\documentclass[english,11pt,a4paper]{article}
\usepackage[T1]{fontenc}
\usepackage[utf8]{inputenc} % this is needed for umlauts
\usepackage{geometry}
\usepackage{color}
\usepackage{babel}
\usepackage{amsmath}
\usepackage{amsthm}
\usepackage{amssymb}
\usepackage{mleftright}
\usepackage[unicode=true,pdfusetitle,
 bookmarks=true,bookmarksnumbered=false,bookmarksopen=false,
 breaklinks=false,pdfborder={0 0 0},pdfborderstyle={},backref=false,colorlinks=true]
 {hyperref}
\hypersetup{
 linkcolor=vblue, citecolor=vblue}

\makeatletter

\numberwithin{equation}{section}

\theoremstyle{plain}
\newtheorem{thm}{\protect\theoremname}[section]
\newtheorem{prop}[thm]{\protect\propositionname}
\newtheorem{lem}[thm]{\protect\lemmaname}
\newtheorem{cor}[thm]{\protect\corollaryname}
\theoremstyle{plain}
\newtheorem*{thm*}{\protect\theoremname}
\theoremstyle{plain}
\newtheorem*{lem*}{\protect\lemmaname}
\theoremstyle{plain}
\newtheorem*{prop*}{\protect\propositionname}

\theoremstyle{definition}
\newtheorem{rmk}{Remark}[section]

\DeclareMathOperator{\vspan}{span}
\DeclareMathOperator{\tr}{tr}
\DeclareMathOperator{\HS}{HS}
\DeclareMathOperator{\Op}{Op}

\DeclareMathOperator{\Area}{Area}
\DeclareMathOperator{\Vol}{Vol}

\DeclareMathOperator{\kin}{kin}
\DeclareMathOperator{\inter}{int}

\DeclareMathOperator{\FS}{FS}

\definecolor{vblue}{RGB}{0,0,255}

\allowdisplaybreaks

\makeatother

\providecommand{\corollaryname}{Corollary}
\providecommand{\lemmaname}{Lemma}
\providecommand{\propositionname}{Proposition}
\providecommand{\theoremname}{Theorem}

\begin{document}
\title{The Random Phase Approximation for Interacting Fermi Gases in the
Mean-Field Regime}
\author{Martin Ravn Christiansen, Christian Hainzl, Phan Th\`anh Nam\\
\\
{\footnotesize{}Department of Mathematics, Ludwig Maximilian University
of Munich, Germany}\\
{\footnotesize{}Emails: christiansen@math.lmu.de, hainzl@math.lmu.de,
nam@math.lmu.de}}
\maketitle
\begin{abstract}
We present a general approach to justify the random phase approximation
for the homogeneous Fermi gas in three dimensions in the mean-field
scaling regime. We consider a system of $N$ fermions on a torus,
interacting via a two-body repulsive potential proportional to $N^{-\frac{1}{3}}$.
In the limit $N\rightarrow\infty$, we derive the exact leading order
of the correlation energy and the bosonic elementary excitations of the system, which are consistent with the
prediction of the random phase approximation in the physics literature. 
\end{abstract}
\tableofcontents{}

\section{Introduction}

In the 1940s, experiments on the cohesive energy and specific heat
of alkali atoms\footnote{When calculated in the Hartree--Fock approximation, the cohesive energy of metals is off by an order of magnitude compared to experiments on alkali metals, as described in \cite[p. 80]{Pines-99}. The same is true for the specific heat, as theoretically calculated in \cite{Bardeen-36}. 
%, which led to a $T \log T$ term, not seen experimentally. and the failure was due to the negligence of correlations among electrons with opposite spin. 
%The calculations for jellium with characteristic parameters concerning typical electron densities for the corresponding alkali metals agrees well with experiments as described in \cite{Pines-99}.
} 
showed a large discrepancy with theoretical calculations
based solely on the Hartree--Fock approximation \cite{Bardeen-36},
further complicated by the fact that second-order perturbation theory
failed because it yielded infinities. Motivated by this unfortunate situation, Bohm and Pines in four seminal
papers \cite{BohPin-51,BohPin-52,BohPin-53, Pines-53} introduced
the random phase approximation (RPA) as a useful tool for studying
the properties of a high-density electron gas moving in a background
of uniform positive charge, called jellium. In the Bohm--Pines RPA
approach, the electron gas could be decoupled into collective plasmon
excitations and quasi-electrons that interacted via a screened Coulomb
interaction. The latter fact justified the independent particle approach
commonly used for many-body fermion systems. Their work was also in
good agreement with experimental data, the culmination of which was
the experimental detection of plasmons \cite{Wat-56,Fer-57}.

\medskip

The microscopic derivation of the RPA has led to notable work by theoretical
physicists since the 1950s. In 1957, Gell-Mann and Brueckner \cite{GelBru-57}
derived the correlation energy of the electron gas in the high density
limit by using a formal summation of a particular class of Feynman
diagrams. Although each diagram is divergent in itself, it turned
out that the sum is finite. This diagrammatic picture further suggested
that the main contribution to the ground-state energy came from the
interaction of pairs of fermions, one from inside and one from outside
the Fermi ball. Shortly thereafter, Sawada \cite{Sawada-57} and Sawada--Brueckner--Fukuda--Brout
\cite{SawBruFukBro-57} interpreted these pairs as bosons and obtained
the correlation energy by diagonalizing an effective Hamiltonian which
is quadratic with respect to the bosonic particle pairs. Since then
the random phase approximation has become a cornerstone in the physics
of condensed matter and nuclear physics \cite{RKNR-15}, also playing
a significant role in bosonic field theory \cite{HCDS-02}, the quark-gluon
plasma \cite{Wal-04}, and especially in computational chemistry and
materials science. Although originally proposed for an electron gas,
it is applicable to a wide variety of fermionic systems.

\medskip

The complete derivation of the RPA from first principles, namely from
the microscopic Schr\"odinger equation, has however long been a major
open problem in mathematical physics. Recently, some rigorous results
on the correlation energy have been derived in the mean-field regime
for small interaction potentials by Hainzl--Porta--Rexze \cite{HaiPorRex-20}
(perturbative results) and by Benedikter--Nam--Porta--Schlein--Seiringer
\cite{Benedikter-20,BNPSS-20,BNPSS-21} (non-perturbative results). 

\medskip

The aim of the present paper is to justify the RPA for a large class
of interaction potentials in the mean-field regime, addressing not only the ground state energy but also the excitation spectrum. As we will explain below, the correlation structure of Fermi gases
can indeed be described correctly by treating appropriate pairs of
fermions as bosons. The corresponding bosonic Hamiltonian can be handled
by Bogolubov's diagonalization method, thus putting the description
in the physics literature \cite{GelBru-57,Sawada-57,SawBruFukBro-57}
on a firm mathematical footing. Although this general point of view
has been employed in \cite{HaiPorRex-20,BNPSS-20,BNPSS-21}, we will provide a new bosonization approach to fermionic
systems which enables us to not only extend the study on the ground state energy initiated in \cite{HaiPorRex-20,BNPSS-20,BNPSS-21}, but also obtain all bosonic elementary excitations predicted in the physics literature, thus justifying the RPA in the mean-field regime. In the long run, we expect that the tools developed in our work will pave the way towards the Coulomb gas in the thermodynamic limit.

%In the long-term perspective, we expect that the tools developed in our work will be helpful in paving the way toward the Coulomb gas in the thermodynamic limit. 

%?? ADD SOMETHING ABOUT "PAVING THE WAY TOWARD THE COULOMB GAS IN THE THERMODYNAMIC LIMIT" ??

%, thus allowing to extend the existing rigorous results in a substantial way. 

%not only extend the existing rigorous results in a substantial way,
%but also provide a completely new bosonization approach to fermionic
%systems.

%The aim of the present paper is to justify the RPA for a large class
%of interaction potentials in the mean-field regime, thus completing
%the research initiated in \cite{HaiPorRex-20,BNPSS-20,BNPSS-21}.
%As we will explain below, the correlation structure of Fermi gases
%can indeed be described correctly by treating appropriate pairs of
%fermions as bosons. The corresponding bosonic Hamiltonian can be handled
%by Bogolubov's diagonalization method, thus putting the description
%in the physics literature \cite{GelBru-57,Sawada-57,SawBruFukBro-57}
%on a firm mathematical footing. Although this general point of view
%has been employed in \cite{HaiPorRex-20,BNPSS-20,BNPSS-21}, we will
%not only extend the existing rigorous results in a substantial way,
%but also provide a completely new bosonization approach to fermionic
%systems.

\subsection{Model}

We consider a system of $N$ (spinless) fermions on the torus $\mathbb{T}^{3}=\left[0,2\pi\right]^{3}$ (with periodic boundary conditions), interacting via a bounded potential $V:\mathbb{T}^{3}\rightarrow\mathbb{R}$. The system is described by the Hamiltonian 
\begin{equation} \label{eq:intro-HN}
H_{N}=H_{\text{kin}}+k_{F}^{-1}H_{\text{int}} = \sum_{i=1}^{N} (-\Delta_{i}) + k_F^{-1}\sum_{1\leq i<j\leq N}V\left(x_{i}-x_{j}\right)
\end{equation}
which acts on the fermionic space
\begin{equation}
\mathcal{H}_{N}=\bigwedge^{N}\mathfrak{h}, \quad \mathfrak{h}=L^{2}\left(\mathbb{T}^{3}\right). 
\end{equation}
Here the coupling constant $k_F^{-1}>0$ corresponds to the interaction strength. We will  focus on the mean-field regime $k_F^{-1}\sim N^{-\frac 1 3}$, where the kinetic and interaction energies are comparable. More precisely,  we assume that  
\begin{equation} \label{eq:intro-N}
N=\left|B_{F}\right| = \frac{4\pi}{3} k_F^3 (1+ o(1)_{k_F\to \infty}), \quad B_{F}=\overline{B}\left(0,k_{F}\right)\cap\mathbb{Z}^{3},
\end{equation}
namely the Fermi ball $B_F$ is completely filled by $N$ integer points. In this case, the kinetic operator $H_{\text{kin}}$ has a unique, non-degenerate ground state which is the Fermi state
\begin{equation} \label{eq:intro-FS}
\psi_{{\rm FS}}=\bigwedge_{p\in B_{F}}u_{p}, \quad u_{p}\left(x\right)=\left(2\pi\right)^{-\frac{3}{2}}e^{ip\cdot x}. 
\end{equation}
More generally, the eigenstates of $H_{\text{kin}}$ can be written explicitly in terms of the plane waves $\left(u_{p}\right)_{p\in\mathbb{Z}^{3}}$. On the other hand, the spectrum of the interacting operator $H_N$ is highly nontrivial and its computation often requires suitable approximations. 
%
%. For practical applications, the spectrum is often computed using suitable approximations. 

We assume that $V$ is of positive type, namely its Fourier transform satisfies $\hat V \ge 0$ with   
\begin{equation}
V\left(x\right)=\frac{1}{\left(2\pi\right)^{3}}\sum_{k\in\mathbb{Z}^{3}}\hat{V}_{k}e^{ik\cdot x}\quad\text{with}\quad\hat{V}_{k}=\int_{\mathbb{T}^{3}}V\left(x\right)e^{-ik\cdot x}\,dx.
\end{equation}
Under our assumption, $H_{N}$ is 
a self-adjoint operator on $\mathcal{H}_{N}$ with domain 
$D\left(H_{N}\right)=D\left(H_{\text{kin}}\right)=\bigwedge^{N}H^{2}\left(\mathbb{T}^{3}\right).$ Moreover, $H_N$ is bounded from below and has compact resolvent. We are interested in the asymptotic behavior of the low-lying spectrum of $H_N$ when $N\to \infty$ and $k_F\to \infty$. 

\medskip
One of the most famous approximations for fermions is the Hartree--Fock
theory, where one restricts the states under consideration to the
set of all Slater determinants $g_{1}\wedge g_{2}\cdots\wedge g_{N}$
with $\left\{ g_{i}\right\} _{i=1}^{N}$ orthonormal in $L^{2}\left(\mathbb{T}^{3}\right)$.
The precision of the Hartree--Fock energy is an interesting subject,
which has been studied for Coulomb systems by Bach \cite{Bach-92}
and Graf--Solovej \cite{GraSol-94}. In general, the Hartree--Fock minimizer could be different from the Fermi
state $\psi_{{\rm FS}}$; see \cite{GonHaiLew-19} for an estimate for Coulomb systems. However, in the mean-field model that we are considering here, the Hartree--Fock minimizer coincides with $\psi_{{\rm FS}}$; see \cite[Theorem A.1]{BNPSS-21} for a precise statement. Thus to obtain the correction to the ansatz of plane waves we have
to understand the correlation
structure of the system\footnote{The Slater determinants are the least correlated states among all
fermionic wave functions (they are eigenfunctions of non-interacting Hamiltonians).}.

\medskip

To go beyond the ansatz of plane waves, the first step is the extraction of the energy of the Fermi state.  For computational purposes, it is convenient to use the second quantization language.  For every $p\in \mathbb{Z}^3$, we denote by  $c_{p}^{\ast}=c^*(u_p)$,
$c_{p}=c(u_p)$ the fermionic creation and annihilation operators associated
to the plane-wave state $u_{p}$. These operators act on the fermionic Fock space 
\begin{equation}
\mathcal{F}^{-}\left(\mathfrak{h}\right)=\bigoplus_{N=0}^{\infty}\bigwedge^{N}\mathfrak{h}
\end{equation}
and obey  the canonical anticommutation relations (CAR)
\begin{equation} \label{eq:CAR}
\left\{ c_{p},c_{q}\right\} =\left\{ c_{p}^{\ast},c_{q}^{\ast}\right\} =0,\quad\left\{ c_{p},c_{q}^{\ast}\right\} =\delta_{p,q},\quad p,q\in\mathbb{Z}^{3},
\end{equation}
where $\left\{ A,B\right\} =AB+BA$. The
Hamiltonian operator $H_{N}$ in \eqref{eq:intro-HN} 
can be expressed as
\begin{equation}
H_{N}=H_{\text{kin}}+k_{F}^{-1}H_{\text{int}} = \sum_{p\in\mathbb{Z}^{3}}\left|p\right|^{2}c_{p}^{\ast}c_{p}+\frac{k_{F}^{-1}}{2\left(2\pi\right)^{3}}\sum_{k\in\mathbb{Z}^{3}}\sum_{p,q\in\mathbb{Z}^{3}}\hat{V}_{k}c_{p+k}^{\ast}c_{q-k}^{\ast}c_{q}c_{p}.\label{eq:IntroductionSecondQuantizedHamiltonian}
\end{equation}

Thanks to the CAR \eqref{eq:CAR} it is straightforward to see that the Fermi state obeys, for all $p\in\mathbb{Z}^{3}$,
\begin{equation}
c_{p}^{\ast}c_{p}\psi_{{\rm FS}}=1_{B_{F}}\left(p\right)\psi_{{\rm FS}}=\begin{cases}
\psi_{{\rm FS}} & p\in B_{F}\\
0 & p\in B_{F}^{c}
\end{cases}\label{eq:FermiStateAnnihilation}
\end{equation}
where $1_{B_{F}}\left(\cdot\right)$ denotes the indicator function
of the Fermi ball $B_{F}$. Thus the kinetic energy of the Fermi state is 
\begin{equation}
\left\langle \psi_{{\rm FS}},H_{\text{kin}}\psi_{{\rm FS}}\right\rangle =\sum_{p\in\mathbb{Z}^{3}}\left|p\right|^{2}\left\langle \psi_{{\rm FS}},c_{p}^{\ast}c_{p}\psi_{{\rm FS}}\right\rangle =\sum_{p\in\mathbb{Z}^{3}}1_{B_{F}}\left(p\right)\left|p\right|^{2}\left\Vert \psi_{{\rm FS}}\right\Vert ^{2}=\sum_{p\in B_{F}}\left|p\right|^{2}.
\end{equation}

Hence, we can define the \textit{localized kinetic operator} $H_{\kin}^{\prime}:D\left(H_{\text{kin}}\right)\subset\mathcal{H}_{N}\rightarrow\mathcal{H}_{N}$
by
\begin{equation} \label{eq:Hkin'}
H_{\kin}^{\prime}=H_{\text{kin}}  - \left\langle \psi_{{\rm FS}},H_{\text{kin}}\psi_{{\rm FS}}\right\rangle = \sum_{p\in B_{F}^{c}}\left|p\right|^{2}c_{p}^{\ast}c_{p}-\sum_{p\in B_{F}}\left|p\right|^{2}c_{p}c_{p}^{\ast}.  
\end{equation}
We refer to this operator as being "localized" since extracting $\langle \psi_\mathrm{FS} , H_\mathrm{kin} \psi_\mathrm{FS}\rangle$ in this manner can be seen as changing the point of reference from the vacuum state $\Omega$ to the Fermi state $\psi_\mathrm{FS}$, so $H_\mathrm{kin}^\prime$ can be seen as a kind of expansion of $ H_\mathrm{kin}$ around $\psi_\mathrm{FS}$.

Note that it is clear from the first identity in \eqref{eq:Hkin'} that $H_{\kin}^{\prime}$ is nonnegative 
since $\psi_{{\rm FS}}$ is the ground state of $H_{\text{kin}}$. On the other hand, the positivity of $H_{\kin}^{\prime}$ is unclear from the second identity in  \eqref{eq:Hkin'} since the difference of two operators which are nonnegative may not have a sign. The resolution of this apparent paradox lies in the underlying Hilbert space: In the $N$-body space $\mathcal{H}_{N}$  we always have 
\begin{align}
N = \sum_{p\in\mathbb{Z}^{3}}c_{p}^{\ast}c_{p}=\sum_{p\in B_{F}} (1- c_{p} c_{p}^{\ast}) +\sum_{p\in B_{F}^{c}}c_{p}^{\ast}c_{p}= |B_F|- \sum_{p\in B_{F}}c_{p}c_{p}^{\ast}+\sum_{p\in B_{F}^{c}}c_{p}^{\ast}c_{p}. 
\end{align}
Therefore, the assumption $|B_F|=N$ implies the \textit{particle-hole symmetry}
\begin{equation}
\mathcal{N}_{E}=\sum_{p\in B_{F}^{c}}c_{p}^{\ast}c_{p}=\sum_{p\in B_{F}}c_{p}c_{p}^{\ast}\quad\text{on }\mathcal{H}_{N},\label{eq:ParticleHoleSymmetry}
\end{equation}
namely the \textit{excitation number operator} (which counts the number of particles outside the Fermi state) coincides with the \textit{hole number operator} (which counts the number of holes inside the Fermi state). Consequently, the kinetic operator in \eqref{eq:Hkin'} can be rewritten as    
\begin{align}\label{eq:ManifestlyNonNegativeHKin}
H_{\kin}^{\prime}   =\sum_{p\in B_{F}^{c}}\vert\left|p\right|^{2}-\zeta \vert\,c_{p}^{\ast}c_{p}+\sum_{p\in B_{F}}\vert\left|p\right|^{2}-\zeta\vert\,c_{p}c_{p}^{\ast}
\end{align}
for any $\zeta\in[\sup_{p\in B_{F}}\left|p\right|^{2},\inf_{p\in B_{F}^{c}}\left|p\right|^{2}]$, which is clearly nonnegative. 

\medskip

For the interaction operator, it is convenient to use the factorized form  
\begin{align}
H_{\text{int}}  =\frac{1}{2\left(2\pi\right)^{3}}\sum_{k\in\mathbb{Z}^{3}}\sum_{p,q\in\mathbb{Z}^{3}}\hat{V}_{k}c_{p+k}^{\ast}c_{q-k}^{\ast}c_{q}c_{p} 
  =\frac{1}{2\left(2\pi\right)^{3}}\sum_{k\in\mathbb{Z}^{3}}\hat{V}_{k}\left(\text{d}\Gamma\left(e^{-ik\cdot x}\right)^{\ast}\text{d}\Gamma\left(e^{-ik\cdot x}\right)-N\right)\label{eq:HintFactorizedForm}
\end{align}
where
\begin{equation}
\text{d}\Gamma\left(e^{-ik\cdot x}\right)=\sum_{p,q\in\mathbb{Z}^{3}}\left\langle u_{p},e^{-ik\cdot x}u_{q}\right\rangle c_{p}^{\ast}c_{q}=\sum_{p,q\in\mathbb{Z}^{3}}\delta_{p,q-k}c_{p}^{\ast}c_{q}=\sum_{p\in\mathbb{Z}^{3}}c_{p}^{\ast}c_{p+k}. 
\end{equation}
Note that for  any $k\in\mathbb{Z}_{\ast}^{3}=\mathbb{Z}^{3}\backslash\left\{ 0\right\}$, we have
\begin{equation}
\text{d}\Gamma\left(e^{-ik\cdot x}\right)\psi_{{\rm FS}}=\sum_{p\in\mathbb{Z}^{3}}c_{p}^{\ast}c_{p+k}\psi_{{\rm FS}}=\sum_{p\in L_{-k}}c_{p}^{\ast}c_{p+k}\psi_{{\rm FS}}\label{eq:rhokOnFermiState}
\end{equation}
since the summand $c_{p}^{\ast}c_{p+k}\psi_{{\rm FS}}$ in (\ref{eq:rhokOnFermiState}) does not vanish if and only if $p\in L_{-k}$ where the {\em lune} 
\begin{equation} \label{eq:lune}
L_k = B_{F}^{c}\cap\left(B_{F}+k\right)= \left\{ p\in\mathbb{Z}^{3}\mid\left|p-k\right|\leq k_{F}<\left|p\right|\right\}
\end{equation}
will play an important role in our analysis. In particular, using  (\ref{eq:FermiStateAnnihilation}) and the CAR again we find that for all $k\in \mathbb{Z}_{\ast}^{3}$
\begin{equation}
\left\Vert \text{d}\Gamma\left(e^{-ik\cdot x}\right)\psi_{{\rm FS}}\right\Vert ^{2}=\sum_{p\in L_{-k}}\left\Vert c_{p}^{\ast}c_{p+k}\psi_{{\rm FS}}\right\Vert ^{2}=\sum_{p\in L_{-k}}1=\left|L_{-k}\right|=\left|L_{k}\right|.
\end{equation}
Thus the interaction energy of the Fermi state is  given by
\begin{equation}
\left\langle \psi_{{\rm FS}},H_{\text{int}}\psi_{{\rm FS}}\right\rangle =\frac{N\left(N-1\right)}{2\left(2\pi\right)^{3}}\hat{V}_{0}+\frac{1}{2\left(2\pi\right)^{3}}\sum_{k\in\mathbb{Z}_{\ast}^{3}}\hat{V}_{k}\left(\left|L_{k}\right|-N\right)
\end{equation}
where we see the direct and exchange energies (involving $\hat V_0$ and $\{\hat V_k\}_{k\ne 0}$, respectively). We can define the \textit{localized interaction operator}  
\begin{equation} \label{eq:LocalizedInteractionEnergy}
H_{\text{int}}^{\prime}=H_{\text{int}}- \left\langle \psi_{{\rm FS}},H_{\text{int}}\psi_{{\rm FS}}\right\rangle =  \frac{1}{2\left(2\pi\right)^{3}}\sum_{k\in\mathbb{Z}_{\ast}^{3}}\hat{V}_{k}\left(\text{d}\Gamma\left(e^{-ik\cdot x}\right)^{\ast}\text{d}\Gamma\left(e^{-ik\cdot x}\right)-\left|L_{k}\right|\right).
\end{equation}
In summary, with  $H_{\kin}^{\prime}$ and $H_{\inter}^{\prime}$ defined in  \eqref{eq:Hkin'} and \eqref{eq:LocalizedInteractionEnergy}  we can write
\begin{align} \label{eq:HN-localized-1}
H_{N}  =E_{\rm FS} +H_{\kin}^{\prime}+k_{F}^{-1}H_{\inter}^{\prime}, \quad E_{\rm FS} = \left\langle \psi_{\FS},H_{N}\psi_{\FS}\right\rangle .  
\end{align}
Note that in the prior works \cite{HaiPorRex-20,BNPSS-20,BNPSS-21} the localization procedure was carried out by employing
what is known as the particle-hole transformation, which maps the Fermi state $\psi_{\rm FS}$ to the vacuum; see e.g. \cite[Eq. (1.20)]{BNPSS-21} for an analogue of \eqref{eq:HN-localized-1}. However, in the present paper we do not follow this approach since we prefer to  work  on the $N$-body Hilbert space. 
 
\subsection{Random Phase Approximation}

In this subsection we explain the ideas of the bosonization approach to the random phase approximation. On the one hand, in the original approach \cite{BohPin-51,BohPin-52,BohPin-53, Pines-53}, Bohm and Pines considered fluctuations of density in the momentum representation where the plasma momenta and the effective particle momenta of different wavelengths $k,l$ are coupled by phases $e^{i(k-l) \cdot x_j}$, summing over the "random" particle positions $x_j$. The assumption that the phases average toward zero for a large number of particles is originally called the ``random phase approximation''. On the other hand, after the work of Sawada \cite{Sawada-57} and Sawada--Brueckner--Fukuda--Brout \cite{SawBruFukBro-57}, the term RPA has been widely used in the physics literature in the context of a quasi-bosonic Hamiltonian, where a quasi-boson consists of a particle-hole pair. The quasi-bosonic approach is used not only for Coulomb gases, but also in a much broader context, especially in nuclear matter (for a standard textbook, see \cite[p. 156]{Fetter-Walecka-71} for Coulomb gases and \cite[pp. 540-543]{Fetter-Walecka-71} for nuclear matter). 

\medskip

In the present paper, we will focus on building a mathematical  formulation of the quasi-bosonic approach for general potentials and eventually apply this theory to regular potentials. In the long run, we hope that this general theory will also be helpful for singular potentials, in particular for Coulomb gases where the next-order correction to the bosonization picture matters (in \cite{CHN-22} we used the formulation provided in the present paper to find the analogue of the Gell-Mann--Brueckner formula for the mean-field Coulomb gas, which shows how important it is to carry the non-bosonic part in the calculation at least to the leading order).

\medskip

Now let us explain the bosonization argument in detail. 
%Roughly speaking, the RPA suggests that the fermionic correlation can be described by a quadratic Hamiltonian in a bosonic interpretation.
Roughly speaking, the RPA suggests that the fermionic correlation can be described by a  Hamiltonian which is quadratic in suitable \textit{bosonic} creation and annihilation operators.   To explain the heuristic bosonization argument, let us decompose further the interaction terms in \eqref{eq:LocalizedInteractionEnergy} by defining, for  every $k\in\mathbb{Z}_{\ast}^{3}$, 
\begin{align}
\text{d}\Gamma\left(e^{-ik\cdot x}\right)=\text{d}\Gamma\left(\left(P_{B_{F}}+P_{B_{F}^{c}}\right)e^{-ik\cdot x}\left(P_{B_{F}}+P_{B_{F}^{c}}\right)\right) = \tilde{B}_{k}+\tilde{B}_{-k}^{\ast}+D_{k} \label{eq:Localizedrhok}
\end{align}
where $P_{B_F}$ and $P_{B_F^c}$ are projections in the one-fermion Hilbert space and
\begin{align} 
\tilde{B}_{k}&=\text{d}\Gamma\left(P_{B_{F}}e^{-ik\cdot x}P_{B_{F}^{c}}\right)=\sum_{p,q\in\mathbb{Z}^{3}}\left\langle u_{p}, P_{B_{F}}e^{-ik\cdot x}P_{B_{F}^{c}} u_{q}\right\rangle c_{p}^{\ast}c_{q} = \sum_{p\in L_{k}}c_{p-k}^{\ast}c_{p}, \label{eq:def-Bk}  \\
D_k &= \text{d}\Gamma\left(P_{B_{F}}e^{-ik\cdot x}P_{B_{F}}\right) + \text{d}\Gamma\left(P_{B_{F}^c}e^{-ik\cdot x}P_{B_{F}^{c}}\right) =  \sum_{p\in  B_F \cap (B_F+k)} c_{p-k}^{\ast}c_{p} +  \sum_{p\in B_F^c \cap (B_F^c+k) } c_{p-k}^{\ast}c_{p}. \nonumber
\end{align}
Note that for all $k\in \mathbb{Z}^3_*$ we have $D_k^*=D_{-k}$ and 
\begin{equation}
 [\tilde{B}_k,\tilde{B}_{-k}]= [\tilde{B}_{-k},D_{k}]=[\tilde{B}_{k}^{\ast},D_{k}]=0
 \end{equation}
which can be seen from the identity $[{\rm d} \Gamma (X),{\rm d} \Gamma (Y) ] = {\rm d} \Gamma ([X,Y])$ and \eqref{eq:def-Bk}.  Due to the symmetry between $k$ and $-k$, it is convenient to introduce the set\footnote{The exact definition of $\mathbb{Z}_{+}^{3}$ is not important, only
that it satisfies $\mathbb{Z}_{+}^{3}\cup\left(-\mathbb{Z}_{+}^{3}\right)=\mathbb{Z}_{\ast}^{3}$
and $\mathbb{Z}_{+}^{3}\cap\left(-\mathbb{Z}_{+}^{3}\right)=\emptyset$.}
\begin{equation} \label{eq:intro-def-Z+}
\mathbb{Z}_{+}^{3}=\left(\left\{ x_{1}>0\right\} \cup\left\{ x_{1}=0,x_{2}>0\right\} \cup\left\{ x_{1}=x_{2}=0,x_{3}>0\right\} \right)\cap\mathbb{Z}_{\ast}^{3}.
\end{equation}
such that
\begin{equation}
\mathbb{Z}_{+}^{3}\cup\left(-\mathbb{Z}_{+}^{3}\right)=\mathbb{Z}_{\ast}^{3}, \quad \mathbb{Z}_{+}^{3}\cap\left(-\mathbb{Z}_{+}^{3}\right)=\emptyset.
\end{equation}
Using this notation and the assumption $\hat V_k=\hat V_{-k}$, we can rewrite the interaction operator in \eqref{eq:LocalizedInteractionEnergy} as
\begin{align}\label{eq:localizedInteractionBandDForm}
k_{F}^{-1} H_{\text{int}}^{\prime} & =\frac{k_F^{-1}}{2\left(2\pi\right)^{3}}\sum_{k\in\mathbb{Z}_{\ast}^{3}} \hat{V}_{k}\left(\left(\tilde{B}_{k}+\tilde{B}_{-k}^{\ast}+D_{k}\right)^{\ast}\left(\tilde{B}_{k}+\tilde{B}_{-k}^{\ast}+D_{k}\right)-\left|L_{k}\right|\right) \\
 & = \sum_{k\in\mathbb{Z}_{+}^{3}}\left(H_{\text{int}}^{k}-\frac{\hat{V}_{k}k_{F}^{-1}}{\left(2\pi\right)^{3}}\left|L_{k}\right|\right)
 +\frac{k_{F}^{-1}}{\left(2\pi\right)^{3}}\sum_{k\in\mathbb{Z}_{\ast}^{3}}\hat{V}_{k}\left(\tilde{B}_{k}^{\ast}D_{k}+D_{k}^{\ast}\tilde{B}_{k}+\frac{1}{2}D_{k}^{\ast}D_{k}\right) \nonumber
\end{align}
where for each $k\in\mathbb{Z}_{+}^{3}$ we denote 
\begin{align}
H_{\text{int}}^{k} & =\frac{\hat{V}_{k}k_{F}^{-1}}{2\left(2\pi\right)^{3}}\left(\left(\tilde{B}_{k}+\tilde{B}_{-k}^{\ast}\right)^{\ast}\left(\tilde{B}_{k}+\tilde{B}_{-k}^{\ast}\right)+\left(\tilde{B}_{-k}+\tilde{B}_{k}^{\ast}\right)^{\ast}\left(\tilde{B}_{-k}+\tilde{B}_{k}^{\ast}\right)\right) \nonumber  \\
 & =\frac{\hat{V}_{k}k_{F}^{-1}}{2\left(2\pi\right)^{3}} \left( \{ \tilde{B}_{k}^{\ast}, \tilde{B}_{k}\} + \{\tilde{B}_{-k}^{\ast}, \tilde{B}_{-k}\} + 2 \tilde{B}_{k}^{\ast} \tilde{B}_{-k}^{\ast}  +2   \tilde{B}_{-k} \tilde{B}_{k}  \right). \label{eq:HintkBBForm} 
\end{align}

Now let us introduce the  quasi-bosonicity. From the CAR \eqref{eq:CAR} it is straightforward to see that   
\begin{equation}
\left[\tilde{B}_{k},\tilde{B}_{l}\right]=\left[\tilde{B}_{k}^{\ast},\tilde{B}_{l}^{\ast}\right]=0,\quad \left[\tilde{B}_{k},\tilde{B}_{l}^{\ast}\right]= \left|L_{k}\right|\delta_{k,l}-\sum_{p\in L_{k}\cap L_{l}}c_{p-l}c_{p-k}^{\ast}-\sum_{p\in L_{k}\cap\left(L_{l}-l+k\right)}c_{p-k+l}^{\ast}c_{p}
\end{equation}
for all $k,l\in \mathbb{Z}^3_*$, where $[A,B]= AB-BA$. Hence, on states with few excitations, e.g. the expectation value of $\mathcal{N}_E$ is much smaller than $|L_k| \sim \min\{|k| k_F^2,k_F^3\}$, then the rescaled operators  $\tilde{B}_{k}^{\prime}=\left|L_{k}\right|^{-\frac{1}{2}}\tilde{B}_{k}$ obey the commutation
relations
\begin{align}
\left[\tilde{B}_{k}^{\prime},\tilde{B}_{l}^{\prime}\right]  =\left[\left(\tilde{B}_{k}^{\prime}\right)^{\ast},\left(\tilde{B}_{l}^{\prime}\right)^{\ast}\right]=0, \quad \left[\tilde{B}_{k}^{\prime},\left(\tilde{B}_{l}^{\prime}\right)^{\ast}\right] \approx \delta_{k,l} 
\end{align}
for all $k,l\in\mathbb{Z}_{\ast}^{3}$, in direct analogy with the canonical commutation
relations (CCR) obeyed by a set of \textit{bosonic} creation and annihilation
operators $a_{k}^{\ast}$, $a_{k}$ indexed by $\mathbb{Z}_{\ast}^{3}$,
\begin{align} \label{eq:CCR}
\left[a_{k},a_{l}\right]  =\left[a_{k}^{\ast},a_{l}^{\ast}\right]=0, \quad \left[a_{k},a_{l}^{\ast}\right]  =\delta_{k,l}. 
\end{align}
Since  the relation $[\tilde{B}_{k}^{\prime}, (\tilde{B}_{l}^{\prime})^{\ast}] \approx \delta_{k,l}$
is only approximate, we call  these
operators  \textit{quasi}-bosonic.

In view of the quasi-bosonicity of these operators, in  the form (\ref{eq:localizedInteractionBandDForm}) of
$H_{\text{int}}'$, we call the first sum on the right-hand side of this equation the \textit{bosonizable
terms}, while the second sum constitutes the \textit{non-bosonizable
terms} which are regarded as error terms. The bosonizable part $H_{\text{int}}^{k}$ can be viewed as a quadratic
Hamiltonian in the bosonic setting, which can be diagonalized
by Bogolubov transformations. This is the spirit of what we will do,
but there is a catch: The kinetic operator $H_{\kin}^{\prime}$ cannot be written in terms of $\tilde{B}_{k}$. The solution is to further decompose the operators $\tilde{B}_{k}$ by defining the \textit{excitation operators}  
\begin{equation} \label{eq:def-bkp}
b_{k,p}=c_{p-k}^{\ast}c_{p},\quad b_{k,p}^{\ast}=c_{p}^{\ast}c_{p-k},\quad k\in\mathbb{Z}_{\ast}^{3},\,p\in L_{k}.
\end{equation}
The name is due to the fact that the action of $b_{k,p}^{\ast}$ is
to create a state
at momentum $p\in B_{F}^{c}$ and annihilate a state at momentum $p-k\in B_{F}$. 

Since $H_{\text{int}}^{k}$ is quadratic in terms of $\tilde{B}_{k}$, it is also quadratic in terms of $b_{k,p}^{\ast}$, namely
\begin{align} \label{eq:Hint-k}
H_{\text{int}}^{k} & =\sum_{p,q\in L_{k}}\frac{\hat{V}_{k}k_{F}^{-1}}{2\left(2\pi\right)^{3}}\left(b_{k,p}^{\ast}b_{k,q}+b_{k,q}b_{k,p}^{\ast}\right)+\sum_{p,q\in L_{-k}}\frac{\hat{V}_{k}k_{F}^{-1}}{2\left(2\pi\right)^{3}}\left(b_{-k,p}^{\ast}b_{-k,q}+b_{-k,q}b_{-k,p}^{\ast}\right)\\
 & +\sum_{p\in L_{k}}\sum_{q\in L_{-k}}\frac{\hat{V}_{k}k_{F}^{-1}}{2\left(2\pi\right)^{3}}\left(b_{k,p}^{\ast}b_{-k,q}^{\ast}+b_{-k,q}b_{k,p}\right)+\sum_{p\in L_{-k}}\sum_{q\in L_{k}}\frac{\hat{V}_{k}k_{F}^{-1}}{2\left(2\pi\right)^{3}}\left(b_{-k,p}^{\ast}b_{k,q}^{\ast}+b_{k,q}b_{-k,p}\right).\nonumber 
\end{align}

The reason that the operators $b_{k,p}$ are preferable to the operators
$\tilde{B}_{k}$ is that they satisfy the following commutation relation with the kinetic operator (see \eqref{LocalizedKineticOperatorCommutator} below)
\begin{equation} \label{eq:intro-Hkin-com}
\left[H_{\kin}^{\prime},b_{k,p}^{\ast}\right]= 2 \lambda_{k,p} b_{k,p}^{\ast}, \quad \lambda_{k,p}=\frac{1}{2}(\left|p\right|^{2}-\left|p-k\right|^{2}). 
\end{equation}
Note that $\lambda_{k,p}\ge \frac{1}{2}$ (first $\lambda_{k,p}>0$ since $p\in L_k$; moreover $|p|^2- |p-k|^2$ is an integer as $p,k\in \mathbb{Z}^3$). This is to be compared with the bosonic setting: If the operators
 $a_{k}$ obey the CCR \eqref{eq:CCR}, then 
\begin{equation}
\left[\sum_{l}\varepsilon_{l}a_{l}^{\ast}a_{l},a_{k}^{\ast}\right]=\varepsilon_{k}a_{k}^{\ast}.
\end{equation}
Therefore, viewing $b_{k,p}^{\ast}$ as being analogous to a bosonic creation
operator we get 
%$H_{\kin}^{\prime}$ as being analogous
%to
\begin{equation} \label{eq:intro-kinetic-bosonic}
H_{\kin}^{\prime}\approx \sum_{k\in\mathbb{Z}_{\ast}^{3}}\sum_{p\in L_{k}} 2 \lambda_{k,p}  b_{k,p}^{\ast}b_{k,p} =\sum_{k\in\mathbb{Z}_{+}^{3}} \Big( \sum_{p\in L_{k}} 2 \lambda_{k,p}  b_{k,p}^{\ast}b_{k,p}  + \sum_{p\in L_{-k}} 2 \lambda_{-k,p}  b_{-k,p}^{\ast}b_{-k,p} \Big) .
\end{equation}
Combining \eqref{eq:Hint-k} and \eqref{eq:intro-kinetic-bosonic} we arrive at a Hamiltonian quadratic in terms of the operators $b_{k,p}$ which could be treated in the  bosonic interpretation. Note that $b_{k,p}\Psi_{\rm FS}=0$ for all $k\in \mathbb{Z}_*^3,p\in L_k$, and hence the Fermi state plays the role of the bosonic vacuum.

\subsubsection*{Overview of the Heuristic Assumptions behind the Random Phase Approximation}

In the physics literature \cite{Sawada-57,SawBruFukBro-57}, the RPA entails two assumptions: % three assumptions:

\textbf{1.} That the excitation operators $b_{k,p}^{\ast}$, $b_{k,p}$ in \eqref{eq:def-bkp} 
can be treated as \textit{bosonic} creation and annihilation operators,
and that the operators $b_{k,p}$ and $b_{l,q}$ with $k\neq l$ can
be considered as acting on independent Fock spaces. Mathematically
we thus expect that the approximate canonical commutation relations
(CCR)
\begin{equation}
\left[b_{k,p},b_{l,q}\right]=\left[b_{k,p}^{\ast},b_{l,q}^{\ast}\right]=0,\quad\left[b_{k,p},b_{l,q}^{\ast}\right]\approx\delta_{k,l}\delta_{p,q},\label{eq:IntroductionApproximateCCR}
\end{equation}
should hold in an appropriate sense.

\textbf{2.} That the operator in (\ref{eq:HN-localized-1})
can be approximated by an effective Hamiltonian which is quadratic
in terms of $b_{k,p}^{\ast}$ and $b_{k,p}$. This is already true
for the interaction part $\sum_{k\in\mathbb{Z}_{+}^{3}}H_{\text{int}}^{k}$ in \eqref{eq:Hint-k},
and in the RPA the \textit{non-bosonizable terms}
\begin{equation}
\frac{k_{F}^{-1}}{\left(2\pi\right)^{3}}\sum_{k\in\mathbb{Z}_{\ast}^{3}}\hat{V}_{k}\left(\tilde{B}_{k}^{\ast}D_{k}+D_{k}^{\ast}\tilde{B}_{k}+\frac{1}{2}D_{k}^{\ast}D_{k}\right)
\end{equation}
are simply dropped. Moreover, the kinetic operator $H_{\kin}^{\prime}$ is not exactly of the desired form, but it can be replaced by the right side of \eqref{eq:intro-kinetic-bosonic}.  All this leads to  
the effective Hamiltonian
\begin{equation}
\sum_{k\in\mathbb{Z}_{+}^{3}}H_{\text{Bog},k}=\sum_{k\in\mathbb{Z}_{+}^{3}}\left(2\sum_{p\in L_{k}}\lambda_{k,p}b_{k,p}^{\ast}b_{k,p}+2\sum_{p\in L_{-k}}\lambda_{-k,p}b_{-k,p}^{\ast}b_{-k,p}+H_{\text{int}}^{k} - \frac{\hat{V}_{k}k_{F}^{-1}}{\left(2\pi\right)^{3}}\left|L_{k}\right| \right)\label{eq:IntroductonEffectiveOperator}
\end{equation}
acting on the bosonic Fock space $\bigoplus_{k\in\mathbb{Z}_{+}^{3}}\mathcal{F}^{+}\left(\ell^{2}\left(L_{k}\cup L_{-k}\right)\right)$.

%\textbf{3.} That the  effective bosonic Hamiltonian does indeed describe
%the fermionic system correctly. 

Consequently, since the operators $b_{k,p}$ and $b_{l,q}$ with $k\neq l$ are considered as acting independently, we can diagonalize separately each quadratic bosonic Hamiltonian $H_{\text{Bog},k}$ by a Bogolubov transformation $\mathcal{U}_{k}$ on $\mathcal{F}^{+}\left(\ell^{2}\left(L_{k}\cup L_{-k}\right)\right)$ such that 
%\begin{align}
%\mathcal{U}_{k}H_{\text{Bog},k}\mathcal{U}_{k}^{\ast} &=2\,\text{tr}\left(\widetilde{E}_{k}-h_{k}\right) - \frac{\hat{V}_{k}k_{F}^{-1}}{\left(2\pi\right)^{3}}\left|L_{k}\right| \nonumber\\
%&\quad +2\sum_{p\in L_{k}\cup L_{-k}}\left\langle e_{p},\widetilde{E}_{k}e_{q}\right\rangle b_{k,p}^{\ast}b_{k,q}+2\sum_{p\in L_{-k}}\left\langle e_{p},\widetilde{E}_{-k}e_{q}\right\rangle b_{-k,p}^{\ast}b_{-k,q}
%\end{align}
\begin{align}
\mathcal{U}_{k}H_{\text{Bog},k}\mathcal{U}_{k}^{\ast} =2\,\text{tr}\left(\widetilde{E}_{k}-h_{k}\right) - \frac{\hat{V}_{k}k_{F}^{-1}}{\left(2\pi\right)^{3}}\left|L_{k}\right|  +2\sum_{p\in L_{k}\cup L_{-k}}\left\langle e_{p},\widetilde{E}_{k}e_{q}\right\rangle b_{k,p}^{\ast}b_{k,q}
\end{align}
where for every $k\in\mathbb{Z}_{\ast}^{3}$ we denote the following quantities on $\ell^{2}\left(L_{k}\right)$: 
\begin{equation}
\widetilde{E}_{k}= (h_{k}^{\frac{1}{2}}\left(h_{k}+2P_{v_{k}}\right)h_{k}^{\frac{1}{2}})^{\frac{1}{2}}, \quad h_{k}e_{p}=\lambda_{k,p}e_{p}, \quad  P_{v_k}  =|v_{k} \rangle \langle v_k|,  \quad v_{k}=\sqrt{\frac{\hat{V}_{k}k_{F}^{-1}}{2\left(2\pi\right)^{3}}}\sum_{p\in L_{k}}e_{p}\label{eq:IntroductionOneBodyOperators}
\end{equation}
with $\left(e_{p}\right)_{p\in L_{k}}$ the standard orthonormal basis of $\ell^{2}\left(L_{k}\right)$.
 
Summing over $k$ we obtain the {\em correlation energy}  (see Proposition \ref{thm:ExplicitKAktBktEstimates})
\begin{equation} \label{eq:correlation-energy}
E_{\rm corr}  = \sum_{k\in \mathbb{Z}^3_+} \left(  \text{tr}\left(\widetilde{E}_{k}-h_{k}\right)-\frac{\hat{V}_{k}k_{F}^{-1}}{2\left(2\pi\right)^{3}}\left|L_{k}\right| \right)=\sum_{k\in \mathbb{Z}^3_*}  \frac{1}{\pi}\int_{0}^{\infty}F\left(\frac{\hat{V}_{k}k_{F}^{-1}}{\left(2\pi\right)^{3}}\sum_{p\in L_{k}}\frac{\lambda_{k,p}}{\lambda_{k,p}^{2}+t^{2}}\right)dt
\end{equation}
where $F\left(x\right)=\log\left(1+x\right)-x$. All in all the RPA  thus suggests that up to a unitary transformation 
%Moreover, up to a unitary transformation 
we expect that 
\begin{align} \label{eq:IntroductionFullRPA}
 H_{N}  \approx E_{\rm FS}  + E_{\rm corr} +2\sum_{k\in\mathbb{Z}_{\ast}^{3}}\sum_{p,q\in L_{k}}\left\langle e_{p},\widetilde{E}_{k}e_{q}\right\rangle b_{k,p}^{\ast}b_{k,p},
\end{align}
at least on states with few excitations. 
%Equation (\ref{eq:IntroductionFullRPA}) amounts to what is often referred to as the RPA in the physics literature.

\subsubsection*{Prediction of the Correlation Energy and  the Excitation Spectrum}

Equation (\ref{eq:IntroductionFullRPA}) leads immediately to the following approximation for the ground state energy 
\begin{equation} \label{eq:intro-GSE}
\inf\sigma\left(H_{N}\right)\approx E_{\rm FS}  +E_{\rm corr}
\end{equation} 
which coincides with \cite[Eq. (34)]{SawBruFukBro-57}\footnote{Provided one replaces $\left(2\pi\right)^{3}$ with the volume $\Omega$
of the box, includes a spin factor and inserts the Coulomb potential,
$\hat{V}_{k}k_{F}^{-1}= 4\pi e^{2} |k|^{-2}$.}, where the authors derived it from the effective operator of equation
(\ref{eq:IntroductonEffectiveOperator}) and also explained the connection
to the original work of Gell-Mann--Brueckner \cite{GelBru-57}. See
also \cite[Eq. (9.54)]{Raimes-72} and \cite[Eq. (12.53)]{Fetter-Walecka-71} for this expression of the ground state energy. 

%More importantly, (\ref{eq:IntroductionFullRPA}) also suggests how the  excitation spectrum of $H_{N}$  should behave: the {\em elementary excitations} should be given by the eigenvalues of $2\widetilde{E}_{k}$, which can be explicitly described. 
\medskip

More importantly, (\ref{eq:IntroductionFullRPA}) also suggests that the  excitation spectrum of $H_{N}$  could be described in terms of the eigenvalues of $2\widetilde{E}_{k}$, which correspond to the {\em bosonic elementary excitations} and can be explicitly computed. 

Indeed, for every eigenvalue $\epsilon$ of  $\widetilde{E}_{k}$, we may find an eigenvector $w\in\ell^{2}\left(L_{k}\right)$
such that
\begin{equation}
\epsilon^{2}w=\widetilde{E}_{k}^2 w = h_{k}^{\frac{1}{2}}\left(h_{k}+2P_{v_{k}}\right)h_{k}^{\frac{1}{2}} w =h_{k}^{2}w+2\langle h_{k}^{\frac{1}{2}}v_{k},w\rangle h_{k}^{\frac{1}{2}}v_{k}. 
\end{equation}
But either $\epsilon$ is also an eigenvalue of $h_{k}$ or $\epsilon^{2}-h_{k}^{2}$
is invertible. In the latter case we can write
\begin{equation}
w=2\langle h_{k}^{\frac{1}{2}}v_{k},w\rangle\left(\epsilon^{2}-h_{k}^{2}\right)^{-1}h_{k}^{\frac{1}{2}}v_{k},
\end{equation}
and taking the inner product with $h_{k}^{\frac{1}{2}}v_{k}$ and
cancelling the factors of $\langle h_{k}^{\frac{1}{2}}v_{k},w\rangle$
yields
\begin{equation}
1=2\langle v_{k},h_{k}\left(\epsilon^{2}-h_{k}^{2}\right)^{-1}v_{k}\rangle=\frac{\hat{V}_{k}k_{F}^{-1}}{\left(2\pi\right)^{3}}\sum_{p\in L_{k}}\frac{\lambda_{k,p}}{\epsilon^{2}-\lambda_{k,p}^{2}},
\end{equation}
which appears in \cite[Eq. (6)]{SawBruFukBro-57}. 
%By manipulating the sum this can be rewritten as
The sum can be rewritten as
\begin{equation} \label{eq:ev-Ek}
1=\frac{\hat{V}_{k}k_{F}^{-1}}{2\left(2\pi\right)^{3}}\sum_{p\in B_{F}}\frac{\left|k\right|^{2}}{\left(\epsilon-k\cdot p\right)^{2}-\left(\frac{1}{2}\left|k\right|^2\right)^{2}}.
\end{equation}
The formula \eqref{eq:ev-Ek} allows to compute all eigenvalues of $\widetilde{E}_{k}$ outside the spectrum of $h_k$. 

In the physically relevant case of the Coulomb potential where $\hat{V}_{k}k_{F}^{-1}$ is replaced by $4\pi e^{2}\left|k\right|^{-2}$
one can immediately derive the famous plasmon frequency from \eqref{eq:ev-Ek}: for $|k|\ll k_F^{1/2}$, the largest eigenvalue $\epsilon$ is proportional to $k_F^{3/2}$ (see \cite[Eq. (2.27)--(2.54)]{CHN-22-LMP} for a detailed explanation), and its leading order behavior can be computed easily in the thermodynamic limit %and assuming $\left|k\right|^2 \ll k_{F}\ll\epsilon$ 
(including also a factor of $2$ for the electron spin states)
\begin{equation}
\epsilon^{2}=\frac{4\pi e^{2}}{\left(2\pi\right)^{3}}\int_{\overline{B}\left(0,k_{F}\right)}\frac{\epsilon^{2}}{\left(\epsilon-k\cdot p\right)^{2}-\left(\frac{1}{2}\left|k\right|^2\right)^{2}}\,dp\approx\frac{2e^{2}}{\left(2\pi\right)^{2}}\text{Vol}\left(\overline{B}\left(0,k_{F}\right)\right)=\frac{2e^{2}}{3\pi}k_{F}^{3}=2\pi ne^{2}
\end{equation}
where $n=\frac{N}{\mathcal{V}}=\frac{1}{3\pi^{2}}k_{F}^{3}$ is the
number density of the system. Recalling that the relevant operator
is $2\widetilde{E}_{k}$ rather than $\widetilde{E}_{k}$ and that $\frac{\hbar^{2}}{2m}=1$,
this yields an excitation energy of
\begin{equation} \label{eq:plasmon-fre-intro}
2\epsilon\approx2\sqrt{2\pi ne^{2}}=\hbar\sqrt{\frac{4\pi ne^{2}}{m}} = \hbar \omega_{\text{plasmon}}
\end{equation}
where $\omega_{\text{plasmon}}=\sqrt{ {4\pi ne^{2}}m^{-1}}$ is called the plasmon frequency in  \cite[Eq. (3-90)]{Pines-99} and \cite[Eq.  (15.16) - (15.18)]{Fetter-Walecka-71}. Note that the Coulomb potential is special as it makes the right-hand side of  \eqref{eq:plasmon-fre-intro} independent of $k$. See also \cite{Benedikter-20,CHN-22-LMP} where \eqref{eq:plasmon-fre-intro} was discussed. 

Establishing the above heuristic computation is a longstanding problem
in mathematical physics. In the present paper, we will give a rigorous formulation for the operator approximation \eqref{eq:IntroductionFullRPA},  and then use this to justify the prediction of the correlation energy and the bosonic elementary excitations for a wide class of bounded potentials in the mean-field regime.
%
%consider the random phase approximation in a form similar to equation
%(\ref{eq:IntroductionFullRPA}), and we will return to this in the
%near future.

\subsection{Main Results}

Our first  result is the following rigorous formulation of the operator approximation \eqref{eq:IntroductionFullRPA}. 
\begin{thm}[Operator Formulation of the RPA] 
\label{them:OperatorStatement}
Let $V:\mathbb{T}^{3}\rightarrow\mathbb{R}$
obey $\hat{V}_{k}\geq0$ and $\hat{V}_{-k}=\hat{V}_{k}$ for all $k\in\mathbb{Z}^{3}$,
and assume furthermore that $\sum_{k\in\mathbb{Z}^{3}}\hat{V}_{k}\left|k\right|<\infty$.  Consider the Hamiltonian $H_N$ given in \eqref{eq:intro-HN} with $N=|B_F|$. Let the operators $H_{\rm kin}'$, $\mathcal{N}_E$,  $\widetilde{E}_{k}-h_k$  be defined in  \eqref{eq:Hkin'}, \eqref{eq:ParticleHoleSymmetry},   \eqref{eq:IntroductionOneBodyOperators}. Let the energies $E_{\rm FS}$, $E_{\rm corr}$ be defined in \eqref{eq:HN-localized-1}, \eqref{eq:correlation-energy}.  Then there exists a unitary transformation $\mathcal{U}:\mathcal{H}_{N}\rightarrow\mathcal{H}_{N}$ such that
\begin{align} \label{eq:intro-operator-statement}
\mathcal{U}H_{N}\mathcal{U}^{\ast}  = E_{\rm FS}  +E_{\rm corr}  +H_{\rm eff} +\mathcal{E}_{\mathcal{U}}
\end{align}
where  the effective operator $H_{\text{eff}}: \mathcal{H}_{N}\rightarrow\mathcal{H}_{N}$ is 
\begin{align} \label{eq:intro-Heff}
H_{\rm eff}=H_{\kin}^{\prime}+2\sum_{k\in\mathbb{Z}^{3}_* }\sum_{p,q\in L_{k}}\left\langle e_{p},\left(\widetilde{E}_{k}-h_{k}\right)e_{q}\right\rangle b_{k,p}^{\ast}b_{k,q} 
\end{align}
and the error operator $\mathcal{E}_{\mathcal{U}}:\mathcal{H}_{N}\rightarrow\mathcal{H}_{N}$
obeys the operator inequality:   For every constant $\epsilon>0$,
\begin{align} \label{eq:intro-E-op}
\pm\mathcal{E}_{\mathcal U}\leq Ck_F^{-\frac 1 {94}+\epsilon} \left(k_{F}^{-1}\mathcal{N}_{E}H_{\kin}^{\prime} + H_{\kin}^{\prime}+ k_F \right),\quad k_{F}\rightarrow\infty.
\end{align}
\end{thm}

The unitary operator in Theorem \ref{them:OperatorStatement} is given explicitly as $\mathcal{U}=e^{\mathcal{J}}e^{\mathcal{K}}$ where $\mathcal{K}$ and $\mathcal{J}$ are given in \eqref{eq:cK-def-intro} and \eqref{eq:cJ-def-intro}, respectively (the transformations $e^{\mathcal{K}}$ and $e^{\mathcal{J}}$ are studied in detail in Sections \ref{sec:quasi-bosonic-Bogolubov} and \ref{sec:TheSecondTransformationandKineticEstimates}). 

\begin{rmk}
The operator $\mathcal{N}_E H_{\rm kin}'$  on the right hand side of \eqref{eq:intro-E-op} is nothing but the ``bosonic kinetic operator'', due to the following remarkable identity (see Proposition \ref{lem:Heff})
\begin{equation}
2\sum_{k\in\mathbb{Z}_{\ast}^{3}}\sum_{p\in L_{k}}\lambda_{k,p}b_{k,p}^{\ast}b_{k,p}=\mathcal{N}_{E}H_{\kin}^{\prime},\label{eq:intro-RemarkableKineticIdentity}
\end{equation}
Thus in Theorem \ref{them:OperatorStatement} we control the error in the random phase approximation using only the fermionic and bosonic kinetic operators which is very natural. 
\end{rmk}

\begin{rmk}
In the expansion \eqref{eq:intro-operator-statement}, $E_{\rm FS}$ is of order $k_F^5$ and $E_{\rm corr}$ is of order $k_F$. As we will argue below, when we apply this to the low-lying eigenstates with energy $E_{\rm FS}+ O(k_F)$, the expectation of the effective Hamiltonian $H_{\rm eff}$ in \eqref{eq:intro-Heff} is of order $k_F$ while the error term $\mathcal{E}_{\mathcal{U}}$ in \eqref{eq:intro-E-op} is of order $O(k_F^{1-\frac 1 {94}+\epsilon})=o(k_F)$. 
\end{rmk}

In order to put Theorem \ref{them:OperatorStatement} to good use, we need some a-priori estimate on the low-lying eigenstates of the Hamiltonian $H_N$. We have 

\begin{thm}[A-priori Estimate for Eigenstates]
\label{thm:Kinetic-estimate} Let $V$ and $\mathcal{U}$ be as in Theorem \ref{them:OperatorStatement}. Let $\Psi\in D\left(H_{\kin}^{\prime}\right)$ be a normalized eigenstate of $H_{N}$ with energy $\left\langle \Psi,H_{N}\Psi\right\rangle \le E_{\rm FS} + \kappa k_F$ for some constant $\kappa>0$ independent of $k_F$. Then  
\[
\left\langle \Psi, \left(H_{\kin}^{\prime}+k_{F}^{-1}\mathcal{N}_{E}H_{\kin}^{\prime}\right) \Psi\right\rangle \leq C (\kappa +1)^2 k_F 
\]
for a constant $C>0$ depending only on $V$. The same bound holds with $\Psi$ replaced by $\mathcal{U} \Psi$. 
\end{thm}

\begin{rmk}
Thanks to the inequality $\mathcal{N}_{E}\leq H_{\kin}^{\prime}$ (see \cite[Lemma 2.4]{BNPSS-21} and also Proposition \ref{prop:Zeta0RepresentationofHKin} below), Theorem \ref{thm:Kinetic-estimate} implies that for an eigenstate $\Psi$ of $H_{N}$ with energy $\left\langle \Psi,H_{N}\Psi\right\rangle \le  E_{\rm FS} + O(k_F)$,  we have
\begin{align} \label{eq:intro-kinetic-weak}
\left\langle \Psi,\mathcal{N}_{E} \Psi\right\rangle \le \left\langle \Psi, H_{\kin}^{\prime}\Psi\right\rangle = O(k_F).
\end{align}
Thus the number of excitations is much smaller than the total number of particles ($k_F\sim N^{1/3}\ll N$). While \eqref{eq:intro-kinetic-weak} has been derived in \cite{HaiPorRex-20,BNPSS-21} for every state with energy $\left\langle \Psi,H_{N}\Psi\right\rangle \le  E_{\rm FS} + O(k_F)$ (at least for a class of potentials $V$), the improved bound in Theorem \ref{thm:Kinetic-estimate} is deeper and  the  eigenstate assumption plays a crucial role in the proof. 
\end{rmk}

From Theorems \ref{them:OperatorStatement} and \ref{thm:Kinetic-estimate}, we can deduce immediately the asymptotic formula \eqref{eq:intro-GSE} on the ground state energy up to an error $o(k_F)$. Indeed, the energy upper bound is given by the trial state $\mathcal{U}^* \Psi_{\rm FS}$, while the energy lower bound follows from the obvious operator inequality $\widetilde{E}_{k} \ge h_k$.  Moreover, our approach is quantitative and we can derive \eqref{eq:intro-GSE} with explicit error estimates. 

\begin{thm}[Ground State Energy]  \label{thm:intro-GSE} Let $V$ be as in Theorem \ref{them:OperatorStatement}.
Then for all $\epsilon>0$,  
\[
\inf\sigma\left(H_{N}\right)= E_{\rm FS}  +E_{\rm corr} +O(k_{F}^{1-\frac{1}{94} +\epsilon}),\quad k_{F}\rightarrow\infty.
\]
\end{thm}

Here are some remarks concerning Theorem \ref{thm:intro-GSE}. 

%\begin{rmk} 
%A result similar to ours, namely the bound \eqref{eq:MainResultWithRiemannIntegralReplaced} for all potentials satisfying $\sum_{k\in\mathbb{Z}^{3}}\hat{V}_{k}\left|k\right|<\infty$, has been independently obtained in \cite{BPSS-21}, based on a refinement of the method in \cite{BNPSS-20,BNPSS-21}.
%\end{rmk}
%
%\medskip

\begin{rmk} The method of our proof can be adapted to give the upper bound under the weaker condition $\sum_{k\in\mathbb{Z}^{3}}\hat{V}_{k}^{2}\left|k\right|<\infty$ (see  \cite[Appendix A]{BPSS-21} for a derivation of the upper bound under this weaker condition). Additionally, under this condition it can  be shown that
\begin{equation}
\frac{1}{\pi}\sum_{k\in\mathbb{Z}^{3}_* }\int_{0}^{\infty}F\left(\frac{\hat{V}_{k}k_{F}^{-1}}{\left(2\pi\right)^{3}}\sum_{p\in L_{k}}\frac{\lambda_{k,p}}{\lambda_{k,p}^{2}+t^{2}}\right)dt=\frac{k_{F}}{\pi}\sum_{k\in\mathbb{Z}^{3}_* }\left|k\right|\int_{0}^{\infty}F\left(\frac{\hat{V}_{k}}{\left(2\pi\right)^{2}}I\left(t\right)\right)dt+o\left(k_{F}\right),\label{eq:CorrelationEnergyRiemannIntegralReplacement}
\end{equation}
where $F\left(x\right)=\log\left(1+x\right)-x$ and $I\left(t\right)=1-t\tan^{-1}\left(t^{-1}\right)$ (this essentially amounts to replacing the Riemann sum by the integral and can be done by following either the proof of \cite[Eq. (5.15)]{BNPSS-20} or the analysis in Appendix \ref{sec:AnalysisofRiemannSums}; the condition $\sum \hat{V}_{k}^{2}\left|k\right|<\infty$ ensures that the main contribution comes from $|k|\sim O(1)$). Hence, Theorem \ref{thm:intro-GSE} implies that 
\begin{equation} 
\inf\sigma\left(H_{N}\right)= E_{\rm FS}  +\frac{k_{F}}{\pi}\sum_{k\in\mathbb{Z}^{3}_* }\left|k\right|\int_{0}^{\infty}F\left(\frac{\hat{V}_{k}}{\left(2\pi\right)^{2}}I\left(t\right)\right)dt+o\left(k_{F}\right).\label{eq:MainResultWithRiemannIntegralReplaced}
\end{equation}
A result similar to ours, namely the bound \eqref{eq:MainResultWithRiemannIntegralReplaced} for all
potentials satisfying $\sum_k \hat V_k |k|<\infty$, has been independently obtained in  \cite{BPSS-21}, based on a
refinement of the method in \cite{BNPSS-20,BNPSS-21}\footnote{Note that the conventions of the Fourier transform and scaling of
$H_{N}$ in \cite{BNPSS-20,BNPSS-21,BPSS-21} differ from ours.}. The bound \eqref{eq:MainResultWithRiemannIntegralReplaced} was proved earlier in \cite{BNPSS-20,BNPSS-21}, under the additional assumption that the Fourier coefficients $\hat{V}_{k}$
be finitely supported and that $\Vert\hat{V}\Vert_{\ell^{1}}$ be
sufficiently small. For small $\hat{V}_{k}$ the logarithm
of equation (\ref{eq:MainResultWithRiemannIntegralReplaced}) can
be expanded for
\begin{equation}
\sigma\left(H_{N}\right)= E_{\rm FS}  -\frac{1 - \log\left(2\right)}{6\left(2\pi\right)^{4}}k_{F}\sum_{k\in\mathbb{Z}^{3}_* }\hat{V}_{k}^{2}\left|k\right|\left(1+O\left(\hat{V}_{k}\right)\right)+o\left(k_{F}\right)
\end{equation}
which was first proved in \cite{HaiPorRex-20}. %Compared to these prior results our approach applies to a significantly larger class of potentials. 
\end{rmk}

\begin{rmk} A further refinement of our method allows a derivation of a rigorous energy upper bound for all potentials satisfying $\sum_{k\ne 0} \hat V_k^2<\infty$, see   \cite{CHN-22}. This covers the case of the Coulomb potential $\hat{V}_{k} =  4\pi e^2 \left|k\right|^{-2}$, where the correlation energy is given by the left-hand side of (\ref{eq:CorrelationEnergyRiemannIntegralReplacement}) which is of order  $k_F \log k_{F}$ plus a correlation exchange correction of order $k_F$ (the correlation exchange contribution comes from the fact that the purely bosonic picture is not exact; it is different from the exchange energy which is part of $E_{\rm FS}$). In particular, for the Coulomb potential, the right-hand side of  (\ref{eq:CorrelationEnergyRiemannIntegralReplacement})
diverges, whereas the left-hand side does not, and hence the discrete form in (\ref{eq:CorrelationEnergyRiemannIntegralReplacement}) is arguably more fundamental than the continuous form. It is interesting that in our method the discrete version of the correlation energy always appears naturally. 
\end{rmk}

Besides containing the information of the ground state energy, another decisive consequence of the operator statement in Theorem \ref{them:OperatorStatement} is that it  allows us to obtain {\em all bosonic elementary excitations} predicted in the physics literature. We have

\begin{thm}[Bosonic elementary excitations] \label{thm:intro-excitations} 
Let $V$ and  $\mathcal{U}$ be as in Theorem \ref{them:OperatorStatement}. Let  $\Psi\in \mathcal{H}_{N}$ be a normalized wave function such that $ \mathcal{N}_{E} \Psi= \Psi$ and $\langle  \Psi, H_{\kin}^{\prime} \Psi \rangle = O(k_F)$. Then  for all $\epsilon>0$ we have 
\begin{align*}
\langle \Psi, \mathcal{U}H_{N}\mathcal{U}^{\ast} \Psi\rangle  = E_{\rm FS}  + E_{\rm corr} + \langle \Psi, \left.H_{\rm eff}\right|_{\mathcal{N}_{E}=1} \Psi\rangle + O(k_{F}^{1-\frac{1}{94} +\epsilon})
\end{align*} 
where  
 \begin{equation} \label{eq:intro-excitation-NE1}
\left.H_{\rm eff}\right|_{\mathcal{N}_{E}=1}=2\sum_{k\in\mathbb{Z}_{\ast}^{3}}\sum_{p,q\in L_{k}}\left\langle e_{p},\widetilde{E}_{k}e_{q}\right\rangle b_{k,p}^{\ast}b_{k,q} =  \tilde{U} \left(  \bigoplus_{k\in\mathbb{Z}_{\ast}^{3}}2\widetilde{E}_{k} \right)  \tilde{U}^{\ast}
\end{equation}
on the space $\left\{ \Psi\in\mathcal{H}_{N}\mid\mathcal{N}_{E}\Psi=\Psi\right\}$, and  
\begin{equation} 
\tilde{U}:\bigoplus_{k\in\mathbb{Z}_{\ast}^{3}}L^{2}\left(L_{k}\right)\rightarrow\left\{ \Psi\in\mathcal{H}_{N}\mid\mathcal{N}_{E}\Psi=\Psi\right\} 
\end{equation}
is a unitary isomorphism defined by
\begin{equation} \label{eq:intro-tilde-U}
\tilde{U}\bigoplus_{k\in\mathbb{Z}_{\ast}^{3}}\varphi_{k}=\sum_{k\in\mathbb{Z}_{\ast}^{3}}b_{k}^{\ast}\left(\varphi_{k}\right)\psi_{\rm FS} = \sum_{k\in\mathbb{Z}_{\ast}^{3}}\sum_{p\in L_k} \langle e_p, \varphi_k\rangle b_{k,p}^{\ast}\psi_{\rm FS}. 
\end{equation}
\end{thm}

%Recall that all of eigenvalues of $\widetilde{E}_{k}$ can be computed explicitly using \eqref{eq:ev-Ek}. Thus from Theorem \ref{thm:intro-excitations}, we may  say that up to the unitary transformation $\mathcal{U}$, the RPA is exact for the $\{\mathcal{N}_{E}=1\}$
%eigenspace of the effective Hamiltonian $H_{\rm eff}$. To our knowledge, this is the first rigorous derivation of the bosonic elementary  excitations from first principles. 
%
%\medskip
%
%Note that in the limit $k_F\to \infty$, the eigenvalues of $\widetilde{E}_{k}$ can be arbitrarily small, and this implies heuristically the fact that the spectrum of $H_N$ is continuous. Therefore, extracting useful information by analyzing the full spectrum of $H_N$ is very difficult. The significance of Theorem \eqref{thm:intro-excitations}  is to offer a non-trivial statement on the bosonic excitations by analyzing exactly the spectrum of the effective Hamiltonian. 

Recall that all of eigenvalues of $\widetilde{E}_{k}$ can be computed explicitly from the spectrum of $h_k$ and \eqref{eq:ev-Ek}. From Theorem \ref{them:OperatorStatement} and Theorem \ref{thm:intro-excitations}, we may say that  up to the unitary transformation $\mathcal{U}$, the RPA is exact for the $\{\mathcal{N}_{E}=1\}$
eigenspace of the effective Hamiltonian $H_{\rm eff}$. To our knowledge, this is the first rigorous derivation of the bosonic elementary  excitations from first principles.

\begin{rmk} For every fixed $k\in \mathbb{Z}^3_*$, in the limit $k_F\to \infty$, most of eigenvalues of $\widetilde{E}_{k}$ are of order $k_F$ but the lowest eigenvalue of $\widetilde{E}_{k}$ is of order $o(k_F)$. This absence of a one-body spectral gap corresponds to the expected fact that the excitation spectrum of $k_F^{-1}H_N$ becomes continuous in the limit $k_F\to \infty$. Therefore, in principle, it is very difficult to extract useful information by analyzing the full spectrum of $H_N$. The significance of Theorem \ref{thm:intro-excitations}  is to offer a non-trivial statement on the bosonic excitations by analyzing exactly the spectrum of the effective Hamiltonian instead of looking directly at the spectrum of $H_N$. 
\end{rmk}

%
%
%We believe that the statetment ``the true spectrum of $H_N$ contains eigenvalues close to the expected ones'' is correct, but it is well-known to be trivial since there is essentially no gap on the bosonic elementary excitations of $H_N$: by adding $M$ bosonic excitations of energy $\epsilon$ to $\Psi_{\rm FS}$, we should be able to show that $H_N$ contains an eigenvalue very close to $E_{\rm FS}+M \epsilon$, and hence the spectrum of $H_N$ cover almost every number in  $[E_{\rm FS}, E_{\rm FS}+O(k_F)]$ up to an error of order $\epsilon$ which can be chosen arbitrarily small in the limit $k_F\to \infty$. 
%
%We agree that the above statement does not follow from our estimate in Theorem 1.1, since this requires a norm estimate rather than a quadratic form estimate. Although we are confident that it can be achieved within our method, we are not motivated to proving that since the statement is conceptually trivial. The significance of Theorem 1.4, in our opinion, is to offer a non-trivial statement on the bosonic elementary excitations by analyzing exactly the spectrum of the effective Hamiltonian, instead of looking at the spectrum of $H_N$ which is continuous in the limit $k_F\to \infty$ and extracting helpful information from this is very difficult. 

\begin{rmk} 
%Let us explain the necessity of restricting to the $\mathcal{N}_{E}=1$
%eigenspace  in Theorem \ref{thm:intro-excitations}. 
In Theorem \ref{thm:intro-excitations}, the restriction to the $\mathcal{N}_{E}=1$ eigenspace is important. Obviously, the effective Hamiltonian \eqref{eq:intro-Heff} does not coincide with that in the heuristic formula \eqref{eq:IntroductionFullRPA}. Hence, it is natural to ask  what to make of the assumption of the
RPA that the effective Hamiltonian should behave like a diagonalized
bosonic Hamiltonian. To approach this question we note that, using  \eqref{eq:intro-RemarkableKineticIdentity} 
we can rewrite the effective Hamiltonian in \eqref{eq:intro-Heff} as 
\begin{align}
H_{\text{eff}}=2\sum_{k\in\mathbb{Z}_{\ast}^{3}}\sum_{p,q\in L_{k}}\left\langle e_{p},\widetilde{E}_{k}e_{q}\right\rangle b_{k,p}^{\ast}b_{k,q}-\left(\mathcal{N}_{E}-1\right)H_{\kin}^{\prime}.
\end{align}
Since this operator commutes with $\mathcal{N}_{E}$,  we can restrict $H_{\text{eff}}$ to the eigenspaces of $\mathcal{N}_{E}$.
Doing so, we see that the trivial eigenspace  $\left\{ \mathcal{N}_{E}=0\right\} =\vspan\left(\psi_{{\rm FS}}\right)$ exactly corresponds to the ground state energy which is already addressed in  Theorem \ref{thm:intro-GSE}. For the first nontrivial eigenspace  $\{\mathcal{N}_{E}=1\}$ we do
indeed obtain the expected operator 
\begin{equation}
\left.H_{\text{eff}}\right|_{\mathcal{N}_{E}=1}=2\sum_{k\in\mathbb{Z}_{\ast}^{3}}\sum_{p,q\in L_{k}}\left\langle e_{p},\widetilde{E}_{k}e_{q}\right\rangle b_{k,p}^{\ast}b_{k,q},
\end{equation}
as in the heuristic formula \eqref{eq:IntroductionFullRPA}. Moreover, the second identity in \eqref{eq:intro-excitation-NE1} tells us that $\left.H_{\text{eff}}\right|_{\mathcal{N}_{E}=1}$ can be diagonalized explicitly on $\{\mathcal{N}_{E}=1\}$, which is important for applications. 
\end{rmk}

More generally, we can also consider the higher excitation sectors $\left\{ \mathcal{N}_{E}=M\right\} $ for $M\in\mathbb{N}$. 

%However, for $M\ge 2$ we meet additional complications since the operator $\left.H_{\text{eff}}\right|_{\mathcal{N}_{E}=M}$ in \eqref{eq:RestrictedEffectiveOperator-2} cannot be diagonalized explicitly as in \eqref{eq:intro-excitation-NE1}

\begin{thm}[Higher Excitations] \label{thm:intro-excitations-II} 
Let $V$ and $\mathcal{U}$ be as in Theorem \ref{them:OperatorStatement}. Let $1\le M\le O(k_F)$. Let  $\Psi\in \mathcal{H}_{N}$ be a normalized wave function such that $ \mathcal{N}_{E} \Psi= M\Psi$  and $\langle  \Psi, H_{\kin}^{\prime} \Psi \rangle \le O(k_F)$. Then  for all $\epsilon>0$ we have 
\begin{align*}
\langle \Psi, \mathcal{U}H_{N}\mathcal{U}^{\ast} \Psi\rangle  = E_{\rm FS}  + E_{\rm corr} + \langle \Psi, \left.H_{\rm eff}\right|_{\mathcal{N}_{E}=M} \Psi\rangle + O(k_{F}^{1-\frac{1}{94} +\epsilon})
\end{align*}
where  
\begin{equation*}
\left.H_{\rm{eff}}\right|_{\mathcal{N}_{E}=M}=2\sum_{k\in\mathbb{Z}_{\ast}^{3}}\sum_{p,q\in L_{k}}\left\langle e_{p},\left(\widetilde{E}_{k}-\left(1-M^{-1}\right)h_{k}\right)e_{q}\right\rangle b_{k,p}^{\ast}b_{k,q}.\label{eq:RestrictedEffectiveOperator-2}
\end{equation*}
\end{thm}

%\begin{align}
%delete\\
%delete
%\end{align}

\begin{rmk} 
%Theorem \ref{thm:intro-excitations-II} suggests that treating $H_{\text{eff}}$ as a quasi-bosonic 
%operator can perhaps be justified, albeit with $M$-dependent one-body
%operators $\widetilde{E}_{k}-\left(1-M^{-1}\right)h_{k}$. 
%it is also possible to consider the higher excitation sectors $\left\{ \mathcal{N}_{E}=M\right\} $ for $M\in\mathbb{N}$, but 
%In  Theorem \ref{thm:intro-excitations-II}, 
%the operator $\left.H_{\rm{eff}}\right|_{\mathcal{N}_{E}=M}$ for $M\ge 2$ is substantially more complicated than the case $M=1$. In particular, it is unlikely diagonalizable Nevertheless, 
For $M\ge 2$, the operator $\left.H_{\text{eff}}\right|_{\mathcal{N}_{E}=M}$ in Theorem \ref{thm:intro-excitations-II} cannot be diagonalized explicitly as in \eqref{eq:intro-excitation-NE1}. The quasi-bosonic property is insufficient to guarantee that it is diagonalizable, even approximately. Understanding the behaviour of $H_{\text{eff}}$ on higher eigenspaces
and reconciling the RPA thus 
appears to be an interesting but non-trivial task. Some progress in this direction was done in \cite{CHN-22-LMP} where the norm $\|(H_{\rm{eff}} - M \epsilon) \Psi\|$ was estimated for suitable trial states. 
\end{rmk}

\subsection{Proof Strategy}

Now let us explain some key ingredients of the proof. Following \cite{SawBruFukBro-57}, our approach
consists of studying pair-excitations $b_{k,p}^{\ast}=c_{p}^{\ast}c_{p-k}$,
where $c_{p-k}$ annihilates a particle with momentum $p-k$, i.e.
creates a hole in the Fermi ball, and $c_{p}^{\ast}$ creates a particle
outside the Fermi ball. These operators $b_{k,p}$, $b_{k,p}^{\ast}$ satisfy the bosonic commutation relations in an appropriate
sense. 
%More precisely, for states having few excitations relative to the Fermi state one might expect the overall
%contribution of the non-bosonic error terms 
%to be small. 
This enables the use of a quasi-bosonic Bogolubov transformation to diagonalize the original fermionic operator. 
A main achievement of the present work is the analytical
elaboration of this bosonic picture.

\medskip

In \cite{BNPSS-20,BNPSS-21} a different, collective bosonization approach was developed by averaging
the pair-excitations $b_{k,p}^{\ast}$ on ``patches'' near the surface of the Fermi ball, thus realizing
strengthened versions of the bosonic commutation relations which make the comparison with the purely bosonic computation significantly easier. In the present paper we show that the bosonization idea can be implemented directly for pairs of fermions without such an averaging procedure. In our opinion this new approach is conceptually closer to the physics of the problem and more transparent for applications.  In particular, it allows us to obtain all bosonic elementary excitations as in Theorem \ref{thm:intro-excitations}. Moreover, the new method is potentially applicable to Coulomb systems, where the correlation exchange correction  to the purely bosonic computation plays an important role; see \cite{CHN-22} for a rigorous ground state energy upper bound.

\medskip
In the context of interacting Bose gases, Bogolubov transformations based on another approximate CCR have been used to study the excitation spectrum, see e.g. \cite{Seiringer-11,GreSei-13,BCSS-19,HST-21}. However, for the fermionic problem considered in the present paper, the approximate CCR holds in a very different setting and requires distinct estimation techniques.

%the bosonic and fermionic problems require completely different settings of estimates. 
%
%
%the approximate CCR holds in a very different context

%an averaging technique was instead implemented
%directly into the bosonization method, by distributing the particles
%into ``patches'' near the surface of the Fermi ball. Our approach can be contrasted to \cite{BNPSS-20,BNPSS-21} where
%a different, collective bosonization approach was developed by averaging
%the pair-excitations $b_{k,p}^{\ast}$ on ``patches'', thus realizing
%averaged versions of the bosonic commutation relations.

\medskip

%{\bf add comments on the different ... potentially apply to Coulomb...} In the present paper we show that the bosonization idea can be implemented
%directly for pairs of fermions without such an averaging procedure.
% In particular, while the two approaches are both useful for deriving the ground state energy, the new approach seems to be more suitable to obtain the bosonic elementary excitations as in Theorem \ref{thm:intro-excitations}. 

%In the context of bosonic systems, 
%
%{\bf add comparison to approximate CCR for bosons, completely different...} 

Now let us provide further details. 

\subsubsection*{Bosonization Method}

The driving concept of the random phase approximation is the bosonization
of fermionic pairs. We must therefore argue why the excitation operators
\begin{equation}
b_{k,p}=c_{p-k}^{\ast}c_{p},\quad b_{k,p}^{\ast}=c_{p}^{\ast}c_{p-k},\quad p\in L_{k}=\left(B_{F}+k\right)\backslash B_{F},
\end{equation}
obey an approximate CCR. Consider for simplicity the case $k=l$:
Then computation shows that for any $p,q\in L_{k}$, $\left[b_{k,p},b_{k,q}\right]=[b_{k,p}^{\ast},b_{k,q}^{\ast}]=0$
but
\begin{equation} \label{eq:intro-comm-b-b*}
\left[b_{k,p},b_{k,q}^{\ast}\right]=\delta_{p,q}-\delta_{p,q}\left(c_{p}^{\ast}c_{p}+c_{p-k}c_{p-k}^{\ast}\right).
\end{equation}
In general, thanks to Pauli's exclusion principle ($c_{p}^{\ast}c_{p},\,c_{p}c_{p}^{\ast}\leq1$), the error term in \eqref{eq:intro-comm-b-b*} satisfies the  simple bound
$\delta_{p,q}\left(c_{p}^{\ast}c_{p}+c_{p-k}c_{p-k}^{\ast}\right)\leq2\delta_{p,q}$, but this is even bigger than the leading term $\delta_{p,q}$.
%$-\delta_{p,q}\left(c_{p}^{\ast}c_{p}+c_{p-k}c_{p-k}^{\ast}\right)$
%can generally not be considered to be small - the general estimates
%$c_{p}^{\ast}c_{p},\,c_{p}c_{p}^{\ast}\leq1$, which are valid for
%fermionic creation and annihilation operators, imply the simple bound
%$\delta_{p,q}\left(c_{p}^{\ast}c_{p}+c_{p-k}c_{p-k}^{\ast}\right)\leq2\delta_{p,q}$,
%but this is even bigger than the leading $\delta_{p,q}$.
The key observation %which justifies treating the excitation operators $b_{k,p}$ as being bosonic 
is that although these errors terms can
not be considered to be small individually, they are so on average.
For instance
\begin{equation}
\sum_{p,q\in L_{k}}\delta_{p,q}\left(c_{p}^{\ast}c_{p}+c_{p-k}c_{p-k}^{\ast}\right)=\sum_{p\in L_{k}}c_{p}^{\ast}c_{p}+\sum_{p\in L_{k}}c_{p-k}c_{p-k}^{\ast}\leq2\,\mathcal{N}_{E}
\end{equation}
where $\mathcal{N}_{E}$ is the ``excitation number operator'' defined in \eqref{eq:ParticleHoleSymmetry}. 
Thus for states where the expectation value of $\mathcal{N}_{E}$ is much
smaller than $\sum_{p,q\in L_{k}}\delta_{p,q}=\left|L_{k}\right|  \sim \min\{|k| k_F^2,k_F^3\}$
one may expect that the contribution of the non-bosonic error terms are also smaller than the leading bosonic behaviour.
Justifying this idea rigorously is one of the main results of this
paper.

\medskip

Note that unlike the works \cite{HaiPorRex-20,BNPSS-20,BNPSS-21}
we do not employ the ``particle-hole transformation'' $R$, which
maps $\psi_{{\rm FS}}$ to the vacuum, so that we always work directly
on the space $\mathcal{H}_{N}$. % (see subsection \ref{subsec:RemarksontheLocalizationProcedure} for further details on this aspect).

\subsubsection*{A Priori Estimates}

As explained above, to apply the bosonization method we  need to show that the expectation of $\mathcal{N}_E$ against  low-lying eigenstates of $H_N$ is much smaller than $\left|L_{k}\right|\sim\min\left\{ k_{F}^{2}\left|k\right|,k_{F}^{3}\right\}$. 
%,
%so if $\Psi_{\text{GS}}$ is a ground state for $H_{N}$ we must control
%$\left\langle \Psi_{\text{GS}},\mathcal{N}_{E}\Psi_{\text{GS}}\right\rangle $
%much better than this.

Using the condition $\sum_{k\in\mathbb{Z}^{3}}\hat{V}_{k}\left|k\right|<\infty$
and a variant of Onsager's lemma, 
we can prove that 
\begin{equation}
H_{N} \ge E_{\rm FS} + H_{\kin}^{\prime} - Ck_{F}.
\end{equation}
Consequently, if $\Psi$ is any eigenstate for $H_N$ satisfying $\left\langle \Psi,H_{N}\Psi\right\rangle \leq E_{\rm FS}  +Ck_{F}$, 
then %we obtain the kinetic bound
\begin{equation} \label{eq:IntroductionStrongerAPriori-0}
\left\langle \Psi,H_{\kin}^{\prime}\Psi\right\rangle \leq Ck_{F}.
\end{equation}

Since $H_{\kin}^{\prime}\ge \mathcal{N}_{E}$, 
which was already explained in \cite{BNPSS-21}, this implies that $\langle \Psi, \mathcal{N}_E \Psi\rangle\le Ck_F \ll |L_k|$. For $V$ sufficiently small this bound was first proved in \cite{HaiPorRex-20} 
(by a different method), and it was also used in \cite{BNPSS-21}. In practice we will also need a stronger a priori estimate, namely
\begin{equation}\label{eq:IntroductionStrongerAPriori}
\left\langle \Psi, k_F^{-1}\mathcal{N}_{E}H_{\kin}^{\prime}\Psi\right\rangle \leq Ck_{F} 
\end{equation}
as stated in Theorem \ref{thm:Kinetic-estimate}. This we will obtain 
by employing a bootstrapping argument for eigenstates, inspired by
the ``improved condensation'' in the context of Bose gases in \cite{Seiringer-11,GreSei-13,Nam-17,NamNap-21}.
%Indeed, we prove that any eigenstate $\Psi$ of $H_{N}$ obeys
%\begin{equation}
%\left\langle \Psi,\mathcal{N}_{E}H_{\kin}^{\prime}\Psi\right\rangle \leq C\max\left\{ \left\langle \Psi,H_{N}\Psi\right\rangle - E_{\rm FS}  ,k_{F}\right\} ^{2}.\label{eq:IntroductionStrongerAPriori}
%\end{equation}
In \cite{BNPSS-21}, an analogue of equation (\ref{eq:IntroductionStrongerAPriori})
was proved for a modified ground state by using a ``localization
in Fock space'' technique. In comparison, our estimate of equation
(\ref{eq:IntroductionStrongerAPriori}) is obtained in a far more
direct fashion, and yields a uniform bound for all low-lying eigenstates. In particular, thanks to \eqref{eq:IntroductionStrongerAPriori-0} and \eqref{eq:IntroductionStrongerAPriori}, the operator estimate in Theorem \ref{them:OperatorStatement} leads to direct consequences on the ground state energy and the excitation spectrum of $H_N$.

\subsubsection*{Removing the Non-Bosonizable Terms}

An important ingredient of the RPA is that the non-bosonizable terms
\begin{equation} \label{eq:intro-nonbosonizable-terms}
\frac{k_{F}^{-1}}{2\left(2\pi\right)^{3}}\sum_{k\in\mathbb{Z}_{\ast}^{3}}\hat{V}_{k}\left(2\,{\rm Re}\left(\tilde{B}_{k}^{\ast}+\tilde{B}_{-k}\right)D_{k}+D_{k}^{\ast}D_{k}\right)
\end{equation}
are negligible to the leading order of the correlation energy. Here we offer a direct estimate for these terms,
which is simpler than the strategy proposed in \cite{BNPSS-21} and
does not require a smallness condition on $V$. More precisely, in Theorem \ref{them:ReductiontoBosonizableTerms}  we will prove that the non-bosonizable terms are bounded by $o(1) (k_F^{-1}\mathcal{N}_E H_{\kin}' + H_{\kin}'+k_F)$, and hence the expectation against the low-lying eigenstates of $H_N$ is of order $o(k_F)$ due to the a-priori estimates mentioned before. 

\subsubsection*{Bosonization of the Kinetic Operator and the Excitation Number Operator}

Concerning the bosonizable terms, while the interaction terms can be interpreted directly as a quadratic Hamiltonian in the quasi-bosonic picture as in \eqref{eq:Hint-k}, the treatment of the kinetic operator is more subtle. In fact, \eqref{eq:intro-kinetic-bosonic} does not hold as a direct operator approximation. 
Instead we will justify it by appealing to the commutator relation
\begin{equation}
\left[H_{\kin}^{\prime},b_{l,q}^{\ast}\right]\approx\left[2\sum_{k\in\mathbb{Z}_{\ast}^{3}}\sum_{p\in L_{k}}\lambda_{k,p}b_{k,p}^{\ast}b_{k,p},b_{l,q}^{\ast}\right].\label{eq:IntroductionApproximatelyEqualCommutators}
\end{equation}
This commutator relation ensures that the difference
\begin{equation}
H_{\kin}^{\prime}-2\sum_{k\in\mathbb{Z}_{\ast}^{3}}\sum_{p\in L_{k}}\lambda_{k,p}b_{k,p}^{\ast}b_{k,p}
\end{equation}
is essentially invariant under the Bogolubov transformations introduced later, which is sufficient for our purpose.  The approximation \eqref{eq:IntroductionApproximatelyEqualCommutators} is a consequence of the exact commutation relation \eqref{eq:intro-Hkin-com}: 
%\begin{prop}
%\label{LocalizedKineticOperatorCommutator}For any $k\in\mathbb{Z}_{\ast}^{3}$
%and $p\in L_{k}$ the operators $b_{k,p}^{\ast}$ obey
%\[
%\left[H_{\kin}^{\prime},b_{k,p}^{\ast}\right]=\left(\left|p\right|^{2}-\left|p-k\right|^{2}\right)b_{k,p}^{\ast}.
%\]
%\end{prop}
For every $p\in L_k= B_F^c \cap (B_F+k)$, by the CAR  we have 
\begin{align} \label{LocalizedKineticOperatorCommutator}
\left[H_{\kin}^{\prime},b_{k,p}^{\ast}\right] &= \sum_{q\in B_{F}^{c}} \left|q\right|^{2} \left[c_{q}^{\ast}c_{q},c_{p}^* c_{p-k} \right] - \sum_{q\in B_{F}}\left|q\right|^{2}  \left[c_{q}c_{q}^{\ast}, c_{p}^* c_{p-k} \right] \nonumber\\
&=\sum_{q\in B_{F}^{c}} \left|q\right|^{2} \left[c_{q}^{\ast}c_{q},c_{p}^*\right]  c_{p-k}  - \sum_{q\in B_{F}}\left|q\right|^{2}  c_{p}^* \left[c_{q}c_{q}^{\ast},  c_{p-k} \right]  \nonumber \\
&=\sum_{q\in B_{F}^{c}} \left|q\right|^{2} \delta_{q,p} c_{q}^{\ast} c_{p-k}  - \sum_{q\in B_{F}}\left|q\right|^{2}  \delta_{q,p-k} c_{p}^* c_{p-k} \nonumber  \\
 & =\left|p\right|^{2}  c_{p}^{\ast} c_{p-k} - \left|p-k\right|^{2} c_{p}^* c_{p-k} =   \left(\left|p\right|^{2}-\left|p-k\right|^{2}\right)b_{k,p}^{\ast}.
\end{align}
A similar strategy was used in \cite{BNPSS-21}, although the analysis
there is more complicated due to the averaging technique of the ``patches''.
In particular, the operators on ``patches'' in \cite{BNPSS-21} do not obey the exact commutator relation $\left[H_{\kin}^{\prime},b_{k,p}^{\ast}\right]=2\lambda_{k,p}b_{k,p}^{\ast}$, and so the kinetic operator has to be handled by an additional
linearization argument.

Note that in the same manner of the dispersion relation in  \eqref{LocalizedKineticOperatorCommutator}, we also have 
\begin{align} \label{eq:intro-comm-NE-b}
\left[\mathcal{N}_{E},b_{k,p}\right] & =\sum_{q\in B_{F}^{c}}\left[c_{q}^{\ast}c_{q},c_{p-k}^{\ast}c_{p}\right]=\sum_{q\in B_{F}^{c}}\left(c_{q}^{\ast}\left[c_{q},c_{p-k}^{\ast}c_{p}\right]+\left[c_{q}^{\ast},c_{p-k}^{\ast}c_{p}\right]c_{q}\right)\\
 & =\sum_{q\in B_{F}^{c}}\left(c_{q}^{\ast}\left(-c_{p-k}^{\ast}\left\{ c_{q},c_{p}\right\} +\left\{ c_{q},c_{p-k}^{\ast}\right\} c_{p}\right)+\left(-c_{p-k}^{\ast}\left\{ c_{q}^{\ast},c_{p}\right\} +\left\{ c_{q}^{\ast},c_{p-k}^{\ast}\right\} c_{p}\right)c_{q}\right)\nonumber \\
 & =-\sum_{q\in B_{F}^{c}}\left(\delta_{q,p}c_{p-k}^{\ast}\right)c_{q}=-c_{p-k}^{\ast}c_{p}=-b_{k,p}\nonumber 
\end{align}
for all $k\in\mathbb{Z}_{\ast}^{3}$ and $p\in L_{k}$. This means that $\mathcal{N}_{E}$ plays the same role as the number operator in the bosonic picture.

\subsubsection*{Bogolubov Transformation I}

%To diagonalize the effective operator in  (\ref{eq:IntroductonEffectiveOperator}), first for practical reasons we define a cut-off set
%\begin{equation}
%S_{C}=\overline{B}\left(0,k_{F}^{\gamma}\right)\cap\mathbb{Z}_{+}^{3},
%\end{equation}
%where $\gamma \in (0,1]$ will be optimized later. 

We will estimate the contribution of high momenta separately, and only diagonalize the effective operator in  (\ref{eq:IntroductonEffectiveOperator}) for low momenta. For this reason, we define a cut-off set
\begin{equation} \label{eq:SC}
S_{C}=\overline{B}\left(0,k_{F}^{\gamma}\right)\cap\mathbb{Z}_{+}^{3},
\end{equation}
where $\gamma \in (0,1]$ will be optimized later. For a given $k_{F}$ we then diagonalize only  
\begin{equation} \label{eq:Recall1-1}
H_{\text{eff}}^{\prime}=\sum_{k\in S_{C}}\left(2\sum_{p\in L_{k}}\lambda_{k,p}b_{k,p}^{\ast}b_{k,p}+2\sum_{p\in L_{-k}}\lambda_{-k,p}b_{-k,p}^{\ast}b_{-k,p}+H_{\text{int}}^{k}\right),
\end{equation}
and treat the remaining terms with $k\in\mathbb{Z}_{+}^{3}\backslash S_{C}$
as an error term. As $\overline{B}\left(0,k_{F}^{\gamma}\right)\cap\mathbb{Z}_{+}^{3}$
forms an exhaustion of $\mathbb{Z}_{+}^{3}$ all terms are thus nonetheless
diagonalized in the limit $k_{F}\rightarrow\infty$.

Inspired by the exact bosonic diagonalization (see Theorem \ref{thm:Bog-diag} for details) we take the diagonalizing Bogolubov transformation to be of the form
$e^{\mathcal{K}}$ for a generator $\mathcal{K}:\mathcal{H}_{N}\rightarrow\mathcal{H}_{N}$
defined by
\begin{equation} \label{eq:cK-def-intro}
\mathcal{K}=\sum_{k\in S_{C}}\left(\sum_{p\in L_{k}}\sum_{q\in L_{-k}}\left\langle e_{p},K_{k}e_{-q}\right\rangle \left(b_{k,p}b_{-k,q}-b_{-k,q}^{\ast}b_{k,p}^{\ast}\right)\right),
\end{equation}
where the transformation kernels $K_{k}:\ell^{2}\left(L_{k}\right)\rightarrow\ell^{2}\left(L_{k}\right)$,
$k\in\mathbb{Z}_{+}^{3}$, are defined by
\begin{equation}
K_{k}=-\frac{1}{2}\log\left(h_{k}^{-\frac{1}{2}}\left(h_{k}^{\frac{1}{2}}\left(h_{k}+2P_{v_{k}}\right)h_{k}^{\frac{1}{2}}\right)^{\frac{1}{2}}h_{k}^{-\frac{1}{2}}\right)
\end{equation}
with $h_{k},P_{v_{k}}$ as defined in equation (\ref{eq:IntroductionOneBodyOperators}).
With this choice we find that
\begin{equation} \label{eq:Recall1-2}
e^{\mathcal{K}}H_{\text{eff}}^{\prime}e^{-\mathcal{K}}\approx\sum_{k\in S_{C}\cup\left(-S_{C}\right)}\left(\text{tr}\left(E_{k}-h_{k}\right)+2\sum_{p,q\in L_{k}}\left\langle e_{p},E_{k}e_{q}\right\rangle b_{k,p}^{\ast}b_{k,q}\right)
\end{equation}
for  
\begin{align}
E_{k}=e^{-K_{k}}h_{k}e^{-K_{k}} %?=\left(\left(h_{k}+2P_{v_{k}}\right)^{\frac 1 2} h_{k} \left(h_{k}+2P_{v_{k}}\right)^{\frac{1}{2}} \right)^{\frac{1}{2}}
\end{align}
and by the commutation relation
of equation (\ref{eq:IntroductionApproximatelyEqualCommutators}),
that
\begin{equation} \label{eq:Recall1-3}
e^{\mathcal{K}}\left(H_{\kin}^{\prime}-2\sum_{k\in S_{C}\cup\left(-S_{C}\right)}\sum_{p\in L_{k}}\lambda_{k,p}b_{k,p}^{\ast}b_{k,p}\right)e^{-\mathcal{K}}\approx H_{\kin}^{\prime}-2\sum_{k\in S_{C}\cup\left(-S_{C}\right)}\sum_{p\in L_{k}}\lambda_{k,p}b_{k,p}^{\ast}b_{k,p}
\end{equation}
so by the equations \eqref{eq:Recall1-1}, \eqref{eq:Recall1-2} and \eqref{eq:Recall1-3}, noting also that $\left\langle e_{p},h_{k}e_{q}\right\rangle =\delta_{p,q}\lambda_{k,p}$,
\begin{align}  \label{eq:intro-result-first-Bog-trans}
 & e^{\mathcal{K}}\left(H_{\kin}^{\prime}+\sum_{k\in S_{C}\cup\left(-S_{C}\right)}H_{\text{int}}^{k}\right)e^{-\mathcal{K}}\\
 & \approx H_{\kin}^{\prime} +\sum_{k\in S_{C}\cup\left(-S_{C}\right)}\left(\text{tr}\left(E_{k}-h_{k}\right)+2\sum_{p,q\in L_{k}}\left\langle e_{p},\left(E_{k}-h_{k}\right)e_{q}\right\rangle b_{k,p}^{\ast}b_{k,q}\right).\nonumber 
\end{align}
%After justifying this approximation we can then apply a trial state
%to conclude the upper bound on the correlation energy.

On the right side of \eqref{eq:intro-result-first-Bog-trans}, the constant $\sum_{k\in S_{C}\cup\left(-S_{C}\right)}\text{tr}\left(E_{k}-h_{k}\right)$ captures correctly the leading order of the correlation energy $E_{\rm corr}$. On the other hand, although $E_{k}$ is isospectral to
\begin{equation}
\widetilde{E}_k=h_{k}^{\frac{1}{2}}e^{-2K_{k}}h_{k}^{\frac{1}{2}}=(h_{k}^{\frac{1}{2}}\left(h_{k}+2P_{v_{k}}\right)h_{k}^{\frac{1}{2}})^{\frac{1}{2}}\geq h_{k},
\end{equation}
the operator $E_{k}-h_{k}$ is not non-negative. Thus the term $2\sum_{p,q\in L_{k}}\left\langle e_{p},\left(E_{k}-h_{k}\right)e_{q}\right\rangle b_{k,p}^{\ast}b_{k,q}$
- a kind of second quantization of $E_{k}-h_{k}$ - cannot be ignored
for the lower bound.

The Bogolubov transformation used in this part is analogous to that
of \cite{BNPSS-21}. It was proved in \cite{BNPSS-21} that if $V$
is small, then the quantization of $E_{k}-h_{k}$ can be controlled
by $H_{\kin}^{\prime}$, leading to the desired lower bound on the ground state energy. 
In order to treat an arbitrary potential we will instead utilize a
second Bogolubov transformation which effectively replaces $E_k$ by $\widetilde{E}_k$ in \eqref{eq:intro-result-first-Bog-trans}.

\subsubsection*{Bogolubov Transformation II}

We define the second Bogolubov transformation $e^{\mathcal{J}}$
for a generator $\mathcal{J}:\mathcal{H}_{N}\rightarrow\mathcal{H}_{N}$
defined by
\begin{equation} \label{eq:cJ-def-intro}
\mathcal{J}=\sum_{k\in S_{C}\cup\left(-S_{C}\right)}\sum_{p,q\in L_{k}}\left\langle e_{p},J_{k}e_{q}\right\rangle b_{k,p}^{\ast}b_{k,q},
\end{equation}
where $J_{k}=\log\left(U_{k}\right)$ denotes the (principal) logarithm
of the unitary transformation $U_{k}:\ell^{2}\left(L_{k}\right)\rightarrow\ell^{2}\left(L_{k}\right)$
defined by
\begin{equation}
U_{k}=\left(h_{k}^{\frac{1}{2}}e^{-2K_{k}}h_{k}^{\frac{1}{2}}\right)^{\frac{1}{2}}h_{k}^{-\frac{1}{2}}e^{K_{k}}.
\end{equation}
This is precisely the unitary transformation which satisfies
\begin{equation}
U_{k}E_{k}U_{k}^{\ast}=h_{k}^{\frac{1}{2}}e^{-2K_{k}}h_{k}^{\frac{1}{2}}=(h_{k}^{\frac{1}{2}}\left(h_{k}+2P_{v_{k}}\right)h_{k}^{\frac{1}{2}})^{\frac{1}{2}} = \widetilde{E}_k,
\end{equation}
as is easily verified. This transformation acts such that
\begin{equation}
e^{\mathcal{J}}\left(\sum_{p,q\in L_{k}}\left\langle e_{p},E_{k}e_{q}\right\rangle b_{k,p}^{\ast}b_{k,q}\right)e^{-\mathcal{J}}\approx\sum_{p,q\in L_{k}}\left\langle e_{p},\widetilde{E}_k e_{q}\right\rangle b_{k,p}^{\ast}b_{k,q},
\end{equation}
and thanks to the relation of equation (\ref{eq:IntroductionApproximatelyEqualCommutators})
also
\begin{equation}
e^{\mathcal{J}}\left(H_{\kin}^{\prime}-2\sum_{k\in S_{C}\cup\left(-S_{C}\right)}\sum_{p\in L_{k}}\lambda_{k,p}b_{k,p}^{\ast}b_{k,p}\right)e^{-\mathcal{J}}\approx H_{\kin}^{\prime}-2\sum_{k\in S_{C}\cup\left(-S_{C}\right)}\sum_{p\in L_{k}}\lambda_{k,p}b_{k,p}^{\ast}b_{k,p},
\end{equation}
so all in all
\begin{align}
 & e^{\mathcal{J}}e^{\mathcal{K}}\left(H_{\kin}^{\prime}+\sum_{k\in S_{C}\cup\left(-S_{C}\right)}H_{\text{int}}^{k}\right)e^{-\mathcal{K}}e^{-\mathcal{J}}  \\
 &\approx\sum_{k\in S_{C}\cup\left(-S_{C}\right)}\text{tr}\left(E_{k}-h_{k}\right)+H_{\kin}^{\prime} +\sum_{k\in S_{C}\cup\left(-S_{C}\right)}\sum_{p,q\in L_{k}}\left\langle e_{p},(\widetilde{E}_k - h_k)  e_{q}\right\rangle b_{k,p}^{\ast}b_{k,q}.\nonumber 
\end{align}
As $\widetilde{E}_k-h_{k}\geq0$
the last term can now be dropped and the energy lower bound concluded. The cut-off $S_C$ can be removed at the end without serious difficulties. On the technical level, the second
Bogolubov transformation is an important new tool to remove the smallness
condition of \cite{BNPSS-21}, thus enabling us to work with a significantly larger
class of interaction potentials. In the independent work \cite{BPSS-21}, the idea of using the second Bogolubov transformation has also been introduced to refine the method in \cite{BNPSS-20,BNPSS-21}.

\subsubsection*{Elementary Excitations} 

The key ingredient to obtain all bosonic elementary excitations is the formula \eqref{eq:intro-excitation-NE1}  in Theorem \ref{thm:intro-excitations}. To prove this, note that  $\left.H_{\rm eff}\right|_{\mathcal{N}_{E}=1}$ commutes with $\mathcal{N}_{E}$ and the total momentum $P= \sum_{p\in \mathbb{Z}^3_*} p c_p^* c_p$, so we may restrict $H_{\text{eff}}$ to the simultanous eigenspaces
of $\mathcal{N}_{E}$ and $P$, which are 
\begin{equation}
\left\{ \Psi\in\mathcal{H}_{N}\mid\mathcal{N}_{E}\Psi=\Psi,\,P\Psi=k\Psi\right\} =\vspan\left(b_{k,p}^{\ast}\psi_{{\rm FS}}\right)_{p\in L_{k}}=\left\{ b_{k}^{\ast}\left(\varphi\right)\psi_{{\rm FS}}\mid\varphi\in L^{2}\left(L_{k}\right)\right\} .
\end{equation}
It turns out that the mapping $U_k:L^{2}\left(L_{k}\right)\rightarrow\left\{ \Psi\in\mathcal{H}_{N}\mid\mathcal{N}_{E}\Psi=\Psi,\,P\Psi=k\Psi\right\} $
defined by
\begin{equation}
U_k\varphi=b_{k}^{\ast}\left(\varphi\right)\psi_{\FS},\quad\varphi\in L^{2}\left(L_{k}\right),
\end{equation}
is a unitary isomorphism with the property that  
\begin{equation}
\left.H_{\text{eff}}\right|_{\mathcal{N}_{E}=1} = U_k (2\widetilde{E}_{k} ) U_k^{\ast}
\end{equation}
Summing over  different momenta $k$'s, we obtain the transformation $\tilde{U}$ introduced in \eqref{eq:intro-tilde-U}.

\medskip

In summary, our approach is different from the previous works \cite{HaiPorRex-20,BNPSS-20,BNPSS-21}
in many aspects.  On the conceptual level, our direct bosonization method (i.e. working
directly with the operators $b_{k,p}$ instead of averaging them on
``patches'') allows us to stick closely to the heuristic argument of the
physics literature, and to obtain not only the ground state energy but also all bosonic elementary excitations, thus leading to the first complete justification of the RPA in the mean-field regime.

Although our general ideas are very transparent, to realize the whole
procedure on a rigorous basis we will need to develop several new
estimates to justify all of the approximations made. In the rest of the paper we will show how to implement the proof strategy
rigorously.

\medskip

\textbf{Outline of the Paper.} In Section \ref{sec:Kinetic} we prove some general estimates involving the kinetic operator $H_{\kin}$ and bound the non-bosonizable terms. In Section \ref{sec:BosonicBogolubovTransformations} we review the
theory of bosonic Bogolubov transformations; in particular we review how one may
explicitly define a Bogolubov transformation which diagonalizes a
given positive-definite quadratic Hamiltonian. We then apply the bosonic theory to our study of the Fermi gas where we implement
the diagonalization procedure in the quasi-bosonic framework. This 
is done by introducing the quasi-bosonic quadratic Hamiltonian in Section \ref{sec:TransformingtheHamiltonian} and the quasi-bosonic Bogolubov transformation $e^{\mathcal{K}}$ in Section \ref{sec:quasi-bosonic-Bogolubov} (these notations mirror the
exact bosonic ones as closely as possible such that the bosonic theory
is easily transferred to the quasi-bosonic setting). In this way the
quasi-bosonic analysis reduces to that of a collection of exact bosonic
quadratic Hamiltonians plus correlation exchange terms - error terms which arise
due to the deviation from the exact CCR. In Section \ref{sec:Analysis of Exchange Terms} we  estimate
the exchange terms,  reducing the analysis of these to the
associated one-body operators of the bosonic problem. The one-body operators are studied separately in Section \ref{sec:AnalysisoftheOne-BodyOperators}. In this part, we will need several estimates of Riemann sums, which are collected in the Appendix. We complete the analysis of the transformation $e^{\mathcal{K}}$ in Section \ref{sec:Gronwall Estimates for the Bogolubov Transformation}, where we prove that $H_{\kin}'$ and $\mathcal{N}_E$ are stable under the transformation $e^{\mathcal{K}}$. In Section \ref{sec:TheSecondTransformationandKineticEstimates} we introduce the second unitary transformation $e^{\mathcal{J}}$. The analysis of this transformation is essentially similar to the first one, except that we require new one-body operator estimates which are somewhat more difficult. 
Finally we conclude the proofs of the main theorems in Section \ref{sec:prof-main-theorems}.

\medskip
\textbf{Acknowledgements.} PTN thanks Niels Benedikter, Marcello Porta,
Benjamin Schlein, and Robert Seiringer for helpful discussions. We thank the referees for constructive remarks and suggestions. MRC
and PTN acknowledge the support from the Deutsche Forschungsgemeinschaft
(DFG project Nr. 426365943).

%%%%%%%%%%%%%%%%%%%%%%%%%%%%%%%%%%%%
%%%%%%%%%%%%%%%%%%%%%%%%%%%%%%%%%%%%

\section{Removal of the Non-Bosonizable Terms} \label{sec:Kinetic}

In this section we collect several basic estimates concerning the operator $H_N$ which can be obtained without using Bogolubov transformations. Recall the decomposition \eqref{eq:HN-localized-1}:
\begin{equation}
H_{N}  =E_{\rm FS} +H_{\kin}^{\prime}+k_{F}^{-1}H_{\inter}^{\prime}, \quad E_{\rm FS} = \left\langle \psi_{\FS},H_{N}\psi_{\FS}\right\rangle .  
\end{equation}
We will bound the interaction operator $H_{\inter}^{\prime}$ in terms of the kinetic operator $H_{\kin}^{\prime}$, and then prove a-priori estimates for eigenstates of $H_N$ which are parts of Theorem \ref{thm:Kinetic-estimate}. 

Recall the following result from \cite[Lemma 2.4]{BNPSS-21} concerning the kinetic operator $H_{\rm kin}'$ in \eqref{eq:Hkin'}. 

\begin{prop}
\label{prop:Zeta0RepresentationofHKin} We have $H_{\kin}^{\prime} \ge \mathcal{N}_{E}$ with $\mathcal{N}_{E}$ given in \eqref{eq:ParticleHoleSymmetry}. 
\end{prop}

\textbf{Proof:} Since  $|p|^2$ is an integer for $p\in \mathbb{Z}^3$, our assumption $|B_F|=N$ implies that 
\begin{equation} 
\inf_{p\in B_{F}^{c}}\left|p\right|^{2} -  \sup_{p\in B_{F}}\left|p\right|^{2} \ge 1.
\end{equation}
Therefore, in \eqref{eq:ManifestlyNonNegativeHKin} we can choose $\zeta$ such that $\vert\left|p\right|^{2}-\zeta\vert\,\geq 1/2$ for all $p\in \mathbb{Z}^3$. %  and obtain the desired bound. 
$\hfill\square$

Next, we consider the bosonizable terms in $H_{\inter}^{\prime}$. The following result is a minor extension of \cite[Lemma 4.7]{HaiPorRex-20}  (see also \cite[Appendix B]{BNPSS-21} for a simplified proof).  

\begin{prop}
\label{prop:BkBkAPriori}For all $k\in\mathbb{Z}_{\ast}^{3}$, the operator $\tilde{B}_{k}$ in \eqref{eq:def-Bk} satisfies  
that
\begin{align*}
\tilde{B}_{k}^{\ast}\tilde{B}_{k}\leq Ck_{F}H_{\kin}^{\prime}, \quad \tilde{B}_{k}\tilde{B}_{k}^{\ast} \leq Ck_{F} ( H_{\kin}^{\prime} + |k| k_F)
\end{align*} 
where the constant $C>0$ is independent of $k$ and $k_{F}$.
\end{prop}

\textbf{Proof:} As argued in \cite{HaiPorRex-20,BNPSS-21}, for any $\Psi\in\mathcal{H}_{N}$ it follows from
the triangle and Cauchy-Schwarz inequalities that
\begin{align}
\left\Vert \tilde{B}_{k}\Psi\right\Vert  =\left\Vert \sum_{p\in L_{k}}c_{p-k}^{\ast}c_{p}\Psi\right\Vert \leq\sum_{p\in L_{k}}\left\Vert c_{p-k}^{\ast}c_{p}\Psi\right\Vert  \leq\sqrt{\sum_{p\in L_{k}}\lambda_{k,p}^{-1}}\sqrt{\sum_{p\in L_{k}}\lambda_{k,p}\left\Vert c_{p-k}^{\ast}c_{p}\Psi\right\Vert ^{2}}  
\end{align}
where $\lambda_{k,p}=\frac{1}{2}\left(\left|p\right|^{2}-\left|p-k\right|^{2}\right)$. Using \eqref{eq:ManifestlyNonNegativeHKin} and Pauli's exclusion principle $\|c_p\|_{\rm op}\le 1$, $\|c^*_p\|_{\rm op}\le 1$, we find that
\begin{align}
\sum_{p\in L_{k}}\lambda_{k,p}\left\Vert c_{p-k}^{\ast}c_{p}\Psi\right\Vert ^{2}  & =\frac{1}{2}\sum_{p\in L_{k}}\left(\vert\left|p\right|^{2}-\zeta\vert+\vert\left|p-k\right|^{2}-\zeta\vert\right)\left\Vert c_{p-k}^{\ast}c_{p}\Psi\right\Vert ^{2}\label{eq:BasicKineticArgument}\\
 & \leq\frac{1}{2}\sum_{p\in L_{k}}\vert\left|p\right|^{2}-\zeta\vert\left\Vert c_{p}\Psi\right\Vert ^{2}+\frac{1}{2}\sum_{p\in L_{k}}\vert\left|p-k\right|^{2}-\zeta\vert\left\Vert c_{p-k}^{\ast}\Psi\right\Vert ^{2}\nonumber \\
 & \leq \frac{1}{2}\sum_{p\in B_F^c}\vert\left|p\right|^{2}-\zeta\vert\left\Vert c_{p}\Psi\right\Vert ^{2}+\frac{1}{2}\sum_{p\in B_F}\vert\left|p\right|^{2}-\zeta\vert\left\Vert c_{p}^{\ast}\Psi\right\Vert ^{2}\nonumber =\frac{1}{2}\left\langle \Psi,H_{\kin}^{\prime}\Psi\right\rangle. \nonumber 
\end{align}
Thus it remains to show that $\sum_{p\in L_{k}}\lambda_{k,p}^{-1}\leq Ck_{F}$. For $|k|\sim O(1)$, this bound was already proved in  \cite{HaiPorRex-20,BNPSS-21}.  For completeness, we will establish this bound for all $k\in \mathbb{Z}^3_*$ in the Appendix (Proposition \ref{coro:CompletelyUniformRiemannSumBound}). Thus in summary, 
\begin{align}
\tilde{B}_{k}^{\ast}\tilde{B}_{k}\leq\frac{1}{2}\left(\sum_{p\in L_{k}}\lambda_{k,p}^{-1}\right)H_{\kin}^{\prime} \le Ck_{F}H_{\kin}^{\prime}.
\end{align} 
Then the bound for $\tilde{B}_{k}\tilde{B}_{k}^{\ast}$ follows from the fact that 
\begin{equation}
\left[\tilde{B}_{k},\tilde{B}_{k}^{\ast}\right]=\left|L_{k}\right|-\sum_{p\in L_{k}}c_{p}^{\ast}c_{p}-\sum_{p\in L_{k}}c_{p-k}c_{p-k}^{\ast} \le |L_k| \le C|k| k_F^2 \label{eq:BTildeAjointCommutator}
\end{equation}
In the last estimate we used  $\left|L_{k}\right|\leq Ck_{F}^{2}\left|k\right|$ for all $k\in\mathbb{Z}_{\ast}^{3}$ (see Proposition \ref{prop:NonSingularRiemannSums} for details).  
$\hfill\square$

For the non-bosonizable terms in $H_{\inter}^{\prime}$, it was proved in \cite[Eq. (5.1)]{BNPSS-20} that 
\begin{equation} \label{eq:DkDkAPriori}
D_{k}^{\ast}D_{k}\leq4\,\mathcal{N}_{E}^{2}.
\end{equation}
However, this bound is not optimal for low-lying eigenfunctions (for which $\mathcal{N}_{E}\sim k_F$). In order to remove the non-bosonizable terms completely, we need the following improvement. 

%In contrast, Proposition \ref{prop:DkDKFirstStep} is sharper thanks to the appearance of the kinetic operator $H_{\rm kin}'$.

%The following  new bound is important to remove the non-bosonizable terms. 
\begin{prop}
\label{prop:DkDKFirstStep} For all $k\in\mathbb{Z}_{\ast}^{3}$ and
any $0<\lambda\le \frac{1}{6}k_F^2$, the operator $D_k$ in \eqref{eq:Localizedrhok} satisfies 
\[
D_{k}^{\ast}D_{k}\leq C\left(\left|k\right|^{-1}\lambda +\left|k\right|^{3+\frac{2}{3}}(\log k_{F})^{\frac{2}{3}}k_{F}^{\frac{2}{3}}\right)\left(\lambda +\left|k\right|\right) \mathcal{N}_{E}+C\lambda^{-\frac{1}{2}}\mathcal{N}_{E}H_{\kin}^{\prime}
\]
for a constant $C>0$ independent of $k$, $k_{F}$ and $\lambda$. 
\end{prop}

In applications, we will eventually choose $\lambda=k_F^{2\gamma}/|k|^{4}$ for some constant $\gamma \in (0,1/9)$.

\textbf{Proof:} For $k\in\mathbb{Z}_{\ast}^{3}$, we write $D_{k}=D_{k}^{1}+D_{k}^{2}$ as in \eqref{eq:def-Bk}, namely
%\begin{align}
%D_{k}^{1}&=\text{d}\Gamma\left(P_{B_{F}}e^{-ik\cdot x}P_{B_{F}}\right)=\sum_{q\in B_{F}\cap\left(B_{F}+k\right)}c_{q-k}^{\ast}c_{q},\\
%D_{k}^{2}&=\text{d}\Gamma\left(P_{B_{F}^{c}}e^{-ik\cdot x}P_{B_{F}^{c}}\right)= \sum_{q\in B_{F}^c \cap\left(B_{F}^c+k\right)}c_{q-k}^{\ast}c_{q}. \nonumber
%\end{align}
\begin{align}
D_{k}^{1}=\sum_{q\in B_{F}\cap\left(B_{F}+k\right)}c_{q-k}^{\ast}c_{q},\quad D_{k}^{2}= \sum_{q\in B_{F}^c \cap\left(B_{F}^c+k\right)}c_{q-k}^{\ast}c_{q}. 
\end{align}
By the Cauchy--Schwarz inequality, 
\begin{align}
D_{k}^{\ast}D_{k}\leq2\left(\left(D_{k}^{1}\right)^{\ast}D_{k}^{1}+\left(D_{k}^{2}\right)^{\ast}D_{k}^{2}\right).
\end{align}
We will estimate $\left(D_{k}^{1}\right)^{\ast}D_{k}^{1}$ in detail, the estimate of $\left(D_{k}^{2}\right)^{\ast}D_{k}^{2}$
being similar. We have
%By the definition
%\begin{align}
%D_{k}^{1}  =\sum_{p,q\in B_{F}}\left\langle u_{p},e^{-ik\cdot x}u_{q}\right\rangle c_{p}^{\ast}c_{q}=\sum_{q\in B_{F}\cap\left(B_{F}+k\right)}c_{q-k}^{\ast}c_{q} ,\nonumber 
%\end{align}
%we can write  
\begin{align} \label{eq:Dk1astDk1Identity} 
\left(D_{k}^{1}\right)^{\ast}D_{k}^{1}  &=\sum_{p,q\in B_{F}\cap\left(B_{F}+k\right)} c_p^* c_{p-k} c_{q-k}^{\ast} c_{q} 
  =\sum_{p,q \in B_{F}\cap\left(B_{F}+k\right)} ( \delta_{p,q}c_{p-k}c_{p-k}^{\ast} - c_{p-k}c_{q}c_{p}^{\ast}c_{q-k}^{\ast} ) \nonumber \\
  &= \sum_{p \in B_{F}\cap\left(B_{F}+k\right)}  c_{p-k}c_{p-k}^{\ast} - \frac{1}{2}\sum_{p,q\in B_{F}\cap\left(B_{F}+k\right)} ( c_{p-k}c_{q}c_{p}^{\ast}c_{q-k}^{\ast} + h.c.). 
\end{align} 
Here we used $k\ne 0$ so that $c_{p-k}$ and $c^*_p$ anti-commute. 
By the definition of $\mathcal{N}_E$ in \eqref{eq:ParticleHoleSymmetry}, 
\begin{align}
\sum_{p\in B_{F}\cap\left(B_{F}+k\right)} c_{p-k}c_{p-k}^{\ast} \le \sum_{p\in B_{F}} c_{p}c_{p}^{\ast} = \mathcal{N}_E.
\end{align} 
Moreover, by the Cauchy--Schwarz inequality, for all $\epsilon_p>0$ we get 
\begin{align}
&\pm \frac{1}{2}\sum_{p,q \in B_{F}\cap\left(B_{F}+k\right)} ( c_{p-k}c_{q}c_{p}^{\ast}c_{q-k}^{\ast} + h.c. ) \nonumber\\
  & \le \frac{1}{2} \sum_{p,q \in B_{F}\cap\left(B_{F}+k\right)} ( \epsilon_p c_{p-k}c_{q} c_q^* c_{p-k}^* + \epsilon_p^{-1} c_{q-k} c_p c_{p}^{\ast}c_{q-k}^{\ast})  \nonumber\\
& \le  \frac{1}{2} \sum_{p,q \in B_{F}\cap\left(B_{F}+k\right)} ( \epsilon_p c_{p-k}c_{p-k}^* c_{q} c_q^*  + \epsilon_p^{-1} c_{q-k} c_{q-k}^{\ast} c_p c_{p}^{\ast})    \\
&\le \frac{1}{2} \sum_{p \in B_{F}\cap\left(B_{F}+k\right)} ( \epsilon_p c_{p-k}c_{p-k}^*  + \epsilon_p^{-1}c_p c_{p}^{\ast} ) \mathcal{N}_E  \nonumber
\end{align}
By taking $\epsilon_p\equiv 1$, we obtain immediately $\left(D_{k}^{1}\right)^{\ast}D_{k}^{1} \le \mathcal{N}_E^2$ which, together with a similar bound for $D_{k}^{2}$,  leads to \eqref{eq:DkDkAPriori}. To improve on this, we have to choose $\epsilon_p$ differently. 

Recall that in \eqref{eq:ManifestlyNonNegativeHKin}  we can choose $\zeta\in[\sup_{p\in B_{F}}\left|p\right|^{2},\inf_{p\in B_{F}^{c}}\left|p\right|^{2}]$ such that $\vert\left|p\right|^{2}-\zeta\vert\,\geq 1/2$ for all $p\in \mathbb{Z}^3$. For any $\lambda>0$ we can split  
\begin{align}
B_{F}\cap\left(B_{F}+k\right)=S_{k,\lambda}^{1}\cup S_{k,\geq\lambda}^{1}
\end{align}
where 
\begin{align} \label{eq:def-Sk1}
S_{k,\lambda}^{1}  &=\left\{ p\in B_{F}\cap\left(B_{F}+k\right)\mid \max\left\{ \vert\left|p\right|^{2}-\zeta\vert,\vert\left|p-k\right|^{2}-\zeta\vert\right\} <\lambda\right\} ,\\
 S_{k,\geq\lambda}^{1} &=\left\{ p\in B_{F}\cap\left(B_{F}+k\right)\mid\max\left\{ \vert\left|p\right|^{2}-\zeta\vert,\vert\left|p-k\right|^{2}-\zeta\vert\right\} \geq\lambda\right\} \nonumber
\end{align}
Choosing  $\epsilon_p=1$ for $p\in S_{k,\lambda}^{1}$ and using $\|c_p^*\|_{\rm op} \le 1$ we get 
\begin{equation}
\frac 1 2\sum_{p \in S_{k,\lambda}^{1}} ( \epsilon_p c_{p-k}c_{p-k}^*  + \epsilon_p^{-1}c_p c_{p}^{\ast} ) \le |S_{k,\lambda}^{1}|. 
\end{equation}
Choosing $\epsilon_p =   \sqrt{ | |p-k|^{2}-\zeta|}  / \sqrt{ | |p|^2 -\zeta|} $ for $p\in S_{k,\geq\lambda}^{1}$ we have
\begin{align}
& \sum_{p \in S_{k,\geq\lambda}^{1}} ( \epsilon_p c_{p-k}c_{p-k}^*  + \epsilon_p^{-1}c_p c_{p}^{\ast} ) \nonumber\\
&= \sum_{p \in S_{k,\geq\lambda}^{1}} \frac{1}{ \sqrt{| |p|^2 -\zeta| \cdot | |p-k|^{2}-\zeta|  }}(  | |p-k|^{2}-\zeta| c_{p-k}c_{p-k}^*  +  | |p|^2 -\zeta| c_p c_{p}^{\ast} ) \\
&\le \sum_{p \in S_{k,\geq\lambda}^{1}} \frac{1}{  \sqrt{\lambda/2}  }(  | |p-k|^{2}-\zeta| c_{p-k}c_{p-k}^*  +  | |p|^2 -\zeta| c_p c_{p}^{\ast} ) \le \frac{2\sqrt{2}}{\sqrt{\lambda}}  H_{\rm kin}'. \nonumber
\end{align}
Here we used that among two factors $| |p|^2 -\zeta|$ and $| |p-k|^{2}-\zeta| $ there is at least one $\ge \lambda$ due to the assumption $p\in S_{k,\geq\lambda}^{1}$, and the other one is trivially $\ge 1/2$. In summary, 
\begin{equation}
 \left(D_{k}^{1}\right)^{\ast}D_{k}^{1} \leq\left|S_{k,\lambda}^{1}\right|  \mathcal{N}_{E}  +C\lambda^{-\frac{1}{2}} H_{\kin}^{\prime}\mathcal{N}_{E}. 
\end{equation}
Similarly, we have
\begin{equation}
 \left(D_{k}^{2}\right)^{\ast}D_{k}^{2} \leq\left|S_{k,\lambda}^{2}\right|  \mathcal{N}_{E}  +C\lambda^{-\frac{1}{2}} H_{\kin}^{\prime}\mathcal{N}_{E}\end{equation}
 where
 \begin{equation}\label{eq:def-Sk2}
 S_{k,\lambda}^{2}  =\left\{ p\in B_{F}^{c}\cap\left(B_{F}^{c}+k\right)\mid\max\left\{ \vert\left|p\right|^{2}-\zeta\vert,\vert\left|p-k\right|^{2}-\zeta\vert\right\} <\lambda\right\} .
\end{equation}
The desired conclusion of $D_k^* D_k$ follows from the bound 
 \begin{equation}\label{eq:EstimationofSklambda}
\left|S_{k,\lambda}^{1}\right|+\left|S_{k,\lambda}^{2}\right|\leq C\left(\left|k\right|^{-1}\lambda +\left|k\right|^{3+\frac{2}{3}}(\log k_{F})^{\frac{2}{3}}k_{F}^{\frac{2}{3}}\right)\left(\lambda +\left|k\right|\right) 
\end{equation}
whose proof can be found in Proposition \ref{prop:SklambdaPointsEstimate} in the Appendix. $\hfill\square$

%%%%%%%%%%%%%%%%%%%%%%%%%%%%%%%%%%%%
%%%%%%%%%%%%%%%%%%%%%%%%%%%%%%%%%%%%

\subsection{Estimation of the Non-Bosonizable Terms}  \label{sec:EstimationoftheNon-BosonizableTerms}

Now we are ready to remove the non-bosonizable terms, namely the terms involving operators $D_k$ in the decomposition \eqref{eq:localizedInteractionBandDForm} of the  interaction operator: 
\begin{align} 
k_{F}^{-1} H_{\text{int}}^{\prime}  = \sum_{k\in\mathbb{Z}_{+}^{3}}\left(H_{\text{int}}^{k}-\frac{\hat{V}_{k}k_{F}^{-1}}{\left(2\pi\right)^{3}}\left|L_{k}\right|\right)
 +\frac{k_{F}^{-1}}{\left(2\pi\right)^{3}}\sum_{k\in\mathbb{Z}_{\ast}^{3}}\hat{V}_{k}\left(\tilde{B}_{k}^{\ast}D_{k}+D_{k}^{\ast}\tilde{B}_{k}+\frac{1}{2}D_{k}^{\ast}D_{k}\right) 
\end{align}
where   $H_{\text{int}}^{k}$ is defined in \eqref{eq:HintkBBForm}.
% that 
%\begin{align}
%H_{\text{int}}^{k}  =\frac{\hat{V}_{k}k_{F}^{-1}}{2\left(2\pi\right)^{3}} \left( \{ \tilde{B}_{k}^{\ast}, \tilde{B}_{k}\} + \{\tilde{B}_{-k}^{\ast}, \tilde{B}_{-k}\} + 2 \tilde{B}_{k}^{\ast} \tilde{B}_{-k}^{\ast}  +2   \tilde{B}_{-k} \tilde{B}_{k}  \right). 
%\end{align}
Moreover, for technical reasons, we will also impose a momentum cut-off in the bosonizable terms. 
%Let us introduce the set  
%\begin{equation} \label{eq:def-SC}
%S_{C}=\overline{B}\left(0,k_{F}^{\gamma}\right)\cap\mathbb{Z}_{+}^{3},
%\end{equation}
%where $\gamma \in (0,1]$ is an exponent which is to be optimized over at the end.  
Recall the set $S_C$ in \eqref{eq:SC}. 
%\begin{align}
%delete
%\end{align}
Define 
\begin{equation} \label{eq:ENB}
\mathcal{E}_{\rm NB} = k_{F}^{-1} H_{\text{int}}^{\prime} - \sum_{k\in S_C}  \left(H_{{\rm int}}^{k}-\frac{\hat{V}_{k}k_{F}^{-1}}{\left(2\pi\right)^{3}}\left|L_{k}\right|\right) .
\end{equation}

%We will prove

\begin{prop}
\label{them:ReductiontoBosonizableTerms}Let $\sum_{k\in\mathbb{Z}^{3}}\hat{V}_{k}\left|k\right|<\infty$. Then for all $\gamma \in (0,1/9)$ in $S_C$ we have
\begin{align*}
\pm \mathcal{E}_{\rm NB}   
\le C k_F^{-\gamma/2} \Big(H_{\rm kin}' + k_F^{-1} \mathcal{N}_E  H_{\rm kin}'  + k_F\Big).  
\end{align*}
Here the constant $C>0$ depends only on $V$ (in particular, it is independent of $k,k_F$ and $\lambda$). 
\end{prop}

We write $\pm X\le Y$ for two operator inequalities $X\le Y$ and $-X\le Y$.

{\bf Proof:} For the bosonizable terms, by \eqref{eq:BTildeAjointCommutator}, Proposition \ref{prop:BkBkAPriori}  and Proposition \ref{prop:Zeta0RepresentationofHKin}, we  can bound 
\begin{equation} \label{eq:2.24-pre}
\pm \Big( \{\tilde{B}_{k}^{\ast},\tilde{B}_{k}\} - |L_k| \Big) = \pm \Big(  2 \tilde{B}_{k}^{\ast}\tilde{B}_{k} - \sum_{p\in L_{k}}c_{p}^{\ast}c_{p}-\sum_{p\in L_{k}}c_{p-k}c_{p-k}^{\ast} \Big) \le 2 \tilde{B}_{k}^{\ast}\tilde{B}_{k} + \mathcal{N}_E\le C k_F H_{\rm kin}'
\end{equation}
for all $k\in \mathbb{Z}^3_*$.  Moreover, by the Cauchy--Schwarz inequality,  
\begin{equation}\label{eq:2.25-pre}
\pm \Big(\tilde{B}_{k}^* \tilde{B}_{-k}^* + \tilde{B}_{-k}\tilde{B}_{k}  \Big) \le |k|^{-1/2} \tilde{B}_{-k} \tilde{B}_{-k}^* + |k|^{1/2} \tilde{B}_{k}^* \tilde{B}_{k} \le  C |k|^{1/2} k_F  (H_{\rm kin}' + k_F) 
\end{equation}
for all $k\in \mathbb{Z}^3_*$. Combining \eqref{eq:2.24-pre} and \eqref{eq:2.25-pre} we find that
\begin{align} \label{eq:bos-cut-off}
\pm \sum_{k \in \mathbb{Z}_{+}^{3} \backslash S_C}  \left(H_{{\rm int}}^{k}-\frac{\hat{V}_{k}k_{F}^{-1}}{\left(2\pi\right)^{3}}\left|L_{k}\right|\right) &\le C  (H_{\rm kin}' + k_F) \sum_{k \in \mathbb{Z}_{+}^{3} \backslash S_C} \hat V_k |k|^{1/2}\nonumber\\
& \le C  (H_{\rm kin}' + k_F)   k_F^{-\gamma/2} \sum_{k \in \mathbb{Z}^{3}}  \hat V_k |k|.
\end{align}
For the non-bosonizable terms, by the Cauchy--Schwarz inequality and Proposition \ref{prop:BkBkAPriori}   we have
\begin{equation}
\pm  \left(  \tilde{B}_{k}^{\ast}D_{k}+D_{k}^{\ast}\tilde{B}_{k} \right) \le k_F^{-\gamma/2} |k| \tilde{B}_{k}^{\ast} \tilde{B}_{k} + k_F^{\gamma/2} |k|^{-1} D_{k}^{\ast} D_{k} \le C k_F^{1-\gamma/2} |k|  H_{\rm kin}' + k_F^{\gamma/2} |k|^{-1} D_{k}^{\ast} D_{k} 
\end{equation}
and hence
\begin{align} \label{eq:non-bos-cut-off-1}
\pm  \sum_{k\in\mathbb{Z}_{\ast}^{3}}\hat{V}_{k} \left(  \tilde{B}_{k}^{\ast}D_{k}+D_{k}^{\ast}\tilde{B}_{k} + D_{k}^{\ast} D_{k}  \right) \le C  k_F^{1-\gamma/2}  H_{\rm kin}' +  \sum_{k\in\mathbb{Z}_{\ast}^{3}}\hat{V}_{k} ( k_F^{\gamma/2}   |k|^{-1}  +1 ) D_{k}^{\ast} D_{k}.  
\end{align}
Let us decompose the sum on the right-hand side of \eqref{eq:non-bos-cut-off-1} into the high-momenta $|k|> k_F^{\gamma/2}$ and the low-momenta $|k|\le k_F^{\gamma/2}$. For the high-momenta, from the simple bound  \eqref{eq:DkDkAPriori} we get
\begin{align} \label{eq:non-bos-cut-off-2}
 \sum_{k\in\mathbb{Z}_{\ast}^{3}, |k|> k_F^{\gamma/2}}\hat{V}_{k} ( k_F^{\gamma/2}   |k|^{-1}  +1 ) D_{k}^{\ast} D_{k} \le C k_F^{-\gamma/2} \mathcal{N}_E^2  \sum_{k\in\mathbb{Z}^{3}}\hat{V}_{k} |k|.  
\end{align}
For the low-momenta, using Proposition \ref{prop:DkDKFirstStep} with $\lambda=k_F^{\gamma}/|k|^{2}$ we have
%\begin{align}
%D_{k}^{\ast}D_{k} &\leq C\left(\left|k\right|^{-1}\lambda +\left|k\right|^{3+\frac{2}{3}}(\log k_{F})^{\frac{2}{3}}k_{F}^{\frac{2}{3}}\right)\left(\lambda +\left|k\right|\right) \mathcal{N}_{E}+C\lambda^{-\frac{1}{2}}\mathcal{N}_{E}H_{\kin}^{\prime} \\
%&\le  C\left( k_F^{2\gamma} + k_F^\gamma |k|^{2/3} (\log k_F)^{\frac 2 3} k_F^{2/3} + |k|^{3+2/3} (\log k_F)^{\frac 2 3} k_F^{2/3}\right) |k|\mathcal{N}_E +Ck_F^{-\gamma/2}  |k| \mathcal{N}_{E}H_{\kin}^{\prime}, \nonumber
%\end{align}
\begin{align}
D_{k}^{\ast}D_{k} \le  C\left( k_F^{2\gamma} + k_F^\gamma |k|^{2/3} (\log k_F)^{\frac 2 3} k_F^{2/3} + |k|^{3+2/3} (\log k_F)^{\frac 2 3} k_F^{2/3}\right) |k|\mathcal{N}_E +Ck_F^{-\gamma/2}  |k| \mathcal{N}_{E}H_{\kin}^{\prime}, 
\end{align}
and hence
\begin{align} \label{eq:non-bos-cut-off-3}
 \sum_{k\in\mathbb{Z}_{\ast}^{3}, |k|\le k_F^{\gamma/2}}\hat{V}_{k} D_{k}^{\ast}D_{k} &\le  C \left( \sum_{k\in\mathbb{Z}^{3}}\hat{V}_{k} |k| \right)   \left( k_F^{(3+\frac 2 3) \frac \gamma 2+ \frac 2 3} (\log k_F)^{\frac 2 3} \mathcal{N}_E + k_F^{-\frac \gamma 2} \mathcal{N}_{E}H_{\kin}^{\prime}\right) \nonumber\\
 &\le C k_F^{-\gamma/2} \left( k_F \mathcal{N}_E +\mathcal{N}_{E}H_{\kin}^{\prime}\right)
 \end{align}
 for all $\gamma\in (0,1/7)$. Moreover, using Proposition \ref{prop:DkDKFirstStep} with $\lambda=k_F^{2\gamma}/|k|^{4}$ we have
%\begin{align}
%& k_F^{\gamma/2}   |k|^{-1} D_{k}^{\ast}D_{k} \leq Ck_F^{\gamma/2}   |k|^{-1} \left[ \left(\left|k\right|^{-1}\lambda +\left|k\right|^{3+\frac{2}{3}}(\log k_{F})^{\frac{2}{3}}k_{F}^{\frac{2}{3}}\right)\left(\lambda +\left|k\right|\right) \mathcal{N}_{E}+ \lambda^{-\frac{1}{2}}\mathcal{N}_{E}H_{\kin}^{\prime} \right] \nonumber \\
%&\le  C\left( k_F^{4\gamma} + k_F^{(2+\frac 1 2)\gamma} (\log k_F)^{\frac 2 3} k_F^{2/3} + k_F^{\frac \gamma 2}|k|^{2+2/3} (\log k_F)^{\frac 2 3} k_F^{2/3}\right) |k|\mathcal{N}_E +Ck_F^{-\gamma/2}  |k| \mathcal{N}_{E}H_{\kin}^{\prime}, 
%\end{align}
\begin{align}
k_F^{\gamma/2}   |k|^{-1} D_{k}^{\ast}D_{k} &\leq  C\left( k_F^{4\gamma} + k_F^{(2+\frac 1 2)\gamma} (\log k_F)^{\frac 2 3} k_F^{2/3} + k_F^{\frac \gamma 2}|k|^{2+2/3} (\log k_F)^{\frac 2 3} k_F^{2/3}\right) |k|\mathcal{N}_E \nonumber\\
&\qquad +Ck_F^{-\gamma/2}  |k| \mathcal{N}_{E}H_{\kin}^{\prime}, 
\end{align}
and hence 
\begin{align} \label{eq:non-bos-cut-off-4}
 \sum_{k\in\mathbb{Z}_{\ast}^{3}, |k|\le k_F^{\gamma/2}}\hat{V}_{k} k_F^{\gamma/2}   |k|^{-1}  D_{k}^{\ast}D_{k} &\le  C \left( \sum_{k\in\mathbb{Z}^{3}}\hat{V}_{k} |k| \right)   \left( k_F^{(2+\frac 1 2)  \gamma + \frac 2 3} (\log k_F)^{\frac 2 3} \mathcal{N}_E + k_F^{-\frac \gamma 2} \mathcal{N}_{E}H_{\kin}^{\prime}\right) \nonumber\\
 &\le C k_F^{-\gamma/2} \left( k_F \mathcal{N}_E +\mathcal{N}_{E}H_{\kin}^{\prime}\right)
 \end{align}
  for all $\gamma\in (0,1/9)$. Inserting \eqref{eq:non-bos-cut-off-2}, \eqref{eq:non-bos-cut-off-3} and \eqref{eq:non-bos-cut-off-4} in \eqref{eq:non-bos-cut-off-1} and using Proposition \ref{prop:Zeta0RepresentationofHKin}, we conclude that 
  \begin{align} \label{eq:non-bos-cut-off-final}
\pm \frac{k_{F}^{-1}}{\left(2\pi\right)^{3}}  \sum_{k\in\mathbb{Z}_{\ast}^{3}}\hat{V}_{k} \left(  \tilde{B}_{k}^{\ast}D_{k}+D_{k}^{\ast}\tilde{B}_{k} + D_{k}^{\ast} D_{k}  \right) \le C  k_F^{-\gamma/2}  \Big( H_{\rm kin}' +  k_F^{-1}\mathcal{N}_{E}H_{\kin}^{\prime} \Big)
\end{align}
  for all $\gamma\in (0,1/9)$.   The conclusion follows from \eqref{eq:bos-cut-off} and \eqref{eq:non-bos-cut-off-final}. 
 $\hfill\square$

%%%%%%%%%%%%%%%%%%%%%%%%%%%%%%%%%%%%%%%%%%%%
%%%%%%%%%%%%%%%%%%%%%%%%%%%%%%%%%%%%%%%%%%%%

%%%%%%%%%%%%%%%%%%%%%%%%%%%%%%%%%%%%
%%%%%%%%%%%%%%%%%%%%%%%%%%%%%%%%%%%%

\section{Overview of Bosonic Bogolubov Transformations}\label{sec:BosonicBogolubovTransformations}

In this section we review the general theory of quadratic
Hamiltonians and Bogolubov transformations in the {\em exact} bosonic setting. Later, in the remainder of the paper, the  analysis here will be adapted to handle the {\em quasi-bosonic} case where error terms have to be estimated carefully.  

\medskip

The study of bosonic quadratic Hamiltonians goes back to Bogolubov\textquoteright s
1947 paper \cite{Bogolubov-47} where he proposed an effective Hamiltonian
to describe the excitation spectrum of weakly interacting Bose gases.
An important property of quadratic Hamiltonians is that they can be
diagonalized by suitable Bogolubov transformations, see e.g. \cite{BacBru-16,NamNapSol-16,Derezinski-17}
for recent results in the infinite dimensional cases. For our application we will only focus on the situation where the
one-body Hilbert space is  {\em real} and {\em finite dimensional}. Historically, the diagonalization problem in finite dimensions can be solved abstractly by using  Williamson\textquoteright s theorem \cite{Williamson-36}. We refer to \cite{Hor-95} and \cite[Section 2]{Derezinski-17} for systematic discussions on the finite dimensional case. 

\medskip
In the present paper, we will need an explicit construction of the diagonalizing transformations so that we can adapt this
to the quasi-bosonic operators. Such an explicit construction can
be found in \cite{GreSei-13}, which was also used in the fermionic
context in \cite{BNPSS-20,BNPSS-21} and will be recalled below. Here we will offer a slightly different treatment of Bogolubov transformations,
in that we will view \textit{quadratic operators} on Fock spaces as
the fundamental object of study rather than the creation and annihilation
operators. % themselves. 

\medskip
{\bf Notation.} We will denote
by $V$ a finite-dimensional real Hilbert space and let
$n=\dim\left(V\right)$. The  bosonic Fock space associated to $V$  is 
\begin{equation}
\mathcal{F}^{+}\left(V\right)=\bigoplus_{N=0}^{\infty}\bigotimes_{\text{Sym}}^{N}V
\end{equation}
where $\bigotimes_{\text{Sym}}^{N}V$ denotes the space of symmetric
$N$-fold tensor products of $V$. To any element $\varphi\in V$
there are associated two operators on $\mathcal{F}^{+}\left(V\right)$:
The annihilation operator $a\left(\varphi\right)$ and the creation
operator $a^{\ast}\left(\varphi\right)$. These are (formal) adjoints
of one another and obey the canonical commutation relations (CCR):
For any $\varphi,\psi\in V$
\begin{align}
\left[a\left(\varphi\right),a\left(\psi\right)\right]  =\left[a^{\ast}\left(\varphi\right),a^{\ast}\left(\psi\right)\right]=0, \quad \left[a\left(\varphi\right),a^{\ast}\left(\psi\right)\right]  =\left\langle \varphi,\psi\right\rangle. \label{eq:CanonicalCommutationRelations}
\end{align}
Additionally, the mappings $\varphi\mapsto a\left(\varphi\right)$,
$\varphi\mapsto a^{\ast}\left(\varphi\right)$ are linear\footnote{If $V$ is a complex Hilbert space space, the mapping $\varphi\mapsto a(\varphi)$ is anti-linear which complicates the exposition. In our quasi-bosonic application, although the relevant Hilbert spaces are complex, all relevant operators have real matrix elements and hence it suffices to restrict to the case of real spaces as in this section.}. 

\subsection{Quadratic Hamiltonians}

Similarly to how we can to any $\varphi\in V$ associate the two operators
$a\left(\varphi\right)$ and $a^{\ast}\left(\varphi\right)$ we may
also associate two types of symmetric operators on $\mathcal{F}^{+}\left(V\right)$
to any symmetric operator on $V$. For the definition we let $\left(e_{i}\right)_{i=1}^{n}$
denote an orthonormal basis of $V$. Given any symmetric operator
$A:V\rightarrow V$ we then define the operator $Q_{1}\left(A\right)$
on $\mathcal{F}^{+}\left(V\right)$ by
\begin{equation}
Q_{1}\left(A\right)=\sum_{i,j=1}^{n}\left\langle e_{i},Ae_{j}\right\rangle \left(a^{\ast}\left(e_{i}\right)a\left(e_{j}\right)+a\left(e_{j}\right)a^{\ast}\left(e_{i}\right)\right)\label{eq:Q1AFullForm}
\end{equation}
and, likewise, for any symmetric operator $B:V\rightarrow V$ we define
the operator $Q_{2}\left(B\right)$ by
\begin{equation}
Q_{2}\left(B\right)=\sum_{i,j=1}^{n}\left\langle e_{i},Be_{j}\right\rangle \left(a^{\ast}\left(e_{i}\right)a^{\ast}\left(e_{j}\right)+a\left(e_{j}\right)a\left(e_{i}\right)\right).\label{eq:Q2BFullForm}
\end{equation}
These definitions are independent of the basis chosen 
and we can write equivalently 
\begin{align}
Q_{1}\left(A\right) & =\sum_{i=1}^{n}\left(a^{\ast}\left(Ae_{i}\right)a\left(e_{i}\right)+a\left(e_{i}\right)a^{\ast}\left(Ae_{i}\right)\right)\label{eq:Q1Q2GoodDefinition}\\
Q_{2}\left(B\right) & =\sum_{i=1}^{n}\left(a^{\ast}\left(Be_{i}\right)a^{\ast}\left(e_{i}\right)+a\left(e_{i}\right)a\left(Be_{i}\right)\right).\nonumber 
\end{align}
Thus for real, symmetric $A,B:V\rightarrow V$ we can define a quadratic Hamiltonian on $\mathcal{F}^{+}\left(V\right)$ by 
%to be an operator of the form
\begin{equation}
H=Q_{1}\left(A\right)+Q_{2}\left(B\right).  %=\sum_{i=1}^{n}\left(a^{\ast}\left(Ae_{i}\right)a\left(e_{i}\right)+a\left(e_{i}\right)a^{\ast}\left(Ae_{i}\right)\right)+\sum_{i=1}^{n}\left(a^{\ast}\left(Be_{i}\right)a^{\ast}\left(e_{i}\right)+a\left(e_{i}\right)a\left(Be_{i}\right)\right)
\end{equation}

Note that by the CCR, we may express $Q_{1}\left(A\right)$ as
\begin{align}
Q_{1}\left(A\right)   =2\sum_{i,j=1}^{n}\left\langle e_{i},Ae_{j}\right\rangle a^{\ast}\left(e_{i}\right)a\left(e_{j}\right)+\text{tr}\left(A\right)=2\,\text{d}\Gamma\left(A\right)+\text{tr}\left(A\right) 
\end{align}
where $\text{d}\Gamma\left(A\right)$ denotes the second quantization
of $A:V\rightarrow V$. Sometimes in the literature, in particular in infinite dimensions, quadratic Hamiltonians are defined by $\text{d}\Gamma\left(A\right)+Q_{2}\left(B\right)$, which is the same to our definition up to the constant $\text{tr}\left(A\right)$. 
Here we prefer to use $Q_1(A)$ instead of  $\text{d}\Gamma\left(\cdot\right)$; the reason for this is that the relations of Proposition \ref{prop:BogolubovQKCommutators} below are symmetric in the $Q$'s.  % since it is more compatible to Bogolubov transformations.  

Note that the basis-independence is a nice property of the real space setting. In general, if $V$ is a complex Hilbert space and $B$ is symmetric, then the definition of $Q_{2}\left(B\right)$ in \eqref{eq:Q2BFullForm} may depend on the basis. In fact, we can obtain a basis-independent formulation in the complex case, but the mapping $B\mapsto Q_{2}\left(B\right)$ is not to be defined for
symmetric linear operators $B$, but rather symmetric \textit{anti-linear}
operators $B$ to make up for the fact that in the complex case the
assignment $\varphi\mapsto a\left(\varphi\right)$ is also anti-linear. 
This is unimportant for our application, which is why we only consider
real Hilbert spaces in this section, for the sake of simplicity.

\subsection{Bogolubov Transformations}

In this subsection, we review an explicit construction of a \textit{Bogolubov transformation} \, $\mathcal{U}:\mathcal{F}^{+}\left(V\right)\rightarrow\mathcal{F}^{+}\left(V\right)$ that diagonalizes the quadratic Hamiltonian $H= Q_1(A)+Q_2(B)$, namely
%\begin{equation}
%\mathcal{U}H\mathcal{U}^{\ast}=\mathcal{U}\left(Q_{1}\left(A\right)+Q_{2}\left(B\right)\right)\mathcal{U}^{\ast}=Q_{1}\left(E\right)
%\end{equation}
\begin{equation}
\mathcal{U}H\mathcal{U}^{\ast}=Q_{1}\left(E\right)
\end{equation}
for a real, symmetric operator $E:V\rightarrow V$. Such a construction is well-known, see e.g. \cite{Derezinski-17} for a recent review. %More precisely,
We consider a unitary  transformation $\mathcal{U}=e^{\mathcal{K}}$ where $\mathcal{K}$ is an anti-symmetric operator on $\mathcal{F}^{+}\left(V\right)$ of the following form: 
\begin{equation}
\mathcal{K}=\frac{1}{2}\sum_{i,j=1}^{n}\left\langle e_{i},Ke_{j}\right\rangle \left(a\left(e_{i}\right)a\left(e_{j}\right)-a^{\ast}\left(e_{j}\right)a^{\ast}\left(e_{i}\right)\right)=\frac{1}{2}\sum_{i=1}^{n}\left(a\left(Ke_{i}\right)a\left(e_{i}\right)-a^{\ast}\left(e_{i}\right)a^{\ast}\left(Ke_{i}\right)\right).\label{eq:BogolubovGeneratorDefinition}
\end{equation}
Here $K:V\rightarrow V$ is a  symmetric operator (called the \textit{transformation kernel}) and $\left(e_{i}\right)_{i=1}^{n}$ denotes any orthonormal basis
of $V$ (as with $Q_{1}\left(\cdot\right)$ and $Q_{2}\left(\cdot\right)$
this definition is independent of the basis). 

\medskip

In this subsection, we will discuss 
\begin{thm}\label{thm:Bog-diag} Let $A,B:V\rightarrow V$ be real, symmetric
operators such that $A\pm B>0$ (namely $A+B>0$ and $A-B>0$). Consider the Bogolubov transformation  $e^{\mathcal{K}}$ where $\mathcal{K}$ is given in \eqref{eq:BogolubovGeneratorDefinition} with
\[
K=-\frac{1}{2}\log\left(\left(A-B\right)^{-\frac{1}{2}}\left(\left(A-B\right)^{\frac{1}{2}}\left(A+B\right)\left(A-B\right)^{\frac{1}{2}}\right)^{\frac{1}{2}}\left(A-B\right)^{-\frac{1}{2}}\right).
\]
Then 
$$
e^{\mathcal{K}} (Q_1(A)+Q_2(B))e^{-\mathcal{K}}=Q_{1}\left(E\right)=2\,\mathrm{d}\Gamma\left(E\right)+{\rm tr}\left(E\right)
$$
where
$$
E=e^{K}\left(A+B\right)e^{K}=e^{-K}\left(A-B\right)e^{-K}.
$$
Moreover, the diagonalizing $K$ is uniquely determined by this. 
\end{thm}

In the following we will prove Theorem \ref{thm:Bog-diag} by using a generalization and simplification of the argument used in \cite{GreSei-13,BNPSS-20}. We will first discuss the action of the Bogolubov transformation with a general kernel $K$, and then explain where the diagonalization condition comes from.

\medskip

Let us start with some basic properties of $\mathcal{K}$. %  has the following properties which follow from a simple computation. 
\begin{prop}
\label{prop:BogolubovCAKCommutators}For any symmetric operator $K:V\rightarrow V$, 
the operator $\mathcal{K}$ defined by  (\ref{eq:BogolubovGeneratorDefinition})
is an anti-symmetric operator on $\mathcal{F}^{+}\left(V\right)$ and obeys the commutators:  
\begin{align*}
\left[\mathcal{K},a\left(\varphi\right)\right]  =a^{\ast}\left(K\varphi\right), \quad \left[\mathcal{K},a^{\ast}\left(\varphi\right)\right] =a\left(K\varphi\right), \quad \forall \varphi\in V. 
\end{align*}
\end{prop}

Thus $\left[\mathcal{K},\cdot\right]$  acts on the creation and annihilation
operators by ``swapping'' each type into the other and applying
the operator $K$ to their arguments. From this one can now deduce
that the unitary transformation $e^{\mathcal{K}}$ acts on the creation
and annihilation operators according to
\begin{align}
e^{\mathcal{K}}a\left(\varphi\right)e^{-\mathcal{K}} & =a\left(\cosh\left(K\right)\varphi\right)+a^{\ast}\left(\sinh\left(K\right)\varphi\right)\label{eq:BogolubovCAAction}\\
e^{\mathcal{K}}a^{\ast}\left(\varphi\right)e^{-\mathcal{K}} & =a^{\ast}\left(\cosh\left(K\right)\varphi\right)+a\left(\sinh\left(K\right)\varphi\right),\nonumber 
\end{align}
since by the Baker-Campbell-Hausdorff formula
\begin{align}
e^{\mathcal{K}}a\left(\varphi\right)e^{-\mathcal{K}} & =a\left(\varphi\right)+\frac{1}{1!}\left[\mathcal{K},a\left(\varphi\right)\right]+\frac{1}{2!}\left[\mathcal{K},\left[\mathcal{K},a\left(\varphi\right)\right]\right]+\frac{1}{3!}\left[\mathcal{K},\left[\mathcal{K},\left[\mathcal{K},a\left(\varphi\right)\right]\right]\right]+\cdots\nonumber \\
 & =a\left(\varphi\right)+\frac{1}{1!}a^{\ast}\left(K\varphi\right)+\frac{1}{2!}a\left(K^{2}\varphi\right)+\frac{1}{3!}a^{\ast}\left(K^{3}\varphi\right)+\cdots\\
 & =a\left(\varphi+\frac{1}{2!}K^{2}\varphi+\cdots\right)+a^{\ast}\left(\frac{1}{1!}K\varphi+\frac{1}{3!}K^{3}\varphi+\cdots\right)\nonumber \\
 & =a\left(\cosh\left(K\right)\varphi\right)+a^{\ast}\left(\sinh\left(K\right)\varphi\right),\nonumber 
\end{align}
and the identity for $e^{\mathcal{K}}a^{\ast}\left(\varphi\right)e^{-\mathcal{K}}$
then follows immediately by taking the adjoint.

Now let us consider $e^{\mathcal{K}}Q_{1}\left(\cdot\right)e^{-\mathcal{K}}$
and $e^{\mathcal{K}}Q_{2}\left(\cdot\right)e^{-\mathcal{K}}$. For this we will first make an observation on their structure
which will greatly simplify computations: Namely, we note that the
operators $Q_{1}\left(A\right)$ and $Q_{2}\left(B\right)$ 
%\begin{align}
%Q_{1}\left(A\right) & =\sum_{i=1}^{n}\left(a^{\ast}\left(Ae_{i}\right)a\left(e_{i}\right)+a\left(e_{i}\right)a^{\ast}\left(Ae_{i}\right)\right)\\
%Q_{2}\left(B\right) & =\sum_{i=1}^{n}\left(a^{\ast}\left(Be_{i}\right)a^{\ast}\left(e_{i}\right)+a\left(e_{i}\right)a\left(Be_{i}\right)\right)\nonumber 
%\end{align}
are both of a ``trace-form'' in the sense that we can write, say,
$Q_{1}\left(A\right)=\sum_{i=1}^{n}q\left(e_{i},Ae_{i}\right)$ where
\begin{equation}
q\left(x,y\right)=a^{\ast}\left(y\right)a\left(x\right)+a\left(x\right)a^{\ast}\left(y\right)
\end{equation}
defines a bilinear mapping from $V\times V$ into the space of operators
on $\mathcal{F}^{+}\left(V\right)$, similar to how the trace of an
operator $T$ is $\text{tr}\left(T\right)=\sum_{i=1}^{n}q\left(e_{i},Te_{i}\right)$
for $q\left(x,y\right)=\left\langle x,y\right\rangle $. This abstract
viewpoint is worth noting because all such expressions are both basis-independent
and obey an additional property, which for the trace is just the familiar
cyclicity property. Since we will encounter such ``trace-form''
expressions repeatedly during computations throughout this paper we
state this property in full generality. In the following we take sesquilinear
to mean anti-linear in the first argument and linear in the second (we note that in the present real case a sesquilinear mapping is of
course just a bilinear mapping, but stating it in this generality
will prove useful later). 
\begin{lem}
\label{lemma:TraceFormLemma}Let $\left(V,\left\langle \cdot,\cdot\right\rangle \right)$
be an $n$-dimensional Hilbert space and let $q:V\times V\rightarrow W$
be a sesquilinear mapping into a vector space $W$. Let $\left(e_{i}\right)_{i=1}^{n}$
be an orthonormal basis for $V$. Then for any linear operators $S,T:V\rightarrow V$
it holds that
\[
\sum_{i=1}^{n}q\left(Se_{i},Te_{i}\right)=\sum_{i=1}^{n}q\left(ST^{\ast}e_{i},e_{i}\right).
\]
As a consequence the expression $\sum_{i=1}^{n}q\left(e_{i},e_{i}\right)$
is independent of the chosen basis.
\end{lem}

\textbf{Proof:} By orthonormal expansion we find that
\begin{align}
\sum_{i=1}^{n}q\left(Se_{i},Te_{i}\right) & =\sum_{i=1}^{n}q\left(Se_{i},\sum_{j=1}^{n}\left\langle e_{j},Te_{i}\right\rangle e_{j}\right) =\sum_{j=1}^{n}q\left(\sum_{i=1}^{n}\left\langle Te_{i},e_{j}\right\rangle Se_{i},e_{j}\right) \nonumber\\%=\sum_{i,j=1}^{n}\left\langle e_{j},Te_{i}\right\rangle q\left(Se_{i},e_{j}\right)\nonumber \\
 & =\sum_{j=1}^{n}q\left(S\sum_{i=1}^{n}\left\langle e_{i},T^{\ast}e_{j}\right\rangle e_{i},e_{j}\right)=\sum_{i=1}^{n}q\left(ST^{\ast}e_{i},e_{i}\right)
 %& =\sum_{j=1}^{n}q\left(ST^{\ast}e_{j},e_{j}\right)=\sum_{i=1}^{n}q\left(ST^{\ast}e_{i},e_{i}\right).\nonumber 
\end{align}
%The basis independence now follows by noting that if $\left(e_{i}^{\prime}\right)_{i=1}^{n}$
%is any other orthonormal basis then with $U:V\rightarrow V$ denoting
%the unitary transformation obeying $Ue_{i}=e_{i}^{\prime}$ for $1\leq i\leq n$
%we find that
%\begin{equation}
%\sum_{i=1}^{n}q\left(e_{i}^{\prime},e_{i}^{\prime}\right)=\sum_{i=1}^{n}q\left(Ue_{i},Ue_{i}\right)=\sum_{i=1}^{n}q\left(UU^{\ast}e_{i},e_{i}\right)=\sum_{i=1}^{n}q\left(e_{i},e_{i}\right).
%\end{equation}
The basis independence  follows from the fact that for all unitary transformation $U:V\rightarrow V$, 
\begin{equation}
\sum_{i=1}^{n}q\left(Ue_{i},Ue_{i}\right)=\sum_{i=1}^{n}q\left(UU^{\ast}e_{i},e_{i}\right)=\sum_{i=1}^{n}q\left(e_{i},e_{i}\right).
\end{equation}
$\hfill\square$

The lemma thus allows us to move a mapping from one argument to the
other when under a sum, which will be immensely useful when simplifying
expressions. As mentioned this can indeed be seen as a generalization
of the cyclicity property of the trace, since  the lemma implies
\begin{equation}
\text{tr}\left(ST\right)=\sum_{i=1}^{n}\left\langle e_{i},STe_{i}\right\rangle =\sum_{i=1}^{n}\left\langle S^{\ast}e_{i},Te_{i}\right\rangle =\sum_{i=1}^{n}\left\langle S^{\ast}T^{\ast}e_{i},e_{i}\right\rangle =\sum_{i=1}^{n}\left\langle e_{i},TSe_{i}\right\rangle =\text{tr}\left(TS\right),
\end{equation}
but it is important to note that cyclicity is not a general property
of trace-form sums - the assignments $A\mapsto Q_{1}\left(A\right)$
and $B\mapsto Q_{2}\left(B\right)$ do not obey such a property.

With the lemma we can now easily derive the commutator of $\mathcal{K}$
with $Q_{1}\left(\cdot\right)$ and $Q_{2}\left(\cdot\right)$:
\begin{prop}
\label{prop:BogolubovQKCommutators}For any real, symmetric operators $A,B,K:V\rightarrow V$, the operator 
$\mathcal{K}$ defined by equation (\ref{eq:BogolubovGeneratorDefinition})
obeys the following commutators on $\mathcal{F}^+(V)$:
\begin{align*}
\left[\mathcal{K},Q_{1}\left(A\right)\right] & =Q_{2}\left(\left\{ K,A\right\} \right)\\
\left[\mathcal{K},Q_{2}\left(B\right)\right] & =Q_{1}\left(\left\{ K,B\right\} \right).
\end{align*}
\end{prop}

\textbf{Proof:} We compute using the commutators of Proposition \ref{prop:BogolubovCAKCommutators}
that
\begin{align}
\left[\mathcal{K},Q_{1}\left(A\right)\right] & =\sum_{i=1}^{n}\left(\left[\mathcal{K},a^{\ast}\left(Ae_{i}\right)a\left(e_{i}\right)\right]+\left[\mathcal{K},a\left(e_{i}\right)a^{\ast}\left(Ae_{i}\right)\right]\right)\\
 & =\sum_{i=1}^{n}\left(a^{\ast}\left(Ae_{i}\right)\left[\mathcal{K},a\left(e_{i}\right)\right]+\left[\mathcal{K},a^{\ast}\left(Ae_{i}\right)\right]a\left(e_{i}\right)+a\left(e_{i}\right)\left[\mathcal{K},a^{\ast}\left(Ae_{i}\right)\right]+\left[\mathcal{K},a\left(e_{i}\right)\right]a^{\ast}\left(Ae_{i}\right)\right)\nonumber \\
 & =\sum_{i=1}^{n}\left(a^{\ast}\left(Ae_{i}\right)a^{\ast}\left(Ke_{i}\right)+a\left(KAe_{i}\right)a\left(e_{i}\right)+a\left(e_{i}\right)a\left(KAe_{i}\right)+a^{\ast}\left(Ke_{i}\right)a^{\ast}\left(Ae_{i}\right)\right).\nonumber 
\end{align}
As the assignments $\varphi,\psi\mapsto a\left(\varphi\right)a\left(\psi\right),a^{\ast}\left(\varphi\right)a^{\ast}\left(\psi\right)$
are bilinear we can apply Lemma \ref{lemma:TraceFormLemma} to see
that
\begin{align}
\left[\mathcal{K},Q_{1}\left(A\right)\right] & =\sum_{i=1}^{n}\left(a^{\ast}\left(AK^{\ast}e_{i}\right)a^{\ast}\left(e_{i}\right)+a\left(e_{i}\right)a\left(\left(KA\right)^{\ast}e_{i}\right)+a\left(e_{i}\right)a\left(KAe_{i}\right)+a^{\ast}\left(KA^{\ast}e_{i}\right)a^{\ast}\left(e_{i}\right)\right)\nonumber \\
 & =\sum_{i=1}^{n}\left(a^{\ast}\left(AKe_{i}\right)a^{\ast}\left(e_{i}\right)+a\left(e_{i}\right)a\left(AKe_{i}\right)+a\left(e_{i}\right)a\left(KAe_{i}\right)+a^{\ast}\left(KAe_{i}\right)a^{\ast}\left(e_{i}\right)\right)\nonumber \\
 & =\sum_{i=1}^{n}\left(a^{\ast}\left(\left(AK+KA\right)e_{i}\right)a^{\ast}\left(e_{i}\right)+a\left(e_{i}\right)a\left(\left(AK+KA\right)e_{i}\right)\right)=Q_{2}\left(\left\{ K,A\right\} \right)
\end{align}
where we also used that $A$ and $K$ are symmetric. The computation of $\left[\mathcal{K},Q_{2}\left(B\right)\right]$ is similar.
$\hfill\square$

%\begin{align}
%delete\\
%delete
%\end{align}

Note the similarity between this result and that of Proposition \ref{prop:BogolubovCAKCommutators}
- again we see that that $\left[\mathcal{K},\cdot\right]$ acts by
``swapping the types and applying $K$ to the argument'', although
now the relevant types are $Q_{1}\left(\cdot\right)$ and $Q_{2}\left(\cdot\right)$
and the application of $K$ is taking the anticommutator.

We can now appeal to the Baker-Campbell-Hausdorff formula again to
conclude that
\begin{align}
&e^{\mathcal{K}}Q_{1}\left(A\right)e^{-\mathcal{K}}  =Q_{1}\left(A\right)+\frac{1}{1!}\left[\mathcal{K},Q_{1}\left(A\right)\right]+\frac{1}{2!}\left[\mathcal{K},\left[\mathcal{K},Q_{1}\left(A\right)\right]\right]+\frac{1}{3!}\left[\mathcal{K},\left[\mathcal{K},\left[\mathcal{K},Q_{1}\left(A\right)\right]\right]\right]+\cdots \nonumber\\
 & =Q_{1}\left(A\right)+\frac{1}{1!}Q_{2}\left(\left\{ K,A\right\} \right)+\frac{1}{2!}Q_{1}\left(\left\{ K,\left\{ K,A\right\} \right\} \right)+\frac{1}{3!}Q_{2}\left(\left\{ K,\left\{ K,\left\{ K,A\right\} \right\} \right\} \right)+\cdots \nonumber\\
 & =Q_{1}\left(A+\frac{1}{2!}\left\{ K,\left\{ K,A\right\} \right\} +\cdots\right)+Q_{2}\left(\frac{1}{1!}\left\{ K,A\right\} +\frac{1}{3!}\left\{ K,\left\{ K,\left\{ K,A\right\} \right\} \right\} +\cdots\right)
\end{align}
but to succeed we must identify the sums of these iterated anticommutators.
First we note that we can rephrase this in a manner closer to that
of equation (\ref{eq:BogolubovCAAction}) for $e^{\mathcal{K}}a\left(\varphi\right)e^{-\mathcal{K}}$:
One may view the anticommutator with $K$ as a linear mapping $A\mapsto\left\{ K,A\right\} $
on the space of operators on $V$, $\mathcal{B}\left(V\right)$ -
denote this mapping by $\mathcal{A}_{K}:\mathcal{B}\left(V\right)\rightarrow\mathcal{B}\left(V\right)$,
i.e. $\mathcal{A}_{K}\left(\cdot\right)=\left\{ K,\cdot\right\} $.
Then we may phrase the above identity as
\begin{equation}
e^{\mathcal{K}}Q_{1}\left(A\right)e^{-\mathcal{K}}=Q_{1}\left(\cosh\left(\mathcal{A}_{K}\right)\left(A\right)\right)+Q_{2}\left(\sinh\left(\mathcal{A}_{K}\right)\left(A\right)\right)
\end{equation}
and likewise
\begin{equation}
e^{\mathcal{K}}Q_{2}\left(B\right)e^{-\mathcal{K}}=Q_{2}\left(\cosh\left(\mathcal{A}_{K}\right)\left(B\right)\right)+Q_{1}\left(\sinh\left(\mathcal{A}_{K}\right)\left(B\right)\right)
\end{equation}
so that the arguments again involve hyperbolic functions of linear
operators, but now acting on $\mathcal{B}\left(V\right)$ rather than
$V$ itself. We then note the following ``anticommutator Baker-Campbell-Hausdorff
formula'':
\begin{prop}
\label{prop:anticommutatorBCH}Let $\left(V,\left\langle \cdot,\cdot\right\rangle \right)$
be an $n$-dimensional Hilbert space, let $K:V\rightarrow V$ be a
self-adjoint operator and let $\mathcal{A}_{K}\left(\cdot\right)=\left\{ K,\cdot\right\} :\mathcal{B}\left(V\right)\rightarrow\mathcal{B}\left(V\right)$
denote the anticommutator with $K$. Then for any linear operator
$T:V\rightarrow V$
\[
e^{\mathcal{A}_{K}}\left(T\right)=\sum_{m=0}^{\infty}\frac{1}{m!}\mathcal{A}_{K}^{m}\left(T\right)=e^{K}Te^{K}.
\]
Consequently,
\begin{align*}
\cosh\left(\mathcal{A}_{K}\right)\left(T\right) & =\frac{1}{2}\left(e^{K}Te^{K}+e^{-K}Te^{-K}\right),\\
\sinh\left(\mathcal{A}_{K}\right)\left(T\right) & =\frac{1}{2}\left(e^{K}Te^{K}-e^{-K}Te^{-K}\right).
\end{align*}
\end{prop}

\textbf{Proof:} Let $\left(x_{i}\right)_{i=1}^{n}$ be an eigenbasis for $K$ with
associated eigenvalues $\left(\lambda_{i}\right)_{i=1}^{n}$. Denote $P_{i,j}=|x_j\rangle \langle x_i|$, namely $P_{i,j}x=\left\langle x_{i},x\right\rangle x_{j}$ for all $x\in V$. It is well-known that for any orthonormal basis $\left(x_{i}\right)_{i=1}^{n}$
of $V$ the collection $\left(P_{i,j}\right)_{i,j=1}^{n}$ form an
orthonormal basis for $\left(\mathcal{B}\left(V\right),\left\langle \cdot,\cdot\right\rangle _{\text{HS}}\right)$. Moreover, for any $x\in V$ and $1\leq i,j\leq n$,
by self-adjointness of $K$,
\begin{align}
\mathcal{A}_{K}\left(P_{i,j}\right)x & =\left\{ K,P_{i,j}\right\} x=\left\langle x_{i},x\right\rangle Kx_{j}+\left\langle x_{i},Kx\right\rangle x_{j}=\left\langle x_{i},x\right\rangle \lambda_{j}x_{j}+\left\langle \lambda_{i}x_{i},x\right\rangle x_{j}\\
 & =\left(\lambda_{i}+\lambda_{j}\right)\left\langle x_{i},x\right\rangle x_{j}=\left(\lambda_{i}+\lambda_{j}\right)P_{i,j}x.\nonumber 
\end{align}
Thus $\{P_{i,j}\}_{i,j=1}^n$ an eigenbasis for $\mathcal{A}_{K}$ with associated eigenvalues $\left(\lambda_{i}+\lambda_{j}\right)_{i,j=1}^{n}$. 

Hence, it suffices to verify the identity $e^{\mathcal{A}_{K}}\left(T\right)=e^{K}Te^{K}$ with the  eigenbasis $\left(P_{i,j}\right)_{i,j=1}^{n}$:
\begin{align}
e^{\mathcal{A}_{K}}\left(P_{i,j}\right) x & =e^{\lambda_{i}+\lambda_{j}}P_{i,j}=e^{\lambda_{i}+\lambda_{j}}\left\langle x_{i},x\right\rangle x_{j}=\left\langle e^{\lambda_{i}}x_{i},x\right\rangle e^{\lambda_{j}}x_{j}=\left\langle e^{K}x_{i},x\right\rangle e^{K}x_{j}\\
 & =\left\langle x_{i},e^{K}x\right\rangle e^{K}x_{j}=e^{K}P_{i,j}e^{K}x.\nonumber 
\end{align}

The statements regarding $\cosh\left(\mathcal{A}_{K}\right)$
and $\sinh\left(\mathcal{A}_{K}\right)$ follow from the identities 
\begin{equation}
\cosh\left(x\right)=\frac{1}{2}\left(e^{x}+e^{-x}\right), \quad \sinh\left(x\right)=\frac{1}{2}\left(e^{x}-e^{-x}\right), \quad \text{and }\left(-\mathcal{A}_{K}\right)=\mathcal{A}_{-K}.
\end{equation}
$\hfill\square$

By these formulas we thus deduce the quadratic operator analogue of
equation (\ref{eq:BogolubovCAAction}):
\begin{align}
e^{\mathcal{K}}Q_{1}\left(A\right)e^{-\mathcal{K}} & =\frac{1}{2}Q_{1}\left(e^{K}Ae^{K}+e^{-K}Ae^{-K}\right)+\frac{1}{2}Q_{2}\left(e^{K}Ae^{K}-e^{-K}Ae^{-K}\right)\\
e^{\mathcal{K}}Q_{2}\left(B\right)e^{-\mathcal{K}} & =\frac{1}{2}Q_{1}\left(e^{K}Be^{K}-e^{-K}Be^{-K}\right)+\frac{1}{2}Q_{2}\left(e^{K}Be^{K}+e^{-K}Be^{-K}\right).\nonumber 
\end{align}

\subsubsection*{Diagonalization Condition}

We can now finally describe how to diagonalize a quadratic Hamiltonian
using a Bogolubov transformation of the form $e^{\mathcal{K}}$. By
the transformation identities above we find that under $e^{\mathcal{K}}$
the quadratic Hamiltonian $H=Q_{1}\left(A\right)+Q_{2}\left(B\right)$
transforms as
\begin{align}
&e^{\mathcal{K}}He^{-\mathcal{K}}  =\frac{1}{2}Q_{1}\left(e^{K}Ae^{K}+e^{-K}Ae^{-K}\right)+\frac{1}{2}Q_{2}\left(e^{K}Ae^{K}-e^{-K}Ae^{-K}\right)\nonumber \\
 &\qquad \qquad +\frac{1}{2}Q_{1}\left(e^{K}Be^{K}-e^{-K}Be^{-K}\right)+\frac{1}{2}Q_{2}\left(e^{K}Be^{K}+e^{-K}Be^{-K}\right)\\
 & =\frac{1}{2}Q_{1}\left(e^{K}\left(A+B\right)e^{K}+e^{-K}\left(A-B\right)e^{-K}\right)+\frac{1}{2}Q_{2}\left(e^{K}\left(A+B\right)e^{K}-e^{-K}\left(A-B\right)e^{-K}\right).\nonumber 
\end{align}
Therefore, the \textit{diagonalization condition}
on $K$ is  
\begin{equation}
e^{K}\left(A+B\right)e^{K}=e^{-K}\left(A-B\right)e^{-K}.\label{eq:DiagonalizationCondition}
\end{equation}
If we can find such a $K$, then 
\begin{equation}
e^{\mathcal{K}}He^{-\mathcal{K}}=Q_{1}\left(E\right)=2\,\text{d}\Gamma\left(E\right)+\text{tr}\left(E\right)
\end{equation}
where
\begin{equation}
E=e^{K}\left(A+B\right)e^{K}=e^{-K}\left(A-B\right)e^{-K}.
\end{equation}

There remains the question of existence and uniqueness of such a $K$:

%\begin{prop}
%\label{prop:AnExactDiagonalizer}Let $A,B:V\rightarrow V$ be real, symmetric
%operators such that $A\pm B>0$. Then
%\[
%K=-\frac{1}{2}\log\left(\left(A-B\right)^{-\frac{1}{2}}\left(\left(A-B\right)^{\frac{1}{2}}\left(A+B\right)\left(A-B\right)^{\frac{1}{2}}\right)^{\frac{1}{2}}\left(A-B\right)^{-\frac{1}{2}}\right)
%\]
%is the unique real, symmetric solution of
%\[
%e^{K}\left(A+B\right)e^{K}=e^{-K}\left(A-B\right)e^{-K}.
%\]
%\end{prop}

\medskip

\textbf{Conclusion of the proof of Theorem \ref{thm:Bog-diag}:} 
%This holds if  $A\pm B>0$ (namely $A+B>0$ and $A-B>0$), and moreover in this case a diagonalizing $K$ is unique. The following statement is a generalization
%and simplification of the argument used in \cite{GreSei-13,BNPSS-20}. 
Write $A_{\pm}=A\pm B>0$ for brevity. Then we may write
the diagonalization condition as
\begin{equation}
e^{-2K}A_{-}e^{-2K}=A_{+}.
\end{equation}
Multiplying by $A_{-}^{\frac{1}{2}}$ on both sides yields
\begin{equation}
\Big(A_{-}^{\frac{1}{2}}e^{-2K}A_{-}^{\frac{1}{2}}\Big)^{2}=A_{-}^{\frac{1}{2}}e^{-2K}A_{-}e^{-2K}A_{-}^{\frac{1}{2}}=A_{-}^{\frac{1}{2}}A_{+}A_{-}^{\frac{1}{2}}
\end{equation}
which is equivalent to 
%For any real, symmetric $K$ the operator $e^{-2K}$ is positive, hence
%$A_{-}^{\frac{1}{2}}e^{-2K}A_{-}^{\frac{1}{2}}$ is positive. This
%equation thus states that $A_{-}^{\frac{1}{2}}e^{-2K}A_{-}^{\frac{1}{2}}$
%is a positive square root of $A_{-}^{\frac{1}{2}}A_{+}A_{-}^{\frac{1}{2}}$,
%so as positive square roots are unique we must have equality with
%$\Big(A_{-}^{\frac{1}{2}}A_{+}A_{-}^{\frac{1}{2}}\Big)^{\frac{1}{2}}$,
%i.e. we may conclude that
\begin{equation}
A_{-}^{\frac{1}{2}}e^{-2K}A_{-}^{\frac{1}{2}}=\left(A_{-}^{\frac{1}{2}}A_{+}A_{-}^{\frac{1}{2}}\right)^{\frac{1}{2}}, \, \text{ namely } e^{-2K}=A_{-}^{-\frac{1}{2}}\left(A_{-}^{\frac{1}{2}}A_{+}A_{-}^{\frac{1}{2}}\right)^{\frac{1}{2}}A_{-}^{-\frac{1}{2}}.
\end{equation}
%or
%\begin{equation}
%e^{-2K}=A_{-}^{-\frac{1}{2}}\left(A_{-}^{\frac{1}{2}}A_{+}A_{-}^{\frac{1}{2}}\right)^{\frac{1}{2}}A_{-}^{-\frac{1}{2}},
%\end{equation}
This implies the existence and uniqueness of the diagonalizing $K$ as the operator exponential is a bijection
between the real, symmetric operators and the real, symmetric, positive-definite
operators.
$\hfill\square$

%In summary, we conclude that for real, symmetric operators $A,B$ on a finite-dimensional 
%Hilbert space satisfying $A\pm B>0$, the quadratic Hamiltonian $H=Q_{1}\left(A\right)+Q_{2}\left(B\right)$ on the bosonic Fock space can be explicitly diagonalized by a Bogolubov transformation of the form $e^{\mathcal{K}}$, where $\mathcal{K}$ is given in \eqref{eq:BogolubovGeneratorDefinition} with $K$ defined in Theorem \ref{thm:Bog-diag}. 

%\subsection{Bogolubov Transformation? To be filled ...} 

%%%%%%%%%%%%%%%%%%%%%%%%%%%%%%
%%%%%%%%%%%%%%%%%%%%%%%%%%%%%%

\section{The Quasi-Bosonic Quadratic Hamiltonian}\label{sec:TransformingtheHamiltonian}

Now we turn to the quasi-bosonic setting.  We start by casting the bosonizable terms
$
H_{\kin}^{\prime} + \sum_{k\in S_C} H^k_{\rm int}$, which we encountered in Section \ref{sec:EstimationoftheNon-BosonizableTerms},
into a form which closely mirrors the form of the bosonic quadratic
Hamiltonians that we considered in the preceding section.

\subsection{Quadratic Hamiltonian}

Let us define the pair excitation operators 
\begin{equation} \label{eq:pair-excitation}
b_{k,p}=c_{p-k}^{\ast}c_{p},\quad b_{k,p}^{\ast}=c_{p}^{\ast}c_{p-k},\quad k\in\mathbb{Z}_{\ast}^{3},\quad p\in L_{k}.
\end{equation}
We remark that in contrast to the bosonic case, the  fermionic creation and
annihilation operators are bounded (in fact $\left\Vert c_{p,\sigma}\right\Vert _{\text{Op}}=\left\Vert c_{p,\sigma}^{\ast}\right\Vert _{\text{Op}}=1$), therefore so are the operators $b_{k,p}^{\ast}$, $b_{k,p}$.

%Once the terms $H_{\kin}^{\prime}+\sum_{k\in\mathbb{Z}_{+}^{3}}H_{\text{int}}^{k}$
%have been expressed in this way it will then be clear how to define
%operators $e^{\mathcal{K}}$ - the quasi-bosonic Bogolubov transformations
%- which emulate the properties of the transformation of the previous
%section. We will then be able to repeat the calculations of that section
%(now also keeping in mind the quasi-bosonic error term) to determine
%the action of $e^{\mathcal{K}}$ on our Hamiltonian.
%
%Once the general transformation identities have been derived we will
%then specify a generator $\mathcal{K}$ such that $e^{\mathcal{K}}$
%will serve to diagonalize the bosonizable terms similarly to the exact
%bosonic case.
%\subsection{Formalizing the Bosonic Analogy}
%We begin by determining what space will play the role of the one-body
%space $V$ of the bosonic case. Recall that we in Section \ref{sec:LocalizingtheHamiltonian}
%found that the bosonizable parts of the Hamiltonian $H_{N}$ are the
%terms $H_{\kin}^{\prime}+\sum_{k\in\mathbb{Z}_{+}^{n}}H_{\text{int}}^{k}$,
%where $H_{\kin}^{\prime}$ behaved in an analogous fashion to
%\begin{equation}
%H_{\kin}^{\prime}\sim\sum_{k\in\mathbb{Z}_{\ast}^{3}}\sum_{p\in L_{k}}\left(\left|p\right|^{2}-\left|p-k\right|^{2}\right)b_{k,p}^{\ast}b_{k,p}\label{eq:KineticEnergyAnalogyReminder}
%\end{equation}
%while 
Then $H_{\text{int}}^{k}$ in \eqref{eq:Hint-k} is exactly given by
\begin{align}
H_{\text{int}}^{k} & =\sum_{p,q\in L_{k}}\frac{\hat{V}_{k}k_{F}^{-1}}{2\left(2\pi\right)^{3}}\left(b_{k,p}^{\ast}b_{k,q}+b_{k,q}b_{k,p}^{\ast}\right)+\sum_{p,q\in L_{-k}}\frac{\hat{V}_{k}k_{F}^{-1}}{2\left(2\pi\right)^{3}}\left(b_{-k,p}^{\ast}b_{-k,q}+b_{-k,q}b_{-k,p}^{\ast}\right)\label{eq:HkintReminder}\\
 & +\sum_{p\in L_{k}}\sum_{q\in L_{-k}}\frac{\hat{V}_{k}k_{F}^{-1}}{2\left(2\pi\right)^{3}}\left(b_{k,p}^{\ast}b_{-k,q}^{\ast}+b_{-k,q}b_{k,p}\right)+\sum_{p\in L_{-k}}\sum_{q\in L_{k}}\frac{\hat{V}_{k}k_{F}^{-1}}{2\left(2\pi\right)^{3}}\left(b_{-k,p}^{\ast}b_{k,q}^{\ast}+b_{k,q}b_{-k,p}\right),\nonumber 
\end{align}
Thus the natural one-body Hilbert space associated to $H_{\text{int}}^{k}$ is $\ell^2 (L_k \cup L_{-k})$. To free us from having to explicitly write sums over $L_{k}$
and $L_{-k}$ separately we introduce some more notation: First we
will denote this union of lunes by
\begin{equation}
L_{k}^{\pm}=L_{k}\cup L_{-k}, \quad \ell^{2}\left(L_{k}^{\pm}\right)=\ell^{2}\left(L_{k}\cup L_{-k}\right)=\ell^{2}\left(L_{k}\right)\oplus\ell^{2}\left(L_{-k}\right), \quad k\in\mathbb{Z}_{+}^{3}.
\end{equation}
Here we used the fact that $L_{k}\cap L_{-k}=\emptyset$ for any $k\in\mathbb{Z}_{+}^{3}$, since if $p\in L_{k}\cap L_{-k}$ then 
$$
2|p|^2 \geq |p-k|^2 + |p+k|^2 = 2|p|^2 + 2|k|^2 > 2|p|^2
$$
which is a contradiction. 
%It is easily seen from the definition of the lunes,
%\begin{equation}
%L_{k}=\left\{ p\in\mathbb{Z}^{3}\mid\left|p-k\right|\leq k_{F}<\left|p\right|\right\} ,
%\end{equation}
%that for any $k$, $L_{k}\cap L_{-k}=\emptyset$, as $p\in L_{k}\cap L_{-k}$
%implies the contradiction
%\begin{equation}
%2k_{F}^{2}\geq\left|p-k\right|^{2}+\left|p+k\right|^{2}=2\left|p\right|^{2}+\left|k\right|^{2}>2k_{F}^{2}+\left|k\right|^{2}\Rightarrow\left|k\right|^{2}<0,
%\end{equation}
%and so we have a natural identification
%\begin{equation}
%\ell^{2}\left(L_{k}^{\pm}\right)=\ell^{2}\left(L_{k}\cup L_{-k}\right)=\ell^{2}\left(L_{k}\right)\oplus\ell^{2}\left(L_{-k}\right).
%\end{equation}
%At the end of this section we will choose the Bogolubov transformation
%such that, once the transformation is carried out, these summands
%will decouple and we may focus on the single space $\ell^{2}\left(L_{k}\right)$
%for the remainder of the paper. Until then we must however consider
%the whole space $\ell^{2}\left(L_{k}^{\pm}\right)$, 
It is also convenient to introduce the ``bar-notation''
\begin{equation} \label{eq:bar-notation}
\overline{k,p}=\begin{cases}
k,p & p\in L_{k}\\
-k,p & p\in L_{-k}
\end{cases},\quad\overline{p-k}=\begin{cases}
p-k & p\in L_{k}\\
p+k & p\in L_{-k}
\end{cases},
\end{equation}
to automatically encode the appropriate sign of $k$ depending on
$p\in L_{k}^{\pm}=L_{k}\cup L_{-k}$ (this will allow us to avoid
expanding all our terms on a case-by-case basis when this is irrelevant).

In analogy with the definitions (\ref{eq:Q1AFullForm})
and (\ref{eq:Q2BFullForm}) we now define, for any $k\in\mathbb{Z}_{+}^{3}$
and symmetric operators $A,B:\ell^{2}\left(L_{k}^{\pm}\right)\rightarrow\ell^{2}\left(L_{k}^{\pm}\right)$,
the quadratic operators $Q_{1}^{k}\left(A\right),Q_{2}^{k}\left(B\right):\mathcal{H}_{N}\rightarrow\mathcal{H}_{N}$
by
\begin{align}
Q_{1}^{k}\left(A\right) & =\sum_{p,q\in L_{k}^{\pm}}\left\langle e_{p},Ae_{q}\right\rangle \left(b_{\overline{k,p}}^{\ast}b_{\overline{k,q}}+b_{\overline{k,q}}b_{\overline{k,p}}^{\ast}\right),\label{eq:Q1kQ2kFirstDefinition}\\
Q_{2}^{k}\left(B\right) & =\sum_{p,q\in L_{k}^{\pm}}\left\langle e_{p},Be_{q}\right\rangle \left(b_{\overline{k,p}}^{\ast}b_{\overline{k,q}}^{\ast}+b_{\overline{k,q}}b_{\overline{k,p}}\right).\nonumber 
\end{align}
In order to cast $H_{\text{int}}^{k}$ as given by equation (\ref{eq:HkintReminder})
into this form we must identify the relevant operators $A$ and $B$. 
%The prefactors of the terms of $H_{\text{int}}^{k}$ are all $\frac{\hat{V}_{k}k_{F}^{-1}}{2\left(2\pi\right)^{3}}$,
%which is independent of $p,q\in L_{k}^{\pm}$. The operator with such
%matrix elements are (un-normalized) one-dimensional projections: Consider
%the vector $v_{k}\in\ell^{2}\left(L_{k}\right)$ (not $\ell^{2}\left(L_{k}^{\pm}\right)$)
%given by
%\begin{equation}
%v_{k}=\sqrt{\frac{\hat{V}_{k}k_{F}^{-1}}{2\left(2\pi\right)^{3}}}\sum_{p\in L_{k}}e_{p},
%\end{equation}
%where $\left(e_{p}\right)_{p\in L_{k}}$ denotes the standard orthonormal
%basis of $\ell^{2}\left(L_{k}\right)$. Let 
%\begin{equation}
%P_{v_k} =|v_k\rangle \langle v_k|
%\end{equation}
%be the rank-one projection on $\ell^2(L_{k})$, namely 
%%and let $P_{v_{k}}:\ell^{2}\left(L_{k}\right)\rightarrow\ell^{2}\left(L_{k}\right)$
%%denote the operator defined by $P_{v_{k}}x=\left\langle v_{k},x\right\rangle v_{k}$
%%for $x\in\ell^{2}\left(L_{k}\right)$. Then the matrix elements of
%%$P_{v_{k}}$ are precisely
%its matrix elements are precisely
%\begin{align}
%\left\langle e_{p},P_{v_{k}}e_{q}\right\rangle   =\left\langle e_{p},v_{k}\right\rangle \left\langle v_{k},e_{q}\right\rangle  =\frac{\hat{V}_{k}k_{F}^{-1}}{2\left(2\pi\right)^{3}},\quad \forall p,q\in L_{k}. 
%\end{align}
Define the (un-normalized) rank-one projection $P_{v_k}: \ell^2(L_k)\to \ell^2(L_k)$ by
\begin{equation}
P_{v_k} =|v_k\rangle \langle v_k|, \quad v_{k}=\sqrt{\frac{\hat{V}_{k}k_{F}^{-1}}{2\left(2\pi\right)^{3}}}\sum_{p\in L_{k}}e_{p} \in \ell^2(L_k)
\end{equation}
where $\left(e_{p}\right)_{p\in L_{k}}$ denotes the standard orthonormal
basis of $\ell^{2}\left(L_{k}\right)$. Put differently, the matrix elements of $P_{v_k}$ are 
$\left\langle e_{p},P_{v_{k}}e_{q}\right\rangle   =\frac{1}{2\left(2\pi\right)^{3}} \hat{V}_{k}k_{F}^{-1}$
for all $p,q\in L_{k}$. Next, we define the operators 
\begin{equation} \label{eq:def-Ak-Bk-plus}
A_{k}^{\oplus},B_{k}^{\oplus}:\ell^{2}\left(L_{k}^{\pm}\right)\rightarrow\ell^{2}\left(L_{k}^{\pm}\right), \quad A_{k}^{\oplus}=\left(\begin{array}{cc}
P_{v_{k}} & 0\\
0 & P_{v_{k}}
\end{array}\right),\quad B_{k}^{\oplus}=\left(\begin{array}{cc}
0 & P_{v_{k}}\\
P_{v_{k}} & 0
\end{array}\right)
\end{equation}
with respect to the decomposition $\ell^{2}\left(L_{k}^{\pm}\right)=\ell^{2}\left(L_{k}\right)\oplus\ell^{2}\left(L_{-k}\right)$
and the identification $\ell^{2}\left(L_{k}\right)\cong\ell^{2}\left(L_{-k}\right)$
(under $e_{p}\mapsto e_{-p}$). 

\medskip
Thus the operator $H_{\text{int}}^{k}$ is concisely expressed as
\begin{equation}
H_{\text{int}}^{k}=Q_{k}^{1}\left(A_{k}^{\oplus}\right)+Q_{k}^{2}\left(B_{k}^{\oplus}\right).
\end{equation}
It remains to consider the kinetic operator. The equality \eqref{LocalizedKineticOperatorCommutator} bids us to think of think of $H_{\kin}^{\prime}$ as it were 
\begin{align}
H_{\kin}^{\prime} &\sim \sum_{k\in\mathbb{Z}_{\ast}^{3}}\sum_{p\in L_{k}}\left(\left|p\right|^{2}-\left|p-k\right|^{2}\right)b_{k,p}^{\ast}b_{k,p} \nonumber\\
&= \sum_{k\in\mathbb{Z}_{+}^{3}} \left(\sum_{p\in L_{k}}\left(\left|p\right|^{2}-\left|p-k\right|^{2}\right)b_{k,p}^{\ast}b_{k,p}+\sum_{p\in L_{-k}}\left(\left|p\right|^{2}-\left|p+k\right|^{2}\right)b_{-k,p}^{\ast}b_{-k,p}\right) \label{eq:KineticEnergyAnalogyReminder}
\end{align}
in an appropriate sense. To put this in the same framework as $H_{\rm int}^k$, let us introduce (for every $k\in\mathbb{Z}_{+}^{3}$) the operator $h_{k}:\ell^{2}\left(L_{k}\right)\rightarrow\ell^{2}\left(L_{k}\right)$
by
\begin{align}
h_{k}e_{p} =\lambda_{k,p}e_{p},\quad\;\;\lambda_{k,p}=\frac{1}{2}\left(\left|p\right|^{2}-\left|p-k\right|^{2}\right). \label{eq:hkvkReminder}
\end{align}
Using again the identification $\ell^{2}\left(L_{k}\right)\cong\ell^{2}\left(L_{-k}\right)$
(under $e_{p}\mapsto e_{-p}$), we define the operators $h_{k}^{\oplus}:\ell^{2}\left(L_{k}^{\pm}\right)\rightarrow\ell^{2}\left(L_{k}^{\pm}\right)$
by
\begin{equation} \label{eq:def-hk-oplus}
h_{k}^{\oplus}=\left(\begin{array}{cc}
h_{{k}} & 0\\
0 & h_{{k}}
\end{array}\right).
\end{equation}
Then we can rewrite \eqref{eq:KineticEnergyAnalogyReminder} as 
\begin{align}
H_{\kin}^{\prime} \sim \sum_{k\in\mathbb{Z}_{+}^{3}}  \Big( Q_1^k (h_k^{\oplus}) - 2\,{\rm tr} (h_k)\Big). 
\end{align}

Recall that $S_{C}=\overline{B}\left(0,k_{F}^{\gamma}\right)\cap\mathbb{Z}_{+}^{3}$ for an exponent $1\ge \gamma>0$ which is to be optimized over at
the end. As far as the lower bound is concerned, we may replace $\sum_{k\in\mathbb{Z}_+}$ by $\sum_{k\in S_C}$ (the upper bound is easier and will be explained separately).  In summary, we arrive at the following quasi-bosonic expression for the bosonizable terms:
\begin{equation} \label{eq:Quasi-quadratic}
H_{\kin}^{\prime} + \sum_{k\in S_C} H_{\rm int}^k \sim  \sum_{k\in S_C}  \Big( Q_1^k (h_k^{\oplus} + A_k^{\oplus}) + Q_2^k (B_k^{\oplus}) - 2\,{\rm tr}(h_k)\Big).
\end{equation}
Note that unlike the bosonic case, the operators on the right side of \eqref{eq:Quasi-quadratic} are bounded.

\subsection{Generalized Pair Operators}

For every $k\in\mathbb{Z}_{+}^{3}$ and $\varphi\in \ell^2(L_k^{\pm})$ we define the operators  
\begin{equation}
b_{k}\left(\varphi\right)=\sum_{p\in L_{k}^{\pm}}\left\langle \varphi,e_{p}\right\rangle b_{\overline{k,p}},\quad b_{k}^{\ast}\left(\varphi\right)=\sum_{p\in L_{k}^{\pm}}\left\langle e_{p},\varphi\right\rangle b_{\overline{k,p}}^{\ast}. 
\end{equation}
They obey the quasi-bosonic commutation relations (for $k,l\in\mathbb{Z}_{+}^{3}$
and $\varphi\in\ell^{2}\left(L_{k}^{\pm}\right)$, $\psi\in\ell^{2}\left(L_{l}^{\pm}\right)$)
\begin{align}  \label{eq:QuasiBosonicCommutationRelations}
\left[b_{k}\left(\varphi\right),b_{l}\left(\psi\right)\right] & =\left[b_{k}^{\ast}\left(\varphi\right),b_{l}^{\ast}\left(\psi\right)\right]=0, \\
\left[b_{k}\left(\varphi\right),b_{l}^{\ast}\left(\psi\right)\right] & =\delta_{k,l}\left\langle \varphi,\psi\right\rangle +\varepsilon_{k,l}\left(\varphi;\psi\right)\nonumber 
\end{align}
where the correction term is
\begin{align}
\varepsilon_{k,l}\left(\varphi;\psi\right) & =\sum_{p\in L_{k}^{\pm}}\sum_{q\in L_{l}^{\pm}}\left\langle \varphi,e_{p}\right\rangle \left\langle e_{q},\psi\right\rangle \varepsilon\left(\overline{k,p};\overline{l,q}\right), \label{eq:ExchangeCorrectionDefinition}\\
\varepsilon\left(\overline{k,p};\overline{l,q}\right) & =-\left(\delta_{p,q}c_{\overline{q-l}}c_{\overline{p-k}}^{\ast}+\delta_{\overline{p-k},\overline{q-l}}c_{q}^{\ast}c_{p}\right).\nonumber 
\end{align}
%The quadratic operators $Q_{1}^{k}\left(\cdot\right)$ and $Q_{2}^{k}\left(\cdot\right)$
%are defined in terms of such operators, and so our first step towards
%estimating quadratic terms will be to estimate these. In the bosonic
%case a practical matter for obtaining estimates on terms of creation
%and annihilation operators is normal ordering, which is to rearrange
%each term so that all creation operators are to the left of all annihilation
%operators, using the canonical commutation relations. The resulting
%normal ordered expression is then typically easy to estimate with
%respect to the number operator $\mathcal{N}$.

We simply have $b_{k}\left(e_{p}\right)=b_{\overline{k,p}}$
and the quadratic operators in \eqref{eq:Q1kQ2kFirstDefinition} can be expressed as
\begin{align}
Q_{1}^{k}\left(A\right) & =\sum_{p\in L_{k}^{\pm}}\left(b_{k}^{\ast}\left(Ae_{p}\right)b_{k}\left(e_{p}\right)+b_{k}\left(e_{p}\right)b_{k}^{\ast}\left(Ae_{p}\right)\right)\\
Q_{2}^{k}\left(B\right) & =\sum_{p\in L_{k}^{\pm}}\left(b_{k}^{\ast}\left(Be_{p}\right)b_{k}^{\ast}\left(e_{p}\right)+b_{k}\left(e_{p}\right)b_{k}\left(Be_{p}\right)\right)\nonumber 
\end{align}
in analogy with equation (\ref{eq:Q1Q2GoodDefinition}). In order to justify the quasi-bosonic interpretation, we need rigorous estimates for the correction term in \eqref{eq:ExchangeCorrectionDefinition}.  Let us start with 
%This is not quite as easy in our case as e.g.
%\begin{equation}
%b_{k}\left(\varphi\right)b_{k}^{\ast}\left(\varphi\right)=b_{k}^{\ast}\left(\varphi\right)b_{k}\left(\varphi\right)+\left\Vert \varphi\right\Vert ^{2}+\varepsilon_{k,k}\left(\varphi;\varphi\right).
%\end{equation}
%In many estimates, 
%The error term $\varepsilon_{k,k}\left(\varphi;\varphi\right)$
%is neither a constant nor of the form $b_{k}^{\ast}\left(\varphi\right)b_{k}\left(\varphi\right)$.
%The first result of this section will however imply that for the purposes
%of estimating $b_{k}\left(\varphi\right)b_{k}^{\ast}\left(\varphi\right)$,
%$\varepsilon_{k,k}\left(\varphi;\varphi\right)$ can be neglected:
\begin{prop}
\label{prop:DirectExchangeCorrectionIsNegative}For all $k\in\mathbb{Z}_{+}^{3}$
and $\varphi\in\ell^{2}\left(L_{k}^{\pm}\right)$, it holds that $\varepsilon_{k,k}\left(\varphi,\varphi\right)\leq0$, namely
$$
b_{k}\left(\varphi\right)b_{k}^{\ast}\left(\varphi\right) \le b_{k}^{\ast}\left(\varphi\right)b_{k}\left(\varphi\right)+\left\Vert \varphi\right\Vert ^{2}. 
$$
\end{prop}

Note that the observation of the error term $\varepsilon_{k,k}$ being non-positive also appeared in \cite[Proof of Lemma 4.2]{BNPSS-20} in the context of
different bosonic operators. 

\medskip
\textbf{Proof:} We expand the term
\begin{align}
\varepsilon_{k,k}\left(\varphi;\varphi\right) & =\sum_{p,q\in L_{k}^{\pm}}\left\langle \varphi,e_{p}\right\rangle \left\langle e_{q},\varphi\right\rangle \varepsilon\left(\overline{k,p};\overline{k,q}\right) \nonumber\\
&=-\sum_{p,q\in L_{k}^{\pm}}\left\langle \varphi,e_{p}\right\rangle \left\langle e_{q},\varphi\right\rangle \left(\delta_{p,q}c_{\overline{q-k}}c_{\overline{p-k}}^{\ast}+\delta_{\overline{p-k},\overline{q-k}}c_{q}^{\ast}c_{p}\right)\nonumber \\
 & =-\sum_{p\in L_{k}^{\pm}}\left|\left\langle e_{p},\varphi\right\rangle \right|^{2}c_{\overline{p-k}}c_{\overline{p-k}}^{\ast}-\sum_{p,q\in L_{k}^{\pm}}\delta_{\overline{p-k},\overline{q-k}}\left\langle \varphi,e_{p}\right\rangle \left\langle e_{q},\varphi\right\rangle c_{q}^{\ast}c_{p} \\
 &\le - \sum_{p,q\in L_{k}^{\pm}}\delta_{\overline{p-k},\overline{q-k}}\left\langle \varphi,e_{p}\right\rangle \left\langle e_{q},\varphi\right\rangle c_{q}^{\ast}c_{p}.\nonumber 
\end{align}
We treat the terms of the last sum on a case-by-case basis according
to which of $L_{k}$ and $L_{-k}$, $p$ and $q$ lie in: If $p$ and
$q$ lie in the same lune then $\delta_{\overline{p-k},\overline{q-k}}=\delta_{p\mp k,q\mp k}=\delta_{p,q}$
and so
\begin{align}
A  =\left(\sum_{p,q\in L_{k}}+\sum_{p,q\in L_{-k}}\right)\delta_{\overline{p-k},\overline{q-k}}\left\langle \varphi,e_{p}\right\rangle \left\langle e_{q},\varphi\right\rangle c_{q}^{\ast}c_{p}=\sum_{p\in L_{k}^{\pm}}\left|\left\langle e_{p},\varphi\right\rangle \right|^{2}c_{p}^{\ast}c_{p} \ge 0. 
\end{align}
On the other hand, by the Cauchy--Schwarz inequality
\begin{align}
&\pm \left( \sum_{p\in L_{k}}\sum_{q\in L_{-k}} + \sum_{p\in L_{-k}}\sum_{q\in L_{k}} \right)  \delta_{\overline{p-k},\overline{q-k}}\left\langle \varphi,e_{p}\right\rangle \left\langle e_{q},\varphi\right\rangle c_{q}^{\ast}c_{p}\nonumber\\
&\le \left( \sum_{p\in L_{k}}\sum_{q\in L_{-k}} + \sum_{p\in L_{-k}}\sum_{q\in L_{k}} \right)  \frac{1}{2}\Big( \delta_{\overline{p-k},\overline{q-k}} |\left\langle \varphi,e_{p}\right\rangle|^2 c_p^* c_p +  |\left\langle e_{q},\varphi\right\rangle|^2 c_{q}^{\ast}c_{q} \Big) \\
&\le \sum_{p\in L_{k}^{\pm}}\left|\left\langle e_{p},\varphi\right\rangle \right|^{2}c_{p}^{\ast}c_{p} = A. \nonumber
\end{align}

We thus conclude that $\varepsilon_{k,k}\left(\varphi,\varphi\right)\leq0$ as claimed.
$\hfill\square$

Next we have
\begin{prop}
\label{prop:GeneralExcitationOperatorEstimate}For all $k\in\mathbb{Z}_{+}^{3}$,
$\varphi\in\ell^{2}\left(L_{k}^{\pm}\right)$ and $\Psi\in\mathcal{H}_{N}$
it holds that
\[
\left\Vert b_{k}\left(\varphi\right)\Psi\right\Vert \leq\left\Vert \varphi\right\Vert \sqrt{\left\langle \Psi,\mathcal{N}_{E}\Psi\right\rangle },\quad\left\Vert b_{k}^{\ast}\left(\varphi\right)\Psi\right\Vert \leq\left\Vert \varphi\right\Vert \sqrt{\left\langle \Psi,\left(1+\mathcal{N}_{E}\right)\Psi\right\rangle }.
\]
\end{prop}

The bounds here are similar to \cite[Lemma 4.2]{BNPSS-20}. Recall that in our quasi-bosonic setting the excitation number operator 
\begin{equation}
\mathcal{N}_{E}=\sum_{p\in B_{F}^{c}}c_{p}^{\ast}c_{p}=\sum_{p\in B_{F}}c_{p}c_{p}^{\ast}
\end{equation}
 play
the role that the usual number operator $\mathcal{N}$ does in the
exact bosonic case. Thus Proposition \ref{prop:GeneralExcitationOperatorEstimate} is the analogue of the well-known bosonic estimate 
\begin{equation}
\left\Vert a\left(\varphi\right)\Psi\right\Vert \leq\left\Vert \varphi\right\Vert \sqrt{\left\langle \Psi,\mathcal{N}\Psi\right\rangle },\quad\left\Vert a^{\ast}\left(\varphi\right)\Psi\right\Vert \leq\left\Vert \varphi\right\Vert \sqrt{\left\langle \Psi,\left(1+\mathcal{N}\right)\Psi\right\rangle }.
\end{equation}
%and the preceding proposition will now let us conclude the analogous
%relations for $b_{k}\left(\varphi\right)$ and $b_{k}^{\ast}\left(\varphi\right)$:

\textbf{Proof:}
%For a single operator $b_{\overline{k,p}}=c_{\overline{p-k}}^{\ast}c_{p}$ with some $p\in L_{k}^{\pm}$, the bound follows trivially from the bound $\left\Vert c_{p}^{\ast}\right\Vert _{\text{Op}}=1$. For a general element $\varphi\in\ell^{2}\left(L_{k}^{\pm}\right)$, 
By the Cauchy-Schwarz inequality 
\begin{align}
\left\Vert b_{k}\left(\varphi\right)\Psi\right\Vert  & = \left\Vert \sum_{p\in L_{k}^{\pm}}\left\langle \varphi,e_{p}\right\rangle b_{\overline{k,p}}\Psi\right\Vert \leq\sqrt{\sum_{p\in L_{k}^{\pm}}\left|\left\langle \varphi,e_{p}\right\rangle \right|^{2}}\sqrt{\sum_{p\in L_{k}^{\pm}}\left\Vert b_{\overline{k,p}}\Psi\right\Vert ^{2}}\nonumber \\
 & \leq\left\Vert \varphi\right\Vert \sqrt{\sum_{p\in L_{k}^{\pm}}\left\Vert c_{p}\Psi\right\Vert ^{2}}\leq\left\Vert \varphi\right\Vert \sqrt{\left\langle \Psi,\mathcal{N}_{E}\Psi\right\rangle }. \label{eq:bkvarphiEstimateEquation}
\end{align}
The second bound follows from the first and Proposition \ref{prop:DirectExchangeCorrectionIsNegative}.
%
%as it is implied by the inequality
%\begin{equation}
%b_{k}\left(\varphi\right)b_{k}^{\ast}\left(\varphi\right)=b_{k}^{\ast}\left(\varphi\right)b_{k}\left(\varphi\right)+\left\Vert \varphi\right\Vert ^{2}+\varepsilon_{k,k}\left(\varphi;\varphi\right)\leq\left\Vert \varphi\right\Vert ^{2}\mathcal{N}_{E}+\left\Vert \varphi\right\Vert ^{2}+0=\left\Vert \varphi\right\Vert ^{2}\left(1+\mathcal{N}_{E}\right).
%\end{equation}
$\hfill\square$

We remark that the above estimate is also valid for $\Psi\in\mathcal{H}_{M}$
when $M\neq N$, provided $\mathcal{N}_{E}$ is understood as $\sum_{p\in B_{F}^{c}}c_{p}^{\ast}c_{p}$
acting on $\mathcal{H}_{M}$ (in \eqref{eq:bkvarphiEstimateEquation} we used $L_k^{\pm} \subset B_F^c$) - one must be precise here as the identity
$\mathcal{N}_{E}=\sum_{p\in B_{F}}c_{p}c_{p}^{\ast}$ does not hold
on $\mathcal{H}_{M}$. In fact the estimate also holds if $\mathcal{N}_{E}$
is understood as $\sum_{p\in B_{F}}c_{p}c_{p}^{\ast}$, up to an additional
factor of $\sqrt{2}$ due to  the necessary overcounting of the holes\footnote{While $L_{k}\cap L_{-k}=\emptyset$ it is
generally the case that $\left(L_{k}-k\right)\cap\left(L_{-k}+k\right)\neq\emptyset$
(a single hole state may be ``shared'' by both lunes), so when estimating
in terms of a single sum over $p\in B_{F}$ a factor of $2$ is often
necessary.}, namely from $\left\Vert b_{\overline{k,p}}\Psi\right\Vert =\left\Vert c_{\overline{p-k}}^{\ast}c_{p}\Psi\right\Vert \leq\left\Vert c_{\overline{p-k}}^{\ast}\Psi\right\Vert $ with $\overline{p-k} \in B_F$ we get 
\begin{align} \label{eq:over-counting-used-first}
\left\Vert b_{k}\left(\varphi\right)\Psi\right\Vert  \leq\left\Vert \varphi\right\Vert \sqrt{\sum_{p\in L_{k}^{\pm}}\left\Vert c_{\overline{p-k}}^{\ast}\Psi\right\Vert ^{2}}\leq\sqrt{2}\left\Vert \varphi\right\Vert \sqrt{\left\langle \Psi,\left(\sum_{p\in B_{F}}c_{p}c_{p}^{\ast}\right)\Psi\right\rangle }. 
\end{align}
This is a point that we must consider, since below we will also encounter
expressions such as $\left\Vert b_{k}\left(\varphi\right)c_{p}\Psi\right\Vert $
for $\Psi\in\mathcal{H}_{N}$ (so that $c_{p}\Psi\in\mathcal{H}_{N-1}$).
For this we denote by $\mathcal{N}_{E}^{\left(-1\right)}:\mathcal{H}_{N-1}\rightarrow\mathcal{H}_{N-1}$
and $\mathcal{N}_{E}^{\left(+1\right)}:\mathcal{H}_{N+1}\rightarrow\mathcal{H}_{N+1}$
the operators
\begin{equation}
\mathcal{N}_{E}^{\left(-1\right)}=\sum_{p\in B_{F}^{c}}c_{p}^{\ast}c_{p},\quad\mathcal{N}_{E}^{\left(+1\right)}=\sum_{p\in B_{F}}c_{p}c_{p}^{\ast}. 
\end{equation}
This choice is motivated by the following identities:
\begin{lem}
\label{lemma:ExcitationNumberOperatorCommutators}For all $p\in B_{F}^{c}$
and $q\in B_{F}$ it holds that
\begin{align*}
\mathcal{N}_{E}c_{p}^{\ast} &=c_{p}^{\ast}\mathcal{N}_{E}^{\left(-1\right)}+c_{p}^{\ast}, \quad c_p\mathcal{N}_{E}^{(-1)}c_{p}^{\ast} \le \mathcal{N}_E,\\
\mathcal{N}_{E}c_{q}  &=c_{q}\mathcal{N}_{E}^{\left(+1\right)}+c_{q}, \quad c_{q}^{\ast}\mathcal{N}_{E}^{\left(+1\right)} c_{q} \le \mathcal{N}_E.
\end{align*}
Consequently 
\[
\sum_{p\in B_{F}^{c}}c_{p}^{\ast}\mathcal{N}_{E}^{\left(-1\right)}c_{p}=\mathcal{N}_{E}^{2}-\mathcal{N}_{E}=\sum_{p\in B_{F}}c_{p}\mathcal{N}_{E}^{\left(+1\right)}c_{p}^{\ast}.
\]
\end{lem}

\textbf{Proof:} This follows directly by the CAR, as for all $p\in B_{F}^{c}$
\begin{equation}
\mathcal{N}_{E}c_{p}^{\ast}=\sum_{q\in B_{F}^{c}}c_{q}^{\ast}c_{q}c_{p}^{\ast}=\sum_{q\in B_{F}^{c}}c_{p}^{\ast}c_{q}^{\ast}c_{q}+\sum_{q\in B_{F}^{c}}\left(c_{q}^{\ast}\left\{ c_{q},c_{p}^{\ast}\right\} -\left\{ c_{q}^{\ast},c_{p}^{\ast}\right\} c_{q}\right)=c_{p}^{\ast}\mathcal{N}_{E}^{\left(-1\right)}+c_{p}^{\ast}.
\end{equation}
Consequently, using $\|c_p\|_{\rm Op}=1$ and $[\mathcal{N}_{E},c_{p}^{\ast}c_p]=0$ we have
\begin{equation}
\mathcal{N}_{E} \ge \mathcal{N}_{E}c_{p}^{\ast}c_p = c_{p}^{\ast}\mathcal{N}_{E}^{\left(-1\right)}c_p + c_{p}^{\ast} c_p \ge c_{p}^{\ast}\mathcal{N}_{E}^{\left(-1\right)}c_p.
\end{equation}
Likewise, for all $q\in B_{F}$
\begin{equation}
\mathcal{N}_{E}c_{q}=\sum_{p\in B_{F}}c_{p}c_{p}^{\ast}c_{q}=\sum_{p\in B_{F}}c_{q}c_{p}c_{p}^{\ast}+\sum_{p\in B_{F}}\left(c_{p}\left\{ c_{p}^{\ast},c_{q}\right\} -\left\{ c_{p},c_{q}\right\} c_{p}^{\ast}\right)=c_{q}\mathcal{N}_{E}^{\left(+1\right)}+c_{q}
\end{equation}
and hence $\mathcal{N}_{E} \ge c_{q}^{\ast}\mathcal{N}_{E}^{\left(+1\right)} c_{q}$. Moreover,  
\begin{equation}
\sum_{p\in B_{F}^{c}}c_{p}^{\ast}\mathcal{N}_{E}^{\left(-1\right)}c_{p}=\sum_{p\in B_{F}^{c}}\left(\mathcal{N}_{E}c_{p}^{\ast}-c_{p}^{\ast}\right)c_{p}=\mathcal{N}_{E}^{2}-\mathcal{N}_{E}=\sum_{p\in B_{F}}\left(\mathcal{N}_{E}c_{p}-c_{p}\right)c_{p}^{\ast}=\sum_{p\in B_{F}}c_{p}\mathcal{N}_{E}^{\left(+1\right)}c_{p}^{\ast}.
\end{equation}
$\hfill\square$

In some cases, it is important to refine error estimates by using the kinetic operator $H_{\kin}^{\prime}$
rather than $\mathcal{N}_{E}$. We can  implement  the kinetic estimate of Proposition \ref{prop:BkBkAPriori}
in the generalized setting: 
%Obtaining these kinetic estimates
%is the subject of the remainder of this section, and we start by deriving
%estimates for the ``pure'' quasi-bosonic operators which we must
%consider. First we implement the kinetic estimate of Proposition \ref{prop:BkBkAPriori}
%in the generalized setting:
\begin{prop}
\label{prop:GeneralizedKineticEstimates}For all $k\in\mathbb{Z}_{+}^{3}$,
$\varphi\in\ell^{2}\left(L_{k}^{\pm}\right)$ and $\Psi\in D\left(H_{\kin}^{\prime}\right)$
it holds that
\[
\left\Vert b_{k}\left(\varphi\right)\Psi\right\Vert \leq\left\Vert \left(h_{k}^{\oplus}\right)^{-\frac{1}{2}}\varphi\right\Vert \sqrt{\left\langle \Psi,H_{\kin}^{\prime}\Psi\right\rangle },\quad\left\Vert b_{k}^{\ast}\left(\varphi\right)\Psi\right\Vert \leq\left\Vert \left(h_{k}^{\oplus}\right)^{-\frac{1}{2}}\varphi\right\Vert \sqrt{\left\langle \Psi,H_{\kin}^{\prime}\Psi\right\rangle }+\left\Vert \varphi\right\Vert \left\Vert \Psi\right\Vert .
\]
\end{prop}

\textbf{Proof:} We start by applying the Cauchy-Schwarz inequality  
\begin{align}
\left\Vert b_{k}\left(\varphi\right)\Psi\right\Vert   =\left\Vert \sum_{p\in L_{k}^{\pm}}\left\langle \varphi,e_{p}\right\rangle b_{\overline{k,p}}\Psi\right\Vert %\leq\sum_{p\in L_{k}^{\pm}}\frac{\sqrt{\lambda_{\overline{k,p}}}}{\sqrt{\lambda_{\overline{k,p}}}}\left|\left\langle \varphi,e_{p}\right\rangle \right|\left\Vert b_{\overline{k,p}}\Psi\right\Vert \\
 \leq\sqrt{\sum_{p\in L_{k}^{\pm}}\lambda_{\overline{k,p}}^{-1}\left|\left\langle \varphi,e_{p}\right\rangle \right|^{2}}\sqrt{\sum_{p\in L_{k}^{\pm}}\lambda_{\overline{k,p}}\left\Vert b_{\overline{k,p}}\Psi\right\Vert ^{2}}. 
\end{align}
As the vectors $\left(e_{p}\right)_{p\in L_{k}^{\pm}}$ obey $h_{k}^{\oplus}e_{p}=\lambda_{\overline{k,p}}e_{p}$
we recognize the first sum on the right-hand side as
\begin{equation}
\sum_{p\in L_{k}^{\pm}}\lambda_{\overline{k,p}}^{-1}\left|\left\langle \varphi,e_{p}\right\rangle \right|^{2}=\left\langle \varphi,\left(h_{k}^{\oplus}\right)^{-1}\varphi\right\rangle =\left\Vert \left(h_{k}^{\oplus}\right)^{-\frac{1}{2}}\varphi\right\Vert ^{2}.
\end{equation}
For the second sum we have by equation (\ref{eq:BasicKineticArgument})
that
\begin{equation}
\sum_{p\in L_{k}^{\pm}}\lambda_{\overline{k,p}}\left\Vert b_{\overline{k,p}}\Psi\right\Vert ^{2}=\sum_{p\in L_{k}}\lambda_{k,p}\left\Vert c_{p-k}^{\ast}c_{p}\Psi\right\Vert ^{2}+\sum_{p\in L_{-k}}\lambda_{-k,p}\left\Vert c_{p+k}^{\ast}c_{p}\Psi\right\Vert ^{2}\leq\left\langle \Psi,H_{\kin}^{\prime}\Psi\right\rangle \label{eq:BasicKineticArgument2}
\end{equation}
which implies the first claim. The second bound follows from the first and Proposition
\ref{prop:DirectExchangeCorrectionIsNegative}:
\begin{equation}
\left\Vert b_{k}^{\ast}\left(\varphi\right)\Psi\right\Vert \leq\sqrt{\left\langle \Psi,\left(\left\Vert \left(h_{k}^{\oplus}\right)^{-\frac{1}{2}}\varphi\right\Vert ^{2}H_{\kin}^{\prime}+\left\Vert \varphi\right\Vert ^{2}\right)\Psi\right\rangle }\leq\left\Vert \left(h_{k}^{\oplus}\right)^{-\frac{1}{2}}\varphi\right\Vert \sqrt{\left\langle \Psi,H_{\kin}^{\prime}\Psi\right\rangle }+\left\Vert \varphi\right\Vert \left\Vert \Psi\right\Vert .
\end{equation}
$\hfill\square$

\subsection{Preliminary Estimates for Quadratic Operators} \label{sec:Preliminary Estimates for Quadratic Operators}

In this subsection, we provide some basic bounds on the quadratic operators   $Q_{1}^{k}\left(A\right)$ and 
$Q_{1}^{k}\left(B\right)$ defined in \eqref{eq:Q1kQ2kFirstDefinition} for any $k\in\mathbb{Z}_{+}^{3}$. 
%In the previous section we defined the quadratic operators $Q_{1}^{k}\left(A\right)$,
%$Q_{1}^{k}\left(B\right)$, for any $k\in\mathbb{Z}_{+}^{3}$ and
%symmetric $A,B:\ell^{2}\left(L_{k}^{\pm}\right)\rightarrow\ell^{2}\left(L_{k}^{\pm}\right)$,
%by
%\begin{align}
%Q_{1}^{k}\left(A\right) & =\sum_{p\in L_{k}^{\pm}}\left(b_{k}^{\ast}\left(Ae_{p}\right)b_{k}\left(e_{p}\right)+b_{k}\left(e_{p}\right)b_{k}^{\ast}\left(Ae_{p}\right)\right)\\
%Q_{2}^{k}\left(B\right) & =\sum_{p\in L_{k}^{\pm}}\left(b_{k}^{\ast}\left(Be_{p}\right)b_{k}^{\ast}\left(e_{p}\right)+b_{k}\left(e_{p}\right)b_{k}\left(Be_{p}\right)\right)\nonumber 
%\end{align}
%and we now derive some general estimates for these operators.
First, for $Q_{1}^{k}\left(A\right)$ we can normal order: 
\begin{align}
Q_{1}^{k}\left(A\right) & =\sum_{p\in L_{k}^{\pm}}\left(2\,b_{k}^{\ast}\left(Ae_{p}\right)b_{k}\left(e_{p}\right)+\left[b_{k}\left(e_{p}\right),b_{k}^{\ast}\left(Ae_{p}\right)\right]\right)\nonumber \\
 & =2\sum_{p\in L_{k}^{\pm}}b_{k}^{\ast}\left(Ae_{p}\right)b_{k}\left(e_{p}\right)+\sum_{p\in L_{k}^{\pm}}\left\langle e_{p},Ae_{p}\right\rangle +\sum_{p\in L_{k}^{\pm}}\varepsilon_{k,k}\left(e_{p};Ae_{p}\right)\label{eq:NormalOrderedQ1A}\\
 & =2\,\tilde{Q}_{1}^{k}\left(A\right)+\text{tr}\left(A\right)+\varepsilon_{k}\left(A\right)\nonumber 
\end{align}
where for brevity, we  have defined the notation
\begin{equation} \label{eq:def-Q1-tilde}
\tilde{Q}_{1}^{k}\left(A\right)=\sum_{p\in L_{k}^{\pm}}b_{k}^{\ast}\left(Ae_{p}\right)b_{k}\left(e_{p}\right),\quad\varepsilon_{k}\left(A\right)=\sum_{p\in L_{k}^{\pm}}\varepsilon_{k,k}\left(e_{p};Ae_{p}\right).
\end{equation}
The term $\tilde{Q}_{1}^{k}\left(A\right)$ plays the same role of $\text{d}\Gamma\left(A\right)$ in the exact bosonic case, whereas $\varepsilon_{k}\left(A\right)$
is a correction term in the quasi-bosonic case. 
%In the exact bosonic case $\tilde{Q}_{1}^{k}\left(A\right)$ would
%simply be $\text{d}\Gamma\left(A\right)$, 
%whereas $\varepsilon_{k}\left(A\right)$
%is a purely quasi-bosonic correction term.
%The basis-independence guaranteed by Lemma \ref{lemma:TraceFormLemma}
%now allows us to estimate both of these terms without much difficulty.
%First we have $\tilde{Q}_{1}^{k}\left(A\right)$:
\begin{prop}
\label{prop:Q1TildeNumberEstimate}For all $k\in\mathbb{Z}_{+}^{3}$,
symmetric $A:\ell^{2}\left(L_{k}^{\pm}\right)\rightarrow\ell^{2}\left(L_{k}^{\pm}\right)$
and $\Psi\in\mathcal{H}_{N}$ it holds that
\begin{align*}
\left|\left\langle \Psi,\tilde{Q}_{1}^{k}\left(A\right)\Psi\right\rangle \right| &\leq \left\Vert A\right\Vert _{\Op}\left\langle \Psi,\mathcal{N}_{E}\Psi\right\rangle ,\\
\left|\left\langle \Psi,\varepsilon_{k}\left(A\right)\Psi\right\rangle \right| &\leq 3\left\Vert A\right\Vert _{\Op}\left\langle \Psi,\mathcal{N}_{E}\Psi\right\rangle .
\end{align*}
If furthermore $A\geq0$ then also $\tilde{Q}_{1}^{k}\left(A\right)\geq0$.
\end{prop}

\textbf{Proof:} Let $\left(x_{i}\right)_i$
be an eigenbasis for $A$ with eigenvalues $\left(\lambda_{i}\right)_i$.
Noting that the mapping $x,y\mapsto b_{k}^{\ast}\left(Ax\right)b_{k}\left(y\right)$
is bilinear we may invoke Lemma \ref{lemma:TraceFormLemma} (the part of basis independence) to write
\begin{equation}
\tilde{Q}_{1}^{k}\left(A\right)=\sum_{i}b_{k}^{\ast}\left(Ax_{i}\right)b_{k}\left(x_{i}\right)=\sum_{i}\lambda_{i}b_{k}^{\ast}\left(x_{i}\right)b_{k}\left(x_{i}\right).
\end{equation}
Clearly if $A\ge 0$, then all $\lambda_i\ge 0$ and hence $\tilde{Q}_{1}^{k}\left(A\right)\ge 0$. In general, we always have $|\lambda_i | \le  \left\Vert A\right\Vert _{\Op}$ for all $i$. Hence,  using Lemma \ref{lemma:TraceFormLemma} again and $b_{\overline{k,p}}^* b_{\overline{k,p}} \le c_p^* c_p$ we have
\begin{align}
\pm \tilde{Q}_{1}^{k}\left(A\right) \le \left\Vert A\right\Vert _{\text{Op}} \sum_{i}b_{k}^{\ast}\left(x_{i}\right)b_{k}\left(x_{i}\right)= \left\Vert A\right\Vert _{\text{Op}} \sum_{p\in L_{k}^{\pm}}b_{\overline{k,p}}^{\ast}b_{\overline{k,p}} \le  \left\Vert A\right\Vert _{\text{Op}} \sum_{p\in L_{k}^{\pm}} c_p^* c_p \le   \left\Vert A\right\Vert _{\text{Op}} \mathcal{N}_{E}. 
\end{align}
%By Lemma \ref{lemma:TraceFormLemma} again we see that $\sum_{i}b_{k}^{\ast}\left(x_{i}\right)b_{k}\left(x_{i}\right)=\sum_{p\in L_{k}^{\pm}}b_{\overline{k,p}}^{\ast}b_{\overline{k,p}}$
%and so this inequality implies
%\begin{equation}
%\left|\left\langle \Psi,\tilde{Q}_{1}^{k}\left(A\right)\Psi\right\rangle \right|\leq\left\Vert A\right\Vert _{\text{Op}}\sum_{p\in L_{k}^{\pm}}\left\Vert b_{\overline{k,p}}\Psi\right\Vert ^{2}\leq\left\Vert A\right\Vert _{\text{Op}}\sum_{p\in L_{k}^{\pm}}\left\langle \Psi,c_{p}^{\ast}c_{p}\Psi\right\rangle \leq\left\Vert A\right\Vert _{\text{Op}}\left\langle \Psi,\mathcal{N}_{E}\Psi\right\rangle 
%\end{equation}
%
%Using $|\lambda_i | \le  \left\Vert A\right\Vert _{\Op}$, Lemma \ref{lemma:TraceFormLemma} again and $\left\Vert b_{\overline{k,p}}\Psi\right\Vert \leq\left\Vert c_{p}\Psi\right\Vert $, 
%where we also used that $\left\Vert b_{\overline{k,p}}\Psi\right\Vert \leq\left\Vert c_{p}\Psi\right\Vert $.
%

%Now $\varepsilon_{k}\left(A\right)$:
Similarly,
\begin{align}
\pm \varepsilon_{k}\left(A\right)&= \pm \sum_{i}\varepsilon_{k,k}\left(x_{i};Ax_{i}\right)= \pm \sum_{i}\lambda_{i}\varepsilon_{k,k}\left(x_{i};x_{i}\right) \nonumber\\
&\le - \|A\|_{\rm Op}\sum_{i} \varepsilon_{k,k}\left(x_{i};x_{i}\right) = - \|A\|_{\rm Op} \sum_{p\in L_{k}^{\pm}}\varepsilon_{k,k}\left(e_{p};e_{p}\right)
\end{align}
where in the first inequality we used the fact that $\varepsilon_{k,k}\left(x_{i};x_{i}\right)\le 0$ as shown in the proof of Proposition \ref{prop:DirectExchangeCorrectionIsNegative}. Using $\varepsilon_{k,k}\left(e_{p};e_{p}\right)=\varepsilon\left(\overline{k,p};\overline{l,q}\right)$ and the definition (\ref{eq:ExchangeCorrectionDefinition}), we get
%\footnote{Note that in the final estimate the sum over the hole states was expanded
%to avoid undercounting - while $L_{k}\cap L_{-k}=\emptyset$ it is
%generally the case that $\left(L_{k}-k\right)\cap\left(L_{-k}+k\right)\neq\emptyset$
%(a single hole state may be ``shared'' by both lunes), so when estimating
%in terms of a single sum over $p\in B_{F}$ a factor of $2$ is often
%necessary.}
\begin{align}
- \sum_{p\in L_{k}^{\pm}}\varepsilon\left(\overline{k,p};\overline{k,p}\right)=\sum_{p\in L_{k}^{\pm}}\left(c_{\overline{p-k}}c_{\overline{p-k}}^{\ast}+c_{p}^{\ast}c_{p}\right)
 \leq 2 \sum_{p\in B_{F}} c_{p}c_{p}^{\ast}  +\sum_{p\in B_{F}^{c}} c_{p}^{\ast}c_{p} =3 \mathcal{N}_{E}   
\end{align}
which implies the desired claim.
$\hfill\square$

From these results and  equation (\ref{eq:NormalOrderedQ1A}) we immediately obtain 
%the following estimate on $Q_{1}^{k}\left(A\right)$,
%which we used in the previous section to justify Proposition \ref{prop:QuasiBosonicQuadraticOperatorTransformation}:
\begin{prop}
\label{prop:Q1kAEstimate}For all $k\in\mathbb{Z}_{+}^{3}$, symmetric
$A:\ell^{2}\left(L_{k}^{\pm}\right)\rightarrow\ell^{2}\left(L_{k}^{\pm}\right)$
and $\Psi\in\mathcal{H}_{N}$ it holds that
\[
\left|\left\langle \Psi,\left(Q_{1}^{k}\left(A\right)-\tr\left(A\right)\right)\Psi\right\rangle \right|\leq5\left\Vert A\right\Vert _{\Op}\left\langle \Psi,\mathcal{N}_{E}\Psi\right\rangle .
\]
\end{prop}

Next, we turn to $Q_{2}^{k}\left(B\right)$. 

\begin{prop}
\label{prop:Q2kBEstimate}For all $k\in\mathbb{Z}_{+}^{3}$, symmetric
$B:\ell^{2}\left(L_{k}^{\pm}\right)\rightarrow\ell^{2}\left(L_{k}^{\pm}\right)$
and $\Psi\in\mathcal{H}_{N}$ it holds that
\[
\left|\left\langle \Psi,Q_{2}^{k}\left(B\right)\Psi\right\rangle \right|\leq2\left\Vert B\right\Vert _{\HS}\sqrt{\left\langle \Psi,\left(1+\mathcal{N}_{E}\right)\Psi\right\rangle \left\langle \Psi,\mathcal{N}_{E}\Psi\right\rangle }\leq2\left\Vert B\right\Vert _{\HS}\left\langle \Psi,\left(1+\mathcal{N}_{E}\right)\Psi\right\rangle .
\]
\end{prop}

\textbf{Proof:} We have (using that the $b_{k}$ operators commute)
\begin{align}
\left\langle \Psi,Q_{2}^{k}\left(B\right)\Psi\right\rangle  &=\sum_{p\in L_{k}^{\pm}}\left\langle \Psi,\left(b_{k}^{\ast}\left(Be_{p}\right)b_{k}^{\ast}\left(e_{p}\right)+b_{k}\left(e_{p}\right)b_{k}\left(Be_{p}\right)\right)\Psi\right\rangle \nonumber\\
&=2\sum_{p\in L_{k}^{\pm}}{\rm Re} \left\langle b_{k}^{\ast}\left(Be_{p}\right)\Psi,b_{k}\left(e_{p}\right)\Psi\right\rangle 
\end{align}
so using the estimates of Proposition \ref{prop:GeneralExcitationOperatorEstimate}
and the Cauchy-Schwarz inequality we conclude that
%\begin{align}
% & \quad\;\left|\left\langle \Psi,Q_{2}^{k}\left(B\right)\Psi\right\rangle \right|\leq2\sum_{p\in L_{k}^{\pm}}\left|\left\langle b_{k}^{\ast}\left(Be_{p}\right)\Psi,b_{k}\left(e_{p}\right)\Psi\right\rangle \right|\leq2\sum_{p\in L_{k}^{\pm}}\left\Vert b_{k}^{\ast}\left(Be_{p}\right)\Psi\right\Vert \left\Vert b_{k}\left(e_{p}\right)\Psi\right\Vert \nonumber \\
% & \leq2\sqrt{\left\langle \Psi,\left(1+\mathcal{N}_{E}\right)\Psi\right\rangle }\sum_{p\in L_{k}^{\pm}}\left\Vert Be_{p}\right\Vert \left\Vert b_{\overline{k,p}}\Psi\right\Vert \leq2\sqrt{\left\langle \Psi,\left(1+\mathcal{N}_{E}\right)\Psi\right\rangle }\sqrt{\sum_{p\in L_{k}^{\pm}}\left\Vert Be_{p}\right\Vert ^{2}}\sqrt{\sum_{p\in L_{k}^{\pm}}\left\Vert b_{\overline{k,p}}\Psi\right\Vert ^{2}}\nonumber \\
% & \leq2\left\Vert B\right\Vert _{\text{HS}}\sqrt{\left\langle \Psi,\left(1+\mathcal{N}_{E}\right)\Psi\right\rangle }\sqrt{\sum_{p\in L_{k}^{\pm}}\left\Vert c_{p}\Psi\right\Vert ^{2}}\leq2\left\Vert B\right\Vert _{\text{HS}}\sqrt{\left\langle \Psi,\left(1+\mathcal{N}_{E}\right)\Psi\right\rangle }\sqrt{\left\langle \Psi,\mathcal{N}_{E}\Psi\right\rangle } \nonumber  \\
% & \leq2\left\Vert B\right\Vert _{\text{HS}}\left\langle \Psi,\left(1+\mathcal{N}_{E}\right)\Psi\right\rangle 
%\end{align}
\begin{align}
 & \quad\;\left|\left\langle \Psi,Q_{2}^{k}\left(B\right)\Psi\right\rangle \right| \leq2\sum_{p\in L_{k}^{\pm}}\left\Vert b_{k}^{\ast}\left(Be_{p}\right)\Psi\right\Vert \left\Vert b_{k}\left(e_{p}\right)\Psi\right\Vert  \leq2\sqrt{\left\langle \Psi,\left(1+\mathcal{N}_{E}\right)\Psi\right\rangle }\sum_{p\in L_{k}^{\pm}}\left\Vert Be_{p}\right\Vert \left\Vert b_{\overline{k,p}}\Psi\right\Vert \nonumber\\
 & \leq2\sqrt{\left\langle \Psi,\left(1+\mathcal{N}_{E}\right)\Psi\right\rangle }\sqrt{\sum_{p\in L_{k}^{\pm}}\left\Vert Be_{p}\right\Vert ^{2}}\sqrt{\sum_{p\in L_{k}^{\pm}}\left\Vert b_{\overline{k,p}}\Psi\right\Vert ^{2}} \leq2\left\Vert B\right\Vert _{\text{HS}}\left\langle \Psi,\left(1+\mathcal{N}_{E}\right)\Psi\right\rangle 
\end{align}
where we again used that $\left\Vert b_{\overline{k,p}}\Psi\right\Vert \leq\left\Vert c_{p}\Psi\right\Vert $.
$\hfill\square$

\subsubsection*{Kinetic Estimates for Quadratic Operators}

Finally let us improve the estimates in this subsection by using the kinetic operator $H_{\kin}^{\prime}$ instead of the number operator $\mathcal{N}_E$. 

\begin{prop}
\label{prop:KineticQ1AEstimate}For all $k\in\mathbb{Z}_{+}^{3}$, symmetric
 $A:\ell^{2}\left(L_{k}^{\pm}\right)\rightarrow\ell^{2}\left(L_{k}^{\pm}\right)$
and $\Psi\in D\left(H_{\kin}^{\prime}\right)$ it holds that
\[
\left|\left\langle \Psi,\tilde{Q}_{1}^{k}\left(A\right)\Psi\right\rangle \right|\leq\left\Vert \left(h_{k}^{\oplus}\right)^{-\frac{1}{2}}A\left(h_{k}^{\oplus}\right)^{-\frac{1}{2}}\right\Vert _{\Op}\left\langle \Psi,H_{\kin}^{\prime}\Psi\right\rangle .
\]

\end{prop}

\textbf{Proof:} Let $\left(x_{i}\right)_{i}$
be an eigenbasis for $\left(h_{k}^{\oplus}\right)^{-\frac{1}{2}}A\left(h_{k}^{\oplus}\right)^{-\frac{1}{2}}$
with eigenvalues $\left(\mu_{i}\right)_i$.
By Lemma \ref{lemma:TraceFormLemma} we then see that we may write
$\tilde{Q}_{1}^{k}\left(A\right)$ as
\begin{align}
\tilde{Q}_{1}^{k}\left(A\right) & =\sum_{i}b_{k}^{\ast}\left(Ax_{i}\right)b_{k}\left(x_{i}\right)=\sum_{i}b_{k}^{\ast}\left(\left(h_{k}^{\oplus}\right)^{\frac{1}{2}}\left(h_{k}^{\oplus}\right)^{-\frac{1}{2}}A\left(h_{k}^{\oplus}\right)^{-\frac{1}{2}}\left(h_{k}^{\oplus}\right)^{\frac{1}{2}}x_{i}\right)b_{k}\left(x_{i}\right)\nonumber \\
 & =\sum_{i}b_{k}^{\ast}\left(\left(h_{k}^{\oplus}\right)^{\frac{1}{2}}\left(h_{k}^{\oplus}\right)^{-\frac{1}{2}}A\left(h_{k}^{\oplus}\right)^{-\frac{1}{2}}x_{i}\right)b_{k}\left(\left(h_{k}^{\oplus}\right)^{\frac{1}{2}}x_{i}\right)\\
 & =\sum_{i}\mu_{i}b_{k}^{\ast}\left(\left(h_{k}^{\oplus}\right)^{\frac{1}{2}}x_{i}\right)b_{k}\left(\left(h_{k}^{\oplus}\right)^{\frac{1}{2}}x_{i}\right),\nonumber 
\end{align}
and so we can estimate
\begin{align}
\left|\left\langle \Psi,\tilde{Q}_{1}^{k}\left(A\right)\Psi\right\rangle \right|  &\leq\left(\max_{1\leq i\leq\left|L_{k}^{\pm}\right|}\left|\mu_{i}\right|\right)\sum_{i}\left\langle \Psi,b_{k}^{\ast}\left(\left(h_{k}^{\oplus}\right)^{\frac{1}{2}}x_{i}\right)b_{k}\left(\left(h_{k}^{\oplus}\right)^{\frac{1}{2}}x_{i}\right)\Psi\right\rangle \\
 & =\left\Vert \left(h_{k}^{\oplus}\right)^{-\frac{1}{2}}A\left(h_{k}^{\oplus}\right)^{-\frac{1}{2}}\right\Vert _{\text{Op}}\left\langle \Psi,\sum_{i}b_{k}^{\ast}\left(\left(h_{k}^{\oplus}\right)^{\frac{1}{2}}x_{i}\right)b_{k}\left(\left(h_{k}^{\oplus}\right)^{\frac{1}{2}}x_{i}\right)\Psi\right\rangle .\nonumber 
\end{align}
Applying Lemma \ref{lemma:TraceFormLemma} again we also see that
\begin{equation}
\sum_{i}b_{k}^{\ast}\left(\left(h_{k}^{\oplus}\right)^{\frac{1}{2}}x_{i}\right)b_{k}\left(\left(h_{k}^{\oplus}\right)^{\frac{1}{2}}x_{i}\right)=\sum_{p\in L_{k}^{\pm}}b_{k}^{\ast}\left(\left(h_{k}^{\oplus}\right)^{\frac{1}{2}}e_{p}\right)b_{k}\left(\left(h_{k}^{\oplus}\right)^{\frac{1}{2}}e_{p}\right)=\sum_{p\in L_{k}^{\pm}}\lambda_{\overline{k,p}}b_{\overline{k,p}}^{\ast}b_{\overline{k,p}}
\end{equation}
so by equation (\ref{eq:BasicKineticArgument2}) we obtain the desired
bound of
\begin{equation}
\left|\left\langle \Psi,\tilde{Q}_{1}^{k}\left(A\right)\Psi\right\rangle \right|\leq\left\Vert \left(h_{k}^{\oplus}\right)^{-\frac{1}{2}}A\left(h_{k}^{\oplus}\right)^{-\frac{1}{2}}\right\Vert _{\text{Op}}\left\langle \Psi,H_{\kin}^{\prime}\Psi\right\rangle .
\end{equation}
%For the second statement we simply note that if $\left(x_{i}\right)_{i}$
%instead denotes an eigenbasis for $A\geq0$ (with eigenvalues $\left(\mu_{i}\right)_i$)
%then
%\begin{equation}
%\tilde{Q}_{1}^{k}\left(A\right)=\sum_{i}b_{k}^{\ast}\left(Ax_{i}\right)b_{k}\left(x_{i}\right)=\sum_{i}\mu_{i}b_{k}^{\ast}\left(x_{i}\right)b_{k}\left(x_{i}\right)\geq0
%\end{equation}
%as $\mu_{i}\geq0$ for all $1\leq i\leq\left|L_{k}^{\pm}\right|$.
%
$\hfill\square$

Next are the $\varepsilon_{k}\left(A\right)$ terms. These we can
not estimate in terms of $H_{\kin}^{\prime}$, but for $A$
of diagonal form we can still control them strongly:
\begin{prop}
\label{prop:epskAEstimate}For all $k\in\mathbb{Z}_{+}^{3}$, symmetric
$A^{\oplus}=\left(\begin{array}{cc}
A & 0\\
0 & A
\end{array}\right):\ell^{2}\left(L_{k}^{\pm}\right)\rightarrow\ell^{2}\left(L_{k}^{\pm}\right)$ and $\Psi\in\mathcal{H}_{N}$ it holds that
\[
\left|\left\langle \Psi,\varepsilon_{k}\left(A^{\oplus}\right)\Psi\right\rangle \right|\leq3\left(\max_{p\in L_{k}}\left|\left\langle e_{p},Ae_{p}\right\rangle \right|\right)\left\langle \Psi,\mathcal{N}_{E}\Psi\right\rangle .
\]
\end{prop}

\textbf{Proof:} By the assumed form of $A^{\oplus}$ we may write
$\varepsilon_{k}\left(A^{\oplus}\right)$ as
\begin{align}
\varepsilon_{k}\left(A^{\oplus}\right) & =\sum_{p\in L_{k}^{\pm}}\varepsilon_{k,k}\left(e_{p};A^{\oplus}e_{p}\right)=\sum_{p,q\in L_{k}^{\pm}}\left\langle e_{q},A^{\oplus}e_{p}\right\rangle \varepsilon_{k,k}\left(\overline{k,p};\overline{k,q}\right)\nonumber \\
 & =-\sum_{p,q\in L_{k}^{\pm}}\left\langle e_{q},A^{\oplus}e_{p}\right\rangle \left(\delta_{p,q}c_{\overline{q-k}}c_{\overline{p-k}}^{\ast}+\delta_{\overline{p-k},\overline{q-k}}c_{q}^{\ast}c_{p}\right)\nonumber \\
 & =-\sum_{p,q\in L_{k}}\left\langle e_{q},Ae_{p}\right\rangle \left(\delta_{p,q}c_{q-k}c_{p-k}^{\ast}+\delta_{p-k,q-k}c_{q}^{\ast}c_{p}\right)\\
 & -\sum_{p,q\in L_{-k}}\left\langle e_{-q},Ae_{-p}\right\rangle \left(\delta_{p,q}c_{q+k}c_{p+k}^{\ast}+\delta_{p+k,q+k}c_{q}^{\ast}c_{p}\right)\nonumber \\
 & =-\sum_{p\in L_{k}}\left\langle e_{p},Ae_{p}\right\rangle \left(c_{p-k}c_{p-k}^{\ast}+c_{p}^{\ast}c_{p}\right)-\sum_{p\in L_{-k}}\left\langle e_{-p},Ae_{-p}\right\rangle \left(c_{p+k}c_{p+k}^{\ast}+c_{p}^{\ast}c_{p}\right)\nonumber 
\end{align}
since the terms with $p\in L_{k},q\in L_{-k}$ or $p\in L_{-k},q\in L_{k}$
 vanish (because $L_k\cap L_{-k}=\emptyset$ and there are $\delta_{p,q}$, $\delta_{p-k,q-k}$ in the summand). We can thus estimate
\begin{align}
\left|\left\langle \Psi,\varepsilon_{k}\left(A^{\oplus}\right)\Psi\right\rangle \right| & \leq\sum_{p\in L_{k}}\left|\left\langle e_{p},Ae_{p}\right\rangle \right|\left\langle \Psi,\left(c_{p-k}c_{p-k}^{\ast}+c_{p}^{\ast}c_{p}\right)\Psi\right\rangle \nonumber \\
 & +\sum_{p\in L_{-k}}\left|\left\langle e_{-p},Ae_{-p}\right\rangle \right|\left\langle \Psi,\left(c_{p+k}c_{p+k}^{\ast}+c_{p}^{\ast}c_{p}\right)\Psi\right\rangle \\
 & \leq\left(\max_{p\in L_{k}}\left|\left\langle e_{p},Ae_{p}\right\rangle \right|\right)\left\langle \Psi,\left(\sum_{p\in L_{k}^{\pm}}c_{p}^{\ast}c_{p}+\sum_{p\in L_{k}-k}c_{p}c_{p}^{\ast}+\sum_{p\in L_{-k}+k}c_{p}c_{p}^{\ast}\right)\Psi\right\rangle \nonumber \\
 & \leq3\left(\max_{p\in L_{k}}\left|\left\langle e_{p},Ae_{p}\right\rangle \right|\right)\left\langle \Psi,\mathcal{N}_{E}\Psi\right\rangle .\nonumber 
\end{align}
$\hfill\square$

Lastly we consider the $Q_{2}^{k}\left(B\right)$ terms:
\begin{prop}
\label{prop:KineticQ2BEstimate}For all $k\in\mathbb{Z}_{+}^{3}$, 
symmetric $B:\ell^{2}\left(L_{k}^{\pm}\right)\rightarrow\ell^{2}\left(L_{k}^{\pm}\right)$
and $\Psi\in D\left(H_{\kin}^{\prime}\right)$ it holds that
\[
\left|\left\langle \Psi,Q_{2}^{k}\left(B\right)\Psi\right\rangle \right|\leq2\left\Vert \left(h_{k}^{\oplus}\right)^{-\frac{1}{2}}B\left(h_{k}^{\oplus}\right)^{-\frac{1}{2}}\right\Vert _{\HS}\left\langle \Psi,H_{\kin}^{\prime}\Psi\right\rangle +2\left\Vert B\left(h_{k}^{\oplus}\right)^{-\frac{1}{2}}\right\Vert _{\HS}\sqrt{\left\langle \Psi,H_{\kin}^{\prime}\Psi\right\rangle }\left\Vert \Psi\right\Vert .
\]
\end{prop}

\textbf{Proof:} By the Cauchy-Schwarz inequality and Proposition \ref{prop:GeneralizedKineticEstimates}
we have  
\begin{align}
& \left|\left\langle \Psi,Q_{2}^{k}\left(B\right)\Psi\right\rangle \right|  =\left|2\sum_{p\in L_{k}^{\pm}}{\rm Re}\left(\left\langle \Psi,b_{k}\left(Be_{p}\right)b_{k}\left(e_{p}\right)\Psi\right\rangle \right)\right|\leq2\sum_{p\in L_{k}^{\pm}}\left\Vert b_{k}^{\ast}\left(Be_{p}\right)\Psi\right\Vert \left\Vert b_{k}\left(e_{p}\right)\Psi\right\Vert \nonumber \\
 & \leq2\sum_{p\in L_{k}^{\pm}}\left(\left\Vert \left(h_{k}^{\oplus}\right)^{-\frac{1}{2}}Be_{p}\right\Vert \sqrt{\left\langle \Psi,H_{\kin}^{\prime}\Psi\right\rangle }+\left\Vert Be_{p}\right\Vert \left\Vert \Psi\right\Vert \right)\left\Vert b_{k}\left(e_{p}\right)\Psi\right\Vert \\
 & \leq2\sqrt{\left\langle \Psi,H_{\kin}^{\prime}\Psi\right\rangle }\sum_{p\in L_{k}^{\pm}}\left\Vert \left(h_{k}^{\oplus}\right)^{-\frac{1}{2}}Be_{p}\right\Vert \left\Vert b_{k}\left(e_{p}\right)\Psi\right\Vert +2\left\Vert \Psi\right\Vert \sum_{p\in L_{k}^{\pm}}\left\Vert Be_{p}\right\Vert \left\Vert b_{k}\left(e_{p}\right)\Psi\right\Vert .\nonumber 
\end{align}
For the first sum we can again apply the Cauchy-Schwarz inequality
and  (\ref{eq:BasicKineticArgument2}): 
\begin{align}
\sum_{p\in L_{k}^{\pm}}\left\Vert \left(h_{k}^{\oplus}\right)^{-\frac{1}{2}}Be_{p}\right\Vert \left\Vert b_{k}\left(e_{p}\right)\Psi\right\Vert  & \leq\sqrt{\sum_{p\in L_{k}^{\pm}}\lambda_{\overline{k,p}}^{-1}\left\Vert \left(h_{k}^{\oplus}\right)^{-\frac{1}{2}}Be_{p}\right\Vert ^{2}}\sqrt{\sum_{p\in L_{k}^{\pm}}\lambda_{\overline{k,p}}\left\Vert b_{\overline{k,p}}\Psi\right\Vert ^{2}}\\
 & \leq\sqrt{\sum_{p\in L_{k}^{\pm}}\left\Vert \left(h_{k}^{\oplus}\right)^{-\frac{1}{2}}B\left(h_{k}^{\oplus}\right)^{-\frac{1}{2}}e_{p}\right\Vert ^{2}}\sqrt{\left\langle \Psi,H_{\kin}^{\prime}\Psi\right\rangle },\nonumber 
\end{align}
and we likewise estimate the second sum as
\begin{align}
\sum_{p\in L_{k}^{\pm}}\left\Vert Be_{p}\right\Vert \left\Vert b_{k}\left(e_{p}\right)\Psi\right\Vert  & \leq\sqrt{\sum_{p\in L_{k}^{\pm}}\lambda_{\overline{k,p}}^{-1}\left\Vert Be_{p}\right\Vert ^{2}}\sqrt{\sum_{p\in L_{k}^{\pm}}\lambda_{\overline{k,p}}\left\Vert b_{\overline{k,p}}\Psi\right\Vert ^{2}}\\
 & \leq\sqrt{\sum_{p\in L_{k}^{\pm}}\left\Vert B\left(h_{k}^{\oplus}\right)^{-\frac{1}{2}}e_{p}\right\Vert ^{2}}\sqrt{\left\langle \Psi,H_{\kin}^{\prime}\Psi\right\rangle }.\nonumber 
\end{align}
The claim now follows by recognizing the Hilbert-Schmidt norms.
$\hfill\square$

\section{The Quasi-Bosonic Bogolubov Transformation} \label{sec:quasi-bosonic-Bogolubov}

Now we are prepared to define the quasi-bosonic Bogolubov transformation that will approximately diagonalize the Hamiltonian in \eqref{eq:Quasi-quadratic}, 
\begin{align}
\sum_{k\in S_C}  \Big( Q_1^k (h_k^{\oplus} + A_k^{\oplus}) + Q_2^k (B_k^{\oplus}) - 2\, {\rm tr}(h_k)\Big),
\end{align}
where $h_k^{\oplus}$, $A_k^{\oplus}$, $B_k^{\oplus}$ are defined in \eqref{eq:def-hk-oplus} and \eqref{eq:def-Ak-Bk-plus}. 

%(recall that $S_{C}=\overline{B}\left(0,k_{F}^{\gamma}\right)\cap\mathbb{Z}_{+}^{3}$ for an exponent $1\ge \gamma>0$ which is to be optimized over at
%the end). 
We define the generator $\mathcal{K}:\mathcal{H}_{N}\rightarrow\mathcal{H}_{N}$
of the Bogolubov transformation as follows: Let $\left(K_{k}^{\oplus}\right)_{k\in S_{C}}$
be a collection of symmetric operators $K_{k}^{\oplus}:\ell^{2}\left(L_{k}^{\pm}\right)\rightarrow\ell^{2}\left(L_{k}^{\pm}\right)$.
Then we define
\begin{align} \label{eq:cK}
\mathcal{K} & =\frac{1}{2}\sum_{k\in S_{C}}\sum_{p,q\in L_{k}^{\pm}}\left\langle e_{p},K_{k}^{\oplus}e_{q}\right\rangle \left(b_{\overline{k,p}}b_{\overline{k,q}}-b_{\overline{k,q}}^{\ast}b_{\overline{k,p}}^{\ast}\right)\\
 & =\frac{1}{2}\sum_{k\in S_{C}}\sum_{p\in L_{k}^{\pm}}\left(b_{k}\left(K_{k}^{\oplus}e_{p}\right)b_{k}\left(e_{p}\right)-b_{k}^{\ast}\left(e_{p}\right)b_{k}^{\ast}\left(K_{k}^{\oplus}e_{p}\right)\right)\nonumber 
\end{align}
in analogy with equation (\ref{eq:BogolubovGeneratorDefinition}).
As in the bosonic case $\mathcal{K}$ is seen to be a skew-symmetric
operator\footnote{In the case of complex spaces, $\mathcal{K}$ is skew-symmetric if the $K_{k}^{\oplus}$'s are symmetric and $\left\langle e_{p},K_{k}^{\oplus}e_{q}\right\rangle$ are real. In our application, all relevant operators have real matrix elements, and hence we can think of the case of real spaces.}. Moreover, unlike the bosonic case, $\mathcal{K}$ is now a bounded operator by the same argument that
$Q_{1}^{k}\left(\cdot\right)$ and $Q_{2}^{k}\left(\cdot\right)$
are. Therefore, $\mathcal{K}$ generates
a unitary transformation $e^{\mathcal{K}}:\mathcal{H}_{N}\rightarrow\mathcal{H}_{N}$,
which is the quasi-bosonic Bogolubov transformation.
%The exact choice of the kernels $K_{k}^{\oplus}$ is obtained by just pretending one has the CCR and applying Section \ref{sec:BosonicBogolubovTransformations}.   But let us first discuss

The specific kernels $K_k^\oplus$ we will use are those which diagonalize the corresponding bosonic Hamiltonian exactly, but first we will consider the action of $e^\mathcal{K}$ on quadratic operators and the localized kinetic operator more generally.

\subsection{Transformation of Quadratic Operators}

By exploiting the similarity of our quasi-bosonic definitions with
the exact bosonic case we can now easily deduce the analogues of Propositions
\ref{prop:BogolubovCAKCommutators} and \ref{prop:BogolubovQKCommutators}:
\begin{prop}
\label{prop:QuasiBosonicExcitationKCommutators}For all $k\in S_{C}$,
$\varphi\in\ell^{2}\left(L_{k}^{\pm}\right)$ and symmetric operators
$\left(K_{l}^{\oplus}\right)_{l\in S_{C}}$ it holds that
\begin{align*}
\left[\mathcal{K},b_{k}\left(\varphi\right)\right] & =b_{k}^{\ast}\left(K_{k}^{\oplus}\varphi\right)+\mathcal{E}_{k}\left(\varphi\right),\\
\left[\mathcal{K},b_{k}^{\ast}\left(\varphi\right)\right] & =b_{k}\left(K_{k}^{\oplus}\varphi\right)+\mathcal{E}_{k}\left(\varphi\right)^{\ast}
\end{align*}
where
\[
\mathcal{E}_{k}\left(\varphi\right)=\frac{1}{2}\sum_{l\in S_{C}}\sum_{q\in L_{l}^{\pm}}\left\{ b_{l}^{\ast}\left(K_{l}^{\oplus}e_{q}\right),\varepsilon_{k,l}\left(\varphi;e_{q}\right)\right\} .
\]
\end{prop}

\textbf{Proof:} We calculate using the commutation relations of (\ref{eq:QuasiBosonicCommutationRelations})
that
\begin{align}
\left[\mathcal{K},b_{k}\left(\varphi\right)\right] & =\frac{1}{2}\sum_{l\in S_{C}}\sum_{q\in L_{l}^{\pm}}\left(\left[b_{l}\left(K_{l}^{\oplus}e_{q}\right)b_{l}\left(e_{p}\right),b_{k}\left(\varphi\right)\right]-\left[b_{l}^{\ast}\left(e_{q}\right)b_{l}^{\ast}\left(K_{l}^{\oplus}e_{q}\right),b_{k}\left(\varphi\right)\right]\right)\nonumber \\
 & =\frac{1}{2}\sum_{l\in S_{C}}\sum_{q\in L_{l}^{\pm}}\left(b_{l}^{\ast}\left(e_{q}\right)\left[b_{k}\left(\varphi\right),b_{l}^{\ast}\left(K_{l}^{\oplus}e_{q}\right)\right]+\left[b_{k}\left(\varphi\right),b_{l}^{\ast}\left(e_{q}\right)\right]b_{l}^{\ast}\left(K_{l}^{\oplus}e_{q}\right)\right)\nonumber \\
 & =\frac{1}{2}\sum_{l\in S_{C}}\sum_{q\in L_{l}^{\pm}}b_{l}^{\ast}\left(e_{q}\right)\left(\delta_{k,l}\left\langle \varphi,K_{l}^{\oplus}e_{q}\right\rangle +\varepsilon_{k,l}\left(\varphi;K_{l}^{\oplus}e_{q}\right)\right)\\
 & +\frac{1}{2}\sum_{l\in S_{C}}\sum_{q\in L_{l}^{\pm}}\left(\delta_{k,l}\left\langle \varphi,e_{q}\right\rangle +\varepsilon_{k,l}\left(\varphi;e_{q}\right)\right)b_{l}^{\ast}\left(K_{l}^{\oplus}e_{q}\right)\nonumber \\
 & =\frac{1}{2}b_{k}^{\ast}\left(\sum_{q\in L_{k}^{\pm}}\left\langle \varphi,K_{k}^{\oplus}e_{q}\right\rangle e_{q}\right)+\frac{1}{2}b_{k}^{\ast}\left(K_{k}^{\oplus}\sum_{q\in L_{k}^{\pm}}\left\langle \varphi,e_{q}\right\rangle e_{q}\right)+\mathcal{E}_{k}\left(\varphi\right)\nonumber \\
 & =b_{k}^{\ast}\left(K_{k}^{\oplus}\varphi\right)+\mathcal{E}_{k}\left(\varphi\right)\nonumber 
\end{align}
for $\mathcal{E}_{k}\left(\varphi\right)$ given by
\begin{align}
\mathcal{E}_{k}\left(\varphi\right) & =\frac{1}{2}\sum_{l\in S_{C}}\sum_{q\in L_{l}^{\pm}}\left(b_{l}^{\ast}\left(e_{q}\right)\varepsilon_{k,l}\left(\varphi;K_{l}^{\oplus}e_{q}\right)+\varepsilon_{k,l}\left(\varphi;e_{q}\right)b_{l}^{\ast}\left(K_{l}^{\oplus}e_{q}\right)\right)\\
 & =\frac{1}{2}\sum_{l\in S_{C}}\sum_{q\in L_{l}^{\pm}}\left\{ b_{l}^{\ast}\left(K_{l}^{\oplus}e_{q}\right),\varepsilon_{k,l}\left(\varphi;e_{q}\right)\right\} \nonumber 
\end{align}
where we used Lemma \ref{lemma:TraceFormLemma} to simplify the expression
(as $x,y\mapsto b_{l}^{\ast}\left(x\right)\varepsilon_{k,l}\left(\varphi;y\right)$
is bilinear for fixed $\varphi$ and $K_{k}^{\oplus}$ is symmetric).
The commutator $\left[\mathcal{K},b_{k}^{\ast}\left(\varphi\right)\right]$
follows by taking the adjoint.
$\hfill\square$

From this we easily deduce the commutator of $\mathcal{K}$ with quadratic
operators:
\begin{prop}
\label{prop:QuasiBosonicQuadraticCommutators}For all $k\in S_{C}$
and symmetric operators $A,B:\ell^{2}\left(L_{k}^{\pm}\right)\rightarrow\ell^{2}\left(L_{k}^{\pm}\right)$
it holds that
\begin{align*}
\left[\mathcal{K},Q_{1}^{k}\left(A\right)\right] & =Q_{2}^{k}\left(\left\{ K_{k}^{\oplus},A\right\} \right)+\mathcal{E}_{1}^{k}\left(A\right)\\
\left[\mathcal{K},Q_{2}^{k}\left(B\right)\right] & =Q_{1}^{k}\left(\left\{ K_{k}^{\oplus},B\right\} \right)+\mathcal{E}_{2}^{k}\left(B\right)
\end{align*}
where
\begin{align*}
\mathcal{E}_{1}^{k}\left(A\right) & =\frac{1}{2}\sum_{l\in S_{C}}\sum_{p\in L_{k}^{\pm}}\sum_{q\in L_{l}^{\pm}}\left(\left\{ b_{k}^{\ast}\left(Ae_{p}\right),\left\{ b_{l}^{\ast}\left(K_{l}^{\oplus}e_{q}\right),\varepsilon_{k,l}\left(e_{p};e_{q}\right)\right\} \right\} \right.\\
 & \qquad\qquad\qquad\quad\,\left.+\left\{ \left\{ \varepsilon_{l,k}\left(e_{q};e_{p}\right),b_{l}\left(K_{l}^{\oplus}e_{q}\right)\right\} ,b_{k}\left(Ae_{p}\right)\right\} \right)\\
\mathcal{E}_{2}^{k}\left(B\right) & =\frac{1}{2}\sum_{l\in S_{C}}\sum_{p\in L_{k}^{\pm}}\sum_{q\in L_{l}^{\pm}}\left(\left\{ b_{k}^{\ast}\left(Be_{p}\right),\left\{ b_{l}\left(K_{l}^{\oplus}e_{q}\right),\varepsilon_{l,k}\left(e_{q};e_{p}\right)\right\} \right\} \right.\\
 & \qquad\qquad\qquad\quad\,\left.+\left\{ \left\{ \varepsilon_{k,l}\left(e_{p};e_{q}\right),b_{l}^{\ast}\left(K_{l}^{\oplus}e_{q}\right)\right\} ,b_{k}\left(Be_{p}\right)\right\} \right).
\end{align*}
\end{prop}

\textbf{Proof:} We compute using the commutators of the previous proposition
(and Lemma \ref{lemma:TraceFormLemma}, to simplify the resulting
expressions) that
\begin{align}
\left[\mathcal{K},Q_{1}^{k}\left(A\right)\right] & =\sum_{p\in L_{k}^{\pm}}\left(\left[\mathcal{K},b_{k}^{\ast}\left(Ae_{p}\right)b_{k}\left(e_{p}\right)\right]+\left[\mathcal{K},b_{k}\left(e_{p}\right)b_{k}^{\ast}\left(Ae_{p}\right)\right]\right)\nonumber \\
 & =\sum_{p\in L_{k}^{\pm}}\left(b_{k}^{\ast}\left(Ae_{p}\right)\left[\mathcal{K},b_{k}\left(e_{p}\right)\right]+\left[\mathcal{K},b_{k}^{\ast}\left(Ae_{p}\right)\right]b_{k}\left(e_{p}\right)\right)\nonumber \\
 & +\sum_{p\in L_{k}^{\pm}}\left(b_{k}\left(e_{p}\right)\left[\mathcal{K},b_{k}^{\ast}\left(Ae_{p}\right)\right]+\left[\mathcal{K},b_{k}\left(e_{p}\right)\right]b_{k}^{\ast}\left(Ae_{p}\right)\right)\\
 & =\sum_{p\in L_{k}^{\pm}}\left(b_{k}^{\ast}\left(Ae_{p}\right)\left(b_{k}^{\ast}\left(K_{k}^{\oplus}e_{p}\right)+\mathcal{E}_{k}\left(e_{p}\right)\right)+\left(b_{k}\left(K_{k}^{\oplus}Ae_{p}\right)+\mathcal{E}_{k}\left(Ae_{p}\right)^{\ast}\right)b_{k}\left(e_{p}\right)\right)\nonumber \\
 & +\sum_{p\in L_{k}^{\pm}}\left(b_{k}\left(e_{p}\right)\left(b_{k}\left(K_{k}^{\oplus}Ae_{p}\right)+\mathcal{E}_{k}\left(Ae_{p}\right)^{\ast}\right)+\left(b_{k}^{\ast}\left(K_{k}^{\oplus}e_{p}\right)+\mathcal{E}_{k}\left(e_{p}\right)\right)b_{k}^{\ast}\left(Ae_{p}\right)\right)\nonumber \\
 & =\sum_{p\in L_{k}^{\pm}}\left(b_{k}^{\ast}\left(\left(AK_{k}^{\oplus}+K_{k}^{\oplus}A\right)e_{p}\right)b_{k}^{\ast}\left(e_{p}\right)+b_{k}\left(e_{p}\right)b_{k}\left(\left(K_{k}^{\oplus}A+K_{k}^{\oplus}A\right)e_{p}\right)\right)\nonumber \\
 & +\sum_{p\in L_{k}^{\pm}}\left(b_{k}^{\ast}\left(Ae_{p}\right)\mathcal{E}_{k}\left(e_{p}\right)+\mathcal{E}_{k}\left(e_{p}\right)^{\ast}b_{k}\left(Ae_{p}\right)+b_{k}\left(Ae_{p}\right)\mathcal{E}_{k}\left(e_{p}\right)^{\ast}+\mathcal{E}_{k}\left(e_{p}\right)b_{k}^{\ast}\left(Ae_{p}\right)\right)\nonumber \\
 & =Q_{2}^{k}\left(\left\{ K_{k}^{\oplus},A\right\} \right)+\sum_{p\in L_{k}^{\pm}}\left(\left\{ b_{k}^{\ast}\left(Ae_{p}\right),\mathcal{E}_{k}\left(e_{p}\right)\right\} +\left\{ \mathcal{E}_{k}\left(e_{p}\right)^{\ast},b_{k}\left(Ae_{p}\right)\right\} \right)\nonumber 
\end{align}
and
\begin{align}
 & \qquad\sum_{p\in L_{k}^{\pm}}\left(\left\{ b_{k}^{\ast}\left(Ae_{p}\right),\mathcal{E}_{k}\left(e_{p}\right)\right\} +\left\{ \mathcal{E}_{k}\left(e_{p}\right)^{\ast},b_{k}\left(Ae_{p}\right)\right\} \right)\nonumber \\
 & =\frac{1}{2}\sum_{l\in S_{C}}\sum_{p\in L_{k}^{\pm}}\sum_{q\in L_{l}^{\pm}}\left(\left\{ b_{k}^{\ast}\left(Ae_{p}\right),\left\{ b_{l}^{\ast}\left(K_{l}^{\oplus}e_{q}\right),\varepsilon_{k,l}\left(e_{p};e_{q}\right)\right\} \right\} \right.\\
 & \qquad\qquad\qquad\quad\;\left.+\left\{ \left\{ \varepsilon_{l,k}\left(e_{q};e_{p}\right),b_{l}\left(K_{l}^{\oplus}e_{q}\right)\right\} ,b_{k}\left(Ae_{p}\right)\right\} \right)=\mathcal{E}_{1}^{k}\left(A\right) \nonumber 
% \\
% & =\mathcal{E}_{1}^{k}\left(A\right)\nonumber 
\end{align}
as $\varepsilon_{k,l}\left(e_{p};e_{q}\right)^{\ast}=\varepsilon_{l,k}\left(e_{q};e_{p}\right)$. The computation of $Q_{2}^{k}\left(B\right)$ is similar.
$\hfill\square$

\subsection*{Action of $e^{\mathcal{K}}$ on Quadratic Operators}

With the commutators calculated we are now ready to determine the
full action of $e^{\mathcal{K}}$ on the quadratic operators $Q_{1}^{k}\left(\cdot\right)$
and $Q_{2}^{k}\left(\cdot\right)$. Rather than appeal to the Baker-Campbell-Hausdorff
formula, which would also require describing the commutators $\left[\mathcal{K},\mathcal{E}_{1}^{k}\left(A\right)\right]$,
etc., we will employ a ``Duhamel-type'' argument which allows us
to more selectively expand the operator $e^{\mathcal{K}}$.

\medskip
As in the section \ref{sec:BosonicBogolubovTransformations} we use the notation $\mathcal{A}_{K_{k}^{\oplus}}=\left\{ K_{k}^{\oplus},\cdot\right\}$
for anticommutators with $K_{k}^{\oplus}$. % - this will allow us to keep the equations to a manageable size. 

\medskip

Before stating the proposition we must make a remark: To use these
identities we will need to take limits, and to justify those limits
we need some general estimates on operators of the form $Q_{1}^{k}\left(\cdot\right),Q_{2}^{k}\left(\cdot\right),\mathcal{E}_{1}^{k}\left(\cdot\right),\mathcal{E}_{2}^{k}\left(\cdot\right)$. The Propositions \ref{prop:Q1kAEstimate}, \ref{prop:Q2kBEstimate} establish these for $Q_{1}^{k}\left(\cdot\right)$ and $Q_{2}^{k}\left(\cdot\right)$, while Proposition \ref{prop:ExchangeTermEstimates} will establish these for $\mathcal{E}_{1}^{k}\left(\cdot\right)$ and $\mathcal{E}_{2}^{k}\left(\cdot\right)$.

%(probably Prp’s 4.6, 4.7, 6.4 as mentioned at
%the end of the next proof?)

%For the sake of presentation we will however postpone those estimates
%until the next section.

Now the statement:
\begin{prop}
\label{prop:QuasiBosonicQuadraticOperatorTransformation}For all $k\in S_{C}$
and symmetric $A,B:\ell^{2}\left(L_{k}^{\pm}\right)\rightarrow\ell^{2}\left(L_{k}^{\pm}\right)$
it holds that
\begin{align*}
e^{\mathcal{K}}Q_{1}^{k}\left(A\right)e^{-\mathcal{K}} & =\frac{1}{2}Q_{1}^{k}\left(e^{K_{k}^{\oplus}}Ae^{K_{k}^{\oplus}}+e^{-K_{k}^{\oplus}}Ae^{-K_{k}^{\oplus}}\right)+\frac{1}{2}Q_{2}^{k}\left(e^{K_{k}^{\oplus}}Ae^{K_{k}^{\oplus}}-e^{-K_{k}^{\oplus}}Ae^{-K_{k}^{\oplus}}\right)\\
 & +\int_{0}^{1}e^{t\mathcal{K}}\left(\mathcal{E}_{1}^{k}\left(\cosh\left(\mathcal{A}_{\left(1-t\right)K_{k}^{\oplus}}\right)\left(A\right)\right)+\mathcal{E}_{2}^{k}\left(\sinh\left(\mathcal{A}_{\left(1-t\right)K_{k}^{\oplus}}\right)\left(A\right)\right)\right)e^{-t\mathcal{K}}\,dt\\
e^{\mathcal{K}}Q_{2}^{k}\left(B\right)e^{-\mathcal{K}} & =\frac{1}{2}Q_{1}^{k}\left(e^{K_{k}^{\oplus}}Be^{K_{k}^{\oplus}}-e^{-K_{k}^{\oplus}}Be^{-K_{k}^{\oplus}}\right)+\frac{1}{2}Q_{2}^{k}\left(e^{K_{k}^{\oplus}}Be^{K_{k}^{\oplus}}+e^{-K_{k}^{\oplus}}Be^{-K_{k}^{\oplus}}\right)\\
 & +\int_{0}^{1}e^{t\mathcal{K}}\left(\mathcal{E}_{1}^{k}\left(\sinh\left(\mathcal{A}_{\left(1-t\right)K_{k}^{\oplus}}\right)\left(B\right)\right)+\mathcal{E}_{2}^{k}\left(\cosh\left(\mathcal{A}_{\left(1-t\right)K_{k}^{\oplus}}\right)\left(B\right)\right)\right)e^{-t\mathcal{K}}\,dt,
\end{align*}
the integrals being Riemann integrals of bounded operators. 
%with respect to the norm\footnote{The estimates of the next section which we use are with respect to
%$\mathcal{N}_{E}$, and are uniform estimates of the associated quadratic
%forms of the respective operators on the diagonal of $\mathcal{H}_{N}\times\mathcal{H}_{N}$.
%This does indeed imply the stated norm convergence: Note that for
%fixed $k_{F}$, $\mathcal{N}_{E}$ is in fact a bounded operator by
%the particle-hole symmetry of equation (\ref{eq:ParticleHoleSymmetry}),
%so $\left\langle \Psi,\mathcal{N}_{E}\Psi\right\rangle \leq C\left\Vert \Psi\right\Vert ^{2}$.
%Furthermore, as the operators we estimate are self-adjoint, norm convergence
%follows by the identity $\left\Vert S\right\Vert _{\text{Op}}=\max_{\left\Vert \Psi\right\Vert =1}\left|\left\langle \Psi,S\Psi\right\rangle \right|$
%for self-adjoint $S$.} topology of $\mathcal{B}\left(\mathcal{H}_{N}\right)$.
\end{prop}

\textbf{Proof:} We consider $e^{\mathcal{K}}Q_{1}^{k}\left(A\right)e^{-\mathcal{K}}$,
the argument for $e^{\mathcal{K}}Q_{2}^{k}\left(B\right)e^{-\mathcal{K}}$
being similar. We first claim that for any $n\in\mathbb{N}$
\begin{align}
 & e^{\mathcal{K}}Q_{1}^{k}\left(A\right)e^{-\mathcal{K}}=Q_{1}^{k}\left(\sum_{m=0}^{n_{1}}\frac{1}{\left(2m\right)!}\mathcal{A}_{K_{k}^{\oplus}}^{2m}\left(A\right)\right)+Q_{2}^{k}\left(\sum_{m=0}^{n_{2}}\frac{1}{\left(2m+1\right)!}\mathcal{A}_{K_{k}^{\oplus}}^{2m+1}\left(A\right)\right)\label{eq:DuhamelInductiveStatement}\\
 & +\int_{0}^{1}e^{t\mathcal{K}}\left(\mathcal{E}_{1}^{k}\left(\sum_{m=0}^{n_{1}}\frac{1}{\left(2m\right)!}\mathcal{A}_{\left(1-t\right)K_{k}^{\oplus}}^{2m}\left(A\right)\right)+\mathcal{E}_{2}^{k}\left(\sum_{m=0}^{n_{2}}\frac{1}{\left(2m+1\right)!}\mathcal{A}_{\left(1-t\right)K_{k}^{\oplus}}^{2m+1}\left(A\right)\right)\right)e^{-t\mathcal{K}}\,dt\nonumber \\
 & +\frac{1}{\left(n-1\right)!}\int_{0}^{1}e^{t\mathcal{K}}Q_{\overline{n-1}}^{k}\left(\mathcal{A}_{K_{k}^{\oplus}}^{n}\left(A\right)\right)e^{-t\mathcal{K}}\left(1-t\right)^{n-1}\,dt,\nonumber 
\end{align}
where for brevity $\overline{n-1}=n-1\mod2$ and $n_{1},n_{2}$ are
the largest integers such that $2n_{1}<n$ and $2n_{2}+1<n$, respectively.

We proceed by induction. For $n=1$ we find by the fundamental theorem
of calculus that
\begin{align}
e^{\mathcal{K}}Q_{1}^{k}\left(A\right)e^{-\mathcal{K}} & =Q_{1}^{k}\left(A\right)+\int_{0}^{1}\frac{d}{dt}\left(e^{t\mathcal{K}}Q_{1}^{k}\left(A\right)e^{-t\mathcal{K}}\right)\,dt=Q_{1}^{k}\left(A\right)+\int_{0}^{1}e^{t\mathcal{K}}\left[\mathcal{K},Q_{1}^{k}\left(A\right)\right]e^{-t\mathcal{K}}\,dt\nonumber \\
 & =Q_{1}^{k}\left(A\right)+\int_{0}^{1}e^{t\mathcal{K}}\left(Q_{2}^{k}\left(\left\{ K_{k}^{\oplus},A\right\} \right)+\mathcal{E}_{1}^{k}\left(A\right)\right)e^{-t\mathcal{K}}\,dt\\
 & =Q_{1}^{k}\left(A\right)+\int_{0}^{1}e^{t\mathcal{K}}\mathcal{E}_{1}^{k}\left(A\right)e^{-t\mathcal{K}}\,dt+\int_{0}^{1}e^{t\mathcal{K}}Q_{2}^{k}\left(\mathcal{A}_{K_{k}^{\oplus}}\left(A\right)\right)e^{-t\mathcal{K}}\,dt\nonumber 
\end{align}
by the commutator of Proposition \ref{prop:QuasiBosonicQuadraticCommutators},
which is the statement for $n=1$ (in this case $n_1=0$ and $n_2=-1$, so $\sum_{m=0}^{n_1}$ contains one term and  $\sum_{m=0}^{n_2}$ is empty). 

\medskip

For the inductive step we now assume that case $n$ holds. Integrating
the last term of equation (\ref{eq:DuhamelInductiveStatement}) by
parts we find that
\begin{align}
 & \quad\;\frac{1}{\left(n-1\right)!}\int_{0}^{1}e^{t\mathcal{K}}Q_{\overline{n-1}}^{k}\left(\mathcal{A}_{K_{k}^{\oplus}}^{n}\left(A\right)\right)e^{-t\mathcal{K}}\left(1-t\right)^{n-1}dt\nonumber \\
 & =\frac{1}{\left(n-1\right)!}\left[e^{t\mathcal{K}}Q_{\overline{n-1}}^{k}\left(\mathcal{A}_{K_{k}^{\oplus}}^{n}\left(A\right)\right)e^{-t\mathcal{K}}\left(-\frac{\left(1-t\right)^{n}}{n}\right)\right]_{0}^{1}\nonumber \\
 &  -\frac{1}{\left(n-1\right)!}\int_{0}^{1}e^{t\mathcal{K}}\left[\mathcal{K},Q_{\overline{n-1}}^{k}\left(\mathcal{A}_{K_{k}^{\oplus}}^{n}\left(A\right)\right)\right]e^{-t\mathcal{K}}\left(-\frac{\left(1-t\right)^{n}}{n}\right)dt\\
 & =\frac{1}{n!}Q_{\overline{n-1}}^{k}\left(\mathcal{A}_{K_{k}^{\oplus}}^{n}\left(A\right)\right)+\frac{1}{n!}\int_{0}^{1}e^{t\mathcal{K}}\left(Q_{\overline{n}}^{k}\left(\left\{ K_{k}^{\oplus},\mathcal{A}_{K_{k}^{\oplus}}^{n}\left(A\right)\right\} \right)+\mathcal{E}_{\overline{n-1}}^{k}\left(\mathcal{A}_{K_{k}^{\oplus}}^{n}\left(A\right)\right)\right)e^{-t\mathcal{K}}\left(1-t\right)^{n}dt\nonumber \\
 & =Q_{\overline{n-1}}^{k}\left(\frac{1}{n!}\mathcal{A}_{K_{k}^{\oplus}}^{n}\left(A\right)\right)+\int_{0}^{1}e^{t\mathcal{K}}\mathcal{E}_{\overline{n-1}}^{k}\left(\frac{1}{n!}\mathcal{A}_{\left(1-t\right)K_{k}^{\oplus}}^{n}\left(A\right)\right)e^{-t\mathcal{K}}\,dt\nonumber \\
 &  +\frac{1}{n!}\int_{0}^{1}e^{t\mathcal{K}}Q_{\overline{n}}^{k}\left(\mathcal{A}_{K_{k}^{\oplus}}^{n+1}\left(A\right)\right)e^{-t\mathcal{K}}\left(1-t\right)^{n}dt,\nonumber 
\end{align}
where we also used that 
\begin{align}
\left(1-t\right)^{n}\mathcal{A}_{K_{k}^{\oplus}}^{n}\left(A\right)=\left(\left(1-t\right)\mathcal{A}_{K_{k}^{\oplus}}\left(A\right)\right)^{n}=\mathcal{A}_{\left(1-t\right)K_{k}^{\oplus}}^{n}\left(A\right).
\end{align}
Inserting this into (\ref{eq:DuhamelInductiveStatement})
and collecting like terms yields the statement for case $n+1$.

We now deduce the statement from \eqref{eq:DuhamelInductiveStatement} by taking $n\to \infty$. Recall the identities
\begin{align} 
\cosh\left(\mathcal{A}_{K_{k}^{\oplus}}\right)\left(T\right) & =\frac{1}{2}\left(e^{K_{k}^{\oplus}}Te^{K_{k}^{\oplus}}+e^{-K_{k}^{\oplus}}Te^{-K_{k}^{\oplus}}\right) \label{eq:HyperbolicAnticommutatorIdentities}\\
\sinh\left(\mathcal{A}_{K_{k}^{\oplus}}\right)\left(T\right) & =\frac{1}{2}\left(e^{K_{k}^{\oplus}}Te^{K_{k}^{\oplus}}-e^{-K_{k}^{\oplus}}Te^{-K_{k}^{\oplus}}\right)\nonumber 
\end{align}
from Proposition \ref{prop:anticommutatorBCH} and note that $\left(\left(n-1\right)!\right)^{-1}\mathcal{A}_{K_{k}^{\oplus}}^{n}\left(A\right)\to 0$ as $n\to \infty$. By Proposition \ref{prop:Q1kAEstimate}, 
$$
Q_{1}^{k}\left(\sum_{m=0}^{n_{1}}\frac{1}{\left(2m\right)!}\mathcal{A}_{K_{k}^{\oplus}}^{2m}\left(A\right)\right)\to \frac{1}{2}Q_{1}^{k}\left(e^{K_{k}^{\oplus}}Ae^{K_{k}^{\oplus}}+e^{-K_{k}^{\oplus}}Ae^{-K_{k}^{\oplus}}\right)
$$
and 
$$
\frac{1}{\left(n-1\right)!}\int_{0}^{1}e^{t\mathcal{K}}Q_{1}^{k}\left(\mathcal{A}_{K_{k}^{\oplus}}^{n}\left(A\right)\right)e^{-t\mathcal{K}}\left(1-t\right)^{n-1}\,dt \to 0.
$$
Similar convergence for $Q_2$ are justified by Proposition \ref{prop:Q2kBEstimate}. The convergence for $\mathcal{E}_{1}^{k}$ and $\mathcal{E}_{2}^{k}$ follow from Proposition \ref{prop:ExchangeTermEstimates}.
$\hfill\square$

\subsubsection*{Remark on the Transformation of Excitation Operators}

Let us make a quick remark on why we choose to approach the Bogolubov
transformation from the point of view of quadratic operators rather
than the usual creation and annihilation operator approach. Recall that in the exact bosonic case the creation and annihilation
operators transformed under a Bogolubov transformation as
\begin{align}
e^{\mathcal{K}}a\left(\varphi\right)e^{-\mathcal{K}} & =a\left(\cosh\left(K\right)\varphi\right)+a^{\ast}\left(\sinh\left(K\right)\varphi\right)\\
e^{\mathcal{K}}a^{\ast}\left(\varphi\right)e^{-\mathcal{K}} & =a^{\ast}\left(\cosh\left(K\right)\varphi\right)+a\left(\sinh\left(K\right)\varphi\right).\nonumber 
\end{align}
In the quasi-bosonic setting we can use the commutators of Proposition
\ref{prop:QuasiBosonicExcitationKCommutators} and a similar Duhamel-type
argument to what we just applied to conclude that 
\begin{align}
e^{\mathcal{K}}b_{k}\left(\varphi\right)e^{-\mathcal{K}} & =b_{k}\left(\cosh\left(K_{k}^{\oplus}\right)\varphi\right)+b_{k}^{\ast}\left(\sinh\left(K_{k}^{\oplus}\right)\varphi\right)\label{eq:ExcitationOperatorTransformation}\\
 & +\int_{0}^{1}e^{t\mathcal{K}}\left(\mathcal{E}_{k}\left(\cosh\left(\left(1-t\right)K_{k}^{\oplus}\right)\varphi\right)+\mathcal{E}_{k}\left(\sinh\left(\left(1-t\right)K_{k}^{\oplus}\right)\varphi\right)^{\ast}\right)e^{-t\mathcal{K}}\,dt\nonumber 
\end{align}
with a similar expression for $e^{\mathcal{K}}b_{k}^{\ast}\left(\varphi\right)e^{-\mathcal{K}}$.
This is a more cumbersome expression to work with, and if we were
to describe $e^{\mathcal{K}}Q_{1}^{k}\left(A\right)e^{-\mathcal{K}}$
by transforming the individual terms of $Q_{1}^{k}\left(A\right)$
like this rather than transforming $Q_{1}^{k}\left(A\right)$ as a
whole, the error terms would not only go from being under a single
integral to involving the product of two integrals, it would also
involve cross terms between the bosonic terms and the error terms
of equation (\ref{eq:ExcitationOperatorTransformation}). These cross
terms, in particular, would severely reduce the quality of the final
error estimate. Hence, we prefer the quadratic operator approach in the quasi-bosonic
setting.

\subsection{Transformation of the Kinetic Operator}

There remains the task of describing the action of $e^{\mathcal{K}}$
on the localized kinetic operator $H_{\kin}^{\prime}$. 
%For
%this we must first formulate $H_{\kin}^{\prime}$ - or rather,
%the commutator
%\begin{equation}
%\left[H_{\kin}^{\prime},b_{k,p}^{\ast}\right]=\left(\left|p\right|^{2}-\left|p-k\right|^{2}\right)b_{k,p}^{\ast},\quad k\in\mathbb{Z}_{\ast}^{3},\,p\in L_{k},\label{eq:KineticOperatorCommutatorReminder}
%\end{equation}
%that we calculated in \eqref{LocalizedKineticOperatorCommutator}
%- within the general framework that we have introduced in this section.
%This is an easy task: Recall that in \eqref{eq:def-hk-oplus} we defined the operators $h_{k}^{\oplus}:\ell^{2}\left(L_{k}^{\pm}\right)\rightarrow\ell^{2}\left(L_{k}^{\pm}\right)$ by
%\begin{align}
%h_{k}^{\oplus}e_{p}  =\lambda_{\overline{k,p}}e_{p}, \quad \lambda_{\overline{k,p}} =\frac{1}{2}\left(\left|p\right|^{2}-\left|\overline{p-k}\right|^{2}\right), \quad p\in L_{k}^{\pm}
%\end{align}
%where $\left(e_{p}\right)_{p\in L_{k}^{\pm}}$ denotes
%the standard orthonormal basis of $\ell^{2}\left(L_{k}^{\pm}\right)$. Then equation (\ref{eq:KineticOperatorCommutatorReminder}) can be
%expressed as
%\begin{align}
%delete\\
%delete\\
%delete
%\end{align}
For this we must first formulate $H_{\kin}^{\prime}$ - or rather the commutator $[H_{\kin}^{\prime},b_{k,p}^{\ast}]$ calculated in \eqref{LocalizedKineticOperatorCommutator} - within the general framework that we have introduced in this section.
Recalling  the operators $h_{k}^{\oplus}:\ell^{2}\left(L_{k}^{\pm}\right)\rightarrow\ell^{2}\left(L_{k}^{\pm}\right)$ in \eqref{eq:def-hk-oplus}, then  by (\ref{LocalizedKineticOperatorCommutator}) and linearity it follows that 
%\begin{equation}
%\left[H_{\kin}^{\prime},b_{k}^{\ast}\left(e_{p}\right)\right]=2\lambda_{\overline{k,p}}b_{k}^{\ast}\left(e_{p}\right)=2\,b_{k}^{\ast}\left(h_{k}^{\oplus}e_{p}\right),\quad k\in\mathbb{Z}_{+}^{3},\,p\in L_{k}^{\pm},
%\end{equation}
%and so by linearity (and by taking the adjoint) it follows that with
%respect to the generalized excitation operators $b_{k}\left(\varphi\right)$,
%$b_{k}^{\ast}\left(\varphi\right)$ (for $\varphi\in\ell^{2}\left(L_{k}^{\pm}\right)$)
%the commutator is
\begin{equation}
\left[H_{\kin}^{\prime},b_{k}\left(\varphi\right)\right]=-2\,b_{k}\left(h_{k}^{\oplus}\varphi\right),\quad\left[H_{\kin}^{\prime},b_{k}^{\ast}\left(\varphi\right)\right]=2\,b_{k}^{\ast}\left(h_{k}^{\oplus}\varphi\right)\label{eq:GeneralizedKineticCommutator}
\end{equation}
for all  $\varphi\in\ell^{2}\left(L_{k}^{\pm}\right)$. (The factor of $2$ is introduced here because in the analogy of equation
(\ref{eq:KineticEnergyAnalogyReminder}) $H_{\kin}^{\prime}$
appears like a $\text{d}\Gamma\left(\cdot\right)=\frac{1}{2}Q_{1}\left(\cdot\right)-\frac{1}{2}\text{tr}\left(\cdot\right)$
term rather than a pure $Q_{1}\left(\cdot\right)$ term.)

We now calculate $\left[\mathcal{K},H_{\kin}^{\prime}\right]$
as follows:
\begin{prop}
\label{prop:HKincalKCommutator}$H_{\kin}^{\prime}$ obeys
\[
\left[\mathcal{K},H_{\kin}^{\prime}\right]=\sum_{k\in S_{C}}Q_{2}^{k}\left(\left\{ K_{k}^{\oplus},h_{k}^{\oplus}\right\} \right).
\]
\end{prop}

\textbf{Proof:} We compute, using the commutators of equation (\ref{eq:GeneralizedKineticCommutator})
and Lemma \ref{lemma:TraceFormLemma}, that
\begin{align}
\left[\mathcal{K},H_{\kin}^{\prime}\right] & =\frac{1}{2}\sum_{k\in S_{C}}\sum_{p\in L_{k}^{\pm}}\left(\left[b_{k}\left(K_{k}^{\oplus}e_{p}\right)b_{k}\left(e_{p}\right),H_{\kin}^{\prime}\right]-\left[b_{k}^{\ast}\left(e_{p}\right)b_{k}^{\ast}\left(K_{k}^{\oplus}e_{p}\right),H_{\kin}^{\prime}\right]\right)\nonumber \\
% & =\frac{1}{2}\sum_{k\in S_{C}}\sum_{p\in L_{k}^{\pm}}\left(b_{k}^{\ast}\left(e_{p}\right)\left[H_{\kin}^{\prime},b_{k}^{\ast}\left(K_{k}^{\oplus}e_{p}\right)\right]+\left[H_{\kin}^{\prime},b_{k}^{\ast}\left(e_{p}\right)\right]b_{k}^{\ast}\left(K_{k}^{\oplus}e_{p}\right)\right)\nonumber \\
% & \qquad -\frac{1}{2}\sum_{k\in S_{C}}\sum_{p\in L_{k}^{\pm}}\left(b_{k}\left(K_{k}^{\oplus}e_{p}\right)\left[H_{\kin}^{\prime},b_{k}\left(e_{p}\right)\right]+\left[H_{\kin}^{\prime},b_{k}\left(K_{k}^{\oplus}e_{p}\right)\right]b_{k}\left(e_{p}\right)\right)\nonumber \\
 & =\sum_{k\in S_{C}}\sum_{p\in L_{k}^{\pm}}\left(b_{k}^{\ast}\left(e_{p}\right)b_{k}^{\ast}\left(h_{k}^{\oplus}K_{k}^{\oplus}e_{p}\right)+b_{k}^{\ast}\left(h_{k}^{\oplus}e_{p}\right)b_{k}^{\ast}\left(K_{k}^{\oplus}e_{p}\right)\right)\\
 & +\sum_{k\in S_{C}}\sum_{p\in L_{k}^{\pm}}\left(b_{k}\left(K_{k}^{\oplus}e_{p}\right)b_{k}\left(h_{k}^{\oplus}e_{p}\right)+b_{k}\left(h_{k}^{\oplus}K_{k}^{\oplus}e_{p}\right)b_{k}\left(e_{p}\right)\right)\nonumber \\
% & =\sum_{k\in S_{C}}\sum_{p\in L_{k}^{\pm}}\left(b_{k}^{\ast}\left(\left(K_{k}^{\oplus}h_{k}^{\oplus}+h_{k}^{\oplus}K_{k}^{\oplus}\right)e_{p}\right)b_{k}^{\ast}\left(e_{p}\right)+b_{k}\left(e_{p}\right)b_{k}\left(\left(h_{k}^{\oplus}K_{k}^{\oplus}+K_{k}^{\oplus}h_{k}^{\oplus}\right)e_{p}\right)\right)\nonumber \\
 & =\sum_{k\in S_{C}}Q_{2}^{k}\left(\left\{ K_{k}^{\oplus},h_{k}^{\oplus}\right\} \right).\nonumber 
\end{align}
$\hfill\square$

Note that because the commutator $\left[H_{\kin}^{\prime},b_{k}^{\ast}\left(\varphi\right)\right]=2\,b_{k}^{\ast}\left(h_{k}^{\oplus}\varphi\right)$
exactly mirrors the bosonic case (in that there is no additional error
term) the commutator $\left[\mathcal{K},H_{\kin}^{\prime}\right]$
is likewise ``purely bosonic'', being simply a sum of $Q_{2}^{k}\left(\cdot\right)$
terms without error terms such as those appearing in the statement
of Proposition \ref{prop:QuasiBosonicQuadraticCommutators}. With the groundwork laid we can now easily deduce %$e^{\mathcal{K}}H_{\kin}^{\prime}e^{-\mathcal{K}}$:
\begin{prop}
\label{prop:LocalizedKineticEnergyTransformation}$H_{\kin}^{\prime}$
obeys
\begin{align*}
 & e^{\mathcal{K}}H_{\kin}^{\prime}e^{-\mathcal{K}}=H_{\kin}^{\prime}\\
 & +\sum_{k\in S_{C}}\left(\frac{1}{2}Q_{1}^{k}\left(e^{K_{k}^{\oplus}}h_{k}^{\oplus}e^{K_{k}^{\oplus}}+e^{-K_{k}^{\oplus}}h_{k}^{\oplus}e^{-K_{k}^{\oplus}}-2h_{k}^{\oplus}\right)+\frac{1}{2}Q_{2}^{k}\left(e^{K_{k}^{\oplus}}h_{k}^{\oplus}e^{K_{k}^{\oplus}}-e^{-K_{k}^{\oplus}}h_{k}^{\oplus}e^{-K_{k}^{\oplus}}\right)\right)\\
 & +\sum_{k\in S_{C}}\int_{0}^{1}e^{t\mathcal{K}}\left(\mathcal{E}_{1}^{k}\left(\cosh\left(\mathcal{A}_{\left(1-t\right)K_{k}^{\oplus}}\right)\left(h_{k}^{\oplus}\right)-h_{k}^{\oplus}\right)+\mathcal{E}_{2}^{k}\left(\sinh\left(\mathcal{A}_{\left(1-t\right)K_{k}^{\oplus}}\right)\left(h_{k}^{\oplus}\right)\right)\right)e^{-t\mathcal{K}}\,dt.
\end{align*}
\end{prop}

\textbf{Proof:} By adding and subtracting we have
\begin{equation}
e^{\mathcal{K}}H_{\kin}^{\prime}e^{-\mathcal{K}}=\sum_{k\in S_{C}}e^{\mathcal{K}}Q_{1}^{k}\left(h_{k}^{\oplus}\right)e^{-\mathcal{K}}+e^{\mathcal{K}}\left(H_{\kin}^{\prime}-\sum_{k\in S_{C}}Q_{1}^{k}\left(h_{k}^{\oplus}\right)\right)e^{-\mathcal{K}},
\end{equation}
and the first term on the right-hand side is by Proposition \ref{prop:QuasiBosonicQuadraticOperatorTransformation}
\begin{align}
&\quad\,\sum_{k\in S_{C}}e^{\mathcal{K}}Q_{1}^{k}\left(h_{k}^{\oplus}\right)e^{-\mathcal{K}} \\
& =\sum_{k\in S_{C}}\left(\frac{1}{2}Q_{1}^{k}\left(e^{K_{k}^{\oplus}}h_{k}^{\oplus}e^{K_{k}^{\oplus}}+e^{-K_{k}^{\oplus}}h_{k}^{\oplus}e^{-K_{k}^{\oplus}}\right)+\frac{1}{2}Q_{2}^{k}\left(e^{K_{k}^{\oplus}}h_{k}^{\oplus}e^{K_{k}^{\oplus}}-e^{-K_{k}^{\oplus}}h_{k}^{\oplus}e^{-K_{k}^{\oplus}}\right)\right) \nonumber\\
 &  +\sum_{k\in S_{C}}\int_{0}^{1}e^{t\mathcal{K}}\left(\mathcal{E}_{1}^{k}\left(\cosh\left(\mathcal{A}_{\left(1-t\right)K_{k}^{\oplus}}\right)\left(h_{k}^{\oplus}\right)\right)+\mathcal{E}_{2}^{k}\left(\sinh\left(\mathcal{A}_{\left(1-t\right)K_{k}^{\oplus}}\right)\left(h_{k}^{\oplus}\right)\right)\right)e^{-t\mathcal{K}}\,dt,\nonumber
\end{align}
while the second is calculated using the commutators of the Propositions
\ref{prop:QuasiBosonicQuadraticCommutators} and \ref{prop:HKincalKCommutator}
to be
\begin{align} \label{eq:to-be-compared-BNPSS-21}
&e^{\mathcal{K}}\left(H_{\kin}^{\prime}-\sum_{k\in S_{C}}Q_{1}^{k}\left(h_{k}^{\oplus}\right)\right)e^{-\mathcal{K}} - \left(  H_{\kin}^{\prime}-\sum_{k\in S_{C}}Q_{1}^{k}\left(h_{k}^{\oplus}\right) \right) \\
& =\int_{0}^{1}e^{t\mathcal{K}}\left[\mathcal{K},H_{\kin}^{\prime}-\sum_{k\in S_{C}}Q_{1}^{k}\left(h_{k}^{\oplus}\right)\right]e^{-t\mathcal{K}}\,dt = -\sum_{k\in S_{C}}\int_{0}^{1}e^{t\mathcal{K}}\mathcal{E}_{1}^{k}\left(h_{k}^{\oplus}\right)e^{-t\mathcal{K}}\,dt \nonumber
\end{align}
which yields the claim.
$\hfill\square$

\subsection{Fixing the Transformation Kernels}

With all the transformation identities determined we now choose the
transformation kernels $\left(K_{k}^{\oplus}\right)_{k\in S_{C}}$
such that $H_{\kin}^{\prime}+\sum_{k\in S_{C}}H_{\text{int}}^{k}$
is diagonalized. 
For any choice of $\left(K_{k}^{\oplus}\right)_{k\in S_{C}}$,
the Propositions \ref{prop:QuasiBosonicQuadraticOperatorTransformation}
and \ref{prop:LocalizedKineticEnergyTransformation}  imply that
%$e^{\mathcal{K}}\left(H_{\kin}^{\prime}+\sum_{k\in S_{C}}H_{\text{int}}^{k}\right)e^{-\mathcal{K}}$
%takes the form (omitting the error terms for the moment)
\begin{align}
 & \quad\;\,e^{\mathcal{K}}\left(H_{\kin}^{\prime}+\sum_{k\in S_{C}}H_{\text{int}}^{k}\right)e^{-\mathcal{K}} \\
% & =\frac{1}{2}\sum_{k\in S_{C}}\left(Q_{1}^{k}\left(e^{K_{k}^{\oplus}}h_{k}^{\oplus}e^{K_{k}^{\oplus}}+e^{-K_{k}^{\oplus}}h_{k}^{\oplus}e^{-K_{k}^{\oplus}}-2h_{k}^{\oplus}\right)+Q_{2}^{k}\left(e^{K_{k}^{\oplus}}h_{k}^{\oplus}e^{K_{k}^{\oplus}}-e^{-K_{k}^{\oplus}}h_{k}^{\oplus}e^{-K_{k}^{\oplus}}\right)\right)\nonumber \\
% & +\frac{1}{2}\sum_{k\in S_{C}}\left(Q_{1}^{k}\left(e^{K_{k}^{\oplus}}A_{k}^{\oplus}e^{K_{k}^{\oplus}}+e^{-K_{k}^{\oplus}}A_{k}^{\oplus}e^{-K_{k}^{\oplus}}\right)+Q_{2}^{k}\left(e^{K_{k}^{\oplus}}A_{k}^{\oplus}e^{K_{k}^{\oplus}}-e^{-K_{k}^{\oplus}}A_{k}^{\oplus}e^{-K_{k}^{\oplus}}\right)\right)\nonumber \\
% & +\frac{1}{2}\sum_{k\in S_{C}}\left(Q_{1}^{k}\left(e^{K_{k}^{\oplus}}B_{k}^{\oplus}e^{K_{k}^{\oplus}}-e^{-K_{k}^{\oplus}}B_{k}^{\oplus}e^{-K_{k}^{\oplus}}\right)+Q_{2}^{k}\left(e^{K_{k}^{\oplus}}B_{k}^{\oplus}e^{K_{k}^{\oplus}}+e^{-K_{k}^{\oplus}}B_{k}^{\oplus}e^{-K_{k}^{\oplus}}\right)\right)\\
% & +H_{\kin}^{\prime}+\text{error terms}\nonumber \\
 & =\frac{1}{2}\sum_{k\in S_{C}}Q_{1}^{k}\left(e^{K_{k}^{\oplus}}\left(h_{k}^{\oplus}+A_{k}^{\oplus}+B_{k}^{\oplus}\right)e^{K_{k}^{\oplus}}+e^{-K_{k}^{\oplus}}\left(h_{k}^{\oplus}+A_{k}^{\oplus}-B_{k}^{\oplus}\right)e^{-K_{k}^{\oplus}}-2h_{k}^{\oplus}\right)\nonumber \\
 & +\frac{1}{2}\sum_{k\in S_{C}}Q_{2}^{k}\left(e^{K_{k}^{\oplus}}\left(h_{k}^{\oplus}+A_{k}^{\oplus}+B_{k}^{\oplus}\right)e^{K_{k}^{\oplus}}-e^{-K_{k}^{\oplus}}\left(h_{k}^{\oplus}+A_{k}^{\oplus}-B_{k}^{\oplus}\right)e^{-K_{k}^{\oplus}}\right)+H_{\kin}^{\prime}+\text{error terms}.\nonumber 
\end{align}

In analogy with the bosonic case we consider this expression to be
diagonalized provided the $Q_{2}^{k}\left(\cdot\right)$ terms vanish,
whence the diagonalization condition is that
\begin{equation}
e^{K_{k}^{\oplus}}\left(h_{k}^{\oplus}+A_{k}^{\oplus}+B_{k}^{\oplus}\right)e^{K_{k}^{\oplus}}=e^{-K_{k}^{\oplus}}\left(h_{k}^{\oplus}+A_{k}^{\oplus}-B_{k}^{\oplus}\right)e^{-K_{k}^{\oplus}},
\end{equation}
which we note is the same as the diagonalization condition (equation
(\ref{eq:DiagonalizationCondition})) of the exact bosonic quadratic
Hamiltonian
\begin{equation}
H=Q_{1}\left(h_{k}^{\oplus}+A_{k}^{\oplus}\right)+Q_{2}\left(B_{k}^{\oplus}\right)\quad\text{on }\mathcal{F}^{+}\left(\ell^{2}\left(L_{k}^{\pm}\right)\right).\label{eq:CoupledBosonicHamiltonian}
\end{equation}
Recalling the definitions of $h_k^{\oplus}$, $A_k^{\oplus}$ and $B_k^{\oplus}$ from \eqref{eq:def-hk-oplus} and \eqref{eq:def-Ak-Bk-plus} we have
\begin{equation*} 
h_{k}^{\oplus}+A_{k}^{\oplus}\pm B_{k}^{\oplus} =\left(\begin{array}{cc}
h_k+P_{v_k} & \pm P_{v_{k}}\\
\pm P_k & h_k + P_{v_{k}}\\
\end{array}\right) >0.
\end{equation*}
So by Theorem \ref{thm:Bog-diag} the choice
\begin{align*}
K_{k}^{\oplus} & =-\frac{1}{2}\log\left(\left(h_{k}^{\oplus}+A_{k}^{\oplus}-B_{k}^{\oplus}\right)^{-\frac{1}{2}}\left(\left(h_{k}^{\oplus}+A_{k}^{\oplus}-B_{k}^{\oplus}\right)^{\frac{1}{2}}\left(h_{k}^{\oplus}+A_{k}^{\oplus}+B_{k}^{\oplus}\right)\left(h_{k}^{\oplus}+A_{k}^{\oplus}-B_{k}^{\oplus}\right)^{\frac{1}{2}}\right)^{\frac{1}{2}}\right.\\
 & \qquad\qquad\qquad\qquad\qquad\qquad\qquad\qquad\qquad\qquad\qquad\qquad\qquad\qquad\quad\;\;\;\left.\left(h_{k}^{\oplus}+A_{k}^{\oplus}-B_{k}^{\oplus}\right)^{-\frac{1}{2}}\right)
\end{align*}
is the unique diagonalizing kernel for the Hamiltonian. In this form
it is however not easy to see how $K_{k}^{\oplus}$ acts, so we will
proceed slightly differently: We define $K_{k}^{\oplus}:\ell^{2}\left(L_{k}^{\pm}\right)\rightarrow\ell^{2}\left(L_{k}^{\pm}\right)$
by
%\begin{equation}
%K_{k}^{\oplus}=\left(\begin{array}{cc}
%0 & K_{k}\\
%K_{k} & 0
%\end{array}\right)=\left(\begin{array}{cc}
%0 & \log\left(S_{k}\right)\\
%\log\left(S_{k}\right) & 0
%\end{array}\right)\label{eq:KkoplusDefinition}
%\end{equation}
\begin{equation}
K_{k}^{\oplus}=\left(\begin{array}{cc}
0 & K_{k}\\
K_{k} & 0
\end{array}\right)\label{eq:KkoplusDefinition}
\end{equation}
where the operator $K_{k}:\ell^{2}\left(L_{k}\right)\rightarrow\ell^{2}\left(L_{k}\right)$
is given by
\begin{equation} 
K_{k}=-\frac{1}{2}\log\left(h_{k}^{-\frac{1}{2}}\left(h_{k}^{\frac{1}{2}}\left(h_{k}+2P_{v_{k}}\right)h_{k}^{\frac{1}{2}}\right)^{\frac{1}{2}}h_{k}^{-\frac{1}{2}}\right)=-\frac{1}{2}\log\left(h_{k}^{-\frac{1}{2}}\left(h_{k}^{2}+2P_{h_{k}^{\frac{1}{2}}v_{k}}\right)^{\frac{1}{2}}h_{k}^{-\frac{1}{2}}\right).\label{eq:KkDefinition}
\end{equation}
A kernel similar to $K_k$ also appeared in \cite{BNPSS-20,BNPSS-21}. Note that $K_{k}$ is precisely the diagonalizer of Theorem \ref{thm:Bog-diag} for the exact bosonic quadratic Hamiltonian
\begin{equation}
H=Q_{1}\left(h_{k}^{\oplus}+P_{v_{k}}\right)+Q_{2}\left(P_{v_{k}}\right)\quad\text{on }\mathcal{F}^{+}\left(\ell^{2}\left(L_{k}\right)\right),\label{eq:DecoupledBosonicHamiltonian}
\end{equation}
rather than that of equation (\ref{eq:CoupledBosonicHamiltonian}).
Now we can verify that this $K_{k}^{\oplus}$ is in fact equal to
the diagonalizing kernel:
\begin{prop}
The operator $K_{k}^{\oplus}$ defined by the equations (\ref{eq:KkoplusDefinition})
and (\ref{eq:KkDefinition}) satisfies
\[
e^{K_{k}^{\oplus}}\left(h_{k}^{\oplus}+A_{k}^{\oplus}+B_{k}^{\oplus}\right)e^{K_{k}^{\oplus}}=e^{-K_{k}^{\oplus}}\left(h_{k}^{\oplus}+A_{k}^{\oplus}-B_{k}^{\oplus}\right)e^{-K_{k}^{\oplus}}=\left(\begin{array}{cc}
E_{k} & 0\\
0 & E_{k}
\end{array}\right)
\]
for $E_{k}=e^{-K_{k}}h_{k}e^{-K_{k}}$.
\end{prop}

\textbf{Proof:} It is easily verified that $e^{\pm K_{k}^{\oplus}}$
is given by
\begin{equation}
e^{\pm K_{k}^{\oplus}}=\left(\begin{array}{cc}
\cosh\left(K_{k}\right) & \pm\sinh\left(K_{k}\right)\\
\pm\sinh\left(K_{k}\right) & \cosh\left(K_{k}\right)
\end{array}\right),
\end{equation}

and so
\begin{align}
 & \quad\;\;e^{\pm K_{k}^{\oplus}}\left(h_{k}^{\oplus}+A_{k}^{\oplus}\pm B_{k}^{\oplus}\right)e^{\pm K_{k}^{\oplus}}\\
 & =\left(\begin{array}{cc}
\cosh\left(K_{k}\right) & \pm\sinh\left(K_{k}\right)\\
\pm\sinh\left(K_{k}\right) & \cosh\left(K_{k}\right)
\end{array}\right)\left(\begin{array}{cc}
h_{k}+P_{v_{k}} & \pm P_{v_{k}}\\
\pm P_{v_{k}} & h_{k}+P_{v_{k}}
\end{array}\right)\left(\begin{array}{cc}
\cosh\left(K_{k}\right) & \pm\sinh\left(K_{k}\right)\\
\pm\sinh\left(K_{k}\right) & \cosh\left(K_{k}\right)
\end{array}\right)\nonumber \\
 & =\frac{1}{2}\left(\begin{array}{cc}
e^{K_{k}}\left(h_{k}+2P_{v_{k}}\right)e^{K_{k}}+e^{-K_{k}}h_{k}e^{-K_{k}} & \pm\left(e^{K_{k}}\left(h_{k}+2P_{v_{k}}\right)e^{K_{k}}-e^{-K_{k}}h_{k}e^{-K_{k}}\right)\\
\pm\left(e^{K_{k}}\left(h_{k}+2P_{v_{k}}\right)e^{K_{k}}-e^{-K_{k}}h_{k}e^{-K_{k}}\right) & e^{K_{k}}\left(h_{k}+2P_{v_{k}}\right)e^{K_{k}}+e^{-K_{k}}h_{k}e^{-K_{k}}
\end{array}\right).\nonumber 
\end{align}
The condition 
\begin{equation}
e^{K_{k}^{\oplus}}\left(h_{k}^{\oplus}+A_{k}^{\oplus}+B_{k}^{\oplus}\right)e^{K_{k}^{\oplus}}=e^{-K_{k}^{\oplus}}\left(h_{k}^{\oplus}+A_{k}^{\oplus}-B_{k}^{\oplus}\right)e^{-K_{k}^{\oplus}}
\end{equation}
thus holds if and only if
\begin{equation}
e^{K_{k}}\left(h_{k}+2P_{v_{k}}\right)e^{K_{k}}=e^{-K_{k}}h_{k}e^{-K_{k}}
\end{equation}
which is the diagonalization condition for the bosonic Hamiltonian
of equation (\ref{eq:DecoupledBosonicHamiltonian}). Theorem \ref{thm:Bog-diag}
asserts that this condition is satisfied for our choice of $K_{k}$,
and the claim follows. $\hfill\square$

\subsection{Full Transformation of the Bosonizable Terms}

With the above choice of transformation kernels we thus conclude that  
\begin{align}
 & \quad\;e^{\mathcal{K}}\left(H_{\kin}^{\prime}+\sum_{k\in S_{C}}H_{\text{int}}^{k}\right)e^{-\mathcal{K}}=H_{\kin}^{\prime}+\text{error terms}\nonumber \\
 & +\frac{1}{2}\sum_{k\in S_{C}}Q_{1}^{k}\left(e^{K_{k}^{\oplus}}\left(h_{k}^{\oplus}+A_{k}^{\oplus}+B_{k}^{\oplus}\right)e^{K_{k}^{\oplus}}+e^{-K_{k}^{\oplus}}\left(h_{k}^{\oplus}+A_{k}^{\oplus}-B_{k}^{\oplus}\right)e^{-K_{k}^{\oplus}}-2h_{k}^{\oplus}\right)\\
 & =H_{\kin}^{\prime}+\sum_{k\in S_{C}}Q_{1}^{k}\left(\begin{array}{cc}
E_{k}-h_{k} & 0\\
0 & E_{k}-h_{k}
\end{array}\right)+\text{error terms}\nonumber 
\end{align}
and so we have succeeded in diagonalizing $H_{\kin}^{\prime}+\sum_{k\in S_{C}}H_{\text{int}}^{k}$
while simultanously decoupling the spaces $\ell^{2}\left(L_{\pm k}\right)\subset\ell^{2}\left(L_{k}^{\pm}\right)$
in a symmetric fashion. We still need to determine the exact form
of the error terms, which we record in the following proposition:

\begin{prop} \label{prop:trans-eK-bosonizable-terms} Let $S_{C}=\overline{B}\left(0,k_{F}^{\gamma}\right)\cap\mathbb{Z}_{+}^{3}$ with $\gamma \in (0,1]$. Then the unitary transformation $e^{\mathcal{K}}:\mathcal{H}_{N}\rightarrow\mathcal{H}_{N}$ with $\mathcal{K}$ defined by \eqref{eq:cK},  \eqref{eq:KkoplusDefinition}, \eqref{eq:KkDefinition} satisfies
\begin{align*}
 & \;e^{\mathcal{K}}\left(H_{\kin}^{\prime}+\sum_{k\in S_{C}}H_{\inter}^{k}\right)e^{-\mathcal{K}}=H_{\kin}^{\prime}+\sum_{k\in S_{C}}Q_{1}^{k} (E_k^{\oplus} - h_k^{\oplus})\\
 & +\sum_{k\in S_{C}}\int_{0}^{1}e^{\left(1-t\right)\mathcal{K}}\left(\mathcal{E}_{1}^{k}(A_k^\oplus(t))+\mathcal{E}_{2}^{k} (B_k^\oplus(t))\right)e^{-\left(1-t\right)\mathcal{K}}\,dt
\end{align*}
where %the exchange terms 
$\mathcal{E}_1(\cdot)$,  $\mathcal{E}_2(\cdot)$ are defined in Proposition \ref{prop:QuasiBosonicQuadraticCommutators} and 
$$
E_k^{\oplus} - h_k^{\oplus}= \left(\begin{array}{cc}
E_{k}-h_{k} & 0\\
0 & E_{k}-h_{k}
\end{array}\right),\,\, A_{k}^\oplus (t) = \left(\begin{array}{cc}
A_{k}\left(t\right) & 0\\
0 & A_{k}\left(t\right)
\end{array}\right),\,\, B_{k}^\oplus (t) = \left(\begin{array}{cc}
0 & B_{k}\left(t\right)\\
B_{k}\left(t\right) & 0
\end{array}\right)
$$
with $E_k= e^{-K_k} h_k e^{-K_k}$ and the operators $A_{k}\left(t\right),B_{k}\left(t\right):\ell^{2}\left(L_{k}\right)\rightarrow\ell^{2}\left(L_{k}\right)$ defined by
\begin{align*}
A_{k}\left(t\right) & =\frac{1}{2}\left(e^{tK_{k}}\left(h_{k}+2P_{v_{k}}\right)e^{tK_{k}}+e^{-tK_{k}}h_{k}e^{-tK_{k}}\right)-h_{k}\\
B_{k}\left(t\right) & =\frac{1}{2}\left(e^{tK_{k}}\left(h_{k}+2P_{v_{k}}\right)e^{tK_{k}}-e^{-tK_{k}}h_{k}e^{-tK_{k}}\right).
\end{align*}
\end{prop}

\textbf{Proof:} By the Propositions \ref{prop:QuasiBosonicQuadraticOperatorTransformation}
and \ref{prop:LocalizedKineticEnergyTransformation} the error terms
are
\begin{align*}
 & \quad\sum_{k\in S_{C}}\int_{0}^{1}e^{\left(1-t\right)\mathcal{K}}\left(\mathcal{E}_{1}^{k}\left(\cosh\left(\mathcal{A}_{tK_{k}^{\oplus}}\right)\left(h_{k}^{\oplus}+A_{k}^{\oplus}\right)+\sinh\left(\mathcal{A}_{tK_{k}^{\oplus}}\right)\left(B_{k}^{\oplus}\right)-h_{k}^{\oplus}\right)\right)e^{-\left(1-t\right)\mathcal{K}}\,dt\\
 & +\sum_{k\in S_{C}}\int_{0}^{1}e^{\left(1-t\right)\mathcal{K}}\left(\mathcal{E}_{2}^{k}\left(\sinh\left(\mathcal{A}_{tK_{k}^{\oplus}}\right)\left(h_{k}^{\oplus}+A_{k}^{\oplus}\right)+\cosh\left(\mathcal{A}_{tK_{k}^{\oplus}}\right)\left(B_{k}^{\oplus}\right)\right)\right)e^{-\left(1-t\right)\mathcal{K}}\,dt
\end{align*}
where we have reparametrized the integral by $t\mapsto1-t$ to simplify
the arguments of the $\mathcal{E}_{1}^{k}\left(\cdot\right)$ and
$\mathcal{E}_{2}^{k}\left(\cdot\right)$ operators. By \eqref{eq:HyperbolicAnticommutatorIdentities},  
%\begin{align}
%delete
%\end{align}
% the general
%identities
%\begin{equation}
%\cosh\left(\mathcal{A}_{K}\right)\left(T\right)=\frac{1}{2}\left(e^{K}Te^{K}+e^{-K}Te^{-K}\right),\quad\sinh\left(\mathcal{A}_{K}\right)\left(T\right)=\frac{1}{2}\left(e^{K}Te^{K}-e^{-K}Te^{-K}\right),
%\end{equation}
the arguments of $\mathcal{E}_{1}^{k}$ and $\mathcal{E}_{2}^{k}$
in each term above equal
\begin{equation}
\frac{1}{2}\left(e^{tK_{k}^{\oplus}}\left(h_{k}^{\oplus}+A_{k}^{\oplus}+B_{k}^{\oplus}\right)e^{tK_{k}^{\oplus}}+e^{-tK_{k}^{\oplus}}\left(h_{k}^{\oplus}+A_{k}^{\oplus}-B_{k}^{\oplus}\right)e^{-tK_{k}^{\oplus}}\right)-h_{k}^{\oplus}
\end{equation}
and
\begin{equation}
\frac{1}{2}\left(e^{tK_{k}^{\oplus}}\left(h_{k}^{\oplus}+A_{k}^{\oplus}+B_{k}^{\oplus}\right)e^{tK_{k}^{\oplus}}-e^{-tK_{k}^{\oplus}}\left(h_{k}^{\oplus}+A_{k}^{\oplus}-B_{k}^{\oplus}\right)e^{-tK_{k}^{\oplus}}\right),
\end{equation}
respectively. By the same identities that we used in the preceding
proposition it holds that
\begin{align}
 & \quad\;\,e^{\pm tK_{k}^{\oplus}}\left(h_{k}^{\oplus}+A_{k}^{\oplus}\pm B_{k}^{\oplus}\right)e^{\pm tK_{k}^{\oplus}}\\
 & =\frac{1}{2}\left(\begin{array}{cc}
e^{tK_{k}}\left(h_{k}+2P_{v_{k}}\right)e^{tK_{k}}+e^{-tK_{k}}h_{k}e^{-tK_{k}} & \pm\left(e^{tK_{k}}\left(h_{k}+2P_{v_{k}}\right)e^{tK_{k}}-e^{-tK_{k}}h_{k}e^{-tK_{k}}\right)\\
\pm\left(e^{tK_{k}}\left(h_{k}+2P_{v_{k}}\right)e^{tK_{k}}-e^{-tK_{k}}h_{k}e^{-tK_{k}}\right) & e^{tK_{k}}\left(h_{k}+2P_{v_{k}}\right)e^{tK_{k}}+e^{-tK_{k}}h_{k}e^{-tK_{k}}
\end{array}\right)\nonumber 
\end{align}
and the claim follows.
$\hfill\square$

\section{Analysis of the Exchange Terms} \label{sec:Analysis of Exchange Terms}

In the preceding section we accomplished a major \textit{qualitative} 
goal of this paper, which was diagonalizing the bosonizable terms $H_{\kin}^{\prime}+\sum_{k\in S_C}H_{\text{int}}^{k}$ in an explicit, quasi-bosonic fashion. In this section we begin the {\em quantitative} study of the quasi-bosonic expression in Proposition \ref{prop:trans-eK-bosonizable-terms}.

The aim of this section is to estimate the  $\mathcal{E}_{1}^{k}\left(\cdot\right)$,
$\mathcal{E}_{2}^{k}\left(\cdot\right)$ operators, which enter in
the error terms due to the presence of the exchange correction $\varepsilon_{k,l}\left(\varphi;\psi\right)$
in the quasi-bosonic commutation relations - we will therefore refer
to them as \textit{exchange terms}. Since these expressions are complicated, we thus devote three subsections to the analysis of them:
In the first we carry out a reduction procedure, in which we systematically
consider the type of terms that can appear in the sums defining $\mathcal{E}_{1}^{k}\left(A\right)$
and $\mathcal{E}_{2}^{k}\left(B\right)$ for given $A,B$, and reduce
these to simpler expressions, or \textit{schematic forms}. In doing
so we will see that every term appearing in $\mathcal{E}_{1}^{k}\left(A\right)$
and $\mathcal{E}_{2}^{k}\left(B\right)$ can for the purpose of estimation
be sorted into one of 4 schematic forms. In the second subsection we provide some basic commutator estimates associated with the 4 schematic forms, and in the final subsection
we then carry out the quantitative analysis of these 4 forms to obtain
the desired estimates of $\mathcal{E}_{1}^{k}\left(\cdot\right)$
and $\mathcal{E}_{2}^{k}\left(\cdot\right)$.

\subsection{Reduction to Simpler Expressions} \label{sec:Reduction to Simpler Expressions}

Recall that for $k\in S_C$ and symmetric operators $A,B:\ell^{2}\left(L_{k}^{\pm}\right)\rightarrow\ell^{2}\left(L_{k}^{\pm}\right)$
we already defined $\mathcal{E}_{1}^{k}\left(A\right)$ and $\mathcal{E}_{2}^{k}\left(B\right)$ in Proposition \ref{prop:QuasiBosonicQuadraticCommutators}. 
%\begin{align}
%\mathcal{E}_{1}^{k}\left(A\right) & =\frac{1}{2}\sum_{l\in S_{C}}\sum_{p\in L_{k}^{\pm}}\sum_{q\in L_{l}^{\pm}}\left(\left\{ b_{k}^{\ast}\left(Ae_{p}\right),\left\{ b_{l}^{\ast}\left(K_{l}^{\oplus}e_{q}\right),\varepsilon_{k,l}\left(e_{p};e_{q}\right)\right\} \right\} \right.\nonumber \\
% & \qquad\qquad\qquad\quad\;\left.+\left\{ \left\{ \varepsilon_{l,k}\left(e_{q};e_{p}\right),b_{l}\left(K_{l}^{\oplus}e_{q}\right)\right\} ,b_{k}\left(Ae_{p}\right)\right\} \right)\label{eq:ExchangeTermReminder}\\
%\mathcal{E}_{2}^{k}\left(B\right) & =\frac{1}{2}\sum_{l\in S_{C}}\sum_{p\in L_{k}^{\pm}}\sum_{q\in L_{l}^{\pm}}\left(\left\{ b_{k}^{\ast}\left(Be_{p}\right),\left\{ b_{l}\left(K_{l}^{\oplus}e_{q}\right),\varepsilon_{l,k}\left(e_{q};e_{p}\right)\right\} \right\} \right.\nonumber \\
% & \qquad\qquad\qquad\quad\;\left.+\left\{ \left\{ \varepsilon_{k,l}\left(e_{p};e_{q}\right),b_{l}^{\ast}\left(K_{l}^{\oplus}e_{q}\right)\right\} ,b_{k}\left(Be_{p}\right)\right\} \right).\nonumber 
%\end{align}
Since these expressions are complicated, it is helpful to discuss the
general structure of $\mathcal{E}_{1}^{k}\left(A\right)$ and $\mathcal{E}_{2}^{k}\left(B\right)$.
Consider the first term of $\mathcal{E}_{1}^{k}\left(A\right)$, which
upon expansion is  
\begin{align}
 & \;\;\left\{ b_{k}^{\ast}\left(Ae_{p}\right),\left\{ b_{l}^{\ast}\left(K_{l}^{\oplus}e_{q}\right),\varepsilon_{k,l}\left(e_{p},e_{q}\right)\right\} \right\} \nonumber \\
 & =b_{k}^{\ast}\left(Ae_{p}\right)\left\{ b_{l}^{\ast}\left(K_{l}^{\oplus}e_{q}\right),\varepsilon_{k,l}\left(e_{p},e_{q}\right)\right\} +\left\{ b_{l}^{\ast}\left(K_{l}^{\oplus}e_{q}\right),\varepsilon_{k,l}\left(e_{p},e_{q}\right)\right\} b_{k}^{\ast}\left(Ae_{p}\right)\\
 & =b_{k}^{\ast}\left(Ae_{p}\right)b_{l}^{\ast}\left(K_{l}^{\oplus}e_{q}\right)\varepsilon_{k,l}\left(e_{p},e_{q}\right)+b_{k}^{\ast}\left(Ae_{p}\right)\varepsilon_{k,l}\left(e_{p},e_{q}\right)b_{l}^{\ast}\left(K_{l}^{\oplus}e_{q}\right)\nonumber \\
 & +b_{l}^{\ast}\left(K_{l}^{\oplus}e_{q}\right)\varepsilon_{k,l}\left(e_{p},e_{q}\right)b_{k}^{\ast}\left(Ae_{p}\right)+\varepsilon_{k,l}\left(e_{p},e_{q}\right)b_{l}^{\ast}\left(K_{l}^{\oplus}e_{q}\right)b_{k}^{\ast}\left(Ae_{p}\right).\nonumber 
\end{align}
which we may expand further using 
\begin{equation}
\varepsilon_{k,l}\left(e_{p},e_{q}\right)=\varepsilon\left(\overline{k,p};\overline{l,q}\right)=-\left(\delta_{p,q}c_{\overline{q-l}}c_{\overline{p-k}}^{\ast}+\delta_{\overline{p-k},\overline{q-l}}c_{q}^{\ast}c_{p}\right)
\end{equation}
and then removing the delta on a case-by-case basis. This causes the sums over $p\in L_{k}^{\pm}$ and $q\in L_{l}^{\pm}$ of any of these terms to reduce to one of the schematic forms
\begin{equation}
\sum_{p\in S}b_{k}^{\natural}\left(Te_{p_{1}}\right)b_{l}^{\natural}\left(K_{l}^{\oplus}e_{p_{2}}\right)\tilde{c}_{p_{3}}^{\ast}\tilde{c}_{p_{4}},\quad\sum_{p\in S}b_{k}^{\natural}\left(Te_{p_{1}}\right)\tilde{c}_{p_{3}}^{\ast}\tilde{c}_{p_{4}}b_{l}^{\natural}\left(K_{l}^{\oplus}e_{p_{2}}\right),\quad\sum_{p\in S}\tilde{c}_{p_{3}}^{\ast}\tilde{c}_{p_{4}}b_{k}^{\natural}\left(Te_{p_{1}}\right)b_{l}^{\natural}\left(K_{l}^{\oplus}e_{p_{2}}\right).\label{eq:SecondSchematicForms}
\end{equation}
subject to the following: $S$ is a subset of $L_{k}^{\pm}\cap L_{l}^{\pm}$, $b_{k}^{\natural}$ can denote either $b_{k}$
or $b_{k}^{\ast}$, $\varepsilon_{k,l}\left(e_{p},e_{q}\right)$ may
instead be $\varepsilon_{l,k}\left(e_{q},e_{p}\right)=\varepsilon_{k,l}\left(e_{p},e_{q}\right)^{\ast}$,
$T$ denotes either $A$ or $B$, the terms $b_{k}^{\natural}\left(Te_{p}\right)$
and $b_{l}^{\natural}\left(K_{l}^{\oplus}e_{q}\right)$ may be interchanged, and the 
notation
\begin{align}
\tilde{c}_{p}=\begin{cases}
c_{p} & p\in B_{F}^{c}\\
c_{p}^{\ast} & p\in B_{F}
\end{cases}
\end{align} encodes the correct type of creation/annihilation operator depending
on whether $p$ corresponds to a hole state or an excited state, and
$p_{1},p_{2},p_{3},p_{4}$ denote indices which depend on $p$.

The same decomposition holds for every term appearing in either $\mathcal{E}_{1}^{k}\left(A\right)$ or $\mathcal{E}_{2}^{k}\left(B\right)$, so it the forms of \eqref{eq:SecondSchematicForms} we must consider.

The only important feature of the dependency that the $p_{i}$ have
with respect to $p$ is that regardless of the term, when summing
over $p\in S$, $p_{i}$ ranges either exclusively over excited states
(i.e. $p_{i}\in L_{k}^{\pm}$) or exclusively
over hole states (i.e. $p_{i}\in\left(L_{k}-k\right)\cup\left(L_{-k}+k\right)$
or the analogous set for $L_{l}^{\pm}$), and that the assignments $p\mapsto p_{i}$
(for a given term) are injective. (Additionally, $p_{1}$ and $p_{2}$
will always be excited states.)

Therefore when estimating we can   always expand the sum to either
all of $B_{F}$ or all of $B_{F}^{c}$, which is why the exact identities
of $S$ and the $p_{i}$ are of no importance to the estimation. For example 
\begin{equation}
\left|\sum_{p\in S}\left\langle \Psi,\tilde{c}_{p_{3}}^{\ast}\tilde{c}_{p_{4}}\Psi\right\rangle \right|  \leq\sum_{p\in S}\left\Vert \tilde{c}_{p_{3}}\Psi\right\Vert \left\Vert \tilde{c}_{p_{4}}\Psi\right\Vert \leq\sqrt{\sum_{p\in S}\left\Vert \tilde{c}_{p_{3}}\Psi\right\Vert ^{2}}\sqrt{\sum_{p\in S}\left\Vert \tilde{c}_{p_{4}}\Psi\right\Vert ^{2}} \le \left\langle \Psi,\mathcal{N}_{E}\Psi\right\rangle
\end{equation}
independently of $S$, $p_{3}$ and $p_{4}$. Here the two situations when both $p_{3}$ and $p_{4}$ range over excited states, and when both $p_{3}$ and $p_{4}$ range over hole states, can be treated similarly thanks to the particle-hole symmetry \eqref{eq:ParticleHoleSymmetry}.

\subsubsection*{Discussion of Estimation Strategy}

We conclude that both $\mathcal{E}_{1}^{k}\left(A\right)$ and $\mathcal{E}_{2}^{k}\left(B\right)$
reduce to sums over $l\in S_{C}$ of finitely many terms of the schematic
forms of equation (\ref{eq:SecondSchematicForms}), so it suffices
to estimate these. To this end we must first perform some additional algebraic
manipulation.

To motivate our goal, let us first derive a simple but insufficent
estimate for one of these terms: 
\begin{equation}
\sum_{p\in S}b_{k}^{\ast}\left(Te_{p_{1}}\right)\tilde{c}_{p_{3}}^{\ast}\tilde{c}_{p_{4}}b_{l}\left(K_{l}^{\oplus}e_{p_{2}}\right).
\end{equation}
Using $\left\Vert c_{p}\right\Vert _{\text{Op}}=1$, Proposition
\ref{prop:GeneralizedKineticEstimates} and the Cauchy--Schwarz
inequality we find that 
\begin{align} \label{eq:SimpleInsufficientEstimate}
 & \quad\,\sum_{p\in S}\left|\left\langle \Psi,b_{k}^{\ast}\left(Te_{p_{1}}\right)\tilde{c}_{p_{3}}^{\ast}\tilde{c}_{p_{4}}b_{l}\left(K_{l}^{\oplus}e_{p_{2}}\right)\Psi\right\rangle \right|
 \leq\sum_{p\in S}\left\Vert b_{k}\left(Te_{p_{1}}\right)\Psi\right\Vert \left\Vert b_{l}\left(K_{l}^{\oplus}e_{p_{2}}\right)\Psi\right\Vert \\
 & \leq\sum_{p\in S}\left\Vert \left(h_{k}^{\oplus}\right)^{-\frac{1}{2}}Te_{p_{1}}\right\Vert \left\Vert \left(h_{l}^{\oplus}\right)^{-\frac{1}{2}}K_{l}^{\oplus}e_{p_{2}}\right\Vert \left\langle \Psi,H_{\kin}^{\prime}\Psi\right\rangle \leq\left\Vert \left(h_{k}^{\oplus}\right)^{-\frac{1}{2}}T\right\Vert _{\HS}\left\Vert \left(h_{l}^{\oplus}\right)^{-\frac{1}{2}}K_{l}^{\oplus}\right\Vert _{\HS}\left\langle \Psi,H_{\kin}^{\prime}\Psi\right\rangle \nonumber 
\end{align}
for any $\Psi\in\mathcal{H}_{N}$. To get a feeling for the quality of this estimate we must know what
to expect of the quantities on the right-hand side. We will see in
the next sections that  $\left\Vert \left(h_{l}^{\oplus}\right)^{-\frac{1}{2}}K_{l}^{\oplus}\right\Vert _{\HS} \leq O(k_{F}^{-\frac{1}{3} + \epsilon})$. In general, what will take the place of $T$ will be the $A_{k}\left(t\right)$
and $B_{k}\left(t\right)$ operators we defined in the last section,
but as a simple example we consider 
\begin{align} \label{eq:example-T-Ak}
T= \left(\begin{array}{cc}
P_{v_{k}} & 0\\
0 & P_{v_{k}}
\end{array}\right),\quad P_{v_{k}} = |v_k\rangle \langle v_k|, \quad v_{k}=\sqrt{\frac{\hat{V}_{k}k_{F}^{-1}}{2\left(2\pi\right)^{3}}}\sum_{p\in L_{k}}e_{p} \in \ell^{2}\left(L_{k}\right)
\end{align}
for which
\begin{align} \label{eq:hkT-HS}
\left\Vert \left(h_{k}^{\oplus}\right)^{-\frac{1}{2}}T\right\Vert _{\HS} & =2\left\Vert h_{k}^{-\frac{1}{2}}P_{v_{k}}\right\Vert _{\HS}=2\sqrt{\tr\left(P_{v_{k}}h_{k}^{-1}P_{v_{k}}\right)}\\
 & =2\left\Vert v_{k}\right\Vert \sqrt{\left\langle v_{k},h_{k}^{-1}v_{k}\right\rangle }=\frac{\hat{V}_{k}k_{F}^{-1}}{\left(2\pi\right)^{3}}\left|L_{k}\right|^{\frac 1 2} \sqrt{\sum_{p\in L_{k}}\lambda_{k,p}^{-1}}\leq O(k_{F}^{\frac{1}{2}})\nonumber 
\end{align}
when $|k|\sim 1$. Here we used $|L_k|\le C|k| k_F^2$ and the bound $\sum_{p\in L_{k}}\lambda_{k,p}^{-1}\leq Ck_{F}$ from Proposition \ref{coro:CompletelyUniformRiemannSumBound}. Thus for any state satisfying $\left\langle \Psi,H_{\kin}^{\prime}\Psi\right\rangle \le O(k_F)$ (c.f. Theorem \ref{thm:Kinetic-estimate}), the overall estimate for the right side of \eqref{eq:SimpleInsufficientEstimate} is $O(k_{F}^{\frac{7}{6} + \epsilon})$ which is insufficient as the correlation energy is of order $k_F$.

%To improve (\ref{eq:SimpleInsufficientEstimate}), let us replace the Hilbert-Schmidt estimate \eqref{eq:hkT-HS} by the stronger bound 

The technical issue with the estimation  in  (\ref{eq:SimpleInsufficientEstimate})
lies in only using that $\left\Vert c_{p}\right\Vert _{\text{Op}}=1$,
for we may get better bounds by using $\left\langle \Psi,\mathcal{N}_{E} H'_{\rm kin}\Psi\right\rangle $ instead of $\left\langle \Psi,H'_{\rm kin}\Psi\right\rangle $. For example, 
\begin{align} \label{eq:GoodSchematicEstimate}
&\quad\,\sum_{p\in S}\left|\left\langle \Psi,b_{k}^{\ast}\left(Te_{p_{1}}\right)\tilde{c}_{p_{3}}^{\ast}\tilde{c}_{p_{4}}b_{l}\left(K_{l}^{\oplus}e_{p_{2}}\right)\Psi\right\rangle \right|\\
 &=\sum_{p\in S}\left|\left\langle \Psi,\tilde{c}_{p_{3}}^{\ast}b_{k}^{\ast}\left(Te_{p_{1}}\right)b_{l}\left(K_{l}^{\oplus}e_{p_{2}}\right)\tilde{c}_{p_{4}}\Psi\right\rangle \right|\leq\sum_{p\in S}\left\Vert b_{k}\left(Te_{p_{1}}\right)\tilde{c}_{p_{3}}\Psi\right\Vert \left\Vert b_{l}\left(K_{l}^{\oplus}e_{p_{2}}\right)\tilde{c}_{p_{4}}\Psi\right\Vert \nonumber \\
 & \leq\sum_{p\in S}\left\Vert \left(h_{k}^{\oplus}\right)^{-\frac{1}{2}}Te_{p_{1}}\right\Vert \left\Vert \left(h_{l}^{\oplus}\right)^{-\frac{1}{2}}K_{l}^{\oplus}e_{p_{2}}\right\Vert \sqrt{\left\langle \tilde{c}_{p_{3}}\Psi,H_{\kin}^{\prime\left(\pm1\right)}\tilde{c}_{p_{3}}\Psi\right\rangle \left\langle \tilde{c}_{p_{4}}\Psi,H_{\kin}^{\prime\left(\pm1\right)}\tilde{c}_{p_{4}}\Psi\right\rangle } \\
 & \leq\left(\max_{p\in L_{k}^{\pm}}\left\Vert \left(h_{k}^{\oplus}\right)^{-\frac{1}{2}}Te_{p}\right\Vert \right)\sqrt{\sum_{p\in S}\left\Vert \left(h_{l}^{\oplus}\right)^{-\frac{1}{2}}K_{l}^{\oplus}e_{q}\right\Vert ^{2}}\sqrt{\sum_{p\in S}\left\langle \Psi,\tilde{c}_{p_{3}}^{\ast}H_{\kin}^{\prime\left(\pm1\right)}\tilde{c}_{p_{3}}\Psi\right\rangle }\sqrt{\left\langle \Psi,H_{\kin}^{\prime}\Psi\right\rangle }\nonumber \\
 & \leq\left(\max_{p\in L_{k}^{\pm}}\left\Vert \left(h_{k}^{\oplus}\right)^{-\frac{1}{2}}Te_{p}\right\Vert \right)\left\Vert \left(h_{l}^{\oplus}\right)^{-\frac{1}{2}}K_{l}^{\oplus}\right\Vert _{\HS}\sqrt{\left\langle \Psi,\mathcal{N}_{E}H_{\kin}^{\prime}\Psi\right\rangle \left\langle \Psi,H_{\kin}^{\prime}\Psi\right\rangle }\nonumber 
\end{align}
where we  used that $\left[\tilde{c}_{p},b_{k}\left(\cdot\right)\right]=0$ (as we will see in Proposition \ref{prop:VanishingofCommutators} below), and momentarily looked ahead to the definition \eqref{eq:H-pm1-def} for $H_{\kin}^{\prime\left(\pm1\right)}$ and Lemma \ref{lemma:KineticEstimationSimplification} (we take supremum over $p_4$ and sum over $p_3$ to get the second inequality). 
%to estimate that
%\begin{align}
%\sum_{p\in S}\left\langle \Psi,\tilde{c}_{p_{3}}^{\ast}\mathcal{N}_{E}^{\left(-1\right)}\tilde{c}_{p_{3}}\Psi\right\rangle  & \leq\sum_{p\in B_{F}^{c}}\left\langle \Psi,c_{p}^{\ast}\mathcal{N}_{E}^{\left(-1\right)}c_{p}\Psi\right\rangle =\left\langle \Psi,\left(\mathcal{N}_{E}^{2}-\mathcal{N}_{E}\right)\Psi\right\rangle \label{eq:CleanNumberEstimates}\\
%\sum_{p\in S}\left\langle \Psi,\tilde{c}_{p_{3}}^{\ast}\mathcal{N}_{E}^{\left(+1\right)}\tilde{c}_{p_{3}}\Psi\right\rangle  & \leq\sum_{p\in B_{F}}\left\langle \Psi,c_{p}\mathcal{N}_{E}^{\left(+1\right)}c_{p}^{\ast}\Psi\right\rangle =\left\langle \Psi,\left(\mathcal{N}_{E}^{2}-\mathcal{N}_{E}\right)\Psi\right\rangle \nonumber 
%\end{align}
%depending on whether $p\mapsto p_{3}$ maps into $B_{F}^{c}$ or $B_{F}$,
%respectively. Clearly 
Considering again
the example in \eqref{eq:example-T-Ak} we find
\begin{equation}
\max_{p\in L_{k}^{\pm}}\left\Vert \left(h_{k}^{\oplus}\right)^{-\frac{1}{2}}Te_{p}\right\Vert =\sqrt{\frac{\hat{V}_{k}k_{F}^{-1}}{2\left(2\pi\right)^{3}}}\left\Vert h_{k}^{-\frac{1}{2}}v_{k}\right\Vert =\frac{\hat{V}_{k}k_{F}^{-1}}{2\left(2\pi\right)^{3}}\sqrt{\sum_{k\in L_{k}}\lambda_{k,p}^{-1}}\leq O(k_{F}^{-\frac{1}{2}}).
\end{equation}
Thus for any state satisfying $\left\langle \Psi,H_{\kin}^{\prime}\Psi\right\rangle \le O(k_F)$ and $\left\langle \Psi,\mathcal{N}_{E}H_{\kin}^{\prime}\Psi\right\rangle \le O(k_F^2)$ (c.f. Theorem \ref{thm:Kinetic-estimate}), the right side of \eqref{eq:GoodSchematicEstimate} is thus bounded by  $O(k_{F}^{ \frac{2}{3} + \epsilon})$ which is much smaller than the correlation energy. 
%Note that the left-hand side of \eqref{eq:GoodSchematicEstimate} is essentially the same with the left-hand side of   (\ref{eq:SimpleInsufficientEstimate}), up to some commutators between $b^{\natural}$ and $c^{\natural}$ that we will discuss below. 

% $O(k_{F}^{- \frac{5}{6} + \epsilon}) \sqrt{\left\langle \Psi,\mathcal{N}_{E}H_{\kin}^{\prime}\Psi\right\rangle \left\langle \Psi,H_{\kin}^{\prime}\Psi\right\rangle }$.

%Thus 
%as going from $\left\Vert A_{k}^{\oplus}\right\Vert _{\text{HS}}$
%to $\max_{p\in L_{k}^{\pm}}\left\Vert A_{k}^{\oplus}e_{p}\right\Vert $
%had the effect of introducing a square root to the $\left|L_{k}\right|$
%factor. 
%Compared to the simple estimate this is an improvement by
%a factor of $k_{F}$, at least for the states satisfying $\langle \Psi, \mathcal{N}_E^2 \Psi\rangle \sim 1$. 
%Although this computation is particular
%to the example $T=A_{k}^{\oplus}$ in \eqref{eq:example-T-Ak} the more general $A_{k}\left(t\right)$
%and $B_{k}\left(t\right)$ operators will exhibit a similar improvement. Note that \eqref{eq:GoodSchematicEstimate} would be sufficient for the states with very few excitations, e.g. $\langle \Psi, \mathcal{N}_E^2 \Psi\rangle \sim 1$. For more general states we will need another improvement where $\langle \Psi, \mathcal{N}_E^2 \Psi\rangle$ is replaced by  $\langle \Psi, \mathcal{N}_E k_F \Psi\rangle$, but we will defer this technical detail to the end of the section. 

Our goal is therefore to reduce the schematic forms of equation (\ref{eq:SecondSchematicForms})
to those of the form $\sum_{p\in S}\tilde{c}_{p_{3}}^{\ast}b_{k}^{\natural}\left(Te_{p_{1}}\right)b_{l}^{\natural}\left(K_{l}^{\oplus}e_{p_{2}}\right)\tilde{c}_{p_{4}}$,
which we may then estimate as above. While $\left[\tilde{c}_{p},b_{k}\left(\cdot\right)\right]=0$
it is generally the case that $\left[\tilde{c}_{p},b_{k}^{\ast}\left(\cdot\right)\right]\neq0$,
so this will also introduce additional commutator terms which we must
then estimate separately.

Taking into account whether $b_{k}^{\natural}=b_{k}$ or $b_{k}^{\natural}=b_{k}^{\ast}$,
the schematic forms of equation (\ref{eq:SecondSchematicForms}) are
either of the form (supressing the summation, the arguments and the
subscripts for brevity)
\begin{equation}
b^{\ast}b^{\ast}\tilde{c}^{\ast}\tilde{c},\quad b^{\ast}b\tilde{c}^{\ast}\tilde{c},\quad bb^{\ast}\tilde{c}^{\ast}\tilde{c},\quad bb\tilde{c}^{\ast}\tilde{c},\quad b^{\ast}\tilde{c}^{\ast}\tilde{c}b,\quad b\tilde{c}^{\ast}\tilde{c}b,\quad b\tilde{c}^{\ast}\tilde{c}b^{\ast},
\end{equation}
or reduce to one of these by taking the adjoint, which as we will
estimate $\mathcal{E}_{1}^{k}\left(A\right)$ and $\mathcal{E}_{2}^{k}\left(B\right)$
as bilinear forms does not matter. Using that commutators of the form
$\left[b,\tilde{c}\right]$, $\left[b^{\ast},\tilde{c}^{\ast}\right]$
and $\left[b,\left[b,\tilde{c}^{\ast}\right]\right]$ vanish (verified
below), these schematic forms reduce to
\begin{align}
b^{\ast}b^{\ast}\tilde{c}^{\ast}\tilde{c} & =\tilde{c}^{\ast}b^{\ast}b^{\ast}\tilde{c} ,\nonumber \\
b^{\ast}b\tilde{c}^{\ast}\tilde{c} & =\tilde{c}^{\ast}b^{\ast}b\tilde{c}+\left[b,\tilde{c}^{\ast}\right]b^{\ast}\tilde{c}+\left[b^{\ast},\left[b,\tilde{c}^{\ast}\right]\right]\tilde{c},\nonumber \\
bb^{\ast}\tilde{c}^{\ast}\tilde{c} & =\tilde{c}^{\ast}bb^{\ast}\tilde{c}+\left[b,\tilde{c}^{\ast}\right]b^{\ast}\tilde{c},\nonumber \\
bb\tilde{c}^{\ast}\tilde{c} & =\tilde{c}^{\ast}bb\tilde{c}+\left[b,\tilde{c}^{\ast}\right]b\tilde{c}+\left[b,\tilde{c}^{\ast}\right]b\tilde{c},\\
b^{\ast}\tilde{c}^{\ast}\tilde{c}b & =\tilde{c}^{\ast}b^{\ast}b\tilde{c},\nonumber \\
b\tilde{c}^{\ast}\tilde{c}b & =\tilde{c}^{\ast}bb\tilde{c}+\left[b,\tilde{c}^{\ast}\right]b\tilde{c},\nonumber \\
b\tilde{c}^{\ast}\tilde{c}b^{\ast} & =\tilde{c}^{\ast}bb^{\ast}\tilde{c}+\left[b,\tilde{c}^{\ast}\right]b^{\ast}\tilde{c}+\tilde{c}^{\ast}b\left[\tilde{c},b^{\ast}\right]+\left[b,\tilde{c}^{\ast}\right]\left[\tilde{c},b^{\ast}\right].\nonumber 
\end{align}
%the schematic forms of the first group reduce to
%\begin{align}
%b^{\ast}b^{\ast}\tilde{c}^{\ast}\tilde{c} & =\tilde{c}^{\ast}b^{\ast}b^{\ast}\tilde{c} ,\nonumber \\
%b^{\ast}b\tilde{c}^{\ast}\tilde{c} & =\tilde{c}^{\ast}b^{\ast}b\tilde{c}+\left[b,\tilde{c}^{\ast}\right]b^{\ast}\tilde{c}+\left[b^{\ast},\left[b,\tilde{c}^{\ast}\right]\right]\tilde{c},\\
%bb^{\ast}\tilde{c}^{\ast}\tilde{c} & =\tilde{c}^{\ast}bb^{\ast}\tilde{c}+\left[b,\tilde{c}^{\ast}\right]b^{\ast}\tilde{c},\nonumber \\
%bb\tilde{c}^{\ast}\tilde{c} & =\tilde{c}^{\ast}bb\tilde{c}+\left[b,\tilde{c}^{\ast}\right]b\tilde{c}+\left[b,\tilde{c}^{\ast}\right]b\tilde{c},\nonumber 
%\end{align}
%while those of the second reduce to
%\begin{align}
%b^{\ast}\tilde{c}^{\ast}\tilde{c}b & =\tilde{c}^{\ast}b^{\ast}b\tilde{c},\nonumber \\
%b\tilde{c}^{\ast}\tilde{c}b & =\tilde{c}^{\ast}bb\tilde{c}+\left[b,\tilde{c}^{\ast}\right]b\tilde{c},\\
%b\tilde{c}^{\ast}\tilde{c}b^{\ast} & =\tilde{c}^{\ast}bb^{\ast}\tilde{c}+\left[b,\tilde{c}^{\ast}\right]b^{\ast}\tilde{c}+\tilde{c}^{\ast}b\left[\tilde{c},b^{\ast}\right]+\left[b,\tilde{c}^{\ast}\right]\left[\tilde{c},b^{\ast}\right].\nonumber 
%\end{align}
%\begin{align}
%delete
%\end{align}

Reintroducing the $b^{\natural}$ notation outside the commutators and
using once more our freedom to take adjoints, we find that every term
on the right-hand sides of the two equations above take one of the
four schematic forms
\begin{equation}
\tilde{c}^{\ast}b^{\natural}b^{\natural}\tilde{c},\quad\left[\tilde{c},b^{\ast}\right]^{\ast}b^{\natural}\tilde{c},\quad\left[\left[\tilde{c},b^{\ast}\right],b\right]^{\ast}\tilde{c},\quad\left[\tilde{c},b^{\ast}\right]^{\ast}\left[\tilde{c},b^{\ast}\right].
\end{equation}
These are the final forms which we will explicitly estimate.

\subsection{Preliminary Commutator Estimates}

In addition to the general estimates which we derived at the start
of this section we will also need estimates on the commutator terms
which appear in the schematic forms of equation (\ref{eq:FinalSchematicForms}),
which we now derive. First we must however verify that the commutators
$\left[b,\tilde{c}\right]$, $\left[b^{\ast},\tilde{c}^{\ast}\right]$
and $\left[b,\left[b,\tilde{c}^{\ast}\right]\right]$ vanish, which
we relied upon in our reduction procedure:
\begin{prop}
\label{prop:VanishingofCommutators}For all $k,l\in\mathbb{Z}_{+}^{3}$,
$\varphi\in\ell^{2}\left(L_{k}^{\pm}\right)$, $\psi\in\ell^{2}\left(L_{l}^{\pm}\right)$
and $p\in\mathbb{Z}^{3}$ it holds that
\begin{align*}
\left[b_{k}\left(\varphi\right),\tilde{c}_{p}\right]=\left[b_{k}^{\ast}\left(\varphi\right),\tilde{c}_{p}^{\ast}\right]  =0,\quad 
\left[b_{l}\left(\psi\right),\left[b_{k}\left(\varphi\right),\tilde{c}_{p}^{\ast}\right]\right]  =0.
\end{align*}
\end{prop}

\textbf{Proof:} We compute from the definitions that for any $q\in L_{k}^{\pm}$
\begin{align}
\left[b_{\overline{k,q}},\tilde{c}_{p}\right] & =\left[c_{\overline{q-k}}^{\ast}c_{q},\tilde{c}_{p}\right]=c_{\overline{q-k}}^{\ast}\left\{ c_{q},\tilde{c}_{p}\right\} -\left\{ c_{\overline{q-k}}^{\ast},\tilde{c}_{p}\right\} c_{q}\nonumber \\
 & =\begin{cases}
c_{\overline{q-k}}^{\ast}\left\{ c_{q},c_{p}\right\} -\left\{ c_{\overline{q-k}}^{\ast},c_{p}\right\} c_{q} ,& p\in B_{F}^{c}\\
c_{\overline{q-k}}^{\ast}\left\{ c_{q},c_{p}^{\ast}\right\} -\left\{ c_{\overline{q-k}}^{\ast},c_{p}^{\ast}\right\} c_{q} ,& p\in B_{F}
\end{cases}\\
 & =0\nonumber 
\end{align}
as all anticommutators on the second line vanish either directly by
the CAR or by disjointness of $B_{F}$ and $B_{F}^{c}$. By linearity $\left[b_{k}\left(\varphi\right),\tilde{c}_{p}\right]=0$,
and  $\left[b_{k}^{\ast}\left(\varphi\right),\tilde{c}_{p}^{\ast}\right]=-\left[b_{k}\left(\varphi\right),\tilde{c}_{p}\right]^{\ast}=0$.

%Since a general
%operator $b_{k}\left(\varphi\right)=\sum_{q\in L_{k}^{\pm}}\left\langle \varphi,e_{q}\right\rangle b_{\overline{k,q}}$
%is a linear combination of $b_{\overline{k,q}}$ operators it follows
%that also $\left[b_{k}\left(\varphi\right),\tilde{c}_{p}\right]=0$,
%and furthermore that $\left[b_{k}^{\ast}\left(\varphi\right),\tilde{c}_{p}^{\ast}\right]=-\left[b_{k}\left(\varphi\right),\tilde{c}_{p}\right]^{\ast}=0$.

For the double commutator we first compute $\left[b_{\overline{k,q}},\tilde{c}_{p}^{\ast}\right]$:
As above we find
\begin{align}
\left[b_{\overline{k,q}},\tilde{c}_{p}^{\ast}\right] =\begin{cases}
c_{\overline{q-k}}^{\ast}\left\{ c_{q},c_{p}^{\ast}\right\} -\left\{ c_{\overline{q-k}}^{\ast},c_{p}^{\ast}\right\} c_{q} ,& p\in B_{F}^{c}\\
c_{\overline{q-k}}^{\ast}\left\{ c_{q},c_{p}\right\} -\left\{ c_{\overline{q-k}}^{\ast},c_{p}\right\} c_{q}, & p\in B_{F}
\end{cases}
%=\begin{cases}
%\delta_{q,p}c_{\overline{q-k}}^{\ast} ,& p\in B_{F}^{c}\\
%-\delta_{\overline{q-k},p}c_{q} ,& p\in B_{F}
%\end{cases}\\
  =\begin{cases}
\delta_{q,p}\tilde{c}_{\overline{q-k}} ,& p\in B_{F}^{c}\\
-\delta_{\overline{q-k},p}\tilde{c}_{q}, & p\in B_{F}
\end{cases}\nonumber 
\end{align}
so
\begin{align}
\left[b_{k}\left(\varphi\right),\tilde{c}_{p}^{\ast}\right] & =\sum_{q\in L_{k}^{\pm}}\left\langle \varphi,e_{q}\right\rangle \left[b_{\overline{k,q}},\tilde{c}_{p}^{\ast}\right]=\sum_{q\in L_{k}^{\pm}}\left\langle \varphi,e_{q}\right\rangle \begin{cases}
\delta_{q,p}\tilde{c}_{\overline{q-k}} & p\in B_{F}^{c}\\
-\delta_{\overline{q-k},p}\tilde{c}_{q} & p\in B_{F}
\end{cases}\label{eq:ExplicitSingleCommutator}\\
 & =1_{L_{k}^{\pm}}\left(p\right)\left\langle \varphi,e_{p}\right\rangle \tilde{c}_{\overline{p-k}}-1_{L_{k}-k}\left(p\right)\left\langle \varphi,e_{p+k}\right\rangle \tilde{c}_{p+k}-1_{L_{k}+k}\left(p\right)\left\langle \varphi,e_{p-k}\right\rangle \tilde{c}_{p-k}\nonumber 
\end{align}
where $1_{S}\left(\cdot\right)$ denotes the indicator function of
a set $S$. Observing that $\left[b_{k}\left(\varphi\right),\tilde{c}_{p}^{\ast}\right]$
is a linear combination of $\tilde{c}_{p}$ terms we conclude that
$\left[b_{l}\left(\psi\right),\left[b_{k}\left(\varphi\right),\tilde{c}_{p}^{\ast}\right]\right]=0$
by the first part.

$\hfill\square$

Now to the estimation of the non-vanishing commutators. We begin with
the single commutator - we state the estimate and make a remark:
\begin{prop}
\label{prop:SingleCommutatorEstimates}For all $k\in\mathbb{Z}_{+}^{3}$,
sequences $\left(\varphi_{p}\right)_{p\in\mathbb{Z}^{3}}\in\ell^{2}\left(L_{k}^{\pm}\right)$
and $\Psi\in\mathcal{H}_{N}$ it holds that
\begin{align*}
\sum_{p\in\mathbb{Z}^{3}}\left\Vert \left[\tilde{c}_{p},b_{k}^{\ast}\left(\varphi_{p}\right)\right]\Psi\right\Vert ^{2} & \leq3\left(\sum_{p\in L_{k}^{\pm}}\max_{q\in\mathbb{Z}^{3}}\left|\left\langle e_{p},\varphi_{q}\right\rangle \right|^{2}\right)\left\Vert \Psi\right\Vert ^{2}\\
\sum_{p\in\mathbb{Z}^{3}}\left\Vert \left[\tilde{c}_{p}^{\ast},b_{k}\left(\varphi_{p}\right)\right]\Psi\right\Vert ^{2} & \leq4\left(\max_{p\in L_{k}^{\pm},q\in\mathbb{Z}^{3}}\left|\left\langle e_{p},\varphi_{q}\right\rangle \right|^{2}\right)\left\langle \Psi,\mathcal{N}_{E}\Psi\right\rangle .
\end{align*}
\end{prop}

\begin{rmk} The statement may appear overly general, in that
it involves general sequences $\left(\varphi_{p}\right)_{p\in\mathbb{Z}^{3}}\subset\ell^{2}\left(L_{k}^{\pm}\right)$
rather than the explicit vectors $\left(Te_{p_{1}}\right)_{p\in S}\subset\ell^{2}\left(L_{k}^{\pm}\right)$
that we must consider. The point of the generality is however only
to avoid having to explicitly state the dependencies of the set $S$
and the $p_{i}$'s of each possible schematic form, as independently
of these it is easy to see that a sum such as $\sum_{p\in S}\left\Vert \left[\tilde{c}_{p_{3}},b_{k}^{\ast}\left(Te_{p_{1}}\right)\right]\Psi\right\Vert ^{2}$
can always be cast into the form in the statement.
\end{rmk}

\textbf{Proof:} Taking the adjoint of equation (\ref{eq:ExplicitSingleCommutator})
yields
\begin{equation}
\left[\tilde{c}_{p},b_{k}^{\ast}\left(\varphi\right)\right]=1_{L_{k}^{\pm}}\left(p\right)\left\langle e_{p},\varphi\right\rangle \tilde{c}_{\overline{p-k}}^{\ast}-1_{L_{k}-k}\left(p\right)\left\langle e_{p+k},\varphi\right\rangle \tilde{c}_{p+k}^{\ast}-1_{L_{k}+k}\left(p\right)\left\langle e_{p-k},\varphi\right\rangle \tilde{c}_{p-k}^{\ast}\label{eq:ExplicitSingleCommutator2}
\end{equation}
and so we can for any $\Psi\in\mathcal{H}_{N}$ estimate by the (squared)
triangle inequality, using also that $L_{k}^{\pm}$ and $\left(L_{k}-k\right)\cap\left(L_{-k}+k\right)$
are disjoint and $\left\Vert \tilde{c}_{p}^{\ast}\right\Vert _{\text{Op}}=1$,
that
\begin{align}
 & \qquad\;\sum_{p\in\mathbb{Z}^{3}}\left\Vert \left[\tilde{c}_{p},b_{k}^{\ast}\left(\varphi_{p}\right)\right]\Psi\right\Vert ^{2}\leq\sum_{p\in L_{k}^{\pm}}\left|\left\langle e_{p},\varphi_{p}\right\rangle \right|^{2}\left\Vert \tilde{c}_{\overline{p-k}}^{\ast}\Psi\right\Vert ^{2} +2\sum_{p\in L_{k}-k}\left|\left\langle e_{p+k},\varphi_{p}\right\rangle \right|^{2}\left\Vert \tilde{c}_{p+k}^{\ast}\Psi\right\Vert ^{2}\nonumber \\ 
 &\qquad\qquad\qquad\qquad\qquad\quad\;\;+2\sum_{p\in L_{-k}+k}\left|\left\langle e_{p-k},\varphi_{p}\right\rangle \right|^{2}\left\Vert \tilde{c}_{p-k}^{\ast}\Psi\right\Vert ^{2}\nonumber \\
 & \leq\left(\sum_{p\in L_{k}^{\pm}}\left|\left\langle e_{p},\varphi_{p}\right\rangle \right|^{2}+2\sum_{p\in L_{k}-k}\left|\left\langle e_{p+k},\varphi_{p}\right\rangle \right|^{2}+2\sum_{p\in L_{-k}+k}\left|\left\langle e_{p-k},\varphi_{p}\right\rangle \right|^{2}\right)\left\Vert \Psi\right\Vert ^{2}\\
 & \leq\left(\sum_{p\in L_{k}^{\pm}}\max_{q\in\mathbb{Z}^{3}}\left|\left\langle e_{p},\varphi_{q}\right\rangle \right|^{2}+2\sum_{p\in L_{k}}\max_{q\in\mathbb{Z}^{3}}\left|\left\langle e_{p},\varphi_{q}\right\rangle \right|^{2}+2\sum_{p\in L_{-k}}\max_{q\in\mathbb{Z}^{3}}\left|\left\langle e_{p},\varphi_{q}\right\rangle \right|^{2}\right)\left\Vert \Psi\right\Vert ^{2}\nonumber \\
 & =3\left(\sum_{p\in L_{k}^{\pm}}\max_{q\in\mathbb{Z}^{3}}\left|\left\langle e_{p},\varphi_{q}\right\rangle \right|^{2}\right)\left\Vert \Psi\right\Vert ^{2}\nonumber 
\end{align}
which implies the first estimate. For the second estimate we find
in a similar manner (now directly from equation (\ref{eq:ExplicitSingleCommutator}))
that
\begin{align}
 & \qquad\;\sum_{p\in\mathbb{Z}^{3}}\left\Vert \left[\tilde{c}_{p}^{\ast},b_{k}\left(\varphi_{p}\right)\right]\Psi\right\Vert ^{2}\leq\sum_{p\in L_{k}^{\pm}}\left|\left\langle e_{p},\varphi_{p}\right\rangle \right|^{2}\left\Vert \tilde{c}_{\overline{p-k}}\Psi\right\Vert ^{2} +2\sum_{p\in L_{k}-k}\left|\left\langle e_{p+k},\varphi_{p}\right\rangle \right|^{2}\left\Vert \tilde{c}_{p+k}\Psi\right\Vert ^{2} \nonumber \\ 
 &\qquad\qquad\qquad\qquad\qquad\qquad+2\sum_{p\in L_{-k}+k}\left|\left\langle e_{p-k},\varphi_{p}\right\rangle \right|^{2}\left\Vert \tilde{c}_{p-k}\Psi\right\Vert ^{2}\\
 & \leq\left(\max_{p\in L_{k}^{\pm},q\in\mathbb{Z}^{3}}\left|\left\langle e_{p},\varphi_{q}\right\rangle \right|^{2}\right)\left(\sum_{p\in L_{k}^{\pm}}\left\Vert \tilde{c}_{\overline{p-k}}\Psi\right\Vert ^{2}+2\sum_{p\in L_{k}-k}\left\Vert \tilde{c}_{p+k}\Psi\right\Vert ^{2}+2\sum_{p\in L_{-k}+k}\left\Vert \tilde{c}_{p-k}\Psi\right\Vert ^{2}\right)\nonumber \\
 & =\left(\max_{p\in L_{k}^{\pm},q\in\mathbb{Z}^{3}}\left|\left\langle e_{p},\varphi_{q}\right\rangle \right|^{2}\right)\left(\sum_{p\in L_{k}^{\pm}}\left\Vert c_{\overline{p-k}}^{\ast}\Psi\right\Vert ^{2}+2\sum_{p\in L_{k}}\left\Vert c_{p}\Psi\right\Vert ^{2}+2\sum_{p\in L_{-k}}\left\Vert c_{p}\Psi\right\Vert ^{2}\right)\nonumber \\
 & \leq4\left(\max_{p\in L_{k}^{\pm},q\in\mathbb{Z}^{3}}\left|\left\langle e_{p},\varphi_{q}\right\rangle \right|^{2}\right)\left\langle \Psi,\mathcal{N}_{E}\Psi\right\rangle.\nonumber 
\end{align}
$\hfill\square$

Lastly we estimate the double commutator:
\begin{prop}
\label{prop:DoubleCommutatorEstimate}For all $k,l\in\mathbb{Z}_{+}^{3}$,
sequences $\left(\varphi_{p}\right)_{p\in\mathbb{Z}^{3}}\subset\ell^{2}\left(L_{k}^{\pm}\right)$
and $\left(\psi_{p}\right)_{p\in\mathbb{Z}^{3}}\subset\ell^{2}\left(L_{l}^{\pm}\right)$,
and $\Psi\in\mathcal{H}_{N}$, it holds that
\[
\sum_{p\in\mathbb{Z}^{3}}\left\Vert \left[\left[\tilde{c}_{p},b_{k}^{\ast}\left(\varphi_{p}\right)\right],b_{l}\left(\psi_{p}\right)\right]\Psi\right\Vert ^{2}\leq12\left(\max_{p\in L_{k}^{\pm},q\in\mathbb{Z}^{3}}\left|\left\langle e_{p},\varphi_{q}\right\rangle \right|^{2}\right)\left(\max_{p\in L_{k}^{\pm},q\in\mathbb{Z}^{3}}\left|\left\langle e_{p},\psi_{q}\right\rangle \right|^{2}\right)\left\langle \Psi,\mathcal{N}_{E}\Psi\right\rangle .
\]
\end{prop}

\textbf{Proof:} From (\ref{eq:ExplicitSingleCommutator2}) we have
that
\begin{align}
\left[\left[\tilde{c}_{p},b_{k}^{\ast}\left(\varphi_{p}\right)\right],b_{l}\left(\psi_{p}\right)\right] & =1_{L_{k}^{\pm}}\left(p\right)\left\langle e_{p},\varphi\right\rangle \left[\tilde{c}_{\overline{p-k}}^{\ast},b_{l}\left(\psi_{p}\right)\right]-1_{L_{k}-k}\left(p\right)\left\langle e_{p+k},\varphi\right\rangle \left[\tilde{c}_{p+k}^{\ast},b_{l}\left(\psi_{p}\right)\right]\\
 & -1_{L_{k}+k}\left(p\right)\left\langle e_{p-k},\varphi\right\rangle \left[\tilde{c}_{p-k}^{\ast},b_{l}\left(\psi_{p}\right)\right]\nonumber 
\end{align}
and so, by the triangle inequality and the second estimate of Proposition
\ref{prop:SingleCommutatorEstimates},
\begin{align}
 & \qquad\;\sum_{p\in\mathbb{Z}^{n}}\left\Vert \left[\left[\tilde{c}_{p},b_{k}^{\ast}\left(\varphi_{p}\right)\right],b_{l}\left(\psi_{p}\right)\right]\Psi\right\Vert ^{2}\leq\sum_{p\in L_{k}^{\pm}}\left|\left\langle e_{p},\varphi_{p}\right\rangle \right|^{2}\left\Vert \left[\tilde{c}_{\overline{p-k}}^{\ast},b_{l}\left(\psi_{p}\right)\right]\Psi\right\Vert ^{2}\nonumber \\
 & +2\sum_{p\in L_{k}-k}\left|\left\langle e_{p+k},\varphi_{p}\right\rangle \right|^{2}\left\Vert \left[\tilde{c}_{p+k}^{\ast},b_{l}\left(\psi_{p}\right)\right]\Psi\right\Vert ^{2}+2\sum_{p\in L_{-k}+k}\left|\left\langle e_{p-k},\varphi_{p}\right\rangle \right|^{2}\left\Vert \left[\tilde{c}_{p-k}^{\ast},b_{l}\left(\psi_{p}\right)\right]\Psi\right\Vert ^{2}\nonumber \\
 & \leq\left(\max_{p\in L_{k}^{\pm},q\in\mathbb{Z}^{3}}\left|\left\langle e_{p},\varphi_{q}\right\rangle \right|^{2}\right)\left(\sum_{p\in L_{k}^{\pm}}\left\Vert \left[\tilde{c}_{\overline{p-k}}^{\ast},b_{l}\left(\psi_{p}\right)\right]\Psi\right\Vert ^{2}+2\sum_{p\in L_{k}}\left\Vert \left[\tilde{c}_{p}^{\ast},b_{l}\left(\psi_{p-k}\right)\right]\Psi\right\Vert ^{2}\right.\nonumber \\
 & \qquad\qquad\qquad\qquad\qquad\qquad\qquad\qquad\qquad\qquad\qquad\qquad\left.+2\sum_{p\in L_{-k}}\left\Vert \left[\tilde{c}_{p}^{\ast},b_{l}\left(\psi_{p+k}\right)\right]\Psi\right\Vert ^{2}\right) \nonumber\\
 &\leq12\left(\max_{p\in L_{k}^{\pm},q\in\mathbb{Z}^{3}}\left|\left\langle e_{p},\varphi_{q}\right\rangle \right|^{2}\right)\left(\max_{p\in L_{k}^{\pm},q\in\mathbb{Z}^{3}}\left|\left\langle e_{p},\psi_{q}\right\rangle \right|^{2}\right)\left\langle \Psi,\mathcal{N}_{E}\Psi\right\rangle . 
\end{align}
$\hfill\square$

\subsection{Final Estimation of the Exchange Terms}

Now we are ready to derive bounds for the exchange terms $\mathcal{E}_{1}^{k}\left(A\right)$ and $\mathcal{E}_{2}^{k}\left(B\right)$ defined in Proposition \ref{prop:QuasiBosonicQuadraticCommutators}. Recall that we have reduced the estimation of these complicated operators to the task of obtaining
a uniform estimate for the four explicit forms
\begin{align}
 & \sum_{l\in S_{C}}\sum_{p\in S}\tilde{c}_{p_{3}}^{\ast}b_{k}^{\natural}\left(Te_{p_{1}}\right)b_{l}^{\natural}\left(K_{l}^{\oplus}e_{p_{2}}\right)\tilde{c}_{p_{4}},\qquad\quad\;\;\sum_{l\in S_{C}}\sum_{p\in S}\left[\tilde{c}_{p_{3}},b_{k}^{\ast}\left(Te_{p_{1}}\right)\right]^{\ast}b_{l}^{\natural}\left(K_{l}^{\oplus}e_{p_{2}}\right)\tilde{c}_{p_{4}}\label{eq:FinalSchematicForms}\\
 & \sum_{l\in S_{C}}\sum_{p\in S}\left[\left[\tilde{c}_{p_{3}},b_{k}^{\ast}\left(Te_{p_{1}}\right)\right],b_{l}\left(K_{l}^{\oplus}e_{p_{2}}\right)\right]^{\ast}\tilde{c}_{p_{4}},\quad\sum_{l\in S_{C}}\sum_{p\in S}\left[\tilde{c}_{p_{3}},b_{k}^{\ast}\left(Te_{p_{1}}\right)\right]^{\ast}\left[\tilde{c}_{p_{4}},b_{l}^{\ast}\left(K_{l}^{\oplus}e_{p_{2}}\right)\right],\nonumber 
\end{align}
subject to the following rules: $b_{k}^{\natural}$ denotes either $b_{k}$
or $b_{k}^{\ast}$, $T$ denotes either $A$ or $B$, and $b_{k}^{\natural}\left(Te_{p_{1}}\right)$
and $b_{l}^{\natural}\left(K_{l}^{\oplus}e_{p_{2}}\right)$ may be interchanged.
Furthermore the notation $\tilde{c}_{p}$ denotes either $c_{p}$
or $c_{p}^{\ast}$ as appropriate for $p$ and the set $S$ is such
that the assignments $p\mapsto p_{1},p_{2},p_{3},p_{4}$ are injective
and map exclusively into $B_{F}$ or $B_{F}^{c}$.

Let us start by giving estimates in terms of $\mathcal{N}_E^2$. For the statement we define the $\left\Vert \cdot\right\Vert _{\infty,2}$-norm
of an operator $T:\ell^{2}\left(L_{k}^{\pm}\right)\rightarrow\ell^{2}\left(L_{k}^{\pm}\right)$
by
\begin{equation}
\left\Vert T\right\Vert _{\infty,2}=\sqrt{\sum_{p\in L_{k}^{\pm}}\max_{q\in L_{k}^{\pm}}\left|\left\langle e_{p},Te_{q}\right\rangle \right|^{2}}.
\end{equation}
This is a minor but necessary detail, as unlike the simple estimate of equation
(\ref{eq:GoodSchematicEstimate}) we cannot take the maximum outside
the sum for all schematic terms, so we need this slightly stronger
norm. Note that
\begin{equation}
\max_{p,q\in L_{k}^{\pm}}\left|\left\langle e_{p},Te_{q}\right\rangle \right|\leq\max_{p\in L_{k}^{\pm}}\left\Vert Te_{p}\right\Vert \leq\left\Vert T\right\Vert _{\infty,2}.
\end{equation}
Now the estimate:
\begin{prop}
\label{prop:ExchangeTermEstimates}For all $k\in\mathbb{Z}_{+}^{3}$,
symmetric $T:\ell^{2}\left(L_{k}^{\pm}\right)\rightarrow\ell^{2}\left(L_{k}^{\pm}\right)$
and $\Psi\in\mathcal{H}_{N}$ it holds that
\[
\left|\left\langle \Psi,\mathcal{E}_{i}^{k}\left(T\right)\Psi\right\rangle \right|\leq C\left\Vert T\right\Vert _{\infty,2}\left(\sum_{l\in S_{C}}\left\Vert K_{l}^{\oplus}\right\Vert _{\infty,2}\right)\left\langle \Psi,\left(1+\mathcal{N}_{E}^{2}\right)\Psi\right\rangle 
\]
with $i=1,2$, for a constant $C>0$ independent of all relevant quantities.
\end{prop}

\textbf{Proof:} We estimate each schematic form of 
(\ref{eq:FinalSchematicForms}) using the estimates of the Propositions
\ref{prop:GeneralExcitationOperatorEstimate}, \ref{prop:SingleCommutatorEstimates},
\ref{prop:DoubleCommutatorEstimate} and Lemma \ref{lemma:ExcitationNumberOperatorCommutators}, as well as the Cauchy-Schwarz inequality. First is $\tilde{c}^{\ast}b^{\natural}b^{\natural}\tilde{c}$: 
\begin{align}
 & \qquad\,\sum_{l\in S_{C}}\sum_{p\in S}\left|\left\langle \Psi,\tilde{c}_{p_{3}}^{\ast}b_{k}^{\natural}\left(Te_{p_{1}}\right)b_{l}^{\natural}\left(K_{l}^{\oplus}e_{p_{2}}\right)\tilde{c}_{p_{4}}\Psi\right\rangle \right|\leq\sum_{l\in S_{C}}\sum_{p\in S}\left\Vert b_{k}^\sharp \left(Te_{p_{1}}\right)^\ast\tilde{c}_{p_{3}}\Psi\right\Vert \left\Vert b_{l}^{\natural}\left(K_{l}^{\oplus}e_{p_{2}}\right)\tilde{c}_{p_{4}}\Psi\right\Vert \nonumber \\
 & \leq C\sum_{l\in S_{C}}\sum_{p\in S}\left\Vert Te_{p_{1}}\right\Vert \left\Vert K_{l}^{\oplus}e_{p_{2}}\right\Vert \sqrt{\left\langle \tilde{c}_{p_{3}}\Psi,\left(1+\mathcal{N}_{E}^{\left(\pm1\right)}\right)\tilde{c}_{p_{3}}\Psi\right\rangle \left\langle \tilde{c}_{p_{4}}\Psi,\left(1+\mathcal{N}_{E}^{\left(\pm1\right)}\right)\tilde{c}_{p_{4}}\Psi\right\rangle }\\
 & \leq C\max_{p\in L_{k}^{\pm}}\left\Vert Te_{p}\right\Vert \sum_{l\in S_{C}}\max_{q\in L_{l}^{\pm}}\left\Vert K_{l}^{\oplus}e_{q}\right\Vert \sqrt{\sum_{p\in S}\left\langle \Psi,\left(\tilde{c}_{p_{3}}^{\ast}\mathcal{N}_{E}^{\left(\pm1\right)}\tilde{c}_{p_{3}}+\tilde{c}_{p_{3}}^{\ast}\tilde{c}_{p_{3}}\right)\Psi\right\rangle }\nonumber \\
 & \quad\;\cdot\sqrt{\sum_{p\in S}\left\langle \Psi,\left(\tilde{c}_{p_{4}}^{\ast}\mathcal{N}_{E}^{\left(\pm1\right)}\tilde{c}_{p_{4}}+\tilde{c}_{p_{4}}^{\ast}\tilde{c}_{p_{4}}\right)\Psi\right\rangle }\leq C\left\Vert T\right\Vert _{\infty,2}\left(\sum_{l\in S_{C}}\left\Vert K_{l}^{\oplus}\right\Vert _{\infty,2}\right)\left\langle \Psi,\mathcal{N}_{E}^{2}\Psi\right\rangle . \nonumber
\end{align}
Then $\left[\tilde{c},b^{\ast}\right]^{\ast}b^{\natural}\tilde{c}$: 
\begin{align}
 & \quad\,\sum_{l\in S_{C}}\sum_{p\in S}\left|\left\langle \Psi,\left[\tilde{c}_{p_{3}},b_{k}^{\ast}\left(Te_{p_{1}}\right)\right]^{\ast}b_{l}^{\natural}\left(K_{l}^{\oplus}e_{p_{2}}\right)\tilde{c}_{p_{4}}\Psi\right\rangle \right| \nonumber\\
 &\leq\sum_{l\in S_{C}}\sum_{p\in S}\left\Vert \left[\tilde{c}_{p_{3}},b_{k}^{\ast}\left(Te_{p_{1}}\right)\right]\Psi\right\Vert \left\Vert b_{l}^{\natural}\left(K_{l}^{\oplus}e_{p_{2}}\right)\tilde{c}_{p_{4}}\Psi\right\Vert \nonumber \\
 & \leq C\sum_{l\in S_{C}}\sum_{p\in S}\left\Vert \left[\tilde{c}_{p_{3}},b_{k}^{\ast}\left(Te_{p_{1}}\right)\right]\Psi\right\Vert \left\Vert K_{l}^{\oplus}e_{p_{2}}\right\Vert \sqrt{\left\langle \tilde{c}_{p_{4}}\Psi,\left(1+\mathcal{N}_{E}^{\left(\pm1\right)}\right)\tilde{c}_{p_{4}}\Psi\right\rangle }\nonumber \\
 & \leq C\sum_{l\in S_{C}}\left\Vert K_{l}^{\oplus}\right\Vert _{\infty,2}\sqrt{\sum_{p\in S}\left\Vert \left[\tilde{c}_{p_{3}},b_{k}^{\ast}\left(Te_{p_{1}}\right)\right]\Psi\right\Vert ^{2}}\sqrt{\sum_{p\in S}\left\langle \tilde{c}_{p_{4}}\Psi,\left(1+\mathcal{N}_{E}^{\left(\pm1\right)}\right)\tilde{c}_{p_{4}}\Psi\right\rangle }\label{eq:ExchangeTermEstimatesEquation2}\\
 & \leq C\sum_{l\in S_{C}}\left\Vert K_{l}^{\oplus}\right\Vert _{\infty,2}\sqrt{\sum_{p\in L_{k}^{\pm}}\max_{q\in L_{k}^{\pm}}\left|\left\langle e_{p},Te_{q}\right\rangle \right|^{2}\left\Vert \Psi\right\Vert ^{2}}\sqrt{\left\langle \Psi,\mathcal{N}_{E}^{2}\Psi\right\rangle }\nonumber \\
 & \leq C\left\Vert T\right\Vert _{\infty,2}\left(\sum_{l\in S_{C}}\left\Vert K_{l}^{\oplus}\right\Vert _{\infty,2}\right)\left\Vert \Psi\right\Vert \sqrt{\left\langle \Psi,\mathcal{N}_{E}^{2}\Psi\right\rangle }.\nonumber 
\end{align}
Now $\left[\left[\tilde{c},b^{\ast}\right],b\right]^{\ast}\tilde{c}$: 
\begin{align}
 & \quad\,\sum_{l\in S_{C}}\sum_{p\in S}\left|\left\langle \Psi,\left[\left[\tilde{c}_{p_{3}},b_{k}^{\ast}\left(Te_{p_{1}}\right)\right],b_{l}\left(K_{l}^{\oplus}e_{p_{2}}\right)\right]^{\ast}\tilde{c}_{p_{4}}\Psi\right\rangle \right|\nonumber \\
 & \leq\sum_{l\in S_{C}}\sum_{p\in S}\left\Vert \left[\left[\tilde{c}_{p_{3}},b_{k}^{\ast}\left(Te_{p_{1}}\right)\right],b_{l}\left(K_{l}^{\oplus}e_{p_{2}}\right)\right]\Psi\right\Vert \left\Vert \tilde{c}_{p_{4}}\Psi\right\Vert \nonumber \\
 & \leq\sum_{l\in S_{C}}\sqrt{\sum_{p\in S}\left\Vert \left[\left[\tilde{c}_{p_{3}},b_{k}^{\ast}\left(Te_{p_{1}}\right)\right],b_{l}\left(K_{l}^{\oplus}e_{p_{2}}\right)\right]\Psi\right\Vert ^{2}}\sqrt{\sum_{p\in S}\left\Vert \tilde{c}_{p_{4}}\Psi\right\Vert ^{2}}\\
 & \leq C\sum_{l\in S_{C}}\sqrt{\left(\max_{p,q\in L_{k}^{\pm}}\left|\left\langle e_{p},Te_{q}\right\rangle \right|^{2}\right)\left(\max_{p,q\in L_{l}^{\pm}}\left|\left\langle e_{p},K_{l}^{\oplus}e_{q}\right\rangle \right|^{2}\right)\left\langle \Psi,\mathcal{N}_{E}\Psi\right\rangle }\sqrt{\left\langle \Psi,\mathcal{N}_{E}\Psi\right\rangle }\nonumber \\
 & \leq C\left\Vert T\right\Vert _{\infty,2}\sum_{l\in S_{C}}\left\Vert K_{l}^{\oplus}\right\Vert _{\infty,2}\left\langle \Psi,\mathcal{N}_{E}\Psi\right\rangle. \nonumber 
\end{align}
And finally $\left[\tilde{c},b^{\ast}\right]^{\ast}\left[\tilde{c},b^{\ast}\right]$: 
\begin{align}
 & \quad\,\sum_{l\in S_{C}}\sum_{p\in S}\left|\left\langle \Psi,\left[\tilde{c}_{p_{3}},b_{k}^{\ast}\left(Te_{p_{1}}\right)\right]^{\ast}\left[\tilde{c}_{p_{4}},b_{l}^{\ast}\left(K_{l}^{\oplus}e_{p_{2}}\right)\right]\Psi\right\rangle \right|\nonumber \\
 & \leq\sum_{l\in S_{C}}\sum_{p\in S}\left\Vert \left[\tilde{c}_{p_{3}},b_{k}^{\ast}\left(Te_{p_{1}}\right)\right]\Psi\right\Vert \left\Vert \left[\tilde{c}_{p_{4}},b_{l}^{\ast}\left(K_{l}^{\oplus}e_{p_{2}}\right)\right]\Psi\right\Vert \nonumber \\
 & \leq\sum_{l\in S_{C}}\sqrt{\sum_{p\in S}\left\Vert \left[\tilde{c}_{p_{3}},b_{k}^{\ast}\left(Te_{p_{1}}\right)\right]\Psi\right\Vert ^{2}}\sqrt{\sum_{p\in S}\left\Vert \left[\tilde{c}_{p_{4}},b_{l}^{\ast}\left(K_{l}^{\oplus}e_{p_{2}}\right)\right]\Psi\right\Vert ^{2}}\label{eq:ExchangeTermEstimatesEquation4}\\
 & \leq C\sum_{l\in S_{C}}\sqrt{\sum_{p\in L_{k}^{\pm}}\max_{q\in L_{k}^{\pm}}\left|\left\langle e_{p},Te_{q}\right\rangle \right|^{2}\left\Vert \Psi\right\Vert ^{2}}\sqrt{\sum_{p\in L_{l}^{\pm}}\max_{q\in L_{l}^{\pm}}\left|\left\langle e_{p},K_{l}^{\oplus}e_{q}\right\rangle \right|^{2}\left\Vert \Psi\right\Vert ^{2}}\nonumber \\
 & \leq C\left\Vert T\right\Vert _{\infty,2}\sum_{l\in S_{C}}\left\Vert K_{l}^{\oplus}\right\Vert _{\infty,2}\left\Vert \Psi\right\Vert ^{2}.\nonumber 
\end{align}
%All of the expressions on the right-hand sides are bounded as in the 
%further by \\
%$C\left\Vert T\right\Vert _{\infty,2}\left(\sum_{l\in S_{C}}\left\Vert K_{l}^{\oplus}\right\Vert _{\infty,2}\right)\left\langle \Psi,\left(1+\mathcal{N}_{E}^{2}\right)\Psi\right\rangle $,
%which proves the claim.
$\hfill\square$

%\subsection*{Kinetic Estimates}

Now we derive a kinetic bound: 
\begin{prop}
\label{prop:KineticExchangeTermEstimates}For all $k\in\mathbb{Z}_{+}^{3}$,
symmetric $T:\ell^{2}\left(L_{k}^{\pm}\right)\rightarrow\ell^{2}\left(L_{k}^{\pm}\right)$
and $\Psi\in D\left(H_{\kin}^{\prime}\right)$, 
\begin{align*}
 & \quad\,\left|\left\langle \Psi,\mathcal{E}_{i}^{k}\left(T\right)\Psi\right\rangle \right| \le C \sum_{l\in S_{C}}\left(\left\Vert \left(h_{l}^{\oplus}\right)^{-\frac{1}{2}}K_{l}^{\oplus}\right\Vert _{\HS}+\left\Vert K_{l}^{\oplus}\right\Vert _{\infty,2}\right) \times \\
 &\qquad\qquad\qquad\qquad\qquad \times \Bigg[ \left(\max_{p\in L_{k}^{\pm}}\left\Vert \left(h_{k}^{\oplus}\right)^{-\frac{1}{2}}Te_{p}\right\Vert \right)\sqrt{\left\langle \Psi,H_{\kin}^{\prime}\Psi\right\rangle \left\langle \Psi,\mathcal{N}_{E}H_{\kin}^{\prime}\Psi\right\rangle }\\
 &\qquad\qquad\qquad\qquad\qquad\qquad\qquad +\left\Vert T\right\Vert _{\infty,2} \left(\left\langle \Psi,\left(1+H_{\kin}^{\prime}\right)\Psi\right\rangle +\left\Vert \Psi\right\Vert \sqrt{\left\langle \Psi,\mathcal{N}_{E}H_{\kin}^{\prime}\Psi\right\rangle }\right) \Bigg]
\end{align*}
for $i=1,2$, for a constant $C>0$ independent of all relevant quantities.
\end{prop}

As a technical preparation, let us observe that from \eqref{eq:ManifestlyNonNegativeHKin} we may associate to $H_{\rm kin}'$ the operators 
\begin{equation} \label{eq:H-pm1-def}
H_{\text{kin}}^{\prime\left(\pm1\right)}=\sum_{p\in B_{F}^{c}}\vert\left|p\right|^{2}-\zeta\vert\,c_{p}^{\ast}c_{p}+\sum_{p\in B_{F}}\vert\left|p\right|^{2}-\zeta\vert\,c_{p}c_{p}^{\ast}
\end{equation}
acting on $\mathcal{H}_{N\pm1}$ (the expressions of $H_{\text{kin}}^{\prime\left(+1\right)}$ and $H_{\text{kin}}^{\prime\left(-1\right)}$ are the same, but the domains are different). With this interpretation, we have the following lemma (c.f. Lemma \ref{lemma:ExcitationNumberOperatorCommutators}): 
\begin{lem}
\label{lemma:KineticEstimationSimplification}It holds that 
$$\tilde{c}_{p}^{\ast}H_{\kin}^{\prime\left(\pm1\right)}\tilde{c}_{p}\leq H_{\kin}^{\prime}$$
for all $p\in\mathbb{Z}^{3}$ and 
\[
\sum_{p\in B_{F}^{c}}c_{p}^{\ast}H_{\kin}^{\prime\left(-1\right)}c_{p} \le \mathcal{N}_{E}H_{\kin}^{\prime}, \quad \sum_{p\in B_{F}}c_{p}H_{\kin}^{\prime\left(+1\right)}c_{p}^{\ast}\leq\mathcal{N}_{E}H_{\kin}^{\prime}.
\]
%and for all $p\in\mathbb{Z}^{3}$
%\[
%\tilde{c}_{p}^{\ast}H_{\kin}^{\prime\left(\pm1\right)}\tilde{c}_{p}\leq H_{\kin}^{\prime}.
%\]
\end{lem}

\textbf{Proof:} By the CAR we have  
that
\begin{align}
 & \quad\quad\sum_{p\in B_{F}^{c}}c_{p}^{\ast}H_{\text{kin}}^{\prime\left(-1\right)}c_{p}=\sum_{p\in B_{F}^{c}}c_{p}^{\ast}\left(\sum_{q\in B_{F}^{c}}\vert\left|q\right|^{2}-\zeta_{0}\vert\,c_{q}^{\ast}c_{q}+\sum_{q\in B_{F}}\vert\left|q\right|^{2}-\zeta_{0}\vert\,c_{q}c_{q}^{\ast}\right)c_{p}\\
 & =\left(\sum_{p\in B_{F}^{c}}c_{p}^{\ast}c_{p}\right)\left(\sum_{q\in B_{F}^{c}}\vert\left|q\right|^{2}-\zeta_{0}\vert\,c_{q}^{\ast}c_{q}+\sum_{q\in B_{F}}\vert\left|q\right|^{2}-\zeta_{0}\vert\,c_{q}c_{q}^{\ast}\right)+\sum_{p\in B_{F}^{c}}c_{p}^{\ast}\left[\sum_{q\in B_{F}^{c}}\vert\left|q\right|^{2}-\zeta_{0}\vert\,c_{q}^{\ast}c_{q},c_{p}\right]\nonumber \\
 & =\mathcal{N}_{E}H_{\kin}^{\prime}-\sum_{p,q\in B_{F}^{c}}\vert\left|q\right|^{2}-\zeta_{0}\vert\,\delta_{p,q}c_{p}^{\ast}c_{q} \leq\mathcal{N}_{E}H_{\kin}^{\prime}\nonumber 
\end{align}
and the inequality for $c_{p}H_{\text{kin}}^{\prime\left(+1\right)}c_{p}^{\ast}$ can be derived similarly. 
That $\tilde{c}_{p}^{\ast}H_{\text{kin}}^{\prime\left(\pm1\right)}\tilde{c}_{p}\leq H_{\kin}^{\prime}$
follows exactly as the inequality $\tilde{c}_{p}^{\ast}\mathcal{N}_{E}^{\left(\pm1\right)}\tilde{c}_{p}\leq\mathcal{N}_{E}$
did in Lemma \ref{lemma:ExcitationNumberOperatorCommutators}.
$\hfill\square$

\medskip
Now we are ready to give the

\textbf{Proof of Proposition \ref{prop:KineticExchangeTermEstimates}:} For all schematic forms except 
\begin{equation} \label{eq:kinetic-exchange-missing}
\sum_{l\in S_{C}}\sum_{p\in S}\tilde{c}_{p_{3}}^{\ast}b_{k}^{\natural}\left(Te_{p_{1}}\right)b_{l}^{\natural}\left(K_{l}^{\oplus}e_{p_{2}}\right)\tilde{c}_{p_{4}}
\end{equation} 
we can use the estimates derived in Proposition \ref{prop:ExchangeTermEstimates},
specifically the equations (\ref{eq:ExchangeTermEstimatesEquation2})
through (\ref{eq:ExchangeTermEstimatesEquation4}), and the fact that $\mathcal{N}_{E}\leq H_{\kin}^{\prime}$. For the schematic form in \eqref{eq:kinetic-exchange-missing}, 
%as
%\begin{align}
%\sum_{l\in S_{C}}\sum_{p\in S}\left|\left\langle \Psi,\left[\tilde{c}_{p_{3}},b_{k}^{\ast}\left(Te_{p_{1}}\right)\right]^{\ast}b_{l}^{\natural}\left(K_{l}^{\oplus}e_{p_{2}}\right)\tilde{c}_{p_{4}}\Psi\right\rangle \right| & \leq C\left\Vert T\right\Vert _{\infty,2}\left(\sum_{l\in S_{C}}\left\Vert K_{l}^{\oplus}\right\Vert _{\infty,2}\right)\left\Vert \Psi\right\Vert \sqrt{\left\langle \Psi,\mathcal{N}_{E}^{2}\Psi\right\rangle }\nonumber \\
%\sum_{l\in S_{C}}\sum_{p\in S}\left|\left\langle \Psi,\left[\left[\tilde{c}_{p_{3}},b_{k}^{\ast}\left(Te_{p_{1}}\right)\right],b_{l}\left(K_{l}^{\oplus}e_{p_{2}}\right)\right]^{\ast}\tilde{c}_{p_{4}}\Psi\right\rangle \right| & \leq C\left\Vert T\right\Vert _{\infty,2}\left(\sum_{l\in S_{C}}\left\Vert K_{l}^{\oplus}\right\Vert _{\infty,2}\right)\left\langle \Psi,\mathcal{N}_{E}\Psi\right\rangle \nonumber \\
%\sum_{l\in S_{C}}\sum_{p\in S}\left|\left\langle \Psi,\left[\tilde{c}_{p_{3}},b_{k}^{\ast}\left(Te_{p_{1}}\right)\right]^{\ast}\left[\tilde{c}_{p_{4}},b_{l}^{\ast}\left(K_{l}^{\oplus}e_{p_{2}}\right)\right]\Psi\right\rangle \right| & \leq C\left\Vert T\right\Vert _{\infty,2}\left(\sum_{l\in S_{C}}\left\Vert K_{l}^{\oplus}\right\Vert _{\infty,2}\right)\left\Vert \Psi\right\Vert ^{2}
%\end{align}
%and since $\mathcal{N}_{E}\leq H_{\kin}^{\prime}$ these estimates
%are all accounted for by the final line in the statement of the proposition.
%
%For $\sum_{l\in S_{C}}\sum_{p\in S}\tilde{c}_{p_{3}}^{\ast}b_{k}^{\natural}\left(Te_{p_{1}}\right)b_{l}^{\natural}\left(K_{l}^{\oplus}e_{p_{2}}\right)\tilde{c}_{p_{4}}$
we can by Proposition \ref{prop:GeneralizedKineticEstimates} estimate
that
\begin{align}
 & \quad\,\sum_{l\in S_{C}}\sum_{p\in S}\left|\left\langle \Psi,\tilde{c}_{p_{3}}^{\ast}b_{k}^{\natural}\left(Te_{p_{1}}\right)b_{l}^{\natural}\left(K_{l}^{\oplus}e_{p_{2}}\right)\tilde{c}_{p_{4}}\Psi\right\rangle \right|\leq\sum_{l\in S_{C}}\sum_{p\in S}\left\Vert b_{k}^{\natural}\left(Te_{p_{1}}\right)\tilde{c}_{p_{3}}\Psi\right\Vert \left\Vert b_{l}^{\natural}\left(K_{l}^{\oplus}e_{p_{2}}\right)\tilde{c}_{p_{4}}\Psi\right\Vert \nonumber \\
 & \leq\sum_{l\in S_{C}}\sum_{p\in S}\left\Vert \left(h_{k}^{\oplus}\right)^{-\frac{1}{2}}Te_{p_{1}}\right\Vert \left\Vert \left(h_{l}^{\oplus}\right)^{-\frac{1}{2}}K_{l}^{\oplus}e_{p_{2}}\right\Vert \sqrt{\left\langle \tilde{c}_{p_{3}}\Psi,H_{\text{kin}}^{\prime\left(\pm1\right)}\tilde{c}_{p_{3}}\Psi\right\rangle \left\langle \tilde{c}_{p_{4}}\Psi,H_{\text{kin}}^{\prime\left(\pm1\right)}\tilde{c}_{p_{4}}\Psi\right\rangle }\nonumber \\
 & +\sum_{l\in S_{C}}\sum_{p\in S}\left\Vert \left(h_{k}^{\oplus}\right)^{-\frac{1}{2}}Te_{p_{1}}\right\Vert \left\Vert K_{l}^{\oplus}e_{p_{2}}\right\Vert \sqrt{\left\langle \tilde{c}_{p_{3}}\Psi,H_{\text{kin}}^{\prime\left(\pm1\right)}\tilde{c}_{p_{3}}\Psi\right\rangle }\left\Vert \tilde{c}_{p_{4}}\Psi\right\Vert \\
 & +\sum_{l\in S_{C}}\sum_{p\in S}\left\Vert Te_{p_{1}}\right\Vert \left\Vert \left(h_{l}^{\oplus}\right)^{-\frac{1}{2}}K_{l}^{\oplus}e_{p_{2}}\right\Vert \left\Vert \tilde{c}_{p_{3}}\Psi\right\Vert \sqrt{\left\langle \tilde{c}_{p_{4}}\Psi,H_{\text{kin}}^{\prime\left(\pm1\right)}\tilde{c}_{p_{4}}\Psi\right\rangle }\nonumber \\
 & +\sum_{l\in S_{C}}\sum_{p\in S}\left\Vert Te_{p_{1}}\right\Vert \left\Vert K_{l}^{\oplus}e_{p_{2}}\right\Vert \left\Vert \tilde{c}_{p_{3}}\Psi\right\Vert \left\Vert \tilde{c}_{p_{4}}\Psi\right\Vert \nonumber \\
 & =:A_{1}+A_{2}+A_{3}+A_{4}.\nonumber 
\end{align}
The terms $A_{1}$ through $A_{4}$ can be estimated by the Cauchy-Schwarz
inequality, Lemma \ref{lemma:KineticEstimationSimplification}, the
inequality $\mathcal{N}_{E}\leq H_{\kin}^{\prime}$ and the
fact that $\max_{p\in L_{k}^{\pm}}\left\Vert Te_{p}\right\Vert \leq\left\Vert T\right\Vert _{\infty,2}$
as
\begin{align}
A_{1} & \leq\left(\max_{p\in L_{k}^{\pm}}\left\Vert \left(h_{k}^{\oplus}\right)^{-\frac{1}{2}}Te_{p}\right\Vert \right)\left(\sum_{l\in S_{C}}\left\Vert \left(h_{l}^{\oplus}\right)^{-\frac{1}{2}}K_{l}^{\oplus}\right\Vert _{\text{HS}}\right)\sqrt{\left\langle \Psi,H_{\kin}^{\prime}\Psi\right\rangle \left\langle \Psi,\mathcal{N}_{E}H_{\kin}^{\prime}\Psi\right\rangle },\nonumber \\
A_{2} & \leq\left(\max_{p\in L_{k}^{\pm}}\left\Vert \left(h_{k}^{\oplus}\right)^{-\frac{1}{2}}Te_{p}\right\Vert \right)\left(\sum_{l\in S_{C}}\left\Vert K_{l}^{\oplus}\right\Vert _{\infty,2}\right)\sqrt{\left\langle \Psi,H_{\kin}^{\prime}\Psi\right\rangle \left\langle \Psi,\mathcal{N}_{E}H_{\kin}^{\prime}\Psi\right\rangle },\\
A_{3} & \leq\left\Vert T\right\Vert _{\infty,2}\left(\sum_{l\in S_{C}}\left\Vert \left(h_{l}^{\oplus}\right)^{-\frac{1}{2}}K_{l}^{\oplus}\right\Vert _{\text{HS}}\right)\left\langle \Psi,H_{\kin}^{\prime}\Psi\right\rangle \nonumber, \\
A_{4} & \leq\left\Vert T\right\Vert _{\infty,2}\left(\sum_{l\in S_{C}}\left\Vert K_{l}^{\oplus}\right\Vert _{\infty,2}\right)\left\langle \Psi,H_{\kin}^{\prime}\Psi\right\rangle ,\nonumber 
\end{align}
all of which are also accounted for by the statement.
$\hfill\square$

%%%%%%%%%%%%%%%%%%%%%%%%%%%%%%%%%%%%
%%%%%%%%%%%%%%%%%%%%%%%%%%%%%%%%%%%%

\section{Analysis of the One-Body
Operators $K$, $A\left(t\right)$ and $B\left(t\right)$} \label{sec:AnalysisoftheOne-BodyOperators}

In this section we study the one-body operators on $\ell^2(L_k)$ defined in Section \ref{sec:quasi-bosonic-Bogolubov}, including $K_k$ introduced in \eqref{eq:KkDefinition} and $A_k,B_k$ defined in Proposition \ref{prop:trans-eK-bosonizable-terms}: 
\begin{align}\label{eq:KernelDefinitionReminder}
K_{k}&=-\frac{1}{2}\log\left(h_{k}^{-\frac{1}{2}}\left(h_{k}^{2}+2P_{h_{k}^{\frac{1}{2}}v_{k}}\right)^{\frac{1}{2}}h_{k}^{-\frac{1}{2}}\right), \\
A_{k}\left(t\right) & =\frac{1}{2}\left(e^{tK_{k}}\left(h_{k}+2P_{v_{k}}\right)e^{tK_{k}}+e^{-tK_{k}}h_{k}e^{-tK_{k}}\right)-h_{k},\nonumber\\
B_{k}\left(t\right) & =\frac{1}{2}\left(e^{tK_{k}}\left(h_{k}+2P_{v_{k}}\right)e^{tK_{k}}-e^{-tK_{k}}h_{k}e^{-tK_{k}}\right)\nonumber 
\end{align}
where 
\begin{align}\label{eq:hkvkReminder}
h_{k}e_{p} =\lambda_{k,p}e_{p},\quad\;\;\lambda_{k,p}=\frac{1}{2}\left(\left|p\right|^{2}-\left|p-k\right|^{2}\right), \quad P_{v_{k}}  = |v_k \rangle \langle v_k|, \quad v_{k}=\sqrt{\frac{\hat{V}_{k}k_{F}^{-1}}{2\left(2\pi\right)^{3}}}\sum_{p\in L_{k}}e_{p},
\end{align}
and $\left(e_{p}\right)_{p\in L_{k}}$ is the standard orthonormal
basis of $\ell^{2}\left(L_{k}\right)$. We will need precise estimates on these operators to control the quasi-bosonic
Bogolubov transformation $e^{\mathcal{K}}$ diagonalizing the
bosonizable terms. In particular, we will prove the following bounds. 

\begin{prop}[Trace formulas]
\label{thm:ExplicitKAktBktEstimates} For all $k\in\mathbb{Z}_{\ast}^{3}$
it holds that $K_k\le 0$ and 
$$
\tr\left(K_{k}\right) = -\frac{1}{4}\log\left(1+2\hat{V}_{k}\left(\frac{k_{F}^{-1}}{2\left(2\pi\right)^{3}}\sum_{p\in L_{k}}\lambda_{k,p}^{-1}\right)  \right) \ge - C\hat{V}_{k}. 
$$
Moreover, with $E_k= e^{-K_k} h_k e^{-K_k}$ we have 
\[
\tr\left(E_{k}-h_{k}\right)-\frac{\hat{V}_{k}k_{F}^{-1}}{2\left(2\pi\right)^{3}}\left|L_{k}\right|=\frac{1}{\pi}\int_{0}^{\infty}F\left(\frac{\hat{V}_{k}k_{F}^{-1}}{\left(2\pi\right)^{3}}\sum_{p\in L_{k}}\frac{\lambda_{k,p}}{\lambda_{k,p}^{2}+t^{2}}\right)dt,
\]
with $F\left(x\right)=\log\left(1+x\right)-x$, and 
\[
\left|\tr\left(E_{k}-h_{k}\right)-\frac{\hat{V}_{k}k_{F}^{-1}}{2\left(2\pi\right)^{3}}\left|L_{k}\right|\right|\leq Ck_{F}\hat{V}_{k}^{2}\left|k\right|,\quad k_{F}\rightarrow\infty.
\]
Here $C>0$ is a constant independent of $k$ and $k_{F}$.
\end{prop}

%Check label of Theorem \ref{thm:ExplicitKAktBktEstimates} and \ref{thm:ExplicitKAktBktEstimates-3}

\begin{prop} [Matrix element estimates] 
\label{thm:ExplicitKAktBktEstimates-2}For all $k\in\overline{B}\left(0,2k_{F}\right)\cap\mathbb{Z}_{\ast}^{3}$ 
it holds that
\begin{align*}
\left\Vert K_{k}\right\Vert _{\infty,2} & \leq C\hat{V}_{k}\log\left(k_{F}\right)^{\frac{1}{3}}k_{F}^{-\frac{2}{3}}\left|k\right|^{1+\frac{5}{6}}
\end{align*}
and for all $t\in\left[0,1\right]$ that
\[
\left\Vert A_{k}\left(t\right)\right\Vert _{\infty,2},\,\left\Vert B_{k}\left(t\right)\right\Vert _{\infty,2}\leq C\hat{V}_{k}\left|k\right|^{\frac{1}{2}}\left(1+\hat{V}_{k}\right).
\]
Moreover, with $E_k= e^{-K_k} h_k e^{-K_k}$ we have
$$
\max_{p\in L_{k}}\left|\left\langle e_{p},\left(E_k -h_{k}\right)e_{p}\right\rangle \right|  \leq Ck_{F}^{-1}\hat{V}_{k}\left(1+\hat{V}_{k}\right).
$$
Here $C>0$ is a constant  independent of $k$ and $k_{F}$.
\end{prop}

\begin{prop} [Kinetic estimates] \label{thm:ExplicitKAktBktEstimates-3} 
For all $k\in\overline{B}\left(0,2k_{F}\right)$ it holds as $k_{F}\rightarrow\infty$
that
\begin{align*}
\left\Vert h_{k}^{-\frac{1}{2}}K_{k}\right\Vert _{\HS} & \leq C(\log k_{F})^{\frac{2}{3}}k_{F}^{-\frac{1}{3}}\hat{V}_{k}\left|k\right|^{3+\frac{2}{3}}\\
\left\Vert \left\{ K_{k},h_{k}\right\} h_{k}^{-\frac{1}{2}}\right\Vert _{\HS} & \leq Ck_{F}^{\frac{1}{2}}\hat{V}_{k}\left|k\right|^{\frac{1}{2}}\\
\left\Vert h_{k}^{-\frac{1}{2}}\left\{ K_{k},h_{k}\right\} h_{k}^{-\frac{1}{2}}\right\Vert _{\HS} & \leq C\hat{V}_{k}
\end{align*}
and for all $t\in\left[0,1\right]$
\[
\max_{p\in L_{k}}\left\Vert h_{k}^{-\frac{1}{2}}A_{k}\left(t\right)e_{p}\right\Vert ,\,\max_{p\in L_{k}}\left\Vert h_{k}^{-\frac{1}{2}}B_{k}\left(t\right)e_{p}\right\Vert \leq Ck_{F}^{-\frac{1}{2}}\hat{V}_{k}\left(1+\hat{V}_{k}^{2}\right).
\]
Here $C>0$ is a constant independent of $k$ and $k_{F}$.
\end{prop}

{\bf Notation.} In order to simplify the notation, we will throughout
this section let $h:V\rightarrow V$ denote any positive self-adjoint
operator acting on an $n$-dimensional Hilbert space $V$, let $\left(x_{i}\right)_{i=1}^{n}$
be an eigenbasis for $h$ with eigenvalues $\left(\lambda_{i}\right)_{i=1}^{n}$
and let $v\in V$ be any vector satisfying $\left\langle x_{i},v\right\rangle \geq0$ for all $1\leq i\leq n$. 
% (this last condition can always be ensured
%by multiplying the basis vectors $\left(x_{i}\right)_{i=1}^{n}$ by
%appropriate phases). @Nam: I think we should avoid mentioning the phase, since we think of V as a real space
We will establish general results for the operators (c.f. \eqref{eq:KernelDefinitionReminder})
\begin{align} \label{eq:CleanKDefinition}
K&=-\frac{1}{2}\log\left(h^{-\frac{1}{2}}\left(h^{2}+2P_{h^{\frac{1}{2}}v}\right)^{\frac{1}{2}}h^{-\frac{1}{2}}\right),\\
A\left(t\right) & =\frac{1}{2}\left(e^{tK}\left(h+2P_{v}\right)e^{tK}+e^{-tK}he^{-tK}\right)-h,\nonumber\\
B\left(t\right) & =\frac{1}{2}\left(e^{tK}\left(h+2P_{v}\right)e^{tK}-e^{-tK}he^{-tK}\right),\nonumber 
\end{align}
and then at the end insert the specific choice \eqref{eq:hkvkReminder} to get explicit estimates.

%For any $k\in\mathbb{Z}_{+}^{3}$ the operators $h_{k}$ and $P_{v_{k}}$
%are of similar forms: All the $h_{k}$ operators are positive and
%diagonalized by the basis $\left(e_{p}\right)_{p\in L_{k}}$, and
%all the $P_{v_{k}}$ operators are (unnormalized) rank one projections
%onto vectors $v_{k}$ which satisfy $\left\langle e_{p},v_{k}\right\rangle \geq0$
%for all $p\in L_{k}$. All other particular details are irrelevant
%for the results of this section, so for simplicity we will throughout
%this section let $h:V\rightarrow V$ denote any positive self-adjoint
%operator acting on an $n$-dimensional Hilbert space $V$, let $\left(x_{i}\right)_{i=1}^{n}$
%be an eigenbasis for $h$ with eigenvalues $\left(\lambda_{i}\right)_{i=1}^{n}$
%and let $v\in V$ be any vector satisfying $\left\langle x_{i},v\right\rangle \geq0$
%for all $1\leq i\leq n$ (this last condition can always be ensured
%by multiplying the basis vectors $\left(x_{i}\right)_{i=1}^{n}$ by
%appropriate phases).

We will prove the trace formulas first. Then we  derive general estimates for the matrix elements
of the operators $e^{-2K}$ and $e^{2K}$ in terms of a single, simpler
operator $T$. 
%We then show that these estimates are optimal by introducing
%a small parameter $g$ to the problem, as we then conclude that $-2K=T+O\left(g^{2}\right)$
%in the small $g$ limit, and observe that this limit is in fact also
%relevant for our particular problem.
This allows us to show that all matrix
elements of $K$ are non-negative, which in turn implies that all matrix elements
of $e^{-tK}$, $\sinh\left(-tK\right)$ and $\cosh\left(-tK\right)$
are convex with respect to $t$. With these estimates we can then obtain the desired estimates of $K$, $A(t)$ and $B(t)$.

%and $\left\Vert B\left(t\right)\right\Vert _{\infty,2}$, and we end
%the section by inserting the particular $h_{k}$'s and $v_{k}$'s
%of our problem into the general estimates of the section for the following
%result:
%\begin{thm}
%\label{them:KAktBktEstimates}For all $k\in\mathbb{Z}_{+}^{3}$ it
%holds that
%\begin{align*}
%\tr\left(E_{k}-h_{k}\right) & =\frac{1}{\pi}\int_{0}^{\infty}\log\left(1+\frac{\hat{V}_{k}k_{F}^{-1}}{\left(2\pi\right)^{3}}\sum_{p\in L_{k}}\frac{\lambda_{k,p}}{\lambda_{k,p}^{2}+t^{2}}\right)dt\\
%\tr\left(\left|K_{k}\right|\right) & \leq C\hat{V}_{k}k_{F}^{-1}\sum_{p\in L_{k}}\frac{1}{\lambda_{k,p}}\\
%\left\Vert K_{k}\right\Vert _{\infty,2} & \leq C\hat{V}_{k}k_{F}^{-1}\sqrt{\sum_{p\in L_{k}}\frac{1}{\lambda_{k,p}^{2}}}
%\end{align*}
%and for all $t\in\left[0,1\right]$
%\[
%\left\Vert A_{k}\left(t\right)\right\Vert _{\infty,2},\,\left\Vert B_{k}\left(t\right)\right\Vert _{\infty,2}\leq C\hat{V}_{k}k_{F}^{-1}\sqrt{\left|L_{k}\right|}\left(1+\hat{V}_{k}k_{F}^{-1}\sum_{p\in L_{k}}\frac{1}{\lambda_{k,p}}\right)
%\]
%for a constant $C>0$ independent of all quantities.
%\end{thm}
%
%The remaining unkown quantities are seen to be Riemann sums over the
%lunes $L_{k}$, the estimation of which is the subject of the next
%section.
%

\subsection{Trace Formulas} \label{sec:Trace-formulas}

In this section we prove Proposition \ref{thm:ExplicitKAktBktEstimates}. We will prove some general results using the notation in \eqref{eq:CleanKDefinition}, and then we insert the special choice of $h_k$, $v_k$ in \eqref{eq:hkvkReminder} to conclude. Let us start with

%We define a self-adjoint operator $K:V\rightarrow V$ as in equation
%(\ref{eq:KernelDefinitionReminder}) by
%\begin{equation}
%K=-\frac{1}{2}\log\left(h^{-\frac{1}{2}}\left(h^{2}+2P_{h^{\frac{1}{2}}v}\right)^{\frac{1}{2}}h^{-\frac{1}{2}}\right).\label{eq:CleanKDefinition}
%\end{equation}

\begin{prop}\label{prop:KTrace} The operator $K$ in \eqref{eq:CleanKDefinition} satisfies $K\le 0$ and 
\[
\tr\left(K\right)=-\frac{1}{4}\log\left(1+2\left\langle v,h^{-1}v \right\rangle \right). 
\]
\end{prop}

\textbf{Proof:} Since $h^{2}+2P_{h^{\frac{1}{2}}v} \ge h^2>0$ and $A\mapsto A^{\frac{1}{2}}$ is operator monotone, we find that 
\begin{equation}
h^{-\frac{1}{2}}\left(h^{2}+2P_{h^{\frac{1}{2}}v}\right)^{\frac{1}{2}}h^{-\frac{1}{2}}\geq h^{-\frac{1}{2}}\left(h^{2}\right)^{\frac{1}{2}}h^{-\frac{1}{2}}=1.
\end{equation}
Hence $K$ is well-defined and $K\le 0$. By the identity $\text{tr}\left(\log\left(A\right)\right)=\log\left(\det\left(A\right)\right)$
and multiplicativity of the determinant we find
\begin{align}
\text{tr}\left(K\right) & =-\frac{1}{2}\log\left(\det\left(h^{-\frac{1}{2}}\left(h^{2}+2P_{h^{\frac{1}{2}}v}\right)^{\frac{1}{2}}h^{-\frac{1}{2}}\right)\right)\\
&=-\frac{1}{4}\log\left(\det\left(h\right)^{-1}\det\left(h^{2}+2P_{h^{\frac{1}{2}}v}\right)\det\left(h\right)^{-1}\right)\nonumber \\
 & =-\frac{1}{4}\log\left(\det\left(h^{-1}\left(h^{2}+2P_{h^{\frac{1}{2}}v}\right)h^{-1}\right)\right)=-\frac{1}{4}\log\left(\det\left(1+2P_{h^{-\frac{1}{2}}v}\right)\right), \nonumber 
\end{align}
and by Sylvester's determinant theorem \cite{Sylvester-1883}, $\det\left(1+\alpha P_{x}\right)=1+\alpha\left\Vert x\right\Vert ^{2}$
for any $\alpha\in\mathbb{C}$, hence
\begin{equation}
\text{tr}\left(K\right)=-\frac{1}{4}\log\left(1+2\left\Vert h^{-\frac{1}{2}}v\right\Vert ^{2}\right)=-\frac{1}{4}\log\left(1+2\left\langle v,h^{-1}v\right\rangle \right).
\end{equation}
$\hfill\square$

Another exact trace formula which we will need is the following integral representation
of the square root of a rank one perturbation, first presented in \cite{BNPSS-20}.  
\begin{prop}
\label{prop:IntegralFormulaForRankOnePerturbation}Let $\left(H,\left\langle \cdot,\cdot\right\rangle \right)$
be a Hilbert space and let $A:H\rightarrow H$ be a positive self-adjoint
operator. Then for any $x\in H$ and $g\in\mathbb{R}$ such that $A+gP_{x}>0$
it holds that
\[
\left(A+gP_{x}\right)^{\frac{1}{2}}=A^{\frac{1}{2}}+\frac{2g}{\pi}\int_{0}^{\infty}\frac{t^{2}}{1+g\left\langle x,\left(A+t^{2}\right)^{-1}x\right\rangle }P_{\left(A+t^{2}\right)^{-1}x}\,dt
\]
and
\[
\tr\left(\left(A+gP_{x}\right)^{\frac{1}{2}}\right)=\tr\left(A^{\frac{1}{2}}\right)+\frac{1}{\pi}\int_{0}^{\infty}\log\left(1+g\left\langle x,\left(A+t^{2}\right)^{-1}x\right\rangle \right)dt.
\]
\end{prop}

Note that Proposition \ref{prop:IntegralFormulaForRankOnePerturbation} follows from the Sherman--Morrison formula \cite{SheMor-50} 
\begin{equation}  \label{eq:SM-formula}
\left(A+gP_{x,y}\right)^{-1}=A^{-1}-\frac{g}{1+g\left\langle x,A^{-1}y\right\rangle }P_{\left(A^{\ast}\right)^{-1}x,A^{-1}y}.
\end{equation}
with $P_{x,y}= |y\rangle \langle x| = \left\langle x,\cdot\right\rangle y$, and the functional calculus 
\begin{equation} \label{eq:integral-sqrtA}
\sqrt{A}=\frac{2}{\pi}\int_{0}^{\infty}\frac{A}{A+t^{2}}\,dt=\frac{2}{\pi}\int_{0}^{\infty}\left(1-\frac{t^{2}}{A+t^{2}}\right)dt
\end{equation}
for every self-adjoint non-negative operator $A$. 
Using this we conclude the following:

\begin{prop}
\label{thm:ExplicitKAktBktEstimates-1}The trace of $E-h$ where $E=e^{-K}he^{-K}$
is given by
\[
\tr\left(E-h\right)=\frac{1}{\pi}\int_{0}^{\infty}\log\left(1+2\left\langle v,h\left(h^{2}+t^{2}\right)^{-1}v\right\rangle \right)dt.
\]
\end{prop}

\textbf{Proof:} By cyclicity of the trace and the definition of $K$
\begin{equation}
\text{tr}\left(e^{-K}he^{-K}\right)=\text{tr}\left(he^{-2K}\right)=\text{tr}\left(h\left(h^{-\frac{1}{2}}\left(h^{2}+2P_{h^{\frac{1}{2}}v}\right)^{\frac{1}{2}}h^{-\frac{1}{2}}\right)\right)=\text{tr}\left(h^{2}+2P_{h^{\frac{1}{2}}v}\right)^{\frac{1}{2}},
\end{equation}
so applying Proposition \ref{prop:IntegralFormulaForRankOnePerturbation}
with $A=h^{2}$, $x=h^{\frac{1}{2}}v$ and $g=2$ 
%we find that
%\begin{align}
%\text{tr}\left(e^{-K}he^{-K}\right) %& =\text{tr}\left(\left(h^{2}\right)^{\frac{1}{2}}\right)+\frac{1}{\pi}\int_{0}^{\infty}\log\left(1+2\left\langle h^{\frac{1}{2}}v,\left(h^{2}+t^{2}\right)^{-1}h^{\frac{1}{2}}v\right\rangle \right)dt\\
%  =\text{tr}\left(h\right)+\frac{1}{\pi}\int_{0}^{\infty}\log\left(1+2\left\langle v,h\left(h^{2}+t^{2}\right)^{-1}v\right\rangle \right)dt
%\end{align}
%which implies the claim.
we get the claim.
$\hfill\square$

%\begin{align}
%delete
%\end{align}

{\bf Proof of Proposition \ref{thm:ExplicitKAktBktEstimates}:} By inserting $h_{k}$ and $v_{k}$ in Proposition \ref{prop:KTrace}, we get $K_k\le 0$ and 
\begin{equation}
\tr\left(K_k\right)=-\frac{1}{4}\log\left(1+2\left\langle v_k,h_k^{-1}v_k \right\rangle \right). 
\end{equation}
With the choice of $h_{k}$ and $v_{k}$ in \eqref{eq:hkvkReminder} we have
\begin{equation} \label{eq:v-h-1-v}
0\le \left\langle v_{k},h_{k}^{-1}v_{k}\right\rangle =\frac{\hat{V}_{k}k_{F}^{-1}}{2\left(2\pi\right)^{3}}\sum_{p,q\in L_{k}}\left\langle e_{p},h_{k}^{-1}e_{q}\right\rangle =\frac{\hat{V}_{k} k_{F}^{-1}}{2\left(2\pi\right)^{3}}\sum_{p\in L_{k}}\lambda_{k,p}^{-1}  \le C \hat V_k 
\end{equation}
where the last inequality is taken from Proposition \ref{coro:CompletelyUniformRiemannSumBound} in the Appendix. Combining with the bound 
$
\log (1+x) \le x
$ with $x>0$, we find that 
\begin{equation}
\tr\left(K_k\right)=-\frac{1}{4}\log\left(1+2 \hat{V}_{k}\left(\frac{k_{F}^{-1}}{2\left(2\pi\right)^{3}}\sum_{p\in L_{k}}\lambda_{k,p}^{-1}\right)    \right) \ge - C  \hat V_k .
\end{equation}
Next, using Proposition \ref{thm:ExplicitKAktBktEstimates-1} and the identity (c.f. \eqref{eq:integral-sqrtA})
\begin{equation}
\left|L_{k}\right|=\sum_{p\in L_{k}}1=\frac{2}{\pi}\int_{0}^{\infty}\sum_{p\in L_{k}}\frac{\lambda_{k,p}}{\lambda_{k,p}^{2}+t^{2}}\,dt
\end{equation}
we conclude that 
\begin{equation}
\tr\left(E_{k}-h_{k}\right)-\frac{\hat{V}_{k}k_{F}^{-1}}{2\left(2\pi\right)^{3}}\left|L_{k}\right|=\frac{1}{\pi}\int_{0}^{\infty}F\left(\frac{\hat{V}_{k}k_{F}^{-1}}{\left(2\pi\right)^{3}}\sum_{p\in L_{k}}\frac{\lambda_{k,p}}{\lambda_{k,p}^{2}+t^{2}}\right)dt
\end{equation}
with $F\left(x\right)=\log\left(1+x\right)-x$. Since $\left|F\left(x\right)\right|\leq\frac{1}{2}x^{2}$
we have
\begin{align}
\left|\text{tr}\left(E_{k}-h_{k}\right)-\frac{\hat{V}_{k}k_{F}^{-1}}{2\left(2\pi\right)^{3}}\left|L_{k}\right|\right| & \leq\frac{1}{\pi}\int_{0}^{\infty}\frac{1}{2}\left(\frac{\hat{V}_{k}k_{F}^{-1}}{\left(2\pi\right)^{3}}\sum_{p\in L_{k}}\frac{\lambda_{k,p}}{\lambda_{k,p}^{2}+t^{2}}\right)^{2}dt\label{eq:CorrelationEnergyContributionEstimate}\\
 & =\frac{\hat{V}_{k}^{2}k_{F}^{-2}}{\left(2\pi\right)^{7}}\sum_{p,q\in L_{k}}\int_{0}^{\infty}\frac{\lambda_{k,p}}{\lambda_{k,p}^{2}+t^{2}}\frac{\lambda_{k,q}}{\lambda_{k,q}^{2}+t^{2}}\,dt\nonumber 
\end{align}
and by the integral identity 
\begin{align} \label{eq:integral-t2-a2-t2-b2}
\int_{0}^{\infty}\frac{a}{a^{2}+t^{2}}\frac{b}{b^{2}+t^{2}}\,dt=\frac{\pi}{2}\frac{1}{a+b},\quad a,b>0,
\end{align}
it holds that
\begin{equation}
\sum_{p,q\in L_{k}}\int_{0}^{\infty}\frac{\lambda_{k,p}}{\lambda_{k,p}^{2}+t^{2}}\frac{\lambda_{k,q}}{\lambda_{k,q}^{2}+t^{2}}\,dt=\frac{\pi}{2}\sum_{p,q\in L_{k}}\frac{1}{\lambda_{k,p}+\lambda_{k,q}}\leq\frac{\pi}{2}\sum_{p,q\in L_{k}}\frac{1}{\sqrt{\lambda_{k,p}}\sqrt{\lambda_{k,q}}}=\frac{\pi}{2}\left(\sum_{p\in L_{k}}\lambda_{k,p}^{-\frac{1}{2}}\right)^{2}.
\end{equation}
By Proposition \ref{prop:NonSingularRiemannSums} we have for any
$k\in\mathbb{Z}_{\ast}^{3}$ that
\begin{equation}
\sum_{p\in L_{k}}\lambda_{k,p}^{-\frac{1}{2}}\leq C\begin{cases}
k_{F}^{\frac{3}{2}}\sqrt{\left|k\right|} & \left|k\right|<2k_{F}\\
k_{F}^{3}\left|k\right|^{-1} & \left|k\right|\geq2k_{F}
\end{cases}\leq Ck_{F}^{\frac{3}{2}}\sqrt{\left|k\right|},\quad k_{F}\rightarrow\infty,
\end{equation}
for a constant $C>0$ independent of $k$ and $k_{F}$, so we get the desired bound
\begin{equation}
\left|\text{tr}\left(E_{k}-h_{k}\right)-\frac{\hat{V}_{k}k_{F}^{-1}}{2\left(2\pi\right)^{3}}\left|L_{k}\right|\right|\leq C\hat{V}_{k}^{2}k_{F}^{-2}\left(k_{F}^{\frac{3}{2}}\sqrt{\left|k\right|}\right)^{2}=Ck_{F}\hat{V}_{k}^{2}\left|k\right|. 
\end{equation}
$\hfill\square$

\subsection{Preliminary Estimates for $e^{-2K}$ and $e^{2K}$}

The square root formula also yields the following exact representations
of $e^{-2K}$ and $e^{2K}$:
\begin{prop}
\label{prop:EKRepresentations} The operator $K$ in \eqref{eq:CleanKDefinition} satisfies 
\begin{align*}
e^{-2K} & =1+\frac{4}{\pi}\int_{0}^{\infty}\frac{t^{2}}{1+2\left\langle v,h \left(h^{2}+t^{2}\right)^{-1}v\right\rangle }P_{\left(h^{2}+t^{2}\right)^{-1}v}\,dt\\
e^{2K} & =1-\frac{4}{\pi}\int_{0}^{\infty}\frac{t^{2}}{1+2\left\langle v,h^{-1}\left(h^{-2}+t^{2}\right)^{-1}v\right\rangle t^{2}}P_{h^{-1}\left(h^{-2}+t^{2}\right)^{-1}v}\,dt.
\end{align*}
\end{prop}

\textbf{Proof:} Let us consider 
\begin{equation}
e^{-2K}=h^{-\frac{1}{2}}\left(h^{2}+2P_{h^{\frac{1}{2}}v}\right)^{\frac{1}{2}}h^{-\frac{1}{2}}.
\end{equation}
first. Applying Proposition \ref{prop:IntegralFormulaForRankOnePerturbation}
with $A=h^{2}$, $x=h^{\frac{1}{2}}v$ and $g=2$ again we find
\begin{align} \label{eq:sqrt-h2-P}
\left(h^{2}+2P_{h^{\frac{1}{2}}v}\right)^{\frac{1}{2}} & =\left(h^{2}\right)^{\frac{1}{2}}+\frac{4}{\pi}\int_{0}^{\infty}\frac{t^{2}}{1+2\left\langle h^{\frac{1}{2}}v,\left(h^{2}+t^{2}\right)^{-1}h^{\frac{1}{2}}v\right\rangle }P_{\left(h^{2}+t^{2}\right)^{-1}h^{\frac{1}{2}}v}\,dt\\
 & =h+\frac{4}{\pi}\int_{0}^{\infty}\frac{t^{2}}{1+2\left\langle v,h\left(h^{2}+t^{2}\right)^{-1}v\right\rangle }P_{h^{\frac{1}{2}}\left(h^{2}+t^{2}\right)^{-1}v}\,dt\nonumber 
\end{align}
whence
\begin{align}
e^{-2K} & =h^{-\frac{1}{2}}\left(h+\frac{4}{\pi}\int_{0}^{\infty}\frac{t^{2}}{1+2\left\langle v,h\left(h^{2}+t^{2}\right)^{-1}v\right\rangle }P_{h^{\frac{1}{2}}\left(h^{2}+t^{2}\right)^{-1}v}\,dt\right)h^{-\frac{1}{2}}\\
 & =1+\frac{4}{\pi}\int_{0}^{\infty}\frac{t^{2}}{1+2\left\langle v,h\left(h^{2}+t^{2}\right)^{-1}v\right\rangle }P_{\left(h^{2}+t^{2}\right)^{-1}v}\,dt.\nonumber 
\end{align}
For $e^{2K}=h^{\frac{1}{2}}\left(h^{2}+2P_{h^{\frac{1}{2}}v}\right)^{-\frac{1}{2}}h^{\frac{1}{2}}$ we first use \eqref{eq:SM-formula}  
%the formula 
%\begin{align}
%\left(A+gP_{x}\right)^{-1}=A^{-1}-\frac{g}{1+g\left\langle x,A^{-1}x\right\rangle }P_{A^{-1}x}
%\end{align}
%for self-adjoint operator $A>0$ 
to write
%\begin{align}
%delete
%\end{align}
\begin{equation}
\left(h^{2}+2P_{h^{\frac{1}{2}}v}\right)^{-1}=\left(h^{2}\right)^{-1}-\frac{2}{1+2\left\langle h^{\frac{1}{2}}v,\left(h^{2}\right)^{-1}h^{\frac{1}{2}}v\right\rangle }P_{\left(h^{2}\right)^{-1}h^{\frac{1}{2}}v}=h^{-2}-\frac{2}{1+2\left\langle v,h^{-1}v\right\rangle }P_{h^{-\frac{3}{2}}v}.
\end{equation}
As this is an equality the right-hand side is in fact positive (as
the left-hand side is), so we may apply Proposition \ref{prop:IntegralFormulaForRankOnePerturbation}
with $A=h^{-2}$, $x=h^{-\frac{3}{2}}v$ and $g=-2\left(1+2\left\langle v,h^{-1}v\right\rangle \right)^{-1}$
for
\begin{align}
 & \quad\left(h^{2}+2P_{h^{\frac{1}{2}}v}\right)^{-\frac{1}{2}}=\left(h^{-2}-\frac{2}{1+2\left\langle v,h^{-1}v\right\rangle }P_{h^{-\frac{3}{2}}v}\right)^{\frac{1}{2}}\nonumber \\
 & =\left(h^{-2}\right)^{\frac{1}{2}}-\frac{2}{1+2\left\langle v,h^{-1}v\right\rangle }\frac{2}{\pi}\int_{0}^{\infty}\frac{t^{2}}{1-\frac{2}{1+2\left\langle v,h^{-1}v\right\rangle }\left\langle h^{-\frac{3}{2}}v,\left(h^{-2}+t^{2}\right)^{-1}h^{-\frac{3}{2}}v\right\rangle }P_{\left(h^{-2}+t^{2}\right)^{-1}h^{-\frac{3}{2}}v}\,dt\nonumber \\
 & =h^{-1}-\frac{4}{\pi}\int_{0}^{\infty}\frac{t^{2}}{1+2\left\langle v,h^{-1}v\right\rangle -2\left\langle v,h^{-3}\left(h^{-2}+t^{2}\right)^{-1}v\right\rangle }P_{h^{-\frac{3}{2}}\left(h^{-2}+t^{2}\right)^{-1}v}\,dt\\
 & =h^{-1}-\frac{4}{\pi}\int_{0}^{\infty}\frac{t^{2}}{1+2\left\langle v,h^{-1}\left(h^{-2}+t^{2}\right)^{-1}v\right\rangle t^{2}}P_{h^{-\frac{3}{2}}\left(h^{-2}+t^{2}\right)^{-1}v}\,dt,\nonumber 
\end{align}
hence
\begin{align}
e^{2K} & =h^{\frac{1}{2}}\left(h^{-1}-\frac{4}{\pi}\int_{0}^{\infty}\frac{t^{2}}{1+2\left\langle v,h^{-1}\left(h^{-2}+t^{2}\right)^{-1}v\right\rangle t^{2}}P_{h^{-\frac{3}{2}}\left(h^{-2}+t^{2}\right)^{-1}v}\,dt\right)h^{\frac{1}{2}}\\
 & =1-\frac{4}{\pi}\int_{0}^{\infty}\frac{t^{2}}{1+2\left\langle v,h^{-1}\left(h^{-2}+t^{2}\right)^{-1}v\right\rangle t^{2}}P_{h^{-1}\left(h^{-2}+t^{2}\right)^{-1}v}\,dt.\nonumber 
\end{align}
$\hfill\square$

These exact formulas now allow us to derive some simple estimates
for $e^{-2K}-1$ and $1-e^{2K}$. To state these estimates we first
define a new operator $T$ on $\ell^2 (L_k)$ with
matrix elements
\begin{equation} \label{eq:def-T}
\left\langle x_{i},T x_{j}\right\rangle =2\frac{\left\langle x_{i},v \right\rangle \left\langle v, x_{j}\right\rangle }{\lambda_{i}+\lambda_{j}},\quad \forall 1\le i,j \le n.
\end{equation}
Recall that $(x_i)_{i=1}^n$ are an eigenbasis of $h$ with eigenvalues $\lambda_{i}$'s and $\langle x_i,v\rangle\ge 0$ for all $1\le i\le n$.

\begin{prop}
\label{prop:EKEstimates} For $K$ in \eqref{eq:CleanKDefinition} and $T$ in \eqref{eq:def-T}, we have both the operator estimates
\begin{align*}
0&\leq e^{-2K} -1 \le T \leq ( 1+2\left\langle v ,h^{-1}v\right\rangle )  \left(e^{-2K}-1\right),\\
0&\leq 1- e^{2K}  \le T \leq ( 1+2\left\langle v ,h^{-1}v\right\rangle )  \left(1-e^{2K}\right),
\end{align*}
and for all $1\le i,j \le n$ the elementwise estimates
\begin{align*}
0&\leq \left\langle x_{i}, (e^{-2K} -1) x_j \right\rangle \le \left\langle x_{i}, T x_j \right\rangle    \le ( 1+2\left\langle v ,h^{-1}v\right\rangle )  \left\langle x_{i},\left(e^{-2K}-1\right)x_{j}\right\rangle,\\
0&\leq \left\langle x_{i}, (1-e^{2K}) x_j \right\rangle \le \left\langle x_{i}, T x_j \right\rangle  \le ( 1+2\left\langle v ,h^{-1}v\right\rangle )   \left\langle x_{i},\left(  1-e^{2K}\right)x_{j}\right\rangle.
\end{align*}
\end{prop}

\textbf{Proof:} We first prove the bound $0\le e^{-2K}-1\leq T$. Obviously $0\le e^{-2K}-1$ since $K\le 0$. Noting
that $\left\langle v,h\left(h^{2}+t^{2}\right)^{-1}v\right\rangle \geq0$
and $P_{\left(h^{2}+t^{2}\right)^{-1}v}\geq0$ for all $t\in\left[0,\infty\right)$
we have by the first identity of Proposition \ref{prop:EKRepresentations}
that
\begin{equation}
e^{-2K}-1=\frac{4}{\pi}\int_{0}^{\infty}\frac{t^{2}}{1+2\left\langle v,h\left(h^{2}+t^{2}\right)^{-1}v\right\rangle }P_{\left(h^{2}+t^{2}\right)^{-1}v}\,dt\leq\frac{4}{\pi}\int_{0}^{\infty}t^{2}P_{\left(h^{2}+t^{2}\right)^{-1}v}\,dt.
\end{equation}
We claim that the right-hand side is precisely $T$. To see this we
compute the matrix elements with respect to $\left(x_{i}\right)_{i=1}^{n}$:
For any $1\leq i,j\leq n$ we have
\begin{align}
 & \left\langle x_{i},\left(\frac{4}{\pi}\int_{0}^{\infty}t^{2}P_{\left(h^{2}+t^{2}\right)^{-1}v}\,dt\right)x_{j}\right\rangle =\frac{4}{\pi}\int_{0}^{\infty}t^{2}\left\langle x_{i},\left(h^{2}+t^{2}\right)^{-1}v\right\rangle \left\langle \left(h^{2}+t^{2}\right)^{-1}v,x_{j}\right\rangle dt\nonumber \\
 & =\frac{4}{\pi}\int_{0}^{\infty}t^{2}\frac{\left\langle x_{i},v\right\rangle }{\lambda_{i}^{2}+t^{2}}\frac{\left\langle v,x_{j}\right\rangle }{\lambda_{j}^{2}+t^{2}}\,dt=\left\langle x_{i},v\right\rangle \left\langle v,x_{j}\right\rangle \left(\frac{4}{\pi}\int_{0}^{\infty}\frac{t^{2}}{\left(\lambda_{i}^{2}+t^{2}\right)\left(\lambda_{j}^{2}+t^{2}\right)}\,dt\right)\\
 & =\left\langle x_{i},v\right\rangle \left\langle v,x_{j}\right\rangle \left(\frac{4}{\pi}\frac{\pi}{2}\frac{1}{\lambda_{i}+\lambda_{j}}\right)=2\frac{\left\langle x_{i},v\right\rangle \left\langle v,x_{j}\right\rangle }{\lambda_{i}+\lambda_{j}}=\left\langle x_{i},Tx_{j}\right\rangle \nonumber 
\end{align}
where we used that $\left(x_{i}\right)_{i=1}^{n}$ is an
eigenbasis for $h$ as well as the integral identity \eqref{eq:integral-t2-a2-t2-b2}.

The lower bound $T\leq \left(1+2\left\langle v,h^{-1}v\right\rangle \right) (e^{-2K}-1)$
follows by the same argument as 
\begin{equation}
\left\langle v,h\left(h^{2}+t^{2}\right)^{-1}v\right\rangle \leq\left\langle v,h\left(h^{2}\right)^{-1}v\right\rangle =\left\langle v,h^{-1}v\right\rangle 
\end{equation}
for all $t\in\left[0,\infty\right)$, so
\begin{equation}
e^{-2K}-1\geq\frac{1}{1+2\left\langle v,h^{-1}v\right\rangle }\left(\frac{4}{\pi}\int_{0}^{\infty}t^{2}P_{\left(h^{2}+t^{2}\right)^{-1}v}\,dt\right)=\frac{1}{1+2\left\langle v,h^{-1}v\right\rangle }T.
\end{equation}
The bounds 
\begin{equation}
0\le 1-e^{2K}\leq T \le  \left(1+2\left\langle v,h^{-1}v\right\rangle \right) (1-e^{2K})
\end{equation}
follow by exactly the same argument, starting from the second identity
of Proposition \ref{prop:EKRepresentations}, using that 
\begin{equation}
0\leq\left\langle v,h^{-1}\left(h^{-2}+t^{2}\right)^{-1}v\right\rangle t^{2}\leq\left\langle v,h^{-1}v\right\rangle 
\end{equation}
for all $t\in\left[0,\infty\right)$ as well as the integral identity \eqref{eq:integral-t2-a2-t2-b2}.

The matrix element estimates likewise follow by the same argument
as e.g.
\begin{align}
 0&\le \;\left\langle x_{i},\left(e^{-2K}-1\right)x_{j}\right\rangle =\frac{4}{\pi}\int_{0}^{\infty}\frac{t^{2}}{1+2\left\langle v,h\left(h^{2}+t^{2}\right)^{-1}v\right\rangle }\frac{\left\langle x_{i},v\right\rangle }{\lambda_{i}^{2}+t^{2}}\frac{\left\langle v,x_{j}\right\rangle }{\lambda_{j}^{2}+t^{2}}\,dt\nonumber \\
 & =\left\langle x_{i},v\right\rangle \left\langle v,x_{j}\right\rangle \left(\frac{4}{\pi}\int_{0}^{\infty}\frac{1}{1+2\left\langle v,h\left(h^{2}+t^{2}\right)^{-1}v\right\rangle }\frac{t^{2}}{\left(\lambda_{i}^{2}+t^{2}\right)\left(\lambda_{j}^{2}+t^{2}\right)}\,dt\right)\\
 & \leq\left\langle x_{i},v\right\rangle \left\langle v,x_{j}\right\rangle \left(\frac{4}{\pi}\int_{0}^{\infty}\frac{t^{2}}{\left(\lambda_{i}^{2}+t^{2}\right)\left(\lambda_{j}^{2}+t^{2}\right)}\,dt\right)=2\frac{\left\langle x_{i},v\right\rangle \left\langle v,x_{j}\right\rangle }{\lambda_{i}+\lambda_{j}}=\left\langle x_{i},Tx_{j}\right\rangle \nonumber 
\end{align}
by the assumption that the inner products $\left\langle x_{i},v\right\rangle $
and $\left\langle v,x_{j}\right\rangle $ are non-negative. % (which were stated before \eqref{eq:CleanKDefinition}).  
$\hfill\square$

%We remark that by using the estimate 
%$$1-e^{2K}\geq\left(1+2\left\langle v,h^{-1}v\right\rangle \right)^{-1}T$$
%rather than simply $1-e^{2K}\geq0$ one may derive the stronger estimate
%$$\cosh\left(-2K\right)-1\leq\frac{1}{2}\frac{\left\langle v,h^{-1}v\right\rangle }{1+\left\langle v,h^{-1}v\right\rangle }T$$
%(and likewise for the elementwise estimate), but for our application
%this is not a notable improvement so we stick with the simpler $\cosh\left(-2K\right)-1\leq\frac{1}{2}T$.

\begin{rmk}[Optimality of the Estimates] We may observe that the estimates for $e^{-2K}$, $e^{2K}$  are in general optimal. To see this, let us add
a small parameter $g\geq0$ to the problem by substituting $\sqrt{g}v$
for $v$ in equation (\ref{eq:CleanKDefinition}), i.e. defining
\begin{equation}
K_{g}=-\frac{1}{2}\log\left(h^{-\frac{1}{2}}\left(h^{2}+2gP_{h^{\frac{1}{2}}v}\right)^{\frac{1}{2}}h^{-\frac{1}{2}}\right),\quad T_{g}=gT.
\end{equation}
Then the general bounds of the corollary read for $K_{g}$ that
\begin{equation}
\frac{1}{1+2g\left\langle v,h^{-1}v\right\rangle }T_{g}\leq-2K_{g}\leq T_{g}.
\end{equation}
Hence, 
\begin{equation}
0\geq-2K_{g}-T_{g}\geq-\left(1-\frac{1}{1+2g\left\langle v,h^{-1}v\right\rangle }\right)T_{g}=-\left(\frac{2\left\langle v,h^{-1}v\right\rangle }{1+2g\left\langle v,h^{-1}v\right\rangle }T\right)g^{2}\geq-Cg^{2}
\end{equation}
which by self-adjointness of the operators involved implies that
\begin{equation}
-2K_{g}=T_{g}+O\left(g^{2}\right)=gT+O\left(g^{2}\right)
\end{equation}
with respect to, say, operator norm. This shows that the operator
$T_{g}=gT$ is in fact the first-order expansion of $K_{g}$ with
respect to the parameter $g$, which is then also the case for $e^{-2K_{g}}-1$,
$1-e^{2K_{g}}$ as e.g. $e^{-2K_{g}}-1=-2K_{g}+O\left(g^{2}\right)=T_{g}+O\left(g^{2}\right)$.
The estimate
\begin{equation}
\left\langle x_{i},\left(e^{-2K_{g}}-1\right)x_{j}\right\rangle \leq\left\langle x_{i},T_{g}x_{j}\right\rangle 
\end{equation}
is therefore (asymptotically) optimal since $T_{g}$ is precisely
the small $g$  limit of $e^{-2K_{g}}-1$.

This is relevant for our application, for although we do not have
an explicit parameter $g$ to consider we do have $\hat V_k$ as an effective one. More precisely, the summability condition of $\hat V_k$ ensures that essentially all but finitely many coefficients
$\hat{V}_{k}$ are small, even when the coefficients $(\hat{V}_{k})_{k\in\mathbb{Z}^{3}}$
are not finitely supported. 
\end{rmk}

\subsection{Matrix Element Estimates for $K$, $A(t)$, $B(t)$}

In this section we prove Proposition \ref{thm:ExplicitKAktBktEstimates}. As before, we will prove some general results using the notation from \eqref{eq:CleanKDefinition}, and then we insert $h_k$, $v_k$ from \eqref{eq:hkvkReminder} at the end. Recall that $(x_i)_{i}$ is an eigenbasis of $h$. We start with

\begin{prop} \label{prop:key-pointwise-positive-convex}
For all $1\leq i,j\leq n$, we have $\langle x_i, -K x_j \rangle \ge 0$ 
and the functions
\[
t\mapsto\left\langle x_{i},\left(e^{-tK}-1\right)x_{j}\right\rangle ,\,\left\langle x_{i},\sinh\left(-tK\right)x_{j}\right\rangle ,\,\left\langle x_{i},\left(\cosh\left(-tK\right)-1\right)x_{j}\right\rangle 
\]
are non-negative and convex for $t\in\left[0,\infty\right)$.
\end{prop}

\textbf{Proof:} By Proposition \ref{prop:EKEstimates}, the operator $S=1-e^{2K}$ satisfies that $0\le S<1$ and $\langle x_i, S x_j\rangle \ge 0$ for all $1\le i,j\le n$. By writing   
\begin{equation}
-2K = - \log(1-S)= S+ \frac{S^2}{2}+ \frac{S^3}{3}+ \frac{S^4}{4} + ...
\end{equation}
we find that $-2K$ also has non-negative matrix elements.
%, namely 
%\begin{equation}
%\langle x_i, (-2K) x_j\rangle = \sum_{m=1}^\infty \frac{\langle x_i, S^m x_j\rangle}{m} \ge 0, \quad \forall 1\le i,j\le n. 
%\end{equation}
%\begin{align}
%delete
%\end{align}
By using the series expansion again we see that for any $1\leq i,j\leq n$ and $t\in\left[0,\infty\right)$, 
\begin{align}
\left\langle x_{i},\left(e^{-tK}-1\right)x_{j}\right\rangle  & =\sum_{m=1}^{\infty}\frac{\left\langle x_{i},\left(-K\right)^{m}x_{j}\right\rangle }{m!}t^{m}\geq0,\\
\frac{d^{2}}{dt^{2}}\left\langle x_{i},\left(e^{-tK}-1\right)x_{j}\right\rangle  & =\sum_{m=3}^{\infty}\frac{\left\langle x_{i},\left(-K\right)^{m}x_{j}\right\rangle }{\left(m-2\right)!}t^{m-2}\geq0,\nonumber 
\end{align}
yielding the claim for $t\mapsto\left\langle x_{i},\left(e^{-tK}-1\right)x_{j}\right\rangle$. The functions $
t\mapsto \left\langle x_{i},\sinh\left(-tK\right)x_{j}\right\rangle$ and $t\mapsto \left\langle x_{i},\left(\cosh\left(-tK\right)-1\right)x_{j}\right\rangle$ can be treated similarly. 

$\hfill\square$

Next we have the key matrix element bounds. 
\begin{prop}
\label{prop:SuperGeneralElementBounds}For all $1\leq i,j\leq n$
and $t\in\left[0,1\right]$ we have the elementwise estimates
\begin{align*}
 & \left|\left\langle x_{i},Kx_{j}\right\rangle \right|,\,\left|\left\langle x_{i},\left(e^{-tK}-1\right)x_{j}\right\rangle \right|,\,\left|\left\langle x_{i},\left(1-e^{tK}\right)x_{j}\right\rangle \right|,\\
 & \qquad\;\;\left|\left\langle x_{i},\sinh\left(-tK\right)x_{j}\right\rangle \right|,\,\left|\left\langle x_{i},\left(\cosh\left(-tK\right)-1\right)x_{j}\right\rangle \right|\leq \frac{\left\langle x_{i},v\right\rangle \left\langle v,x_{j}\right\rangle }{\lambda_{i}+\lambda_{j}}.
\end{align*}
\end{prop}

\textbf{Proof:} The arguments for $e^{-tK}-1$, $\sinh\left(-tK\right)$ and $\cosh\left(-tK\right)-1$
are again the same, so we focus on $e^{-tK}-1$: By the convexity of Proposition \ref{prop:key-pointwise-positive-convex}  and the elementwise estimate of Proposition \ref{prop:EKEstimates}  we
find for all $t\in\left[0,1\right]$ that
\begin{align}
0 & \leq\left\langle x_{i},\left(e^{-tK}-1\right)x_{j}\right\rangle   \leq\left(1-\frac{t}{2}\right)\left\langle x_{i},\left(e^{-0\cdot K}-1\right)x_{j}\right\rangle +\frac{t}{2}\left\langle x_{i},\left(e^{-2K}-1\right)x_{j}\right\rangle \\
 & =\frac{t}{2}\left\langle x_{i},\left(e^{-2K}-1\right)x_{j}\right\rangle \leq\frac{1}{2}\left|\left\langle x_{i},\left(e^{-2K}-1\right)x_{j}\right\rangle \right|\leq\frac{1}{2}\left\langle x_{i},Tx_{j}\right\rangle =\frac{\left\langle x_{i},v\right\rangle \left\langle v,x_{j}\right\rangle }{\lambda_{i}+\lambda_{j}}.\nonumber 
\end{align}
This also gives us the estimate for $K$ as
\begin{equation}
0\leq\left\langle x_{i},\left(-K\right)x_{j}\right\rangle \leq\sum_{m=1}^{\infty}\frac{1}{m!}\left\langle x_{i},\left(-K\right)^{m}x_{j}\right\rangle =\left\langle x_{i},\left(e^{-K}-1\right)x_{j}\right\rangle \leq\frac{\left\langle x_{i},v\right\rangle \left\langle v,x_{j}\right\rangle }{\lambda_{i}+\lambda_{j}}
\end{equation}
where we used again the positivity of $\langle x_i, -K x_j\rangle$ from Proposition \ref{prop:key-pointwise-positive-convex}. Finally the estimate for $1-e^{-tK}$ is deduced from that of $\sinh\left(-tK\right)$
and $\cosh\left(-tK\right)-1$ as
\begin{align}
\left|\left\langle x_{i},\left(1-e^{tK}\right)x_{j}\right\rangle \right| & =\left|\left\langle x_{i},\sinh\left(-tK\right)x_{j}\right\rangle -\left\langle x_{i},\left(\cosh\left(-tK\right)-1\right)x_{j}\right\rangle \right|\\
 & \leq\max\left\{ \left|\left\langle x_{i},\sinh\left(-tK\right)x_{j}\right\rangle \right|,\left|\left\langle x_{i},\left(\cosh\left(-tK\right)-1\right)x_{j}\right\rangle \right|\right\} \leq\frac{\left\langle x_{i},v\right\rangle \left\langle v,x_{j}\right\rangle }{\lambda_{i}+\lambda_{j}}\nonumber 
\end{align}
where we also used the positivity of 
$\left\langle x_{i},\left(\cosh\left(-tK\right)-1\right)x_{j}\right\rangle$ and $\left\langle x_{i},\sinh\left(-tK\right)x_{j}\right\rangle $
from Proposition \ref{prop:key-pointwise-positive-convex} to justify the first inequality.

$\hfill\square$

As a simple application of these estimates we can  easily obtain 

\begin{prop}
\label{prop:Infty2NormofK}It holds that
\[
\left\Vert K\right\Vert _{\infty,2}\leq\alpha\sqrt{\left\langle v,h^{-2}v\right\rangle }
\]
where $\alpha=\max_{1\leq j\leq n}\left\langle v,x_{j}\right\rangle $.
\end{prop}

\textbf{Proof:} We estimate using Proposition \ref{prop:SuperGeneralElementBounds}
that
\begin{align}
\left\Vert K\right\Vert _{\infty,2}^{2} & =\sum_{i=1}^{n}\max_{1\leq j\leq n}\left|\left\langle x_{i},Kx_{j}\right\rangle \right|^{2}\leq\sum_{i=1}^{n}\max_{1\leq j\leq n}\left(\frac{\left\langle x_{i},v\right\rangle \left\langle v,x_{j}\right\rangle }{\lambda_{i}+\lambda_{j}}\right)^{2}\\
 & \leq\left(\max_{1\leq j\leq n}\left|\left\langle v,x_{j}\right\rangle \right|^{2}\right)\sum_{i=1}^{n}\frac{\left|\left\langle x_{i},v\right\rangle \right|^{2}}{\lambda_{i}^{2}}=\alpha^{2}\left\langle v,h^{-2}v\right\rangle . \nonumber
\end{align}
$\hfill\square$

Now we consider  $A\left(t\right)$ and $B\left(t\right)$, which can be written as  
\begin{equation} \label{eq:A49} 
A\left(t\right)=A_{h}\left(t\right)+e^{tK}P_{v}e^{tK},\quad B\left(t\right)=B_{h}\left(t\right)+e^{tK}P_{v}e^{tK},
\end{equation}
for
\begin{align}
A_{h}\left(t\right) & =\frac{1}{2}\left(e^{tK}he^{tK}+e^{-tK}he^{-tK}\right)-h\nonumber \\
 & =\cosh\left(-tK\right)h\cosh\left(-tK\right)+\sinh\left(-tK\right)h\sinh\left(-tK\right)-h\label{eq:AhtDefinition}\\
 & =\sinh\left(-tK\right)h\sinh\left(-tK\right)+\left(\cosh\left(-tK\right)-1\right)h\left(\cosh\left(-tK\right)-1\right)+\left\{ h,\cosh\left(-tK\right)-1\right\} \nonumber 
\end{align}
and
\begin{align} %\label{eq:A51}
B_{h}\left(t\right) & =\frac{1}{2}\left(e^{tK}he^{tK}-e^{-tK}he^{-tK}\right)\nonumber \\
 & =-\left(\cosh\left(-tK\right)h\sinh\left(-tK\right)+\sinh\left(-tK\right)h\cosh\left(-tK\right)\right)\label{eq:BhtDefinition}\\
 & =-\left(\left(\cosh\left(-tK\right)-1\right)h\sinh\left(-tK\right)+\sinh\left(-tK\right)h\left(\cosh\left(-tK\right)-1\right)+\left\{ h,\sinh\left(-tK\right)\right\} \right).\nonumber 
\end{align}
Specifically we must estimate the $\left\Vert \cdot\right\Vert _{\infty,2}$
norms of $A\left(t\right)$ and $B\left(t\right)$ with respect to
$\left(x_{i}\right)_{i=1}^{n}$. We begin with the $e^{tK}P_{v}e^{tK}$
term:
\begin{prop}
\label{prop:Infty2NormofetKPvetK}It holds for all $t\in\left[0,1\right]$
that
\[
\left\Vert e^{tK}P_{v}e^{tK}\right\Vert _{\infty,2}\leq\alpha\left(1+\left\langle v,h^{-1}v\right\rangle \right)\left\Vert v\right\Vert % \le C \alpha \|v\|
\]
where $\alpha=\max_{1\leq j\leq n}\left\langle v,x_{j}\right\rangle $.
\end{prop}

\textbf{Proof:} We first observe that
\begin{align}
\left\Vert e^{tK}P_{v}e^{tK}\right\Vert _{\infty,2}^{2} & =\sum_{i=1}^{n}\max_{1\leq j\leq n}\left|\left\langle x_{i},e^{tK}P_{v}e^{tK}x_{j}\right\rangle \right|^{2}=\sum_{i=1}^{n}\max_{1\leq j\leq n}\left|\left\langle x_{i},e^{tK}v\right\rangle \right|^{2}\left|\left\langle v,e^{tK}x_{j}\right\rangle \right|^{2}\label{eq:PvFirstEstimate}\\
 & =\left(\max_{1\leq j\leq n}\left|\left\langle v,e^{tK}x_{j}\right\rangle \right|^{2}\right)\left\Vert e^{tK}v\right\Vert ^{2}\leq\left(\max_{1\leq j\leq n}\left|\left\langle v,e^{tK}x_{j}\right\rangle \right|^{2}\right)\left\Vert v\right\Vert ^{2}\nonumber 
\end{align}
where we used that by monotonicity of $e^{x}$ and the fact that $K\leq0$,
$\left\Vert e^{tK}v\right\Vert ^{2}=\left\langle v,e^{2tK}v\right\rangle \leq\left\Vert v\right\Vert ^{2}$.
For the remaining factor we first write
\begin{equation}
\left\langle v,e^{tK}x_{j}\right\rangle =\left\langle v,x_{j}\right\rangle +\left\langle v,\left(e^{tK}-1\right)x_{j}\right\rangle =\left\langle v,x_{j}\right\rangle +\sum_{i=1}^{n}\left\langle v,x_{i}\right\rangle \left\langle x_{i},\left(e^{tK}-1\right)x_{j}\right\rangle ,\quad1\leq j\leq n,
\end{equation}
and estimate using Proposition \ref{prop:SuperGeneralElementBounds}
that
\begin{align}
\left|\sum_{i=1}^{n}\left\langle v,x_{i}\right\rangle \left\langle x_{i},\left(e^{tK}-1\right)x_{j}\right\rangle \right| & \leq\sum_{i=1}^{n}\left|\left\langle v,x_{i}\right\rangle \right|\left|\left\langle x_{i},\left(e^{tK}-1\right)x_{j}\right\rangle \right|\leq\sum_{i=1}^{n}\left|\left\langle v,x_{i}\right\rangle \right|\frac{\left\langle x_{i},v\right\rangle \left\langle v,x_{j}\right\rangle }{\lambda_{i}+\lambda_{j}}\nonumber \\
 & \leq\left\langle v,x_{j}\right\rangle \sum_{i=1}^{n}\frac{\left|\left\langle x_{i},v\right\rangle \right|^{2}}{\lambda_{i}}=\left\langle v,x_{j}\right\rangle \left\langle v,h^{-1}v\right\rangle ,\quad1\leq j\leq n,
\end{align}
hence
\begin{equation}
\left|\left\langle v,e^{tK}x_{j}\right\rangle \right|\leq\left\langle v,x_{j}\right\rangle +\left|\sum_{i=1}^{n}\left\langle v,x_{i}\right\rangle \left\langle x_{i},\left(e^{tK}-1\right)x_{j}\right\rangle \right|\leq\left\langle v,x_{j}\right\rangle \left(1+\left\langle v,h^{-1}v\right\rangle \right),\quad1\leq j\leq n,
\end{equation}
so returning to equation (\ref{eq:PvFirstEstimate}) we conclude that
\begin{equation}
\left\Vert e^{tK}P_{v}e^{tK}\right\Vert _{\infty}^{2}\leq\left(\max_{1\leq j\leq n}\left|\left\langle v,x_{j}\right\rangle \left(1+\left\langle v,h^{-1}v\right\rangle \right)\right|^{2}\right)\left\Vert v\right\Vert ^{2}=\alpha^{2}\left(1+\left\langle v,h^{-1}v\right\rangle \right)^{2}\left\Vert v\right\Vert ^{2}
\end{equation}
implying the claim. %Recall $\langle v, h^{-1}v\rangle$ is bounded by \eqref{eq:v-h-1-v}. 

$\hfill\square$

For $A_{h}\left(t\right)$ and $B_{h}\left(t\right)$ we estimate
the matrix elements of the operators appearing in the equations (\ref{eq:AhtDefinition})
and (\ref{eq:BhtDefinition}):
\begin{prop}
\label{prop:AhtBhtComponentBounds}It holds for all $1\leq i,j\leq n$
and $t\in\left[0,1\right]$ that, for $C_{t}=\cosh\left(-tK\right)-1$
and $S_{t}=\sinh\left(-tK\right)$,
\begin{equation}
\left|\left\langle x_{i},C_{t}hC_{t}x_{j}\right\rangle \right|,\left|\left\langle x_{i},C_{t}hS_{t}x_{j}\right\rangle \right|,\left|\left\langle x_{i},S_{t}hS_{t}x_{j}\right\rangle \right|\leq\left\langle x_{i},v\right\rangle \left\langle v,x_{j}\right\rangle \left\langle v,h^{-1}v\right\rangle 
\end{equation}
and
\begin{equation}
\left|\left\langle x_{i},\left\{ h,C_{t}\right\} x_{j}\right\rangle \right|,\left|\left\langle x_{i},\left\{ h,S_{t}\right\} x_{j}\right\rangle \right|\leq\left\langle x_{i},v\right\rangle \left\langle v,x_{j}\right\rangle .
\end{equation}
\end{prop}

\textbf{Proof:} The arguments for the elements of the two groups are
the same so we focus on particular representatives. For the first
we have by the estimates of Proposition \ref{prop:SuperGeneralElementBounds}
that
\begin{align}
 & \quad\,\left|\left\langle x_{i},\sinh\left(-tK\right)h\sinh\left(-tK\right),x_{j}\right\rangle \right|=\left|\sum_{k=1}^{n}\lambda_{k}\left\langle x_{i},\sinh\left(-tK\right),x_{k}\right\rangle \left\langle x_{k},\sinh\left(-tK\right),x_{j}\right\rangle \right|\nonumber \\
 & \leq\sum_{k=1}^{n}\lambda_{k}\frac{\left\langle x_{i},v\right\rangle \left\langle v,x_{k}\right\rangle }{\lambda_{i}+\lambda_{k}}\frac{\left\langle x_{k},v\right\rangle \left\langle v,x_{j}\right\rangle }{\lambda_{k}+\lambda_{j}}\leq\left\langle x_{i},v\right\rangle \left\langle v,x_{j}\right\rangle \sum_{k=1}^{n}\frac{\left|\left\langle x_{k},v\right\rangle \right|^{2}}{\lambda_{k}} = \left\langle x_{i},v\right\rangle \left\langle v,x_{j}\right\rangle \left\langle v,h^{-1}v\right\rangle
\end{align}
and for the second that
\begin{align}
\left|\left\langle x_{i},\left\{ h,\sinh\left(-tK\right)\right\} ,x_{j}\right\rangle \right| & =\left(\lambda_{i}+\lambda_{j}\right)\left|\left\langle x_{i},\sinh\left(-tK\right),x_{j}\right\rangle \right|\\
 & \leq\left(\lambda_{i}+\lambda_{j}\right)\frac{\left\langle x_{i},v\right\rangle \left\langle v,x_{j}\right\rangle }{\lambda_{i}+\lambda_{j}}=\left\langle x_{i},v\right\rangle \left\langle v,x_{j}\right\rangle .\nonumber 
\end{align}
$\hfill\square$

We can now obtain the desired estimate:
\begin{prop}
\label{prop:AtBtFinalEstimates}It holds for all $t\in\left[0,1\right]$
that
\[
\left\Vert A\left(t\right)\right\Vert _{\infty,2},\,\left\Vert B\left(t\right)\right\Vert _{\infty,2}\leq3\alpha\left(1+\left\langle v,h^{-1}v\right\rangle \right)\left\Vert v\right\Vert 
\]
where $\alpha=\max_{1\leq j\leq n}\left\langle v,x_{j}\right\rangle $.
\end{prop}

\textbf{Proof:} Again the arguments for $A\left(t\right)$ and $B\left(t\right)$
are the same so we focus on $A\left(t\right)$. Using that $\left\Vert \cdot\right\Vert _{\infty,2}$
is indeed a norm, hence obeys the triangle inequality, we have for
any $t\in\left[0,1\right]$ that
\begin{align}
\left\Vert A\left(t\right)\right\Vert _{\infty,2} & \leq\left\Vert e^{tK}P_{v}e^{tK}\right\Vert _{\infty,2}+\left\Vert A_{h}\left(t\right)\right\Vert _{\infty,2}\leq\left\Vert e^{tK}P_{v}e^{tK}\right\Vert _{\infty,2}+\left\Vert \sinh\left(-tK\right)h\sinh\left(-tK\right)\right\Vert _{\infty,2}\nonumber \\
 & +\left\Vert \left(\cosh\left(-tK\right)-1\right)h\left(\cosh\left(-tK\right)-1\right)\right\Vert _{\infty,2}+\left\Vert \left\{ h,\cosh\left(-tK\right)-1\right\} \right\Vert _{\infty,2}.
\end{align}
We estimate $\left\Vert \sinh\left(-tK\right)h\sinh\left(-tK\right)\right\Vert _{\infty,2}$
using Proposition \ref{prop:AhtBhtComponentBounds} as
\begin{align}
\left\Vert \sinh\left(-tK\right)h\sinh\left(-tK\right)\right\Vert _{\infty,2}^{2} & =\sum_{i=1}^{n}\max_{1\leq j\leq n}\left|\left\langle x_{i},\sinh\left(-tK\right)h\sinh\left(-tK\right)x_{j}\right\rangle \right|^{2}\nonumber \\
 & \leq\sum_{i=1}^{n}\max_{1\leq j\leq n}\left|\left\langle x_{i},v\right\rangle \left\langle v,x_{j}\right\rangle \left\langle v,h^{-1}v\right\rangle \right|^{2}\\
 & =\alpha^{2}\left\langle v,h^{-1}v\right\rangle ^{2}\sum_{i=1}^{n}\left|\left\langle x_{i},v\right\rangle \right|^{2}=\alpha^{2}\left\langle v,h^{-1}v\right\rangle ^{2}\left\Vert v\right\Vert ^{2},\nonumber 
\end{align}
the same bound holding also for $\left\Vert \left(\cosh\left(-tK\right)-1\right)h\left(\cosh\left(-tK\right)-1\right)\right\Vert _{\infty,2}^{2}$.
We likewise find
\begin{align}
\left\Vert \left\{ h,\cosh\left(-tK\right)-1\right\} \right\Vert _{\infty,2}^{2} & =\sum_{i=1}^{n}\max_{1\leq j\leq n}\left|\left\langle x_{i},\left\{ h,\cosh\left(-tK\right)-1\right\} x_{j}\right\rangle \right|^{2}\\
 & \leq\sum_{i=1}^{n}\max_{1\leq j\leq n}\left|\left\langle x_{i},v\right\rangle \left\langle v,x_{j}\right\rangle \right|^{2}=\alpha^{2}\left\Vert v\right\Vert ^{2},\nonumber 
\end{align}
so recalling the estimate of Proposition \ref{prop:Infty2NormofetKPvetK}
we conclude that
\begin{equation}
\left\Vert A\left(t\right)\right\Vert _{\infty,2}\leq\alpha\left(1+\left\langle v,h^{-1}v\right\rangle \right)\left\Vert v\right\Vert +2\alpha\left\langle v,h^{-1}v\right\rangle \left\Vert v\right\Vert +\alpha\left\Vert v\right\Vert \leq3\alpha\left(1+\left\langle v,h^{-1}v\right\rangle \right)\left\Vert v\right\Vert .
\end{equation}
$\hfill\square$

Now we come to the last ingredient of Proposition \ref{thm:ExplicitKAktBktEstimates}. 

\begin{prop}
\label{prop:new-E0MatrixElementEstimate} Let $E=e^{-K}he^{-K}$. For all $1\leq i,j\leq n$ it
holds that
\[
\left|\left\langle x_{i},\left(E-h\right)x_{j}\right\rangle \right|\leq\left(1+\left\langle v,h^{-1}v\right\rangle \right)\left\langle x_{i},v\right\rangle \left\langle v,x_{j}\right\rangle .
\]
\end{prop}

\textbf{Proof:} Using the identity 
\begin{equation}
e^{-K}he^{-K}-h=\left\{ h,e^{-K}-1\right\} +\left(e^{-K}-1\right)h\left(e^{-K}-1\right)
\end{equation}
we can write 
\begin{equation}
\left\langle x_{i},\left(e^{-K}he^{-K}-h\right)x_{j}\right\rangle =\left(\lambda_{i}+\lambda_{j}\right)\left\langle x_{i},\left(e^{-K}-1\right)x_{j}\right\rangle +\left\langle x_{i},\left(e^{-K}-1\right)h\left(e^{-K}-1\right)x_{j}\right\rangle .
\end{equation}
We can apply Proposition \ref{prop:SuperGeneralElementBounds}
to estimate the first term of this equation as
\begin{equation}
\left|\left(\lambda_{i}+\lambda_{j}\right)\left\langle x_{i},\left(e^{-K}-1\right)x_{j}\right\rangle \right|\leq\left(\lambda_{i}+\lambda_{j}\right)\frac{\left\langle x_{i},v\right\rangle \left\langle v,x_{j}\right\rangle }{\lambda_{i}+\lambda_{j}}=\left\langle x_{i},v\right\rangle \left\langle v,x_{j}\right\rangle 
\end{equation}
and the second term as
\begin{align}
\left|\left\langle x_{i},\left(e^{-K}-1\right)h\left(e^{-K}-1\right)x_{j}\right\rangle \right| & =\left|\sum_{k=1}^{n}\lambda_{k}\left\langle x_{i},\left(e^{-K}-1\right)x_{k}\right\rangle \left\langle x_{k},\left(e^{-K}-1\right)x_{j}\right\rangle \right|\nonumber \\
 & \leq\sum_{k=1}^{n}\lambda_{k}\frac{\left\langle x_{i},v\right\rangle \left\langle v,x_{k}\right\rangle }{\lambda_{i}+\lambda_{k}}\frac{\left\langle x_{k},v\right\rangle \left\langle v,x_{j}\right\rangle }{\lambda_{k}+\lambda_{j}}\\
 & \leq\left\langle x_{i},v\right\rangle \left\langle v,x_{j}\right\rangle \sum_{k=1}^{n}\frac{\left|\left\langle x_{k},v\right\rangle \right|^{2}}{\lambda_{k}}=\left\langle v,h^{-1}v\right\rangle \left\langle x_{i},v\right\rangle \left\langle v,x_{j}\right\rangle \nonumber 
\end{align}
which implies the claim.

$\hfill\square$

{\bf Proof of Proposition \ref{thm:ExplicitKAktBktEstimates-2}:} Now we insert $h_{k}$ and $v_{k}$ to conclude. 
Using Proposition \ref{prop:Infty2NormofK}, and noting that ``$\alpha$'' of our problem is simply
the constant
\begin{align} \label{eq:alpha-uniform}
\max_{p\in L_{k}}\left\langle v_{k},e_{p}\right\rangle =\sqrt{\frac{\hat{V}_{k}k_{F}^{-1}}{2\left(2\pi\right)^{3}}}
\end{align}
we find that 
\begin{equation}
\left\Vert K_{k}\right\Vert _{\infty,2} \leq\sqrt{\frac{\hat{V}_{k}k_{F}^{-1}}{2\left(2\pi\right)^{3}}}\sqrt{\left\langle v_{k},h_{k}^{-2}v_{k}\right\rangle }=\sqrt{\frac{\hat{V}_{k}k_{F}^{-1}}{2\left(2\pi\right)^{3}}}\sqrt{\frac{\hat{V}_{k}k_{F}^{-1}}{2\left(2\pi\right)^{3}}\sum_{p\in L_{k}}\frac{1}{\lambda_{k,p}^{2}}}\leq C\hat{V}_{k}k_{F}^{-1}\sqrt{\sum_{p\in L_{k}}\frac{1}{\lambda_{k,p}^{2}}}.
\end{equation}
The desired upper bound 
\begin{align}
\left\Vert K_{k}\right\Vert _{\infty,2} & \leq C\hat{V}_{k}\log\left(k_{F}\right)^{\frac{1}{3}}k_{F}^{-\frac{2}{3}}\left|k\right|^{1+\frac{5}{6}}
\end{align}
then follows from an estimate from Proposition \ref{prop:MoreSingularRiemannSums} in the Appendix: 
\[
\sum_{p\in L_{k}}\frac{1}{\lambda_{k,p}^{2}} \leq C\left|k\right|^{3+\frac{2}{3}}(\log k_{F})^{\frac{2}{3}}k_{F}^{\frac{2}{3}},\quad k_{F}\rightarrow\infty,
\]

On the other hand, by Proposition \ref{prop:AtBtFinalEstimates} and \eqref{eq:v-h-1-v} we conclude that 
\begin{align}
\left\Vert A_{k}\left(t\right)\right\Vert _{\infty,2},\,\left\Vert B_{k}\left(t\right)\right\Vert _{\infty,2} & \leq3\sqrt{\frac{\hat{V}_{k}k_{F}^{-1}}{2\left(2\pi\right)^{3}}}\left(1+\left\langle v_{k},h_{k}^{-1}v_{k}\right\rangle \right)\sqrt{\frac{\hat{V}_{k}k_{F}^{-1}}{2\left(2\pi\right)^{3}}\left|L_{k}\right|}\\
 & \leq C\hat{V}_{k}k_{F}^{-1}\sqrt{\left|L_{k}\right|}\left(1+\hat{V}_{k}k_{F}^{-1}\sum_{p\in L_{k}}\frac{1}{\lambda_{k,p}}\right) \leq C\hat{V}_{k}\left|k\right|^{\frac{1}{2}}\left(1+\hat{V}_{k}\right) \nonumber 
\end{align}
where we used 
\begin{align} \label{eq:Lk-lambda-1}
|L_k| \le C|k| k_F^2, \quad \sum_{p\in L_{k}}\frac{1}{\lambda_{k,p}} \le C k_F
\end{align}
from Proposition \ref{prop:NonSingularRiemannSums} and Proposition \ref{coro:CompletelyUniformRiemannSumBound}. Finally from Proposition \ref{prop:new-E0MatrixElementEstimate} we have 
\begin{align}
\max_{p\in L_{k}} \left|\left\langle e_p,\left(E_k-h_k\right)e_p\right\rangle \right|  & \leq \left(1+\left\langle v_k,h_k^{-1}v_k\right\rangle \right) \sup_{p\in L_k} |\left\langle e_p,v\right\rangle|^2 \\
&= \left(1+\hat{V}_{k}k_{F}^{-1}\sum_{p\in L_{k}}\frac{1}{\lambda_{k,p}}\right)   \frac{\hat{V}_{k}k_{F}^{-1}}{2\left(2\pi\right)^{3}} \le  Ck_{F}^{-1}\hat{V}_{k}\left(1+\hat{V}_{k}\right) \nonumber
\end{align}
where we  used \eqref{eq:alpha-uniform} and \eqref{eq:Lk-lambda-1} again in the last estimate. 
$\hfill\square$
%%%%%%%%%%%%%%%%%%%%%%%%%%%%%%%%%%%%
%%%%%%%%%%%%%%%%%%%%%%%%%%%%%%%%%%%%

\subsection{\label{subsec:EstimatesofAkandBkfortheLowerBound}Kinetic Estimates}

Now we prove Proposition \ref{thm:ExplicitKAktBktEstimates-3}. Again let us start with the notation \eqref{eq:CleanKDefinition}. We have

\begin{prop} Under the notation \eqref{eq:CleanKDefinition}, it holds that
\begin{align*}
\left\Vert h^{-\frac{1}{2}}K\right\Vert _{\HS} & \leq\left\langle v,h^{-\frac{3}{2}}v\right\rangle ,\\
\left\Vert \left\{ K,h\right\} h^{-\frac{1}{2}}\right\Vert _{\HS} & \leq2\left\Vert v\right\Vert \sqrt{\left\langle v,h^{-1}v\right\rangle },\\
\left\Vert h^{-\frac{1}{2}}\left\{ K,h\right\} h^{-\frac{1}{2}}\right\Vert _{\HS} & \leq2\left\langle v,h^{-1}v\right\rangle .
\end{align*}
\end{prop}

\textbf{Proof:} Using Proposition \ref{prop:SuperGeneralElementBounds}
we estimate
\begin{equation}
\left\Vert h^{-\frac{1}{2}}K\right\Vert _{\text{HS}}^{2}=\sum_{i,j=1}^{n}\frac{1}{\lambda_{i}}\left|\left\langle x_{i},Kx_{j}\right\rangle \right|^{2}\leq\sum_{i,j=1}^{n}\frac{1}{\lambda_{i}}\left|\frac{\left\langle x_{i},v\right\rangle \left\langle v,x_{j}\right\rangle }{\lambda_{i}+\lambda_{j}}\right|^{2}\leq\left(\sum_{i=1}^{n}\frac{\left|\left\langle x_{i},v\right\rangle \right|^{2}}{\lambda_{i}^{\frac{3}{2}}}\right)^{2}=\left\langle v,h^{-\frac{3}{2}}v\right\rangle ^{2},
\end{equation}
and for $\left\Vert \left\{ K,h\right\} h^{-\frac{1}{2}}\right\Vert _{\text{HS}}$
use that $\left\Vert \left\{ K,h\right\} h^{-\frac{1}{2}}\right\Vert _{\text{HS}}\leq\left\Vert Kh^{\frac{1}{2}}\right\Vert _{\text{HS}}+\left\Vert hKh^{-\frac{1}{2}}\right\Vert _{\text{HS}}$
to estimate
\begin{align}
\left\Vert Kh^{\frac{1}{2}}\right\Vert _{\text{HS}}^{2} & =\sum_{i,j=1}^{n}\lambda_{j}\left|\left\langle x_{i},Kx_{j}\right\rangle \right|^{2}\leq\sum_{i,j=1}^{n}\left|\left\langle x_{i},v\right\rangle \right|^{2}\frac{\left|\left\langle x_{j},v\right\rangle \right|^{2}}{\lambda_{j}}=\left\Vert v\right\Vert ^{2}\left\langle v,h^{-1}v\right\rangle \\
\left\Vert hKh^{-\frac{1}{2}}\right\Vert _{\text{HS}}^{2} & =\sum_{i,j=1}^{n}\frac{\lambda_{i}^{2}}{\lambda_{j}}\left|\left\langle x_{i},Kx_{j}\right\rangle \right|^{2}\leq\sum_{i,j=1}^{n}\left|\left\langle x_{i},v\right\rangle \right|^{2}\frac{\left|\left\langle x_{j},v\right\rangle \right|^{2}}{\lambda_{j}}=\left\Vert v\right\Vert ^{2}\left\langle v,h^{-1}v\right\rangle \nonumber 
\end{align}
for the claimed $\left\Vert \left\{ K,h\right\} h^{-\frac{1}{2}}\right\Vert _{\text{HS}}\leq2\left\Vert v\right\Vert \sqrt{\left\langle v,h^{-1}v\right\rangle }$.
We likewise have that
\begin{equation}
\left\Vert h^{-\frac{1}{2}}\left\{ K,h\right\} h^{-\frac{1}{2}}\right\Vert _{\text{HS}}\leq\left\Vert h^{-\frac{1}{2}}Kh^{\frac{1}{2}}\right\Vert _{\text{HS}}+\left\Vert h^{\frac{1}{2}}Kh^{-\frac{1}{2}}\right\Vert _{\text{HS}}=2\left\Vert h^{-\frac{1}{2}}Kh^{\frac{1}{2}}\right\Vert _{\text{HS}}
\end{equation}
so the bound
\begin{equation}
\left\Vert h^{-\frac{1}{2}}Kh^{\frac{1}{2}}\right\Vert _{\text{HS}}^{2}=\sum_{i,j=1}^{n}\frac{\lambda_{i}}{\lambda_{j}}\left|\left\langle x_{i},Kx_{j}\right\rangle \right|^{2}\leq\left(\sum_{i=1}^{n}\frac{\left|\left\langle x_{i},v\right\rangle \right|^{2}}{\lambda_{i}}\right)^{2}=\left\langle v,h^{-1}v\right\rangle ^{2}
\end{equation}
implies the final claim.

$\hfill\square$

For $A\left(t\right)$ and $B\left(t\right)$ we recall the decompositions  \eqref{eq:A49}-\eqref{eq:BhtDefinition}. 
%\begin{align}
%delete\\
%delete
%\end{align}
%\begin{equation}
%A\left(t\right)=A_{h}\left(t\right)+e^{tK}P_{v}e^{tK},\quad B\left(t\right)=B_{h}\left(t\right)+e^{tK}P_{v}e^{tK},
%\end{equation}
%where
%\begin{align}
%A_{h}\left(t\right) & =\sinh\left(-tK\right)h\sinh\left(-tK\right)+\left(\cosh\left(-tK\right)-1\right)h\left(\cosh\left(-tK\right)-1\right)+\left\{ h,\cosh\left(-tK\right)-1\right\} \nonumber \\
%B_{h}\left(t\right) & =\sinh\left(-tK\right)h\left(\cosh\left(-tK\right)-1\right)+\left(\cosh\left(-tK\right)-1\right)h\sinh\left(-tK\right)-\left\{ h,\sinh\left(-tK\right)\right\} .
%\end{align}
Recall also that $(x_i)_i$ is an eigenbasis of $h$ and $\langle x_i,v\rangle\ge 0$ for all $1\le i\le n$. We first estimate the $e^{tK}P_{v}e^{tK}$ term:
\begin{prop}
\label{prop:etKPvetKfortheLowerBound}For all $t\in\left[0,1\right]$
it holds that
\[
\max_{1\leq j\leq n}\left\Vert h^{-\frac{1}{2}}e^{tK}P_{v}e^{tK}x_{j}\right\Vert \leq\alpha\left(1+\left\langle v,h^{-1}v\right\rangle \right)^{2}\sqrt{\left\langle v,h^{-1}v\right\rangle }
\]
where $\alpha=\max_{1\leq j\leq n}\left\langle v,x_{j}\right\rangle $.
\end{prop}

\textbf{Proof:} We write $e^{tK}P_{v}e^{tK}$ as
\begin{equation}
e^{tK}P_{v}e^{tK}=P_{v}+\left(e^{tK}-1\right)P_{v}+P_{v}\left(e^{tK}-1\right)+\left(e^{tK}-1\right)P_{v}\left(e^{tK}-1\right)
\end{equation}
and estimate each term separately. By the definition of $P_{v}$ the
first term is simply
\begin{equation}
\left\Vert h^{-\frac{1}{2}}P_{v}x_{j}\right\Vert =\left|\left\langle v,x_{j}\right\rangle \right|\left\Vert h^{-\frac{1}{2}}v\right\Vert \leq\alpha\sqrt{\left\langle v,h^{-1}v\right\rangle }.
\end{equation}
For the remaining terms we use Proposition \ref{prop:SuperGeneralElementBounds}
to estimate that
\begin{align}
\left\Vert h^{-\frac{1}{2}}\left(e^{tK}-1\right)P_{v}x_{j}\right\Vert ^{2} & =\sum_{i=1}^{n}\frac{1}{\lambda_{i}}\left|\sum_{k=1}^{n}\left\langle x_{i},\left(e^{tK}-1\right)x_{k}\right\rangle \left\langle x_{k},P_{v}x_{j}\right\rangle \right|^{2}\nonumber \\
 & \leq\sum_{i=1}^{n}\frac{1}{\lambda_{i}}\left|\sum_{k=1}^{n}\frac{\left\langle x_{i},v\right\rangle \left\langle v,x_{k}\right\rangle }{\lambda_{i}+\lambda_{k}}\left\langle x_{k},v\right\rangle \left\langle v,x_{j}\right\rangle \right|^{2}\\
 & \leq\left|\left\langle v,x_{j}\right\rangle \right|^{2}\left(\sum_{i=1}^{n}\frac{\left|\left\langle x_{i},v\right\rangle \right|^{2}}{\lambda_{i}}\right)^{3}\leq\alpha^{2}\left\langle v,h^{-1}v\right\rangle ^{3},
 \end{align}
 and
 \begin{align}
\left\Vert h^{-\frac{1}{2}}P_{v}\left(e^{tK}-1\right)x_{j}\right\Vert ^{2} & =\sum_{i=1}^{n}\frac{1}{\lambda_{i}}\left|\sum_{k=1}^{n}\left\langle x_{i},P_{v}x_{k}\right\rangle \left\langle x_{k},\left(e^{tK}-1\right)x_{j}\right\rangle \right|^{2}\nonumber \\
 & \leq\sum_{i=1}^{n}\frac{1}{\lambda_{i}}\left|\sum_{k=1}^{n}\left\langle x_{i},v\right\rangle \left\langle v,x_{k}\right\rangle \frac{\left\langle x_{k},v\right\rangle \left\langle v,x_{j}\right\rangle }{\lambda_{k}+\lambda_{j}}\right|^{2}\\
 & \leq\left|\left\langle v,x_{j}\right\rangle \right|^{2}\left(\sum_{i=1}^{n}\frac{\left|\left\langle x_{i},v\right\rangle \right|^{2}}{\lambda_{i}}\right)^{3}\leq\alpha^{2}\left\langle v,h^{-1}v\right\rangle ^{3}\nonumber 
\end{align}
and
\begin{align}
 & \,\,\left\Vert h^{-\frac{1}{2}}\left(e^{tK}-1\right)P_{v}\left(e^{tK}-1\right)x_{j}\right\Vert ^{2}\nonumber\\
 &=\sum_{i=1}^{n}\frac{1}{\lambda_{i}}\left|\sum_{k,l=1}^{n}\left\langle x_{i},\left(e^{tK}-1\right)x_{k}\right\rangle \left\langle x_{k},P_{v}x_{l}\right\rangle \left\langle x_{l},\left(e^{tK}-1\right)x_{j}\right\rangle \right|^{2}\nonumber \\
 & \leq\sum_{i=1}^{n}\frac{1}{\lambda_{i}}\left|\sum_{k,l=1}^{n}\frac{\left\langle x_{i},v\right\rangle \left\langle v,x_{k}\right\rangle }{\lambda_{i}+\lambda_{k}}\left\langle x_{k},v\right\rangle \left\langle v,x_{l}\right\rangle \frac{\left\langle x_{l},v\right\rangle \left\langle v,x_{j}\right\rangle }{\lambda_{l}+\lambda_{j}}\right|^{2}\\
 & \leq\left|\left\langle v,x_{j}\right\rangle \right|^{2}\sum_{i=1}^{n}\frac{\left|\left\langle x_{i},v\right\rangle \right|^{2}}{\lambda_{i}}\left(\sum_{k,l=1}^{n}\frac{\left|\left\langle x_{k},v\right\rangle \right|^{2}}{\lambda_{k}}\frac{\left|\left\langle x_{l},v\right\rangle \right|^{2}}{\lambda_{l}}\right)^{2}\leq\alpha^{2}\left\langle v,h^{-1}v\right\rangle ^{5}\nonumber 
\end{align}
which imply the claim.

$\hfill\square$

Finally, the full estimates on $A\left(t\right)$ and $B\left(t\right)$ are
now easily obtained:
\begin{prop}
It holds for all $t\in\left[0,1\right]$ that
\[
\max_{1\leq j\leq n}\left\Vert h^{-\frac{1}{2}}A\left(t\right)x_{j}\right\Vert ,\,\max_{1\leq j\leq n}\left\Vert h^{-\frac{1}{2}}B\left(t\right)x_{j}\right\Vert \leq2\alpha\left(1+\left\langle v,h^{-1}v\right\rangle \right)^{2}\sqrt{\left\langle v,h^{-1}v\right\rangle }
\]
where $\alpha=\max_{1\leq j\leq n}\left\langle v,x_{j}\right\rangle $.
\end{prop}

\textbf{Proof:} The estimates for $A\left(t\right)$ and $B\left(t\right)$
are similar so we focus on $A\left(t\right)$. We have
\begin{align}
&\left\Vert h^{-\frac{1}{2}}A\left(t\right)x_{j}\right\Vert  \nonumber\\
& \leq\left\Vert h^{-\frac{1}{2}}\sinh\left(-tK\right)h\sinh\left(-tK\right)x_{j}\right\Vert +\left\Vert h^{-\frac{1}{2}}\left(\cosh\left(-tK\right)-1\right)h\left(\cosh\left(-tK\right)-1\right)x_{j}\right\Vert \\
 & +\left\Vert h^{-\frac{1}{2}}\left\{ h,\cosh\left(-tK\right)-1\right\} x_{j}\right\Vert +\left\Vert h^{-\frac{1}{2}}e^{tK}P_{v}e^{tK}x_{j}\right\Vert \nonumber
\end{align}
and by Proposition \ref{prop:AhtBhtComponentBounds} we can estimate that
\begin{align}
\left\Vert h^{-\frac{1}{2}}\sinh\left(-tK\right)h\sinh\left(-tK\right)x_{j}\right\Vert ^{2} & =\sum_{i=1}^{n}\frac{1}{\lambda_{i}}\left|\left\langle x_{i},\sinh\left(-tK\right)h\sinh\left(-tK\right)x_{j}\right\rangle \right|^{2}\\
 & \leq\left|\left\langle v,x_{j}\right\rangle \right|^{2}\left\langle v,h^{-1}v\right\rangle ^{2}\sum_{i=1}^{n}\frac{\left|\left\langle x_{i},v\right\rangle \right|^{2}}{\lambda_{i}}\leq\alpha^{2}\left\langle v,h^{-1}v\right\rangle ^{3},\nonumber 
\end{align}
the same estimate holding also for $\left\Vert h^{-\frac{1}{2}}\left(\cosh\left(-tK\right)-1\right)h\left(\cosh\left(-tK\right)-1\right)x_{j}\right\Vert $,
and
\begin{align}
\left\Vert h^{-\frac{1}{2}}\left\{ h,\cosh\left(-tK\right)-1\right\} x_{j}\right\Vert ^{2} & =\sum_{i=1}^{n}\frac{1}{\lambda_{i}}\left|\left\langle x_{i},\left\{ h,\cosh\left(-tK\right)-1\right\} x_{j}\right\rangle \right|^{2}\\
 & \leq\left|\left\langle v,x_{j}\right\rangle \right|^{2}\sum_{i=1}^{n}\frac{\left|\left\langle x_{i},v\right\rangle \right|^{2}}{\lambda_{i}}\leq\alpha^{2}\left\langle v,h^{-1}v\right\rangle .\nonumber 
\end{align}
Inserting also the estimate of Proposition \ref{prop:etKPvetKfortheLowerBound}
we thus obtain
\begin{align}
\max_{1\leq j\leq n}\left\Vert h^{-\frac{1}{2}}A\left(t\right)x_{j}\right\Vert  & \leq2\alpha\left\langle v,h^{-1}v\right\rangle ^{\frac{3}{2}}+\alpha\sqrt{\left\langle v,h^{-1}v\right\rangle }+\alpha\left(1+\left\langle v,h^{-1}v\right\rangle \right)^{2}\sqrt{\left\langle v,h^{-1}v\right\rangle }\\
 & \leq2\alpha\left(1+\left\langle v,h^{-1}v\right\rangle \right)^{2}\sqrt{\left\langle v,h^{-1}v\right\rangle }.\nonumber
\end{align}
$\hfill\square$

{\bf Proof of Proposition \ref{thm:ExplicitKAktBktEstimates-3}:} The desired bounds follow from applying the general estimates of this section to $h_k$ and $v_k$, plus using the uniform bound on $\alpha$ in \eqref{eq:alpha-uniform} and the estimates
\begin{equation}
\left\Vert v\right\Vert ^{2}\leq C\hat{V}_{k}\left|k\right|k_{F},\quad\left\langle v_{k},h_{k}^{-1}v_{k}\right\rangle \leq C\hat{V}_{k},\quad\left\langle v_{k},h_{k}^{-\frac{3}{2}}v_{k}\right\rangle \leq C\hat{V}_{k}\left|k\right|^{3+\frac{2}{3}}(\log k_{F})^{\frac{2}{3}}k_{F}^{-\frac{1}{3}},
\end{equation}
which hold for all $k\in\overline{B}\left(0,2k_{F}\right)$ due to Propositions \ref{prop:NonSingularRiemannSums}, \ref{coro:CompletelyUniformRiemannSumBound}, and \ref{prop:MoreSingularRiemannSums}.
$\hfill\square$

\section{Gronwall Estimates for the Bogolubov Transformation} \label{sec:Gronwall Estimates for the Bogolubov Transformation}

In the previous sections,  we have bounded several error terms using the operators $H_{\rm kin}'$ and $\mathcal{N}_E$. In this section, we control the propagation of these operators under the Bogolubov transformation $e^{-\mathcal{K}}$ defined in Section \ref{sec:quasi-bosonic-Bogolubov}. We  have the following  Gronwall-type estimates.

\begin{prop}
\label{prop:KineticGronwall}Let $\sum_{k\in\mathbb{Z}^{3}}\hat{V}_{k}\left|k\right|<\infty$.
Then for all $\Psi\in D\left(H_{\kin}^{\prime}\right)$ and $\left|t\right|\leq1$
it holds that
\begin{align*}
\left\langle e^{-t\mathcal{K}}\Psi, (H_{\kin}^{\prime} + k_F) e^{-t\mathcal{K}}\Psi\right\rangle &\leq C \left\langle \Psi, ( H_{\kin}^{\prime} + k_F) \Psi\right\rangle,\\
\left\langle e^{-t\mathcal{K}}\Psi, (k_F^{-1} \mathcal{N}_E H_{\kin}^{\prime} + H_{\kin}^{\prime} + k_F) e^{-t\mathcal{K}}\Psi\right\rangle &\leq C\left\langle \Psi,(k_F^{-1} \mathcal{N}_E H_{\kin}^{\prime} + H_{\kin}^{\prime} + k_F)  \Psi\right\rangle 
\end{align*}
for a constant $C>0$ independent of $k_{F}$.
\end{prop}

%\begin{prop}
%\label{prop:KineticGronwall}Let $\sum_{k\in\mathbb{Z}^{3}}\hat{V}_{k}\left|k\right|<\infty$.
%Then for all $\Psi\in D\left(H_{\kin}^{\prime}\right)$ and $\left|t\right|\leq1$
%it holds that
%\[
%\left\langle e^{t\mathcal{K}}\Psi,\mathcal{N}_{E}H_{\kin}^{\prime}e^{t\mathcal{K}}\Psi\right\rangle \leq C\left(\left\langle \Psi,\mathcal{N}_{E}H_{\kin}^{\prime}\Psi\right\rangle +k_{F}\left\langle \Psi,\left(1+H_{\kin}^{\prime}\right)\Psi\right\rangle \right)
%\]
%for a constant $C>0$ independent of $k_{F}$.
%\end{prop}

As a preparation, let us first prove  
\begin{lem} \label{lem:double-commutator} Let $X,Y,Z$ be self-adjoint operators on a Hilbert space such that 
$$X,Z >0, \quad [X,Z]=0,  \quad \pm \left[\left[Y,X\right],X\right] \le Z.$$
Then 
$$
\pm \left[\left[Y,\sqrt{X}\right],\sqrt{X}\right] \le \frac{Z}{4X} .
$$
\end{lem}

{\bf Proof of Lemma \ref{lem:double-commutator}:} Using \eqref{eq:integral-sqrtA}  we can write 
\begin{equation} 
\left[Y,\sqrt{X}\right]=\frac{2}{\pi}\int_{0}^{\infty} \left[Y,  \frac{X}{X+t^{2}}\right]\,dt= \frac{2}{\pi}\int_{0}^{\infty} \left[Y,  \frac{-t^2}{X+t^{2}}\right]\,dt = \frac{2}{\pi}\int_{0}^{\infty} \frac{1}{X+t^{2}} \left[Y, X\right] \frac{1}{X+t^{2}}\, t^2 dt, 
\end{equation}
and applying this identity twice we get 
\begin{align} 
\left[\left[Y,\sqrt{X}\right],\sqrt{X}\right] %&= \frac{2}{\pi}\int_{0}^{\infty} \frac{1}{X+t^{2}} \left[\left[Y, X\right],\sqrt{X} \right] \frac{1}{X+t^{2}}\, t^2 dt \\
= \left(\frac{2}{\pi} \right)^2 \int_{0}^{\infty} \int_0^\infty \frac{1}{X+t^{2}} \frac{1}{X+s^{2}} \left[\left[Y, X\right],X \right] \frac{1}{X+s^{2}} \frac{1}{X+t^{2}}\, s^2 t^2 ds dt.  
\end{align}
Therefore, the assumptions $\pm \left[\left[Y,X\right],X\right] \le Z$ and $[X,Z]=0$ imply that 
\begin{align} 
\pm \left[\left[Y,\sqrt{X}\right],\sqrt{X}\right] \le  \left(\frac{2}{\pi} \right)^2 \int_{0}^{\infty} \int_0^\infty \frac{1}{X+t^{2}} \frac{1}{X+s^{2}} Z \frac{1}{X+s^{2}} \frac{1}{X+t^{2}}\, s^2 t^2 ds dt = \frac{Z}{4X}. 
\end{align}
$\hfill\square$

Now we give the 

\textbf{Proof of Proposition \ref{prop:KineticGronwall}} Write $\Psi_{t}=e^{t\mathcal{K}} \Psi $ for brevity. Recalling Proposition \ref{prop:HKincalKCommutator},
we see that
\begin{equation} \label{eq:apply-prop:HKincalKCommutator}
\frac{d}{dt}\left\langle \Psi_{t},(H_{\kin}^{\prime}+k_F)\Psi_{t}\right\rangle =\left\langle \Psi_{t},\left[\mathcal{K},H_{\kin}^{\prime}\right]\Psi_{t}\right\rangle =\sum_{k\in S_{C}}\left\langle \Psi_{t},Q_{2}^{k}\left(\left\{ K_{k}^{\oplus},h_{k}^{\oplus}\right\} \right)\Psi_{t}\right\rangle .
\end{equation}
The right-hand side can be bounded by using Propositions \ref{prop:KineticQ2BEstimate}
and \ref{thm:ExplicitKAktBktEstimates-3} 
as
\begin{align}
 & \sum_{k\in S_{C}}\left|\left\langle \Psi_{t},Q_{2}^{k}\left(\left\{ K_{k}^{\oplus},h_{k}^{\oplus}\right\} \right)\Psi_{t}\right\rangle \right| 
  \leq2\sum_{k\in S_{C}}\left\Vert \left(h_{k}^{\oplus}\right)^{-\frac{1}{2}}\left\{ K_{k}^{\oplus},h_{k}^{\oplus}\right\} \left(h_{k}^{\oplus}\right)^{-\frac{1}{2}}\right\Vert _{\text{HS}}\left\langle \Psi_{t},H_{\kin}^{\prime}\Psi_{t}\right\rangle \nonumber \\
 & \qquad\qquad\qquad\qquad\qquad\qquad\qquad+2\sum_{k\in S_{C}}\left\Vert \left\{ K_{k}^{\oplus},h_{k}^{\oplus}\right\} \left(h_{k}^{\oplus}\right)^{-\frac{1}{2}}\right\Vert _{\text{HS}}\sqrt{\left\langle \Psi_{t},H_{\kin}^{\prime}\Psi_{t}\right\rangle }\left\Vert \Psi_{t}\right\Vert \nonumber \\
 & \leq C\left(\sum_{k\in S_{C}}\hat{V}_{k}\right)\left\langle \Psi_{t},H_{\kin}^{\prime}\Psi_{t}\right\rangle +Ck_{F}^{\frac{1}{2}}\left(\sum_{k\in S_{C}}\hat{V}_{k}\left|k\right|^{\frac{1}{2}}\right)\sqrt{\left\langle \Psi_{t},H_{\kin}^{\prime}\Psi_{t}\right\rangle }\left\Vert \Psi_t\right\Vert \nonumber \\
 & \leq C\left\langle \Psi_{t}, (H_{\kin}^{\prime} + k_F) \Psi_{t}\right\rangle \label{eq:FirstKineticGronwall}
\end{align}
%for some $C>0$ depending only on $V$, 
where we also used the Cauchy--Schwarz inequality in the last step. Thus the first estimate of Proposition \ref{prop:KineticGronwall} follows by Gronwall's lemma. 
%\begin{equation}
%\left|\frac{d}{dt}\left\langle \Psi_{t},(H_{\kin}^{\prime} + k_F) \Psi_{t}\right\rangle \right|\leq C\left\langle \Psi_{t}, (H_{\kin}^{\prime} +k_F) \Psi_{t}\right\rangle, \end{equation}
%so by Gronwall's lemma
%\begin{equation}
%\left\langle \Psi_{t},(H_{\kin}^{\prime}+k_F) \Psi_{t}\right\rangle \leq C \left\langle \Psi,(H_{\kin}^{\prime}+k_F) \Psi\right\rangle, \quad \forall |t|\le 1.
%\end{equation}
%\begin{align}
%delete\\
%delete
%\end{align}
For the second bound of Proposition \ref{prop:KineticGronwall}, let us denote 
\begin{align}
 X_1 = \mathcal{N}_E +k_F \ge k_F, \quad X_2= k_F+ H_{\kin}^{\prime}  \ge k_F, \quad Y_1= [\mathcal{K}, X_2]  \quad \text{ and }\quad Y_2= [\mathcal{K}, X_1] .
 \end{align}
 Note that $Y_1,Y_2$ are symmetric since $X_1,X_2$ are symmetric and $\mathcal{K}$ is skew-symmetric. Moreover, since $[X_1,X_2]=0$, $[\mathcal{K}, X_1X_2]$ is also symmetric and we can write 
 \begin{align} \label{eq:comm-K-X1X2}
 2[\mathcal{K}, X_1 X_2] &= 2\Big( X_1 [\mathcal{K},  X_2] + [\mathcal{K},  X_1] X_2\Big) =   2( X_1 Y_1+ Y_2X_2)    \nonumber\\
 &= \sum_{i=1}^2 ( X_i Y_i+ Y_iX_i) =  \sum_{i=1}^2   \left(  2 \sqrt{X_i} Y_i \sqrt{X_i}  + [[ Y_i, \sqrt{X_i}], \sqrt{X_i}] \right) . 
 \end{align}
 For $i=1$, arguing similarly to \eqref{eq:apply-prop:HKincalKCommutator} and \eqref{eq:FirstKineticGronwall} we have
 \begin{align} \label{eq:comm-K-X1X2-a}
 \pm Y_1 = \pm [\mathcal{K}, H_{\kin}']= \pm  \sum_{k\in S_{C}} Q_{2}^{k}\left(\left\{ K_{k}^{\oplus},h_{k}^{\oplus}\right\} \right) \le C X_2, \quad  \pm \sqrt{X_1} Y_1 \sqrt{X_1} \le C X_1 X_2. 
 \end{align}
 Here we used $[X_1,X_2]=0$ in the last estimate. To apply Lemma \ref{lem:double-commutator}, let us compute $[[Y_1, X_1], X_1]$.  Note that for every symmetric operator $B$ on $\ell^2(L_k^{\pm})$, we deduce from \eqref{eq:intro-comm-NE-b}  that 
%\begin{align}
%\left[ Q_{2}^{k}\left(B\right) ,\mathcal{N}_E\right] & = \sum_{p,q\in L_{k}^{\pm}}  \left\langle e_{p},Be_{q}\right\rangle \left[  \left(b_{\overline{k,p}}^{\ast}b_{\overline{k,q}}^{\ast}+b_{\overline{k,q}}b_{\overline{k,p}}\right),\mathcal{N}_E\right] \nonumber\\
%&= 2 \sum_{p,q\in L_{k}^{\pm}}   \left\langle e_{p},Be_{q}\right\rangle \left(- b_{\overline{k,p}}^{\ast}b_{\overline{k,q}}^{\ast}+b_{\overline{k,q}}b_{\overline{k,p}}\right) \nonumber\\
% \label{eq:comm-Q2-NE-NE}
%\left[ \left[ Q_{2}^{k}\left(B\right) ,\mathcal{N}_E\right],\mathcal{N}_E\right]  & = 2 \sum_{p,q\in L_{k}^{\pm}}  \left\langle e_{p},Be_{q}\right\rangle   \left[\left(- b_{\overline{k,p}}^{\ast}b_{\overline{k,q}}^{\ast}+b_{\overline{k,q}}b_{\overline{k,p}}\right),\mathcal{N}_E\right] \\
%&= 4 \sum_{p,q\in L_{k}^{\pm}}   \left\langle e_{p},Be_{q}\right\rangle \left( b_{\overline{k,p}}^{\ast}b_{\overline{k,q}}^{\ast}+b_{\overline{k,q}}b_{\overline{k,p}}\right) = 4 Q_{2}^{k}(B).\nonumber
%\end{align}
\begin{align}
\left[ Q_{2}^{k}\left(B\right) ,\mathcal{N}_E\right] & = 2 \sum_{p,q\in L_{k}^{\pm}}   \left\langle e_{p},Be_{q}\right\rangle \left(- b_{\overline{k,p}}^{\ast}b_{\overline{k,q}}^{\ast}+b_{\overline{k,q}}b_{\overline{k,p}}\right), \label{eq:comm-Q2-NE-NE} \\
\left[ \left[ Q_{2}^{k}\left(B\right) ,\mathcal{N}_E\right],\mathcal{N}_E\right]  &= 4 \sum_{p,q\in L_{k}^{\pm}}   \left\langle e_{p},Be_{q}\right\rangle \left( b_{\overline{k,p}}^{\ast}b_{\overline{k,q}}^{\ast}+b_{\overline{k,q}}b_{\overline{k,p}}\right) = 4 Q_{2}^{k}(B).  \nonumber
\end{align}
Using \eqref{eq:comm-Q2-NE-NE} and \eqref{eq:comm-K-X1X2-a} we  have
% \begin{align}
%  \pm [[Y_1, X_1], X_1] = \pm \sum_{k\in S_{C}} \left[ \left[ Q_{2}^{k}\left(\left\{ K_{k}^{\oplus},h_{k}^{\oplus}\right\} \right), \mathcal{N}_E\right], \mathcal{N}_E\right] =  \pm 4 \sum_{k\in S_{C}} Q_{2}^{k}\left(\left\{ K_{k}^{\oplus},h_{k}^{\oplus}\right\} \right) \le C X_2  
% \end{align}
  \begin{align}
  \pm [[Y_1, X_1], X_1] =  \pm 4 \sum_{k\in S_{C}} Q_{2}^{k}\left(\left\{ K_{k}^{\oplus},h_{k}^{\oplus}\right\} \right) \le C X_2  
 \end{align}
  which implies by Lemma \ref{lem:double-commutator} that 
   \begin{align}\label{eq:comm-K-X1X2-b}
  \pm [[Y_1, \sqrt{X_1}], \sqrt{X_1}] \le C X_2  X_1^{-1}. 
  \end{align}
 Next, we consider the terms of $i=2$ in \eqref{eq:comm-K-X1X2}. Let us compute the commutator $Y_2=\left[\mathcal{K},\mathcal{N}_{E}\right]$. By linearity we deduce from \eqref{eq:intro-comm-NE-b} that $\left[b_{k}\left(\varphi\right),\mathcal{N}_{E}\right]=b_{k}\left(\varphi\right)$ for any $\varphi\in\ell^{2}\left(L_{k}^{\pm}\right)$, and hence from the definition of $\mathcal{K}$ in \eqref{eq:cK}, 
%\begin{equation}
%\left[b_{k}\left(\varphi\right),\mathcal{N}_{E}\right]=\sum_{p\in L_{k}^{\pm}}\left\langle \varphi,e_{p}\right\rangle \left[b_{\overline{k,p}},\mathcal{N}_{E}\right]=\sum_{p\in L_{k}^{\pm}}\left\langle \varphi,e_{p}\right\rangle b_{\overline{k,p}}=b_{k}\left(\varphi\right),\label{eq:NumberExcitationCommutator}
%\end{equation}
%and so $\left[b_{k}^{\ast}\left(\varphi\right),\mathcal{N}_{E}\right]=-b_{k}^{\ast}\left(\varphi\right)$ by taking the adjoint. This in turn yields 
\begin{align} \label{eq:calKNECommutator}
Y_2=\left[\mathcal{K},\mathcal{N}_{E}\right] %& =\frac{1}{2}\sum_{k\in S_{C}}\sum_{p\in L_{k}^{\pm}}\left(\left[b_{k}\left(K_{k}^{\oplus}e_{p}\right)b_{k}\left(e_{p}\right),\mathcal{N}_{E}\right]-\left[b_{k}^{\ast}\left(e_{p}\right)b_{k}^{\ast}\left(K_{k}^{\oplus}e_{p}\right),\mathcal{N}_{E}\right]\right)  \\
% & =\frac{1}{2}\sum_{k\in S_{C}}\sum_{p\in L_{k}^{\pm}}\left(b_{k}^{\ast}\left(e_{p}\right)\left[\mathcal{N}_{E},b_{k}^{\ast}\left(K_{k}^{\oplus}e_{p}\right)\right]+\left[\mathcal{N}_{E},b_{k}^{\ast}\left(e_{p}\right)\right]b_{k}^{\ast}\left(K_{k}^{\oplus}e_{p}\right)\right)\nonumber \\
% & \quad -\frac{1}{2}\sum_{k\in S_{C}}\sum_{p\in L_{k}^{\pm}}\left(b_{k}\left(K_{k}^{\oplus}e_{p}\right)\left[\mathcal{N}_{E},b_{k}\left(e_{p}\right)\right]+\left[\mathcal{N}_{E},b_{k}\left(K_{k}^{\oplus}e_{p}\right)\right]b_{k}\left(e_{p}\right)\right)\label{eq:calKNECommutator}\\
% & =\frac{1}{2}\sum_{k\in S_{C}}\sum_{p\in L_{k}^{\pm}} \Big(b_{k}\left(K_{k}^{\oplus}e_{p}\right)b_{k}\left(e_{p}\right) +b_{k}\left(K_{k}^{\oplus}e_{p}\right)b_{k}\left(e_{p}\right) \nonumber\\
% &\qquad \qquad \qquad \qquad + b_{k}^{\ast}\left(e_{p}\right)b_{k}^{\ast}\left(K_{k}^{\oplus}e_{p}\right)+b_{k}^{\ast}\left(e_{p}\right)b_{k}^{\ast}\left(K_{k}^{\oplus}e_{p}\right) \Big) \nonumber \\
  =\sum_{k\in S_{C}}\sum_{p\in L_{k}^{\pm}}\left(b_{k}^{\ast}\left(e_{p}\right)b_{k}^{\ast}\left(K_{k}^{\oplus}e_{p}\right)+b_{k}\left(K_{k}^{\oplus}e_{p}\right)b_{k}\left(e_{p}\right)\right)=\sum_{k\in S_{C}}Q_{2}^{k}\left(K_{k}^{\oplus}\right). 
\end{align}
Note that 
\begin{equation}
K_{k}^{\oplus}= \left(\begin{array}{cc}
0 & K_{k}\\
K_{k} & 0
\end{array}\right)=\left(\begin{array}{cc}
\frac{1}{\sqrt{2}} & \frac{1}{\sqrt{2}}\\
\frac{1}{\sqrt{2}} & -\frac{1}{\sqrt{2}}
\end{array}\right)\left(\begin{array}{cc}
K_{k} & 0\\
0 & -K_{k}
\end{array}\right)\left(\begin{array}{cc}
\frac{1}{\sqrt{2}} & \frac{1}{\sqrt{2}}\\
\frac{1}{\sqrt{2}} & -\frac{1}{\sqrt{2}}
\end{array}\right),
\end{equation}
and hence by Proposition \ref{thm:ExplicitKAktBktEstimates} we obtain  
\begin{equation} \label{eq:GronwallSummationEstimate}
\sum_{k\in S_{C}}\left\Vert K_{k}^{\oplus}\right\Vert _{\text{HS}}\leq\sum_{k\in S_{C}} {\rm tr} ( |K_{k}^{\oplus}| )  =2\sum_{k\in S_{C}}\text{tr}\left(\left|K_{k}\right|\right) \le C \sum_{k\in \mathbb{Z}^3} \hat V_k.
\end{equation} 

Therefore, by Proposition \ref{prop:Q2kBEstimate} 
\begin{equation} \label{eq:comm-K-X1X2-c}
\pm Y_2 \leq2\left(\sum_{k\in S_{C}}\left\Vert K_{k}^{\oplus}\right\Vert _{\text{HS}}\right) \left(1+\mathcal{N}_{E}\right) \le C X_1, \quad   \pm \sqrt{X_2} Y_2 \sqrt{X_2} \le C X_1 X_2. 
\end{equation}

Finally, consider 
  \begin{align} \label{eq:comm-Y1-X2-X2}
 [[Y_2, X_2], X_2] =  \sum_{k\in S_{C}} \left[ \left[ Q_{2}^{k}\left(K_{k}^{\oplus} \right), H_{\kin}' \right], H_{\kin}'\right].  
  \end{align}
  For every symmetric operator $B$ on $\ell^2(L_k^{\pm})$, by  \eqref{LocalizedKineticOperatorCommutator} we  compute 
  \begin{align}
\left[ Q_{2}^{k}\left(B\right) , H_{\kin}' \right] & = \sum_{p,q\in L_{k}^{\pm}}  \left\langle e_{p},Be_{q}\right\rangle \left[  \left(b_{\overline{k,p}}^{\ast}b_{\overline{k,q}}^{\ast}+b_{\overline{k,q}}b_{\overline{k,p}}\right),H_{\kin}' \right]  \nonumber\\
&=  \sum_{p,q\in L_{k}^{\pm}}   \left\langle e_{p},Be_{q}\right\rangle (\lambda_{\overline{k,p}} +\lambda_{ \overline{k,q}} ) \left(- b_{\overline{k,p}}^{\ast}b_{\overline{k,q}}^{\ast}+b_{\overline{k,q}}b_{\overline{k,p}}\right), \nonumber  \\ 
\left[ \left[ Q_{2}^{k}\left(B\right) ,H_{\kin}' \right],H_{\kin}' \right]  & = \sum_{p,q\in L_{k}^{\pm}}   \left\langle e_{p},Be_{q}\right\rangle (\lambda_{\overline{k,p}} +\lambda_{ \overline{k,q}} ) \left[ \left(- b_{\overline{k,p}}^{\ast}b_{\overline{k,q}}^{\ast}+b_{\overline{k,q}}b_{\overline{k,p}}\right) , H_{\kin}' \right]   \nonumber \\
&= \sum_{p,q\in L_{k}^{\pm}}   \left\langle e_{p},Be_{q}\right\rangle (\lambda_{\overline{k,p}} +\lambda_{ \overline{k,q}} )^2   \left( b_{\overline{k,p}}^{\ast}b_{\overline{k,q}}^{\ast}+b_{\overline{k,q}}b_{\overline{k,p}}\right).
\end{align}
By the Cauchy--Schwarz inequality we can estimate
  \begin{align} \label{eq:comm-Q-H-H-a}
\pm \left[ \left[ Q_{2}^{k}\left(B\right) ,H_{\kin}' \right],H_{\kin}' \right]  \le  \sum_{p,q\in L_{k}^{\pm}} \left( \epsilon \Big( \lambda_{\overline{k,p}} +\lambda_{\overline{k,q}} \Big)^4    b_{\overline{k,p}}^{\ast}  b_{\overline{k,q}}^{\ast} b_{\overline{k,q}} b_{\overline{k,p}} + \epsilon^{-1} |\left\langle e_{p},Be_{q}\right\rangle|^2 \right)  
\end{align}
for all $\epsilon>0$. From Propositions \ref{prop:Q1TildeNumberEstimate}, \ref{prop:KineticQ1AEstimate} and the commutation relations \eqref{LocalizedKineticOperatorCommutator}, \eqref{eq:intro-comm-NE-b} we have
\begin{align}
0&\le \sum_{p,q\in L_k^{\pm}} \lambda_{\overline{k,p}}  b_{\overline{k,p}}^{\ast}  b_{\overline{k,q}}^{\ast} b_{\overline{k,q}} b_{\overline{k,p}} \le  \sum_{p\in L_k^{\pm}} \lambda_{\overline{k,p}}  b_{\overline{k,p}}^{\ast}  \mathcal{N}_E b_{\overline{k,p}}  \le  H_{\kin}' \mathcal{N}_E,\\
0&\le \sum_{p,q\in L_k^{\pm}} \lambda_{\overline{k,q}}  b_{\overline{k,p}}^{\ast}  b_{\overline{k,q}}^{\ast} b_{\overline{k,q}} b_{\overline{k,p}} \le  \sum_{p\in L_k^{\pm}} b_{\overline{k,p}}^{\ast} H_{\kin}' b_{\overline{k,p}}  \le  H_{\kin}' \mathcal{N}_E. \nonumber 
\end{align} 
Moreover, when $k\in S_{C}=\overline{B}\left(0,k_{F}^{\gamma}\right)\cap\mathbb{Z}_{+}^{3}$ with $1\ge \gamma>0$ we have 
\begin{equation}
|\lambda_{\overline{k,p}}| \le  C|k| k_F,\quad  |\lambda_{\overline{k,p}}|^3 \le  C |k|^3 k_F^3 \le C |k|^2 k_F^4, \quad \forall p\in L_k^{\pm}. 
\end{equation}
Hence, we conclude from \eqref{eq:comm-Q-H-H-a} that 
   \begin{align} \label{eq:comm-Q-H-H}
\pm \left[ \left[ Q_{2}^{k}\left(B\right) ,H_{\kin}' \right],H_{\kin}' \right]  \le  C \Big( \epsilon |k|^2 k_F^4 H_{\kin}' \mathcal{N}_E +   \epsilon^{-1} \|B\|_{\rm HS}^2 \Big) 
\end{align}
for all $\epsilon>0$. Optimizing over $\epsilon$ gives 
   \begin{align} \label{eq:comm-Q-H-H}
\pm \left[ \left[ Q_{2}^{k}\left(B\right) ,H_{\kin}' \right],H_{\kin}' \right]  \le  C \|B\|_{\rm HS} |k|  k_F \Big(  H_{\kin}' \mathcal{N}_E +  k_F^2 \Big)
\end{align}
for all symmetric operators $B$ on $\ell^2(L_k^{\pm})$. Inserting this in \eqref{eq:comm-Y1-X2-X2} and using 
\begin{equation} \label{eq:GronwallSummationEstimate-22}
\sum_{k\in S_{C}} |k| \left\Vert K_{k}^{\oplus}\right\Vert _{\text{HS}}\leq\sum_{k\in S_{C}} |k| {\rm tr} (|K_{k}^{\oplus}|) =2 \sum_{k\in S_{C}} |k| \text{tr}\left(\left|K_{k}\right|\right) \le C \sum_{k\in \mathbb{Z}^3} |k| \hat V_k 
\end{equation} 
(which is similar to \eqref{eq:GronwallSummationEstimate}), we find that
  \begin{align} \label{eq:comm-Y1-X2-X2-final}
\pm [[Y_1, X_2], X_2] = \pm  \sum_{k\in S_{C}} \left[ \left[ Q_{2}^{k}\left(K_{k}^{\oplus} \right), H_{\kin}' \right], H_{\kin}'\right] \le C k_F \Big(  H_{\kin}' \mathcal{N}_E +  k_F^2 \Big)   \le C X_1 X_2^2. 
  \end{align}
Applying Lemma \ref{lem:double-commutator} we obtain 
   \begin{align}\label{eq:comm-K-X1X2-d}
  \pm [[Y_1, \sqrt{X_2}], \sqrt{X_2}] \le  C  X_1 X_2. 
  \end{align}
Putting together \eqref{eq:comm-K-X1X2-a}, \eqref{eq:comm-K-X1X2-b}, \eqref{eq:comm-K-X1X2-c} and \eqref{eq:comm-K-X1X2-d}, we conclude from \eqref{eq:comm-K-X1X2} that 
 \begin{align}
 \pm [\mathcal{K}, X_1 X_2] \le C X_1 X_2.
 \end{align}
Thus
\begin{equation}
\left|\frac{d}{dt}\left\langle \Psi_{t},X_1X_2  \Psi_{t}\right\rangle \right| =\left| \left\langle \Psi_{t}, [\mathcal{K}, X_1X_2]  \Psi_{t}\right\rangle \right| \le C \left\langle \Psi_{t},X_1X_2  \Psi_{t}\right\rangle. 
\end{equation}
By Gronwall's lemma, we have
\begin{equation}
\left\langle \Psi_{t},X_1 X_2 \Psi_{t}\right\rangle \leq C \left\langle \Psi, X_1 X_2 \Psi\right\rangle, \quad \forall |t|\le 1.
\end{equation}
This implies the desired bound since $\frac{1}{2}X_1X_2  \le  \mathcal{N}_E H_{\kin}^{\prime} + k_F H_{\kin}^{\prime} + k_F^2  \le  X_1X_2$. Here we used again Proposition \ref{prop:Zeta0RepresentationofHKin}. 
$\hfill\square$

\section{The Second Bogolubov Transformation}\label{sec:TheSecondTransformationandKineticEstimates}

Recall that after the conjugation by $e^{\mathcal{K}}$, up to negligible error terms, we obtain the correlation energy and the operator  
\begin{align} \label{eq:Heff-after-first-transformation}
H_{\kin}^{\prime}+2\sum_{k\in S_{C}}\tilde{Q}_{1}^{k}\left(E_{k}^{\oplus}-h_{k}^{\oplus}\right).
\end{align}
In the bosonic analogy, where we informally consider $H_{\kin}^{\prime}\sim2\sum_{k\in\mathbb{Z}_{+}^{3}}\tilde{Q}_{1}^{k}\left(h_{k}^{\oplus}\right)$,
this expression would be manifestly non-negative as $H_{\kin}^{\prime}$
cancels the negative terms $2\sum_{k\in S_{C}}\tilde{Q}_{1}^{k}\left(-h_{k}^{\oplus}\right)$
(and $E_{k}^{\oplus}>0$ as $E_{k}=e^{-K_{k}}h_{k}e^{-K_{k}}>0$),
so this term could be neglected for the lower bound. 
%This analogy
%is only informal, however, so this is not a valid argument. 
This analogy
is only formal, however. One might
still hope that $E_{k}^{\oplus}-h_{k}^{\oplus}\geq0$ since $E_{k}$ 
is isospectral to $\widetilde{E}_k$ and $\widetilde{E}_k\ge h_k$, 
% h_{k}^{\frac{1}{2}}e^{-2K_{k}}h_{k}^{\frac{1}{2}}$
%and
%\begin{equation}
%h_{k}^{\frac{1}{2}}e^{-2K_{k}}h_{k}^{\frac{1}{2}}-h_{k}=h_{k}^{\frac{1}{2}}\left(e^{-2K_{k}}-1\right)h_{k}^{\frac{1}{2}}\geq0,
%\end{equation}
%as $e^{-2K_{k}}\geq1$ by the one-body operator estimates of Section
%\ref{sec:AnalysisoftheOne-BodyOperators}, 
but this fails too - it
can be shown that $E_{k}-h_{k}$ is indefinite. While these two ideas - the bosonic analogy and the fact that $E_k-h_{k}\geq0$
- fail on their own we will overcome this issue by combining them. In this section, we will carry out another unitary transformation which effectively replaces $E_k$ by $\widetilde{E}_k$ in  \eqref{eq:Heff-after-first-transformation}.

%Heuristically, In the bosonic analogy, where we informally take \eqref{eq:intro-kinetic-bosonic} into account, then 
%
%$H_{\kin}^{\prime}\sim2\sum_{k\in\mathbb{Z}_{+}^{3}}\tilde{Q}_{1}^{k}\left(h_{k}^{\oplus}\right)$,
%this expression would be manifestly non-negative as $H_{\kin}^{\prime}$
%cancels the negative terms $2\sum_{k\in S_{C}}\tilde{Q}_{1}^{k}\left(-h_{k}^{\oplus}\right)$
%(and $E_{k}^{\oplus}>0$ as $E_{k}=e^{-K_{k}}h_{k}e^{-K_{k}}>0$),
%so this term could be neglected for the lower bound. This analogy
%is only informal, however, so this is not a valid argument. One might
%still hope that $E_{k}^{\oplus}-h_{k}^{\oplus}\geq0$ since $E_{k}$ 
%is isospectral to $\widetilde{E}_k$ and $\widetilde{E}_k\ge h_k$, but 
% h_{k}^{\frac{1}{2}}e^{-2K_{k}}h_{k}^{\frac{1}{2}}$
%and
%\begin{equation}
%h_{k}^{\frac{1}{2}}e^{-2K_{k}}h_{k}^{\frac{1}{2}}-h_{k}=h_{k}^{\frac{1}{2}}\left(e^{-2K_{k}}-1\right)h_{k}^{\frac{1}{2}}\geq0,
%\end{equation}
%as $e^{-2K_{k}}\geq1$ by the one-body operator estimates of Section
%\ref{sec:AnalysisoftheOne-BodyOperators}, 
%but this fails too - it
%can be shown that $E_{k}-h_{k}$ is indefinite.

%
%
%
%While these two ideas - the bosonic analogy and the fact that $E_k-h_{k}\geq0$
%- fail on their own we will overcome this issue by combining them.
%What we will do is carry out another unitary transformation which effectively replaces $E_k$ by $\widetilde{E}_k$.  
%

%While these two ideas - the bosonic analogy and the fact that $h_{k}^{\frac{1}{2}}e^{-2K_{k}}h_{k}^{\frac{1}{2}}-h_{k}\geq0$
%- fail on their own we will overcome this issue by combining them.

Consider the unitary transformation  
$e^{\mathcal{J}}:\mathcal{H}_{N}\rightarrow\mathcal{H}_{N}$, where
$\mathcal{J}:\mathcal{H}_{N}\rightarrow\mathcal{H}_{N}$ is now of
the form
\begin{equation} \label{eq:def-cJ}
\mathcal{J}=\sum_{k\in S_{C}}\sum_{p,q\in L_{k}^{\pm}}\left\langle e_{p},J_{k}^{\oplus}e_{q}\right\rangle b_{k}^{\ast}\left(e_{p}\right)b_{k}\left(e_{q}\right)=\sum_{k\in S_{C}}\sum_{p\in L_{k}^{\pm}}b_{k}^{\ast}\left(J_{k}^{\oplus}e_{p}\right)b_{k}\left(e_{p}\right)
\end{equation}
where $S_{C}=\overline{B}\left(0,k_{F}^{\gamma}\right)\cap\mathbb{Z}_{+}^{3}$ with $1\ge \gamma>0$ and 
\begin{equation} \label{eq:def-Jk}
J_{k}^{\oplus}=\left(\begin{array}{cc}
J_{k} & 0\\
0 & J_{k}
\end{array}\right), \quad J_{k}=\log\left(U_{k}\right), \quad U_{k}=\left(h_{k}^{\frac{1}{2}}e^{-2K_{k}}h_{k}^{\frac{1}{2}}\right)^{\frac{1}{2}}h_{k}^{-\frac{1}{2}}e^{K_{k}}.
\end{equation}
Here $U_{k}:\ell^{2}\left(L_{k}\right)\rightarrow\ell^{2}\left(L_{k}\right)$ is the unitary transformation which takes $E_{k}$ to $\widetilde E_k$, namely 
\begin{align} \label{eq:Uk-tEk-Uk*}
U_{k} E_{k}U_{k}^{\ast}  =\left(h_{k}^{\frac{1}{2}}e^{-2K_{k}}h_{k}^{\frac{1}{2}}\right)^{\frac{1}{2}}\left(h_{k}^{\frac{1}{2}}e^{-2K_{k}}h_{k}^{\frac{1}{2}}\right)^{\frac{1}{2}}=h_{k}^{\frac{1}{2}}e^{-2K_{k}}h_{k}^{\frac{1}{2}}=\widetilde E_k,  
\end{align}
and  $J_k$ is the (principal) logarithm of $U_{k}$, so that $e^{J_k}= U_k$.   
%%\begin{align}
%%U_{k}E_{k}U_{k}^{\ast} & =\left(h_{k}^{\frac{1}{2}}e^{-2K_{k}}h_{k}^{\frac{1}{2}}\right)^{\frac{1}{2}}h_{k}^{-\frac{1}{2}}e^{K_{k}}e^{-K_{k}}h_{k}e^{-K_{k}}e^{K_{k}}h_{k}^{-\frac{1}{2}}\left(h_{k}^{\frac{1}{2}}e^{-2K_{k}}h_{k}^{\frac{1}{2}}\right)^{\frac{1}{2}}\\
%% & =\left(h_{k}^{\frac{1}{2}}e^{-2K_{k}}h_{k}^{\frac{1}{2}}\right)^{\frac{1}{2}}\left(h_{k}^{\frac{1}{2}}e^{-2K_{k}}h_{k}^{\frac{1}{2}}\right)^{\frac{1}{2}}=h_{k}^{\frac{1}{2}}e^{-2K_{k}}h_{k}^{\frac{1}{2}}.\nonumber 
%%\end{align}
%\begin{equation} \label{eq:Uk-tEk-Uk*}
%U_{k}\widetilde E_{k}U_{k}^{\ast}= U_{k}\left(e^{-K_k} h_k e^{-K_k} \right) U_{k}^{\ast} =  h_{k}^{\frac{1}{2}}e^{-2K_{k}}h_{k}^{\frac{1}{2}}=(h_{k}^{\frac{1}{2}}\left(h_{k}+2P_{v_{k}}\right)h_{k}^{\frac{1}{2}})^{\frac{1}{2}}=  E_k. 
%\end{equation}
Since $J_k$ is skew-symmetric so are $J_{k}^{\oplus}$ and $\mathcal{J}$, and hence $e^{\mathcal{J}}$ is a unitary operator on $\mathcal{H}_{N}$.

%In the exact bosonic case, for every skew-symmetric operator $J: V\to V$, the unitary operator $e^{\mathcal{J}}$  with  $\mathcal{J}= \sum_{i} a^{\ast}\left(Je_{i}\right)a\left(e_{i}\right)$ is a Bogolubov transformation on $\mathcal{F}^+(V)$ which acts on creation and annihilation operators according to
%\begin{equation}
%e^{\mathcal{J}}a\left(\varphi\right)e^{-\mathcal{J}}=a\left(e^{J}\varphi\right),\quad e^{\mathcal{J}}a^{\ast}\left(\varphi\right)e^{-\mathcal{J}}=a^{\ast}\left(e^{J}\varphi\right).
%\end{equation}
%%i.e. it applies the unitary operator $e^{J}$ to the arguments of
%%these but does not mix the two types of operator. 
%Indeed, it is not
%difficult to see that $\mathcal{J}=\text{d}\Gamma\left(J\right)$
%and $e^{\mathcal{J}}=\Gamma\left(e^{J}\right)=\bigoplus_{N=0}^{\infty}e^{J}\otimes\cdots\otimes e^{J}$,
%and that the action of $e^{\mathcal{J}}$ on a second-quantized operator
%is simply
%\begin{equation}
%e^{\mathcal{J}}\text{d}\Gamma\left(A\right)e^{-\mathcal{J}}=\text{d}\Gamma\left(e^{J}Ae^{-J}\right).
%\end{equation}

In the exact bosonic case, it is not
difficult to see that for every skew-symmetric operator $J: V\to V$, the unitary operator $e^{\mathcal{J}}$  with  $\mathcal{J}= \text{d}\Gamma\left(J\right) = \sum_{i} a^{\ast}\left(Je_{i}\right)a\left(e_{i}\right)$ is a Bogolubov transformation on $\mathcal{F}^+(V)$ which acts on 
 a second-quantized operator as
\begin{equation}
e^{\mathcal{J}}\text{d}\Gamma\left(A\right)e^{-\mathcal{J}}=\text{d}\Gamma\left(e^{J}Ae^{-J}\right).
\end{equation}

Returning to the quasi-bosonic case, we will show that 
\begin{align}
  \quad\;e^{\mathcal{J}} \left( \sum_{k\in S_{C}}\tilde{Q}_{1}^{k} (  E_{k}^{\oplus} ) \right)e^{-\mathcal{J}} \approx \sum_{k\in S_{C}}\tilde{Q}_{1}^{k}\left(e^{J_{k}^{\oplus}}E_{k}^{\oplus}e^{-J_{k}^{\oplus}}\right) = \sum_{k\in S_{C}}\tilde{Q}_{1}^{k}\left( \widetilde {E}_{k}^{\oplus}\right)  
\end{align}
up to error terms which are similar to the exchange terms coming from
the first transformation. 
Moreover, although $H_{\kin}^{\prime} \sim 2\sum_{k\in\mathbb{Z}_{+}^{3}}\tilde{Q}_{1}^{k}\left(h_{k}^{\oplus}\right)$
does not hold precisely, it is valid from the point of view of commutators as explained in \eqref{eq:IntroductionApproximatelyEqualCommutators},
which results in $H_{\kin}^{\prime}-2\sum_{k\in\mathbb{Z}_{+}^{3}}\tilde{Q}_{1}^{k}\left(h_{k}^{\oplus}\right)$ being essentially invariant under the Bogolubov transformation $e^{\mathcal{J}}$. 
The overall transformation then takes the form
\begin{align} \label{eq:overall-second-transformation}
e^{\mathcal{J}}\left(H_{\kin}^{\prime}+2\sum_{k\in S_{C}}\tilde{Q}_{1}^{k}\left( E_{k}^{\oplus}-h_{k}^{\oplus}\right)\right)e^{-\mathcal{J}} \approx  H_{\kin}^{\prime}+2\sum_{k\in S_{C}}\tilde{Q}_{1}^{k}\left(\widetilde  {E}_{k}^{\oplus}-h_{k}^{\oplus}\right) 
\end{align}
and we now have the desired  non-negative operator $\widetilde {E}_{k}^{\oplus}-h_{k}^{\oplus}\ge 0$ on the right-hand side.

%operator 
%\begin{align}
%\widetilde {E}_{k}^{\oplus}-h_{k}^{\oplus} = \left(\begin{array}{cc}
%h_{k}^{\frac{1}{2}}e^{-2K_{k}}h_{k}^{\frac{1}{2}}-h_{k} & 0\\
%0 & h_{k}^{\frac{1}{2}}e^{-2K_{k}}h_{k}^{\frac{1}{2}}-h_{k}
%\end{array}\right)  \ge 0.
%\end{align}

\medskip

While the error terms in \eqref{eq:overall-second-transformation} are similar to those coming from the first
transformation, they are in practice more difficult to estimate, for
although we derived simple, optimal estimates for the transformation
kernels $\left(K_{k}\right)_{k\in S_{C}}$ in Section \ref{sec:AnalysisoftheOne-BodyOperators}
we cannot obtain the same for the transformation kernels $\left(J_{k}\right)_{k\in S_{C}}$.
The justification that the second transformation works as claimed
will therefore take more effort than was needed for the first transformation. 

\subsection{Actions  on the Bosonizable Terms}

The first step of justifying \eqref{eq:overall-second-transformation} is to prove the following exact equality.

\begin{prop}
\label{prop:ApplicationoftheSecondTransformation} The unitary transformation $e^{\mathcal{J}}:\mathcal{H}_{N}\rightarrow\mathcal{H}_{N}$ given in \eqref{eq:def-cJ}-\eqref{eq:def-Jk} satisfies 
\begin{align*}
  &e^{\mathcal{J}}\left(H_{\kin}^{\prime}+2\sum_{k\in S_{C}}\tilde{Q}_{1}^{k}\left(  E_{k}^{\oplus}-h_{k}^{\oplus}\right)\right)e^{-\mathcal{J}}
 \\
 & =H_{\kin}^{\prime}+2\sum_{k\in S_{C}}\tilde{Q}_{1}^{k}\left(\widetilde{E}_{k}^{\oplus}-h_{k}^{\oplus}\right)   +2\sum_{k\in S_{C}}\int_{0}^{1}e^{\left(1-t\right)\mathcal{J}} {\mathcal{E}}_{3}^{k}(F_{k}^{\oplus}(t))e^{-\left(1-t\right)\mathcal{J}}dt
\end{align*}
where for all $k\in\mathbb{Z}_{\ast}^{3}$ and $t\in\left[0,1\right]$ we defined the operator $F_{k}^{\oplus}(t):\ell^{2}\left(L_{k}^{\pm}\right)\rightarrow\ell^{2}\left(L_{k}^{\pm}\right)$  by
\[
F_{k}^{\oplus} (t) = \left(\begin{array}{cc}
e^{tJ_{k}} E_{k}e^{-tJ_{k}} - h_k   & 0\\
0 & e^{tJ_{k}}E_{k}e^{-tJ_{k}} - h_k
\end{array}\right),
\]
and for symmetric $A:\ell^{2}\left(L_{k}^{\pm}\right)\rightarrow\ell^{2}\left(L_{k}^{\pm}\right)$ we defined the {\em new exchange operator} 
\begin{align*} % \label{eq:def-E3-exchange-term}
\mathcal{E}_{3}^{k}\left(A\right) & =2\sum_{l\in S_{C}}\sum_{p\in L_{k}^{\pm}}\sum_{q\in L_{l}^{\pm}}{\rm Re}\left(b_{k}^{\ast}\left(Ae_{p}\right)\varepsilon_{k,l}\left(e_{p};e_{q}\right)b_{l}\left(J_{l}^{\oplus}e_{q}\right)\right)\\
 & =2\sum_{l\in S_{C}}\sum_{p\in L_{k}^{\pm}}\sum_{q\in L_{l}^{\pm}}{\rm Re}\left(b_{k}^{\ast}\left(Ae_{p}\right)\left(\delta_{p,q}c_{\overline{q-k}}c_{\overline{p-k}}^{\ast}+\delta_{\overline{p-k},\overline{q-k}}c_{q}^{\ast}c_{p}\right)b_{l}\left(J_{l}^{\oplus}e_{q}\right)\right).\nonumber 
\end{align*}
%where $J_{k}$ denotes the principal logarithm of $U_{k}=\left(h_{k}^{\frac{1}{2}}e^{-2K_{k}}h_{k}^{\frac{1}{2}}\right)^{\frac{1}{2}}h_{k}^{-\frac{1}{2}}e^{K_{k}}$.
\end{prop}

We will follow the same strategy as we did when we considered
the action of the quasi-bosonic Bogolubov transformation on the $Q_{1}^{k}\left(A\right)$
and $Q_{2}^{k}\left(B\right)$ terms. First we calculate the commutator:
\begin{prop}
\label{prop:JQ1kCommutator}For all $k\in S_{C}$ and symmetric $A:\ell^{2}\left(L_{k}^{\pm}\right)\rightarrow\ell^{2}\left(L_{k}^{\pm}\right)$
it holds that
\[
\left[\mathcal{J},\tilde{Q}_{1}^{k}\left(A\right)\right]=\tilde{Q}_{1}^{k}\left(\left[J_{k}^{\oplus},A\right]\right)+\mathcal{E}_{3}^{k}\left(A\right).
\]
%where
%\[
%\mathcal{E}_{3}^{k}\left(A\right)=2\sum_{l\in S_{C}}\sum_{p\in L_{k}^{\pm}}\sum_{q\in L_{l}^{\pm}}\re\left(b_{k}^{\ast}\left(Ae_{p}\right)\varepsilon_{k,l}\left(e_{p};e_{q}\right)b_{l}\left(J_{l}^{\oplus}e_{q}\right)\right).
%\]
\end{prop}

\textbf{Proof:} We first calculate, using the commutation relations
of the excitation operators $b_{k}\left(\varphi\right)$ and $b_{k}^{\ast}\left(\varphi\right)$,
that for any $k\in S_{C}$ and $\varphi\in\ell^{2}\left(L_{k}^{\pm}\right)$
\begin{align}
\left[\mathcal{J},b_{k}\left(\varphi\right)\right] 
 & =-\sum_{l\in S_{C}}\sum_{q\in L_{l}^{\pm}}\left(b_{l}^{\ast}\left(J_{l}^{\oplus}e_{q}\right)\left[b_{k}\left(\varphi\right),b_{l}\left(e_{q}\right)\right]+\left[b_{k}\left(\varphi\right),b_{l}^{\ast}\left(J_{l}^{\oplus}e_{q}\right)\right]b_{l}\left(e_{q}\right)\right)\nonumber \\
 & =-\sum_{l\in S_{C}}\sum_{q\in L_{l}^{\pm}}\left(\delta_{k,l}\left\langle \varphi,J_{l}^{\oplus}e_{q}\right\rangle +\varepsilon_{k,l}\left(\varphi;J_{l}^{\oplus}e_{q}\right)\right)b_{l}\left(e_{q}\right)\\
 & =-\sum_{q\in L_{k}^{\pm}}\left\langle \varphi,J_{k}^{\oplus}e_{q}\right\rangle b_{k}\left(e_{q}\right)-\sum_{l\in S_{C}}\sum_{q\in L_{l}^{\pm}}\varepsilon_{k,l}\left(\varphi;J_{l}^{\oplus}e_{q}\right)b_{l}\left(e_{q}\right)\nonumber \\
 & =\sum_{q\in L_{k}^{\pm}}\left\langle J_{k}^{\oplus}\varphi,e_{q}\right\rangle b_{k}\left(e_{q}\right)+\sum_{l\in S_{C}}\sum_{q\in L_{l}^{\pm}}\varepsilon_{k,l}\left(\varphi;e_{q}\right)b_{l}\left(J_{l}^{\oplus}e_{q}\right)\nonumber \\
 & =b_{k}\left(J_{k}^{\oplus}\varphi\right)+\mathcal{E}_{k}^{\mathcal{J}}\left(\varphi\right)\nonumber 
\end{align}
for
\begin{equation}
\mathcal{E}_{k}^{\mathcal{J}}\left(\varphi\right)=\sum_{l\in S_{C}}\sum_{q\in L_{l}^{\pm}}\varepsilon_{k,l}\left(\varphi;e_{q}\right)b_{l}\left(J_{l}^{\oplus}e_{q}\right),
\end{equation}
where we used the skew-symmetry of $J_{k}^{\oplus}$, anti-linearity
of $\varphi\mapsto b_{k}\left(\varphi\right)$, and Lemma \ref{lemma:TraceFormLemma}.
Consequently we compute for $\tilde{Q}_{1}^{k}\left(A\right)$ that
\begin{align}
&\left[\mathcal{J},\tilde{Q}_{1}^{k}\left(A\right)\right]  =\sum_{p\in L_{k}^{\pm}}\left[\mathcal{J},b_{k}^{\ast}\left(Ae_{p}\right)b_{k}\left(e_{p}\right)\right]=\sum_{p\in L_{k}^{\pm}}\left(b_{k}^{\ast}\left(Ae_{p}\right)\left[\mathcal{J},b_{k}\left(e_{p}\right)\right]+\left[\mathcal{J},b_{k}\left(Ae_{p}\right)\right]^{\ast}b_{k}\left(e_{p}\right)\right)\nonumber \\
 & =\sum_{p\in L_{k}^{\pm}}\left(b_{k}^{\ast}\left(Ae_{p}\right)\left(b_{k}\left(J_{k}^{\oplus}e_{p}\right)+\mathcal{E}_{k}^{\mathcal{J}}\left(e_{p}\right)\right)+\left(b_{k}\left(J_{k}^{\oplus}Ae_{p}\right)+\mathcal{E}_{k}^{\mathcal{J}}\left(Ae_{p}\right)\right)^{\ast}b_{k}\left(e_{p}\right)\right)\nonumber \\
 & =\sum_{p\in L_{k}^{\pm}}\left(b_{k}^{\ast}\left(Ae_{p}\right)b_{k}\left(J_{k}^{\oplus}e_{p}\right)+b_{k}^{\ast}\left(J_{k}^{\oplus}Ae_{p}\right)b_{k}\left(e_{p}\right)\right)\\
 & +\sum_{p\in L_{k}^{\pm}}\left(b_{k}^{\ast}\left(Ae_{p}\right)\mathcal{E}_{k}^{\mathcal{J}}\left(e_{p}\right)+\left(b_{k}^{\ast}\left(e_{p}\right)\mathcal{E}_{k}^{\mathcal{J}}\left(Ae_{p}\right)\right)^{\ast}\right)\nonumber \\
 & =\sum_{p\in L_{k}^{\pm}}b_{k}^{\ast}\left(\left(J_{k}^{\oplus}A-AJ_{k}^{\oplus}\right)e_{p}\right)b_{k}\left(e_{p}\right)+2\sum_{l\in S_{C}}\sum_{p\in L_{k}^{\pm}}\sum_{q\in L_{l}^{\pm}}{\rm Re}\left(b_{k}^{\ast}\left(Ae_{p}\right)\varepsilon_{k,l}\left(e_{p};e_{q}\right)b_{l}\left(J_{l}^{\oplus}e_{q}\right)\right)\nonumber \\
 & =\tilde{Q}_{1}^{k}\left(\left[J_{k}^{\oplus},A\right]\right)+\mathcal{E}_{3}^{k}\left(A\right).\nonumber 
\end{align}
$\hfill\square$

To derive an expression for $e^{\mathcal{J}}\tilde{Q}_{1}^{k}\left(A\right)e^{-\mathcal{J}}$
%we will similarly to the notation $\mathcal{A}_{K_{k}^{\oplus}}\left(A\right)=\left\{ K_{k}^{\oplus},A\right\} $
%of Section \ref{sec:TransformingtheHamiltonian} write $\mathcal{C}_{J_{k}^{\oplus}}\left(A\right)=\left[J_{k}^{\oplus},A\right]$
%for the commutator with $J_{k}^{\oplus}$. The Baker-Campbell-Hausdorff
%formula then takes the form
%\begin{equation}
%\exp\left(\mathcal{C}_{J_{k}^{\oplus}}\right)\left(A\right)=\sum_{m=0}^{\infty}\frac{1}{m!}\mathcal{C}_{J_{k}^{\oplus}}^{m}\left(A\right)=e^{J_{k}^{\oplus}}Ae^{-J_{k}^{\oplus}},\label{eq:BCHCondensedNotation}
%\end{equation}
we will use the Baker-Campbell-Hausdorff
formula 
\begin{equation}
\exp\left(\mathcal{C}_{J_{k}^{\oplus}}\right)\left(A\right)=\sum_{m=0}^{\infty}\frac{1}{m!}\mathcal{C}_{J_{k}^{\oplus}}^{m}\left(A\right)=e^{J_{k}^{\oplus}}Ae^{-J_{k}^{\oplus}},\quad \text{with } \mathcal{C}_{J_{k}^{\oplus}}\left(A\right)=\left[J_{k}^{\oplus},A\right].\label{eq:BCHCondensedNotation}
\end{equation}
Imitating the proof of Proposition \ref{prop:QuasiBosonicQuadraticOperatorTransformation}
we deduce the following:
\begin{prop}
\label{prop:SecondTransformationQ1kAction}For all $k\in S_{C}$ and
symmetric $A:\ell^{2}\left(L_{k}^{\pm}\right)\rightarrow\ell^{2}\left(L_{k}^{\pm}\right)$
it holds that
\[
e^{\mathcal{J}}\tilde{Q}_{1}^{k}\left(A\right)e^{-\mathcal{J}}=\tilde{Q}_{1}^{k}\left(e^{J_{k}^{\oplus}}Ae^{-J_{k}^{\oplus}}\right)+\int_{0}^{1}e^{t\mathcal{J}}\mathcal{E}_{3}^{k}\left(e^{\left(1-t\right)J_{k}^{\oplus}}Ae^{-\left(1-t\right)J_{k}^{\oplus}}\right)e^{-t\mathcal{J}}dt,
\]
the integrals being Riemann integrals of bounded operators. 
\end{prop}

\textbf{Proof:} We claim that for any $n\in\mathbb{N}$ it holds that
\begin{align}
e^{\mathcal{J}}\tilde{Q}_{1}^{k}\left(A\right)e^{-\mathcal{J}} & =\tilde{Q}_{1}^{k}\left(\sum_{m=0}^{n-1}\frac{1}{m!}\mathcal{C}_{J_{k}^{\oplus}}^{m}\left(A\right)\right)+\int_{0}^{1}e^{t\mathcal{J}}\mathcal{E}_{3}^{k}\left(\sum_{m=0}^{n-1}\frac{1}{m!}\mathcal{C}_{\left(1-t\right)J_{k}^{\oplus}}^{m}\left(A\right)\right)e^{-t\mathcal{J}}dt\label{eq:SecondTransformationInduction}\\
 & +\frac{1}{\left(n-1\right)!}\int_{0}^{1}e^{t\mathcal{J}}\tilde{Q}_{1}^{k}\left(\mathcal{C}_{J_{k}^{\oplus}}^{n}\left(A\right)\right)e^{-t\mathcal{J}}\left(1-t\right)^{n-1}dt.\nonumber 
\end{align}
We proceed by induction. For $n=1$ we have by the fundamental theorem
of calculus and Proposition \ref{prop:JQ1kCommutator} that
\begin{align}
e^{\mathcal{J}}\tilde{Q}_{1}^{k}\left(A\right)e^{-\mathcal{J}} & =\tilde{Q}_{1}^{k}\left(A\right)+\int_{0}^{1}e^{t\mathcal{J}}\left[\mathcal{J},\tilde{Q}_{1}^{k}\left(A\right)\right]e^{-t\mathcal{J}}\,dt\\
 & =\tilde{Q}_{1}^{k}\left(A\right)+\int_{0}^{1}e^{t\mathcal{J}}\tilde{Q}_{1}^{k}\left(\left[J_{k}^{\oplus},A\right]\right)e^{-t\mathcal{J}}\,dt+\int_{0}^{1}e^{t\mathcal{J}}\mathcal{E}_{3}^{k}\left(A\right)e^{-t\mathcal{J}}\,dt\nonumber 
\end{align}
which is the claim. For the inductive step we assume that case $n$
holds and integrate the last term of equation (\ref{eq:SecondTransformationInduction})
by parts:
\begin{align}
 & \quad\;\frac{1}{\left(n-1\right)!}\int_{0}^{1}e^{t\mathcal{J}}\tilde{Q}_{1}^{k}\left(\mathcal{C}_{J_{k}^{\oplus}}^{n}\left(A\right)\right)e^{-t\mathcal{J}}\left(1-t\right)^{n-1}dt\nonumber \\
 & =\frac{1}{\left(n-1\right)!}\left[e^{t\mathcal{J}}\tilde{Q}_{1}^{k}\left(\mathcal{C}_{J_{k}^{\oplus}}^{n}\left(A\right)\right)e^{-t\mathcal{J}}\left(-\frac{\left(1-t\right)^{n}}{n}\right)\right]_{0}^{1}\nonumber \\
 &  -\frac{1}{\left(n-1\right)!}\int_{0}^{1}e^{t\mathcal{J}}\left[\mathcal{J},\tilde{Q}_{1}^{k}\left(\mathcal{C}_{J_{k}^{\oplus}}^{n}\left(A\right)\right)\right]e^{-t\mathcal{J}}\left(-\frac{\left(1-t\right)^{n}}{n}\right)dt\\
 & =\frac{1}{n!}\tilde{Q}_{1}^{k}\left(\mathcal{C}_{J_{k}^{\oplus}}^{n}\left(A\right)\right)+\frac{1}{n!}\int_{0}^{1}e^{t\mathcal{J}}\left(\tilde{Q}_{1}^{k}\left(\left[J_{k}^{\oplus},\mathcal{C}_{J_{k}^{\oplus}}^{n}\left(A\right)\right]\right)+\mathcal{E}_{3}^{k}\left(\mathcal{C}_{J_{k}^{\oplus}}^{n}\left(A\right)\right)\right)e^{-t\mathcal{J}}\left(1-t\right)^{n}dt\nonumber \\
 & =\tilde{Q}_{1}^{k}\left(\frac{1}{n!}\mathcal{C}_{J_{k}^{\oplus}}^{n}\left(A\right)\right)+\int_{0}^{1}e^{t\mathcal{J}}\mathcal{E}_{3}^{k}\left(\frac{1}{n!}\mathcal{C}_{\left(1-t\right)J_{k}^{\oplus}}^{n}\left(A\right)\right)e^{-t\mathcal{J}}\,dt \nonumber\\
 & +\frac{1}{n!}\int_{0}^{1}e^{t\mathcal{J}} \tilde{Q}_1^k \left( \mathcal{C}_{J_{k}^{\oplus}}^{n+1}\left(A\right) \right)e^{-t\mathcal{J}}\left(1-t\right)^{n}dt.\nonumber 
\end{align}
Insertion of this identity into equation (\ref{eq:SecondTransformationInduction})
yields the statement for case $n+1$, so our claim \eqref{eq:SecondTransformationInduction} holds. 
%We can now take $n\rightarrow\infty$ and appeal to equation (\ref{eq:BCHCondensedNotation})
%to see that
%\begin{equation}
%e^{\mathcal{J}}\tilde{Q}_{1}^{k}\left(A\right)e^{-\mathcal{J}}=\tilde{Q}_{1}^{k}\left(e^{J_{k}^{\oplus}}Ae^{-J_{k}^{\oplus}}\right)+\int_{0}^{1}e^{t\mathcal{J}}\mathcal{E}_{3}^{k}\left(e^{\left(1-t\right)J_{k}^{\oplus}}Ae^{-\left(1-t\right)J_{k}^{\oplus}}\right)e^{-t\mathcal{J}}dt.
%\end{equation}
We can now take $n\rightarrow\infty$ and appeal to equation (\ref{eq:BCHCondensedNotation}) to get the claim.
%to see that
%\begin{equation}
%e^{\mathcal{J}}\tilde{Q}_{1}^{k}\left(A\right)e^{-\mathcal{J}}=\tilde{Q}_{1}^{k}\left(e^{J_{k}^{\oplus}}Ae^{-J_{k}^{\oplus}}\right)+\int_{0}^{1}e^{t\mathcal{J}}\mathcal{E}_{3}^{k}\left(e^{\left(1-t\right)J_{k}^{\oplus}}Ae^{-\left(1-t\right)J_{k}^{\oplus}}\right)e^{-t\mathcal{J}}dt.
%\end{equation}
$\hfill\square$
%
%\begin{align}
%delete
%\end{align}

Proposition \ref{prop:JQ1kCommutator} also allows us to describe
the action of $e^{\mathcal{J}}$ on $H_{\kin}^{\prime}$:
\begin{prop}
\label{prop:SecondTransformationKineticAction}It holds that
\[
e^{\mathcal{J}}\left(H_{\kin}^{\prime}-2\sum_{k\in S_{C}}\tilde{Q}_{1}^{k}\left(h_{k}^{\oplus}\right)\right)e^{-\mathcal{J}}=H_{\kin}^{\prime}-2\sum_{k\in S_{C}}\tilde{Q}_{1}^{k}\left(h_{k}^{\oplus}\right)-2\sum_{k\in S_{C}}\int_{0}^{1}e^{t\mathcal{J}}\mathcal{E}_{3}^{k}\left(h_{k}^{\oplus}\right)e^{-t\mathcal{J}}dt.
\]
\end{prop}

\textbf{Proof:} By the fundemental theorem of calculus and the fact that $\partial_t (e^{tA}B e^{-tA})= e^{tA}[A,B]e^{-tA}$, the left side is equal to 
\begin{align}
H_{\kin}^{\prime}-2\sum_{k\in S_{C}}\tilde{Q}_{1}^{k}\left(h_{k}^{\oplus}\right)  +\int_{0}^{1}e^{t\mathcal{J}}\left[\mathcal{J},H_{\kin}^{\prime}-2\sum_{k\in S_{C}}\tilde{Q}_{1}^{k}\left(h_{k}^{\oplus}\right)\right]e^{-t\mathcal{J}}dt.\nonumber 
\end{align}
Recalling \eqref{LocalizedKineticOperatorCommutator}, 
%\begin{equation}
%\left[H_{\kin}^{\prime},b_{k}\left(\varphi\right)\right]=-2b_{k}\left(h_{k}^{\oplus}\right),\quad\left[H_{\kin}^{\prime},b_{k}^{\ast}\left(\varphi\right)\right]=2b_{k}^{\ast}\left(h_{k}^{\oplus}\right),
%\end{equation}
%of Proposition \ref{LocalizedKineticOperatorCommutator}, 
we may compute
using Lemma \ref{lemma:TraceFormLemma} that
\begin{align}
\left[\mathcal{J},H_{\kin}^{\prime}\right] & =-\sum_{k\in S_{C}}\sum_{p\in L_{k}^{\pm}}\left[H_{\kin}^{\prime},b_{k}^{\ast}\left(J_{k}^{\oplus}e_{p}\right)b_{k}\left(e_{p}\right)\right]\nonumber \\
 & -2\sum_{k\in S_{C}}\sum_{p\in L_{k}^{\pm}}\left(-b_{k}^{\ast}\left(J_{k}^{\oplus}e_{p}\right)b_{k}\left(h_{k}^{\oplus}e_{p}\right)+b_{k}^{\ast}\left(h_{k}^{\oplus}J_{k}^{\oplus}e_{p}\right)b_{k}\left(e_{p}\right)\right)\label{eq:JHkinCommutator}\\
 & =2\sum_{k\in S_{C}}\sum_{p\in L_{k}^{\pm}}b_{k}^{\ast}\left(\left(J_{k}^{\oplus}h_{k}^{\oplus}-h_{k}^{\oplus}J_{k}^{\oplus}\right)e_{p}\right)b_{k}\left(e_{p}\right)=2\sum_{k\in S_{C}}\tilde{Q}_{1}^{k}\left(\left[J_{k}^{\oplus},h_{k}^{\oplus}\right]\right).\nonumber 
\end{align}
Combining with  Proposition \ref{prop:JQ1kCommutator}
we have that
\begin{equation}
\left[\mathcal{J},H_{\kin}^{\prime}-2\sum_{k\in S_{C}}\tilde{Q}_{1}^{k}\left(h_{k}^{\oplus}\right)\right]=\left[\mathcal{J},H_{\kin}^{\prime}\right]-2\sum_{k\in S_{C}}\left[\mathcal{J},\tilde{Q}_{1}^{k}\left(h_{k}^{\oplus}\right)\right]=-2\sum_{k\in S_{C}}\mathcal{E}_{3}^{k}\left(h_{k}^{\oplus}\right)
\end{equation}
which implies the claim.

$\hfill\square$

We can now conclude 

{\bf Proof of Proposition \ref{prop:ApplicationoftheSecondTransformation}:} 
By the Propositions \ref{prop:SecondTransformationQ1kAction} and
\ref{prop:SecondTransformationKineticAction} we see that
\begin{align}
 & \quad\;e^{\mathcal{J}}\left(H_{\kin}^{\prime}- 2\sum_{k\in S_{C}}\tilde{Q}_{1}^{k}\left( h_{k}^{\oplus}\right) + 2\sum_{k\in S_{C}}\tilde{Q}_{1}^{k}\left( E_{k}^{\oplus}\right)\right)e^{-\mathcal{J}}\nonumber \\
 & =H_{\kin}^{\prime}-2\sum_{k\in S_{C}}\tilde{Q}_{1}^{k}\left(h_{k}^{\oplus}\right)-2\sum_{k\in S_{C}}\int_{0}^{1}e^{t\mathcal{J}}\mathcal{E}_{3}^{k}\left(h_{k}^{\oplus}\right)e^{-t\mathcal{J}}\,dt\label{eq:SecondTransformationAction}\\
 & +2\sum_{k\in S_{C}}\tilde{Q}_{1}^{k}\left(e^{J_{k}^{\oplus}} E_{k}^{\oplus}e^{-J_{k}^{\oplus}}\right)+2\int_{0}^{1}e^{t\mathcal{J}}\mathcal{E}_{3}^{k}\left(e^{\left(1-t\right)J_{k}^{\oplus}} E_{k}^{\oplus}e^{-\left(1-t\right)J_{k}^{\oplus}}\right)e^{-t\mathcal{J}}dt\nonumber \\
 & =H_{\kin}^{\prime}+2\sum_{k\in S_{C}}\tilde{Q}_{1}^{k}\left(e^{J_{k}^{\oplus}} E_{k}^{\oplus}e^{-J_{k}^{\oplus}}-h_{k}^{\oplus}\right)+2\int_{0}^{1}e^{\left(1-t\right)\mathcal{J}}\mathcal{E}_{3}^{k}\left(e^{tJ_{k}^{\oplus}} E_{k}^{\oplus}e^{-tJ_{k}^{\oplus}}-h_{k}^{\oplus}\right)e^{-\left(1-t\right)\mathcal{J}}dt,\nonumber 
\end{align}
where we also reparametrized the integral. From the choice of $J_k^{\oplus}$ in \eqref{eq:def-Jk}, we have
\begin{equation}
e^{tJ_{k}^{\oplus}} E_{k}^{\oplus}e^{-tJ_{k}^{\oplus}} - h_k^{\oplus} = \left(\begin{array}{cc}
e^{tJ_{k}}  E_k e^{-tJ_k} - h_k & 0\\
0 & e^{tJ_{k}}   E_k e^{-tJ_k} -h_k
\end{array}\right) ={E}_k^{\oplus}(t) 
\end{equation} 
for all $t\in [0,1]$. Moreover, using $e^{J_k}=U_k$ and \eqref{eq:Uk-tEk-Uk*} we get $e^{J_{k}^{\oplus}} E_{k}^{\oplus}e^{-tJ_{k}^{\oplus}}=\widetilde{E}_k^{\oplus}$.
$\hfill\square$

\subsection{Estimates for the Exchange Terms}

Now we estimate the new exchange term $\mathcal{E}_3$ in Proposition \ref{prop:ApplicationoftheSecondTransformation}. We have

\begin{prop}
\label{prop:KineticcalEEstimate}For all $k\in\mathbb{Z}_{+}^{3}$,
symmetric $E:\ell^{2}\left(L_{k}^{\pm}\right)\rightarrow\ell^{2}\left(L_{k}^{\pm}\right)$
and $\Psi\in D\left(H_{\kin}^{\prime}\right)$ it holds that
\[
\left|\left\langle \Psi,\mathcal{E}_{3}^{k}\left(E\right)\Psi\right\rangle \right|\leq C\left(\max_{p\in L_{k}^{\pm}}\left\Vert \left(h_{k}^{\oplus}\right)^{-\frac{1}{2}}Ee_{p}\right\Vert \right)\left(\sum_{l\in S_{C}}\left\Vert \left(h_{l}^{\oplus}\right)^{-\frac{1}{2}}J_{l}^{\oplus}\right\Vert _{\HS}\right)\sqrt{\left\langle \Psi,H_{\kin}^{\prime}\Psi\right\rangle \left\langle \Psi,\mathcal{N}_{E}H_{\kin}^{\prime}\Psi\right\rangle }
\]
for a constant $C>0$ independent of all quantities.
\end{prop}

\textbf{Proof of Proposition \ref{prop:KineticcalEEstimate}:} We can follow the analysis in Section \ref{sec:Analysis of Exchange Terms}. In particular, the same reduction in Section \ref{sec:Reduction to Simpler Expressions} applies to $\mathcal{E}_{3}^{k}\left(E\right)$,
but in this case it is significantly simpler: By definition, up to taking adjoints every term of $\mathcal{E}_{3}^{k}\left(E\right)$
immediately reduces to the schematic form
\begin{equation}
\sum_{l\in S_{C}}\sum_{p\in S}b_{k}^{\ast}\left(Ee_{p_{1}}\right)\tilde{c}_{p_{3}}^{\ast}\tilde{c}_{p_{4}}b_{l}\left(J_{l}^{\oplus}e_{p_{2}}\right),
\end{equation}
and recalling that commutators of the forms $\left[\tilde{c}_{p},b_{k}\left(\varphi\right)\right]$
and $\left[\tilde{c}_{p}^{\ast},b_{k}^{\ast}\left(\varphi\right)\right]$
also vanish we may normal-order this schematic form without introducing
additional terms. Controlling $\mathcal{E}_{3}^{k}\left(E\right)$
thus reduces entirely to the estimation of the single schematic form
\begin{equation}
\sum_{l\in S_{C}}\sum_{p\in S}\tilde{c}_{p_{3}}^{\ast}b_{k}^{\ast}\left(Ee_{p_{1}}\right)b_{l}\left(J_{l}^{\oplus}e_{p_{2}}\right)\tilde{c}_{p_{4}}.\label{eq:Exchange3SchematicForm}
\end{equation}

We estimate the schematic form of equation (\ref{eq:Exchange3SchematicForm})
using Proposition \ref{prop:GeneralizedKineticEstimates}, Lemma \ref{lemma:KineticEstimationSimplification},
and the Cauchy-Schwarz inequality:
\begin{align}
 & \quad\;\sum_{l\in S_{C}}\sum_{p\in S}\left|\left\langle \Psi,\tilde{c}_{p_{3}}^{\ast}b_{k}^{\ast}\left(Ee_{p_{1}}\right)b_{l}\left(J_{l}^{\oplus}e_{p_{2}}\right)\tilde{c}_{p_{4}}\Psi\right\rangle \right|\leq\sum_{l\in S_{C}}\sum_{p\in S}\left\Vert b_{k}\left(Ee_{p_{1}}\right)\tilde{c}_{p_{3}}\Psi\right\Vert \left\Vert b_{l}\left(J_{l}^{\oplus}e_{p_{2}}\right)\tilde{c}_{p_{4}}\Psi\right\Vert \nonumber \\
 & \leq\sum_{l\in S_{C}}\sum_{p\in S}\left\Vert \left(h_{k}^{\oplus}\right)^{-\frac{1}{2}}Ee_{p_{1}}\right\Vert \left\Vert \left(h_{l}^{\oplus}\right)^{-\frac{1}{2}}J_{l}^{\oplus}e_{p_{2}}\right\Vert \sqrt{\left\langle \tilde{c}_{p_{3}}\Psi,H_{\text{kin}}^{\prime\left(\pm1\right)}\tilde{c}_{p_{3}}\Psi\right\rangle \left\langle \tilde{c}_{p_{4}}\Psi,H_{\text{kin}}^{\prime\left(\pm1\right)}\tilde{c}_{p_{4}}\Psi\right\rangle }\\
 & \leq\left(\max_{p\in L_{k}^{\pm}}\left\Vert \left(h_{k}^{\oplus}\right)^{-\frac{1}{2}}Ee_{p}\right\Vert \right)\sqrt{\left\langle \Psi,H_{\kin}^{\prime}\Psi\right\rangle }\sum_{l\in S_{C}}\sqrt{\sum_{p\in S}\left\Vert \left(h_{l}^{\oplus}\right)^{-\frac{1}{2}}J_{l}^{\oplus}e_{p_{2}}\right\Vert ^{2}}\sqrt{\sum_{p\in S}\left\langle \Psi,\tilde{c}_{p_{4}}^{\ast}H_{\text{kin}}^{\prime\left(\pm1\right)}\tilde{c}_{p_{4}}\Psi\right\rangle }\nonumber \\
 & \leq\left(\max_{p\in L_{k}^{\pm}}\left\Vert \left(h_{k}^{\oplus}\right)^{-\frac{1}{2}}Ee_{p_{1}}\right\Vert \right)\left(\sum_{l\in S_{C}}\left\Vert \left(h_{l}^{\oplus}\right)^{-\frac{1}{2}}J_{l}^{\oplus}\right\Vert _{\text{HS}}\right)\sqrt{\left\langle \Psi,H_{\kin}^{\prime}\Psi\right\rangle \left\langle \Psi,\mathcal{N}_{E}H_{\kin}^{\prime}\Psi\right\rangle }.\nonumber 
\end{align}
%where we estimated $\left\langle \tilde{c}_{p_{3}}\Psi,H_{\text{kin}}^{\prime\left(\pm1\right)}\tilde{c}_{p_{3}}\Psi\right\rangle =\left\langle \Psi,\tilde{c}_{p_{3}}^{\ast}H_{\text{kin}}^{\prime\left(\pm1\right)}\tilde{c}_{p_{3}}\Psi\right\rangle \leq\left\langle \Psi,H_{\kin}^{\prime}\Psi\right\rangle $
%in order to Cauchy-Schwarz only the $\left\Vert \left(h_{l}^{\oplus}\right)^{-\frac{1}{2}}J_{l}^{\oplus}e_{p_{2}}\right\Vert $
%and $\left\langle \tilde{c}_{p_{4}}\Psi,H_{\text{kin}}^{\prime\left(\pm1\right)}\tilde{c}_{p_{4}}\Psi\right\rangle $
%terms.
$\hfill\square$

%(Applying the Cauchy-Schwarz inequality to the $\left\Vert \left(h_{l}^{\oplus}\right)^{-\frac{1}{2}}J_{l}^{\oplus}e_{p_{2}}\right\Vert $
%and $\left\langle \tilde{c}_{p_{4}}\Psi,H_{\text{kin}}^{\prime\left(\pm1\right)}\tilde{c}_{p_{4}}\Psi\right\rangle $
%terms rather than both of the $\left\langle \tilde{c}_{p_{4}}\Psi,H_{\text{kin}}^{\prime\left(\pm1\right)}\tilde{c}_{p_{4}}\Psi\right\rangle $
%terms is motivated by our later estimates of the quantities, which
%show that this yields the optimal final error.)

%%%%%%%%%%%%%%%%%%%%%%%%%%%%%%%%%%%%
%%%%%%%%%%%%%%%%%%%%%%%%%%%%%%%%%%%%

\subsection{\label{sec:OneBodyOperatorEstimatesfortheSecondTransformation}One-Body
Operator Estimates}

In this subsection we derive estimates on the one-body quantities 
%operators which
%we encountered in the previous section, specifically the quantities
\begin{equation}
\max_{p\in L_{k}}\left\Vert h_{k}^{-\frac{1}{2}}E_{k}\left(t\right)e_{p}\right\Vert ,\quad\left\Vert h_{k}^{-\frac{1}{2}}J_{k}\right\Vert _{\text{HS}},\quad \quad\text{\ensuremath{\left\Vert h_{k}^{-\frac{1}{2}}\left[J_{k},h_{k}\right]h_{k}^{-\frac{1}{2}}\right\Vert _{\text{HS}}}}, \quad {\rm tr} ( h_k^{-1/2} (\widetilde{E}_k  - h_k) h_k^{-1/2}) 
\end{equation}
The first two quantities arise from the analysis of the exchange terms in the previous subsection, while the third quantity will be needed in order to derive Gronwall-type estimates for the kinetic operator and the last one is useful to remove the cut-off $S_C$ on the right hand side of \eqref{eq:overall-second-transformation} at the end. The estimates we will establish
are the following:
\begin{prop}
\label{them:SecondTransformationOneBodyEstimates} Assume $\sum_{k\in \mathbb{Z}^3_*}\hat V_k |k|<\infty$. Then for all $k\in \mathbb{Z}^3_*$ we have
\begin{align*}
{\rm tr} \Big( h_k^{-1/2} (\widetilde{E}_k  - h_k) h_k^{-1/2}\Big) \le C \hat V_k.
\end{align*}
Moreover, if $k\in\overline{B}\left(0,k_{F}^{\gamma}\right)\cap\mathbb{Z}_{\ast}^{3}$, $0<\gamma<\frac{1}{47}$, and $t\in\left[0,1\right]$ it holds that
\begin{align*}
\max_{p\in L_{k}}\left\Vert h_{k}^{-\frac{1}{2}}E_{k}\left(t\right) e_{p}\right\Vert  & \leq Ck_{F}^{-\frac{1}{2}}\left(\hat{V}_{k}+\hat{V}_{k}^{3}\left|k\right|^{6}\log\left(k_{F}\right)\right),\\
\left\Vert h_{k}^{-\frac{1}{2}}J_{k}\right\Vert _{\HS} & \leq C(\log k_{F})^{\frac{2}{3}}k_{F}^{-\frac{1}{3}}\hat{V}_{k},\\
\left\Vert h_{k}^{-\frac{1}{2}}\left[J_{k},h_{k}\right]h_{k}^{-\frac{1}{2}}\right\Vert _{\HS} & \leq C\hat{V}_{k}.
\end{align*}
Here the constant $C>0$ is independent of $k$ and $k_{F}$.
\end{prop}

Proposition \ref{them:SecondTransformationOneBodyEstimates} is the main source of the technical restriction $\gamma<\frac 1 {47}$ which comes from the use of the first bound in Proposition \ref{prop:MoreSingularRiemannSums} (we need $\gamma<\frac{4+3\beta}{8-3\beta}$ with $\beta=-\frac 5 4$).

\medskip

%We will also need estimates on $\max_{p\in L_{k}}\left\Vert h_{k}^{-\frac{1}{2}}A_{k}\left(t\right)e_{p}\right\Vert $,
%$\left\Vert h_{k}^{-\frac{1}{2}}K_{k}\right\Vert _{\text{HS}}$ etc.
%in order to control the first transformation by our kinetic estimates,
%but these quantities can be bounded in exactly the same fashion as
%the corresponding non-kinetic estimates of Section \ref{sec:AnalysisoftheOne-BodyOperators}.
%We therefore leave these in appendix section \ref{subsec:EstimatesofAkandBkfortheLowerBound},
%``Estimates on $K$ Quantities for the Lower Bound''.

As in Section \ref{sec:AnalysisoftheOne-BodyOperators}, in order to simplify the notation  let $h:V\rightarrow V$ denote a self-adjoint
operator acting on an $n$-dimensional Hilbert space $V$, let $\left(x_{i}\right)_{i=1}^{n}$
denote an eigenbasis for $h$ with eigenvalues $\left(\lambda_{i}\right)_{i=1}^{n}$,
and let $v\in V$ be any vector such that $\left\langle v,x_{i}\right\rangle \geq0$
for all $1\leq i\leq n$. As before, we take
\begin{align}
K=-\frac{1}{2}\log\left(h^{-\frac{1}{2}}\left(h^{2}+2P_{h^{\frac{1}{2}}v}\right)^{\frac{1}{2}}h^{-\frac{1}{2}}\right). 
\end{align}
We will establish general estimates for the operators  
\begin{align} \label{eq:CleanKDefinition_U-J-E}
U=\left(h^{\frac{1}{2}}e^{-2K}h^{\frac{1}{2}}\right)^{\frac{1}{2}}h^{-\frac{1}{2}}e^{K},\quad J = \log(U),\quad E\left(t\right)=e^{tJ}e^{-K}he^{-K}e^{-tJ} -h,
\end{align}
and then at the end insert the explicit choice \eqref{eq:hkvkReminder} to get the desired estimates.

Unlike the case in Section \ref{sec:AnalysisoftheOne-BodyOperators}
we will now also take $V$ to be a complex Hilbert space - this is
not a strictly necessary assumption but it allows us to streamline
the presentation significantly, as it implies that the unitary operator
$U$ is diagonalizable and so lets us describe the operators $J=\log\left(U\right)$
and $e^{tJ}$ solely in terms of eigenvectors of $U$.

%The difficult part starts from the derivation of a matrix element estimate for $U-1$. We do this
%by utilizing the same argument which we used to obtain matrix element
%estimates for $e^{-2K}-1$ and $1-e^{2K}$ in Section \ref{sec:AnalysisoftheOne-BodyOperators},
%but unlike that case we cannot generalize these to the operators
%$J$ and $e^{tJ}$. In order to estimate the quantities involving
%these operators we will therefore utilize a technique which effectively
%lets us replace instances of these by $U-1$ and powers thereof, by
%exploiting the diagonalizability of $U$. Having obtained the necessary
%estimates in this fashion we end the section by applying the general
%estimates to our particular problem in order to conclude Proposition \ref{them:SecondTransformationOneBodyEstimates}.

\medskip
The main difficulty of the proof of Proposition \ref{them:SecondTransformationOneBodyEstimates} is that we cannot extend the argument leading to matrix element
estimates for $e^{-2K}-1$ and $1-e^{2K}$ in Section \ref{sec:AnalysisoftheOne-BodyOperators} to handle the operators
$J$ and $e^{tJ}$. Instead, we will utilize a technique which effectively
lets us replace relevant quantities of $J$ by these of $U-1$, by exploiting the diagonalizability of $U$. 

%The difficult part starts from the derivation of a matrix element estimate for $U-1$. We do this
%by utilizing the same argument which we used to obtain matrix element
%estimates for $e^{-2K}-1$ and $1-e^{2K}$ in Section \ref{sec:AnalysisoftheOne-BodyOperators},
%but unlike that case we cannot generalize these to the operators
%$J$ and $e^{tJ}$. In order to estimate the quantities involving
%these operators we will therefore utilize a technique which effectively
%lets us replace instances of these by $U-1$ and powers thereof, by
%exploiting the diagonalizability of $U$. Having obtained the necessary
%estimates in this fashion we end the section by applying the general
%estimates to our particular problem in order to conclude Proposition \ref{them:SecondTransformationOneBodyEstimates}.

%\subsubsection*{Estimate for $h^{-1/2} (\widetilde{E}  - h) h^{-1/2}$}

\medskip

We start with the easy part of Proposition \ref{them:SecondTransformationOneBodyEstimates}. 

\begin{prop}
\label{prop:tE-h} With $\widetilde{E} = \left(h^{2}+2P_{h^{\frac{1}{2}}v}\right)^{\frac{1}{2}}$ we have
$$ {\rm tr} \Big(h^{-1/2} (\widetilde{E}  - h) h^{-1/2}\Big) \le  \langle v, h^{-1}v\rangle.$$
\end{prop}

{\bf Proof:} Using \eqref{eq:sqrt-h2-P} for $\widetilde{E} = \left(h^{2}+2P_{h^{\frac{1}{2}}v}\right)^{\frac{1}{2}}$ we can write  
\begin{align}
h^{-1/2} (\widetilde{E}  - h) h^{-1/2} = \frac{4}{\pi}\int_{0}^{\infty}\frac{t^{2}}{1+2\left\langle v,h\left(h^{2}+t^{2}\right)^{-1}v\right\rangle }P_{\left(h^{2}+t^{2}\right)^{-1}v}\,dt \le \frac{4}{\pi}\int_{0}^{\infty} P_{\left(h^{2}+t^{2}\right)^{-1}v} t^2\,dt 
\end{align}
Taking the trace and using \eqref{eq:integral-sqrtA} we complete the proof.
% is complete
%Consequently,
%\begin{align}
% {\rm tr} ( h^{-1/2} (\widetilde{E}  - h) h^{-1/2}) \le \frac{4}{\pi}\int_{0}^{\infty}  \langle v, \left(h^{2}+t^{2}\right)^{-2}v \rangle t^{2} dt  = \langle v, h^{-1}v\rangle.  
%\end{align}
$\hfill\square$

%\begin{align}
%delete
%\end{align}

\subsubsection*{Estimates for $U$}

Let us consider the unitary operator $U:V\rightarrow V$ defined by 
\begin{equation}
U=\left(h^{\frac{1}{2}}e^{-2K}h^{\frac{1}{2}}\right)^{\frac{1}{2}}h^{-\frac{1}{2}}e^{K}=\left(h^{2}+2P_{h^{\frac{1}{2}}v}\right)^{\frac{1}{4}}h^{-\frac{1}{2}}e^{K}. 
\end{equation}
%where we also used that from the definition of $K=-\frac{1}{2}\log\left(h^{-\frac{1}{2}}\left(h^{2}+2P_{h^{\frac{1}{2}}v}\right)^{\frac{1}{2}}h^{-\frac{1}{2}}\right)$
%\begin{equation}
%h^{\frac{1}{2}}e^{-2K}h^{\frac{1}{2}}=h^{\frac{1}{2}}h^{-\frac{1}{2}}\left(h^{2}+2P_{h^{\frac{1}{2}}v}\right)^{\frac{1}{2}}h^{-\frac{1}{2}}h^{\frac{1}{2}}=\left(h^{2}+2P_{h^{\frac{1}{2}}v}\right)^{\frac{1}{2}}.
%\end{equation}

First, the analysis of $\left(h^{2}+2P_{h^{\frac{1}{2}}v}\right)^{\frac{1}{2}}$ in Section \ref{sec:AnalysisoftheOne-BodyOperators} can be extended to $\left(h^{2}+2P_{h^{\frac{1}{2}}v}\right)^{\frac{1}{4}}$. We have 
%rather than $a^{\frac{1}{2}}=\frac{2}{\pi}\int_{0}^{\infty}\left(1-\frac{t^{2}}{a+t^{2}}\right)dt$.

\begin{prop} \label{lem:hP1/4}
For all $1\leq i,j\leq n$ it holds that
\[
\left|\left\langle x_{i},\left(\left(h^{2}+2P_{h^{\frac{1}{2}}v}\right)^{\frac{1}{4}}-h^{\frac{1}{2}}\right)x_{j}\right\rangle \right|\leq\frac{2\sqrt{\lambda_{i}\lambda_{j}}}{\sqrt{\lambda_{i}}+\sqrt{\lambda_{j}}}\frac{\left\langle x_{i},v\right\rangle \left\langle v,x_{j}\right\rangle }{\lambda_{i}+\lambda_{j}}.
\]
\end{prop}

Note that by using the integral identity
\begin{align}
A^{\frac{1}{4}}=\frac{2\sqrt{2}}{\pi}\int_{0}^{\infty}\left(1-\frac{t^{4}}{A+t^{4}}\right)dt
\end{align}
for every self-adjoint non-negative operator $A$ instead of \eqref{eq:integral-sqrtA}, we obtain the following analogue
of Proposition \ref{prop:IntegralFormulaForRankOnePerturbation}:
\begin{prop}
\label{prop:QuarticRootFormula}Let $\left(H,\left\langle \cdot,\cdot\right\rangle \right)$
be a Hilbert space and let $A:H\rightarrow H$ be a positive self-adjoint
operator. Then for any $x\in H$ and $g\in\mathbb{R}$ such that $A+gP_{x}>0$
it holds that
\[
\left(A+gP_{x}\right)^{\frac{1}{4}}=A^{\frac{1}{4}}+\frac{2\sqrt{2}g}{\pi}\int_{0}^{\infty}\frac{t^{4}}{1+g\left\langle v,\left(A+t^{4}\right)^{-1}v\right\rangle }P_{\left(A+t^{4}\right)^{-1}v}\,dt.
\]
\end{prop}

\textbf{Proof of Proposition \ref{lem:hP1/4}:} Applying Proposition \ref{prop:QuarticRootFormula}
with $A=h^{2}$, $x=h^{\frac{1}{2}}v$ and $g=2$ we find
\begin{align}
\left(h^{2}+2P_{h^{\frac{1}{2}}v}\right)^{\frac{1}{4}} & =\left(h^{2}\right)^{\frac{1}{4}}+\frac{4\sqrt{2}}{\pi}\int_{0}^{\infty}\frac{t^{4}}{1+2\left\langle h^{\frac{1}{2}}v,\left(h^{2}+t^{4}\right)^{-1}h^{\frac{1}{2}}v\right\rangle }P_{\left(h^{2}+t^{4}\right)^{-1}h^{\frac{1}{2}}v}\,dt\\
 & =h^{\frac{1}{2}}+\frac{4\sqrt{2}}{\pi}\int_{0}^{\infty}\frac{t^{4}}{1+2\left\langle v,h\left(h^{2}+t^{4}\right)^{-1}v\right\rangle }P_{h^{\frac{1}{2}}\left(h^{2}+t^{4}\right)^{-1}v}\,dt,\nonumber 
\end{align}
and so we can estimate that
\begin{align}
 0 & \le \left\langle x_{i},\left(\left(h^{2}+2P_{h^{\frac{1}{2}}v}\right)^{\frac{1}{4}}-h^{\frac{1}{2}}\right)x_{j}\right\rangle \nonumber \\
 & =\frac{4\sqrt{2}}{\pi}\int_{0}^{\infty}\frac{t^{4}}{1+2\left\langle v,h\left(h^{2}+t^{4}\right)^{-1}v\right\rangle }\left\langle x_{i},P_{h^{\frac{1}{2}}\left(h^{2}+t^{4}\right)^{-1}v}x_{j}\right\rangle dt\nonumber \\
 & =\frac{4\sqrt{2}}{\pi}\left\langle x_{i},v\right\rangle \left\langle v,x_{j}\right\rangle \int_{0}^{\infty}\frac{t^{4}}{1+2\left\langle v,h\left(h^{2}+t^{4}\right)^{-1}v\right\rangle }\frac{\sqrt{\lambda_{i}}}{\lambda_{i}^{2}+t^{4}}\frac{\sqrt{\lambda_{j}}}{\lambda_{j}^{2}+t^{4}}\,dt\label{eq:QuarticRootEquality}\\
 & \leq\frac{4\sqrt{2}}{\pi}\left\langle x_{i},v\right\rangle \left\langle v,x_{j}\right\rangle \int_{0}^{\infty}\frac{\sqrt{\lambda_{i}}}{\lambda_{i}^{2}+t^{4}}\frac{\sqrt{\lambda_{j}}}{\lambda_{j}^{2}+t^{4}}t^{4}\,dt=\frac{2\sqrt{\lambda_{i}\lambda_{j}}}{\sqrt{\lambda_{i}}+\sqrt{\lambda_{j}}}\frac{\left\langle x_{i},v\right\rangle \left\langle v,x_{j}\right\rangle }{\lambda_{i}+\lambda_{j}},\nonumber 
\end{align}
where we also applied the integral identity
\begin{equation}
\int_{0}^{\infty}\frac{\sqrt{a}}{a^{2}+t^{4}}\frac{\sqrt{b}}{b^{2}+t^{4}}t^{4}\,dt=\frac{\pi}{2\sqrt{2}}\frac{\sqrt{ab}}{\sqrt{a}+\sqrt{b}}\frac{1}{a+b},\quad a,b>0.
\end{equation}
%The equality of equation (\ref{eq:QuarticRootEquality}) also shows
%that $\left\langle x_{i},\left(\left(h^{2}+2P_{h^{\frac{1}{2}}v}\right)^{\frac{1}{4}}-h^{\frac{1}{2}}\right)x_{j}\right\rangle \geq0$
%and the claim follows.
$\hfill\square$

%We recall the estimates on $e^{-tK}-1$ and $1-e^{tK}$ of Proposition
%\ref{prop:SuperGeneralElementBounds}:
%\begin{prop}
%\label{prop:SuperGeneralElementBounds}For all $1\leq i,j\leq n$
%and $t\in\left[0,1\right]$ it holds that
%\[
%\left|\left\langle x_{i},\left(e^{-tK}-1\right)x_{j}\right\rangle \right|,\left|\left\langle x_{i},\left(1-e^{tK}\right)x_{j}\right\rangle \right|\leq\frac{\left\langle x_{i},v\right\rangle \left\langle v,x_{j}\right\rangle }{\lambda_{i}+\lambda_{j}}.
%\]
%\end{prop}

We may then conclude the following:
\begin{prop}
\label{prop:Uminus1MatrixElements}For all $1\leq i,j\leq n$ it holds
that
\[
\left|\left\langle x_{i},\left(U-1\right)x_{j}\right\rangle \right|,\,\left|\left\langle x_{i},\left(U^{\ast}-1\right)x_{j}\right\rangle \right|\leq3\left(1+\left\langle v,h^{-1}v\right\rangle \right)\frac{\left\langle x_{i},v\right\rangle \left\langle v,x_{j}\right\rangle }{\lambda_{i}+\lambda_{j}}.
\]
\end{prop}

\textbf{Proof:} As $\left|\left\langle x_{i},\left(U-1\right)x_{j}\right\rangle \right|=\left|\left\langle x_{j},\left(U^{\ast}-1\right)x_{i}\right\rangle \right|$
and the claimed estimate is symmetric with respect to $i$ and $j$
it suffices to consider $U-1$. We write  
\begin{align}
U-1 & =\left(h^{2}+2P_{h^{\frac{1}{2}}v}\right)^{\frac{1}{4}}h^{-\frac{1}{2}}e^{K}-1=\left(\left(h^{2}+2P_{h^{\frac{1}{2}}v}\right)^{\frac{1}{4}}-h^{\frac{1}{2}}\right)h^{-\frac{1}{2}}e^{K}+e^{K}-1\\
 & =e^{K}-1+\left(\left(h^{2}+2P_{h^{\frac{1}{2}}v}\right)^{\frac{1}{4}}-h^{\frac{1}{2}}\right)h^{-\frac{1}{2}}+\left(\left(h^{2}+2P_{h^{\frac{1}{2}}v}\right)^{\frac{1}{4}}-h^{\frac{1}{2}}\right)h^{-\frac{1}{2}}\left(e^{K}-1\right)\nonumber 
\end{align}
and estimate each term separately. The first is directly
covered by Proposition \ref{prop:SuperGeneralElementBounds},
with
\begin{equation}
\left|\left\langle x_{i},\left(e^{K}-1\right)x_{j}\right\rangle \right|\leq\frac{\left\langle x_{i},v\right\rangle \left\langle v,x_{j}\right\rangle }{\lambda_{i}+\lambda_{j}}.
\end{equation}
For the second term we can by Proposition \ref{lem:hP1/4}
estimate that
\begin{align}
\left|\left\langle x_{i},\left(\left(h^{2}+2P_{h^{\frac{1}{2}}v}\right)^{\frac{1}{4}}-h^{\frac{1}{2}}\right)h^{-\frac{1}{2}}x_{j}\right\rangle \right| & =\frac{1}{\sqrt{\lambda_{j}}}\left|\left\langle x_{i},\left(\left(h^{2}+2P_{h^{\frac{1}{2}}v}\right)^{\frac{1}{4}}-h^{\frac{1}{2}}\right)x_{j}\right\rangle \right|\\
 & \leq\frac{1}{\sqrt{\lambda_{j}}}\frac{2\sqrt{\lambda_{i}\lambda_{j}}}{\sqrt{\lambda_{i}}+\sqrt{\lambda_{j}}}\frac{\left\langle x_{i},v\right\rangle \left\langle v,x_{j}\right\rangle }{\lambda_{i}+\lambda_{j}}\leq2\frac{\left\langle x_{i},v\right\rangle \left\langle v,x_{j}\right\rangle }{\lambda_{i}+\lambda_{j}}.\nonumber 
\end{align}
For the final term we carry out an orthonormal expansion and apply
the previous two estimates to see that
\begin{align}
 & \quad\left|\left\langle x_{i},\left(\left(h^{2}+2P_{h^{\frac{1}{2}}v}\right)^{\frac{1}{4}}-h^{\frac{1}{2}}\right)h^{-\frac{1}{2}}\left(e^{K}-1\right)x_{j}\right\rangle \right|\nonumber \\
 & \leq\sum_{k=1}^{n}\left|\left\langle x_{i},\left(\left(h^{2}+2P_{h^{\frac{1}{2}}v}\right)^{\frac{1}{4}}-h^{\frac{1}{2}}\right)h^{-\frac{1}{2}}x_{k}\right\rangle \right|\left|\left\langle x_{k},\left(e^{K}-1\right)x_{j}\right\rangle \right|\\
 & \leq2\sum_{k=1}^{n}\frac{\left\langle x_{i},v\right\rangle \left\langle v,x_{k}\right\rangle }{\lambda_{i}+\lambda_{k}}\frac{\left\langle x_{k},v\right\rangle \left\langle v,x_{j}\right\rangle }{\lambda_{k}+\lambda_{j}}=2\frac{\left\langle x_{i},v\right\rangle \left\langle v,x_{j}\right\rangle }{\lambda_{i}+\lambda_{j}}\sum_{k=1}^{n}\frac{\lambda_{i}+\lambda_{j}}{\left(\lambda_{i}+\lambda_{k}\right)\left(\lambda_{k}+\lambda_{j}\right)}\left|\left\langle x_{k},v\right\rangle \right|^{2}\nonumber \\
 & \leq2\frac{\left\langle x_{i},v\right\rangle \left\langle v,x_{j}\right\rangle }{\lambda_{i}+\lambda_{j}}\sum_{k=1}^{n}\frac{\left|\left\langle x_{k},v\right\rangle \right|^{2}}{\lambda_{k}}=2\left\langle v,h^{-1}v\right\rangle \frac{\left\langle x_{i},v\right\rangle \left\langle v,x_{j}\right\rangle }{\lambda_{i}+\lambda_{j}}\nonumber 
\end{align}
where we also applied the elementary inequality 
\begin{equation}
\frac{a+b}{\left(a+c\right)\left(c+b\right)}=\frac{a+b}{c(a+b)+ ab + c^2} <  \frac{1}{c},\quad \forall a,b,c>0.
\end{equation}
%$a,b,c>0$, which follows by verifying that $\left(a+b\right)c<\left(a+c\right)\left(c+b\right)$.
Combining the estimates now yields the claim.
$\hfill\square$

\subsubsection*{Estimates for $J$}

Recall that we defined $J:V\rightarrow V$ to be the principal logarithm of $U$.  Since $U$ is a unitary operator on the finite-dimensional complex Hilbert space
$V$,  by the spectral theorem it is diagonalizable, i.e. there
exists an orthonormal basis $\left(w_{j}\right)_{j=1}^{n}$ for $V$
of eigenstates of $U$ with eigenvalues $\left(e^{i\theta_{j}}\right)_{j=1}^{n}$, $\left(\theta_{j}\right)_{j=1}^{n}\subset\left(-\pi,\pi\right]$, 
%
%. By unitarity it also follows that each eigenvalue
%has unit modulus, and so there exists unique real numbers $\left(\theta_{j}\right)_{j=1}^{n}\subset\left(-\pi,\pi\right]$
%such that the eigenvalues of $U$ can be expressed as $\left(e^{i\theta_{j}}\right)_{j=1}^{n}$,
i.e. $Uw_{j}=e^{i\theta_{j}}w_{j}$ for all $1\leq j\leq n$. Thus $J$ can be explicitly written as 
\begin{equation}
Jw_{j}=i\theta_{j}w_{j},\quad1\leq j\leq n.
\end{equation}
%such that $e^{J}=U$.

To estimate the quantity $\left\Vert h^{-\frac{1}{2}}J\right\Vert _{\text{HS}}$
we will apply the following:
\begin{prop}
It holds that
\[
JJ^{\ast}\leq\frac{\pi^{2}}{4}\left(U-1\right)^{\ast}\left(U-1\right).
\]
\end{prop}

\textbf{Proof:} We note the elementary inequality
\begin{equation}
\left|x\right|\leq\frac{\pi}{2}\sqrt{2\left(1-\cos\left(x\right)\right)}=\frac{\pi}{2}\left|e^{ix}-1\right|,\quad x\in\left[-\pi,\pi\right],
\end{equation}
which can be deduced from the fact that $x\mapsto\left|e^{ix}-1\right|$
is an even function and concave on $x\in\left[0,\pi\right]$. As the
eigenbasis $\left(w_{j}\right)_{j=1}^{n}$ obeys
\begin{equation}
Uw_{j}=e^{i\theta_{j}}w_{j},\quad U^{\ast}w_{j}=e^{-i\theta_{j}}w_{j},\quad Jw_{j}=i\theta_{j}w_{j},\quad J^{\ast}w_{j}=-i\theta_{j}w_{j},
\end{equation}
we can for any $w\in V$ perform an orthonormal expansion in terms
of $\left(w_{j}\right)_{j=1}^{n}$ to see that
\begin{align}
\left\langle w,JJ^{\ast}w\right\rangle  & =\left\Vert J^{\ast}w\right\Vert ^{2}=\sum_{j=1}^{n}\left|\theta_{j}\right|^{2}\left|\left\langle w_{j},w\right\rangle \right|^{2}\leq\sum_{j=1}^{n}\left(\frac{\pi}{2}\left|e^{i\theta_{j}}-1\right|\right)^{2}\left|\left\langle w_{j},w\right\rangle \right|^{2}\\
 & =\frac{\pi^{2}}{4}\sum_{j=1}^{n}\left|\left\langle \left(U^{\ast}-1\right)w_{j},w\right\rangle \right|^{2}=\frac{\pi^{2}}{4}\left\Vert \left(U-1\right)w\right\Vert ^{2}=\frac{\pi^{2}}{4}\left\langle w,\left(U-1\right)^{\ast}\left(U-1\right)w\right\rangle \nonumber 
\end{align}
which is the claim. $\hfill\square$

\begin{cor}
\label{coro:hJHSEstimate}  There exists a universal constant $C>0$ such that
\[
\left\Vert h^{-\frac{1}{2}}J\right\Vert _{\HS}\leq C\left(1+\left\langle v,h^{-1}v\right\rangle \right)\left\langle v,h^{-\frac{3}{2}}v\right\rangle .
\]
%for a constant $C>0$ independent of all quantities.
\end{cor}

\textbf{Proof:} By cyclicity of the trace and the estimate of the
previous proposition we have that
\begin{align}
\left\Vert h^{-\frac{1}{2}}J\right\Vert _{\text{HS}}^{2} & =\text{tr}\left(J^{\ast}h^{-1}J\right)=\text{tr}\left(h^{-\frac{1}{2}}JJ^{\ast}h^{-\frac{1}{2}}\right)\\
 & \leq\frac{\pi^{2}}{4}\text{tr}\left(h^{-\frac{1}{2}}\left(U-1\right)^{\ast}\left(U-1\right)h^{-\frac{1}{2}}\right)=\frac{\pi^{2}}{4}\left\Vert \left(U-1\right)h^{-\frac{1}{2}}\right\Vert _{\text{HS}}^{2},\nonumber 
\end{align}
and by the matrix element estimate of Proposition \ref{prop:Uminus1MatrixElements}, 
\begin{align}
 & \quad\;\left\Vert \left(U-1\right)h^{-\frac{1}{2}}\right\Vert _{\text{HS}}^{2}=\sum_{i,j=1}^{n}\left|\left\langle x_{i},\left(U-1\right)h^{-\frac{1}{2}}x_{j}\right\rangle \right|^{2}=\sum_{i,j=1}^{n}\frac{1}{\lambda_{j}}\left|\left\langle x_{i},\left(U-1\right)x_{j}\right\rangle \right|^{2}\\
 & \leq C\left(1+\left\langle v,h^{-1}v\right\rangle \right)^{2}\sum_{i,j=1}^{n}\frac{1}{\lambda_{j}}\left|\frac{\left\langle x_{i},v\right\rangle \left\langle v,x_{j}\right\rangle }{\lambda_{i}+\lambda_{j}}\right|^{2}\leq C\left(1+\left\langle v,h^{-1}v\right\rangle \right)^{2}\sum_{i,j=1}^{n}\frac{1}{\lambda_{j}}\left|\frac{\left\langle x_{i},v\right\rangle \left\langle v,x_{j}\right\rangle }{\lambda_{i}^{\frac{3}{4}}\lambda_{j}^{\frac{1}{4}}}\right|^{2}\nonumber \\
 & =C\left(1+\left\langle v,h^{-1}v\right\rangle \right)^{2}\left(\sum_{i=1}^{n}\frac{\left|\left\langle x_{i},v\right\rangle \right|^{2}}{\lambda_{j}^{\frac{3}{2}}}\right)^{2}=C\left(1+\left\langle v,h^{-1}v\right\rangle \right)^{2}\left\langle v,h^{-\frac{3}{2}}v\right\rangle ^{2}\nonumber 
\end{align}
which gives the claim.
$\hfill\square$

\medskip

Next, consider $\left\Vert h^{-\frac{1}{2}}\left[J,h\right]h^{-\frac{1}{2}}\right\Vert _{\text{HS}}$. By the triangle inequality, it suffices to bound $\left\Vert h^{-\frac{1}{2}}Jh^{\frac{1}{2}}\right\Vert _{\text{HS}}$. Unlike $\left\Vert h^{-\frac{1}{2}}J\right\Vert _{\text{HS}}$ this is more involved as the presence of factors of $h$ on both sides
of $J$ prevents us from combining $J$ and $J^{\ast}$ in $\left\Vert h^{-\frac{1}{2}}Jh^{\frac{1}{2}}\right\Vert _{\text{HS}}^{2}=\text{tr}\left(J^{\ast}h^{-1}Jh\right)$,
and so we need to proceed differently. First we note the following
elementary estimate:
\begin{lem}
\label{lemma:SelfEstimateforJ}There exists a constant $C>0$ such
that
\[
\left|i\theta-\frac{1}{2}\left(e^{i\theta}-e^{-i\theta}\right)\right|\leq C\left|e^{i\theta}-1\right|^{3},\quad\theta\in\left[-\pi,\pi\right].
\]
\end{lem}

{\bf Proof:} The left-hand side is  $\left|\theta-\sin\left(\theta\right)\right|=O\left(\left|\theta\right|^3\right)$, while $ |\theta| \ge \left|e^{i\theta}-1\right| \ge C^{-1}\theta$. $\hfill\square$
%,
%so the claim follows by compactness of $\left[-\pi,\pi\right]$.)

%We can apply this to estimate $\left\Vert h^{-\frac{1}{2}}\left[J,h\right]h^{-\frac{1}{2}}\right\Vert _{\text{HS}}$
%as follows:
\begin{prop}
\label{prop:hJhhHSEstimate} There exists a universal constant $C>0$ such that
\[
\left\Vert h^{-\frac{1}{2}}\left[J,h\right]h^{-\frac{1}{2}}\right\Vert _{\HS}\leq C\left(1+\left\langle v,h^{-1}v\right\rangle \right)^{3}\left(\left\langle v,h^{-1}v\right\rangle +\left\langle v,h^{-\frac{1}{2}}v\right\rangle \left\langle v,h^{-\frac{5}{4}}h\right\rangle ^{2}\right).
\]
\end{prop}

\textbf{Proof:} It suffices to bound  $\left\Vert h^{-\frac{1}{2}}Jh^{\frac{1}{2}}\right\Vert _{\text{HS}}$. By writing
\begin{equation} \label{eq:expansion-J}
J=\frac{1}{2}\left(U-1\right)+\frac{1}{2}\left(1-U^{\ast}\right) + \widetilde{J}, \quad \widetilde{J} = J-\frac{1}{2}\left(U-U^{\ast}\right) 
\end{equation}
we see by the triangle inequality that
\begin{equation}
\left\Vert h^{-\frac{1}{2}}Jh^{\frac{1}{2}}\right\Vert _{\text{HS}}\leq\frac{1}{2}\left\Vert h^{-\frac{1}{2}}\left(U-1\right)h^{\frac{1}{2}}\right\Vert _{\text{HS}}+\frac{1}{2}\left\Vert h^{-\frac{1}{2}}\left(1-U^{\ast}\right)h^{\frac{1}{2}}\right\Vert _{\text{HS}}+\left\Vert h^{-\frac{1}{2}} \widetilde{J} h^{\frac{1}{2}}\right\Vert _{\text{HS}}.
\end{equation}
By Proposition \ref{prop:Uminus1MatrixElements}, we have
\begin{align}
 & \quad\;\left\Vert h^{-\frac{1}{2}}\left(U-1\right)h^{\frac{1}{2}}\right\Vert _{\text{HS}}^{2}=\sum_{i,j=1}^{n}\left|\left\langle x_{i},h^{-\frac{1}{2}}\left(U-1\right)h^{\frac{1}{2}}x_{j}\right\rangle \right|^{2}=\sum_{i,j=1}^{n}\frac{\lambda_{j}}{\lambda_{i}}\left|\left\langle x_{i},\left(U-1\right)x_{j}\right\rangle \right|^{2}\nonumber \\
 & \leq C\left(1+\left\langle v,h^{-1}v\right\rangle \right)^{2}\sum_{i,j=1}^{n}\frac{\lambda_{j}}{\lambda_{i}}\left|\frac{\left\langle x_{i},v\right\rangle \left\langle v,x_{j}\right\rangle }{\lambda_{i}+\lambda_{j}}\right|^{2}\leq C\left(1+\left\langle v,h^{-1}v\right\rangle \right)^{2}\left(\sum_{i=1}^{n}\frac{\left|\left\langle x_{i},v\right\rangle \right|^{2}}{\lambda_{i}}\right)^{2}\\
 & =C\left(1+\left\langle v,h^{-1}v\right\rangle \right)^{2}\left\langle v,h^{-1}v\right\rangle ^{2}\nonumber 
\end{align}
and likewise for $\left\Vert h^{-\frac{1}{2}}\left(1-U^{\ast}\right)h^{\frac{1}{2}}\right\Vert _{\text{HS}}^{2}$. For $h^{-\frac{1}{2}}\widetilde{J} h^{\frac{1}{2}}$
we instead apply Lemma \ref{lemma:SelfEstimateforJ} and the Cauchy-Schwarz
inequality to see that for any $1\leq i,j\leq n$
\begin{align}
 & \quad\,\left|\left\langle x_{i},h^{-\frac{1}{2}} \widetilde{J} h^{\frac{1}{2}}x_{j}\right\rangle \right|^2\nonumber \\
 & =\left|\sum_{k=1}^{n}\left(i\theta_{k}-\frac{1}{2}\left(e^{i\theta_{k}}-e^{-i\theta_{k}}\right)\right)\left\langle h^{-\frac{1}{2}}x_{i},w_{k}\right\rangle \left\langle w_{k},h^{\frac{1}{2}}x_{j}\right\rangle \right|^2 \nonumber \\
 & \leq C \left( \sum_{k=1}^{n}\left|e^{i\theta_{k}}-1\right|^{3}\left|\left\langle h^{-\frac{1}{2}}x_{i},w_{k}\right\rangle \right|\left|\left\langle w_{k},h^{\frac{1}{2}}x_{j}\right\rangle \right| \right)^2\\
 & \leq C \left(\sum_{k=1}^{n}\left|e^{i\theta_{k}}-1\right|^{4}\left|\left\langle w_{k},h^{-\frac{1}{2}}x_{i}\right\rangle \right|^{2}\right) \left( \sum_{k=1}^{n}\left|e^{i\theta_{k}}-1\right|^{2}\left|\left\langle w_{k},h^{\frac{1}{2}}x_{j}\right\rangle \right|^{2} \right)\nonumber \\
 & =C \left( \sum_{k=1}^{n}\left|\left\langle \left(U^{\ast}-1\right)^{2}w_{k},h^{-\frac{1}{2}}x_{i}\right\rangle \right|^{2} \right) \left( \sum_{k=1}^{n}\left|\left\langle \left(U^{\ast}-1\right)w_{k},h^{\frac{1}{2}}x_{j}\right\rangle \right|^{2} \right) \nonumber \\
 & =C\left\Vert \left(U-1\right)^{2}h^{-\frac{1}{2}}x_{i}\right\Vert^2 \left\Vert \left(U-1\right)h^{\frac{1}{2}}x_{j}\right\Vert^2.\nonumber 
\end{align}
Summing over $i,j$ we obtain 
\begin{align} \label{eq:tildeJ-HS}
\left\Vert h^{-\frac{1}{2}} \widetilde{J} h^{\frac{1}{2}}\right\Vert _{\text{HS}}^{2} \le C\left\Vert \left(U-1\right)^{2}h^{-\frac{1}{2}}\right\Vert _{\text{HS}}^{2}\left\Vert \left(U-1\right)h^{\frac{1}{2}}\right\Vert _{\text{HS}}^{2}. 
\end{align}
We can now again apply Proposition \ref{prop:Uminus1MatrixElements}
to estimate that
\begin{align}
&  \left\Vert \left(U-1\right)h^{\frac{1}{2}}\right\Vert _{\text{HS}}^{2} =\sum_{i,j=1}^{n}\left|\left\langle x_{i},\left(U-1\right)h^{\frac{1}{2}}x_{j}\right\rangle \right|^{2}=\sum_{i,j=1}^{n}\lambda_{j}\left|\left\langle x_{i},\left(U-1\right)x_{j}\right\rangle \right|^{2}\nonumber \\
 & \leq C\left(1+\left\langle v,h^{-1}v\right\rangle \right)^{2}\sum_{i,j=1}^{n}\lambda_{j}\left|\frac{\left\langle x_{i},v\right\rangle \left\langle v,x_{j}\right\rangle }{\lambda_{i}+\lambda_{j}}\right|^{2}\label{eq:U-1hEstimate}\\
 & \leq C\left(1+\left\langle v,h^{-1}v\right\rangle \right)^{2}\sum_{i,j=1}^{n}\lambda_{j}\left|\frac{\left\langle x_{i},v\right\rangle \left\langle v,x_{j}\right\rangle }{\lambda_{i}^{\frac{1}{4}}\lambda_{j}^{\frac{3}{4}}}\right|^{2}=C\left(1+\left\langle v,h^{-1}v\right\rangle \right)^{2}\left\langle v,h^{-\frac{1}{2}}v\right\rangle ^{2}\nonumber 
\end{align}
and
\begin{align}
 & \quad\;\left\Vert \left(U-1\right)^{2}h^{-\frac{1}{2}}\right\Vert _{\text{HS}}^{2}=\sum_{i,j=1}^{n}\left|\left\langle x_{i},\left(U-1\right)^{2}h^{-\frac{1}{2}}x_{j}\right\rangle \right|^{2}\nonumber \\
 & =\sum_{i,j=1}^{n}\frac{1}{\lambda_{j}}\left|\sum_{k=1}^{n}\left\langle x_{i},\left(U-1\right)x_{k}\right\rangle \left\langle x_{k},\left(U-1\right)x_{j}\right\rangle \right|^{2}\nonumber \\
 & \leq C\left(1+\left\langle v,h^{-1}v\right\rangle \right)^{4}\sum_{i,j=1}^{n}\frac{1}{\lambda_{j}}\left(\sum_{k=1}^{n}\frac{\left\langle x_{i},v\right\rangle \left\langle v,x_{k}\right\rangle }{\lambda_{i}+\lambda_{k}}\frac{\left\langle x_{k},v\right\rangle \left\langle v,x_{j}\right\rangle }{\lambda_{k}+\lambda_{j}}\right)^{2}\label{eq:U-1sqhEstimate}\\
 & \leq C\left(1+\left\langle v,h^{-1}v\right\rangle \right)^{4}\sum_{i,j=1}^{n}\left|\left\langle x_{i},v\right\rangle \right|^{2}\frac{\left|\left\langle x_{j},v\right\rangle \right|^{2}}{\lambda_{j}}\left(\sum_{k=1}^{n}\frac{\left\langle v,x_{k}\right\rangle }{\lambda_{i}^{\frac{5}{8}}\lambda_{k}^{\frac{3}{8}}}\frac{\left\langle x_{k},v\right\rangle }{\lambda_{k}^{\frac{7}{8}}\lambda_{j}^{\frac{1}{8}}}\right)^{2}\nonumber \\
 & =C\left(1+\left\langle v,h^{-1}v\right\rangle \right)^{4}\left(\sum_{i=1}^{n}\frac{\left|\left\langle x_{i},v\right\rangle \right|^{2}}{\lambda_{i}^{\frac{5}{4}}}\right)^{4}=C\left(1+\left\langle v,h^{-1}v\right\rangle \right)^{4}\left\langle v,h^{-\frac{5}{4}}v\right\rangle ^{4}\nonumber 
\end{align}
so
\begin{equation}
\left\Vert h^{-\frac{1}{2}} \widetilde{J} h^{\frac{1}{2}}\right\Vert _{\text{HS}}\leq\left(1+\left\langle v,h^{-1}v\right\rangle \right)^{3}\left\langle v,h^{-\frac{1}{2}}v\right\rangle \left\langle v,h^{-\frac{5}{4}}v\right\rangle ^{2}.
\end{equation}
Combining the estimates yields the claim.
$\hfill\square$

\begin{rmk}[Remarks on the estimation technique]

As we will use the same approach to obtain estimates on $E\left(t\right)$,
let us consider the technique of the proof in detail. The idea is
that, as we have a good estimate for the matrix elements of $U-1$
and $U^{\ast}-1$, we should attempt to express our operator solely
in terms of these. The first step is therefore to decompose $J$ as in \eqref{eq:expansion-J}. The error term $\widetilde{J}=J-\frac{1}{2}\left(U-U^{\ast}\right) $ cannot be simplified further in terms of $U$ but
by orthonormal expansion and Lemma \ref{lemma:SelfEstimateforJ} we
can nonetheless estimate it solely in terms of $U-1$, despite being
unable to apply an operator inequality, as we did for $\| h^{-\frac{1}{2}}J\|_{\text{HS}}$,
to ``substitute'' $U-1$ for $J$ directly. The utility of the estimate \eqref{eq:tildeJ-HS} 
%\begin{equation}
%\left\Vert h^{-\frac{1}{2}}\left(J-\frac{1}{2}\left(U-U^{\ast}\right)\right)h^{\frac{1}{2}}\right\Vert _{\text{HS}}\leq C\left\Vert \left(U-1\right)^{2}h^{-\frac{1}{2}}\right\Vert _{\text{HS}}\left\Vert \left(U-1\right)h^{\frac{1}{2}}\right\Vert _{\text{HS}}
%\end{equation}
is thus that it allows us to replace the unknown error operator with
factors of $U-1$, which we can estimate well. The downside to this
is that it simultanously ``decouples'' the $h^{-\frac{1}{2}}$ and
$h^{\frac{1}{2}}$ factors, which prevents us from exploiting the
cancellation between these.

This decoupling is also the reason why it is important that in \eqref{eq:tildeJ-HS}  we distribute
two factors of $U-1$ to $h^{-\frac{1}{2}}$ rather than only one:
One can by the same argument estimate that
\begin{align} \label{eq:tildeJ-HS-bad}
\left\Vert h^{-\frac{1}{2}} \widetilde{J} h^{\frac{1}{2}}\right\Vert _{\text{HS}} & \leq C\left\Vert \left(U-1\right)h^{-\frac{1}{2}}\right\Vert _{\text{HS}}\left\Vert \left(U-1\right)^{2}h^{\frac{1}{2}}\right\Vert _{\text{HS}}\\
 & \leq C\left(1+\left\langle v,h^{-3}v\right\rangle \right)^{3}\left\langle v,h^{-\frac{3}{4}}v\right\rangle ^{2}\left\langle v,h^{-\frac{3}{2}}v\right\rangle \nonumber 
\end{align}
but in Proposition \ref{prop:MoreSingularRiemannSums} we only have the good estimates $\left\langle v_{k},h_{k}^{\alpha}v_{k}\right\rangle \sim Ck_{F}^{1+\alpha}$
for $\alpha>-\frac{4}{3}$, which makes \eqref{eq:tildeJ-HS-bad} a worse estimate due
to the $\left\langle v,h^{-\frac{3}{2}}v\right\rangle $ factor. There
is therefore a limit to how low the exponent $\alpha$ can be without
affecting our estimates, and so it is advantageous to distribute the
factors of $U-1$ such that the overall minimal exponent is not too
small.
\end{rmk}

\subsubsection*{Estimation of $E\left(t\right)$}

We now estimate $\max_{j}\left\Vert h^{-\frac{1}{2}}E\left(t\right)x_{j}\right\Vert $
using the technique outlined above. First we decompose 
\begin{equation}
E\left(t\right) =e^{tJ}e^{-K}he^{-K}e^{-tJ}-h=\left(e^{tJ}he^{-tJ}-h\right)+e^{tJ}\left(e^{-K}he^{-K}-h\right)e^{-tJ}=:E_{1}\left(t\right)+E_{2}\left(t\right)
\end{equation}
and using the algebraic identity
\begin{equation}
ABC=B+\left(A-1\right)B+B\left(C-1\right)+\left(A-1\right)B\left(C-1\right)\label{eq:ABCIdentity}
\end{equation}
with $A=e^{tJ}$, $B=h$ and $C=e^{-tJ}$ further decompose $E_{1}\left(t\right)$
as
\begin{align}
E_{1}\left(t\right)=e^{tJ}he^{-tJ}-h & =\left(\left(e^{tJ}-1\right)h+h\left(e^{-tJ}-1\right)\right)+\left(e^{tJ}-1\right)h\left(e^{-tJ}-1\right)\\
 & =:E_{1,1}\left(t\right)+E_{1,2}\left(t\right).\nonumber 
\end{align}
Defining $E_{0}=E\left(0\right)=e^{-K}he^{-K}-h$ we likewise decompose
$E_{2}\left(t\right)$ according to
\begin{align}
E_{2}\left(t\right)=e^{tJ}E_{0}e^{-tJ} & =E_{0}+\left(\left(e^{tJ}-1\right)E_{0}+E_{0}\left(e^{-tJ}-1\right)\right)+\left(e^{tJ}-1\right)E_{0}\left(e^{-tJ}-1\right)\\
 & =:E_{0}+E_{2,1}\left(t\right)+E_{2,2}\left(t\right).\nonumber 
\end{align}
The $E_{1,1}\left(t\right)$, $E_{1,2}\left(t\right)$ and $E_{2,1}\left(t\right)$,
$E_{2,2}\left(t\right)$ terms differ only in replacing the operator
$h$ by $E_{0}$. We can therefore estimate these terms similarly,
provided we have an estimate on $E_{0}$. This is given by the following:

\begin{prop}
\label{prop:E0MatrixElementEstimate}For all $1\leq i,j\leq n$ it
holds that
\[
\left|\left\langle x_{i},E_{0}x_{j}\right\rangle \right|=\left|\left\langle x_{i},\left(e^{-K}he^{-K}-h\right)x_{j}\right\rangle \right|\leq\left(1+\left\langle v,h^{-1}v\right\rangle \right)\left\langle x_{i},v\right\rangle \left\langle v,x_{j}\right\rangle .
\]
Consequently, \[
\max_{1\leq j\leq n}\left\Vert h^{-\frac{1}{2}}E_{0}x_{j}\right\Vert \leq\alpha\left(1+\left\langle v,h^{-1}v\right\rangle \right)\sqrt{\left\langle v,h^{-1}v\right\rangle }
\]
where $\alpha=\max_{1\leq j\leq n}\left\langle v,x_{j}\right\rangle $.
\end{prop}

\textbf{Proof:} Using the identity of equation (\ref{eq:ABCIdentity})
with $A=e^{-K}=C$ and $B=h$ we have that
\begin{equation}
e^{-K}he^{-K}-h=\left\{ h,e^{-K}-1\right\} +\left(e^{-K}-1\right)h\left(e^{-K}-1\right)
\end{equation}
hence
\begin{equation}
\left\langle x_{i},\left(e^{-K}he^{-K}-h\right)x_{j}\right\rangle =\left(\lambda_{i}+\lambda_{j}\right)\left\langle x_{i},\left(e^{-K}-1\right)x_{j}\right\rangle +\left\langle x_{i},\left(e^{-K}-1\right)h\left(e^{-K}-1\right)x_{j}\right\rangle .
\end{equation}
We can apply Proposition \ref{prop:SuperGeneralElementBounds}
to estimate the first term of this equation as
\begin{equation}
\left|\left(\lambda_{i}+\lambda_{j}\right)\left\langle x_{i},\left(e^{-K}-1\right)x_{j}\right\rangle \right|\leq\left(\lambda_{i}+\lambda_{j}\right)\frac{\left\langle x_{i},v\right\rangle \left\langle v,x_{j}\right\rangle }{\lambda_{i}+\lambda_{j}}=\left\langle x_{i},v\right\rangle \left\langle v,x_{j}\right\rangle 
\end{equation}
and the second term as
\begin{align}
\left|\left\langle x_{i},\left(e^{-K}-1\right)h\left(e^{-K}-1\right)x_{j}\right\rangle \right| & =\left|\sum_{k=1}^{n}\lambda_{k}\left\langle x_{i},\left(e^{-K}-1\right)x_{k}\right\rangle \left\langle x_{k},\left(e^{-K}-1\right)x_{j}\right\rangle \right|\nonumber \\
 & \leq\sum_{k=1}^{n}\lambda_{k}\frac{\left\langle x_{i},v\right\rangle \left\langle v,x_{k}\right\rangle }{\lambda_{i}+\lambda_{k}}\frac{\left\langle x_{k},v\right\rangle \left\langle v,x_{j}\right\rangle }{\lambda_{k}+\lambda_{j}}\\
 & \leq\left\langle x_{i},v\right\rangle \left\langle v,x_{j}\right\rangle \sum_{k=1}^{n}\frac{\left|\left\langle x_{k},v\right\rangle \right|^{2}}{\lambda_{k}}=\left\langle v,h^{-1}v\right\rangle \left\langle x_{i},v\right\rangle \left\langle v,x_{j}\right\rangle \nonumber 
\end{align}
which implies the first claim. Consequently, 
\begin{align}
 & \left\Vert h^{-\frac{1}{2}}E_{0}x_{j}\right\Vert ^{2} =\sum_{i=1}^{n}\left|\left\langle x_{i},h^{-\frac{1}{2}}\left(e^{-K}he^{-K}-h\right)x_{j}\right\rangle \right|^{2}=\sum_{i=1}^{n}\frac{1}{\lambda_{i}}\left|\left\langle x_{i},\left(e^{-K}he^{-K}-h\right)x_{j}\right\rangle \right|^{2}\nonumber \\
 & \leq\left(1+\left\langle v,h^{-1}v\right\rangle \right)^{2}\sum_{i=1}^{n}\frac{1}{\lambda_{i}}\left|\left\langle x_{i},v\right\rangle \left\langle v,x_{j}\right\rangle \right|^{2} \leq\alpha^{2}\left(1+\left\langle v,h^{-1}v\right\rangle \right)^{2}\left\langle v,h^{-1}v\right\rangle .
\end{align}
$\hfill\square$

Now it remains to consider the operators $e^{tJ}-1$ and $e^{-tJ}-1=\left(e^{tJ}-1\right)^{\ast}$. To implement the above estimation technique,
% we must first determine which known operator we will approximate
%these by, and as was the case for Lemma \ref{lemma:SelfEstimateforJ}
%we will need the error term to be cubic with respect to $U-1$. Noting
%the inequality
from the following analogue of Lemma \ref{lemma:SelfEstimateforJ}
\begin{equation}
\left|\left(e^{it\theta}-1\right)-t\left(e^{i\theta}-1\right)+\frac{t\left(1-t\right)}{2}\left(e^{i\theta}+e^{-i\theta}-2\right)\right|\leq C\left|e^{i\theta}-1\right|^{3},\quad t\in\left[0,1\right],\,\theta\in\left[-\pi,\pi\right],\label{eq:FtNumericalEstimate}
\end{equation}
we are motivated in approximating $e^{tJ}-1$ by 
\begin{equation}
F_{t}=t\left(U-1\right)-\frac{t\left(1-t\right)}{2}\left(U+U^{\ast}-2\right),\quad t\in\left[0,1\right],
\end{equation}
with the error term being cubic with respect to $U-1$. We then have the following bounds for $F_{t}$ and the associated error terms:
\begin{prop}
\label{prop:EstimationTechnique}For any $T:V\rightarrow V$, $x\in V$,
$m\in\left\{ 1,2\right\} $ and $t\in\left[0,1\right]$ it holds that
\[
\left\Vert T\left(e^{tJ}-1-F_{t}\right)x\right\Vert ,\,\left\Vert T\left(e^{-tJ}-1-F_{t}^{\ast}\right)x\right\Vert \leq C\left\Vert T\left(U-1\right)^{m}\right\Vert _{\HS}\left\Vert \left(U-1\right)^{3-m}x\right\Vert 
\]
and for all $1\leq i,j\leq n$, $t\in\left[0,1\right]$,
\[
\left|\left\langle x_{i},F_{t}x_{j}\right\rangle \right|,\,\left|\left\langle x_{i},F_{t}^{\ast}x_{j}\right\rangle \right|\leq C\left(1+\left\langle v,h^{-1}v\right\rangle \right)\frac{\left\langle x_{i},v\right\rangle \left\langle v,x_{j}\right\rangle }{\lambda_{i}+\lambda_{j}}
\]
for a constant $C>0$ independent of all quantities.
\end{prop}

\textbf{Proof:} Recall that  $\left(w_{j}\right)_{j=1}^{n}$ is an orthonormal eigenbasis of $J$, namely $e^{tJ}w_{j}=e^{it\theta_{j}} w_j$ for all $1\leq j\leq n$. Using 
 (\ref{eq:FtNumericalEstimate}) and the
Cauchy-Schwarz inequality we have that
\begin{align}
 & \quad\,\left\Vert T\left(e^{tJ}-1-F_{t}\right)x\right\Vert ^{2}=\sum_{j=1}^{n}\left|\left\langle w_{j},T\left(e^{tJ}-1-F_{t}\right)x\right\rangle \right|^{2}\nonumber \\
 & =\sum_{j=1}^{n}\left|\sum_{k=1}^{n}\left(\left(e^{it\theta_{k}}-1\right)-t\left(e^{i\theta_{k}}-1\right)+\frac{t\left(1-t\right)}{2}\left(e^{i\theta_{k}}+e^{-i\theta_{k}}-2\right)\right)\left\langle w_{j},Tw_{k}\right\rangle \left\langle w_{k},x\right\rangle \right|^{2}\nonumber \\
 & \leq C\sum_{j=1}^{n}\left(\sum_{k=1}^{n}\left|e^{i\theta_{k}}-1\right|^{3}\left|\left\langle w_{j},Tw_{k}\right\rangle \right|\left|\left\langle w_{k},x\right\rangle \right|\right)^{2}\\
 & \leq C\sum_{j=1}^{n}\left(\sum_{k=1}^{n}\left|e^{i\theta_{k}}-1\right|^{2m}\left|\left\langle w_{j},Tw_{k}\right\rangle \right|^{2}\right)\left(\sum_{k=1}^{n}\left|e^{i\theta_{k}}-1\right|^{2\left(3-m\right)}\left|\left\langle w_{k},x\right\rangle \right|^{2}\right)\nonumber \\
 & =C\left(\sum_{j,k=1}^{n}\left|\left\langle w_{j},T\left(U-1\right)^{m}w_{k}\right\rangle \right|^{2}\right)\left(\sum_{k=1}^{n}\left|\left\langle \left(U^{\ast}-1\right)^{3-m}w_{k},x\right\rangle \right|^{2}\right)\nonumber \\
 & =C\left\Vert T\left(U-1\right)^{m}\right\Vert _{\text{HS}}^{2}\left\Vert \left(U-1\right)^{3-m}x\right\Vert ^{2},\nonumber 
\end{align}
the same estimate holding also for $\left\Vert T\left(e^{-tJ}-1-F_{t}^{\ast}\right)x\right\Vert $.
For the matrix element estimate of $F_{t}$ we have by Proposition
\ref{prop:Uminus1MatrixElements} that
\begin{align}
\left|\left\langle x_{i},F_{t}x_{j}\right\rangle \right| %& =\left|\left\langle x_{i},t\left(U-1\right)-\frac{t\left(1-t\right)}{2}\left(U+U^{\ast}-2\right)x_{j}\right\rangle \right|\nonumber \\
 & =\left|\left\langle x_{i},\left(\frac{t\left(1+t\right)}{2}\left(U-1\right)-\frac{t\left(1-t\right)}{2}\left(U^{\ast}-1\right)\right)x_{j}\right\rangle \right|\nonumber \\
 & \leq\frac{t\left(1+t\right)}{2}\left|\left\langle x_{i},\left(U-1\right)x_{j}\right\rangle \right|+\frac{t\left(1-t\right)}{2}\left|\left\langle x_{i},\left(U^{\ast}-1\right)x_{j}\right\rangle \right|\\
 & \leq C\left(1+\left\langle v,h^{-1}v\right\rangle \right)\frac{\left\langle x_{i},v\right\rangle \left\langle v,x_{j}\right\rangle }{\lambda_{i}+\lambda_{j}}\nonumber 
\end{align}
as we only consider $t\in\left[0,1\right]$, and likewise for $\left|\left\langle x_{i},F_{t}^{\ast}x_{j}\right\rangle \right|$.
$\hfill\square$

\subsubsection*{Estimation of $E_{1}\left(t\right)$}

We are now ready to estimate $E_{1}\left(t\right)=E_{1,1}\left(t\right)+E_{1,2}\left(t\right)$,
starting with $E_{1,1}\left(t\right)=\left(e^{tJ}-1\right)h+h\left(e^{-tJ}-1\right)$. Recall that $(x_i)_i$ are an eigenbasis of $h$ with $\langle x_i, v\rangle \ge 0$ for all $1\le i\le n$. 
\begin{prop}
\label{prop:E11Estimation}For all $t\in\left[0,1\right]$ it holds
that
\begin{align*}
\max_{1\leq j\leq n}\left\Vert h^{-\frac{1}{2}}E_{1,1}\left(t\right)x_j\right\Vert  & \leq C\alpha \left(1+\left\langle v,h^{-1}v\right\rangle \right)\sqrt{\left\langle v,h^{-1}v\right\rangle }\\
 & + C\alpha \left(1+\left\langle v,h^{-1}v\right\rangle \right)^{3}\left(\left\Vert v\right\Vert \left\langle v,h^{-\frac{5}{4}}h\right\rangle ^{2}+\left\langle v,h^{-\frac{1}{2}}v\right\rangle \left\langle v,h^{-\frac{4}{3}}v\right\rangle ^{\frac{3}{2}}\right)
\end{align*}
where $\alpha=\max_{1\leq j\leq n}\left\langle v,x_{j}\right\rangle $
and $C>0$ is a constant independent of all  quantities.
\end{prop}

\textbf{Proof:} We write 
\begin{equation}
E_{1,1}\left(t\right)=F_{t}h+hF_{t}^{\ast}+\left(e^{tJ}-1-F_{t}\right)h+h\left(e^{-tJ}-1-F_{t}^{\ast}\right)
\end{equation}
so that for
any $1\leq j\leq n$ we can   estimate by Proposition \ref{prop:EstimationTechnique}
\begin{align}
\left\Vert h^{-\frac{1}{2}}E_{1,1}\left(t\right)x_{j}\right\Vert  & \leq\left\Vert h^{-\frac{1}{2}}F_{t}hx_{j}\right\Vert +\left\Vert h^{\frac{1}{2}}F_{t}^{\ast}x_{j}\right\Vert +C\left\Vert h^{-\frac{1}{2}}\left(U-1\right)^{2}\right\Vert _{\text{HS}}\left\Vert \left(U-1\right)hx_{j}\right\Vert \label{eq:E11Estimate}\\
 & +C\left\Vert h^{\frac{1}{2}}\left(U-1\right)\right\Vert _{\text{HS}}\left\Vert \left(U-1\right)^{2}x_{j}\right\Vert .\nonumber 
\end{align}
We consider each term above for the following. By Proposition \ref{prop:EstimationTechnique}
we see that independently of $1\leq j\leq n$
\begin{align}
\left\Vert h^{-\frac{1}{2}}F_{t}hx_{j}\right\Vert ^{2} & =\sum_{i=1}^{n}\frac{\lambda_{j}^{2}}{\lambda_{i}}\left|\left\langle x_{i},F_{t}x_{j}\right\rangle \right|^{2}\leq C\left(1+\left\langle v,h^{-1}v\right\rangle \right)^{2}\sum_{i=1}^{n}\frac{\lambda_{j}^{2}}{\lambda_{i}}\left|\frac{\left\langle x_{i},v\right\rangle \left\langle v,x_{j}\right\rangle }{\lambda_{i}+\lambda_{j}}\right|^{2}\nonumber \\
 & \leq C\left|\left\langle v,x_{j}\right\rangle \right|^{2}\left(1+\left\langle v,h^{-1}v\right\rangle \right)^{2}\sum_{i=1}^{n}\frac{\left|\left\langle x_{i},v\right\rangle \right|^{2}}{\lambda_{i}}\leq C\alpha^{2}\left(1+\left\langle v,h^{-1}v\right\rangle \right)^{2}\left\langle v,h^{-1}v\right\rangle ,\nonumber \\
\left\Vert h^{\frac{1}{2}}F^*_{t}x_{j}\right\Vert ^{2} & =\sum_{i=1}^{n}\lambda_{i}\left|\left\langle x_{i},F^*_{t}x_{j}\right\rangle \right|^{2}\leq C\left(1+\left\langle v,h^{-1}v\right\rangle \right)^{2}\sum_{i=1}^{n}\lambda_{i}\left|\frac{\left\langle x_{i},v\right\rangle \left\langle v,x_{j}\right\rangle }{\lambda_{i}+\lambda_{j}}\right|^{2}\label{eq:E11EasyEstimates}\\
 & \leq C\left|\left\langle v,x_{j}\right\rangle \right|^{2}\left(1+\left\langle v,h^{-1}v\right\rangle \right)^{2}\sum_{i=1}^{n}\frac{\left|\left\langle x_{i},v\right\rangle \right|^{2}}{\lambda_{i}}\leq C\alpha^{2}\left(1+\left\langle v,h^{-1}v\right\rangle \right)^{2}\left\langle v,h^{-1}v\right\rangle .\nonumber 
\end{align}
For the remaining terms of equation (\ref{eq:E11Estimate}) we recall
that we already estimated $\left\Vert h^{-\frac{1}{2}}\left(U-1\right)^{2}\right\Vert _{\text{HS}}$
and $\left\Vert h^{\frac{1}{2}}\left(U-1\right)\right\Vert _{\text{HS}}$
in the equations (\ref{eq:U-1hEstimate}) and (\ref{eq:U-1sqhEstimate})
to be
\begin{align}
\left\Vert h^{\frac{1}{2}}\left(U-1\right)\right\Vert _{\text{HS}} & =\left\Vert \left(U-1\right)h^{\frac{1}{2}}\right\Vert _{\text{HS}}\leq C\left(1+\left\langle v,h^{-1}v\right\rangle \right)\left\langle v,h^{-\frac{1}{2}}v\right\rangle , \label{eq:UsefulhU-1Estimates}\\
\left\Vert h^{-\frac{1}{2}}\left(U-1\right)^{2}\right\Vert _{\text{HS}} & =\left\Vert \left(U-1\right)^{2}h^{-\frac{1}{2}}\right\Vert _{\text{HS}}\leq C\left(1+\left\langle v,h^{-1}v\right\rangle \right)^{2}\left\langle v,h^{-\frac{5}{4}}v\right\rangle ^{2},\nonumber 
\end{align}
the equalities holding by normality of $U$. The only unknown quantities
are thus $\left\Vert \left(U-1\right)hx_{j}\right\Vert $ and $\left\Vert \left(U-1\right)^{2}x_{j}\right\Vert $,
which we estimate using Proposition \ref{prop:Uminus1MatrixElements}
as
\begin{align}
\left\Vert \left(U-1\right)hx_{j}\right\Vert ^{2} & =\sum_{i=1}^{n}\lambda_{j}^{2}\left|\left\langle x_{i},\left(U-1\right)x_{j}\right\rangle \right|^{2}\leq C\left(1+\left\langle v,h^{-1}v\right\rangle \right)^{2}\sum_{i=1}^{n}\lambda_{j}^{2}\left|\frac{\left\langle x_{i},v\right\rangle \left\langle v,x_{j}\right\rangle }{\lambda_{i}+\lambda_{j}}\right|^{2}\\
 & \leq C\left|\left\langle v,x_{j}\right\rangle \right|^{2}\left(1+\left\langle v,h^{-1}v\right\rangle \right)^{2}\sum_{i=1}^{n}\left|\left\langle x_{i},v\right\rangle \right|^{2}\leq C\alpha^{2}\left(1+\left\langle v,h^{-1}v\right\rangle \right)^{2}\left\Vert v\right\Vert ^{2},\nonumber \\
\left\Vert \left(U-1\right)^{2}x_{j}\right\Vert ^{2} & =\sum_{i=1}^{n}\left|\sum_{k=1}^{n}\left\langle x_{i},\left(U-1\right)x_{k}\right\rangle \left\langle x_{k},\left(U-1\right)x_{j}\right\rangle \right|^{2}\nonumber \\
 & \leq C\left(1+\left\langle v,h^{-1}v\right\rangle \right)^{4}\sum_{i=1}^{n}\left|\sum_{k=1}^{n}\frac{\left\langle x_{i},v\right\rangle \left\langle v,x_{k}\right\rangle }{\lambda_{i}+\lambda_{k}}\frac{\left\langle x_{k},v\right\rangle \left\langle v,x_{j}\right\rangle }{\lambda_{k}+\lambda_{j}}\right|^{2}\label{eq:U-1sqxjEstimate}\\
 & \leq C\left|\left\langle v,x_{j}\right\rangle \right|^{2}\left(1+\left\langle v,h^{-1}v\right\rangle \right)^{4}\sum_{i=1}^{n}\left|\left\langle x_{i},v\right\rangle \right|^{2}\left(\sum_{k=1}^{n}\frac{\left|\left\langle x_{k},v\right\rangle \right|^{2}}{\lambda_{i}^{\frac{2}{3}}\lambda_{k}^{\frac{4}{3}}}\right)^{2}\nonumber \\
 & \leq C\alpha^{2}\left(1+\left\langle v,h^{-1}v\right\rangle \right)^{4}\left\langle v,h^{-\frac{4}{3}}v\right\rangle ^{3}.\nonumber 
\end{align}
Thus
\begin{align}
\left\Vert h^{-\frac{1}{2}}\left(U-1\right)^{2}\right\Vert _{\text{HS}}\left\Vert \left(U-1\right)hx_{j}\right\Vert  & \leq C\alpha\left(1+\left\langle v,h^{-1}v\right\rangle \right)^{3}\left\Vert v\right\Vert \left\langle v,h^{-\frac{5}{4}}h\right\rangle ^{2}\\
\left\Vert h^{\frac{1}{2}}\left(U-1\right)\right\Vert _{\text{HS}}\left\Vert \left(U-1\right)^{2}x_{j}\right\Vert  & \leq C\alpha\left(1+\left\langle v,h^{-1}v\right\rangle \right)^{3}\left\langle v,h^{-\frac{1}{2}}v\right\rangle \left\langle v,h^{-\frac{4}{3}}v\right\rangle ^{\frac{3}{2}}\nonumber 
\end{align}
which upon combination with the estimates of equation (\ref{eq:E11EasyEstimates})
imply the claim.
$\hfill\square$

\begin{prop}
For all $t\in\left[0,1\right]$ it holds that
\begin{align*}
 & \quad\,\left(C\alpha\right)^{-1}\max_{1\leq j\leq n}\left\Vert h^{-\frac{1}{2}}E_{1,2}\left(t\right)x_{j}\right\Vert \\
 & \leq\left(1+\left\langle v,h^{-1}v\right\rangle \right)^{2}\left\langle v,h^{-1}v\right\rangle ^{\frac{3}{2}}+\left(1+\left\langle v,h^{-1}v\right\rangle \right)^{6}\left\langle v,h^{-\frac{1}{2}}v\right\rangle ^{2}\left\langle v,h^{-\frac{5}{4}}h\right\rangle ^{2}\left\langle v,h^{-\frac{4}{3}}v\right\rangle ^{\frac{3}{2}}\\
 & +\left(1+\left\langle v,h^{-1}v\right\rangle \right)^{4}\left(\sqrt{\left\langle v,h^{-1}v\right\rangle }\left\langle v,h^{-\frac{2}{3}}v\right\rangle ^{\frac{3}{2}}\left\langle v,h^{-\frac{4}{3}}v\right\rangle ^{\frac{3}{2}}+\left\langle v,h^{-\frac{2}{3}}v\right\rangle ^{\frac{3}{2}}\left\langle v,h^{-\frac{5}{4}}v\right\rangle ^{2}\right)
\end{align*}
where $\alpha=\max_{1\leq j\leq n}\left\langle v,x_{j}\right\rangle $
and $C>0$ is a constant independent of all quantities.
\end{prop}

\textbf{Proof:} We write $E_{1,2}\left(t\right)=\left(e^{tJ}-1\right)h\left(e^{-tJ}-1\right)$ as
\begin{equation}
E_{1,2}\left(t\right)= F_{t}hF_{t}^{\ast}+F_{t}h\left(e^{-tJ}-1-F_{t}^{\ast}\right)+\left(e^{tJ}-1-F_{t}\right)hF_{t}^{\ast}+\left(e^{tJ}-1-F_{t}\right)h\left(e^{-tJ}-1-F_{t}^{\ast}\right)
\end{equation}
and see by Proposition \ref{prop:EstimationTechnique} that
\begin{align}
\left\Vert h^{-\frac{1}{2}}E_{1,2}\left(t\right)x_{j}\right\Vert  & \leq\left\Vert h^{-\frac{1}{2}}F_{t}hF_{t}^{\ast}x_{j}\right\Vert +C\left\Vert h^{-\frac{1}{2}}F_{t}h\left(U-1\right)\right\Vert _{\text{HS}}\left\Vert \left(U-1\right)^{2}x_{j}\right\Vert \nonumber \\
 & +C\left\Vert h^{-\frac{1}{2}}\left(U-1\right)^{2}\right\Vert _{\text{HS}}\left\Vert \left(U-1\right)hF_{t}^{\ast}x_{j}\right\Vert \\
 & +C\left\Vert h^{-\frac{1}{2}}\left(U-1\right)^{2}\right\Vert _{\text{HS}}\left\Vert \left(U-1\right)h\left(U-1\right)\right\Vert _{\text{HS}}\left\Vert \left(U-1\right)^{2}x_{j}\right\Vert .\nonumber 
\end{align}
We estimate by Propositions \ref{prop:Uminus1MatrixElements}
and \ref{prop:EstimationTechnique} that
\begin{align}
 & \quad\;\left\Vert h^{-\frac{1}{2}}F_{t}hF_{t}^{\ast}x_{j}\right\Vert ^{2}\leq C\left(1+\left\langle v,h^{-1}v\right\rangle \right)^{4}\sum_{i=1}^{n}\frac{1}{\lambda_{i}}\left|\sum_{k=1}^{n}\lambda_{k}\frac{\left\langle x_{i},v\right\rangle \left\langle v,x_{k}\right\rangle }{\lambda_{i}+\lambda_{k}}\frac{\left\langle x_{k},v\right\rangle \left\langle v,x_{j}\right\rangle }{\lambda_{k}+\lambda_{j}}\right|^{2}\\
 & \leq C\alpha^{2}\left(1+\left\langle v,h^{-1}v\right\rangle \right)^{4}\sum_{i=1}^{n}\frac{\left|\left\langle x_{i},v\right\rangle \right|^{2}}{\lambda_{i}}\left(\sum_{k=1}^{n}\frac{\left|\left\langle x_{k},v\right\rangle \right|^{2}}{\lambda_{k}}\right)^{2}=C\alpha^{2}\left(1+\left\langle v,h^{-1}v\right\rangle \right)^{4}\left\langle v,h^{-1}v\right\rangle ^{3},\nonumber 
 \end{align}
 and
 \begin{align}
 & \quad\;\left\Vert h^{-\frac{1}{2}}F_{t}h\left(U-1\right)\right\Vert _{\text{HS}}^{2}\leq C\left(1+\left\langle v,h^{-1}v\right\rangle \right)^{4}\sum_{i,j=1}^{n}\frac{1}{\lambda_{i}}\left|\sum_{k=1}^{n}\lambda_{k}\frac{\left\langle x_{i},v\right\rangle \left\langle v,x_{k}\right\rangle }{\lambda_{i}+\lambda_{k}}\frac{\left\langle x_{k},v\right\rangle \left\langle v,x_{j}\right\rangle }{\lambda_{k}+\lambda_{j}}\right|^{2}\nonumber \\
 & \leq C\left(1+\left\langle v,h^{-1}v\right\rangle \right)^{4}\sum_{i,j=1}^{n}\frac{\left|\left\langle x_{i},v\right\rangle \right|^{2}}{\lambda_{i}}\left|\left\langle x_{j},v\right\rangle \right|^{2}\left(\sum_{k=1}^{n}\frac{\left|\left\langle x_{k},v\right\rangle \right|^{2}}{\lambda_{k}^{\frac{2}{3}}\lambda_{j}^{\frac{1}{3}}}\right)^{2}\\
 & =C\left(1+\left\langle v,h^{-1}v\right\rangle \right)^{4}\left\langle v,h^{-1}v\right\rangle \left\langle v,h^{-\frac{2}{3}}v\right\rangle ^{3},\nonumber 
 \end{align}
and
 \begin{align}
 & \quad\;\left\Vert \left(U-1\right)hF_{t}^{\ast}x_{j}\right\Vert ^{2}\leq C\left(1+\left\langle v,h^{-1}v\right\rangle \right)^{4}\sum_{i=1}^{n}\left|\sum_{k=1}^{n}\lambda_{k}\frac{\left\langle x_{i},v\right\rangle \left\langle v,x_{k}\right\rangle }{\lambda_{i}+\lambda_{k}}\frac{\left\langle x_{k},v\right\rangle \left\langle v,x_{j}\right\rangle }{\lambda_{k}+\lambda_{j}}\right|^{2}\\
 & \leq C\alpha^{2}\left(1+\left\langle v,h^{-1}v\right\rangle \right)^{4}\sum_{i=1}^{n}\left|\left\langle x_{i},v\right\rangle \right|^{2}\left(\sum_{k=1}^{n}\frac{\left|\left\langle x_{k},v\right\rangle \right|^{2}}{\lambda_{i}^{\frac{1}{3}}\lambda_{k}^{\frac{2}{3}}}\right)^{2}=C\alpha^{2}\left(1+\left\langle v,h^{-1}v\right\rangle \right)^{4}\left\langle v,h^{-\frac{2}{3}}v\right\rangle ^{3}\nonumber 
 \end{align}
 and 
  \begin{align}
 & \quad\;\left\Vert \left(U-1\right)h\left(U-1\right)\right\Vert _{\text{HS}}^{2}\leq C\left(1+\left\langle v,h^{-1}v\right\rangle \right)^{4}\sum_{i,j=1}^{n}\left|\sum_{k=1}^{n}\lambda_{k}\frac{\left\langle x_{i},v\right\rangle \left\langle v,x_{k}\right\rangle }{\lambda_{i}+\lambda_{k}}\frac{\left\langle x_{k},v\right\rangle \left\langle v,x_{j}\right\rangle }{\lambda_{k}+\lambda_{j}}\right|^{2} \nonumber\\
 & \leq C\left(1+\left\langle v,h^{-1}v\right\rangle \right)^{4}\sum_{i,j=1}^{n}\left|\left\langle x_{i},v\right\rangle \right|^{2}\left|\left\langle x_{j},v\right\rangle \right|^{2}\left(\sum_{k=1}^{n}\frac{\left|\left\langle x_{k},v\right\rangle \right|^{2}}{\lambda_{i}^{\frac{1}{4}}\lambda_{j}^{\frac{1}{4}}\lambda_{k}^{\frac{1}{2}}}\right)^{2} \\
 &=C\left(1+\left\langle v,h^{-1}v\right\rangle \right)^{4}\left\langle v,h^{-\frac{1}{2}}v\right\rangle ^{4}.\nonumber 
\end{align}
Combining these with the estimates of the equations (\ref{eq:UsefulhU-1Estimates})
and (\ref{eq:U-1sqxjEstimate}) yields
\begin{align}
\left\Vert h^{-\frac{1}{2}}F_{t}h\left(U-1\right)\right\Vert _{\text{HS}}\left\Vert \left(U-1\right)^{2}x_{j}\right\Vert  & \leq C\alpha\left(1+\left\langle v,h^{-1}v\right\rangle \right)^{4}\sqrt{\left\langle v,h^{-1}v\right\rangle }\left\langle v,h^{-\frac{2}{3}}v\right\rangle ^{\frac{3}{2}}\left\langle v,h^{-\frac{4}{3}}v\right\rangle ^{\frac{3}{2}},\nonumber \\
\left\Vert h^{-\frac{1}{2}}\left(U-1\right)^{2}\right\Vert _{\text{HS}}\left\Vert \left(U-1\right)hF_{t}^{\ast}x_{j}\right\Vert  & \leq C\alpha\left(1+\left\langle v,h^{-1}v\right\rangle \right)^{4}\left\langle v,h^{-\frac{2}{3}}v\right\rangle ^{\frac{3}{2}}\left\langle v,h^{-\frac{5}{4}}v\right\rangle ^{2}
\end{align}
and
\begin{align}
 & \quad\;\left\Vert h^{-\frac{1}{2}}\left(U-1\right)^{2}\right\Vert _{\text{HS}}\left\Vert \left(U-1\right)h\left(U-1\right)\right\Vert _{\text{HS}}\left\Vert \left(U-1\right)^{2}x_{j}\right\Vert \\
 & \leq C\alpha\left(1+\left\langle v,h^{-1}v\right\rangle \right)^{6}\left\langle v,h^{-\frac{1}{2}}v\right\rangle ^{2}\left\langle v,h^{-\frac{5}{4}}v\right\rangle ^{2}\left\langle v,h^{-\frac{4}{3}}v\right\rangle ^{\frac{3}{2}}\nonumber 
\end{align}
which imply the claim.
$\hfill\square$

\subsubsection*{Estimation of $E_{2}\left(t\right)$}

We now repeat the same steps for $E_{2}\left(t\right)=E_{0}+E_{2,1}\left(t\right)+E_{2,2}\left(t\right)$ where 
\begin{equation}
E_{2,1}\left(t\right)=F_{t}E_{0}+E_{0}F_{t}^{\ast}+\left(e^{tJ}-1-F_{t}\right)E_{0}+E_{0}\left(e^{-tJ}-1-F_{t}^{\ast}\right). 
\end{equation}
\begin{prop}
For all $t\in\left[0,1\right]$ it holds that
\begin{align*}
 &\max_{1\leq j\leq n}\left\Vert h^{-\frac{1}{2}}E_{2,1}\left(t\right)x\right\Vert \\
 & \leq C\alpha \left(1+\left\langle v,h^{-1}v\right\rangle \right)^{2}\left\langle v,h^{-1}v\right\rangle ^{\frac{3}{2}}+ C\alpha \left(1+\left\langle v,h^{-1}v\right\rangle \right)^{4}\left\langle v,h^{-\frac{2}{3}}v\right\rangle ^{\frac{3}{2}}\left\langle v,h^{-\frac{5}{4}}v\right\rangle ^{2}\\
 &  +C\alpha \left(1+\left\langle v,h^{-1}v\right\rangle \right)^{4}\sqrt{\left\langle v,h^{-1}v\right\rangle }\left\langle v,h^{-\frac{2}{3}}v\right\rangle ^{\frac{3}{2}}\left\langle v,h^{-\frac{4}{3}}v\right\rangle ^{\frac{3}{2}}
\end{align*}
where $\alpha=\max_{1\leq j\leq n}\left\langle v,x_{j}\right\rangle $
and $C>0$ is a constant independent of all quantities.
\end{prop}

\textbf{Proof:} By Proposition \ref{prop:EstimationTechnique} we can estimate that
\begin{align}
\left\Vert h^{-\frac{1}{2}}E_{2,1}\left(t\right)x_{j}\right\Vert  & \leq\left\Vert h^{-\frac{1}{2}}F_{t}E_{0}x_{j}\right\Vert +\left\Vert h^{-\frac{1}{2}}E_{0}F_{t}^{\ast}x_{j}\right\Vert +C\left\Vert h^{-\frac{1}{2}}\left(U-1\right)^{2}\right\Vert _{\text{HS}}\left\Vert \left(U-1\right)E_{0}x_{j}\right\Vert \nonumber \\
 & +C\left\Vert h^{-\frac{1}{2}}E_{0}\left(U-1\right)\right\Vert _{\text{HS}}\left\Vert \left(U-1\right)^{2}x_{j}\right\Vert. 
\end{align}
Then let us consider each term separately. By the Propositions \ref{prop:Uminus1MatrixElements},
\ref{prop:E0MatrixElementEstimate}, \ref{prop:EstimationTechnique}
we have that
\begin{align}
 & \quad\;\left\Vert h^{-\frac{1}{2}}F_{t}E_{0}x_{j}\right\Vert ^{2}\leq C\left(1+\left\langle v,h^{-1}v\right\rangle \right)^{4}\sum_{i=1}^{n}\frac{1}{\lambda_{i}}\left|\sum_{k=1}^{n}\frac{\left\langle x_{i},v\right\rangle \left\langle v,x_{k}\right\rangle }{\lambda_{i}+\lambda_{k}}\left\langle x_{k},v\right\rangle \left\langle v,x_{j}\right\rangle \right|^{2}\\
 & \leq C\alpha^{2}\left(1+\left\langle v,h^{-1}v\right\rangle \right)^{4}\sum_{i=1}^{n}\frac{\left|\left\langle x_{i},v\right\rangle \right|^{2}}{\lambda_{i}}\left(\sum_{k=1}^{n}\frac{\left|\left\langle x_{k},v\right\rangle \right|^{2}}{\lambda_{k}}\right)^{2}=C\alpha^{2}\left(1+\left\langle v,h^{-1}v\right\rangle \right)^{4}\left\langle v,h^{-1}v\right\rangle ^{3},\nonumber
 \end{align}
 and
 \begin{align}
 & \quad\;\left\Vert h^{-\frac{1}{2}}E_{0}F_{t}x_{j}\right\Vert ^{2}\leq C\left(1+\left\langle v,h^{-1}v\right\rangle \right)^{4}\sum_{i=1}^{n}\frac{1}{\lambda_{i}}\left|\sum_{k=1}^{n}\left\langle x_{i},v\right\rangle \left\langle v,x_{k}\right\rangle \frac{\left\langle x_{k},v\right\rangle \left\langle v,x_{j}\right\rangle }{\lambda_{k}+\lambda_{j}}\right|^{2}\\
 & \leq C\alpha^{2}\left(1+\left\langle v,h^{-1}v\right\rangle \right)^{4}\sum_{i=1}^{n}\frac{\left|\left\langle x_{i},v\right\rangle \right|^{2}}{\lambda_{i}}\left(\sum_{k=1}^{n}\frac{\left|\left\langle x_{k},v\right\rangle \right|^{2}}{\lambda_{k}}\right)^{2}=C\alpha^{2}\left(1+\left\langle v,h^{-1}v\right\rangle \right)^{4}\left\langle v,h^{-1}v\right\rangle ^{3},\nonumber 
 \end{align}
 and 
 \begin{align}
 & \quad\;\left\Vert \left(U-1\right)E_{0}x_{j}\right\Vert ^{2}\leq C\left(1+\left\langle v,h^{-1}v\right\rangle \right)^{4}\sum_{i=1}^{n}\left|\sum_{k=1}^{n}\frac{\left\langle x_{i},v\right\rangle \left\langle v,x_{k}\right\rangle }{\lambda_{i}+\lambda_{k}}\left\langle x_{k},v\right\rangle \left\langle v,x_{j}\right\rangle \right|^{2}\\
 & \leq C\alpha^{2}\left(1+\left\langle v,h^{-1}v\right\rangle \right)^{4}\sum_{i=1}^{n}\left|\left\langle x_{i},v\right\rangle \right|^{2}\left(\sum_{k=1}^{n}\frac{\left|\left\langle x_{k},v\right\rangle \right|^{2}}{\lambda_{i}^{\frac{1}{3}}\lambda_{k}^{\frac{2}{3}}}\right)^{2}=C\alpha^{2}\left(1+\left\langle v,h^{-1}v\right\rangle \right)^{4}\left\langle v,h^{-\frac{2}{3}}v\right\rangle ^{3},\nonumber
 \end{align}
 and 
 \begin{align}
 & \quad\;\left\Vert h^{-\frac{1}{2}}E_{0}\left(U-1\right)\right\Vert _{\text{HS}}^{2}\leq C\left(1+\left\langle v,h^{-1}v\right\rangle \right)^{4}\sum_{i,j=1}^{n}\frac{1}{\lambda_{i}}\left|\sum_{k=1}^{n}\left\langle x_{i},v\right\rangle \left\langle v,x_{k}\right\rangle \frac{\left\langle x_{k},v\right\rangle \left\langle v,x_{j}\right\rangle }{\lambda_{k}+\lambda_{j}}\right|^{2}\nonumber \\
 & \leq C\left(1+\left\langle v,h^{-1}v\right\rangle \right)^{4}\sum_{i,j=1}^{n}\frac{\left|\left\langle x_{i},v\right\rangle \right|^{2}}{\lambda_{i}}\left|\left\langle x_{j},v\right\rangle \right|^{2}\left(\sum_{k=1}^{n}\frac{\left|\left\langle x_{k},v\right\rangle \right|^{2}}{\lambda_{k}^{\frac{2}{3}}\lambda_{j}^{\frac{1}{3}}}\right)^{2}\\
 & =C\left(1+\left\langle v,h^{-1}v\right\rangle \right)^{4}\left\langle v,h^{-1}v\right\rangle \left\langle v,h^{-\frac{2}{3}}v\right\rangle ^{3}.\nonumber 
\end{align}
Combining these with our prior estimates that
\begin{align}
\left\Vert h^{-\frac{1}{2}}\left(U-1\right)^{2}\right\Vert _{\text{HS}} & \leq C\left(1+\left\langle v,h^{-1}v\right\rangle \right)^{2}\left\langle v,h^{-\frac{5}{4}}v\right\rangle ^{2},\label{eq:UsefulEstimates2}\\
\left\Vert \left(U-1\right)^{2}x_{j}\right\Vert _{\text{HS}} & \leq C\alpha\left(1+\left\langle v,h^{-1}v\right\rangle \right)^{2}\left\langle v,h^{-\frac{4}{3}}v\right\rangle ^{\frac{3}{2}}\nonumber 
\end{align}
we obtain the claim.
$\hfill\square$

\begin{prop}
\label{Prop:E22Estimation}For all $t\in\left[0,1\right]$ it holds
that
\begin{align*}
 & \quad\,\left(C\alpha\right)^{-1}\max_{1\leq j\leq n}\left\Vert h^{-\frac{1}{2}}E_{2,2}\left(t\right)x_{j}\right\Vert \\
 & \leq\left(1+\left\langle v,h^{-1}v\right\rangle \right)^{3}\left\langle v,h^{-1}v\right\rangle ^{\frac{5}{2}}+\left(1+\left\langle v,h^{-1}v\right\rangle \right)^{7}\left\langle v,h^{-\frac{2}{3}}\right\rangle ^{3}\left\langle v,h^{-\frac{5}{4}}v\right\rangle ^{2}\left\langle v,h^{-\frac{4}{3}}v\right\rangle ^{\frac{3}{2}}\\
 & +\left(1+\left\langle v,h^{-1}v\right\rangle \right)^{5}  \left\langle v,h^{-\frac{2}{3}}v\right\rangle ^{\frac{3}{2}}  \left(\left\langle v,h^{-1}v\right\rangle ^{\frac{3}{2}}\left\langle v,h^{-\frac{4}{3}}v\right\rangle ^{\frac{3}{2}}+\left\langle v,h^{-1}v\right\rangle \left\langle v,h^{-\frac{5}{4}}v\right\rangle ^{2}\right)
\end{align*}
where $\alpha=\max_{1\leq j\leq n}\left\langle v,x_{j}\right\rangle $
and $C>0$ is a constant independent of all quantities.
\end{prop}

\textbf{Proof:} We decompose $E_{2,2}\left(t\right)=\left(e^{tJ}-1\right)E_{0}\left(e^{-tJ}-1\right)$ as
\begin{equation}
F_{t}E_{0}F_{t}^{\ast}+F_{t}E_{0}\left(e^{-tJ}-1-F_{t}^{\ast}\right)+\left(e^{tJ}-1-F_{t}\right)E_{0}F_{t}^{\ast}+\left(e^{tJ}-1-F_{t}\right)E_{0}\left(e^{-tJ}-1-F_{t}^{\ast}\right)
\end{equation}
and estimate by Proposition \ref{prop:EstimationTechnique} that
\begin{align}
\left\Vert h^{-\frac{1}{2}}E_{2,2}\left(t\right)x_{j}\right\Vert  & \leq\left\Vert h^{-\frac{1}{2}}F_{t}E_{0}F_{t}^{\ast}x_{j}\right\Vert +C\left\Vert h^{-\frac{1}{2}}F_{t}E_{0}\left(U-1\right)\right\Vert _{\text{HS}}\left\Vert \left(U-1\right)^{2}x_{j}\right\Vert \nonumber \\
 & +C\left\Vert h^{-\frac{1}{2}}\left(U-1\right)^{2}\right\Vert _{\text{HS}}\left\Vert \left(U-1\right)E_{0}F_{t}^{\ast}x_{j}\right\Vert \\
 & +C\left\Vert h^{-\frac{1}{2}}\left(U-1\right)^{2}\right\Vert _{\text{HS}}\left\Vert \left(U-1\right)E_{0}\left(U-1\right)\right\Vert _{\text{HS}}\left\Vert \left(U-1\right)^{2}x_{j}\right\Vert .\nonumber 
\end{align}
We estimate as in the previous proposition that
\begin{align}
 & \quad\;\left\Vert h^{-\frac{1}{2}}F_{t}E_{0}F_{t}^{\ast}x_{j}\right\Vert ^{2} \nonumber\\
 &\leq C\left(1+\left\langle v,h^{-1}v\right\rangle \right)^{6}\sum_{i=1}^{n}\frac{1}{\lambda_{i}}\left|\sum_{k,l=1}^{n}\frac{\left\langle x_{i},v\right\rangle \left\langle v,x_{k}\right\rangle }{\lambda_{i}+\lambda_{k}}\left\langle x_{k},v\right\rangle \left\langle v,x_{l}\right\rangle \frac{\left\langle x_{l},v\right\rangle \left\langle v,x_{j}\right\rangle }{\lambda_{l}+\lambda_{j}}\right|^{2}\nonumber \\
 & \leq C\alpha^{2}\left(1+\left\langle v,h^{-1}v\right\rangle \right)^{6}\sum_{i=1}^{n}\frac{\left|\left\langle x_{i},v\right\rangle \right|^{2}}{\lambda_{i}}\left(\sum_{k=1}^{n}\frac{\left|\left\langle x_{k},v\right\rangle \right|^{2}}{\lambda_{k}}\frac{\left|\left\langle x_{l},v\right\rangle \right|^{2}}{\lambda_{l}}\right)^{2}\\
 & =C\alpha^{2}\left(1+\left\langle v,h^{-1}v\right\rangle \right)^{6}\left\langle v,h^{-1}v\right\rangle ^{5},\nonumber 
\end{align}
and
\begin{align}
 & \quad\;\left\Vert h^{-\frac{1}{2}}F_{t}E_{0}\left(U-1\right)\right\Vert _{\text{HS}}^{2} \nonumber\\
 &\leq C\left(1+\left\langle v,h^{-1}v\right\rangle \right)^{6}\sum_{i,j=1}^{n}\frac{1}{\lambda_{i}}\left|\sum_{k,l=1}^{n}\frac{\left\langle x_{i},v\right\rangle \left\langle v,x_{k}\right\rangle }{\lambda_{i}+\lambda_{k}}\left\langle x_{k},v\right\rangle \left\langle v,x_{l}\right\rangle \frac{\left\langle x_{l},v\right\rangle \left\langle v,x_{j}\right\rangle }{\lambda_{l}+\lambda_{j}}\right|^{2}\nonumber \\
 & \leq C\left(1+\left\langle v,h^{-1}v\right\rangle \right)^{6}\sum_{i,j=1}^{n}\frac{\left|\left\langle x_{i},v\right\rangle \right|^{2}}{\lambda_{i}}\left|\left\langle x_{j},v\right\rangle \right|^{2}\left(\sum_{k,l=1}^{n}\frac{\left|\left\langle x_{k},v\right\rangle \right|^{2}}{\lambda_{k}}\frac{\left|\left\langle x_{l},v\right\rangle \right|^{2}}{\lambda_{l}^{\frac{2}{3}}\lambda_{j}^{\frac{1}{3}}}\right)^{2}\\
 & =C\left(1+\left\langle v,h^{-1}v\right\rangle \right)^{6}\left\langle v,h^{-1}v\right\rangle ^{3}\left\langle v,h^{-\frac{2}{3}}v\right\rangle ^{3},\nonumber 
\end{align}
and
\begin{align}
 & \quad\;\left\Vert \left(U-1\right)E_{0}F_{t}^{\ast}x_{j}\right\Vert ^{2} \nonumber\\
 &\leq C\left(1+\left\langle v,h^{-1}v\right\rangle \right)^{6}\sum_{i=1}^{n}\left|\sum_{k,l=1}^{n}\frac{\left\langle x_{i},v\right\rangle \left\langle v,x_{k}\right\rangle }{\lambda_{i}+\lambda_{k}}\left\langle x_{k},v\right\rangle \left\langle v,x_{l}\right\rangle \frac{\left\langle x_{l},v\right\rangle \left\langle v,x_{j}\right\rangle }{\lambda_{l}+\lambda_{j}}\right|^{2}\nonumber \\
 & \leq C\alpha^{2}\left(1+\left\langle v,h^{-1}v\right\rangle \right)^{6}\sum_{i=1}^{n}\left|\left\langle x_{i},v\right\rangle \right|^{2}\left(\sum_{k,l=1}^{n}\frac{\left|\left\langle x_{k},v\right\rangle \right|^{2}}{\lambda_{i}^{\frac{1}{3}}\lambda_{k}^{\frac{2}{3}}}\frac{\left|\left\langle x_{l},v\right\rangle \right|^{2}}{\lambda_{l}}\right)^{2}\\
 & =C\alpha^{2}\left(1+\left\langle v,h^{-1}v\right\rangle \right)^{6}\left\langle v,h^{-1}v\right\rangle ^{2}\left\langle v,h^{-\frac{2}{3}}v\right\rangle ^{3},\nonumber 
\end{align}
and finally that
\begin{align}
 & \quad\;\left\Vert \left(U-1\right)E_{0}\left(U-1\right)\right\Vert _{\text{HS}}^{2} \nonumber\\
 &\leq C\left(1+\left\langle v,h^{-1}v\right\rangle \right)^{6}\sum_{i,j=1}^{n}\left|\sum_{k,l=1}^{n}\frac{\left\langle x_{i},v\right\rangle \left\langle v,x_{k}\right\rangle }{\lambda_{i}+\lambda_{k}}\left\langle x_{k},v\right\rangle \left\langle v,x_{l}\right\rangle \frac{\left\langle x_{l},v\right\rangle \left\langle v,x_{j}\right\rangle }{\lambda_{l}+\lambda_{j}}\right|^{2}\nonumber \\
 & \leq C\left(1+\left\langle v,h^{-1}v\right\rangle \right)^{6}\sum_{i,j=1}^{n}\left|\left\langle x_{i},v\right\rangle \right|^{2}\left|\left\langle x_{j},v\right\rangle \right|^{2}\left|\sum_{k,l=1}^{n}\frac{\left|\left\langle x_{k},v\right\rangle \right|^{2}}{\lambda_{i}^{\frac{1}{3}}\lambda_{k}^{\frac{2}{3}}}\frac{\left|\left\langle x_{l},v\right\rangle \right|^{2}}{\lambda_{l}^{\frac{2}{3}}\lambda_{j}^{\frac{1}{3}}}\right|^{2}\\
 & =C\left(1+\left\langle v,h^{-1}v\right\rangle \right)^{6}\left\langle v,h^{-\frac{2}{3}}\right\rangle ^{6}\nonumber .
\end{align}
Combining these bounds with equation (\ref{eq:UsefulEstimates2}) yields the
claim.

$\hfill\square$

Combining the estimates from Proposition \ref{prop:E11Estimation} through \ref{Prop:E22Estimation} and the last bound of Proposition \ref{prop:E0MatrixElementEstimate} we obtain
\begin{align}
 & \quad\left(C\alpha\right)^{-1}\max_{1\leq j\leq n}\left\Vert h^{-\frac{1}{2}}E\left(t\right)x_{j}\right\Vert \nonumber\\
 &\leq\left(1+\left\langle v,h^{-1}v\right\rangle \right)\sqrt{\left\langle v,h^{-1}v\right\rangle }+\left(1+\left\langle v,h^{-1}v\right\rangle \right)^{2}\left\langle v,h^{-1}v\right\rangle ^{\frac{3}{2}} \nonumber\\
 & +\left(1+\left\langle v,h^{-1}v\right\rangle \right)^{3}\left(\left\langle v,h^{-1}v\right\rangle ^{\frac{5}{2}}+\left\Vert v\right\Vert \left\langle v,h^{-\frac{5}{4}}v\right\rangle ^{2}+\left\langle v,h^{-\frac{1}{2}}v\right\rangle \left\langle v,h^{-\frac{4}{3}}v\right\rangle ^{\frac{3}{2}}\right) \nonumber\\
 & +\left(1+\left\langle v,h^{-1}v\right\rangle \right)^{4} \left\langle v,h^{-\frac{2}{3}}v\right\rangle ^{\frac{3}{2}} \left(\sqrt{\left\langle v,h^{-1}v\right\rangle }\left\langle v,h^{-\frac{4}{3}}v\right\rangle ^{\frac{3}{2}}+\left\langle v,h^{-\frac{5}{4}}v\right\rangle ^{2}\right) \\
 & +\left(1+\left\langle v,h^{-1}v\right\rangle \right)^{5} \left\langle v,h^{-\frac{2}{3}}v\right\rangle ^{\frac{3}{2}} \left(\left\langle v,h^{-1}v\right\rangle ^{\frac{3}{2}}\left\langle v,h^{-\frac{4}{3}}v\right\rangle ^{\frac{3}{2}}+\left\langle v,h^{-1}v\right\rangle \left\langle v,h^{-\frac{5}{4}}v\right\rangle ^{2}\right)\nonumber\\
 & +\left(1+\left\langle v,h^{-1}v\right\rangle \right)^{6}\left\langle v,h^{-\frac{1}{2}}v\right\rangle ^{2}\left\langle v,h^{-\frac{5}{4}}v\right\rangle ^{2}\left\langle v,h^{-\frac{4}{3}}v\right\rangle ^{\frac{3}{2}} \nonumber\\
 & +\left(1+\left\langle v,h^{-1}v\right\rangle \right)^{7}\left\langle v,h^{-\frac{2}{3}}v\right\rangle ^{3}\left\langle v,h^{-\frac{5}{4}}v\right\rangle ^{2}\left\langle v,h^{-\frac{4}{3}}v\right\rangle ^{\frac{3}{2}}.\nonumber
\end{align}
The right hand can be simplified further using the H\"older estimates 
\begin{align}
\left\langle v, h^{-\frac 1 2}v\right\rangle \le \|v\| \left\langle v, h^{-1}v\right\rangle^{\frac 1 2},\quad \left\langle v, h^{-\frac 2 3}v\right\rangle^{\frac 3 2} \le \|v\| \left\langle v, h^{-1}v\right\rangle.
\end{align}
All this gives the following:

\begin{prop}
\label{prop:CombinedEtEstimate}For all $t\in\left[0,1\right]$ it
holds that
\begin{align*}
 \max_{1\leq j\leq n}\left\Vert h^{-\frac{1}{2}}E\left(t\right)x_{j}\right\Vert \le C \alpha \left(1+\left\langle v,h^{-1}v\right\rangle \right)^{8} \left( \langle v, h^{-1} v\rangle^{\frac 1 2} + \|v\|  \langle v, h^{-\frac 5 4} v\rangle^{2} \right) \left( 1+ \|v\| \langle v, h^{-\frac 4 3} v\rangle^{\frac 3 2}  \right) 
\end{align*}
where $\alpha=\max_{1\leq j\leq n}\left\langle v,x_{j}\right\rangle $
and $C>0$ is a constant independent of all quantities.
\end{prop}

\medskip

\textbf{Conclusion of Proposition \ref{them:SecondTransformationOneBodyEstimates}:} Inserting $h_k$ and $v_k$ in Proposition \ref{prop:tE-h} and \eqref{eq:v-h-1-v} we have immediately 
\begin{align}
{\rm tr} \Big|h_k^{-1/2} (\widetilde{E}_k  - h_k) h_k^{-1/2}\Big| \le \langle v_k, h_k^{-1} v_k\rangle \le C \hat V_k. 
\end{align}
Next, consider the corresponding expressions on the right-hand side of Proposition \ref{prop:CombinedEtEstimate}. Recall that $\alpha_{k}=\max_{p\in L_{k}}\left\langle v_{k},e_{p}\right\rangle \le C(\hat{V}_{k})^{\frac{1}{2}}k_{F}^{-\frac{1}{2}}$. Moreover, by Propositions \ref{prop:NonSingularRiemannSums}, \ref{coro:CompletelyUniformRiemannSumBound} and \ref{prop:MoreSingularRiemannSums} we get 
\begin{align}\label{eq:A-1-2-3-2ndlast}
\left\langle v_k, h_k^{\beta} v_k\right\rangle &\le C \hat V_k k_F^{-1} \sum_{p\in L_k} \lambda_{k,p}^{\beta} \le C \hat V_k (|k| k_F)^{1+\beta}, \quad 0\ge \beta\ge  -\frac 5 4, \nonumber\\
\left\langle v_{k},h_{k}^{\beta}v_{k}\right\rangle^{\frac 3 2} & \le C(\hat{V}_{k})^{\frac 3 2} \left(\left|k\right|k_{F}\right)^{-\frac{1}{2}} \left|k\right|^{6}\log\left(k_{F}\right),\quad \beta\le -\frac 4 3. 
\end{align}
Putting these bounds together, we deduce from Proposition \ref{prop:CombinedEtEstimate} that 
\begin{align}
\max_{p\in L_{k}}\left\Vert h_{k}^{-\frac{1}{2}}E_{k}\left(t\right)e_{p}\right\Vert  & \leq C(\hat{V}_{k})^{\frac{1}{2}}k_{F}^{-\frac{1}{2}} \left(1+\hat{V}_{k}\right)^8 \left( (\hat V_k)^{\frac 1 2} + (\hat V_k)^{\frac 5 2}\right) \left( 1 + \hat V_k^2 |k|^6 \log k_F\right) \nonumber \\
 & \leq Ck_{F}^{-\frac{1}{2}}\left(\hat{V}_{k}+\hat{V}_{k}^{3}\left|k\right|^{6}\log\left(k_{F}\right)\right)
\end{align}
for $\left|k\right|\leq k_{F}^{\gamma}$, as claimed. Similarly, inserting \eqref{eq:A-1-2-3-2ndlast} in Corollary \ref{coro:hJHSEstimate} and Proposition \ref{prop:hJhhHSEstimate}
we  see that
\begin{align}
\left\Vert h^{-\frac{1}{2}}J\right\Vert _{\text{HS}} &\leq C\left(1+\left\langle v_{k},h_{k}^{-1}v_{k}\right\rangle \right)\left\langle v_{k},h_{k}^{-\frac{3}{2}}v_{k}\right\rangle \leq C(\log k_{F})^{\frac{2}{3}}k_{F}^{-\frac{1}{3}}\hat{V}_{k}\left(1+\hat{V}_{k}\right)\left|k\right|^{3+\frac{2}{3}}, \nonumber\\
\left\Vert h^{-\frac{1}{2}}\left[J,h\right]h^{-\frac{1}{2}}\right\Vert _{\text{HS}} & \leq C\left(1+\left\langle v_{k},h_{k}^{-1}v_{k}\right\rangle \right)^{3}\left(\left\langle v_{k},h_{k}^{-1}v_{k}\right\rangle +\left\langle v_{k},h_{k}^{-\frac{1}{2}}v_{k}\right\rangle \left\langle v_{k},h_{k}^{-\frac{5}{4}}v_{k}\right\rangle ^{2}\right)\nonumber \\
 & \leq C\left(1+\hat{V}_{k}\right)^{3}\left(\hat{V}_{k}+\hat{V}_{k}\left(\left|k\right|k_{F}\right)^{\frac{1}{2}}\left(\hat{V}_{k}\left(\left|k\right|k_{F}\right)^{-\frac{1}{4}}\right)^{2}\right) \leq C\hat{V}_{k}.
\end{align}
Here we also note that $\hat V_k$ is uniformly bounded, and hence the constant $C$ may depend on $V$ but it is still independent of $k$ and $k_F$. 
$\hfill\square$

\subsection{Gronwall Estimates for the Kinetic Operator}

We now come to the kinetic Gronwall estimates for the transformation $e^{\mathcal{J}}$.  We have 

\begin{prop}
\label{prop:KineticGronwall3} Assume $\sum_{k\in\mathbb{Z}^{3}}\hat{V}_{k}|k|<\infty$ and $S_C=\mathbb{Z}^3_+ \cap \overline{B}\left(0,k_{F}^{\gamma}\right)$ with $0<\gamma<\frac{1}{47}$. Then for all $\Psi\in D\left(H_{\kin}^{\prime}\right)$ and $\left|t\right|\leq1$
it holds that
\begin{align*}
\left\langle e^{t\mathcal{J}}\Psi,H_{\kin}^{\prime}e^{t\mathcal{J}}\Psi\right\rangle  & \leq C\left\langle \Psi,H_{\kin}^{\prime}\Psi\right\rangle \\
\left\langle e^{t\mathcal{J}}\Psi,\mathcal{N}_{E}H_{\kin}^{\prime}e^{t\mathcal{J}}\Psi\right\rangle  & \leq C\left\langle \Psi,\mathcal{N}_{E}H_{\kin}^{\prime}\Psi\right\rangle 
\end{align*}
for a constant $C>0$ independent of $k_{F}$.
\end{prop}

\textbf{Proof:} Write $\Psi_{t}=e^{t\mathcal{J}}\Psi$ for brevity.
By the commutator in \eqref{eq:JHkinCommutator}, we have
\begin{equation}
-\frac{d}{dt}\left\langle \Psi_{t},H_{\kin}^{\prime}\Psi_{t}\right\rangle =\left\langle \Psi_{t},\left[\mathcal{J},H_{\kin}^{\prime}\right]\Psi_{t}\right\rangle =2\sum_{k\in S_{C}}\left\langle \Psi_{t},\tilde{Q}_{1}^{k}\left(\left[J_{k}^{\oplus},h_{k}^{\oplus}\right]\right)\Psi_{t}\right\rangle 
\end{equation}
with $\tilde{Q}_{1}^{k}$ defined in \eqref{eq:def-Q1-tilde}. 
Moreover, Proposition \ref{prop:KineticQ1AEstimate} allows us to estimate
\begin{equation}
\sum_{k\in S_{C}}\left|\left\langle \Psi_{t},\tilde{Q}_{1}^{k}\left(\left[J_{k}^{\oplus},h_{k}^{\oplus}\right]\right)\Psi_{t}\right\rangle \right|\leq\sum_{k\in S_{C}}\left\Vert \left(h_{k}^{\oplus}\right)^{-\frac{1}{2}}\left[J_{k}^{\oplus},h_{k}^{\oplus}\right]\left(h_{k}^{\oplus}\right)^{-\frac{1}{2}}\right\Vert _{\text{Op}}\left\langle \Psi_{t},H_{\kin}^{\prime}\Psi_{t}\right\rangle .
\end{equation}
Since  \begin{equation}
\left(h_{k}^{\oplus}\right)^{-\frac{1}{2}}\left[J_{k}^{\oplus},h_{k}^{\oplus}\right]\left(h_{k}^{\oplus}\right)^{-\frac{1}{2}}=\left(\begin{array}{cc}
h_{k}^{-\frac{1}{2}}\left[J_{k},h_{k}\right]h_{k}^{-\frac{1}{2}} & 0\\
0 & h_{k}^{-\frac{1}{2}}\left[J_{k},h_{k}\right]h_{k}^{-\frac{1}{2}}
\end{array}\right)
\end{equation}
by Proposition \ref{them:SecondTransformationOneBodyEstimates} we
can estimate further that
\begin{align}
\sum_{k\in S_{C}}\left\Vert \left(h_{k}^{\oplus}\right)^{-\frac{1}{2}}\left[J_{k}^{\oplus},h_{k}^{\oplus}\right]\left(h_{k}^{\oplus}\right)^{-\frac{1}{2}}\right\Vert _{\text{Op}} & =\sum_{k\in S_{C}}\left\Vert h_{k}^{-\frac{1}{2}}\left[J_{k},h_{k}\right]h_{k}^{-\frac{1}{2}}\right\Vert _{\text{Op}}\leq\sum_{k\in S_{C}}\left\Vert h_{k}^{-\frac{1}{2}}\left[J_{k},h_{k}\right]h_{k}^{-\frac{1}{2}}\right\Vert _{\text{HS}}\nonumber \\
 & \leq C\sum_{k\in S_C}\hat{V}_{k}\leq C.
\end{align}
Hence, $\left|\frac{d}{dt}\left\langle \Psi_{t},H_{\kin}^{\prime}\Psi_{t}\right\rangle \right|\leq C\left\langle \Psi_{t},H_{\kin}^{\prime}\Psi_{t}\right\rangle $,
so by Gronwall's lemma
\begin{equation}
\left\langle \Psi_{t},H_{\kin}^{\prime}\Psi_{t}\right\rangle \leq\left\langle \Psi,H_{\kin}^{\prime}\Psi\right\rangle e^{C\left|t\right|}\leq C\left\langle \Psi,H_{\kin}^{\prime}\Psi\right\rangle ,\quad\left|t\right|\leq1.
\end{equation}
For $\left\langle \Psi_{t},\mathcal{N}_{E}H_{\kin}^{\prime}\Psi_{t}\right\rangle$, besides the commutator in \eqref{eq:JHkinCommutator} we also note that  
\begin{align}
\left[\mathcal{J},\mathcal{N}_{E}\right] & =\sum_{k\in S_{C}}\sum_{p\in L_{k}^{\pm}}\left[b_{k}^{\ast}\left(J_{k}^{\oplus}e_{p}\right)b_{k}\left(e_{p}\right),\mathcal{N}_{E}\right]\nonumber \\
 & =\sum_{k\in S_{C}}\sum_{p\in L_{k}^{\pm}}\left(b_{k}^{\ast}\left(J_{k}^{\oplus}e_{p}\right)\left[b_{k}\left(e_{p}\right),\mathcal{N}_{E}\right]+\left[b_{k}^{\ast}\left(J_{k}^{\oplus}e_{p}\right),\mathcal{N}_{E}\right]b_{k}\left(e_{p}\right)\right)\label{eq:JNECommutator}\\
 & =\sum_{k\in S_{C}}\sum_{p\in L_{k}^{\pm}}\left(b_{k}^{\ast}\left(J_{k}^{\oplus}e_{p}\right)b_{k}\left(e_{p}\right)-b_{k}^{\ast}\left(J_{k}^{\oplus}e_{p}\right)b_{k}\left(e_{p}\right)\right)=0.\nonumber 
\end{align}
Here again we used $\left[\mathcal{N}_{E},b_{k}\left(\varphi\right)\right]=-b_{k}$ for all $\varphi\in\ell^{2}\left(L_{k}^{\pm}\right)$, which follows from \eqref{eq:intro-comm-NE-b} and linearity. Hence, 
\begin{align}
-\frac{d}{dt}\left\langle \Psi_{t},\mathcal{N}_{E}H_{\kin}^{\prime}\Psi_{t}\right\rangle   =\left\langle \Psi_{t},\mathcal{N}_{E}\left[\mathcal{J},H_{\kin}^{\prime}\right]\Psi_{t}\right\rangle=2\sum_{k\in S_{C}}\left\langle \Psi_{t},\mathcal{N}_{E}\tilde{Q}_{1}^{k}\left(\left[J_{k}^{\oplus},h_{k}^{\oplus}\right]\right)\Psi_{t}\right\rangle .\nonumber 
\end{align}
Now, it holds that $\left[\mathcal{N}_{E},\tilde{Q}_{1}^{k}\left(\left[J_{k}^{\oplus},h_{k}^{\oplus}\right]\right)\right]=0$
(as can be seen by a computation similar to that of equation (\ref{eq:JNECommutator})),
so we may estimate as above for
\begin{align}
\sum_{k\in S_{C}}\left|\left\langle \Psi_{t},\mathcal{N}_{E}\tilde{Q}_{1}^{k}\left(\left[J_{k}^{\oplus},h_{k}^{\oplus}\right]\right)\Psi_{t}\right\rangle \right| & =\sum_{k\in S_{C}}\left|\left\langle \mathcal{N}_{E}^{\frac{1}{2}}\Psi_{t},\tilde{Q}_{1}^{k}\left(\left[J_{k}^{\oplus},h_{k}^{\oplus}\right]\right)\mathcal{N}_{E}^{\frac{1}{2}}\Psi_{t}\right\rangle \right|\\
 & \leq\sum_{k\in S_{C}}\left\Vert \left(h_{k}^{\oplus}\right)^{-\frac{1}{2}}\left[J_{k}^{\oplus},h_{k}^{\oplus}\right]\left(h_{k}^{\oplus}\right)^{-\frac{1}{2}}\right\Vert _{\text{Op}}\left\langle \mathcal{N}_{E}^{\frac{1}{2}}\Psi_{t},H_{\kin}^{\prime}\mathcal{N}_{E}^{\frac{1}{2}}\Psi_{t}\right\rangle \nonumber \\
 & \leq C\left\langle \Psi_{t},\mathcal{N}_{E}H_{\kin}^{\prime}\Psi_{t}\right\rangle ,\nonumber 
\end{align}
where we also used that $\left[\mathcal{N}_{E},H_{\kin}^{\prime}\right]=0$.
The second claim now follows.
$\hfill\square$

%%%%%%%%%%%%%%%%%%%%%%%%%%%%%%%%%%%%
%%%%%%%%%%%%%%%%%%%%%%%%%%%%%%%%%%%%

\section{Conclusion of the Main Results}\label{sec:prof-main-theorems}

Now we are ready to provide the proof of the main theorems stated in the introduction. 

\subsection{Proof of Theorem \ref{them:OperatorStatement}}

 The proof follows almost immediately by the analysis
we have performed throughout the paper, for we will simply take $\mathcal{U}=e^{\mathcal{J}}e^{\mathcal{K}}$
where $e^{\mathcal{K}}$ is the quasi-bosonic Bogolubov transformation
$e^{\mathcal{K}}$ of Section \ref{sec:TransformingtheHamiltonian} and $e^{\mathcal{J}}$ is the second transformation of Section \ref{sec:TheSecondTransformationandKineticEstimates}.

\medskip
{\bf Step 1:} Let us start from  the decomposition \eqref{eq:HN-localized-1}: 
\begin{equation}\label{eq:Prof-Thm1-0}
H_N - E_{\rm FS} =  H_{\kin}' + k_F^{-1} H_{\rm int}' =  H_{\kin}'  + \sum_{k\in S_C}  \left(H_{{\rm int}}^{k}-\frac{\hat{V}_{k}k_{F}^{-1}}{\left(2\pi\right)^{3}}\left|L_{k}\right|\right) + \mathcal{E}_{\rm NB} 
\end{equation}
where $H_{\text{int}}^{k}$ is given in \eqref{eq:HintkBBForm}, $\mathcal{E}_{\rm NB}$ is given in \eqref{eq:ENB}, and $S_{C}=\overline{B}\left(0,k_{F}^{\gamma}\right)\cap\mathbb{Z}_{+}^{3}$ with $0<\gamma< \frac 1 {47}$. From Proposition \ref{them:ReductiontoBosonizableTerms}, the non-bosonizable term $\mathcal{E}_{\rm NB}$ is estimated as 
\begin{align}
\pm \mathcal{E}_{\rm NB}
\le C k_F^{-\gamma/2} \Big(H_{\rm kin}' + k_F^{-1} \mathcal{N}_E  H_{\rm kin}'  + k_F\Big).
\end{align}
By the Gronwall estimates of Propositions \ref{prop:KineticGronwall}, \ref{prop:KineticGronwall3} and the choice $\mathcal{U}=e^{\mathcal{J}}e^{\mathcal{K}}$, we have
\begin{align}\label{eq:Prof-Thm1-err1}
\pm \mathcal{U} \mathcal{E}_{\rm NB} \mathcal{U}^* \le  C k_F^{-\gamma/2} \Big(H_{\rm kin}' + k_F^{-1} \mathcal{N}_E  H_{\rm kin}'  + k_F\Big).
\end{align}
%Here we also used the fact that $H_{\rm kin}' \ge \mathcal{N}_E$ of Proposition \ref{prop:Zeta0RepresentationofHKin} to simplify the expression. 
Thus it remains to apply  the transformations $e^{\mathcal K}$ and $e^{\mathcal J}$ to the bosonizable terms. 

\medskip
{\bf Step 2:} Now we apply the transformation $e^{\mathcal K}$. By Proposition \ref{prop:trans-eK-bosonizable-terms} we have
\begin{align} \label{eq:Prof-Thm1-2.0}
  \;e^{\mathcal{K}}\left(H_{\kin}^{\prime}+\sum_{k\in S_{C}}H_{\inter}^{k}\right)e^{-\mathcal{K}}&=H_{\kin}^{\prime}+\sum_{k\in S_{C}}Q_{1}^{k} (E_k^\oplus - h_k^\oplus)  \\
 & \qquad+\sum_{k\in S_{C}}\int_{0}^{1}e^{\left(1-t\right)\mathcal{K}} \left(\mathcal{E}_{1}^{k} (A_k^{\oplus}(t)) +\mathcal{E}_{2}^{k} (B_k^\oplus(t))\right)e^{-\left(1-t\right)\mathcal{K}}\,dt. \nonumber
\end{align}
%where the exchange terms $\mathcal{E}_1(\cdot)$,  $\mathcal{E}_2(\cdot)$ are defined in Proposition \ref{prop:QuasiBosonicQuadraticCommutators}. 

We will use the kinetic estimate of Proposition \ref{prop:KineticExchangeTermEstimates} and the Gronwall estimates of Proposition \ref{prop:KineticGronwall}  to bound the exchange terms in \eqref{eq:Prof-Thm1-2.0}. Thanks to the one-body estimates in Propositions  \ref{thm:ExplicitKAktBktEstimates-3}, \ref{thm:ExplicitKAktBktEstimates-2} and our assumption $\sum_{k\in S_{C}}\hat{V}_{k}\left|k\right|<\infty$ we get
\begin{align}
\sum_{k\in S_{C}}\max_{t\in\left[0,1\right]}\left\{ \max_{p\in L_{k}}\left\Vert h_{k}^{-\frac{1}{2}}A_{k}^{\oplus}\left(t\right)e_{p}\right\Vert ,\max_{p\in L_{k}}\left\Vert h_{k}^{-\frac{1}{2}}B_{k}^{\oplus}\left(t\right)e_{p}\right\Vert \right\} 
&\leq C\sum_{k\in S_{C}}k_{F}^{-\frac{1}{2}}\hat{V}_{k}\left(1+\hat{V}_{k}^{2}\right)\leq Ck_{F}^{-\frac{1}{2}}, \nonumber\\
\sum_{k\in S_{C}}\max_{t\in\left[0,1\right]}\left\{ \left\Vert A_{k}^{\oplus}\left(t\right)\right\Vert _{\infty,2},\left\Vert B_{k}^{\oplus}\left(t\right)\right\Vert _{\infty,2}\right\} 
\leq C\sum_{k\in S_{C}}&\hat{V}_{k}\left|k\right|^{\frac{1}{2}}\left(1+\hat{V}_{k}\right)\leq C,  \nonumber\\
\sum_{k\in S_{C}}\left(\left\Vert \left(h_{k}^{\oplus}\right)^{-\frac{1}{2}}K_{k}^{\oplus}\right\Vert _{\text{HS}}+\left\Vert K_{k}^{\oplus}\right\Vert _{\infty,2}\right) 
\leq C(\log k_{F})^{\frac{2}{3}}& k_{F}^{-\frac{1}{3}}\sum_{k\in S_C}\hat{V}_{k}\left|k\right|^{3+\frac{2}{3}}.
\end{align}
All this gives that for every state $\Psi\in D\left(H_{\kin}^{\prime}\right)$ and $\Psi_t=e^{-(1-t)\mathcal{K}}\Psi$, 
\begin{align}
 & \quad\;\sum_{k\in S_{C}}\int_{0}^{1}\left|\left\langle \Psi_t,\left(\mathcal{E}_{1}^{k}  (A^{\oplus}_k(t) )+\mathcal{E}_{2}^{k} (B^{\oplus}_k(t) )\right) \Psi_t \right\rangle \right|dt\label{eq:ExchangeTerm1PreliminaryEstimate}\\
 & \leq C(\log k_{F})^{\frac{2}{3}}\left(\sum_{k\in S_C}\hat{V}_{k}\left|k\right|^{3+\frac{2}{3}}\right)\left(k_{F}^{-\frac{5}{6}}\max_{t\in\left[0,1\right]}\sqrt{\left\langle \Psi_t,H_{\kin}^{\prime}\Psi_t\right\rangle \left\langle \Psi_t,\mathcal{N}_{E}H_{\kin}^{\prime}\Psi_t\right\rangle }\right.\nonumber \\
 & \qquad\qquad\qquad\qquad\qquad\qquad\left.+k_{F}^{-\frac{1}{3}}\max_{t\in\left[0,1\right]}\left\langle \Psi_t,H_{\kin}^{\prime}\Psi_t\right\rangle +k_{F}^{-\frac{1}{3}}\max_{t\in\left[0,1\right]}\sqrt{\left\langle \Psi_t,\mathcal{N}_{E}H_{\kin}^{\prime}\Psi_t\right\rangle }\right)\nonumber \\
 & \leq C(\log k_{F})^{\frac{2}{3}}k_{F}^{-\frac{1}{3}}\left(\sum_{k\in S_C}\hat{V}_{k}\left|k\right|^{3+\frac{2}{3}}\right)\left(\left\langle \Psi,H_{\kin}^{\prime}\Psi\right\rangle +k_{F}^{-1}\left\langle \Psi,\mathcal{N}_{E}H_{\kin}^{\prime}\Psi\right\rangle +k_{F}\left\Vert \Psi\right\Vert ^{2}\right),\nonumber 
\end{align}
where we also used the Cauchy--Schwarz inequality to split the square roots at the end. Thus the exchange terms in \eqref{eq:Prof-Thm1-2.0} can be estimated as 
\begin{align}\label{eq:ECalEstimateEnd} 
 &\pm \sum_{k\in S_{C}}\int_{0}^{1} e^{(1-t)\mathcal{K}}\left(\mathcal{E}_{1}^{k}  (A^{\oplus}_k(t) )+\mathcal{E}_{2}^{k} (B^{\oplus}_k(t) )\right) e^{-(1-t)\mathcal{K}} dt \nonumber\\
 & \leq C(\log k_{F})^{\frac{2}{3}}k_{F}^{-\frac{1}{3}}\left(\sum_{k\in S_C}\hat{V}_{k}\left|k\right|^{3+\frac{2}{3}}\right)\left( H_{\kin}^{\prime}+k_{F}^{-1} \mathcal{N}_{E}H_{\kin}^{\prime} +k_{F}\right).
\end{align}

%
%where the operators $A_{k}\left(t\right),B_{k}\left(t\right):\ell^{2}\left(L_{k}\right)\rightarrow\ell^{2}\left(L_{k}\right)$
%are for $t\in\left[0,1\right]$ defined by
%\begin{align*}
%A_{k}\left(t\right) & =\frac{1}{2}\left(e^{tK_{k}}\left(h_{k}+2P_{v_{k}}\right)e^{tK_{k}}+e^{-tK_{k}}h_{k}e^{-tK_{k}}\right)-h_{k}\\
%B_{k}\left(t\right) & =\frac{1}{2}\left(e^{tK_{k}}\left(h_{k}+2P_{v_{k}}\right)e^{tK_{k}}-e^{-tK_{k}}h_{k}e^{-tK_{k}}\right).
%\end{align*}

It remains to consider the main term $Q_1(E_k^{\oplus} - h_k^{\oplus})$ on the right side of \eqref{eq:Prof-Thm1-2.0}. We use the normal order form in \eqref{eq:NormalOrderedQ1A}: 
\begin{align}
\sum_{k\in S_C} Q_1( E_k^{\oplus} - h_k^{\oplus})= \sum_{k\in S_C} 2\widetilde{Q}_1 ( E_k^{\oplus} - h_k^{\oplus}) + \sum_{k\in S_C} 2{\rm tr}  ( E_k -h_k)+ \sum_{k\in S_C} \varepsilon_k(E_k^{\oplus} -h_k^\oplus).
\end{align}
By Propositions \ref{prop:epskAEstimate}, \ref{thm:ExplicitKAktBktEstimates-2} and \ref{prop:Zeta0RepresentationofHKin}, 
\begin{equation}
\pm\sum_{k\in S_{C}}\varepsilon_{k}\left( E_{k}^{\oplus}-h_{k}^{\oplus}\right)\leq C \sum_{k\in S_C} k_{F}^{-1}\hat{V}_{k}\left(1+\hat{V}_{k}\right) \mathcal{N}_{E} \le Ck_{F}^{-1} H_{\kin}'.
\end{equation} 
Moreover, by Proposition \ref{thm:ExplicitKAktBktEstimates} we have
\begin{align}
\sum_{k\in S_C} \left( 2\tr\left({E}_{k}-h_{k}\right)-\frac{\hat{V}_{k}k_{F}^{-1}}{\left(2\pi\right)^{3}}\left|L_{k}\right| \right) & = \sum_{k\in S_C} \frac{2}{\pi}\int_{0}^{\infty}F\left(\frac{\hat{V}_{k}k_{F}^{-1}}{\left(2\pi\right)^{3}}\sum_{p\in L_{k}}\frac{\lambda_{k,p}}{\lambda_{k,p}^{2}+t^{2}}\right)dt 
%&=  \sum_{k\in \mathbb{Z}_{\ast}^{3}} \frac{1}{\pi}\int_{0}^{\infty}F\left(\frac{\hat{V}_{k}k_{F}^{-1}}{\left(2\pi\right)^{3}}\sum_{p\in L_{k}}\frac{\lambda_{k,p}}{\lambda_{k,p}^{2}+t^{2}}\right)dt + \mathcal{E}_F \nonumber
\end{align}
with $F\left(x\right)=\log\left(1+x\right)-x$.
% and 
%\begin{equation}
%|\mathcal{E}_F| = \left|\frac{1}{\pi}\sum_{k\in\mathbb{Z}_{\ast}^{3}\backslash\overline{B}\left(0,k_{F}^{\gamma}\right)}\int_{0}^{\infty}F\left(\frac{\hat{V}_{k}k_{F}^{-1}}{\left(2\pi\right)^{3}}\sum_{p\in L_{k}}\frac{\lambda_{k,p}}{\lambda_{k,p}^{2}+t^{2}}\right)dt\right|\leq Ck_{F}\sum_{k\in\mathbb{Z}_{\ast}^{3}\backslash\overline{B}\left(0,k_{F}^{\gamma}\right)}\hat{V}_{k}^{2}\left|k\right|.\label{eq:ECalEstimateStart}
%\end{equation}
Thus in summary, we conclude from \eqref{eq:Prof-Thm1-2.0} that
\begin{align}\label{eq:Prof-Thm1-2}
& e^{\mathcal{K}} \left( H_{\kin}'  + \sum_{k\in S_C}  \left(H_{{\rm int}}^{k}-\frac{\hat{V}_{k}k_{F}^{-1}}{\left(2\pi\right)^{3}}\left|L_{k}\right|\right) \right) e^{-\mathcal{K}}  \nonumber\\
&= H_{\kin}^{\prime}+ 2\sum_{k\in S_{C}} \widetilde Q_{1}^{k} (E_{k}^{\oplus}-h_{k}^{\oplus}) + \sum_{k\in S_C} \frac{2}{\pi}\int_{0}^{\infty}F\left(\frac{\hat{V}_{k}k_{F}^{-1}}{\left(2\pi\right)^{3}}\sum_{p\in L_{k}}\frac{\lambda_{k,p}}{\lambda_{k,p}^{2}+t^{2}}\right)dt  + \mathcal{E}_{\mathcal K}
\end{align}
where
\begin{align}\label{eq:cEk-0}
\pm \mathcal{E}_{\mathcal K} \le C(\log k_{F})^{\frac{2}{3}}k_{F}^{-\frac{1}{3}}\left(\sum_{k\in S_C}\hat{V}_{k}\left|k\right|^{3+\frac{2}{3}}\right)\left( H_{\kin}^{\prime}+k_{F}^{-1} \mathcal{N}_{E}H_{\kin}^{\prime} +k_{F}\right).
\end{align}

{\bf Step 3:} Next, we apply the transformation $e^{\mathcal J}$ to the right hand side of \eqref{eq:Prof-Thm1-2}. From \eqref{eq:cEk-0} and the Gronwall estimates of Proposition \ref{prop:KineticGronwall3} we have
\begin{align}\label{eq:Prof-Thm1-err2}
\pm e^{\mathcal{J}} \mathcal{E}_{\mathcal K} e^{-\mathcal{J}} \le C(\log k_{F})^{\frac{2}{3}}k_{F}^{-\frac{1}{3}}\left(\sum_{k\in S_C}\hat{V}_{k}\left|k\right|^{3+\frac{2}{3}}\right)\left( H_{\kin}^{\prime}+k_{F}^{-1} \mathcal{N}_{E}H_{\kin}^{\prime} +k_{F}\right).
\end{align}
For the main terms, by Proposition  \ref{prop:ApplicationoftheSecondTransformation}  
\begin{align} \label{eq:J-main-exchange}
  &e^{\mathcal{J}}\left(H_{\kin}^{\prime}+2\sum_{k\in S_{C}}\tilde{Q}_{1}^{k}\left(E_{k}^{\oplus}-h_{k}^{\oplus}\right)\right)e^{-\mathcal{J}}
\\ 
 & =H_{\kin}^{\prime}+2\sum_{k\in S_{C}}\tilde{Q}_{1}^{k}\left( \widetilde{E}_{k}^{\oplus}-h_{k}^{\oplus}\right)   +2\sum_{k\in S_{C}}\int_{0}^{1}e^{\left(1-t\right)\mathcal{J}} {\mathcal{E}}_{3}^{k}(F_{k}^{\oplus}(t))  e^{-\left(1-t\right)\mathcal{J}}dt .\nonumber
\end{align}
Let us bound the exchange term $\mathcal{E}_3(\cdot)$. For all $k\in\overline{B}\left(0,k_{F}^{\gamma}\right)\cap\mathbb{Z}_{\ast}^{3}$ with 
$0<\gamma<\frac{1}{47}$ and $t\in [0,1]$, by Proposition \ref{them:SecondTransformationOneBodyEstimates} we have
\begin{align}
\max_{p\in L_{k}^{\pm}}\left\Vert \left(h_{k}^{\oplus}\right)^{-\frac{1}{2}}E_k(t) e_{p}\right\Vert &\le Ck_{F}^{-\frac{1}{2}}\left(\hat{V}_{k}+\hat{V}_{k}^{3}\left|k\right|^{6}\log\left(k_{F}\right)\right) \le Ck_{F}^{-\frac{1}{2}}\left(\hat{V}_{k}+\hat{V}_{k}^{3}\left|k\right|^{3} k_F^{\frac 3 {47}}\log\left(k_{F}\right)\right) ,\nonumber\\
\sum_{l\in S_{C}}\left\Vert \left(h_{l}^{\oplus}\right)^{-\frac{1}{2}}J_{l}^{\oplus}\right\Vert _{\HS} &\le C (\log k_{F})^{\frac{2}{3}}k_{F}^{-\frac{1}{3}} \sum_{l\in S_{C}} \hat{V}_{l}\left(1+\hat{V}_{l}\right) \le C (\log k_{F})^{\frac{2}{3}}k_{F}^{-\frac{1}{3}} .
\end{align} 
Hence, using the kinetic estimate of Proposition \ref{prop:KineticcalEEstimate}, Gronwall's bounds of Proposition \ref{prop:KineticGronwall3} and the assumption $\sum_{k\in \mathbb{Z}^3_*} \hat V_k |k| <\infty$, we find that for every state $\Psi\in D\left(H_{\kin}^{\prime}\right)$ and $\Psi_t=e^{-(1-t)\mathcal{J}}\Psi$, 
\begin{align}
&\sum_{k\in S_{C}}\int_{0}^{1} \left| \langle \Psi_t,  {\mathcal{E}}_{3}^{k}(F_{k}^{\oplus}(t))  \Psi_t\rangle \right| dt \\
&\le \sum_{k\in S_{C}} C (\log k_{F})^{\frac{2}{3}}k_{F}^{-\frac{1}{3}}  k_{F}^{-\frac{1}{2}}\left(\hat{V}_{k}+\hat{V}_{k}^{3}\left|k\right|^{3}k_F^{\frac 3 {47}}\log\left(k_{F}\right)\right) \max_{t\in [0,1]}   \sqrt{\left\langle \Psi_t,H_{\kin}^{\prime}\Psi_t\right\rangle \left\langle \Psi_t,\mathcal{N}_{E}H_{\kin}^{\prime}\Psi_t \right\rangle} \nonumber\\
%&\le C (\log k_F)^{\frac 5 3} k_F^{-\frac{1}{3}} k_{F}^{-\frac{1}{2}} \sqrt{\left\langle \Psi,(H_{\kin}^{\prime}+k_F) \Psi \right\rangle \left\langle \Psi, (\mathcal{N}_{E}H_{\kin}^{\prime} + k_F H_{\kin}^{\prime}+k_F^2 )   \Psi \right\rangle}   \nonumber \\
&\le C(\log k_F)^{\frac 5 3} k_F^{-\frac{1}{3}}  \left\langle \Psi, ( k_F^{-1}\mathcal{N}_{E}H_{\kin}^{\prime} +  H_{\kin}^{\prime}+k_F )   \Psi \right\rangle. \nonumber
\end{align}
Here we used $\sum_{k\in \mathbb{Z}^3_*} \hat V_k^3 |k|^3 \le \left( \sum_{k\in \mathbb{Z}^3_*} \hat V_k |k| \right)^3 <\infty$. Consequently, %, the exchange terms in \eqref{eq:J-main-exchange} is bounded as
\begin{align} \label{eq:Prof-Thm1-err3}
\pm \sum_{k\in S_{C}}\int_{0}^{1}e^{\left(1-t\right)\mathcal{J}} {\mathcal{E}}_{3}^{k}(F_{k}^{\oplus}(t))  e^{-\left(1-t\right)\mathcal{J}}dt \le C  (\log k_F)^{\frac 5 3} k_F^{-\frac{1}{3}} (k_F^{-1}\mathcal{N}_{E}H_{\kin}^{\prime} + H_{\kin}^{\prime}+k_F ). 
\end{align}
In summary, we have for $\mathcal{U}= e^{\mathcal{J}} e^{\mathcal{K}}$ and $0<\gamma<\frac 1 {47}$, 
\begin{align} \label{eq:Prof-Thm1-3}
\mathcal{U} H_N \mathcal{U}^* = E_{\rm FS} + H_{\kin}^{\prime}+2\sum_{k\in S_{C}}\tilde{Q}_{1}^{k}\left( \widetilde{E}_{k}^{\oplus}-h_{k}^{\oplus}\right)  +  \sum_{k\in S_C} \frac{2}{\pi}\int_{0}^{\infty}F\left(\frac{\hat{V}_{k}k_{F}^{-1}}{\left(2\pi\right)^{3}}\sum_{p\in L_{k}}\frac{\lambda_{k,p}}{\lambda_{k,p}^{2}+t^{2}}\right)dt  + \mathcal{E}_{\mathcal J}
\end{align}
where the error term is collected from \eqref{eq:Prof-Thm1-err1}, \eqref{eq:Prof-Thm1-err2}, \eqref{eq:Prof-Thm1-err3} which satisfies 
\begin{align}
\pm \mathcal{E}_{\mathcal J} \le C k_F^{-\gamma/2} (k_F^{-1}\mathcal{N}_{E}H_{\kin}^{\prime} + H_{\kin}^{\prime}+k_F ). 
\end{align}

{\bf Step 4:} Finally let us remove the cut-off $S_C= \mathbb{Z}_{+}^{3}\cap \overline{B}\left(0,k_{F}^{\gamma}\right) $ on the right hand side of \eqref{eq:Prof-Thm1-3}. 
%\begin{align}
%delete
%\end{align}
%For the missing tail of 
%\begin{align}
% \sum_{k\in S_C} \frac{2}{\pi}\int_{0}^{\infty}F\left(\frac{\hat{V}_{k}k_{F}^{-1}}{\left(2\pi\right)^{3}}\sum_{p\in L_{k}}\frac{\lambda_{k,p}}{\lambda_{k,p}^{2}+t^{2}}\right)dt =\sum_{k\in\mathbb{Z}_{\ast}^{3}\cap\overline{B}\left(0,k_{F}^{\gamma}\right)}  \frac{1}{\pi}\int_{0}^{\infty}F\left(\frac{\hat{V}_{k}k_{F}^{-1}}{\left(2\pi\right)^{3}}\sum_{p\in L_{k}}\frac{\lambda_{k,p}}{\lambda_{k,p}^{2}+t^{2}}\right)dt 
%\end{align}
By Proposition \ref{thm:ExplicitKAktBktEstimates} we can bound 
\begin{equation}
\left|\frac{1}{\pi}\sum_{k\in\mathbb{Z}_{\ast}^{3}\backslash S_C}\int_{0}^{\infty}F\left(\frac{\hat{V}_{k}k_{F}^{-1}}{\left(2\pi\right)^{3}}\sum_{p\in L_{k}}\frac{\lambda_{k,p}}{\lambda_{k,p}^{2}+t^{2}}\right)dt\right|\leq Ck_{F}\sum_{k\in\mathbb{Z}_{\ast}^{3}\backslash S_C}\hat{V}_{k}^{2}\left|k\right| \le C k_F^{1 - \gamma} .\label{eq:ECalEstimateStart}
\end{equation}
Here we used $\sum_{k\in \mathbb{Z}^3_*} \hat V_k^2 |k|^2 \le \left( \sum_{k\in \mathbb{Z}^3_*} \hat V_k |k| \right)^2 <\infty$.
Moreover, by Propositions
\ref{prop:KineticQ1AEstimate} and \ref{them:SecondTransformationOneBodyEstimates} (together with the fact that the trace norm dominates the operator norm) we can bound
\begin{align}
\pm\tilde{Q}_{1}^{k}\left(\widetilde{E}_{k}^{\oplus}-h_{k}^{\oplus}\right) &\leq \left\Vert \left(h_{k}^{\oplus}\right)^{-\frac{1}{2}}\left(\widetilde{E}_{k}^{\oplus}-h_{k}^{\oplus}\right)\left(h_{k}^{\oplus}\right)^{-\frac{1}{2}}\right\Vert _{\text{Op}}H_{\kin}^{\prime} \nonumber\\
&= \left\Vert h_{k}^{-\frac{1}{2}}\left(\widetilde{E}_{k} -h_{k} \right) h_{k}^{-\frac{1}{2}}\right\Vert _{\text{Op}}H_{\kin}^{\prime} \le C \hat V_k H_{\kin}^{\prime}
\end{align}
for all $k\in \mathbb{Z}^3_+$, and hence
\begin{align}
\pm \sum_{k\in \mathbb{Z}^3_+\backslash S_C} \tilde{Q}_{1}^{k}\left(\widetilde{E}_{k}^{\oplus}-h_{k}^{\oplus}\right)  \le C \left( \sum_{k\in \mathbb{Z}^3_+\backslash S_C}  \hat V_k \right) H_{\kin}^{\prime} \le C k_F^{-\gamma}H_{\kin}^{\prime}. 
\end{align}
Therefore, we can deduce from \eqref{eq:Prof-Thm1-3} that for $\mathcal{U}= e^{\mathcal{J}} e^{\mathcal{K}}$ and $0<\gamma<\frac 1 {47}$, 
\begin{align} \label{eq:Prof-Thm1-4}
\mathcal{U} H_N \mathcal{U}^* =E_{\rm FS} + H_{\kin}^{\prime}+2\sum_{k\in \mathbb{Z}^3_+}\tilde{Q}_{1}^{k}\left( \widetilde{E}_{k}^{\oplus}-h_{k}^{\oplus}\right)  +  \sum_{k\in \mathbb{Z}^3_*} \frac{1}{\pi}\int_{0}^{\infty}F\left(\frac{\hat{V}_{k}k_{F}^{-1}}{\left(2\pi\right)^{3}}\sum_{p\in L_{k}}\frac{\lambda_{k,p}}{\lambda_{k,p}^{2}+t^{2}}\right)dt  + \mathcal{E}_{\mathcal U}
\end{align}
where
\begin{align}
\pm \mathcal{E}_{\mathcal U} \le C k_F^{-\gamma/2} (k_F^{-1}\mathcal{N}_{E}H_{\kin}^{\prime} + H_{\kin}^{\prime}+k_F ). 
\end{align}
The statement of Theorem \ref{them:OperatorStatement} follows by recognizing the identity 
$$
2\sum_{k\in \mathbb{Z}^3_+}\tilde{Q}_{1}^{k}\left( \widetilde{E}_{k}^{\oplus}-h_{k}^{\oplus}\right) = 2\sum_{k\in\mathbb{Z}_{\ast}^{3}}\sum_{p,q\in L_{k}}\left\langle e_{p},\left(\widetilde{E}_{k}-h_{k}\right)e_{q}\right\rangle b_{k,p}^{\ast}b_{k,q},
$$
which follows from the definition of $\tilde{Q}_{1}^{k}$  in \eqref{eq:def-Q1-tilde}. 
$\hfill\square$

\subsection{Proof of Theorem \ref{thm:Kinetic-estimate}}

%In this subsection we prove the part not concerning $\mathcal{U}$ of Theorem \ref{thm:Kinetic-estimate}. 
%
%\begin{thm} \label{thm:Kinetic-estimate-1} Let $V$ be as in Theorem \ref{them:OperatorStatement}. Then we have the operator lower bound 
%\begin{align} \label{eq:HN-lower-bound-1}
%H_N \ge  E_{\rm FS}  + H_{\rm kin}'  - C k_F.
%\end{align}
%Moreover, if $\Psi\in \mathcal{H}_N$ is a normalized eigenstate of $H_{N}$ with energy $\left\langle \Psi,H_{N}\Psi\right\rangle \le  E_{\rm FS}  +\kappa k_F$ for some $\kappa>0$, then for all $s\in [0,2]$ we have 
%\begin{align} \label{eq:HN-lower-bound-2}
%\left\langle \Psi,\mathcal{N}_{E}^{s}H_{\kin}^{\prime}\Psi\right\rangle \leq C (\kappa +1 ) k_F^{1+s}.
%\end{align}
%Here the constant $C>0$ depends only on $V$ via $\sum_{k\in \mathbb{Z}^3_*} \hat V_k |k|$. 
%\end{thm}

Let $\Psi\in D\left(H_{\kin}^{\prime}\right)$ be a normalized eigenstate of $H_{N}$ with energy $\left\langle \Psi,H_{N}\Psi\right\rangle \le E_{\rm FS} + \kappa k_F$ for some $\kappa>0$.  Denoting $\tilde{H}_{N}=H_{N}- E_{\rm FS}$, we have $\tilde{H}_{N}\Psi=E'\Psi$ with $E' \le \kappa k_F$.  Using \eqref{eq:HN-localized-1} and the obvious inequality $A^*A\ge 0$ we obtain the Onsager-type estimate
\begin{align}
\tilde{H}_{N} - H_{\kin}^{\prime} &= \frac{k_{F}^{-1}}{2\left(2\pi\right)^{3}}\sum_{k\in\mathbb{Z}_{\ast}^{3}}\hat{V}_{k}\left(\text{d}\Gamma\left(e^{-ik\cdot x}\right)^{\ast}\text{d}\Gamma\left(e^{-ik\cdot x}\right)-\left|L_{k}\right|\right)\label{eq:OnsagerEstimate}\\
 & \geq -\frac{k_{F}^{-1}}{2\left(2\pi\right)^{3}}\sum_{k\in\mathbb{Z}_{\ast}^{3}}\hat{V}_{k}\left|L_{k}\right| \ge - Ck_{F} \sum_{k\in \mathbb{Z}^3} |k| \hat V_k.\nonumber 
\end{align}
%The first bound \eqref{eq:HN-lower-bound-1} can be deduced by a form of Onsager's lemma. More precisely, 
Here we used  $\left|L_{k}\right|\leq Ck_{F}^{2}\left|k\right|$ for all $k\in\mathbb{Z}_{\ast}^{3}$ (see Proposition \ref{prop:NonSingularRiemannSums}).  From  \eqref{eq:OnsagerEstimate} and the assumption $\tilde{H}_{N}\Psi=E'\Psi$ with $E' \le \kappa k_F$, we deduce immediately  that 
%Next, let  $\Psi$  be a normalized eigenstate of $H_{N}$ such that $\tilde{H}_{N}\Psi=E'\Psi$ with $E' \le \kappa k_F$. From \eqref{eq:HN-lower-bound-1} we obtain immediately that
\begin{equation} \label{eq:Hkin<=kF}
\langle \Psi, H_{\rm kin}' \Psi\rangle \le C (\kappa+1) k_F. 
\end{equation}
To prove the bound for $\mathcal{N}_E H_{\rm kin}'$, we use  the operator inequality 
%
%We will use a bootstrap argument based on the eigenvalue equation  $\tilde{H}_{N}\Psi=E'\Psi$, which is originated from the study of Bose gases in \cite{Seiringer-11,GreSei-13,Nam-17,NamNap-21}. 
\begin{align}
\mathcal{N}_{E}^{2}H_{\kin}^{\prime}&=\mathcal{N}_{E}H_{\kin}^{\prime}\mathcal{N}_{E}\leq\mathcal{N}_{E}\tilde{H}_{N}\mathcal{N}_{E}+Ck_{F}\mathcal{N}_{E}^{2} \nonumber\\
&= \frac{1}{2}\left(\mathcal{N}_{E}^{2}\tilde{H}_{N}+\tilde{H}_{N}\mathcal{N}_{E}^{2}-\left[\mathcal{N}_{E},\left[\mathcal{N}_{E},\tilde{H}_{N}\right]\right]\right)+Ck_{F}\mathcal{N}_{E}^{2}
\end{align}
which follows from \eqref{eq:OnsagerEstimate} and the fact that $\left[\mathcal{N}_{E},H_{\kin}^{\prime}\right]=0$. Thanks to the eigenvalue equation  $\tilde{H}_{N}\Psi=E'\Psi$ with $E'\le \kappa k_F$, we deduce that 
\begin{equation}
\left\langle \Psi,\mathcal{N}_{E}^{2}H_{\kin}^{\prime}\Psi\right\rangle \leq C(\kappa+1) k_{F} \left\langle \Psi,\mathcal{N}_{E}^{2}\Psi\right\rangle -\frac{1}{2}\left\langle \Psi,\left[\mathcal{N}_{E},\left[\mathcal{N}_{E},\tilde{H}_{N}\right]\right]\Psi\right\rangle. \label{eq:FirstEigenstateBound}
\end{equation}
Using $\mathcal{N}_{E} = \sum_{s\in B_F^c} c_{s}^* c_s$ and 
\begin{equation}
[c_{s}^* c_s, c_{p+k}^{\ast}c_{q-k}^{\ast}c_{q}c_{p}] = c_{p+k}^{\ast}c_{q-k}^{\ast}c_{q}c_{p} (\delta_{s,p+k} + \delta_{s,q-k}- \delta_{s,q}- \delta_{s,p})
\end{equation}
we deduce from  \eqref{eq:IntroductionSecondQuantizedHamiltonian} that 
\begin{align}
\left[\mathcal{N}_{E},\left[\mathcal{N}_{E},\tilde{H}_{N}\right]\right] %&= \frac{k_{F}^{-1}}{2\left(2\pi\right)^{3}}\sum_{k\in\mathbb{Z}^{3}}\sum_{p,q\in\mathbb{Z}^{3}}\hat{V}_{k} \left[\mathcal{N}_{E},\left[\mathcal{N}_{E}, c_{p+k}^{\ast}c_{q-k}^{\ast}c_{q}c_{p}\right]\right] \\
=\frac{k_{F}^{-1}}{2\left(2\pi\right)^{3}}\sum_{k\in\mathbb{Z}^{3}}  \sum_{p,q\in\mathbb{Z}^{3}} \hat{V}_{k} c_{p+k}^{\ast}c_{q-k}^{\ast}c_{q}c_{p}  \Big(  \sum_{s\in B_F^c}  (\delta_{s,p+k} + \delta_{s,q-k}- \delta_{s,q}- \delta_{s,p}) \Big)^2. %\nonumber
\end{align}
Using the obvious bound 
\begin{equation}
0\le \Big(  \sum_{s\in B_F^c}  (\delta_{s,p+k} + \delta_{s,q-k}- \delta_{s,q}- \delta_{s,p}) \Big)^2 \le 4
\end{equation}
and the Cauchy--Schwarz inequality, we  estimate 
\begin{align}  \label{eq:SecondEigenstateBound-0a}
&\left| \left\langle \Psi,\left[\mathcal{N}_{E},\left[\mathcal{N}_{E},\tilde{H}_{N}\right]\right]\Psi\right\rangle \right|  \le C k_F^{-1}  \sum_{k\in\mathbb{Z}^{3}} \sum_{p,q\in\mathbb{Z}^{3}}  \hat V_k  \| c_{p+k} c_{q-k} \Psi\| \|c_{q}c_{p} \Psi\|  \nonumber \\
&\le C k_F^{-1}  \sum_{k\in\mathbb{Z}^{3}}  \hat V_k \sum_{p,q\in\mathbb{Z}^{3}}  ( \| c_{p+k} c_{q-k} \Psi\|^2 +  \|c_{q}c_{p} \Psi\|^2)  \le C k_F^{-1}  \sum_{k\in\mathbb{Z}^{3}}  \hat V_k \langle \Psi, \mathcal{N}_E^2 \Psi\rangle.
\end{align}
Since $\hat V$ is summable, \eqref{eq:FirstEigenstateBound} and \eqref{eq:SecondEigenstateBound-0a} imply that
\begin{align}\label{eq:SecondEigenstateBound-0}
\left\langle \Psi,\mathcal{N}_{E}^{2}H_{\kin}^{\prime}\Psi\right\rangle \leq C(\kappa+1) k_{F} \left\langle \Psi,\mathcal{N}_{E}^{2}\Psi\right\rangle.  
\end{align}
Combining with the inequality $H_{\kin}^{\prime} \ge \mathcal{N}_{E}$ from Proposition \ref{prop:Zeta0RepresentationofHKin}, we deduce by H\"older's inequality
\begin{equation}
\left\langle \Psi,\mathcal{N}_{E}^{2}\Psi\right\rangle \le \left\langle \Psi,\mathcal{N}_{E}^{3} \Psi\right\rangle^{2/3} \le \left\langle \Psi,\mathcal{N}_{E}^{2}H_{\kin}^{\prime}\Psi\right\rangle^{2/3} \leq \Big( C(\kappa+1) k_{F} \left\langle \Psi,\mathcal{N}_{E}^{2}\Psi\right\rangle \Big)^{2/3},
\end{equation}
which implies that $\left\langle \Psi,\mathcal{N}_{E}^{2} \Psi\right\rangle \le C (\kappa +1)^2k_F^2$, and hence by \eqref{eq:SecondEigenstateBound-0} again 
%and hence $\left\langle \Psi,\mathcal{N}_{E}^{2} \Psi\right\rangle \le C (\kappa +1)^2$. Thus \eqref{eq:SecondEigenstateBound-0} implies that \eqref{eq:HN-lower-bound-2} holds for $s=2$, namely
\begin{align} \label{eq:HN-lower-bound-3}
\quad \langle \Psi, \mathcal{N}_E^2 H_{\rm kin}' \Psi\rangle \le C (\kappa+1)^3 k_F^3. 
\end{align}
The bound $\left\langle \Psi,\mathcal{N}_{E} H_{\kin}' \Psi\right\rangle \le C(\kappa+1)^2k_F^2$ follows from \eqref{eq:Hkin<=kF} and \eqref{eq:HN-lower-bound-3}.  In summary,  we have
\begin{align}
\left\langle \Psi, (k_F^{-1}\mathcal{N}_{E} H_{\kin}^{\prime} + H_{\kin}^{\prime} + k_F) \Psi\right\rangle \leq C (\kappa +1 )^2 k_F.
\end{align}
By the Gronwall estimates of Propositions \ref{prop:KineticGronwall}, \ref{prop:KineticGronwall3} and the choice $\mathcal{U}=e^{\mathcal{J}}e^{\mathcal{K}}$, we also obtain 
\begin{align}
\left\langle \mathcal{U}  \Psi, (k_F^{-1}\mathcal{N}_{E} H_{\kin}^{\prime} + H_{\kin}^{\prime} + k_F) \mathcal{U}  \Psi\right\rangle \leq C (\kappa +1 )^2 k_F.
\end{align}
$\hfill\square$

\subsection{Proof of Theorem \ref{thm:intro-GSE}}\label{sec:KineticGronwallEstimatesandFinalDetailsfortheLowerBound}

Taking the expectation against  $\Psi_{\rm FS}$ of the operator estimate in Theorem \ref{them:OperatorStatement} we have
\begin{align}
\inf \sigma(H_N) =\inf \sigma(\mathcal{U} H_N\mathcal{U}^*)  \le \left\langle \Psi_{\rm FS},\mathcal{U} H_N\mathcal{U}^* \Psi_{\rm FS}\right\rangle = E_{\rm FS} + E_{\rm corr} + O(k_F^{1-\frac{1}{94}+\epsilon}). 
\end{align}
Here we used the bound on $\mathcal{E}_U$ from Theorem \ref{them:OperatorStatement} and the identities $H_{\kin}' \Psi_{\rm FS} =  H_{\rm eff} \Psi_{\rm FS}  =0$.

To see the lower bound, let  $\Psi_{\rm GS}\in D\left(H_{\kin}^{\prime}\right)$ be the normalized ground state of $H_{N}$. By the definition of  $\Psi_{\rm GS}$ and the above upper bound, we have
\begin{align}
\left\langle \Psi_{\rm GS},H_{N} \Psi_{\rm GS} \right\rangle = \inf \sigma(H_N)\le   E_{\rm FS} + Ck_F, 
\end{align}
and hence Theorem \ref{thm:Kinetic-estimate} implies that the state $\Psi_{\rm GS}'= \mathcal{U} \Psi_{\rm GS}$  satisfies 
\begin{align} \label{eq:kinetic-bound-GS}
\left\langle  \Psi_{\rm GS}', (k_F^{-1}\mathcal{N}_{E} H_{\kin}^{\prime} + H_{\kin}^{\prime} + k_F)   \Psi_{\rm GS}' \right\rangle \leq C  k_F.
\end{align}
Taking the expectation against $\Psi_{\rm GS}'$ of the operator estimate in Theorem \ref{them:OperatorStatement} we conclude that 
\begin{align}
\inf \sigma(H_N) &= \left\langle \Psi_{\rm GS},H_{N} \Psi_{\rm GS} \right\rangle = \left\langle  \Psi_{\rm GS}',\mathcal{U}  H_{N} \mathcal{U} ^*  \Psi_{\rm GS}' \right\rangle\\
&=E_{\rm FS} + E_{\rm corr} +  \left\langle \Psi_{\rm GS}', \left( H_{\kin}^{\prime}+ H_{\rm eff} +\mathcal{E}_{\mathcal U} \right) \Psi_{\rm GS}' \right\rangle   \ge E_{\rm FS} + E_{\rm corr} + O(k_F^{1-\frac{1}{94}+\epsilon}).  \nonumber
\end{align}
Here we used the operator inequalities 
\begin{align}
H_{\kin}^{\prime}\ge 0,\quad H_{\rm eff}\ge 0, \quad \mathcal{E}_{\mathcal U} \ge C k_F^{1-\frac{1}{94}+\epsilon} (k_F^{-1}\mathcal{N}_{E} H_{\kin}^{\prime} + H_{\kin}^{\prime} + k_F) 
\end{align}  
and the a-priori estimate \eqref{eq:kinetic-bound-GS}. This completes the proof of Theorem \ref{thm:intro-GSE}. $\hfill\square$

%%%%%%%%%%%%%%%%%%%%%%%%%%%%%%%%%%%
%%%%%%%%%%%%%%%%%%%%%%%%%%%%%%%%%%%

\subsection{Proof of Theorems \ref{thm:intro-excitations} and \ref{thm:intro-excitations-II}}

In this subsection we study the effective operator $H_{\rm eff}$ in  
Theorem \ref{them:OperatorStatement}  
%\begin{equation}
%H_{\text{eff}}=H_{\kin}^{\prime}+2\sum_{k\in\mathbb{Z}_{\ast}^{3}}\sum_{p,q\in L_{k}}\left\langle e_{p},\left(E_{k}-h_{k}\right)e_{q}\right\rangle b_{k,p}^{\ast}b_{k,q},
%\end{equation}
in more detail.  First we prove the following remarkable fact.

\begin{prop} \label{lem:Heff} We have the operator identity on $D(H_{\kin}^{\prime})$: 
\[
2\sum_{k\in\mathbb{Z}_{\ast}^{3}}\sum_{p\in L_{k}}\lambda_{k,p}b_{k,p}^{\ast}b_{k,p}=\mathcal{N}_{E}H_{\kin}^{\prime}. 
\]
%\[
%2\sum_{k\in\overline{B}\left(0,R\right)\cap\mathbb{Z}_{\ast}^{3}}\sum_{p\in L_{k}}\lambda_{k,p}b_{k,p}^{\ast}b_{k,p}\Psi=\mathcal{N}_{E}H_{\kin}^{\prime}\Psi.
%\]
\end{prop}

%we must address the question of the bosonic analogy for $H_{\kin}^{\prime}$:
%In this paper we have been guided by the idea that there is an indirect
%correspondence between $H_{\kin}^{\prime}$ and the quasi-bosonic
%operator
%\begin{equation}
%2\sum_{k\in\mathbb{Z}_{\ast}^{3}}\sum_{p,q\in L_{k}}\left\langle e_{p},h_{k}e_{q}\right\rangle b_{k,p}^{\ast}b_{k,q}=2\sum_{k\in\mathbb{Z}_{\ast}^{3}}\sum_{p\in L_{k}}\lambda_{k,p}b_{k,p}^{\ast}b_{k,p}=\sum_{k\in\mathbb{Z}_{\ast}^{3}}\sum_{p\in L_{k}}\left(\left|p\right|^{2}-\left|p-k\right|^{2}\right)b_{k,p}^{\ast}b_{k,p},
%\end{equation}
%so we must now ask ourselves if there is a more direct relation between
%these two operators that would allow us to cancel these with each
%other, such that $H_{\text{eff}}$ is only
%\begin{equation}
%H_{\text{eff}}\sim2\sum_{k\in\mathbb{Z}_{\ast}^{3}}\sum_{p,q\in L_{k}}\left\langle e_{p},E_{k}e_{q}\right\rangle b_{k,p}^{\ast}b_{k,q}
%\end{equation}
%as would be expected from the assumptions of the RPA.
%
%There is indeed a direct relation between the operators, but contrary
%to implying a cancellation betwem them, it shows that in general,
%only slight cancellation occurs, as there holds the remarkable identity
%\begin{equation}
%2\sum_{k\in\mathbb{Z}_{\ast}^{3}}\sum_{p\in L_{k}}\lambda_{k,p}b_{k,p}^{\ast}b_{k,p}=\mathcal{N}_{E}H_{\kin}^{\prime}.\label{eq:RemarkableKineticIdentity}
%\end{equation}

\bigskip
\textbf{Proof of Proposition \ref{lem:Heff}:} The idea is simply to interchange the summation
on $k\in\mathbb{Z}_{\ast}^{3}$ and $p\in L_{k}$: By rephrasing the
condition that $p\in L_{k}$, we have the equivalences
\begin{align}
\left(k\in\mathbb{Z}_{\ast}^{3}\right)\wedge\left(p\in L_{k}\right) & \Leftrightarrow\left(k\in\mathbb{Z}_{\ast}^{3}\right)\wedge\left(\left|p-k\right|\leq k_{F}<\left|p\right|\right)\nonumber \\
 & \Leftrightarrow\left(k\in\mathbb{Z}_{\ast}^{3}\right)\wedge\left(k\in\overline{B}\left(p,k_{F}\right)\right)\wedge\left(p\in B_{F}^{c}\right)\label{eq:InterchangementEquivalences}\\
 & \Leftrightarrow\left(p\in B_{F}^{c}\right)\wedge\left(k\in\overline{B}\left(p,k_{F}\right)\cap\mathbb{Z}^{3}\right),\nonumber 
\end{align}
where we could replace $\mathbb{Z}_{\ast}^{3}$ by $\mathbb{Z}^{3}$
in the last line as the conditions $p\in B_{F}^{c}=\mathbb{Z}^{3}\backslash\overline{B}\left(0,k_{F}\right)$
and $k\in\overline{B}\left(p,k_{F}\right)$ exclude $k=0$ automatically.
Recognizing that $\overline{B}\left(p,k_{F}\right)\cap\mathbb{Z}^{3}=B_{F}+p$,
we can now write
\begin{align}
2\sum_{k\in\mathbb{Z}_{\ast}^{3}}\sum_{p\in L_{k}}\lambda_{k,p}b_{k,p}^{\ast}b_{k,p} & =\sum_{k\in\mathbb{Z}_{\ast}^{3}}\sum_{p\in L_{k}}\left(\left|p\right|^{2}-\left|p-k\right|^{2}\right)b_{k,p}^{\ast}b_{k,p}\\
 & =\sum_{p\in B_{F}^{c}}\sum_{k\in\left(B_{F}+p\right)}\left|p\right|^{2}b_{k,p}^{\ast}b_{k,p}-\sum_{p\in B_{F}^{c}}\sum_{k\in\left(B_{F}+p\right)}\left|p-k\right|^{2}b_{k,p}^{\ast}b_{k,p},\nonumber 
\end{align}
and by expanding the excitation operators we find for the first sum
that
\begin{align}
\sum_{p\in B_{F}^{c}}\sum_{k\in\left(B_{F}+p\right)}\left|p\right|^{2}b_{k,p}^{\ast}b_{k,p} & =\sum_{p\in B_{F}^{c}}\sum_{k\in\left(B_{F}+p\right)}\left|p\right|^{2}c_{p}^{\ast}c_{p-k}c_{p-k}^{\ast}c_{p}=\sum_{p\in B_{F}^{c}}\left(\sum_{k\in\left(B_{F}+p\right)}c_{p-k}c_{p-k}^{\ast}\right)\left|p\right|^{2}c_{p}^{\ast}c_{p}\nonumber \\
 & =\sum_{p\in B_{F}^{c}}\left(\sum_{k\in B_{F}}c_{-k}c_{-k}^{\ast}\right)\left|p\right|^{2}c_{p}^{\ast}c_{p}=\mathcal{N}_{E}\sum_{p\in B_{F}^{c}}\left|p\right|^{2}c_{p}^{\ast}c_{p}
\end{align}
as $\sum_{k\in B_{F}}c_{-k}c_{-k}^{\ast}=\sum_{k\in B_{F}}c_{k}c_{k}^{\ast}=\mathcal{N}_{E}$
by the particle-hole symmetry, and similarly
\begin{align}
\sum_{p\in B_{F}^{c}}\sum_{k\in\left(B_{F}+p\right)}\left|p-k\right|^{2}b_{k,p}^{\ast}b_{k,p} & =\sum_{p\in B_{F}^{c}}c_{p}^{\ast}c_{p}\sum_{k\in\left(B_{F}+p\right)}\left|p-k\right|^{2}c_{p-k}c_{p-k}^{\ast}\\
 & =\sum_{p\in B_{F}^{c}}c_{p}^{\ast}c_{p}\sum_{k\in B_{F}}\left|k\right|^{2}c_{k}c_{k}^{\ast}=\mathcal{N}_{E}\sum_{k\in B_{F}}\left|k\right|^{2}c_{k}c_{k}^{\ast}\nonumber 
\end{align}
for the claimed equality of
\begin{equation}
T= 2\sum_{k\in\mathbb{Z}_{\ast}^{3}}\sum_{p\in L_{k}}\lambda_{k,p}b_{k,p}^{\ast}b_{k,p}=\mathcal{N}_{E}\left(\sum_{p\in B_{F}^{c}}\left|p\right|^{2}c_{p}^{\ast}c_{p}-\sum_{p\in B_{F}}\left|p\right|^{2}c_{p}c_{p}^{\ast}\right)=\mathcal{N}_{E}H_{\kin}^{\prime}.
\end{equation}

To complete the proof, let us show that the relevant operators are well-defined on the domain $D(H_{\rm kin}')$. This is clear for $\mathcal{N}_{E}H_{\kin}^{\prime}$  since $\mathcal{N}_{E}$ is a bounded operator ($0\le \mathcal{N}_E\le N$ on $\mathcal{H}_N$). For $T$, we can interchange the summations of $k$ and $p$ using the same observation in \eqref{eq:InterchangementEquivalences}. This gives the quadratic form estimate
\begin{align}
T&=2\sum_{k\in \mathbb{Z}_{\ast}^{3}}\sum_{p\in L_{k}}\lambda_{k,p}b_{k,p}^{\ast}b_{k,p} =\sum_{k\in\mathbb{Z}_{\ast}^{3}}\sum_{p\in L_{k}}\vert\left|p\right|^{2}-\zeta\vert\,b_{k,p}^{\ast}b_{k,p}+\sum_{k\in \mathbb{Z}_{\ast}^{3}}\sum_{p\in L_{k}}\vert\left|p-k\right|^{2}-\zeta\vert\,b_{k,p}^{\ast}b_{k,p} \nonumber\\
&\le \sum_{p\in B_{F}^{c}}\left(\sum_{k\in B_{F}}c_{k}c_{k}^{\ast}\right)\vert\left|p\right|^{2}-\zeta\vert\,c_{p}^{\ast}c_{p}+\sum_{p\in B_{F}^{c}}c_{p}^{\ast}c_{p}\sum_{k\in B_{F}}\vert\left|k\right|^{2}-\zeta\vert\,c_{k}c_{k}^{\ast}\le \mathcal{N}_E H_{\rm kin}'\label{eq:TRIdentity}
\end{align}
%for brevity. Using the identity
%\begin{equation}
%2\lambda_{k,p}=\left|p\right|^{2}-\left|p-k\right|^{2}=\vert\left|p\right|^{2}-\zeta\vert+\vert\left|p-k\right|^{2}-\zeta\vert,
%\end{equation}
where $\zeta>0$ is the constant in \eqref{eq:ManifestlyNonNegativeHKin}. 
%defined by Proposition \ref{prop:Zeta0RepresentationofHKin},
%we may write $T_{R}$ as
%\begin{equation}
%T_{R}=\sum_{k\in\overline{B}\left(0,R\right)\cap\mathbb{Z}_{\ast}^{3}}\sum_{p\in L_{k}}\vert\left|p\right|^{2}-\zeta\vert\,b_{k,p}^{\ast}b_{k,p}+\sum_{k\in\overline{B}\left(0,R\right)\cap\mathbb{Z}_{\ast}^{3}}\sum_{p\in L_{k}}\vert\left|p-k\right|^{2}-\zeta\vert\,b_{k,p}^{\ast}b_{k,p}.
%\end{equation}
Moreover, it is easily seen that $T$ commutes with both $\mathcal{N}_E$ and $H_{\kin}^{\prime}$. Therefore, the above quadratic form estimate also implies the stronger estimate
\begin{equation}
T^{2}\leq\left(\mathcal{N}_{E}H_{\kin}^{\prime}\right)^{2}
\end{equation}
which justifies that $D(T)\subset D(\mathcal{N}_{E}H'_{\kin}) \subset D(H_{\kin}')$. 

$\hfill\square$

Now we are ready to give the 

\medskip

{\bf Proof of Theorem \ref{thm:intro-excitations-II}:} Thanks to Proposition \ref{lem:Heff} and the identity $\langle e_p, h_k e_q\rangle=\lambda_{k,p}\delta_{p,q}$, we have  
\begin{align}
H_{\text{eff}}&=H_{\kin}^{\prime}+2\sum_{k\in\mathbb{Z}^{3}_* }\sum_{p,q\in L_{k}}\left\langle e_{p},\left(\widetilde{E}_{k}-h_{k}\right)e_{q}\right\rangle b_{k,p}^{\ast}b_{k,q}  \nonumber\\
&=2\sum_{k\in\mathbb{Z}_{\ast}^{3}}\sum_{p,q\in L_{k}}\left\langle e_{p},\widetilde{E}_{k}e_{q}\right\rangle b_{k,p}^{\ast}b_{k,q}-\left(\mathcal{N}_{E}-1\right)H_{\kin}^{\prime}.
\end{align}
Since $\left[H_{\text{eff}},\mathcal{N}_{E}\right]=0$, we can restrict $H_{\text{eff}}$ to the eigenspaces of $\mathcal{N}_{E}$: for every $M=\{1,2,...\}$ we can  write the restriction to $\left\{ \mathcal{N}_{E}=M\right\} $
for $M\in\mathbb{N}$ in the quasi-bosonic form
\begin{equation}
\left.H_{\text{eff}}\right|_{\mathcal{N}_{E}=M}=2\sum_{k\in\mathbb{Z}_{\ast}^{3}}\sum_{p,q\in L_{k}}\left\langle e_{p},\left(\widetilde{E}_{k}-\left(1-M^{-1}\right)h_{k}\right)e_{q}\right\rangle b_{k,p}^{\ast}b_{k,q}.\label{eq:RestrictedEffectiveOperator}
\end{equation}
$\hfill\square$

{\bf Proof of Theorem \ref{thm:intro-excitations}:} We only need to verify the statement on the effective operator $\left.H_{\text{eff}}\right|_{\mathcal{N}_{E}=M}$ with $M=1$. In this case it is convenient to introduce the 
total momentum $P=\left(P_{1},P_{2},P_{3}\right)$, where each $P_{j}$
is given by $P_{j}=\sum_{p\in\mathbb{Z}^{3}}p_{j}c_{p}^{\ast}c_{p}.$ 
It is easily checked that $P_{j}$ obeys the commutators
\begin{equation}
\left[P_{j},b_{k,p}\right]=-k_{j}b_{k,p},\quad\left[P_{j},b_{k,p}^{\ast}\right]=k_{j}b_{k,p}^{\ast},
\end{equation}
and additionally $\left[P_{j},H_{\kin}^{\prime}\right]=0$,
whence the effective Hamiltonian $H_{\text{eff}}$ also commutes with
$P_{j}$, $j=1,2,3$. It also holds that $\left[\mathcal{N}_{E},P_{j}\right]=0$,
so we may restrict $H_{\text{eff}}$ to the simultanous eigenspaces
of $\mathcal{N}_{E}$ and $P$. It follows from $\left[P_{j},b_{k,p}^{\ast}\right]=k_{j}b_{k,p}^{\ast}$
that this simultaneous eigenspace is precisely
\begin{equation}
\left\{ \Psi\in\mathcal{H}_{N}\mid\mathcal{N}_{E}\Psi=\Psi,\,P\Psi=k\Psi\right\} =\vspan\left(b_{k,p}^{\ast}\psi_{{\rm FS}}\right)_{p\in L_{k}}=\left\{ b_{k}^{\ast}\left(\varphi\right)\psi_{{\rm FS}}\mid\varphi\in L^{2}\left(L_{k}\right)\right\} .
\end{equation}
In fact the mapping $U:\varphi\mapsto b_{k}^{\ast}\left(\varphi\right)\psi_{{\rm FS}}$
is an isomorphism. To see that, we compute, using the commutation relations of the
excitation operators and the fact that $b_{k}\left(\phi\right)\psi_{{\rm FS}}=0=\varepsilon_{k,k}\left(\phi;\varphi\right)\psi_{{\rm FS}}$
for any $\phi,\varphi\in L^{2}\left(L_{k}\right)$, that
\begin{align}
\left\langle U\phi,U\varphi\right\rangle  & =\left\langle b_{k}^{\ast}\left(\phi\right)\psi_{{\rm FS}},b_{k}^{\ast}\left(\varphi\right)\psi_{{\rm FS}}\right\rangle =\left\langle \psi_{{\rm FS}},\left(b_{k}^{\ast}\left(\varphi\right)b_{k}\left(\phi\right)+\left\langle \phi,\varphi\right\rangle +\varepsilon_{k,k}\left(\phi;\varphi\right)\right)\psi_{{\rm FS}}\right\rangle \\
 & =\left\langle \phi,\varphi\right\rangle \left\langle \psi_{{\rm FS}},\psi_{{\rm FS}}\right\rangle =\left\langle \phi,\varphi\right\rangle ,\nonumber 
\end{align}
so $U$ is a unitary embedding of $L^{2}\left(L_{k}\right)$ into
$\left\{ \Psi\in\mathcal{H}_{N}\mid\mathcal{N}_{E}\Psi=\Psi,\,P\Psi=k\Psi\right\} $
hence an isomorphism for dimensional reasons.

Similarly we find as $\left.H_{\text{eff}}\right|_{\mathcal{N}_{E}=1}=2\sum_{l\in\mathbb{Z}_{\ast}^{3}}\sum_{p,q\in L_{l}}\left\langle e_{p},\widetilde{E}_{l}e_{q}\right\rangle b_{l,p}^{\ast}b_{l,q}$
that for any $\phi,\varphi\in L^{2}\left(L_{k}\right)$
\begin{align}
\left\langle U\phi,H_{\text{eff}}U\varphi\right\rangle  & =2\sum_{l\in\mathbb{Z}_{\ast}^{3}}\sum_{p,q\in L_{l}}\left\langle e_{p},\widetilde{E}_{l}e_{q}\right\rangle \left\langle b_{l,p}b_{k}^{\ast}\left(\phi\right)\psi_{{\rm FS}},b_{l,q}b_{k}^{\ast}\left(\varphi\right)\psi_{{\rm FS}}\right\rangle \\
 & =2\sum_{k\in\mathbb{Z}_{\ast}^{3}}\sum_{p,q\in L_{k}}\left\langle e_{p},\widetilde{E}_{l}e_{q}\right\rangle \delta_{k,l}\left\langle \phi,e_{p}\right\rangle \left\langle e_{q},\varphi\right\rangle =2\left\langle \phi,\widetilde{E}_{k}\varphi\right\rangle, \nonumber 
\end{align}
whence $U^{\ast}H_{\mathrm{eff}}U=2\widetilde E_{k}$. By elaborating the above argument slightly, one finds that
the mapping
\begin{equation}
\tilde{U}:\bigoplus_{k\in\mathbb{Z}_{\ast}^{3}}L^{2}\left(L_{k}\right)\rightarrow\left\{ \Psi\in\mathcal{H}_{N}\mid\mathcal{N}_{E}\Psi=\Psi\right\} 
\end{equation}
defined by
\begin{equation}
\tilde{U}\bigoplus_{k\in\mathbb{Z}_{\ast}^{3}}\varphi_{k}=\sum_{k\in\mathbb{Z}_{\ast}^{3}}b_{k}^{\ast}\left(\varphi_{k}\right)\psi_{{\rm FS}}
\end{equation}
is likewise a unitary isomorphism under which $\tilde{U}^{\ast}H_{\text{eff}}\tilde{U}=\bigoplus_{k\in\mathbb{Z}_{\ast}^{3}} \widetilde E_{k}$. 
%This completes the proof of Theorem \ref{thm:intro-excitations}. 
$\hfill\square$

\appendix

\section{Appendix: Lattice Estimates and Riemann Sums}\label{sec:AnalysisofRiemannSums}

In this appendix, we collect several useful estimates for the lattice points and Riemann sums. In particular, we want to obtain estimates on the sum $\sum_{p\in L_k} \lambda_{k,p}^{\beta}$ 
%\begin{equation}
%\left|L_{k}\right|=\sum_{p\in L_{k}}1,\quad\sum_{p\in L_{k}}\frac{1}{\lambda_{k,p}},\quad\sum_{p\in L_{k}}\frac{1}{\lambda_{k,p}^{2}}\label{eq:ParticularRiemannSums}
%\end{equation}
where $\beta \le 0$ and
$$L_{k}=\left(B_{F}+k\right)\backslash B_{F}=\left(\overline{B}\left(k,k_{F}\right)\backslash\overline{B}\left(0,k_{F}\right)\right)\cap\mathbb{Z}^{3},\quad \lambda_{k,p}=\frac{1}{2}\left(\left|p\right|^{2}-\left|p-k\right|^{2}\right)=k\cdot p-\frac{1}{2}\left|k\right|^{2}.$$

It is natural to expect the sum to be 
%These sums are all of the form $\sum_{p\in L_{k}}f\left(\lambda_{k,p}\right)$, and we recognize that these sums are in fact Riemann sums of integrals
%over $\overline{B}\left(k,k_{F}\right)\backslash\overline{B}\left(0,k_{F}\right)$.
%One therefore expects these to be 
approximated by the corresponding
integrals, i.e.
\begin{equation} \label{eq:r-sum-r-int}
\sum_{p\in L_{k}}f\left(\lambda_{k,p}\right)\sim\int_{\overline{B}\left(k,k_{F}\right)\backslash\overline{B}\left(0,k_{F}\right)}f\left(k\cdot p-\frac{1}{2}\left|k\right|^{2}\right)\,dp,
\end{equation}
with $f(t)=t^{\beta}$. Indeed, when $-1<\beta\le 0$, the Riemann sum is well-behaved 
and using general estimation methods based on \eqref{eq:r-sum-r-int} we have
\begin{prop}
\label{prop:NonSingularRiemannSums}For all $k\in\mathbb{Z}_{\ast}^{3}$
and $-1<\beta\leq0$ it holds that
\begin{align*}
\sum_{p\in L_{k}}\lambda_{k,p}^{\beta} & \leq C\begin{cases}
k_{F}^{2+\beta}\left|k\right|^{1+\beta} & \left|k\right|<2k_{F}\\
k_{F}^{3}\left|k\right|^{2\beta} & \left|k\right|\geq2k_{F}
\end{cases}
\end{align*}
for a constant $C>0$ depending only on $\beta$.
\end{prop}

For $\beta\leq-1$ the summands are however too divergent to obtain
good estimates using only general methods. For example, when $\beta=-1$, using standard estimates based on \eqref{eq:r-sum-r-int} we  obtain 
\begin{equation} \label{eq:r-sum-beta=-1}
\sum_{p\in L_{k}}\lambda_{k,p}^{-1}  \leq C\begin{cases}
\left(1+\left|k\right|^{-1}\log\left(k_{F}\right)\right)k_{F} & \left|k\right|<2k_{F}\\
k_{F}^{3}\left|k\right|^{-2} & \left|k\right|\geq2k_{F}
\end{cases}
\end{equation}
which is non-optimal when $|k|<2k_F$. % (and we get worse results for $\beta<-1$). 
%It is estimating these Riemann sums efficiently that are the main subject of this section. 
To obtain good estimates on the sums $\sum_{p\in L_{k}}f\left(\lambda_{k,p}\right)$
for more singular $f$ we will instead derive a summation formula which reduces the $3$-dimensional
Riemann sum  to two
$1$-dimensional Riemann sums plus an error term. The utility of this
summation formula, apart from reducing the dimensionality of the sums,
is that the $1$-dimensional Riemann sums contain weighting factors
which explicitly cancel the divergent behaviour of the summands. To derive this summation formula, we need to carry 
out a detailed analysis of the structure of the lunes $L_{k}$, which is related to a lattice point counting problem in the plane and can be handled by classical results  from analytic number theory.

%To derive this summation formula we begin this section by carrying
%out a detailed analysis of the structure of the lunes $L_{k}$. Once
%this is done we will see that estimating the sums $\sum_{p\in L_{k}}f\left(\lambda_{k,p}\right)$
%well reduces to a lattice point counting problem in the plane, which
%we analyze using a classical result from analytic number theory.

With the summation formula at our disposal we can improve \eqref{eq:r-sum-beta=-1} to 
\begin{prop}
\label{coro:CompletelyUniformRiemannSumBound}For all $k\in\mathbb{Z}_{\ast}^{3}$
it holds that
\[
\sum_{p\in L_{k}}\lambda_{k,p}^{-1}\leq Ck_{F},\quad k_{F}\rightarrow\infty,
\]
for a constant $C>0$ independent of $k$ and $k_{F}$.
\end{prop}

We refer to \cite[Lemma 4.7]{HaiPorRex-20} and  \cite[Eq. B.1]{BNPSS-21} for results similar to Proposition \ref{coro:CompletelyUniformRiemannSumBound}. However, the $k$-independence of the constant $C$ was not completely clear in these previous results.  

For more singular functions,  we have

%\begin{prop}
%\label{prop:MoreSingularRiemannSums} For all $k\in\overline{B}\left(0,2k_{F}\right)$ it holds that 
%\[
%\sum_{p\in L_{k}}\lambda_{k,p}^{-\frac 4 3}\leq C\left|k\right|^{3+\frac{2}{3}}(\log k_{F})^{\frac{2}{3}}k_{F}^{\frac{2}{3}}.
%\]
%Moreover if $k\in\overline{B}\left(0,2k_{F}\right)$ and $-4/3<\beta<-1$, then 
%\[
%\sum_{p\in L_{k}}\lambda_{k,p}^{\beta}\leq C k_F^{4+3\beta}  \left( \left|k\right|^{3+\frac{2}{3}}(\log k_{F})^{\frac{2}{3}}k_{F}^{\frac{2}{3}} \right)^{-3(1+\beta)} . 
%\]
%Here the constant $C>0$ independent of $k$ and $k_{F}$.
%\end{prop}
%
%In our applications, if $\beta<-4/3$ it suffices to bound $\sum_{p\in L_k} \lambda_{k,p}^{\beta} \le C\sum_{p\in L_k} \lambda_{k,p}^{-\frac 4 3}$ due to the obvious bound $\lambda_{k,p} \ge 1/2$. 
%

\begin{prop}
\label{prop:MoreSingularRiemannSums} 
For $-\frac{4}{3}<\beta<-1$ and  $k\in\overline{B}\left(0,k_{F}^{\gamma}\right)$
with $0<\gamma<\frac{4+3\beta}{8-3\beta}$ we have 
\[
\sum_{p\in L_{k}}\lambda_{k,p}^{\beta}\leq Ck_{F}^{2+\beta}\left|k\right|^{1+\beta},\quad k_{F}\rightarrow\infty.
\]
Moreover, for $\beta\leq-\frac{4}{3}$ and $k\in\overline{B}\left(0,2k_{F}\right)$ we have
\[
\sum_{p\in L_{k}}\lambda_{k,p}^{\beta}\leq C\left|k\right|^{3+\frac{2}{3}}(\log k_{F})^{\frac{2}{3}}k_{F}^{\frac{2}{3}},\quad k_{F}\rightarrow\infty.
\]
Here the constant $C>0$ is independent of $k$ and $k_{F}$.
\end{prop}

In Proposition \ref{prop:MoreSingularRiemannSums} the first bound is optimal in terms of both $k_F^{2+\beta}$ and $|k|^\beta$. The second bound is unlikely to be optimal, but is sufficient in applications if $|k|$ is relatively small.

%For all $k\in\mathbb{Z}_{\ast}^{3}$
%it holds that
%\[
%\sum_{p\in L_{k}}\lambda_{k,p}^{-1}\leq Ck_{F},\quad k_{F}\rightarrow\infty,
%\]
%for a constant $C>0$ independent of $k$ and $k_{F}$.
%\end{cor}

%The first estimate extends that of Proposition \ref{prop:NonSingularRiemannSums},
%with the caveat that it only holds for $k$ with $\left|k\right|$
%sufficiently small compared to $k_{F}$. In practice this limits how
%big we can take the exponent $\gamma>0$, which determines our cut-off
%set $S_{C}=\overline{B}\left(0,k_{F}^{\gamma}\right)\cap\mathbb{Z}_{+}^{3}$,
%to be. For the upper bound, where only the sums of equation (\ref{eq:ParticularRiemannSums})
%enter, this limit will be $\gamma\leq1$, as the $\beta=-1$ cases
%of the Propositions \ref{prop:NonSingularRiemannSums} and \ref{prop:MoreSingularRiemannSums}
%can be combined to obtain the following:
%\begin{cor}
%\label{coro:CompletelyUniformRiemannSumBound}For all $k\in\mathbb{Z}_{\ast}^{3}$
%it holds that
%\[
%\sum_{p\in L_{k}}\lambda_{k,p}^{-1}\leq Ck_{F},\quad k_{F}\rightarrow\infty,
%\]
%for a constant $C>0$ independent of $k$ and $k_{F}$.
%\end{cor}
%
%
%
%When we consider the lower bound we will however also need to estimate
%$\sum_{k\in L_{k}}\lambda_{k,p}^{-\frac{5}{4}}$, which constrains
%the range of $\gamma$ further to 
%$$\gamma<\frac{4+3\left(-\frac{5}{4}\right)}{8-3\left(-\frac{5}{4}\right)}=\frac{1}{47}.$$
%

\medskip

Finally for the kinetic estimate in Proposition \ref{prop:DkDKFirstStep} we  need the following proposition, which can be obtained by the same argument of the above results. 

\begin{prop}
\label{prop:SklambdaPointsEstimate} Let $S_{k,\lambda}^{1},S_{k,\lambda}^{2}$ as in \eqref{eq:def-Sk1},  \eqref{eq:def-Sk2} with $k\in\overline{B}\left(0,k_{F}\right)\cap\mathbb{Z}_{\ast}^{3}$ and  $0<\lambda=\lambda\left(k_{F},k\right)\leq\frac{1}{6}k_{F}^{2}$. Then there exists a constant $C>0$ independent of $k$, $k_{F}$, $\lambda$ such that 
\[
\left|S_{k,\lambda}^{1}\right|+\left|S_{k,\lambda}^{2}\right|\leq C\left(\left|k\right|^{-1}\lambda +\left|k\right|^{3+\frac{2}{3}}(\log k_{F})^{\frac{2}{3}}k_{F}^{\frac{2}{3}}\right)\left(\lambda+\left|k\right|\right),\quad k_{F}\rightarrow\infty. 
\]
\end{prop}

In the rest of the appendix, we will discuss some preliminary results in Sections \ref{subsec:SomeLatticeConcepts} and \ref{sec:app-Plane-Decomposition}, and then turn to the proofs of Propositions \ref{prop:NonSingularRiemannSums}, \ref{coro:CompletelyUniformRiemannSumBound},  \ref{prop:MoreSingularRiemannSums} and \ref{prop:SklambdaPointsEstimate}.  

%%%%%%%%%%%%%%%%%%%%%%%%%%%%%%%%%%%
%%%%%%%%%%%%%%%%%%%%%%%%%%%%%%%%%%%

\subsection{\label{subsec:SomeLatticeConcepts}Some Lattice Concepts}

Let $V$ be a real $n$-dimensional vector space. The lattice $\Lambda \subset V$ generated by $\left(v_{i}\right)_{i=1}^{n}$ is 
%is defined by 
%
%
%  the lattice  generated
%by $\left(v_{i}\right)_{i=1}^{n}$ 
%of $V$, 
%
%A lattice $\Lambda$ in a real
%$n$-dimensional vector space $V$ is defined to be a subset of $V$
%with the following property: There exists a basis $\left(v_{i}\right)_{i=1}^{n}$
%of $V$ such that $\Lambda$ equals the integral span of $\left(v_{i}\right)_{i=1}^{n}$,
%i.e.
\begin{equation} \label{eq:def-lattice-Lambda}
\Lambda=\Lambda(v_1, \cdots ,v_n) =\left\{ \sum_{i=1}^{n}m_{i}v_{i}\mid m_{1},\ldots,m_{n}\in\mathbb{Z}\right\} .
\end{equation}
%Given the basis $\left(v_{i}\right)_{i=1}^{n}$ the right-hand side
%of this equation always defines a lattice, called the lattice generated
%by $\left(v_{i}\right)_{i=1}^{n}$ and denoted $\left\langle v_{1},\ldots,v_{n}\right\rangle $.
Given two bases $\left(v_{i}\right)_{i=1}^{n}$ and $\left(w_{i}\right)_{i=1}^{n}$
it may happen that $\Lambda(v_{1},\ldots,v_{n}) =\Lambda(w_{1},\ldots,w_{n})$
even if the bases are not equal. The following is well-known (see e.g. \cite[p. 4]{MG-02})
\begin{prop}
Let $\left(v_{i}\right)_{i=1}^{n}$ and $\left(w_{i}\right)_{i=1}^{n}$
be bases of $V$. Then $\Lambda(v_{1},\ldots,v_{n}) =\Lambda(w_{1},\ldots,w_{n})$ 
if and only if the transition matrix $T=\left(T_{i,j}\right)_{i,j=1}^{n}$
defined by
\[
w_{i}=\sum_{j=1}^{n}T_{i,j}v_{j},\quad1\leq i\leq n,
\]
has integer entries and determinant $\pm1$.
\end{prop}

This result has an important consequence when $V$ is endowed with
an inner product. 

\begin{prop}
Let $\Lambda$ be a lattice in $\left(V,\left\langle \cdot,\cdot\right\rangle \right)$
and let $\left(v_{i}\right)_{i=1}^{n}$ generate $\Lambda$. Then
the quantity
\[
d\left(\Lambda\right)=\left|\det\left(\begin{array}{ccc}
\left\langle e_{1},v_{1}\right\rangle  & \cdots & \left\langle e_{n},v_{1}\right\rangle \\
\vdots & \ddots & \vdots\\
\left\langle e_{1},v_{n}\right\rangle  & \cdots & \left\langle e_{n},v_{n}\right\rangle 
\end{array}\right)\right|=\sqrt{\det\left(\begin{array}{ccc}
\left\langle v_{1},v_{1}\right\rangle  & \cdots & \left\langle v_{n},v_{1}\right\rangle \\
\vdots & \ddots & \vdots\\
\left\langle v_{1},v_{n}\right\rangle  & \cdots & \left\langle v_{n},v_{n}\right\rangle 
\end{array}\right)}
\]
is independent of the choice of generators
$\left(v_{i}\right)_{i=1}^{n}$. Here $\left(e_{i}\right)_{i=1}^{n}$ is any orthonormal basis for $V$. 
\end{prop}

Here $d\left(\Lambda\right)$ is referred to as the covolume (or simply
determinant) of $\Lambda$. The fact that $d(\Lambda)$ is independent of $\left(e_{i}\right)_{i=1}^{n}$ follows by a standard orthonormal expansion, while the fact that  $d(\Lambda)$ is independent of $\left(v_{i}\right)_{i=1}^{n}$ follows from the previous proposition: if $\left(v_{i}\right)_{i=1}^{n}$ and $\left(w_{i}\right)_{i=1}^{n}$
are two bases with transition matrix $T$ then
\begin{equation}
\left|\det\left(\begin{array}{ccc}
\left\langle e_{1},w_{1}\right\rangle  & \cdots & \left\langle e_{n},w_{1}\right\rangle \\
\vdots & \ddots & \vdots\\
\left\langle e_{1},w_{n}\right\rangle  & \cdots & \left\langle e_{n},w_{n}\right\rangle 
\end{array}\right)\right|=\left|\det\left(T\right)\right|\left|\det\left(\begin{array}{ccc}
\left\langle e_{1},v_{1}\right\rangle  & \cdots & \left\langle e_{n},v_{1}\right\rangle \\
\vdots & \ddots & \vdots\\
\left\langle e_{1},v_{n}\right\rangle  & \cdots & \left\langle e_{n},v_{n}\right\rangle 
\end{array}\right)\right|. 
\end{equation}

%and as a consequence of the previous proposition one sees the following:

Given a lattice $\Lambda$ in an $n$-dimensional inner product space
$V$, one defines the successive minima $\left(\lambda_{i}\right)_{i=1}^{n}$
(relative to the closed unit ball $\overline{B}\left(0,1\right)$)
by
\begin{equation}
\lambda_{i}=\inf \left\{ \lambda\mid\overline{B}\left(0,\lambda\right)\cap\Lambda\text{ contains }i\text{\text{ linearly independent vectors}}\right\},\quad1\leq i\leq n.
\end{equation}
A well-known theorem due to Minkowski provides an inequality relating
the successive minima of a lattice $\text{\ensuremath{\Lambda}}$ to
its covolume:
\begin{thm}[Minkowski's Second Theorem]
Let $\Lambda$ be a lattice in an $n$-dimensional inner product
space $V$. Then it holds that
\[
\frac{2^{n}d\left(\Lambda\right)}{n!\Vol\left(\overline{B}\left(0,1\right)\right)}\leq\lambda_{1}\cdots\lambda_{n}\leq\frac{2^{n}d\left(\Lambda\right)}{\Vol\left(\overline{B}\left(0,1\right)\right)}.
\]
\end{thm}

Note that although $\overline{B}\left(0,\lambda_{n}\right)\cap\Lambda$
contains $n$ linearly independent vectors, it is not ensured that
these $n$ vectors can be chosen to generate $\Lambda$. For $n=2$
this is nonetheless the case:
\begin{cor} \label{cor:to-second-Minkowski}
Let $\Lambda$ be a lattice in a $2$-dimensional inner product space
$V$. Then there exists vectors $v_{1},v_{2}\in\Lambda$ which generate
$\Lambda$ such that
\[
\left|v_{1}\right|\left|v_{2}\right|\leq\frac{4}{\pi}d\left(\Lambda\right).
\]
\end{cor}

\textbf{Proof:} By definition of $\lambda_{2}$ there exists linearly
independent vectors $v_{1},v_{2}\in\Lambda$ such that $\left|v_{1}\right|,\left|v_{2}\right|\leq\lambda_{2}$
and by Minkowski's second theorem $\left|v_{1}\right|\left|v_{2}\right|\leq\frac{4}{\pi}d\left(\Lambda\right)$.
We argue that $v_{1}$ and $v_{2}$ must necessarily generate $\Lambda$. Suppose otherwise, i.e. that there exists a $v\in\Lambda$ such that
$v\neq m_{1}v_{1}+m_{2}v_{2}$ for $m_{1},m_{2}\in\mathbb{Z}$. As
$v_{1}$ and $v_{2}$ are linearly independent and $\dim\left(V\right)=2$
these do nonetheless span $V$, i.e. there must exist $c_{1},c_{2}\in\mathbb{R}$
such that $v=c_{1}v_{1}+c_{2}v_{2}$.

Now we can assume that $\left|c_{1}\right|,\left|c_{2}\right|\leq\frac{1}{2}$,
since as $\Lambda$ is a lattice and $v_{1},v_{2},v\in\Lambda$ we
may subtract multiples of $v_{1}$ and $v_{2}$ from $v$ until this
is the case. Then since $\left|\left\langle v_{1},v_{2}\right\rangle \right|<\left|v_{1}\right|\left|v_{2}\right|$
by the Cauchy-Schwarz inequality (strict inequality being a consequence
of the linear independence of $v_{1}$ and $v_{2}$) we can estimate
that
\begin{align}
\left|v\right|^{2} & =\left|v_{1}\right|^{2}c_{1}^{2}+\left|v_{2}\right|^{2}c_{2}^{2}+2\left\langle v_{1},v_{2}\right\rangle c_{1}c_{2}<\left|v_{1}\right|^{2}c_{1}^{2}+\left|v_{2}\right|^{2}c_{2}^{2}+2\left|v_{1}\right|\left|v_{2}\right|\left|c_{1}\right|\left|c_{2}\right|\\
 & =\left(\left|c_{1}\right|\left|v_{1}\right|+\left|c_{2}\right|\left|v_{2}\right|\right)^{2}\leq\left(\frac{1}{2}\lambda_{2}+\frac{1}{2}\lambda_{2}\right)^{2}=\lambda_{2}^{2},\nonumber 
\end{align}
or $\left|v\right|<\lambda_{2}$. But this contradicts the minimality
of $\lambda_{2}$ as $v\neq0$ and at least one of $\left\{ v_{1},v\right\} $
and $\left\{ v_{2},v\right\} $ must be a linearly independent set,
so such a $v$ cannot exist.
$\hfill\square$

\subsubsection*{The Sublattice Orthogonal to a Vector $k\in\mathbb{Z}^{3}$}

Consider $\mathbb{Z}^{3}$ as a lattice in $\mathbb{R}^{3}$ endowed
with the usual dot product. Let $k=\left(k_{1},k_{2},k_{3}\right)\in\mathbb{Z}^{3}\backslash\left\{ 0\right\} $
be arbitrary and write $\hat{k}=\left|k\right|^{-1}k$. Now we consider the set $\left\{ p\in\mathbb{Z}^{3}\mid\hat{k}\cdot p=0\right\} $, namely the sublattice orthogonal to $k$.  Let us recall the following well-known result.

\begin{thm}
\label{them:SolutionsofLinearDiophantineEquations} For $\left(k_{1},k_{2},k_{3}\right)\in\mathbb{Z}^{3}\backslash\left\{ 0\right\} $
and $c\in\mathbb{Z}$, the linear Diophantine equation
\[
k_{1}m_{1}+k_{2}m_{2}+k_{3}m_{3}=c
\]
is solvable with $\left(m_{1},m_{2},m_{3}\right)\in\mathbb{Z}^{3}$
if and only if $c$ is a multiple of $\gcd\left(k_{1},k_{2},k_{3}\right)$. Moreover, in this case  there exist linearly independent vectors
$v_{1},v_{2}\in\mathbb{Z}^{3}$, which does not depend on $c$, such
that if $\left(m_{1}^{\ast},m_{2}^{\ast},m_{3}^{\ast}\right)$ is
any particular solution of the equation then all solutions are given
by
\[
\left\{ \left(m_{1},m_{2},m_{3}\right)\in\mathbb{Z}^{3}\mid k_{1}m_{1}+k_{2}m_{2}+k_{3}m_{3}=c\right\} =\left(m_{1}^{\ast},m_{2}^{\ast},m_{3}^{\ast}\right)+\left\{ a_{1}v_{1}+a_{2}v_{2}\mid a_{1},a_{2}\in\mathbb{Z}\right\} .
\]
\end{thm}

Note that the second part of the proposition states that (up to translation
by a particular solution) the solution set of a linear Diophantine
equation forms a lattice, much as the solution set of a real-variable
linear equation forms a linear subspace. This result implies the following:
\begin{prop}
\label{prop:PlaneDecompositionofZ3}Let $k=\left(k_{1},k_{2},k_{3}\right)\in\mathbb{Z}^{3}\backslash\left\{ 0\right\} $
be given. Then with $l=\left|k\right|^{-1}\gcd\left(k_{1},k_{2},k_{3}\right)$
the following disjoint union of non-empty sets holds:
\[
\mathbb{Z}^{3}=\bigcup_{m\in\mathbb{Z}}\left\{ p\in\mathbb{Z}^{3}\mid\hat{k}\cdot p=lm\right\} .
\]
Additionally, there exist linearly independent vectors $v_{1},v_{2}\in\mathbb{Z}^{3}$,
which span $\left\{ p\in\mathbb{R}^{3}\mid\hat{k}\cdot p=0\right\} $,
such that for any $m\in\mathbb{Z}$, it holds for all $q\in\left\{ p\in\mathbb{Z}^{3}\mid\hat{k}\cdot p=lm\right\} $
that
\[
\left\{ p\in\mathbb{Z}^{3}\mid\hat{k}\cdot p=lm\right\} =q+\left\{ a_{1}v_{1}+a_{2}v_{2}\mid a_{1},a_{2}\in\mathbb{Z}\right\} .
\]
\end{prop}

\textbf{Proof:} Clearly $\mathbb{Z}^{3}=\bigcup_{t\in\mathbb{R}}\left\{ p\in\mathbb{Z}^{3}\mid\hat{k}\cdot p=t\right\} $
so we must determine for which values of $t$ it holds that $\left\{ p\in\mathbb{Z}^{3}\mid\hat{k}\cdot p=t\right\} \neq\emptyset$. 
The equation $\hat{k}\cdot p=t$ is equivalent to 
\begin{equation}
k_{1}p_{1}+k_{2}p_{2}+k_{3}p_{3}=\left|k\right|t
\end{equation}
where $p=\left(p_{1},p_{2},p_{3}\right)\in\mathbb{Z}^{3}$, and as the left-hand side is an integer, we must have $t=\left|k\right|^{-1}c$ for some
$c\in\mathbb{Z}$. Theorem \ref{them:SolutionsofLinearDiophantineEquations}
now furthermore implies that $c=\gcd\left(k_{1},k_{2},k_{3}\right)\cdot m$ for some $m\in\mathbb{Z}$,
so that $t=\left|k\right|^{-1}\gcd\left(k_{1},k_{2},k_{3}\right)\cdot m=lm$,
and as $p$ was arbitrary we see that $\mathbb{Z}^{3}=\bigcup_{m\in\mathbb{Z}}\left\{ p\in\mathbb{Z}^{3}\mid\hat{k}\cdot p=lm\right\} $
as claimed.

That all the sets $\left\{ p\in\mathbb{Z}^{3}\mid\hat{k}\cdot p=lm\right\} $,
$m\in\mathbb{Z}$, are also non-empty similarly follows from the ``only
if'' part of Theorem \ref{them:SolutionsofLinearDiophantineEquations},
and the representation
\begin{equation}
\left\{ p\in\mathbb{Z}^{3}\mid\hat{k}\cdot p=lm\right\} =q+\left\{ a_{1}v_{1}+a_{2}v_{2}\mid a_{1},a_{2}\in\mathbb{Z}\right\} 
\end{equation}
for linearly independent $v_{1},v_{2}\in\mathbb{Z}^{3}$ is likewise
a simple restatement of the second part of the theorem. Finally, that $v_{1}$ and $v_{2}$ span $\left\{ p\in\mathbb{R}^{3}\mid\hat{k}\cdot p=0\right\} $
follows by noting that $q=\left(0,0,0\right)$ is a particular solution
of $\hat{k}\cdot p=0$, whence by the previous part
\begin{equation}
\left\{ v_{1},v_{2}\right\} \subset q+\left\{ a_{1}v_{1}+a_{2}v_{2}\mid a_{1},a_{2}\in\mathbb{Z}\right\} =\left\{ p\in\mathbb{Z}^{3}\mid\hat{k}\cdot p=0\right\} \subset\left\{ p\in\mathbb{R}^{3}\mid\hat{k}\cdot p=0\right\} 
\end{equation}
so we find that $\text{span}\left(\left\{ v_{1},v_{2}\right\} \right)=\left\{ p\in\mathbb{R}^{3}\mid\hat{k}\cdot p=0\right\}.$
 by linear independence of $\left\{ v_{1},v_{2}\right\} $ and dimensionality consideration.  
%\begin{equation}\text{span}\left(\left\{ v_{1},v_{2}\right\} \right)=\left\{ p\in\mathbb{R}^{3}\mid\hat{k}\cdot p=0\right\} .\end{equation}
$\hfill\square$

%\begin{align}
%delete
%\end{align}

Proposition \ref{prop:PlaneDecompositionofZ3} implies that $\left\{ p\in\mathbb{Z}^{3}\mid\hat{k}\cdot p=0\right\} $ is
a lattice in $\{k\}^{\perp}=\left\{ p\in\mathbb{R}^{3}\mid\hat{k}\cdot p=0\right\}$. 
%of the main text can be stated
%as follows:
%\begin{prop}
%Let $k\in\left(k_{1},k_{2},k_{3}\right)\in\mathbb{Z}^{3}\backslash\left\{ 0\right\} $
%and define $l=\left|k\right|^{-1}\gcd\left(k_{1},k_{2},k_{3}\right)$.
%Then we have the disjoint union of non-empty sets
%\[
%\mathbb{Z}^{3}=\bigcup_{m\in\mathbb{Z}}\left\{ p\in\mathbb{Z}^{3}\mid\hat{k}\cdot p=lm\right\} 
%\]
%and $\left\{ p\in\mathbb{Z}^{3}\mid\hat{k}\cdot p=0\right\} $ is
%a lattice in $\{k\}^{\perp}=\left\{ p\in\mathbb{R}^{3}\mid\hat{k}\cdot p=0\right\} $.
%\end{prop}
Since $\left\{ p\in\mathbb{Z}^{3}\mid\hat{k}\cdot p=0\right\} $ is a
lattice, it has a well-defined covolume 
\begin{equation} \label{eq:A11}
\sqrt{\det\left(\begin{array}{cc}
v_{1}\cdot v_{1} & v_{2}\cdot v_{1}\\
v_{1}\cdot v_{2} & v_{2}\cdot v_{2}
\end{array}\right)}=\sqrt{\left|v_{1}\right|^{2}\left|v_{2}\right|^{2}-\left(v_{1}\cdot v_{2}\right)^{2}}
\end{equation}
for any choice of generators $v_{1}$ and $v_{2}$. This covolume
is explicitly given by the following:

\begin{prop}
\label{prop:GeneratorProperties}For any $v_{1},v_{2}\in\mathbb{Z}^{3}$
generating $\left\{ p\in\mathbb{Z}^{3}\mid\hat{k}\cdot p=0\right\} $
it holds that
\[
\left|v_{1}\right|^{2}\left|v_{2}\right|^{2}-\left(v_{1}\cdot v_{2}\right)^{2}=l^{-2}
\]
with $l=\left|k\right|^{-1}\gcd\left(k_{1},k_{2},k_{3}\right)$. Additionally, $v_{1}$ and $v_{2}$ can be chosen such that
$
\left|v_{1}\right|^{2}+\left|v_{2}\right|^{2}\leq \frac{8}{\pi^2 l^2} . 
$
%for a constant $C>0$ independent of $k$.
\end{prop}

\textbf{Proof:} Let $v_{1}$ and $v_{2}$ generate $\left\{ p\in\mathbb{Z}^{3}\mid\hat{k}\cdot p=0\right\} $
and let $w\in\left\{ p\in\mathbb{Z}^{3}\mid\hat{k}\cdot p=l\right\} $
be arbitrary. By linearity it holds that
\begin{equation}
\left\{ p\in\mathbb{Z}^{3}\mid\hat{k}\cdot p=lm\right\} =mw+\left\{ p\in\mathbb{Z}^{3}\mid\hat{k}\cdot p=0\right\} ,\quad m\in\mathbb{Z},
\end{equation}
so by the Proposition \ref{prop:PlaneDecompositionofZ3}
\begin{equation}
\mathbb{Z}^{3}=\bigcup_{m\in\mathbb{Z}}\left(mw+\left\{ p\in\mathbb{Z}^{3}\mid\hat{k}\cdot p=0\right\} \right)=\left\{ m_{1}v_{1}+m_{2}v_{2}+m_{3}w\mid m_{1},m_{2},m_{3}\in\mathbb{Z}\right\} ,
\end{equation}
i.e. $\left(v_{1},v_{2},w\right)$ is a set of generators for $\mathbb{Z}^{3}$. Now let $\{k\}^{\perp}=\left\{ p\in\mathbb{R}^{3}\mid\hat{k}\cdot p=0\right\}$ be the orthogonal complement of $\left\{ k\right\}$. 
Let $\left(e_{1},e_{2}\right)$ be an orthonormal basis for $\{k\}^{\perp}$
so that $\left(e_{1},e_{2},\hat{k}\right)$ forms an orthonormal basis
for $\mathbb{R}^{3}$. Then $d\left(\mathbb{Z}^{3}\right)$ is equal to 
\begin{align}
& \left|\det\left(\begin{array}{ccc}
e_{1}\cdot v_{1} & e_{2}\cdot v_{1} & \hat{k}\cdot v_{1}\\
e_{1}\cdot v_{2} & e_{2}\cdot v_{2} & \hat{k}\cdot v_{2}\\
e_{1}\cdot w & e_{2}\cdot w & \hat{k}\cdot w
\end{array}\right)\right|=\left|\det\left(\begin{array}{ccc}
e_{1}\cdot v_{1} & e_{2}\cdot v_{1} & 0\\
e_{1}\cdot v_{2} & e_{2}\cdot v_{2} & 0\\
e_{1}\cdot w & e_{2}\cdot w & l
\end{array}\right)\right|=l\left|\det\left(\begin{array}{cc}
e_{1}\cdot v_{1} & e_{2}\cdot v_{1}\\
e_{1}\cdot v_{2} & e_{2}\cdot v_{2}
\end{array}\right)\right|\nonumber \\
 & =l\sqrt{\det\left(\begin{array}{cc}
v_{1}\cdot v_{1} & v_{2}\cdot v_{1}\\
v_{1}\cdot v_{2} & v_{2}\cdot v_{2}
\end{array}\right)}=l\sqrt{\left|v_{1}\right|^{2}\left|v_{2}\right|^{2}-\left(v_{1}\cdot v_{2}\right)^{2}}
\end{align}
but it is also clear that $d\left(\mathbb{Z}^{3}\right)=1$, so the
first result follows. From this result, \eqref{eq:A11} and Corollary \ref{cor:to-second-Minkowski}, we deduce that there exist generators $v_{1}$
and $v_{2}$ such that
\begin{equation}
\left|v_{1}\right|\left|v_{2}\right|\leq\frac{4}{\pi}d\left(\left\{ p\in\mathbb{Z}^{3}\mid\hat{k}\cdot p=0\right\} \right)=\frac{4}{\pi}l^{-1}. 
\end{equation}
Since $v_1,v_2\in \mathbb{Z}^{3}\backslash\left\{ 0\right\}$, we have $|v_1|,|v_2|\ge 1$, and hence 
%\begin{equation}
%\left|v_{1}\right|\leq\frac{4}{\pi}\left|v_{2}\right|^{-1}l^{-1}\leq\frac{4}{\pi}l^{-1},
%\end{equation}
%as all elements $v\in\mathbb{Z}^{3}\backslash\left\{ 0\right\} $
%obey $\left|v\right|\geq1$. Likewise $\left|v_{2}\right|\leq\frac{4}{\pi}l^{-1}$
%hence
\begin{equation}
\left|v_{1}\right|^{2}+\left|v_{2}\right|^{2}\leq 2 |v_1|^2 |v_2|^2 \le \frac{8}{\pi^2} l^{-2}.
\end{equation}
$\hfill\square$

%We now prove the claims of Proposition \ref{prop:GeneratorProperties}.

\subsection{Plane Decomposition of $L_{k}$ and the Summation Formula} \label{sec:app-Plane-Decomposition}

Now we turn to consider the lune $L_{k}=\left\{ p\in\mathbb{Z}^{3}\mid\left|p-k\right|\leq k_{F}<\left|p\right|\right\} $. Throughout this subsection we let $k=\left(k_{1},k_{2},k_{3}\right)\in\mathbb{Z}^{3}\backslash\left\{ 0\right\} $
be fixed and write $\hat{k}=\left|k\right|^{-1}k$ and $l=\left|k\right|^{-1}\gcd\left(k_{1},k_{2},k_{3}\right)$ for the sake of
brevity. The integrands of the Riemann sums we must consider only depend on
the quantity $\lambda_{k,p}=k\cdot p-\frac{1}{2}\left|k\right|^{2}=\left|k\right|\left(\hat{k}\cdot p-\frac{1}{2}\left|k\right|\right)$,
so we begin by decomposing $L_{k}$ along the $\hat{k}\cdot p=\text{constant}$
planes. By the definition of $L_{k}$
it easily follows that
\begin{equation}
L_{k}\subset\left\{ p\in\mathbb{R}^{3}\mid\frac{1}{2}\left|k\right|<\hat{k}\cdot p\leq k_{F}+\left|k\right|\right\}. \label{eq:SimpleLuneContainment}
\end{equation}
%so $L_{k}\cap\left\{ \hat{k}\cdot p=t\right\} =\emptyset$ unless
%$t\in\left(\frac{1}{2}\left|k\right|,k_{F}+\left|k\right|\right]$.
%By the discrete nature of $L_{k}$ only finite many of the intersections
%$L_{k}\cap\left\{ \hat{k}\cdot p=t\right\} $ are non-empty within
%this range, however. To describe exactly when this is the case we
%will need the following well-known result: {\bf Theorem \ref{them:SolutionsofLinearDiophantineEquations}, which implies Prop. \ref{prop:PlaneDecompositionofZ3}} 
%Defining $l=\left|k\right|^{-1}\gcd\left(k_{1},k_{2},k_{3}\right)$
%as in Proposition \ref{prop:PlaneDecompositionofZ3} and 
Letting $m^{\ast}$ be the least integer
and $M^{\ast}$ the greatest integer such that
\begin{equation}
\frac{1}{2}\left|k\right|<lm^{\ast},\quad lM^{\ast}\leq k_{F}+\left|k\right|,
\end{equation}
we see that the lune $L_{k}$ can be expressed as the disjoint union
\begin{equation}
L_{k}=\bigcup_{m=m^{\ast}}^{M^{\ast}}L_{k}^{m}, \quad L_{k}^{m}=\left\{ p\in L_{k}\mid\hat{k}\cdot p=lm\right\}.
\end{equation}
So for any function $f:\mathbb{R}\to \mathbb{R}$ we may express a sum of the form $\sum_{p\in L_{k}}f\left(\lambda_{k,p}\right)$
as
\begin{equation}
\sum_{p\in L_{k}}f\left(\lambda_{k,p}\right)=\sum_{m=m^{\ast}}^{M^{\ast}}\sum_{p\in L_{k}^{m}}f\left(\left|k\right|\left(\hat{k}\cdot p-\frac{1}{2}\left|k\right|\right)\right)=\sum_{m=m^{\ast}}^{M^{\ast}}f\left(\left|k\right|\left(lm-\frac{1}{2}\left|k\right|\right)\right)\left|L_{k}^{m}\right|.\label{eq:SliceSummation}
\end{equation}

\subsubsection*{Rewriting $L_{k}^{m}$}

To proceed we must analyze $\left|L_{k}^{m}\right|$, the number of
points contained in $L_{k}^{m}$. For this we first rewrite 
\begin{align}
L_{k}=\left\{ p\in\mathbb{Z}^{3}\mid\left|p-k\right|\leq k_{F}<\left|p\right|\right\}
=\left\{ p\in\mathbb{Z}^{3}\mid k_{F}^{2}<\left|p\right|^{2}\leq k_{F}^{2}-\left|k\right|^{2}+2k\cdot p\right\} .
\end{align}
Now let $P_{\perp}:\mathbb{R}^{3}\rightarrow \{k\}^{\perp}$
denote the orthogonal projection onto $\{k\}^{\perp}$. Then for any $p\in\mathbb{R}^{3}$,
$\left|p\right|^{2}=\left|P_{\perp}p\right|^{2}+\left(\hat{k}\cdotp p\right)^{2}$,
whence
\begin{align}
L_{k} & =\left\{ p\in\mathbb{Z}^{3}\mid k_{F}^{2}-\left(\hat{k}\cdotp p\right)^{2}<\left|P_{\perp}p\right|^{2}\leq k_{F}^{2}-\left|k\right|^{2}+2k\cdot p-\left(\hat{k}\cdotp p\right)^{2}\right\} \\
 & =\left\{ p\in\mathbb{Z}^{3}\mid k_{F}^{2}-\left(\hat{k}\cdotp p\right)^{2}<\left|P_{\perp}p\right|^{2}\leq k_{F}^{2}-\left(\hat{k}\cdotp p-\left|k\right|\right)^{2}\right\} \nonumber 
\end{align}
and so the sets $L_{k}^{m}=L_{k}\cap\left\{ p\in\mathbb{Z}^{3}\mid\hat{k}\cdotp p=lm\right\} $
may be written as
\begin{align}
L_{k}^{m} & =\left\{ p\in\mathbb{Z}^{3}\mid\hat{k}\cdotp p=lm,\,k_{F}^{2}-\left(lm\right)^{2}<\left|P_{\perp}p\right|^{2}\leq k_{F}^{2}-\left(lm-\left|k\right|\right)^{2}\right\} \label{eq:LkmUsefulRepresentation}\\
 & =\left\{ p\in\mathbb{Z}^{3}\mid\hat{k}\cdotp p=lm,\,\left(R_{1}^{m}\right)^{2}<\left|P_{\perp}p\right|^{2}\leq\left(R_{2}^{m}\right)^{2}\right\} \nonumber 
\end{align}
where the real numbers $R_{1}^{m}$ and $R_{2}^{m}$ are
\begin{equation}
R_{1}^{m}=\sqrt{k_{F}^{2}-\left(lm\right)^{2}},\quad R_{2}^{m}=\sqrt{k_{F}^{2}-\left(lm-\left|k\right|\right)^{2}},\quad m^{\ast}<m\leq M^{\ast},
\end{equation}
which are well-defined by definition of $m^{\ast}$ and $M^{\ast}$.

%Now let $v_{1},v_{2}\in\mathbb{Z}^{3}$ be the generators of $\left\{ p\in\mathbb{Z}^{3}\mid\hat{k}\cdot p=0\right\} $
%which Proposition \ref{prop:PlaneDecompositionofZ3} asserts exist.
%Consider a fixed $m^{\ast}\leq m\leq M^{\ast}$ and let $q\in\left\{ p\in\mathbb{Z}^{3}\mid\hat{k}\cdot p=lm\right\} $
%be arbitrary (existence of such a $q$ is also ensured by Proposition
%\ref{prop:PlaneDecompositionofZ3}). Then by the proposition a vector
%$p\in\mathbb{Z}^{3}$ is an element of $\left\{ p\in\mathbb{Z}^{3}\mid\hat{k}\cdot p=lm\right\} $
%if and only if it can be written as

Now by Proposition \ref{prop:PlaneDecompositionofZ3}, we can find the generators $v_{1},v_{2}\in\mathbb{Z}^{3}$ of $\left\{ p\in\mathbb{Z}^{3}\mid\hat{k}\cdot p=0\right\} $. Moreover, a fixed $m^{\ast}\leq m\leq M^{\ast}$, there exists $q\in\left\{ p\in\mathbb{Z}^{3}\mid\hat{k}\cdot p=lm\right\} $, and any $p\in\mathbb{Z}^{3}$ is an element of $\left\{ p\in\mathbb{Z}^{3}\mid\hat{k}\cdot p=lm\right\} $
if and only if it can be written as
\begin{equation}
p=a_{1}v_{1}+a_{2}v_{2}+q
\end{equation}
for some $a_{1},a_{2}\in\mathbb{Z}$. Since $P_{\perp}q\in \{k\}^{\perp}$
by definition and the proposition likewise asserts that $v_{1}$ and
$v_{2}$ span $\{k\}^{\perp}$ there must also exist $b_{1},b_{2}\in\mathbb{R}$
such that $P_{\perp}q=b_{1}v_{1}+b_{2}v_{2}$. Consequently $P_{\perp}p$
for our arbitrary element $p$ takes the form
\begin{equation}
P_{\perp}p=a_{1}P_{\perp}v_{1}+a_{2}P_{\perp}v_{2}+P_{\perp}q=\left(a_{1}+b_{1}\right)v_{1}+\left(a_{2}+b_{2}\right)v_{2}
\end{equation}
whence
\begin{align}
\left|P_{\perp}p\right|^{2} & =\left(a_{1}+b_{1}\right)^{2}\left|v_{1}\right|^{2}+\left(a_{2}+b_{2}\right)^{2}\left|v_{2}\right|^{2}+2\left(a_{1}+b_{1}\right)\left(a_{2}+b_{2}\right)\left(v_{1}\cdot v_{2}\right),
\end{align}
so by equation (\ref{eq:LkmUsefulRepresentation}) we conclude that  
\begin{align}
 \left|L_{k}^{m}\right| &= \Big|\Big\{ \left(a_{1},a_{2}\right)\in\mathbb{Z}^{2}\mid\left(R_{1}^{m}\right)^{2}<\left(a_{1}+b_{1}\right)^{2}\left|v_{1}\right|^{2}+\left(a_{2}+b_{2}\right)^{2}\left|v_{2}\right|^{2}\nonumber \\
 &\qquad \qquad\qquad\qquad\qquad\qquad\qquad +2\left(a_{1}+b_{1}\right)\left(a_{2}+b_{2}\right)\left(v_{1}\cdot v_{2}\right)\leq\left(R_{2}^{m}\right)^{2}\Big\} \Big|\nonumber \\
% & =\left|\left\{ \left(x,y\right)\in\left(\mathbb{R}^{2}+\left(b_{1},b_{2}\right)\right)\mid\left(R_{1}^{m}\right)^{2}<\left|v_{1}\right|^{2}x^{2}+\left|v_{2}\right|^{2}y^{2}+2\left(v_{1}\cdot v_{2}\right)xy\leq\left(R_{2}^{m}\right)^{2}\right\} \cap\mathbb{Z}^{2}\right|\nonumber \\
 & =\left|\left(E_{2}^{m}\backslash E_{1}^{m}-\left(b_{1},b_{2}\right)\right)\cap\mathbb{Z}^{2}\right|
\end{align}
where the sets $E_{1}^{m}$ and $E_{2}^{m}$, defined by
\begin{equation}
E_{i}^{m}=\left\{ \left(x,y\right)\in\mathbb{R}^{2}\mid\left|v_{1}\right|^{2}x^{2}+\left|v_{2}\right|^{2}y^{2}+2\left(v_{1}\cdot v_{2}\right)xy\leq\left(R_{i}^{m}\right)^{2}\right\} ,\quad i=1,2,\label{eq:EmiDefinition}
\end{equation}
are seen to be (the closed interiors of) ellipses. The analysis of
$\left|L_{k}^{m}\right|$ thus reduces to the estimation of the number
of lattice points enclosed by these.

\subsubsection*{Lattice Point Estimation}\label{sec:app-Lattice-Point-Estimation}

To estimate $\left|L_{k}^{m}\right|=\left|\left(E_{2}^{m}\backslash E_{1}^{m} - \left(b_{1},b_{2}\right)\right)\cap\mathbb{Z}^{2}\right|$
we will use the following result on the number of lattice points contained
in compact, strictly convex regions in the plane:
\begin{thm}[\cite{GelLin-66}]
Let $K\subset\mathbb{R}^{2}$ be a compact, strictly convex set with
$C^{2}$ boundary and let $\partial K$ have minimal and maximal radii
of curvature $0<r_{1}\leq r_{2}$. If $r_{2}\geq1$ then
\[
\left|\left|K\cap\mathbb{Z}^{2}\right|-\Area\left(K\right)\right|\leq C\frac{r_{2}}{r_{1}}r_{2}^{\frac{2}{3}}\log\left(1+2\sqrt{2 r_2}\right)^{\frac{2}{3}}
\]
for a constant $C>0$ independent of $K$, $r_{1}$ and $r_{2}$.
\end{thm}

This result follows from the techniques of Chapter 8 of \cite{GelLin-66}. 
%A detailed derivation is available upon request.
%See also \cite[Theorem 7.7.16]{Hor03} and \cite{Hux-03}  for  related results. 

\medskip

From the theorem we deduce the following practical corollary:
\begin{cor}
Let $E\subset\mathbb{R}^{2}$ be an ellipse with 
radii of curvature $0<r_{1}\leq r_{2}$. Then
\[
\left|\left|E\cap\mathbb{Z}^{2}\right|-\Area\left(E\right)\right|\leq C\left(1+\frac{r_{2}}{r_{1}}r_{2}^{\frac{2}{3}}\log\left(1+2\sqrt{2 r_2}\right)^{\frac{2}{3}}\right)
\]
for a constant $C>0$ independent of $E$, $r_{1}$ and $r_{2}$.
\end{cor}

\textbf{Proof:} The theorem gives the case that $r_{2}\geq1$. If
$r_{2}<1$ then we can circumscribe some disk $D$ of radius $1$
around $E$, and trivially
\begin{equation}
\left|\left|E\cap\mathbb{Z}^{2}\right|-\text{Area}\left(E\right)\right|\leq\max\left(\left|E\cap\mathbb{Z}^{2}\right|,\text{Area}\left(E\right)\right)\leq\max\left(\left|D\cap\mathbb{Z}^{2}\right|,\text{Area}\left(D\right)\right)\leq C
\end{equation}
as the right-hand side is seen to be bounded irrespective of the exact
position of $D$.
$\hfill\square$

This corollary lets us estimate that
\begin{equation}
\left|L_{k}^{m}\right|=\text{Area}\left(E_{2}^{m}\backslash E_{1}^{m}\right)+O\left(1+\frac{r_{2}}{r_{1}}r_{2}^{\frac{2}{3}}\log\left(1+2\sqrt{2 r_2}\right)^{\frac{2}{3}}+\frac{r_{2}^{\prime}}{r_{1}^{\prime}}\left(r_{2}^{\prime}\right)^{\frac{2}{3}}\log\left(1+2\sqrt{2 r_2'}\right)^{\frac{2}{3}}\right)\label{eq:LuneSlicePointEstimate1}
\end{equation}
where $r_{i}$ and $r_{i}^{\prime}$, $i=1,2$, denote the radii of
curvature of $E_{1}^{m}$ and $E_{2}^{m}$, as the translation by
$\left(b_{1},b_{2}\right)$ affects neither the areas nor the radii
of curvature of the ellipses.

To proceed we must obtain some information on the geometry of the
ellipses $E_{i}^{m}$. 
%By consulting a reference on conic sections
%one concludes that by the definition of these (equation (\ref{eq:EmiDefinition}))
%the semi-axes $a_{i}\geq b_{i}>0$ of $E_{i}^{m}$ are given by
By the definition \eqref{eq:EmiDefinition}, the semi-axes $a_{i}\geq b_{i}>0$ of $E_{i}^{m}$ are given by
\begin{align}
a_{i} & =\sqrt{2}R_{i}^{m}\left(\left|v_{1}\right|^{2}+\left|v_{2}\right|^{2}-\sqrt{\left(\left|v_{1}\right|^{2}-\left|v_{2}\right|^{2}\right)^{2}+4\left(v_{1}\cdot v_{2}\right)^{2}}\right)^{-\frac{1}{2}},\label{eq:SemiAxes}\\
b_{i} & =\sqrt{2}R_{i}^{m}\left(\left|v_{1}\right|^{2}+\left|v_{2}\right|^{2}+\sqrt{\left(\left|v_{1}\right|^{2}-\left|v_{2}\right|^{2}\right)^{2}+4\left(v_{1}\cdot v_{2}\right)^{2}}\right)^{-\frac{1}{2}}.\nonumber 
\end{align}
%We now need some information on the vectors $v_{1},v_{2}\in\mathbb{Z}^{3}$,
%which we recall are generators of the solution set $\left\{ p\in\mathbb{Z}^{3}\mid\hat{k}\cdot p=0\right\} $.
%we have the following {\bf Prop. \ref{prop:GeneratorProperties}}

%For the proof see appendix section \ref{subsec:SomeLatticeConcepts},
%``Some Lattice Concepts''.

We can now describe the geometry of the ellipses $E_{i}^{m}$ in terms
of $k$ and $m$:
\begin{prop}
\label{prop:EllipseGeometry}If $\left|k\right|\leq2k_{F}$ then
\[
\Area\left(E_{2}^{m}\backslash E_{1}^{m}\right)=\begin{cases}
2\pi\left|k\right|\left(lm-\frac{1}{2}\left|k\right|\right)l & \text{ if } \, lm^{\ast}\leq lm\leq k_{F},\\
\pi\left(k_{F}^{2}-\left(lm-\left|k\right|\right)^{2}\right)l & \text{ if } \, k_{F}<lm\leq lM^{\ast},
\end{cases}
\]
and the radii of curvature $0<r_{1}\leq r_{2}$
of both $E_{1}^{m}$, $E_{2}^{m}$ obey 
\[
\frac{r_{2}}{r_{1}}\leq Cl^{-3},\quad r_{2}\leq Cl^{-1}k_{F},
\]
for a constant $C>0$ independent of $k$ and $m$.
\end{prop}

(The condition $\left|k\right|\leq2k_{F}$ ensures that the lune does
not degenerate into a ball, in which case the area formula must be
modified.)

\textbf{Proof:} Let $v_{1}$ and $v_{2}$ be the generators given by Proposition \ref{prop:GeneratorProperties}. The area enclosed by an ellipse with semi-axes $a$
and $b$ is $\pi ab$, so as $E_{1}^{m}\subset E_{2}^{m}$ for any
$m^{\ast}\leq m\leq M^{\ast}$ and $E_{1}^{m}\neq\emptyset$ when
$lm\leq k_{F}$ we find in this case that
\begin{align}
\text{Area}\left(E_{2}^{m}\backslash E_{1}^{m}\right) & =\pi\left(a_{2}b_{2}-a_{1}b_{1}\right)=\frac{2\pi\left(\left(R_{2}^{m}\right)^{2}-\left(R_{1}^{m}\right)^{2}\right)}{\sqrt{\left(\left|v_{1}\right|^{2}+\left|v_{2}\right|^{2}\right)^{2}-\left(\left(\left|v_{1}\right|^{2}-\left|v_{2}\right|^{2}\right)^{2}+4\left(v_{1}\cdot v_{2}\right)^{2}\right)}}\nonumber \\
 & =\frac{2\pi\left(k_{F}^{2}-\left(lm-\left|k\right|\right)^{2}-\left(k_{F}^{2}-\left(lm\right)^{2}\right)\right)}{\sqrt{4\left|v_{1}\right|^{2}\left|v_{2}\right|^{2}+4\left(v_{1}\cdot v_{2}\right)^{2}}}
  %=\frac{\pi\left(2lm\left|k\right|-\left|k\right|^{2}\right)}{\sqrt{\left|v_{1}\right|^{2}\left|v_{2}\right|^{2}+\left(v_{1}\cdot v_{2}\right)^{2}}}
 =2\pi\left|k\right|\left(lm-\frac{1}{2}\left|k\right|\right)l %,\nonumber 
\end{align}
and similarly in the case $k_{F}<lm$ that
\begin{equation}
\text{Area}\left(E_{2}^{m}\backslash E_{1}^{m}\right)=\text{Area}\left(E_{2}^{m}\right)=\pi a_{2}b_{2}=\frac{2\pi\left(R_{2}^{m}\right)^{2}}{2l^{-1}}=\pi\left(k_{F}^{2}-\left(lm-\left|k\right|\right)^{2}\right)l.
\end{equation}
%by the first part of Proposition \ref{prop:GeneratorProperties}.
For the radii of curvature we note that for an ellipse with semi-axes
$a\geq b>0$ these are given by $r_{1}=a^{-1}b^{2}$ and $r_{2}=b^{-1}a^{2}$,
respectively, so for the ratio $r_{1}^{-1}r_{2}$ we can for either
of $E_{1}^{m}$ and $E_{2}^{m}$ estimate using equation (\ref{eq:SemiAxes})
that
\begin{align}
\frac{r_{2}}{r_{1}} & =\left(\frac{a_{i}}{b_{i}}\right)^{3}=\left(\frac{\left|v_{1}\right|^{2}+\left|v_{2}\right|^{2}+\sqrt{\left(\left|v_{1}\right|^{2}-\left|v_{2}\right|^{2}\right)^{2}+4\left(v_{1}\cdot v_{2}\right)^{2}}}{\left|v_{1}\right|^{2}+\left|v_{2}\right|^{2}-\sqrt{\left(\left|v_{1}\right|^{2}-\left|v_{2}\right|^{2}\right)^{2}+4\left(v_{1}\cdot v_{2}\right)^{2}}}\right)^{\frac{3}{2}}\nonumber \\
 & =\left(\frac{\left(\left|v_{1}\right|^{2}+\left|v_{2}\right|^{2}+\sqrt{\left(\left|v_{1}\right|^{2}-\left|v_{2}\right|^{2}\right)^{2}+\left(v_{1}\cdot v_{2}\right)^{2}}\right)^{2}}{\left(\left|v_{1}\right|^{2}+\left|v_{2}\right|^{2}\right)^{2}-\left(\left(\left|v_{1}\right|^{2}-\left|v_{2}\right|^{2}\right)^{2}+4\left(v_{1}\cdot v_{2}\right)^{2}\right)}\right)^{\frac{3}{2}}\\
 & \le \left( \frac{\left( 2( \left|v_{1}\right|^{2}+\left|v_{2}\right|^{2}) \right)^2}{4( \left|v_{1}\right|^{2}\left|v_{2}\right|^{2}-\left(v_{1}\cdot v_{2}\right)^{2} )} \right)^{\frac{3}{2}} \le \left( \frac{(Cl^{-2})^2}{l^2}\right)^{3/2} \leq Cl^{-3} \nonumber
 %\left(\left|v_{1}\right|^{2}+\left|v_{2}\right|^{2}\right)^{3}l^{3},\nonumber 
\end{align}
and likewise estimate for $r_{2}$ that
\begin{align}
r_{2} & =\frac{a_{i}^{2}}{b_{i}}=\sqrt{2}R_{i}^{m}\frac{\sqrt{\left|v_{1}\right|^{2}+\left|v_{2}\right|^{2}+\sqrt{\left(\left|v_{1}\right|^{2}-\left|v_{2}\right|^{2}\right)^{2}+4\left(v_{1}\cdot v_{2}\right)^{2}}}}{\left|v_{1}\right|^{2}+\left|v_{2}\right|^{2}-\sqrt{\left(\left|v_{1}\right|^{2}-\left|v_{2}\right|^{2}\right)^{2}+4\left(v_{1}\cdot v_{2}\right)^{2}}}\nonumber \\
 & =\sqrt{2}R_{i}^{m}\frac{\left(\left|v_{1}\right|^{2}+\left|v_{2}\right|^{2}+\sqrt{\left(\left|v_{1}\right|^{2}-\left|v_{2}\right|^{2}\right)^{2}+4\left(v_{1}\cdot v_{2}\right)^{2}}\right)^{\frac{3}{2}}}{\left(\left|v_{1}\right|^{2}+\left|v_{2}\right|^{2}\right)^{2}-\left(\left(\left|v_{1}\right|^{2}-\left|v_{2}\right|^{2}\right)^{2}+4\left(v_{1}\cdot v_{2}\right)^{2}\right)}\\
 & \leq\sqrt{2}R_{i}^{m}\frac{\left(2\left(\left|v_{1}\right|^{2}+\left|v_{2}\right|^{2}\right)\right)^{\frac{3}{2}}}{4( \left|v_{1}\right|^{2}\left|v_{2}\right|^{2}-\left(v_{1}\cdot v_{2}\right)^{2} )} \leq\left(Cl^{-2}\right)^{\frac{3}{2}}l^{2}R_{i}^{m}\leq Cl^{-1}k_{F} .\nonumber 
\end{align}
Here we also used that $R_{1}^{m},R_{2}^{m}\leq k_{F}$
for all $m^{\ast}\leq m\leq M^{\ast}$.
%asserts exist we can therefore estimate
%\begin{equation}
%\frac{r_{2}}{r_{1}}\leq\left(Cl^{-2}\right)^{3}l^{3}\leq Cl^{-3}\quad\text{and}\quad r_{2}\leq\left(Cl^{-2}\right)^{\frac{3}{2}}l^{2}R_{i}^{m}\leq Cl^{-1}k_{F}
%\end{equation}
%as claimed, where we also used that $R_{1}^{m},R_{2}^{m}\leq k_{F}$
%for all $m^{\ast}\leq m\leq M^{\ast}$.
$\hfill\square$

\subsubsection*{The Summation Formula}

We can now present the summation formula that we will use to estimate
the sums $\sum_{p\in L_{k}}f\left(\lambda_{k,p}\right)$. Noting that
the quantity $l=\left|k\right|^{-1}\gcd\left(k_{1},k_{2},k_{3}\right)$
obeys the lower bound $l\geq\left|k\right|^{-1}$ independently of
$k$ we can by equation (\ref{eq:LuneSlicePointEstimate1}) and Proposition
\ref{prop:EllipseGeometry} estimate (provided $\left|k\right|\leq2k_{F}$)
that
\begin{align}
\left|\left|L_{k}^{m}\right|-\text{Area}\left(E_{2}^{m}\backslash E_{1}^{m}\right)\right| & \leq C\left(1+l^{-3}\left(l^{-1}k_{F}\right)^{\frac{2}{3}}\log\left(1+2\sqrt{2}\left(l^{-1}k_{F}\right)^{\frac{1}{2}}\right)^{\frac{2}{3}}\right)\\
 & \leq C\left(1+\left|k\right|^{3+\frac{2}{3}}k_{F}^{\frac{2}{3}}\log\left(1+\sqrt{\left|k\right|k_{F}}\right)^{\frac{2}{3}}\right)\leq C\left|k\right|^{3+\frac{2}{3}}(\log k_{F})^{\frac{2}{3}}k_{F}^{\frac{2}{3}}\nonumber 
\end{align}
as $k_{F}\rightarrow\infty$, for a constant $C>0$ independent of
$k$ and $m$. Inserting the expression for $\text{Area}\left(E_{2}^{m}\backslash E_{1}^{m}\right)$
that we determined in Proposition \ref{prop:EllipseGeometry} we then
have
\begin{equation}
\left|L_{k}^{m}\right|=\begin{cases}
2\pi\left|k\right|\left(lm-\frac{1}{2}\left|k\right|\right)l & lm^{\ast}\leq lm\leq k_{F}\\
\pi\left(k_{F}^{2}-\left(lm-\left|k\right|\right)^{2}\right)l & k_{F}<lm\leq lM^{\ast}
\end{cases}+O\left(\left|k\right|^{3+\frac{2}{3}}(\log k_{F})^{\frac{2}{3}}k_{F}^{\frac{2}{3}}\right).\label{eq:LuneSlicePointEstimate2}
\end{equation}
Letting $M$ denote the greatest integer such that $lM\leq k_{F}$
it now follows from equation (\ref{eq:SliceSummation}) that for any
$f:\left(0,\infty\right)\rightarrow\mathbb{R}$ it holds that
\begin{align}
\sum_{p\in L_{k}}f\left(\lambda_{k,p}\right) & =2\pi\left|k\right|\sum_{m=m^{\ast}}^{M}f\left(\left|k\right|\left(lm-\frac{1}{2}\left|k\right|\right)\right)\left(lm-\frac{1}{2}\left|k\right|\right)l\nonumber \\
 & +\pi\sum_{m=M+1}^{M^{\ast}}f\left(\left|k\right|\left(lm-\frac{1}{2}\left|k\right|\right)\right)\left(k_{F}^{2}-\left(lm-\left|k\right|\right)^{2}\right)l\label{eq:FirstSumFormula}\\
 & +O\left(\left|k\right|^{3+\frac{2}{3}}(\log k_{F})^{\frac{2}{3}}k_{F}^{\frac{2}{3}}\sum_{m=m^{\ast}}^{M^{\ast}}\left|f\left(\left|k\right|\left(lm-\frac{1}{2}\left|k\right|\right)\right)\right|\right),\nonumber 
\end{align}
so the $3$-dimensional Riemann sum $\sum_{p\in L_{k}}f\left(\lambda_{k,p}\right)$
has been reduced to two $1$-dimensional Riemann sums plus an error
term. In fact these two $1$-dimensional Riemann sums are just what
one would expect, since by 3D integrating along the $\hat{k}$ axis it
is not difficult to show that in general
\begin{align}
\int_{\overline{B}\left(k,k_{F}\right)\backslash\overline{B}\left(0,k_{F}\right)}f\left(k\cdot p-\frac{1}{2}\left|k\right|^{2}\right)dp & =2\pi\left|k\right|\int_{\frac{1}{2}\left|k\right|}^{k_{F}}f\left(\left|k\right|\left(t-\frac{1}{2}\left|k\right|\right)\right)\left(t-\frac{1}{2}\left|k\right|\right)dt\label{eq:IntegrationFormula}\\
 & +\pi\int_{k_{F}}^{k_{F}+\left|k\right|}f\left(\left|k\right|\left(t-\frac{1}{2}\left|k\right|\right)\right)\left(k_{F}^{2}-\left(t-\left|k\right|\right)^{2}\right)dt\nonumber 
\end{align}
and the two Riemann sums of equation (\ref{eq:FirstSumFormula}) are
seen to be Riemann sums for the two $1$-dimensional integrals above.

In the statement  in the following proposition, we make a minor
adjustment: We expand the factor $k_{F}^{2}-\left(lm-\left|k\right|\right)^{2}$
as
\begin{equation}
k_{F}^{2}-\left(lm-\left|k\right|\right)^{2}=k_{F}^{2}-\left(lm\right)^{2}-\left|k\right|^{2}+2\left|k\right|lm=\left(k_{F}^{2}-\left(lm\right)^{2}\right)+2\left|k\right|\left(lm-\frac{1}{2}\left|k\right|\right)
\end{equation}
and collect the $2\left|k\right|\left(lm-\frac{1}{2}\left|k\right|\right)$
terms in the first sum. We have the summation formula: 
\begin{prop}
\label{prop:SummationFormula} Let $k=\left(k_{1},k_{2},k_{3}\right)\in\mathbb{Z}^{3}\backslash\left\{ 0\right\}$
with $\left|k\right|\leq2k_{F}$, $f:\left(0,\infty\right)\rightarrow\mathbb{R}$. Let $l=\left|k\right|^{-1}\gcd\left(k_{1},k_{2},k_{3}\right)$ and
$m^{\ast}$ is the least integer and $M$, $M^{\ast}$ the greatest
integers for which
\[
\frac{1}{2}\left|k\right|<lm^{\ast},\quad lM\leq k_{F},\quad lM^{\ast}\leq k_{F}+\left|k\right|.
\]
Then for all functions $f: (0,\infty)\to \mathbb{R}$ it holds that
\begin{align*}
\sum_{p\in L_{k}}f\left(\lambda_{k,p}\right) & =2\pi\left|k\right|\sum_{m=m^{\ast}}^{M^{\ast}}f\left(\left|k\right|\left(lm-\frac{1}{2}\left|k\right|\right)\right)\left(lm-\frac{1}{2}\left|k\right|\right)l\\
 & +\pi\sum_{m=M+1}^{M^{\ast}}f\left(\left|k\right|\left(lm-\frac{1}{2}\left|k\right|\right)\right)\left(k_{F}^{2}-\left(lm\right)^{2}\right)l\\
 & +O\left(\left|k\right|^{3+\frac{2}{3}}(\log k_{F})^{\frac{2}{3}}k_{F}^{\frac{2}{3}}\sum_{m=m^{\ast}}^{M^{\ast}}\left|f\left(\left|k\right|\left(lm-\frac{1}{2}\left|k\right|\right)\right)\right|\right),\quad k_{F}\rightarrow\infty.
\end{align*}
\end{prop}

\subsection{Proof of Proposition \ref{prop:NonSingularRiemannSums}} \label{subsect:RiemannSumEstimatesforBetageq-1} 

Now we prove Proposition \ref{prop:NonSingularRiemannSums} and \eqref{eq:r-sum-beta=-1}. In this part, we do not use Proposition \ref{prop:SummationFormula}.  

%We prove the following:
%\begin{prop*}[\ref{prop:NonSingularRiemannSums}]
%For all $k\in\mathbb{Z}_{\ast}^{3}$ and $-1<\beta\leq0$ it holds
%that
%\begin{align*}
%\sum_{p\in L_{k}}\lambda_{k,p}^{\beta} & \leq C\begin{cases}
%k_{F}^{2+\beta}\left|k\right|^{1+\beta} & \left|k\right|<2k_{F}\\
%k_{F}^{3}\left|k\right|^{2\beta} & \left|k\right|\geq2k_{F}
%\end{cases}\\
%\sum_{p\in L_{k}}\lambda_{k,p}^{-1} & \leq C\begin{cases}
%\left(1+\left|k\right|^{-1}\log\left(k_{F}\right)\right)k_{F} & \left|k\right|<2k_{F}\\
%k_{F}^{3}\left|k\right|^{-2} & \left|k\right|\geq2k_{F}
%\end{cases}
%\end{align*}
%for a constant $C>0$ depending only on $\beta$.
%\end{prop*}

\subsubsection*{Some Riemann Sum Estimation Techniques}
We must first establish some preliminary Riemann
sum estimation results. Let $S\subset\mathbb{R}^{n}$, $n\in\mathbb{N}$,
be given, define for $k\in\mathbb{Z}^{n}$ the translated unit cube
$\mathcal{C}_{k}$ by
\begin{equation}
\mathcal{C}_{k}=\left[-2^{-1},2^{-1}\right]^{n}+k
\end{equation}
and let $\mathcal{C}_{S}=\bigcup_{k\in S\cap\mathbb{Z}^{n}}\mathcal{C}_{k}$
denote the union of the cubes centered at the lattice points contained
in $S$. The first result we will establish is that for a convex function
$f$ the integral $\int_{\mathcal{C}_{S}}f\left(p\right)dp$ always
yields an upper bound to the Riemann sum $\sum_{k\in S\cap\mathbb{Z}^{n}}f\left(k\right)$:
\begin{prop}
\label{prop:ConvexRiemannSumEstimation}Let $f\in C\left(\mathcal{C}_{S}\right)$
be a function which is convex on $\mathcal{C}_{k}$ for all $k\in S\cap\mathbb{Z}^{n}$.
Then
\[
\sum_{k\in S\cap\mathbb{Z}^{n}}f\left(k\right)\leq\int_{\mathcal{C}_{S}}f\left(p\right)dp.
\]
\end{prop}

\textbf{Proof:} As a convex function admits a supporting hyperplane
at every interior point of its domain we see that for every $k\in S\cap\mathbb{Z}^{n}$
there exists a $c\in\mathbb{R}^{n}$ such that
\begin{equation}
f\left(p\right)\geq f\left(k\right)+c\cdot\left(p-k\right),\quad p\in\mathcal{C}_{k},
\end{equation}
which upon integration over $\mathcal{C}_{k}$ yields
\begin{equation}
\int_{\mathcal{C}_{k}}f\left(p\right)dp\geq\int_{\mathcal{C}_{k}}f\left(k\right)dp+\int_{\mathcal{C}_{k}}c\cdot\left(p-k\right)dp=f\left(k\right)
\end{equation}
as $\int_{\mathcal{C}_{S}}f\left(k\right)dp=f\left(k\right)$ since
$\text{Vol}\left(\mathcal{C}_{k}\right)=1$ and $\int_{\mathcal{C}_{S}}c\cdot\left(p-k\right)dp=0$,
as $\mathcal{C}_{k}$ is symmetric with respect to $k$ but the integrand
$p\mapsto c\cdot\left(p-k\right)$ is antisymmetric. Consequently
\begin{equation}
\sum_{k\in S\cap\mathbb{Z}^{n}}f\left(k\right)\leq\sum_{k\in S\cap\mathbb{Z}^{n}}\int_{\mathcal{C}_{k}}f\left(p\right)dp=\int_{\mathcal{C}_{S}}f\left(p\right)dp.
\end{equation}
$\hfill\square$

This proposition lets us replace the sum by an integral, but over
an integration domain $\mathcal{C}_{S}$ which will generally be complicated.
An exception is the $n=1$ case which we record in the following (generalizing
also the statement to any lattice spacing $l$):
\begin{prop}\label{prop:ConvexRiemannSumEstimate}
Let $a,b\in\mathbb{Z}$, $l>0$, and   $f\in C\left(\left[la-\frac{1}{2}l,lb+\frac{1}{2}l\right]\right)$ be a convex function. Then
\[
\sum_{m=a}^{b}f\left(lm\right)l\leq\int_{la-\frac{1}{2}l}^{lb+\frac{1}{2}l}f\left(x\right)dx.
\]
\end{prop}
For $n\neq1$ we instead require an additional result that lets us
replace $\mathcal{C}_{S}$ by a simpler integration domain. We define a subset $S_{+}\subset \mathbb{R}^{n}$ by
\begin{equation}
S_{+} = \left\{ p\in\mathbb{R}^{n}\mid\inf_{q\in S}\left|p-q\right|\leq\frac{\sqrt{n}}{2}\right\}
\end{equation}
and observe the following:
\begin{prop} \label{lem:basic-Gauss}
It holds that $\mathcal{C}_{S}\subset S_{+}$. Consequently,
\[
\left|S\cap\mathbb{Z}^{n}\right|\leq\Vol\left(S_{+}\right).
\]
\end{prop}

\textbf{Proof:} We first note that for
any $p\in\mathbb{R}^{n}$, every point of the translated cube $\left(\left[-2^{-1},2^{-1}\right]+p\right)^{n}$
is a distance of at most $\frac{\sqrt{n}}{2}$ separated from $p$
itself. Now, let $p\in\mathcal{C}_{S}$. Then by definition of $\mathcal{C}_{S}$ and the
previous observation there exists some $k\in S\cap\mathbb{Z}^{n}$
such that $\left|p-k\right|\leq\frac{\sqrt{n}}{2}$, and hence $p\in S_{+}$ since 
\begin{equation}
\inf_{q\in S}\left|p-q\right|\leq\left|p-k\right|\leq\frac{\sqrt{n}}{2}.
\end{equation}

Clearly $\left|S\cap\mathbb{Z}^{n}\right|=\sum_{k\in S\cap\mathbb{Z}^{n}}1=\sum_{k\in S\cap\mathbb{Z}^{n}}\text{Vol}\left(\mathcal{C}_{k}\right)=\text{Vol}\left(\mathcal{C}_{S}\right)$
so the inclusion $\mathcal{C}_{S}\subset S_{+}$ immediately
implies that $\left|S\cap\mathbb{Z}^{n}\right|\leq\Vol\left(S_{+}\right)$. 
$\hfill\square$

\subsubsection*{Lune Geometry}

Returning to Proposition \ref{prop:NonSingularRiemannSums} and \eqref{eq:r-sum-beta=-1}, we now let $k\in\mathbb{Z}_{\ast}^{3}$ and $-1\leq\beta\leq0$ be
fixed. The Riemann sum ranges over $p\in L_{k}=\left(\overline{B}\left(k,k_{F}\right)\backslash\overline{B}\left(0,k_{F}\right)\right)\cap\mathbb{Z}^{3}$,
so in the notation of the above discussion we must consider $S=\overline{B}\left(k,k_{F}\right)\backslash\overline{B}\left(0,k_{F}\right)$.
The relevant integrand,
\begin{equation}
p\mapsto\lambda_{k,p}^{\beta}=\left(\frac{1}{2}\left(\left|p\right|^{2}-\left|p-k\right|^{2}\right)\right)^{\beta}=\left|k\right|^{\beta}\left(\hat{k}\cdot p-\frac{1}{2}\left|k\right|\right)^{\beta},
\end{equation}
is convex on $\left\{ p\in\mathbb{R}^{3}\mid\hat{k}\cdot p>\frac{1}{2}\left|k\right|\right\} $
but singular at $\left\{ p\in\mathbb{R}^{3}\mid\hat{k}\cdot p=\frac{1}{2}\left|k\right|\right\} $.
For this reason we must introduce a cut-off to the Riemann sum $\sum_{p\in L_{k}}\lambda_{k,p}^{\beta}$:
We write $S=S^{1}\cup S^{2}$ 
\begin{equation}
S^{1}=\left\{ p\in S\mid\hat{k}\cdot p\leq\frac{1}{2}\left|k\right|+\frac{2+\sqrt{3}}{2}\right\} ,\quad S^{2}=\left\{ p\in S\mid\hat{k}\cdot p>\frac{1}{2}\left|k\right|+\frac{2+\sqrt{3}}{2}\right\},
\end{equation}
so that likewise $L_{k}=L_{k}^{1}\cup L_{k}^{2}$ where $L_k^1=L_k\cap S^1$, $L_k^2=L_k\cap S^2$. 
%
%\begin{equation}
%L_{k}^{1}=\left\{ p\in L_{k}\mid\hat{k}\cdot p\leq\frac{1}{2}\left|k\right|+\frac{2+\sqrt{3}}{2}\right\} ,\quad L_{k}^{2}=\left\{ p\in L_{k}\mid\hat{k}\cdot p>\frac{1}{2}\left|k\right|+\frac{2+\sqrt{3}}{2}\right\} ,
%\end{equation}
%so that likewise $S=S^{1}\cup S^{2}$ for
%\begin{equation}
%S^{1}=\left\{ p\in S\mid\hat{k}\cdot p\leq\frac{1}{2}\left|k\right|+\frac{2+\sqrt{3}}{2}\right\} ,\quad S^{2}=\left\{ p\in S\mid\hat{k}\cdot p>\frac{1}{2}\left|k\right|+\frac{2+\sqrt{3}}{2}\right\}.
%\end{equation}
Hence, by Proposition \ref{lem:basic-Gauss} 
\begin{align}
\sum_{p\in L_{k}}\lambda_{k,p}^{\beta} =\sum_{p\in L_{k}^{1}}\lambda_{k,p}^{\beta}+\sum_{p\in L_{k}^{2}}\lambda_{k,p}^{\beta}& \leq\left(\inf_{p\in L_{k}}\lambda_{k,p}\right)^{\beta}\left|L_{k}^{1}\right|+\int_{\mathcal{C}_{S^{2}}}\left|k\right|^{\beta}\left(\hat{k}\cdot p-\frac{1}{2}\left|k\right|\right)^{\beta}dp \\
 & \leq\left(\inf_{p\in L_{k}}\lambda_{k,p}\right)^{\beta}\text{Vol}\left(S_{+}^{1}\right)+\left|k\right|^{\beta}\int_{S_{+}^{2}}\left(\hat{k}\cdot p-\frac{1}{2}\left|k\right|\right)^{\beta}dp, \nonumber
\end{align}
where we also used that $p\mapsto\left(\hat{k}\cdot p-\frac{1}{2}\left|k\right|\right)^{\beta}$
is non-negative to expand the integration range of the integral. In order to apply this inequality we will again replace the sets $S_{+}^{1}$,
$S_{+}^{2}$ by ones which are easier to work with. We have 

\begin{prop} \label{prop:SSSSSS}
For all $k\in\mathbb{Z}^{3}$ it holds that
\begin{align*}
S_{+}=\left\{ p\in\mathbb{R}^{3}\mid\inf_{q\in S}\left|p-q\right|\leq\frac{\sqrt{3}}{2}\right\} &\subset \widetilde{S} =\overline{B}\left(k,k_{F}+\frac{\sqrt{3}}{2}\right)\backslash B\left(0,k_{F}-\frac{\sqrt{3}}{2}\right),\\
S_{+}^{1}=\left\{ p\in\mathbb{R}^{3}\mid\inf_{q\in S^{1}}\left|p-q\right|\leq\frac{\sqrt{3}}{2}\right\} &\subset \widetilde{S}^{1} =\left\{ p\in\widetilde{S}\mid-\frac{\sqrt{3}}{2}\leq\hat{k}\cdot p-\frac{1}{2}\left|k\right|\leq1+\sqrt{3}\right\} , \nonumber\\
S_{+}^{2} = \left\{ p\in\mathbb{R}^{3}\mid\inf_{q\in S^{2}}\left|p-q\right|\leq\frac{\sqrt{3}}{2}\right\} &\subset \widetilde{S}^{2}=\left\{ p\in\widetilde{S}\mid\hat{k}\cdot p-\frac{1}{2}\left|k\right|\geq1\right\}.\nonumber
\end{align*}
\end{prop}

\textbf{Proof:} We first show that $S_+\subset \widetilde{S}$. 
%\begin{equation}
%S_{+}=\left\{ p\in\mathbb{R}^{3}\mid\inf_{q\in S}\left|p-q\right|\leq\frac{\sqrt{3}}{2}\right\}\subset\widetilde{S} = \overline{B}\left(k,k_{F}+\frac{\sqrt{3}}{2}\right)\backslash B\left(0,k_{F}-\frac{\sqrt{3}}{2}\right).
%\end{equation}
For every $p\in S_{+}$  by the triangle
inequality we can estimate 
\begin{align}
\left|p\right| & \geq \sup_{q\in S} \Big( \left|q\right|-\left|p-q\right| \Big) >k_{F}- \inf_{q\in S}\left|p-q\right| \ge k_F - \frac{\sqrt{3}}{2},\\
\left|p-k\right| &  \leq \inf_{q\in S} \Big( \left|q-k\right|+\left|p-q\right| \Big) \leq k_{F}+ \inf_{q\in S}\left|p-q\right| \le k_F + \frac{\sqrt{3}}{2},\nonumber 
\end{align}
and hence $p\in\widetilde{S}$. Next, we prove $S_{+}^{1}\subset \widetilde{S}^{1}$: for every $p\in S_{+}^{1}$, we have
% and the Cauchy-Schwarz inequality we can
%for any $q\in S^{1}$ estimate that
\begin{align}
\hat{k}\cdot p-\frac{1}{2}\left|k\right| &= \inf_{q\in S^1} \Big( \hat{k}\cdot q-\frac{1}{2}\left|k\right|+\hat{k}\cdot\left(p-q\right) \Big) \leq\frac{2+\sqrt{3}}{2}+ \inf_{q\in S^1}\left|p-q\right| \le 1 +\sqrt{3},\\
\hat{k}\cdot p-\frac{1}{2}\left|k\right| &= \sup_{q\in S^1}\Big( \hat{k}\cdot q-\frac{1}{2}\left|k\right|+\hat{k}\cdot\left(p-q\right) \Big) \ge  - \inf_{q\in S^1}\left|p-q\right|  \ge -\frac {\sqrt 3} 2\end{align}
and hence $p\in \widetilde{S}^{1}$. Here we used the definition of $S^{1}$ and $S^{1}\subset S\subset\left\{ q\in\mathbb{R}^{3}\mid\hat{k}\cdot q>\frac{1}{2}\left|k\right|\right\} $. That $p\in S_{+}^{2}$ implies $\hat{k}\cdot p-\frac{1}{2}\left|k\right|\geq1$
follows by the same argument.

$\hfill\square$

Thanks to the simple bound  $\lambda_{k,p}\geq\frac{1}{2}$ for all $p\in L_k$,  we can now conclude the inequality
\begin{equation}
\sum_{p\in L_{k}}\lambda_{k,p}^{\beta}\leq2^{-\beta}\,\text{Vol}\left(\widetilde{S}^{1}\right)+\left|k\right|^{\beta}\int_{\widetilde{S}^{2}}\left(\hat{k}\cdot p-\frac{1}{2}\left|k\right|\right)^{\beta}dp.\label{eq:SmallBetaCutoffEstimate}
\end{equation}
Hence, we need only consider the sets $\widetilde{S}^{1}$ and $\widetilde{S}^{2}$,
which consist of ``slices'' of $\widetilde{S}$:
\begin{equation}
\widetilde{S}=\bigcup_t \widetilde{S}_t,\quad \widetilde{S}_t= \left\{ p\in\widetilde{S}\mid\hat{k}\cdot p=t\right\}.
\end{equation}
%\begin{equation}
%\widetilde{S}=\overline{B}\left(k,k_{F}+\frac{\sqrt{3}}{2}\right)\backslash B\left(0,k_{F}-\frac{\sqrt{3}}{2}\right).
%\end{equation}
Recalling the definition of $\widetilde{S}$ from Proposition \ref{prop:SSSSSS} and using elementary trigonometry we can show that
\begin{align}
\text{Area}\left(\widetilde{S}_{t}\right) & =\pi\left(\left(k_{F}+\frac{\sqrt{3}}{2}\right)^{2}-\left(t-\left|k\right|\right)^{2}\right)-\pi\left(\left(k_{F}-\frac{\sqrt{3}}{2}\right)^{2}-\left|t\right|^{2}\right)\\
 & =\pi\left(2\sqrt{3}k_{F}-\left(\left|k\right|^{2}-2\left|k\right|t\right)\right)=2\pi\left(\left|k\right|\left(t-\frac{1}{2}\left|k\right|\right)+\sqrt{3}k_{F}\right)\nonumber 
\end{align}
for $\left|k\right|/2-\sqrt{3}/2 \leq t\leq k_{F}-\sqrt{3}/2$, and that 
%where $t=k_{F}-\frac{\sqrt{3}}{2}$ correponds to the ``upper end''
%of $B\left(0,k_{F}-\frac{\sqrt{3}}{2}\right)$. Thereafter the planes
%$\left\{ p\in\mathbb{R}^{3}\mid\hat{k}\cdot p=t\right\} $ only intersect
%$\overline{B}\left(k,k_{F}+\frac{\sqrt{3}}{2}\right)$, so
\begin{align}
\text{Area}\left(\widetilde{S}_{t}\right) & =\pi\left(\left(k_{F}+\frac{\sqrt{3}}{2}\right)^{2}-\left(t-\left|k\right|\right)^{2}\right)=\pi\left(\left(k_{F}+\frac{\sqrt{3}}{2}\right)^{2}-\left(t^{2}-2\left|k\right|\left(t-\frac{1}{2}\left|k\right|\right)\right)\right)\nonumber \\
 & =2\pi\left(\left|k\right|\left(t-\frac{1}{2}\left|k\right|\right)+\sqrt{3}k_{F}\right)+\pi\left(\left(k_{F}-\frac{\sqrt{3}}{2}\right)^{2}-t^{2}\right) \\
 &\leq2\pi\left(\left|k\right|\left(t-\frac{1}{2}\left|k\right|\right)+\sqrt{3}k_{F}\right) \nonumber
\end{align}
for $k_{F}-\sqrt{3}/2\leq t\leq k_{F}+\sqrt{3}/2+\left|k\right|$. 

With these formulas we can now give the 

%\begin{prop} \label{prop:one-half-NonSingularRiemannSums}
%For all $k\in B\left(0,2k_{F}\right)\cap\mathbb{Z}_{\ast}^{3}$ it
%holds that
%\[
%\sum_{p\in L_{k}}\lambda_{k,p}^{\beta}\leq C\begin{cases}
%k_{F}^{2+\beta}\left|k\right|^{1+\beta} & -1<\beta\leq0\\
%\left(1+\left|k\right|^{-1}\log\left(k_{F}\right)\right)k_{F} & \beta=-1.
%\end{cases}
%\]
%%for a constant $C>0$ depending only on $\beta$.
%\end{prop}

\medskip
\textbf{Proof of the $\left|k\right|<2k_{F}$
case of Proposition \ref{prop:NonSingularRiemannSums} and \eqref{eq:r-sum-beta=-1}:} By equation (\ref{eq:SmallBetaCutoffEstimate}) we
have 
\begin{equation}
\sum_{p\in L_{k}}\lambda_{k,p}^{\beta}\leq2^{-\beta}\,\text{Vol}\left(\widetilde{S}^{1}\right)+\left|k\right|^{\beta}\int_{\widetilde{S}^{2}}\left(\hat{k}\cdot p-\frac{1}{2}\left|k\right|\right)^{\beta}dp
\end{equation}
and we can estimate  
\begin{align}
\text{Vol}\left(\widetilde{S}^{1}\right) & =\int_{\frac{1}{2}\left|k\right|-\frac{\sqrt{3}}{2}}^{\frac{1}{2}\left|k\right|+1+\sqrt{3}}\text{Area}\left(\widetilde{S}_{t}\right)dt=2\pi\int_{\frac{1}{2}\left|k\right|-\frac{\sqrt{3}}{2}}^{\frac{1}{2}\left|k\right|+1+\sqrt{3}}\left(\left|k\right|\left(t-\frac{1}{2}\left|k\right|\right)+\sqrt{3}k_{F}\right)dt\\
 & =2\pi\int_{-\frac{\sqrt{3}}{2}}^{1+\sqrt{3}}\left(\left|k\right|t+\sqrt{3}k_{F}\right)dt\leq C\left(\left|k\right|+k_{F}\right)\leq Ck_{F} = O\left(k_{F}^{2+\beta}\left|k\right|^{1+\beta}\right),\nonumber 
\end{align}
for all $-1\leq\beta\leq0$, and
\begin{align}
& \int_{\widetilde{S}^{2}}\left(\hat{k}\cdot p-\frac{1}{2}\left|k\right|\right)^{\beta}dp  =\int_{\frac{1}{2}\left|k\right|+1}^{k_{F}+\frac{\sqrt{3}}{2}+\left|k\right|}\left(t-\frac{1}{2}\left|k\right|\right)^{\beta}\text{Area}\left(\widetilde{S}_{t}\right)dt\nonumber \\
 & \leq2\pi\int_{\frac{1}{2}\left|k\right|+1}^{k_{F}+\frac{\sqrt{3}}{2}+\left|k\right|}\left(t-\frac{1}{2}\left|k\right|\right)^{\beta}\left(\left|k\right|\left(t-\frac{1}{2}\left|k\right|\right)+\sqrt{3}k_{F}\right)dt\nonumber \\
 & =2\pi\left(\left|k\right|\int_{1}^{k_{F}+\frac{\sqrt{3}}{2}+\frac{1}{2}\left|k\right|}t^{1+\beta}\,dt+\sqrt{3}k_{F}\int_{1}^{k_{F}+\frac{\sqrt{3}}{2}+\frac{1}{2}\left|k\right|}t^{\beta}\,dt\right)\label{eq:luneIntegralEstimate}\\
 & \leq2\pi\left(\frac{\left|k\right|}{2+\beta}\left(k_{F}+\frac{\sqrt{3}}{2}+\frac{1}{2}\left|k\right|\right)^{2+\beta}+\frac{\sqrt{3}}{1+\beta}k_{F}\left(k_{F}+\frac{\sqrt{3}}{2}+\frac{1}{2}\left|k\right|\right)^{1+\beta}\right) \leq Ck_{F}^{2+\beta}\left|k\right|\nonumber 
\end{align}
for $-1<\beta\leq0$, and 
\begin{align}
& \int_{\widetilde{S}^{2}}\left(\hat{k}\cdot p-\frac{1}{2}\left|k\right|\right)^{-1}dp  \leq2\pi\left(\left|k\right|\int_{1}^{k_{F}+\frac{\sqrt{3}}{2}+\frac{1}{2}\left|k\right|}1\,dt+\sqrt{3}k_{F}\int_{1}^{k_{F}+\frac{\sqrt{3}}{2}+\frac{1}{2}\left|k\right|}t^{-1}\,dt\right)  \label{eq:luneIntegralEstimate2}\\
 & \leq C\left(\left|k\right|k_{F}+k_{F}\log\left(k_{F}+\frac{\sqrt{3}}{2}+\frac{1}{2}\left|k\right|\right)\right)  \leq C\left|k\right|\left(1+\left|k\right|^{-1}\log\left(k_{F}\right)\right)k_{F}\nonumber 
\end{align}
for $\beta=-1$. Combining the estimates yields the claim.

$\hfill\square$

{\bf Proof of the $\left|k\right|\protect\geq2k_{F}$ case of Proposition
\ref{prop:NonSingularRiemannSums}:} For $\left|k\right|\geq2k_{F}$ the lune $S=\overline{B}\left(k,k_{F}\right)\backslash\overline{B}\left(0,k_{F}\right)$
degenerates into a ball and so we must adapt our argument. Now it
is simply the case that
\begin{equation}
S_{+}=\widetilde{S}=\overline{B}\left(k,k_{F}+\frac{\sqrt{3}}{2}\right). 
\end{equation}
If $\frac{1}{2}\left|k\right|\geq k_{F}+\frac{2+\sqrt{3}}{2}$, then every $p\in\widetilde{S}$ satisfies $\hat{k}\cdot p-\frac{1}{2}\left|k\right|\geq1$
and the cut-off set $\widetilde{S}^{1}$ is unnecessary. Otherwise, the equation (\ref{eq:SmallBetaCutoffEstimate}),
\begin{equation}
\sum_{p\in L_{k}}\lambda_{k,p}^{\beta}\leq2^{-\beta}\,\text{Vol}\left(\widetilde{S}^{1}\right)+\left|k\right|^{\beta}\int_{\widetilde{S}^{2}}\left(\hat{k}\cdot p-\frac{1}{2}\left|k\right|\right)^{\beta}\,dp,
\end{equation}
still holds for
\begin{equation}
\widetilde{S}^{1}=\left\{ p\in\widetilde{S}\mid\hat{k}\cdot p-\frac{1}{2}\left|k\right|\leq1+\sqrt{3}\right\} ,\quad\widetilde{S}^{2}=\left\{ p\in\widetilde{S}\mid\hat{k}\cdot p-\frac{1}{2}\left|k\right|\geq+1\right\} ,
\end{equation}
where we simplified the description for $\widetilde{S}^{1}$ using that $\hat{k}\cdot p-\frac{1}{2}\left|k\right|\geq-\frac{\sqrt{3}}{2}$
holds for all $p\in\widetilde{S}$ when $\left|k\right|\geq2k_{F}$.
We can then easily estimate $\text{Vol}\left(\widetilde{S}_{1}\right)$,
as it is now seen to be a spherical cap of radius $k_{F}+\frac{\sqrt{3}}{2}$
and height
\begin{equation}
\left(\frac{1}{2}\left|k\right|+1+\sqrt{3}\right)-\left(\left|k\right|-k_{F}-\frac{\sqrt{3}}{2}\right)\leq k_{F}-\frac{1}{2}\left|k\right|+\frac{2+3\sqrt{3}}{2}\leq\frac{2+3\sqrt{3}}{2}
\end{equation}
whence
\begin{equation}
\text{Vol}\left(\widetilde{S}^{1}\right)\leq\frac{\pi}{3}\left(\frac{2+3\sqrt{3}}{2}\right)^{2}\left(3\left(k_{F}+\frac{\sqrt{3}}{2}\right)-\frac{2+3\sqrt{3}}{2}\right)\leq Ck_{F}
\end{equation}
so as $k_{F}=O\left(k_{F}^{3}\left|k\right|^{2\beta}\right)$ for
all $-1\leq\beta\leq0$ when $2k_{F}\leq\left|k\right|\leq2k_{F}+\frac{\sqrt{3}}{2}$
this is again negligible.

We estimate the integrals to conclude the 
%second part of Proposition
%\ref{prop:NonSingularRiemannSums}:
%\begin{prop} \label{prop:second-half-NonSingularRiemannSums}
%For all $k\in\mathbb{Z}_{\ast}^{3}\backslash B\left(0,2k_{F}\right)$
%and $-1\leq\beta\leq0$ it holds that
%\[
%\sum_{p\in L_{k}}\lambda_{k,p}^{\beta}\leq Ck_{F}^{3}\left|k\right|^{2\beta}
%\]
%for a constant $C>0$ depending only on $\beta$.
%\end{prop}

\textbf{Proof of the second part of Proposition
\ref{prop:NonSingularRiemannSums}:} We again note that the area of the slice $\widetilde{S}_{t}$
is given by
\begin{equation}
\text{Area}\left(\widetilde{S}_{t}\right)=\pi\left(\left(k_{F}+\frac{\sqrt{3}}{2}\right)^{2}-\left(t-\left|k\right|\right)^{2}\right),
\end{equation}
now for $\left|k\right|-k_{F}-\frac{\sqrt{3}}{2}\leq t\leq\left|k\right|+k_{F}+\frac{\sqrt{3}}{2}$.
If $\left|k\right|\leq2k_{F}+1+\sqrt{3}$ we just saw that the contribution
coming from the cut-off set $\widetilde{S}^{1}$ is negligible, while
the integral term is
\begin{equation}
\left|k\right|^{\beta}\int_{\widetilde{S}^{2}}\left(\hat{k}\cdot p-\frac{1}{2}\left|k\right|\right)^{\beta}dp=\left|k\right|^{\beta}\int_{\frac{1}{2}\left|k\right|+1}^{k_{F}+\frac{\sqrt{3}}{2}+\left|k\right|}\left(t-\frac{1}{2}\left|k\right|\right)^{\beta}\text{Area}\left(\widetilde{S}_{t}\right)dt\leq Ck_{F}^{2+\beta}\left|k\right|^{1+\beta}
\end{equation}
as calculated in equation (\ref{eq:luneIntegralEstimate}), which
is $O\left(k_{F}^{3}\left|k\right|^{2\beta}\right)$ for $2k_{F}\leq\left|k\right|\leq2k_{F}+1+\sqrt{3}$
(here we also use that for $\beta=-1$, the logarithmic term in the
estimate of equation (\ref{eq:luneIntegralEstimate2}) is negligible
when $\left|k\right|\geq2k_{F}$ due to the additional factor of $\left|k\right|^{-1}$).

If $\left|k\right|>2k_{F}+\frac{2+\sqrt{3}}{2}$ we simply have
\begin{equation}
\sum_{p\in L_{k}}\lambda_{k,p}^{\beta}\leq\left|k\right|^{\beta}\int_{\widetilde{S}}\left(\hat{k}\cdot p-\frac{1}{2}\left|k\right|\right)^{\beta}dp=\left|k\right|^{\beta}\int_{\left|k\right|-k_{F}-\frac{\sqrt{3}}{2}}^{\left|k\right|+k_{F}+\frac{\sqrt{3}}{2}}\left(t-\frac{1}{2}\left|k\right|\right)^{\beta}\text{Area}\left(\widetilde{S}_{t}\right)dt,
\end{equation}
and by writing $\left(t-\left|k\right|\right)^{2}=\left(t-\frac{1}{2}\left|k\right|\right)^{2}-\left|k\right|\left(t-\frac{1}{2}\left|k\right|\right)+\frac{1}{4}\left|k\right|^{2}$
we can furthermore estimate that
\begin{align}
&\text{Area}\left(\widetilde{S}_{t}\right)  =\pi\left(\left(k_{F}+\frac{\sqrt{3}}{2}\right)^{2}-\left(\left(t-\frac{1}{2}\left|k\right|\right)^{2}-\left|k\right|\left(t-\frac{1}{2}\left|k\right|\right)+\frac{1}{4}\left|k\right|^{2}\right)\right)\\
 & =\pi\left(\left|k\right|\left(t-\frac{1}{2}\left|k\right|\right)-\left(\frac{1}{4}\left|k\right|^{2}-\left(k_{F}+\frac{\sqrt{3}}{2}\right)^{2}\right)-\left(t-\frac{1}{2}\left|k\right|\right)^{2}\right)\leq\pi\left|k\right|\left(t-\frac{1}{2}\left|k\right|\right)\nonumber 
\end{align}
so
\begin{align}
\sum_{p\in L_{k}}\lambda_{k,p}^{\beta} & \leq\pi\left|k\right|^{1+\beta}\int_{\left|k\right|-k_{F}-\frac{\sqrt{3}}{2}}^{\left|k\right|+k_{F}+\frac{\sqrt{3}}{2}}\left(t-\frac{1}{2}\left|k\right|\right)^{1+\beta}dt\nonumber \\
 & =\frac{\pi\left|k\right|^{1+\beta}}{2+\beta}\left(\left(\frac{1}{2} \left|k\right|+k_{F}+\frac{\sqrt{3}}{2}\right)^{2+\beta}-\left(\frac{1}{2} \left|k\right|-k_{F}-\frac{\sqrt{3}}{2}\right)^{2+\beta}\right)\\
 & \leq C\left|k\right|^{1+\beta}\left(\frac{1}{2} \left|k\right|+k_{F}+\frac{\sqrt{3}}{2}\right)^{2+\beta}\leq C\left|k\right|^{3+2\beta}.\nonumber 
\end{align}
If additionally $\left|k\right|\leq3k_{F}$ (say) then this is again
$O\left(k_{F}^{3}\left|k\right|^{2\beta}\right)$. If this is not
the case, however, then we can instead trivially estimate that
\begin{align}
\sum_{p\in L_{k}}\lambda_{k,p}^{\beta} & \leq\left|k\right|^{\beta}\int_{\widetilde{S}}\left(\hat{k}\cdot p-\frac{1}{2}\left|k\right|\right)^{\beta}dp\leq\left|k\right|^{\beta}\left(\inf_{p\in\widetilde{S}}\left(\hat{k}\cdot p-\frac{1}{2}\left|k\right|\right)\right)^{\beta}\int_{\widetilde{S}}1\,dp\nonumber \\
 & \leq \left| k \right|^{\beta} \left(\frac{1}{2}\left|k\right|-k_{F}-\frac{\sqrt{3}}{2}\right)^{\beta}\text{Vol}\left(\overline{B}\left(0,k_{F}+\frac{\sqrt{3}}{2}\right)\right)\\
 & \leq Ck_{F}^{3}\left|k\right|^{\beta}\left(\frac{1}{2}\left|k\right|-\frac{1}{3}\left|k\right|-\frac{\sqrt{3}}{2}\right)^{\beta}\leq Ck_{F}^{3}\left|k\right|^{2\beta}. \nonumber
\end{align}
$\hfill\square$

\subsection{Proof of Proposition \ref{coro:CompletelyUniformRiemannSumBound}} \label{sec:beta=-1}

%\begin{prop}
%\label{prop:LambdaInverseEstimate}For all $k\in\overline{B}\left(0,2k_{F}\right)\cap\mathbb{Z}_{\ast}^{3}$
%it holds that
%\[
%\sum_{p\in L_{k}^{\pm}}\frac{1}{\lambda_{k,p}}\leq C\left(1+\left|k\right|^{3+\frac{2}{3}}\log\left(k_{F}\right)^{\frac{5}{3}}k_{F}^{-\frac{1}{3}}\right)k_{F},\quad k_{F}\rightarrow\infty,
%\]
%for a constant $C>0$ independent of $k$ and $k_{F}$.
%\end{prop}

In the cases $|k| \ge 2k_F$ and $2k_F\ge |k| \ge \log(k_F)$, the claim has been proved. 
%the claim follows from Propositions \ref{prop:second-half-NonSingularRiemannSums} and \ref{prop:one-half-NonSingularRiemannSums}, respectively. 
Thus it remains to consider the case $|k|\le \log(k_F)$, for which we will apply the summation formula in Proposition \ref{prop:SummationFormula} to improve \eqref{eq:r-sum-beta=-1}. By Proposition \ref{prop:SummationFormula} we have
\begin{align}
\sum_{p\in L_{k}}\frac{1}{\lambda_{k,p}} & =2\pi\left|k\right|\sum_{m=m^{\ast}}^{M^{\ast}}\frac{lm-\frac{1}{2}\left|k\right|}{\left|k\right|\left(lm-\frac{1}{2}\left|k\right|\right)}l+\pi\sum_{m=M+1}^{M^{\ast}}\frac{k_{F}^{2}-\left(lm\right)^{2}}{\left|k\right|\left(lm-\frac{1}{2}\left|k\right|\right)}l\nonumber \\
 &  +O\left(\left|k\right|^{3+\frac{2}{3}}(\log k_{F})^{\frac{2}{3}}k_{F}^{\frac{2}{3}}\sum_{m=m^{\ast}}^{M^{\ast}}\frac{1}{\left|k\right|\left(lm-\frac{1}{2}\left|k\right|\right)}\right)\\
 & \leq2\pi\sum_{m=m^{\ast}}^{M^{\ast}}l+O\left(\left|k\right|^{2+\frac{2}{3}}(\log k_{F})^{\frac{2}{3}}k_{F}^{\frac{2}{3}}\sum_{m=m^{\ast}}^{M^{\ast}}\frac{1}{lm-\frac{1}{2}\left|k\right|}\right),\quad k_{F}\rightarrow\infty,\nonumber 
\end{align}
where we used that by definition of $M$, $\left(k_{F}^{2}-\left(lm\right)^{2}\right)<0$
for all $m\geq M+1$. As $\left|k\right|\leq2k_{F}$, 
\begin{equation}
\sum_{m=m^{\ast}}^{M^{\ast}}l=l\left(M^{\ast}-m^{\ast}+1\right)\leq k_{F}+\left|k\right|+l\leq Ck_{F},\quad k_{F}\rightarrow\infty,
\end{equation}
where we also used that
\begin{equation}
l=\left|k\right|^{-1}\gcd\left(k_{1},k_{2},k_{3}\right)\leq\frac{\max\left(\left|k_{1}\right|,\left|k_{2}\right|,\left|k_{3}\right|\right)}{\sqrt{k_{1}^{2}+k_{2}^{2}+k_{3}^{2}}}\leq1.
\end{equation}
We now consider the sum $\sum_{m=m^{\ast}}^{M^{\ast}} \Big( lm-\frac{1}{2}\left|k\right| \Big)^{-1}$.
To apply Proposition \ref{prop:ConvexRiemannSumEstimate} we must
estimate the $m=m^{\ast}$ term separately, so that the integration
range does not cross the point $x=\frac{1}{2}\left|k\right|$, where
the integrand diverges. Note that using $
\lambda_{k,p}\geq\frac{1}{2}
$ for all $p\in L_k$, we have
\begin{equation}\label{eq:lemma:LatticeBound}
lm^{\ast}-\frac{1}{2}\left|k\right|=\min_{p\in L_{k}}\left(\hat{k}\cdot p-\frac{1}{2}\left|k\right|\right)=\left|k\right|^{-1}\left(\min_{p\in L_{k}}\left(k\cdot p-\frac{1}{2}\left|k\right|^{2}\right)\right)\geq\frac{1}{2}\left|k\right|^{-1}.
\end{equation}
%
%We will also need Proposition \ref{prop:ConvexRiemannSumEstimate} in Section \ref{subsect:RiemannSumEstimatesforBetageq-1}, which requires a lower bound on $lm^{\ast}-\frac{1}{2}\left|k\right|$. 
%
%\begin{lem}
%\label{lemma:LatticeBound}For all $k\in\mathbb{Z}_{\ast}^{3}$ and
%$p\in L_{k}$ it holds that
%$
%\lambda_{k,p}\geq\frac{1}{2}.
%$ 
%Consequently $m^{\ast}$ obeys
%\[
%lm^{\ast}-\frac{\left|k\right|}{2}\geq\frac{1}{2}\left|k\right|^{-1}.
%\]
%\end{lem}
%
%\textbf{Proof:} For any $k\in\mathbb{Z}_{\ast}^{3}$ and $p\in L_{k}$
%we have as noted in equation (\ref{eq:SimpleLuneContainment}) that
%$\hat{k}\cdot p>\frac{1}{2}\left|k\right|$, hence $\lambda_{k,p}=k\cdot p-\frac{1}{2}\left|k\right|^{2} >0$. 
%%\begin{equation}
%%k\cdot p>\frac{1}{2}\left|k\right|^{2}.
%%\end{equation}
%Since $k,p\in\mathbb{Z}^{3}$, both $k\cdot p$ and $\left|k\right|^{2}$
%are integers, and hence $\lambda_{k,p}\ge \frac 1 2$. 
%% As the above inequality is strict we therefore conclude
%%that $k\cdot p$ is greater than or equal to the least integer which
%%is greater than the right-hand side, and since $\left|k\right|^{2}$
%%is also an integer this is either $\frac{1}{2}\left|k\right|^{2}+1$
%%or $\frac{1}{2}\left|k\right|^{2}+\frac{1}{2}$ depending on whether
%%$\left|k\right|^{2}$ is even or odd. Either way we have that
%%\begin{equation}
%%\lambda_{k,p}=k\cdot p-\frac{1}{2}\left|k\right|^{2}\geq\frac{1}{2}
%%\end{equation}
%%as claimed. 
%The bound on $m^{\ast}$ now follows by noting that
%
%$\hfill\square$
Therefore, 
\begin{align} 
& \sum_{m=m^{\ast}}^{M^{\ast}}\frac{1}{lm-\frac{1}{2}\left|k\right|} \le 2|k| + \sum_{m=m^{\ast}+1}^{M^{\ast}}\frac{1}{lm-\frac{1}{2}\left|k\right|} \leq 2|k|  +\left|k\right|\int_{lm^{\ast}+\frac{1}{2}l}^{lM^{\ast}+\frac{1}{2}l}\frac{1}{x-\frac{1}{2}\left|k\right|}\,dx\nonumber \\
 & \leq C\left|k\right|\left(1+\log\left(\frac{lM^{\ast}+\frac{1}{2}l-\frac{1}{2}\left|k\right|}{lm^{\ast}+\frac{1}{2}l-\frac{1}{2}\left|k\right|}\right)\right)\leq C\left|k\right|\left(1+\log\left(\frac{k_{F}+\left|k\right|+\frac{1}{2}l-\frac{1}{2}\left|k\right|}{\frac{1}{2}l}\right)\right)\label{eq:InverseSumEstimate}\\
 & \leq C\left|k\right|\left(1+\log\left(\left|k\right|k_{F}\right)\right)\leq C\left|k\right|\log\left(k_{F}\right),\quad k_{F}\rightarrow\infty,\nonumber 
\end{align}
yielding the total bound when $|k|\le \log(k_F)$ 
\begin{align} \label{eq:Inverse-lambda-tricky}
\sum_{p\in L_{k}} \frac{1}{\lambda_{k,p}}  \leq C\left( k_F +\left|k\right|^{3+\frac{2}{3}}\log\left(k_{F}\right)^{\frac{5}{3}}k_{F}^{\frac{2}{3}}\right) \le Ck_F.
\end{align}
%which is bounded by $Ck_F$ when $|k|\le \log(k_F)$.
$\hfill\square$

%
%The claim follows from Propositions \ref{prop:second-half-NonSingularRiemannSums} if $|k| \ge 2k_F$, Proposition \ref{prop:one-half-NonSingularRiemannSums} if $\log(k_F) \le |k| \le 2k_F$ 

%The presence of the term $\left|k\right|^{3+\frac{2}{3}}\log\left(k_{F}\right)^{\frac{5}{3}}k_{F}^{-\frac{1}{3}}$
%is unfortunate, but this bound is still good if $|k|$ is not too large. % unavoidable, and the remaining estimates will contain similar terms - this is the source of the constraint on $\gamma$ in Proposition \ref{prop:MoreSingularRiemannSums}.
%
%
%
%The first estimate extends that of Proposition \ref{prop:NonSingularRiemannSums},
%with the caveat that it only holds for $k$ with $\left|k\right|$
%sufficiently small compared to $k_{F}$. In practice this limits how
%big we can take the exponent $\gamma>0$, which determines our cut-off
%set $S_{C}=\overline{B}\left(0,k_{F}^{\gamma}\right)\cap\mathbb{Z}_{+}^{3}$,
%to be. For the upper bound, where only the sums of equation (\ref{eq:ParticularRiemannSums})
%enter, this limit will be $\gamma\leq1$, as the $\beta=-1$ cases
%of the Propositions \ref{prop:NonSingularRiemannSums} and \ref{prop:MoreSingularRiemannSums}
%can be combined to obtain the following:
%\[
%\sum_{p\in L_{k}}\lambda_{k,p}^{-1}\leq Ck_{F},\quad k_{F}\rightarrow\infty,
%\]

\subsection{Proof of Proposition \ref{prop:MoreSingularRiemannSums}}

First, consider the case $-\frac 4 3 \le \beta<-1$ and $k \in \overline{B}(0,2k_F)$. By Proposition \ref{prop:SummationFormula} we can
estimate using the argument leading to \eqref{eq:Inverse-lambda-tricky} that
\begin{align}
&\sum_{p\in L_{k}}\lambda_{k,p}^{\beta}  =2\pi\left|k\right|\sum_{m=m^{\ast}}^{M^{\ast}}\left(\left|k\right|\left(lm-\frac{1}{2}\left|k\right|\right)\right)^{\beta}\left(lm-\frac{1}{2}\left|k\right|\right)l\nonumber \\
 & \qquad\qquad +\pi\sum_{m=M+1}^{M^{\ast}}\left(\left|k\right|\left(lm-\frac{1}{2}\left|k\right|\right)\right)^{\beta}\left(k_{F}^{2}-\left(lm\right)^{2}\right)l\\
 & \qquad\qquad +O\left(\left|k\right|^{3+\frac{2}{3}}(\log k_{F})^{\frac{2}{3}}k_{F}^{\frac{2}{3}}\sum_{m=m^{\ast}}^{M^{\ast}}\left(\left|k\right|\left(lm-\frac{1}{2}\left|k\right|\right)\right)^{\beta}\right)\nonumber \\
 & \leq2\pi\left|k\right|^{1+\beta}\sum_{m=m^{\ast}}^{M^{\ast}}\left(lm-\frac{1}{2}\left|k\right|\right)^{1+\beta}l+O\left(\left|k\right|^{3+\frac{2}{3}+\beta}(\log k_{F})^{\frac{2}{3}}k_{F}^{\frac{2}{3}}\sum_{m=m^{\ast}}^{M^{\ast}}\left(lm-\frac{1}{2}\left|k\right|\right)^{\beta}\right).\nonumber 
\end{align}
Applying Proposition \ref{prop:ConvexRiemannSumEstimate} and \ref{eq:lemma:LatticeBound}  again we have
\begin{align}
 & \qquad\;\sum_{m=m^{\ast}}^{M^{\ast}}\left(lm-\frac{1}{2}\left|k\right|\right)^{1+\beta}l=\left(lm^{\ast}-\frac{1}{2}\left|k\right|\right)^{1+\beta}l+\sum_{m=m^{\ast}+1}^{M^{\ast}}\left(lm-\frac{1}{2}\left|k\right|\right)^{1+\beta}l\nonumber \\
 & \leq\left(\frac{1}{2}\left|k\right|^{-1}\right)^{1+\beta}+\int_{lm^{\ast}+\frac{1}{2}l}^{lM^{\ast}+\frac{1}{2}l}\left(lm-\frac{1}{2}\left|k\right|\right)^{1+\beta}dx\\
 & =\left(\frac{1}{2}\left|k\right|^{-1}\right)^{1+\beta}+\frac{1}{2+\beta}\left(\left(lM^{\ast}+\frac{1}{2}l-\frac{1}{2}\left|k\right|\right)^{2+\beta}-\left(lm^{\ast}+\frac{1}{2}l-\frac{1}{2}\left|k\right|\right)^{2+\beta}\right)\nonumber \\
 & \leq C\left(\left|k\right|^{-(1+\beta)} + \left(k_{F}+\frac{1}{2}l+\frac{1}{2}\left|k\right|\right)^{2+\beta}\right)\leq Ck_{F}^{2+\beta},\quad k_{F}\rightarrow\infty,\nonumber 
\end{align}
and likewise  
\begin{align}
 & \qquad\;\sum_{m=m^{\ast}}^{M^{\ast}}\left(lm-\frac{1}{2}\left|k\right|\right)^{\beta}=\left(lm^{\ast}-\frac{1}{2}\left|k\right|\right)^{\beta}+l^{-1}\sum_{m=m^{\ast}+1}^{M^{\ast}}\left(lm-\frac{1}{2}\left|k\right|\right)^{\beta}l\nonumber \\
 & \leq\left(\frac{1}{2}\left|k\right|^{-1}\right)^{\beta}+\left|k\right|\int_{lm^{\ast}+\frac{1}{2}l}^{lM^{\ast}+\frac{1}{2}l}\left(lm-\frac{1}{2}\left|k\right|\right)^{\beta}dx\label{eq:GeneralErrorTermRiemannEstimate}\\
 & =2^{-\beta}\left|k\right|^{-\beta}+\frac{\left|k\right|}{1+\beta}\left(\left(lm^{\ast}+\frac{1}{2}l-\frac{1}{2}\left|k\right|\right)^{1+\beta}-\left(lM^{\ast}+\frac{1}{2}l-\frac{1}{2}\left|k\right|\right)^{1+\beta}\right)\nonumber \\
 & \leq C\left(\left|k\right|^{-\beta}+\left|k\right|\left(lm^{\ast}+\frac{1}{2}l-\frac{1}{2}\left|k\right|\right)^{1+\beta}\right)\geq C\left(\left|k\right|^{-\beta}+\left|k\right|\left(\frac{1}{2}\left|k\right|^{-1}\right)^{1+\beta}\right)\leq C\left|k\right|^{-\beta}.\nonumber 
\end{align}
Combining these we find that for all $-\frac 4 3 \le \beta<-1$ and $k\in \overline{B}(0,2k_F)$, 
\begin{equation} \label{eq:prop:MoreSingularLambdaEstimate}
\sum_{p\in L_{k}}\lambda_{k,p}^{\beta}\leq C\left(k_{F}^{2+\beta}\left|k\right|^{1+\beta}+\left|k\right|^{3+\frac{2}{3}}(\log k_{F})^{\frac{2}{3}}k_{F}^{\frac{2}{3}}\right),\quad k_{F}\rightarrow\infty. 
\end{equation}
Consequently, if $\beta\le -\frac 4 3$ and $k\in \overline{B}(0,2k_F)$, then using $\lambda_{k,p}\ge \frac 1 2$ we have
\begin{equation}  
\sum_{p\in L_{k}}\lambda_{k,p}^{\beta}\leq C \sum_{p\in L_{k}}\lambda_{k,p}^{- \frac 4 3} \le C \left|k\right|^{3+\frac{2}{3}}(\log k_{F})^{\frac{2}{3}}k_{F}^{\frac{2}{3}}. 
\end{equation}
Moreover, if $-\frac{4}{3}<\beta<-1$ and $\left|k\right|\leq k_{F}^{\gamma}$ with $\gamma<\frac{4+3\beta}{8-3\beta}$, then the right-hand side of\eqref{eq:prop:MoreSingularLambdaEstimate} can be simplified to $C k_{F}^{2+\beta}\left|k\right|^{1+\beta}.$
$\hfill\square$

\subsection{Proof of Proposition \ref{prop:SklambdaPointsEstimate}} \label{subsec:EstimationofSklambda}

In this subsection we prove Proposition \ref{prop:SklambdaPointsEstimate}. 
%\begin{prop}
%\label{prop:SklambdaPointsEstimate}For all $k\in\overline{B}\left(0,k_{F}\right)\cap\mathbb{Z}_{\ast}^{3}$
%and $\lambda\left(k_{F},k\right)$ such that $\lambda\left(k_{F},k\right)\leq\frac{1}{6}k_{F}^{2}$
%it holds that
%\[
%\left|S_{k,\lambda}^{1}\right|+\left|S_{k,\lambda}^{2}\right|\leq C\left(\left|k\right|^{-1}\lambda\left(k_{F},k\right)+\left|k\right|^{3+\frac{2}{3}}(\log k_{F})^{\frac{2}{3}}k_{F}^{\frac{2}{3}}\right)\left(\lambda\left(k_{F},k\right)+\left|k\right|\right),\quad k_{F}\rightarrow\infty,
%\]
%for a constant $C>0$ independent of $k$ and $k_{F}$.
%\end{prop}
We first establish a simple upper bound:
\begin{prop}
\label{prop:Sklambda12Inclusion}For all $k\in\mathbb{Z}_{\ast}^{3}$
and any $\lambda>0$ it holds that $\left|S_{k,\lambda}^{1}\right|+\left|S_{k,\lambda}^{2}\right|\leq\left|S_{k,\lambda}\right|$
where
\[
S_{k,\lambda}=\left\{ p\in\mathbb{Z}^{3}\mid\vert\left|p\right|^{2}-\zeta\vert<\lambda \,\, \text{ and }\,\, \left|\hat{k}\cdot p-\frac{1}{2}\left|k\right|\right|<\frac{1}{2}\left|k\right|^{-1}\lambda\right\} .
\]
\end{prop}

\textbf{Proof:} As $S_{k,\lambda}^{1}\cap S_{k,\lambda}^{2}=\emptyset$
the claim will follow if we can show that $S_{k,\lambda}^{1},S_{k,\lambda}^{2}\subset S_{k,\lambda}$.
Consider an arbitrary $p\in S_{k,\lambda}^{1}:$ By definition of
$S_{k,\lambda}^{1}$
\begin{equation}
\vert\left|p\right|^{2}-\zeta\vert\leq\max\left\{ \vert\left|p\right|^{2}-\zeta\vert,\vert\left|p-k\right|^{2}-\zeta\vert\right\} <\lambda
\end{equation}
so the first condition for $S_{k,\lambda}$ is satisfied. For the
other we note that
\begin{align}
\left|2k\cdot p-\left|k\right|^{2}\right| & =\left|\left|p\right|^{2}-\left|p-k\right|^{2}\right|=\left|\vert\left|p\right|^{2}-\zeta\vert-\vert\left|p-k\right|^{2}-\zeta\vert\right|\\
 & \leq\max\left\{ \vert\left|p\right|^{2}-\zeta\vert,\vert\left|p-k\right|^{2}-\zeta\vert\right\} <\lambda\nonumber 
\end{align}
where in the second equality we used that both $p,\left(p-k\right)\in B_{F}$ if $p\in S_{k,\lambda}$.  This now implies that
$p\in S_{k,\lambda}$ so indeed $S_{k,\lambda}^{1}\subset S_{k,\lambda}$.
The inclusion $S_{k,\lambda}^{2}\subset S_{k,\lambda}$ follows similarly.

$\hfill\square$

The quantity $\left|S_{k,\lambda}\right|$ can in turn be estimated
with exactly the same techniques which we used for the estimation
of Riemann sums in the previous subsections. Let us start by using the arguments from  Proposition \ref{prop:PlaneDecompositionofZ3}. 
%the following:
%\begin{prop*}[\ref{prop:PlaneDecompositionofZ3}]
%Let $k=\left(k_{1},k_{2},k_{3}\right)\in\mathbb{Z}^{3}\backslash\left\{ 0\right\} $
%be given. Then with $l=\left|k\right|^{-1}\gcd\left(k_{1},k_{2},k_{3}\right)$
%the following disjoint union of non-empty sets holds:
%\[
%\mathbb{Z}^{3}=\bigcup_{m\in\mathbb{Z}}\left\{ p\in\mathbb{Z}^{3}\mid\hat{k}\cdot p=lm\right\} .
%\]
%Additionally, there exists linearly independent vectors $v_{1},v_{2}\in\mathbb{Z}^{3}$,
%which span $\left\{ p\in\mathbb{R}^{3}\mid\hat{k}\cdot p=0\right\} $,
%such that for any $m\in\mathbb{Z}$ it holds for all $q\in\left\{ p\in\mathbb{Z}^{3}\mid\hat{k}\cdot p=lm\right\} $
%that
%\[
%\left\{ p\in\mathbb{Z}^{3}\mid\hat{k}\cdot p=lm\right\} =q+\left\{ a_{1}v_{1}+a_{2}v_{2}\mid a_{1},a_{2}\in\mathbb{Z}\right\} .
%\]
%\end{prop*}
Now, the condition that $\vert\left|p\right|^{2}-\zeta\vert<\lambda$
is equivalent with $\zeta-\lambda<\left|p\right|^{2}<\zeta+\lambda$,
and writing $\left|p\right|^{2}=\left(\hat{k}\cdot p\right)^{2}+\left|P_{\perp}p\right|^{2}$
(where $P_{\perp}:\mathbb{R}^{3}\rightarrow\mathbb{R}^{3}$ denotes
the orthogonal projection onto $\left\{ k\right\} ^{\perp}=\left\{ p\in\mathbb{R}^{3}\mid\hat{k}\cdot p=0\right\} $)
this is equivalent with
\begin{equation}
\zeta-\left(\hat{k}\cdot p\right)^{2}-\lambda<\left|P_{\perp}p\right|^{2}<\zeta-\left(\hat{k}\cdot p\right)^{2}+\lambda.\label{eq:SklambdaEquivalentCondition}
\end{equation}
Consequently, if we let $m_{-}$ and $m_{+}$ be the least and greatest
integers, respectively, such that
\begin{equation}
\frac{1}{2}\left(\left|k\right|-\left|k\right|^{-1}\lambda\right)<lm_{-}\quad\text{and}\quad lm_{+}<\frac{1}{2}\left(\left|k\right|+\left|k\right|^{-1}\lambda\right),
\end{equation}
it follows that we can decompose $S_{k,\lambda}=\bigcup_{m=m_{-}}^{m_{+}}S_{k,\lambda}^{m}$ where  
%the sets $S_{k,\lambda}^{m}$  are
%the intersections of $S_{k,\lambda}$ with the $\hat{k}\cdot p=\text{constant}$
%planes, i.e. (by equation (\ref{eq:SklambdaEquivalentCondition}))
\begin{align}
S_{k,\lambda}^{m} & =S_{k,\lambda}\cap\left\{ p\in\mathbb{Z}^{3}\mid\hat{k}\cdot p=lm\right\} =\left\{ p\in\mathbb{Z}^{3}\mid\hat{k}\cdot p=lm,\vert\left|p\right|^{2}-\zeta\vert<\lambda\right\} \nonumber \\
 & =\left\{ p\in\mathbb{Z}^{3}\mid\hat{k}\cdot p=lm,\,\zeta-\left(lm\right)^{2}-\lambda<\left|P_{\perp}p\right|^{2}<\zeta-\left(lm\right)^{2}+\lambda\right\} \\
 & =\left\{ p\in\mathbb{Z}^{3}\mid\hat{k}\cdot p=lm,\,\left(R_{-}^{m}\right)^{2}<\left|P_{\perp}p\right|^{2}<\left(R_{-}^{m}\right)^{2}\right\} \nonumber 
\end{align}
for
\begin{equation}
\left(R_{\pm}^{m}\right)^{2}=\zeta-\left(lm\right)^{2}\pm\lambda,\quad m_{-}\leq m\leq m_{+}.
\end{equation}

We see that the sets $S_{k,\lambda}^{m}$ are of the same form as
the sets $L_{k}^{m}$ which we considered in Section \ref{sec:app-Plane-Decomposition}. 
%  as these were given by
%\begin{equation}
%L_{k}^{m}=\left\{ p\in\mathbb{Z}^{3}\mid\hat{k}\cdot p=lm,\,\left(R_{1}^{m}\right)^{2}<\left|P_{\perp}p\right|^{2}\leq\left(R_{2}^{m}\right)^{2}\right\} 
%\end{equation}
%for certain $R_{1}^{m},R_{2}^{m}\in\mathbb{R}$. 
The arguments which
we used to estimate $\left|L_{k}^{m}\right|$ thus immediately carry
over, provided we can establish some basic estimates on $R_{-}^{m}$
and $R_{+}^{m}$. We have

\begin{prop}
\label{prop:SklambdaRadiusProperties}For all $k\in\overline{B}\left(0,k_{F}\right)\cap\mathbb{Z}_{\ast}^{3}$
and $0<\lambda =\lambda \left(k_{F},k\right)\leq\frac{1}{6}k_{F}^{2}$
it holds that
\begin{equation}
C^{-1}k_F<R_{-}^{m}<R_{+}^{m}\leq Ck_{F},\quad \forall m_{-}\leq m\leq m_{+},
\end{equation}
as $k_{F}\rightarrow\infty$ for a constant $C>0$ independent of
$k$, $k_{F}$ and $\lambda$.
\end{prop}

\textbf{Proof:} First, recall that $\zeta$ is the midpoint of
the interval $I=\left[\sup_{p\in B_{F}}\left|p\right|^{2},\inf_{p\in B_{F}^{c}}\left|p\right|^{2}\right]$. Since $k_{F}^{2}\in I$ by definition of the Fermi ball, we can bound 
\begin{equation}\label{eq:lemma:Zeta0Estimate}
|\zeta-k_F^2|\le \frac{|I|}{2}=\frac{1}{2}\left( \inf_{q\in B_{F}^{c}}\left|p\right|^{2}-\sup_{q\in B_{F}}\left|p\right|^{2} \right) \le k_F+1.
\end{equation}
Here the last inequality can be seen by taking the trial points $p_{-}=\left(\left\lfloor k_{F}\right\rfloor ,0,0\right) \in B_F$ and
$p_{+}=\left(\left\lfloor k_{F}\right\rfloor +1,0,0\right) \in B_F^c$. Combining \eqref{eq:lemma:Zeta0Estimate}, the
definitions of $m_{-},$ $m_{+}$ and the assumptions of the statement
we may estimate independently of $m$ that 
\begin{align}
\left(R_{-}^{m}\right)^{2}&\geq\zeta-\max\left\{ \left(lm_{-}\right)^{2},\left(lm_{+}\right)^{2}\right\} -\lambda\nonumber \\
 & \geq\zeta-\frac{1}{4}\left(\left(\left|k\right|-\left|k\right|^{-1}\lambda\right)^{2}+\left(\left|k\right|+\left|k\right|^{-1}\lambda\right)^{2}\right)-\lambda\\
 & \geq\zeta-\frac{1}{2}\left(\left|k\right|^{2}+\left|k\right|^{-2}\lambda\right)-\lambda\geq\zeta-\frac{1}{2}k_{F}^{2}-\frac{3}{2}\lambda\geq\frac{1}{4}k_{F}^{2}-k_{F}-1,\nonumber 
 \end{align}
 and
 \begin{align}
\left(R_{+}^{m}\right)^{2}=\zeta-\left(lm\right)^{2}+\lambda\leq\zeta+\frac{1}{6}k_{F}^{2}\leq\frac{7}{6}k_{F}^{2}+k_{F}+1.
\end{align}
$\hfill\square$

This allows us to estimate $|S_{k,\lambda}^{m}|$
with the same error term as that of $|L_{k}^{m}|$, which
is to say $C\left|k\right|^{3+\frac{2}{3}}(\log k_{F})^{\frac{2}{3}}k_{F}^{\frac{2}{3}}$.
We can now give: 

\medskip
%\begin{prop*}[\ref{prop:SklambdaPointsEstimate}]
%For all $k\in\overline{B}\left(0,k_{F}\right)\cap\mathbb{Z}_{\ast}^{3}$
%and $\lambda\left(k_{F},k\right)$ such that $\lambda\left(k_{F},k\right)\leq\frac{1}{6}k_{F}^{2}$
%it holds that
%\[
%\left|S_{k,\lambda}\right|\leq C\left(\left|k\right|^{-1}\lambda\left(k_{F},k\right)+\left|k\right|^{3+\frac{2}{3}}(\log k_{F})^{\frac{2}{3}}k_{F}^{\frac{2}{3}}\right)\left(\lambda\left(k_{F},k\right)+\left|k\right|\right),\quad k_{F}\rightarrow\infty,
%\]
%for a constant $C>0$ independent of $k$ and $k_{F}$.
%\end{prop*}

\textbf{Proof of Proposition \ref{prop:SklambdaPointsEstimate}:} By Proposition \ref{prop:SklambdaRadiusProperties}
and the above arguments,  we can
estimate 
\begin{align}
\left|S_{k,\lambda}^{m}\right| &\leq\frac{2\pi\left(\left(R_{+}^{m}\right)^{2}-\left(R_{-}^{m}\right)^{2}\right)}{2l^{-1}}+C\left|k\right|^{3+\frac{2}{3}}(\log k_{F})^{\frac{2}{3}}k_{F}^{\frac{2}{3}}\nonumber\\
&=2\pi\lambda\left(k_{F},k\right)l+C\left|k\right|^{3+\frac{2}{3}}(\log k_{F})^{\frac{2}{3}}k_{F}^{\frac{2}{3}}
\end{align}
for $m_{-}\leq m\leq m_{+}$. By the decomposition $S_{k,\lambda}=\bigcup_{m=m_{-}}^{m_{+}}S_{k,\lambda}^{m}$
we can then estimate further 
\begin{align}
\left|S_{k,\lambda}\right| & =\sum_{m=m_{-}}^{m_{+}}\left|S_{k,\lambda}^{m}\right|\leq2\pi\lambda\sum_{m=m_{-}}^{m_{+}}l+C\left|k\right|^{3+\frac{2}{3}}(\log k_{F})^{\frac{2}{3}}k_{F}^{\frac{2}{3}}\sum_{m=m_{-}}^{m_{+}}1\nonumber \\
 & \le C\left(\lambda+\left|k\right|^{4+\frac{2}{3}}(\log k_{F})^{\frac{2}{3}}k_{F}^{\frac{2}{3}}\right)\left(lm_{+}-lm_{-}+l\right)\\
 & \leq C\left(\lambda+\left|k\right|^{4+\frac{2}{3}}(\log k_{F})^{\frac{2}{3}}k_{F}^{\frac{2}{3}}\right)\left(\frac{1}{2}\left(\left|k\right|+\left|k\right|^{-1}\lambda\right)-\frac{1}{2}\left(\left|k\right|-\left|k\right|^{-1}\lambda\right)+l\right)\nonumber \\
 & \leq C\left(\left|k\right|^{-1}\lambda+\left|k\right|^{3+\frac{2}{3}}(\log k_{F})^{\frac{2}{3}}k_{F}^{\frac{2}{3}}\right)\left(\lambda+\left|k\right|\right)\nonumber 
\end{align}
where we also applied the estimate $\left|k\right|^{-1}\leq l\leq1$.
$\hfill\square$

%%%%%%%%%%%%%%%%%%%%%%%%%%%%%%%%%%%%
%%%%%%%%%%%%%%%%%%%%%%%%%%%%%%%%%%%%

\newpage

\section{Supplementary Note on Lattice Points in Convex Regions of the Plane}
%{\footnotesize{}Supplementary Note to ``The Random Phase Approximation
%for Interacting Fermi Gases in the Mean-Field Regime''}}
%{Martin Ravn Christiansen, Christian Hainzl, Phan Th\`anh Nam}
%\end{center}

\subsection{Introduction}

In this note we follow the arguments of Chapter 8 of \cite{GelLin-66}
to prove the following theorem:
\begin{thm}
\label{them:ConvexLatticePoints}Let $K\subset\mathbb{R}^{2}$ be
a compact, strictly convex set with $C^{2}$ boundary. Let $\partial K$
have minimal and maximal radii of curvature $0<R_{1}\leq R_{2}$.
If $R_{2}\geq1$ then
\[
\left|\left|K\cap\mathbb{Z}^{2}\right|-\mathrm{Area}\mleft(K\mright)\right|\leq C\frac{R_{2}}{R_{1}}R_{2}^{\frac{2}{3}}\log\mleft(1+2\sqrt{2R_{2}}\mright)^{\frac{2}{3}}
\]
for a constant $C>0$ independent of $K$, $R_{1}$ and $R_{2}$.
\end{thm}

\subsubsection*{Rounding Conventions}

For $x\in\mathbb{R}$ we denote by $\left\lfloor x\right\rfloor $
the greatest integer for which $\left\lfloor x\right\rfloor \leq x$
holds, and by $\left\lceil x\right\rceil $ the least integer for
which $x\leq\left\lceil x\right\rceil $. Thus for any $x$
\begin{equation}
\left\lfloor x\right\rfloor \leq x<\left\lfloor x\right\rfloor +1,\quad\left\lceil x\right\rceil -1\leq x\leq\left\lceil x\right\rceil .
\end{equation}
Note that $\left\lfloor x\right\rfloor =x=\left\lceil x\right\rceil $
if and only if $x\in\mathbb{Z}$, and $\left\lceil x\right\rceil =\left\lfloor x\right\rfloor +1$
otherwise.

We define the fractional part of $x$ to be $\left\{ x\right\} =x-\left\lfloor x\right\rfloor $.

With these definitions we have the following:
\begin{lem}
For any interval $\mleft[a,b\mright]\subset\mathbb{R}$, $a\leq b$,
it holds that
\[
\left|\mleft[a,b\mright]\cap\mathbb{Z}\right|=\left\lfloor b\right\rfloor -\left\lceil a\right\rceil +1\quad\text{and}\quad\left|\left|\mleft[a,b\mright]\cap\mathbb{Z}\right|-\mleft(b-a\mright)\right|\leq1.
\]
\end{lem}

\subsection{Lattice Points Bounded by a Graph}

Let $f:\mleft[a,b\mright]\rightarrow\mleft[0,\infty\mright)$ be given,
and consider the set of points $\Gamma_{f}^{\leq}\subset\mathbb{R}^{2}$
bounded by the $x$-axis and the graph of $f$,
\begin{equation}
\Gamma_{f}^{\leq}=\left\{ \mleft(x,y\mright)\in\mathbb{R}^{2}\mid a\leq x\leq b,\,0<y\leq f\mleft(x\mright)\right\} .
\end{equation}
In this section we prove the following estimate on the number of lattice
points in $\Gamma_{f}^{\leq}$, i.e. $\left|\Gamma_{f}^{\leq}\cap\mathbb{Z}^{2}\right|$:
\begin{prop}
\label{prop:LatticePointsBoundedbyaGraph}Let $f\in C^{2}\mleft(\mleft[a,b\mright]\mright)$
be a positive function satisfying
\[
0<\mathcal{C}_{1}\leq\left|f''\mleft(x\mright)\right|\leq\mathcal{C}_{2},\quad x\in\mleft[a,b\mright].
\]
Then if $\mathcal{C}_{1}\leq1$ it holds that
\begin{align*}
 & \quad\;\left|\left|\Gamma_{f}^{\leq}\cap\mathbb{Z}^{2}\right|-\mleft(\int_{a}^{b}f\mleft(x\mright)\,dx+\beta_{f}\mleft(a,b\mright)-\frac{b-a}{2}\mright)\right|\\
 & \leq\frac{1}{2}\max_{x\in\mleft[a,b\mright]}\left|f'\mleft(x\mright)\right|+C\mleft(1+\mathcal{C}_{2}\mleft(b-a+1\mright)\mright)\mathcal{C}_{1}^{-\frac{2}{3}}\log\mleft(1+2\sqrt{2}\mathcal{C}_{1}^{-\frac{1}{2}}\mright)^{\frac{2}{3}}
\end{align*}
for a constant $C>0$ independent of all quantities, where
\[
\beta_{f}\mleft(a,b\mright)=\mleft(\frac{1}{2}+a-\left\lceil a\right\rceil \mright)f\mleft(a\mright)+\mleft(\frac{1}{2}+\left\lfloor b\right\rfloor -b\mright)f\mleft(b\mright).
\]
\end{prop}

Note that in the notation $\sum_{a\leq m\leq b}=\sum_{m\in\mleft[a,b\mright]\cap\mathbb{Z}}$,
$\left|\Gamma_{f}^{\leq}\cap\mathbb{Z}^{2}\right|$ can be expressed
as
\begin{align}
\left|\Gamma_{f}^{\leq}\cap\mathbb{Z}^{2}\right| & =\sum_{a\leq m\leq b}\mleft(\left|\mleft[0,f\mleft(m\mright)\mright]\cap\mathbb{Z}\right|-1\mright)=\sum_{a\leq m\leq b}\left\lfloor f\mleft(m\mright)\right\rfloor \\
 & =\sum_{a\leq m\leq b}f\mleft(m\mright)-\sum_{a\leq m\leq b}\left\{ f\mleft(m\mright)\right\} .\nonumber 
\end{align}
To prove the proposition we analyze the two terms on the right-hand
side separately. The first term is simply a Riemann sum, and we may
derive an integral identity for this as follows: Consider the sawtooth
wave $g:\mathbb{R}\rightarrow\mathbb{R}$ given by
\begin{equation}
g\mleft(x\mright)=\frac{1}{2}-\left\{ x\right\} =\frac{1}{2}+\left\lfloor x\right\rfloor -x.
\end{equation}
Then the following holds:
\begin{prop}
Let $f\in C^{1}\mleft(\mleft[a,b\mright]\mright)$ be given. Then
\[
\int_{a}^{b}f'\mleft(x\mright)g\mleft(x\mright)\,dx=\int_{a}^{b}f\mleft(x\mright)\,dx+\beta_{f}\mleft(a,b\mright)-\sum_{a\leq m\leq b}f\mleft(m\mright)
\]
where
\[
\beta_{f}\mleft(a,b\mright)=\mleft(\frac{1}{2}+a-\left\lceil a\right\rceil \mright)f\mleft(a\mright)+\mleft(\frac{1}{2}+\left\lfloor b\right\rfloor -b\mright)f\mleft(b\mright).
\]
\end{prop}

\textbf{Proof:} We have
\begin{equation}
\int_{a}^{b}f'\mleft(x\mright)g\mleft(x\mright)\,dx=\int_{a}^{\left\lceil a\right\rceil }f'\mleft(x\mright)g\mleft(x\mright)\,dx+\sum_{m=\left\lceil a\right\rceil }^{\left\lfloor b\right\rfloor -1}\int_{m}^{m+1}f'\mleft(x\mright)g\mleft(x\mright)\,dx+\int_{\left\lfloor b\right\rfloor }^{b}f'\mleft(x\mright)g\mleft(x\mright)\,dx
\end{equation}
and for any $m\in\mathbb{Z}$
\begin{align}
\int_{m}^{m+1}f'\mleft(x\mright)g\mleft(x\mright)\,dx & =\lim_{x\rightarrow\mleft(m+1\mright)^{-}}f\mleft(x\mright)g\mleft(x\mright)-\lim_{x\rightarrow m^{+}}f\mleft(x\mright)g\mleft(x\mright)-\int_{m}^{m+1}f\mleft(x\mright)g'\mleft(x\mright)\,dx\\
 & =-\frac{1}{2}f\mleft(m+1\mright)-\frac{1}{2}f\mleft(m\mright)+\int_{m}^{m+1}f\mleft(x\mright)\,dx,\nonumber 
\end{align}
so
\begin{equation}
\sum_{m=\left\lceil a\right\rceil }^{\left\lfloor b\right\rfloor -1}\int_{m}^{m+1}f'\mleft(x\mright)g\mleft(x\mright)\,dx=\int_{\left\lceil a\right\rceil }^{\left\lfloor b\right\rfloor }f\mleft(x\mright)\,dx+\frac{1}{2}f\mleft(\left\lceil a\right\rceil \mright)+\frac{1}{2}f\mleft(\left\lfloor b\right\rfloor \mright)-\sum_{a\leq m\leq b}f\mleft(m\mright).
\end{equation}
If $a$ and $b$ are integers this is the claim. If not, then as above
\begin{align}
\int_{a}^{\left\lceil a\right\rceil }f'\mleft(x\mright)g\mleft(x\mright)\,dx & =-\frac{1}{2}f\mleft(\left\lceil a\right\rceil \mright)-f\mleft(a\mright)g\mleft(a\mright)+\int_{a}^{\left\lceil a\right\rceil }f\mleft(x\mright)\,dx\nonumber \\
 & =-\frac{1}{2}f\mleft(\left\lceil a\right\rceil \mright)-f\mleft(a\mright)\mleft(\frac{1}{2}+\left\lfloor a\right\rfloor -a\mright)+\int_{a}^{\left\lceil a\right\rceil }f\mleft(x\mright)\,dx\nonumber \\
 & =-\frac{1}{2}f\mleft(\left\lceil a\right\rceil \mright)+\mleft(\frac{1}{2}+a-\left\lceil a\right\rceil \mright)f\mleft(a\mright)+\int_{a}^{\left\lceil a\right\rceil }f\mleft(x\mright)\,dx\\
\int_{\left\lfloor b\right\rfloor }^{b}f'\mleft(x\mright)g\mleft(x\mright)\,dx & =f\mleft(b\mright)g\mleft(b\mright)-\frac{1}{2}f\mleft(\left\lfloor b\right\rfloor \mright)+\int_{\left\lfloor b\right\rfloor }^{b}f\mleft(x\mright)\,dx\nonumber \\
 & =-\frac{1}{2}f\mleft(\left\lfloor b\right\rfloor \mright)+\mleft(\frac{1}{2}+\left\lfloor b\right\rfloor -b\mright)f\mleft(b\mright)+\int_{\left\lfloor b\right\rfloor }^{b}f\mleft(x\mright)\,dx\nonumber 
\end{align}
which implies the claim.

$\hfill\square$

Noting that $g$ admits the antiderivative $G:\mathbb{R}\rightarrow\mathbb{R}$
given by
\begin{equation}
G\mleft(x\mright)=\int_{0}^{x}g\mleft(x\mright)\,dx=\frac{1}{2}\left\{ x\right\} \mleft(1-\left\{ x\right\} \mright)
\end{equation}
we can conclude the following:
\begin{prop}
Let $f\in C^{2}\mleft[a,b\mright]$ be convex or concave. Then
\[
\left|\sum_{a\leq m\leq b}f\mleft(m\mright)-\mleft(\int_{a}^{b}f\mleft(x\mright)\,dx+\beta_{f}\mleft(a,b\mright)\mright)\right|\leq\frac{1}{2}\max_{x\in\mleft[a,b\mright]}\left|f'\mleft(x\mright)\right|.
\]
\end{prop}

\textbf{Proof:} By the previous proposition we must estimate $\int_{a}^{b}f'\mleft(x\mright)g\mleft(x\mright)\,dx$.
To do this we integrate by parts:
\begin{align}
 & \quad\;\left|\int_{a}^{b}f'\mleft(x\mright)g\mleft(x\mright)\,dx\right|=\left|\mleft[f'\mleft(x\mright)G\mleft(x\mright)\mright]_{a}^{b}-\int_{a}^{b}f''\mleft(x\mright)G\mleft(x\mright)\,dx\right|\nonumber \\
 & \leq\left|f'\mleft(b\mright)G\mleft(b\mright)\right|+\left|f'\mleft(a\mright)G\mleft(a\mright)\right|+\int_{a}^{b}\left|f''\mleft(x\mright)G\mleft(x\mright)\right|\,dx\\
 & \leq\frac{1}{8}\mleft(\left|f'\mleft(b\mright)\right|+\left|f'\mleft(a\mright)\right|+\left|\int_{a}^{b}f''\mleft(x\mright)\,dx\right|\mright)\leq\frac{1}{8}\mleft(\left|f'\mleft(b\mright)\right|+\left|f'\mleft(a\mright)\right|+\left|f'\mleft(b\mright)-f'\mleft(a\mright)\right|\mright)\nonumber \\
 & \leq\frac{1}{2}\max_{x\in\mleft[a,b\mright]}\left|f'\mleft(x\mright)\right|\nonumber 
\end{align}
where we used that $0\leq G\mleft(x\mright)\leq\frac{1}{8}$ and that
by the assumption on $f$, $\int_{a}^{b}\left|f''\mleft(x\mright)\right|\,dx=\left|\int_{a}^{b}f''\mleft(x\mright)\,dx\right|$.

$\hfill\square$

\subsection{Estimation of the Sum of Fractional Parts}

We now come to the sum $\sum_{a\leq m\leq b}\left\{ f\mleft(m\mright)\right\} $.
To estimate this we will need 3 auxilliary results. The first is a
version of Dirichlet's approximation theorem:
\begin{thm}
\label{them:DirichletApp}Let $a,\tau\in\mathbb{R}$ be given with
$\tau\geq1$. Then there exists coprime integers $p$ and $q$ such
that
\[
\left|a-\frac{p}{q}\right|<\frac{1}{q\tau}\quad\text{with}\quad1\leq q\leq\tau.
\]
\end{thm}

For the second result, we make the following definition: Let $\mleft(n_{1},\ldots,n_{k}\mright)$
be a finite sequence of numbers. Then we define a \textit{consecutive
grouping of $\mleft(n_{1},\ldots,n_{k}\mright)$} to be a partition
$\left\{ Q_{j}\right\} $ of this sequence into subsequences of the
form
\begin{equation}
Q_{j}=\mleft(n_{m_{j}},n_{m_{j}+1},\ldots,n_{m_{j}+l_{j}}\mright)
\end{equation}
for some $1\leq m_{j}\leq k$ and $m_{j}\leq m_{j}+l_{j}\leq k$.
We can now state the following lemma:
\begin{lem}
\label{lemma:GroupingLemma}Let $n_{1},\ldots,n_{k}\in\mathbb{N}$
be given with $1\leq n_{i}\leq N$, $1\leq i\leq k$, for some $N\in\mathbb{N}$.
Then there exists a $k'\geq k-N$ such that $\mleft(n_{1},\ldots,n_{k'}\mright)$
admits a consecutive grouping $\left\{ Q_{j}\right\} _{j=1}^{T}$
with the property that
\[
\max\mleft(Q_{j}\mright)=\left|Q_{j}\right|
\]
for an integer $T\leq\sum_{i=1}^{k}n_{i}^{-1}$.
\end{lem}

Finally there is the following:
\begin{prop}
\label{prop:PerturbedSumEstimate}Let $n\in\mathbb{N}$ and $k\in\mathbb{Z}$
be coprime, and let for some $M\in\mathbb{Z}$, $P:\left\{ M+1,\ldots,M+n\right\} \rightarrow\mathbb{R}$
be a function such that
\[
\max_{M+1\leq l,m\leq M+n}\left|P\mleft(l\mright)-P\mleft(m\mright)\right|\leq C
\]
for some $C>0$. Then
\[
\left|\sum_{m=M+1}^{M+n}\left\{ \frac{km+P\mleft(m\mright)}{n}\right\} -\frac{n}{2}\right|\leq C+\frac{1}{2}.
\]
\end{prop}

We collect the proofs of these results in Section \ref{sec:ProofsoftheAuxilliaryResults}.

We split the analysis of $\sum_{a\leq m\leq b}\left\{ f\mleft(m\mright)\right\} $
into two parts, starting with the following:
\begin{prop}
Let $f\in C^{2}\mleft(\mleft[a,b\mright]\mright)$ satisfy
\[
0<\mathcal{C}_{1}\leq\left|f''\mleft(x\mright)\right|\leq\mathcal{C}_{2},\quad x\in\mleft[a,b\mright].
\]
Then for any real $\tau\geq1$ it holds that
\[
\left|\sum_{a\leq m\leq b}\left\{ f\mleft(m\mright)\right\} -\frac{b-a}{2}\right|\leq\mleft(\frac{1}{\tau}+\frac{1}{2}\mathcal{C}_{2}\tau^{2}\mright)\mleft(\left\lfloor b\right\rfloor -\left\lceil a\right\rceil +1\mright)+\tau+\frac{1}{2}T_{\tau}
\]
for an integer $T_{\tau}$ described below.
\end{prop}

\textbf{Proof:} By Theorem \ref{them:DirichletApp} we can for each
$\left\lceil a\right\rceil \leq m\leq\left\lfloor b\right\rfloor $
find coprime $p_{m}$ and $q_{m}$ with $1\leq q_{m}\leq\tau$ such
that
\begin{equation}
\left|f'\mleft(m\mright)-\frac{p_{m}}{q_{m}}\right|\leq\frac{1}{q_{m}\tau},\quad\left\lceil a\right\rceil \leq m\leq\left\lfloor b\right\rfloor .
\end{equation}
By Lemma \ref{lemma:GroupingLemma} we can by omitting at most $\left\lfloor \tau\right\rfloor $
of the tail values from the sequence $\mleft(q_{\left\lceil a\right\rceil },\ldots,q_{\left\lfloor b\right\rfloor }\mright)$
form consecutive groupings $\left\{ Q_{i}\right\} _{i=1}^{T_{\tau}}$
of the resulting sequence $\mleft(q_{\left\lceil a\right\rceil },\ldots,q_{k}\mright)$
such that $\max\mleft(Q_{i}\mright)=\left|Q_{i}\right|$.

Now focus on a particular grouping $Q_{i}$, of the form $Q_{i}=\mleft(q_{M+1},\ldots,q_{M+\left|Q_{i}\right|}\mright)$.
Let $1\leq m'\leq\left|Q_{i}\right|$ be such that $q_{M+m'}=\max\mleft(Q_{i}\mright)$.
Then
\begin{equation}
f'\mleft(M+m'\mright)=\frac{p_{M+m'}}{q_{M+m'}}+\frac{\theta}{q_{M+m'}\tau}
\end{equation}
for some $\theta\in\mleft(-1,1\mright)$. Then by Taylor's theorem with
Lagrange's remainder
\begin{align}
&\sum_{m=M+1}^{M+\left|Q_{i}\right|}\left\{ f\mleft(m\mright)\right\}  \nonumber\\
& =\sum_{m=M+1}^{M+\left|Q_{i}\right|}\left\{ f\mleft(M+m'\mright)+\mleft(m-\mleft(M+m'\mright)\mright)f'\mleft(M+m'\mright)+\frac{1}{2}\mleft(m-\mleft(M+m'\mright)\mright)^{2}f''\mleft(\xi_{m}\mright)\right\} \nonumber \\
 & =\sum_{m=1-m'}^{\left|Q_{i}\right|-m'}\left\{ f\mleft(M+m'\mright)+mf'\mleft(M+m'\mright)+\frac{m^{2}}{2}f''\mleft(\xi_{M+m'+m}\mright)\right\} \\
 & =\sum_{m=1-m'}^{\left|Q_{i}\right|-m'}\left\{ \frac{p_{M+m'}m+P\mleft(m\mright)}{q_{M+m'}}\right\} \nonumber 
\end{align}
for some $\xi_{m}$'s lying between $M+m'$ and $M+m$, where
\begin{equation}
P\mleft(m\mright)=\left|Q_{i}\right|f\mleft(M+m'\mright)+m\frac{\theta}{\tau}+\frac{\left|Q_{i}\right|m^{2}}{2}f''\mleft(\xi_{M+m'+m}\mright).
\end{equation}
Now, for any $1-m'\leq l,m\leq\left|Q_{i}\right|-m'$ this $P$ obeys
\begin{align}
\left|\frac{P\mleft(l\mright)-P\mleft(m\mright)}{\left|Q_{i}\right|}\right| & =\left|\frac{l-m}{\left|Q_{i}\right|}\frac{\theta}{\tau}+\frac{1}{2}\mleft(l^{2}f''\mleft(\xi_{M+m'+l}\mright)-m^{2}f''\mleft(\xi_{M+m'+m}\mright)\mright)\right|\\
 & \leq\frac{1}{\tau}+\frac{1}{2}\left|l^{2}f''\mleft(\xi_{M+m'+l}\mright)-m^{2}f''\mleft(\xi_{M+m'+m}\mright)\right|\nonumber 
\end{align}
and since $f''$ does not change sign on $\mleft[a,b\mright]$, we can
further estimate
\begin{align}
 & \quad\;\left|l^{2}f''\mleft(\xi_{M+m'+l}\mright)-m^{2}f''\mleft(\xi_{M+m'+m}\mright)\right|\leq\max_{1-m'\leq n\leq\left|Q_{i}\right|-m'}n^{2}\left|f''\mleft(\xi_{M+m'+n}\mright)\right|\\
 & \leq\mleft(\left|Q_{i}\right|-1\mright)^{2}\max_{x\in\mleft[a,b\mright]}\left|f''\mleft(x\mright)\right|\leq\mathcal{C}_{2}\tau^{2},\nonumber 
\end{align}
so by Proposition \ref{prop:PerturbedSumEstimate}
\begin{equation}
\left|\sum_{m=M+1}^{M+\left|Q_{i}\right|}\left\{ f\mleft(m\mright)\right\} -\frac{\left|Q_{i}\right|}{2}\right|\leq\mleft(\frac{1}{\tau}+\frac{1}{2}\mathcal{C}_{2}\tau^{2}\mright)\left|Q_{i}\right|+\frac{1}{2}.
\end{equation}
Summing up we thus find that
\begin{align}
&\left|\sum_{a\leq m\leq b}\left\{ f\mleft(m\mright)\right\} -\frac{\left\lfloor b\right\rfloor -\left\lceil a\right\rceil +1}{2}\right| \nonumber\\
& =\left|\mleft(\sum_{i=1}^{T_{\tau}}\sum_{m=M+1}^{M+\left|Q_{i}\right|}\left\{ f\mleft(m\mright)\right\} +\sum_{m=k+1}^{\left\lfloor b\right\rfloor }\left\{ f\mleft(m\mright)\right\} \mright)-\mleft(\sum_{i=1}^{T_{\tau}}\frac{\left|Q_{i}\right|}{2}+\sum_{m=k+1}^{\left\lfloor b\right\rfloor }\frac{1}{2}\mright)\right|\nonumber \\
 & \leq\sum_{i=1}^{T_{\tau}}\left|\sum_{m=M+1}^{M+\left|Q_{i}\right|}\left\{ f\mleft(m\mright)\right\} -\frac{\left|Q_{i}\right|}{2}\right|+\sum_{m=k+1}^{\left\lfloor b\right\rfloor }\left|\left\{ f\mleft(m\mright)\right\} -\frac{1}{2}\right|\\
 & \leq\sum_{i=1}^{T_{\tau}}\mleft(\mleft(\frac{1}{\tau}+\frac{1}{2}\mathcal{C}_{2}\tau^{2}\mright)\left|Q_{i}\right|+\frac{1}{2}\mright)+\frac{\left\lfloor b\right\rfloor -k}{2}\nonumber \\
 & \leq\mleft(\frac{1}{\tau}+\frac{1}{2}\mathcal{C}_{2}\tau^{2}\mright)\mleft(\left\lfloor b\right\rfloor -\left\lceil a\right\rceil +1\mright)+\frac{T_{\tau}}{2}+\frac{\tau}{2}.\nonumber 
\end{align}
Replacing $\frac{\left\lfloor b\right\rfloor -\left\lceil a\right\rceil +1}{2}$
by $\frac{b-a}{2}$ on the left incurs an error of at most $\frac{1}{2}\leq\frac{\tau}{2}$,
so we have the claim.

$\hfill\square$

Now we estimate $T_{\tau}$:
\begin{prop}
The quantity $T_{\tau}$ of the previous proposition satisfies
\[
T_{\tau}\leq\mathcal{C}_{2}\mleft(\left\lfloor b\right\rfloor -\left\lceil a\right\rceil \mright)\tau+\mleft(2\frac{\mathcal{C}_{2}}{\mathcal{C}_{1}}\frac{\left\lfloor b\right\rfloor -\left\lceil a\right\rceil }{\tau}+3\mright)\log\mleft(1+2\tau\mright)+\frac{\pi^{2}}{\mathcal{C}_{1}\tau}.
\]
\end{prop}

\textbf{Proof:} $T_{\tau}$ is the number of groupings that we form
when we subject the sequence $q_{\left\lceil a\right\rceil },\ldots,q_{\left\lfloor b\right\rfloor }$,
determined by the condition
\begin{equation}
\left|f'\mleft(m\mright)-\frac{p_{m}}{q_{m}}\right|<\frac{1}{q_{m}\tau},
\end{equation}
to the procedure of Lemma \ref{lemma:GroupingLemma}, so
\begin{equation}
T_{\tau}\leq\sum_{m=\left\lceil a\right\rceil }^{\left\lfloor b\right\rfloor }\frac{1}{q_{m}}=\sum_{q=1}^{\left\lfloor \tau\right\rfloor }\frac{1}{q}\left|\left\{ m\mid q_{m}=q\right\} \right|.
\end{equation}
For a given $q$ we expand the quantity on the right as
\begin{equation}
\left|\left\{ m\mid q_{m}=q\right\} \right|=\sum_{p\in\mathbb{Z}}\left|\left\{ m\mid\mleft(p_{m},q_{m}\mright)=\mleft(p,q\mright)\right\} \right|=\sum_{p\in P_{q}}\left|M_{p,q}\right|
\end{equation}
where
\begin{align}
P_{q} & =\left\{ p\mid\exists m\in\left\{ \left\lceil a\right\rceil ,\ldots,\left\lfloor b\right\rfloor \right\} :\,\mleft(p_{m},q_{m}\mright)=\mleft(p,q\mright)\right\} \\
M_{p,q} & =\left\{ m\mid\mleft(p_{m},q_{m}\mright)=\mleft(p,q\mright)\right\} .\nonumber 
\end{align}
Now, observe that by the mean-value theorem, it holds that
\begin{equation}
\mathcal{C}_{1}\leq\left|f'\mleft(m+1\mright)-f'\mleft(m\mright)\right|\leq\mathcal{C}_{2},\quad\left\lceil a\right\rceil \leq m\leq\left\lfloor b\right\rfloor -1,
\end{equation}
so for any interval $\mleft[\alpha,\beta\mright]\subset\mathbb{R}$
(since $m\mapsto f'\mleft(m\mright)$ is also monotone)
\begin{equation}
\left|\left\{ m\mid\alpha\leq f'\mleft(m\mright)\leq\beta\right\} \right|\leq\frac{\beta-\alpha}{\mathcal{C}_{1}}+1;
\end{equation}
furthermore $F_{1}=\min_{\left\lceil a\right\rceil \leq m\leq\left\lfloor b\right\rfloor }f'\mleft(m\mright)$
and $F_{2}=\max_{\left\lceil a\right\rceil \leq m\leq\left\lfloor b\right\rfloor }f'\mleft(m\mright)$
obey
\begin{equation}
F_{2}-F_{1}\leq\mathcal{C}_{2}\mleft(\left\lfloor b\right\rfloor -\left\lceil a\right\rceil \mright).
\end{equation}
We can then estimate $\left|M_{p,q}\right|$ as follows: For each
pair $\mleft(p_{m},q_{m}\mright)$ we have the simultaneous inequalities
\begin{equation}
\frac{1}{q_{m}}\mleft(p_{m}-\frac{1}{\tau}\mright)<f'\mleft(m\mright)<\frac{1}{q_{m}}\mleft(p_{m}+\frac{1}{\tau}\mright),
\end{equation}
so for any pair $\mleft(p,q\mright)$, there can be at most as many
$m$'s for which $\mleft(p_{m},q_{m}\mright)=\mleft(p,q\mright)$ as values
$f'\mleft(m\mright)$ in the interval $\mleft(\frac{1}{q}\mleft(p-\frac{1}{\tau}\mright),\frac{1}{q}\mleft(p+\frac{1}{\tau}\mright)\mright)$,
so
\begin{equation}
\left|M_{p,q}\right|\leq\frac{2}{\mathcal{C}_{1}q\tau}+1.
\end{equation}
For $P_{q}$ we likewise note that each pair $\mleft(p_{m},q_{m}\mright)$
satisfies
\begin{equation}
-\frac{1}{\tau}+F_{1}q_{m}<p_{m}<\frac{1}{\tau}+F_{2}q_{m}
\end{equation}
so for a given $q$ it must be the case that
\begin{equation}
\left|P_{q}\right|\leq\left|\mleft(-\frac{1}{\tau}+F_{1}q,\frac{1}{\tau}+F_{2}q\mright)\cap\mathbb{Z}\right|\leq\frac{2}{\tau}+\mleft(F_{2}-F_{1}\mright)q+1\leq\mathcal{C}_{2}\mleft(\left\lfloor b\right\rfloor -\left\lceil a\right\rceil \mright)q+3.
\end{equation}
We can then estimate that
\begin{align}
\left|\left\{ m\mid q_{m}=q\right\} \right| & =\sum_{p\in P_{q}}\left|M_{p,q}\right|\leq\mleft(\mathcal{C}_{2}\mleft(\left\lfloor b\right\rfloor -\left\lceil a\right\rceil \mright)q+3\mright)\mleft(\frac{2}{\mathcal{C}_{1}q\tau}+1\mright)\\
 & =\mathcal{C}_{2}\mleft(\left\lfloor b\right\rfloor -\left\lceil a\right\rceil \mright)q+\mleft(2\frac{\mathcal{C}_{2}}{\mathcal{C}_{1}}\frac{\left\lfloor b\right\rfloor -\left\lceil a\right\rceil }{\tau}+3\mright)+\frac{6}{\mathcal{C}_{1}\tau}\frac{1}{q}\nonumber 
\end{align}
so we finally find
\begin{align}
T_{\tau} & =\sum_{q=1}^{\left\lfloor \tau\right\rfloor }\frac{1}{q}\left|\left\{ m\mid q_{m}=q\right\} \right|\leq\mathcal{C}_{2}\mleft(\left\lfloor b\right\rfloor -\left\lceil a\right\rceil \mright)\left\lfloor \tau\right\rfloor +\mleft(2\frac{\mathcal{C}_{2}}{\mathcal{C}_{1}}\frac{\left\lfloor b\right\rfloor -\left\lceil a\right\rceil }{\tau}+3\mright)\sum_{q=1}^{\left\lfloor \tau\right\rfloor }\frac{1}{q}+\frac{6}{\mathcal{C}_{1}\tau}\sum_{q=1}^{\left\lfloor \tau\right\rfloor }\frac{1}{q^{2}}\\
 & \leq\mathcal{C}_{2}\mleft(\left\lfloor b\right\rfloor -\left\lceil a\right\rceil \mright)\tau+\mleft(2\frac{\mathcal{C}_{2}}{\mathcal{C}_{1}}\frac{\left\lfloor b\right\rfloor -\left\lceil a\right\rceil }{\tau}+3\mright)\log\mleft(1+2\tau\mright)+\frac{\pi^{2}}{\mathcal{C}_{1}\tau}\nonumber 
\end{align}
since $\sum_{q=1}^{\left\lfloor \tau\right\rfloor }q^{-2}\leq\sum_{q=1}^{\infty}q^{-2}=\frac{\pi^{2}}{6}$
and $\sum_{q=1}^{\left\lfloor \tau\right\rfloor }q^{-1}\leq\int_{\frac{1}{2}}^{\left\lfloor \tau\right\rfloor +\frac{1}{2}}x^{-1}dx=\log\mleft(1+2\left\lfloor \tau\right\rfloor \mright)\leq\log\mleft(1+2\tau\mright)$.

$\hfill\square$

We end our analysis of $\sum_{a\leq m\leq b}\left\{ f\mleft(m\mright)\right\} $
by combining these propositions for the following:
\begin{thm}
Let $f\in C^{2}\mleft(\mleft[a,b\mright]\mright)$ satisfy
\[
0<\mathcal{C}_{1}\leq\left|f''\mleft(x\mright)\right|\leq\mathcal{C}_{2},\quad x\in\mleft[a,b\mright].
\]
Then if $\mathcal{C}_{1}\leq1$ it holds that
\[
\left|\sum_{a\leq m\leq b}\left\{ f\mleft(m\mright)\right\} -\frac{b-a}{2}\right|\leq C\mleft(1+\mathcal{C}_{2}\mleft(b-a+1\mright)\mright)\mathcal{C}_{1}^{-\frac{2}{3}}\log\mleft(1+2\sqrt{2}\mathcal{C}_{1}^{-\frac{1}{2}}\mright)^{\frac{2}{3}}
\]
for a constant $C>0$ independent of all quantities.
\end{thm}

\textbf{Proof:} By the propositions above we have that
\begin{align}
\left|\sum_{a\leq m\leq b}\left\{ f\mleft(m\mright)\right\} -\frac{b-a}{2}\right| & \leq C\mleft(\mleft(\frac{1}{\tau}+\mathcal{C}_{2}\mleft(\tau^{2}+\frac{1}{\mathcal{C}_{1}\tau}\log\mleft(1+2\tau\mright)\mright)\mright)N+\tau+\frac{1}{\mathcal{C}_{1}\tau}\mright)
\end{align}
for all $\tau\geq1$, for a purely numerical constant $C>0$, where
$N=\left\lfloor b\right\rfloor -\left\lceil a\right\rceil +1$ for
brevity.

Now, if $\mathcal{C}_{1}\leq1$ we can take $\tau=\mathcal{C}_{1}^{-\frac{1}{3}}\log\mleft(1+2\sqrt{2}\,\mathcal{C}_{1}^{-\frac{1}{2}}\mright)^{\frac{1}{3}}$
(as this is then greater than $1$). This obeys
\begin{equation}
\mathcal{C}_{1}^{-\frac{1}{3}}\leq\tau\leq\mathcal{C}_{1}^{-\frac{1}{3}}\mleft(2^{\frac{3}{2}}\,\mathcal{C}_{1}^{-\frac{1}{2}}\mright)^{\frac{1}{3}}=\sqrt{2}\,\mathcal{C}_{1}^{-\frac{1}{3}-\frac{1}{6}}=\sqrt{2}\,\mathcal{C}_{1}^{-\frac{1}{2}}
\end{equation}
so
\begin{equation}
\tau^{2}+\frac{1}{\mathcal{C}_{1}\tau}\log\mleft(1+2\tau\mright)\leq2\,\mathcal{C}_{1}^{-\frac{2}{3}}\log\mleft(1+2\sqrt{2}\,\mathcal{C}_{1}^{-\frac{1}{2}}\mright)^{\frac{2}{3}}
\end{equation}
and
\begin{align}
\left|\sum_{a\leq m\leq b}\left\{ f\mleft(m\mright)\right\} -\frac{b-a}{2}\right| & \leq C'\mleft(\mathcal{C}_{1}^{\frac{1}{3}}\mleft(1+2\,\mathcal{C}_{2}\mathcal{C}_{1}^{-1}\log\mleft(1+2\sqrt{2}\,\mathcal{C}_{1}^{-\frac{1}{2}}\mright)^{\frac{2}{3}}\mright)N+\mathcal{C}_{1}^{-\frac{1}{2}}+\mathcal{C}_{1}^{-\frac{2}{3}}\mright)\nonumber \\
 & \leq C''\mleft(\mathcal{C}_{2}N\mathcal{C}_{1}^{-\frac{2}{3}}\log\mleft(1+2\sqrt{2}\,\mathcal{C}_{1}^{-\frac{1}{2}}\mright)^{\frac{2}{3}}+\mathcal{C}_{1}^{-\frac{2}{3}}\mright)\\
 & \leq C'''\mleft(1+\mathcal{C}_{2}N\mright)\mathcal{C}_{1}^{-\frac{2}{3}}\log\mleft(1+2\sqrt{2}\,\mathcal{C}_{1}^{-\frac{1}{2}}\mright)^{\frac{2}{3}}.\nonumber 
\end{align}
Since $N\leq b-a+1$ this implies the claim.

$\hfill\square$

\subsection{Lattice Points in a Convex Region}

We now consider the number of lattice points in a bounded, convex
region $K\subset\mathbb{R}^{2}$.

One defines the discrepancy $\delta\mleft(K\mright)$ of such a set
to be
\begin{equation}
\delta\mleft(K\mright)=\left|K\cap\mathbb{Z}^{2}\right|-\mathrm{Area}\mleft(K\mright).
\end{equation}
Note that since both $K\mapsto\left|K\cap\mathbb{Z}^{2}\right|,\,\mathrm{Area}\mleft(K\mright)$
are additive, the same is true of $\delta\mleft(K\mright)$, i.e. if
$A\cap B=\emptyset$ then
\begin{equation}
\delta\mleft(A\cup B\mright)=\delta\mleft(A\mright)+\delta\mleft(B\mright)
\end{equation}
and if $B\subset A$ then
\begin{equation}
\delta\mleft(A\backslash B\mright)=\delta\mleft(A\mright)-\delta\mleft(B\mright).
\end{equation}
In order to apply the result of the previous section to study $\delta\mleft(K\mright)$,
we make the following observation:
\begin{lem}
\label{fact:CompactStrictlyConvexRegions}Let $K\subset\mathbb{R}^{2}$
be a compact, strictly convex region with $C^{2}$ boundary. Then
there exists $a,b\in\mathbb{R}$ and $f_{-},f_{+}\in C\mleft(\mleft[a,b\mright]\mright)\cap C^{2}\mleft(\mleft(a,b\mright)\mright)$
with $f_{-}$ strictly convex and $f_{+}$ strictly concave, such
that
\[
K=\left\{ \mleft(x,y\mright)\in\mathbb{R}^{2}\mid a\leq x\leq b,\,f_{-}\mleft(x\mright)\leq y\leq f_{+}\mleft(x\mright)\right\} .
\]
Furthermore
\[
f_{-}\mleft(a\mright)=f_{+}\mleft(a\mright),\quad f_{-}\mleft(b\mright)=f_{+}\mleft(b\mright),
\]
and
\begin{align*}
\lim_{x\rightarrow a^{+}}f_{-}^{\prime}\mleft(x\mright) & =-\infty=\lim_{x\rightarrow b^{-}}f_{+}^{\prime}\mleft(x\mright)\\
\lim_{x\rightarrow b^{-}}f_{-}^{\prime}\mleft(x\mright) & =+\infty=\lim_{x\rightarrow a^{+}}f_{+}^{\prime}\mleft(x\mright).
\end{align*}
\end{lem}

Given a bounded, strictly convex $K$ we can then write this as
\begin{equation}
K=\mleft(K_{+}\backslash K_{-}\mright)\backslash\Gamma_{-}
\end{equation}
for
\begin{align}
K_{\pm} & =\left\{ \mleft(x,y\mright)\in\mathbb{R}^{2}\mid a\leq x\leq b,\,2^{-1}<y\leq f_{\pm}\mleft(x\mright)\right\} \\
\Gamma_{-} & =\left\{ \mleft(x,y\mright)\in\mathbb{R}^{2}\mid a\leq x\leq b,\,y=f_{-}\mleft(x\mright)\right\} \nonumber 
\end{align}
where we assume for later convenience, without loss of generality,
that $f_{-}>\frac{1}{2}$ (this can always be ensured by an integral
translation of $K$, which does not affect the discrepancy). Thus
\begin{equation}
\left|\delta\mleft(K\mright)\right|\leq\left|\delta\mleft(K_{+}\mright)-\delta\mleft(K_{-}\mright)\right|+\left|\delta\mleft(\Gamma_{-}\mright)\right|.
\end{equation}

\subsection{Geometric Reformulation of Proposition \ref{prop:LatticePointsBoundedbyaGraph}}

Given a function $f:\mleft[a,b\mright]\rightarrow\mathbb{R}^{2}$, the
mapping $x\mapsto\mleft(x,f\mleft(x\mright)\mright)$ paremetrizes the
graph of $f$,
\begin{equation}
\Gamma_{f}=\left\{ \mleft(x,y\mright)\in\mathbb{R}^{2}\mid a\leq x\leq b,\,y=f\mleft(x\mright)\right\} .
\end{equation}
Recall that for such a graph parametrization, the radius of curvature
at $\mleft(x,f\mleft(x\mright)\mright)$ is given by
\begin{equation}
R\mleft(x\mright)=\mleft(1+f'\mleft(x\mright)^{2}\mright)^{\frac{3}{2}}\left|f''\mleft(x\mright)\right|^{-1}.
\end{equation}
The statement that $\Gamma_{f}$ has minimal and maximal radii of
curvature $0<R_{1}\leq R_{2}$ then amounts to the assertion that
\begin{equation}
0<R_{1}\leq\mleft(1+f'\mleft(x\mright)^{2}\mright)^{\frac{3}{2}}\left|f''\mleft(x\mright)\right|^{-1}\leq R_{2},\quad x\in\mleft(a,b\mright),
\end{equation}
hence
\begin{equation}
0<R_{2}^{-1}\mleft(1+f'\mleft(x\mright)^{2}\mright)^{\frac{3}{2}}\leq\left|f''\mleft(x\mright)\right|\leq R_{1}^{-1}\mleft(1+f'\mleft(x\mright)^{2}\mright)^{\frac{3}{2}},\quad x\in\mleft(a,b\mright).
\end{equation}
To restate Proposition \ref{prop:LatticePointsBoundedbyaGraph} in
geometric terms, we begin with the following:
\begin{prop}
Let $f\in C^{2}\mleft(\mleft[a,b\mright]\mright)$ be such that $\Gamma_{f}$
has minimal and maximal radii of curvature $0<R_{1}\leq R_{2}$, and
assume furthermore that
\[
0<c<f\mleft(x\mright),\quad\left|f'\mleft(x\mright)\right|\leq1,\quad x\in\mleft[a,b\mright],
\]
and consider
\[
\Gamma_{f}^{c\leq}=\left\{ \mleft(x,y\mright)\in\mathbb{R}^{2}\mid a\leq x\leq b,\,c\leq y\leq f\mleft(x\mright)\right\} .
\]
Then if $R_{2}\geq1$ it holds that
\[
\left|\delta\mleft(\Gamma_{f}^{c\leq}\mright)-\alpha_{c}\mleft(a,b\mright)-\beta_{f}\mleft(a,b\mright)\right|\leq C\frac{R_{2}}{R_{1}}R_{2}^{\frac{2}{3}}\log\mleft(1+2\sqrt{2R_{2}}\mright)^{\frac{2}{3}}
\]
for a constant $C>0$ independent of all quantities, where
\begin{align*}
\alpha_{c}\mleft(a,b\mright) & =\mleft(b-a\mright)\mleft(c-\frac{1}{2}\mright)-\mleft(\left\lfloor b\right\rfloor -\left\lceil a\right\rceil +1\mright)\mleft(\left\lceil c\right\rceil -1\mright)\\
\beta_{f}\mleft(a,b\mright) & =\mleft(\frac{1}{2}+a-\left\lceil a\right\rceil \mright)f\mleft(a\mright)+\mleft(\frac{1}{2}+\left\lfloor b\right\rfloor -b\mright)f\mleft(b\mright).
\end{align*}
\end{prop}

\textbf{Proof:} By the assumptions we have that
\begin{equation}
R_{2}^{-1}\leq\left|f''\mleft(x\mright)\right|\leq2\sqrt{2}R_{1}^{-1},\quad x\in\mleft[a,b\mright],
\end{equation}
so in particular
\begin{equation}
b-a\leq R_{2}\int_{a}^{b}\left|f''\mleft(x\mright)\right|dx\leq R_{2}\left|f'\mleft(b\mright)-f'\mleft(a\mright)\right|\leq2R_{2}.
\end{equation}
Furthermore
\begin{align}
\left|\Gamma_{f}^{c\leq}\cap\mathbb{Z}^{2}\right| & =\sum_{a\leq m\leq b}\mleft(\left|\mleft[c,f\mleft(m\mright)\mright]\cap\mathbb{Z}\right|\mright)=\sum_{a\leq m\leq b}\left\lfloor f\mleft(m\mright)\right\rfloor -\sum_{a\leq m\leq b}\mleft(\left\lceil c\right\rceil -1\mright)\\
 & =\left|\Gamma_{f}^{\leq}\cap\mathbb{Z}^{2}\right|-\mleft(\left\lfloor b\right\rfloor -\left\lceil a\right\rceil +1\mright)\mleft(\left\lceil c\right\rceil -1\mright)\nonumber 
\end{align}
while
\begin{equation}
\mathrm{Area}\mleft(\Gamma_{f}^{c\leq}\mright)=\int_{a}^{b}f\mleft(x\mright)dx-\mleft(b-a\mright)c.
\end{equation}
We can then apply Proposition \ref{prop:LatticePointsBoundedbyaGraph}
to see that under the condition $R_{2}\geq1$
\begin{align}
\left|\delta\mleft(\Gamma_{f}^{c\leq}\mright)-\alpha_{c}\mleft(a,b\mright)-\beta_{f}\mleft(a,b\mright)\right| & =\left|\left|\Gamma_{f}^{\leq}\cap\mathbb{Z}^{2}\right|-\mleft(\int_{a}^{b}f\mleft(x\mright)dx+\beta_{f}\mleft(a,b\mright)-\frac{b-a}{2}\mright)\right|\nonumber \\
 & \leq\frac{1}{2}+C\mleft(1+2\sqrt{2}R_{1}^{-1}\mleft(2R_{2}+1\mright)\mright)R_{2}^{\frac{2}{3}}\log\mleft(1+2\sqrt{2R_{2}}\mright)^{\frac{2}{3}}\\
 & \leq C'\frac{R_{2}}{R_{1}}R_{2}^{\frac{2}{3}}\log\mleft(1+2\sqrt{2R_{2}}\mright)^{\frac{2}{3}}.\nonumber 
\end{align}
$\hfill\square$

We can now further dispense with the assumption that $\left|f'\mleft(x\mright)\right|\leq1$
by exploiting the geometric nature of the radius of curvature:
\begin{prop}
Let $f\in C^{2}\mleft(\mleft[a,b\mright]\mright)$ be such that $\Gamma_{f}$
has minimal and maximal radii of curvature $0<R_{1}\leq R_{2}$, and
assume furthermore that $2^{-1}<f\mleft(x\mright)$ for all $x\in\mleft[a,b\mright]$.
Consider
\[
\Gamma_{f}^{2^{-1}\leq}=\left\{ \mleft(x,y\mright)\in\mathbb{R}^{2}\mid a\leq x\leq b,\,2^{-1}\leq y\leq f\mleft(x\mright)\right\} .
\]
Then if $R_{2}\geq1$ it holds that
\[
\left|\delta\mleft(\Gamma_{f}^{c\leq}\mright)-\beta_{f}\mleft(a,b\mright)\right|\leq C\frac{R_{2}}{R_{1}}R_{2}^{\frac{2}{3}}\log\mleft(1+2\sqrt{2R_{2}}\mright)^{\frac{2}{3}}
\]
for a constant $C>0$ independent of all quantities.
\end{prop}

\textbf{Proof:} First note that we can assume without loss of generality
that $2^{-1}\leq a$, by integral translation.

Now, by strict monotonicity of $f'$, $f'\mleft(x_{-}\mright)=1$ and
$f'\mleft(x_{+}\mright)=-1$ hold for at most one $x_{-},x_{+}\in\mleft[a,b\mright]$
each. Assume without loss of generality that $x_{-}<x_{+}$ (set $x_{-}=a$
or $x_{+}=b$ if $\pm1$ is not attained by $f'$), i.e. that $f$
is concave.

Then we can decompose $\Gamma_{f}^{2^{-1}\leq}$ as $\Gamma_{f}^{2^{-1}\leq}=A_{-}\cup A_{0}\cup A_{+}$
for
\begin{align}
A_{-} & =\left\{ \mleft(x,y\mright)\in\mathbb{R}^{2}\mid a\leq x<x_{-},\,2^{-1}\leq y\leq f\mleft(x\mright)\right\} \nonumber \\
A_{0} & =\left\{ \mleft(x,y\mright)\in\mathbb{R}^{2}\mid x_{-}\leq x\leq x_{+},\,2^{-1}\leq y\leq f\mleft(x\mright)\right\} \\
A_{+} & =\left\{ \mleft(x,y\mright)\in\mathbb{R}^{2}\mid x_{+}<x\leq b,\,2^{-1}\leq y\leq f\mleft(x\mright)\right\} ,\nonumber 
\end{align}
hence
\begin{equation}
\delta\mleft(\Gamma_{f}^{c\leq}\mright)=\delta\mleft(A_{-}\mright)+\delta\mleft(A_{0}\mright)+\delta\mleft(A_{+}\mright).
\end{equation}
Now $A_{0}$ satisfies the hypotheses of the previous proposition,
and if $A_{-}=\emptyset=A_{+}$ we are done. If this is not the case,
say if $A_{+}\neq\emptyset$, let $g:\mleft[f\mleft(b\mright),f\mleft(x_{+}\mright)\mright]\rightarrow\mleft[x_{+},b\mright]$
denote the inverse of $\left.f\right\vert _{\mleft[x_{0},b\mright]}$.
Then we can decompose $A_{+}$ as
\[
A_{+}=\mleft(\mleft(x_{+},b\mright]\times\mleft[2^{-1},f\mleft(b\mright)\mright)\mright)\cup\mleft(\left\{ \mleft(x,y\mright)\in\mathbb{R}^{2}\mid x_{+}\leq x\leq b,\,f\mleft(b\mright)\leq y\leq f\mleft(x\mright)\right\} \backslash\mleft(\left\{ x_{+}\right\} \times\mleft[f\mleft(b\mright),f\mleft(x_{+}\mright)\mright]\mright)\mright)
\]
and note that
\begin{align}
B_{+} & :=\left\{ \mleft(x,y\mright)\in\mathbb{R}^{2}\mid x_{+}\leq x\leq b,\,f\mleft(b\mright)\leq y\leq f\mleft(x\mright)\right\} \nonumber \\
 & =\left\{ \mleft(x,y\mright)\in\mathbb{R}^{2}\mid x_{+}\leq x\leq b,\,x\leq g\mleft(y\mright)\leq b\right\} \nonumber \\
 & =\left\{ \mleft(x,y\mright)\in\mathbb{R}^{2}\mid x_{+}\leq x\leq g\mleft(y\mright)\leq b\right\} \\
 & =\left\{ \mleft(x,y\mright)\in\mathbb{R}^{2}\mid x_{+}\leq g\mleft(y\mright)\leq b,\;x_{+}\leq x\leq g\mleft(y\mright)\right\} \nonumber \\
 & =\left\{ \mleft(x,y\mright)\in\mathbb{R}^{2}\mid f\mleft(b\mright)\leq y\leq f\mleft(x_{+}\mright),\;x_{+}\leq x\leq g\mleft(y\mright)\right\} .\nonumber 
\end{align}
Then
\begin{equation}
\delta\mleft(A_{+}\mright)=\delta\mleft(B_{+}\mright)+\delta\mleft(\mleft(x_{+},b\mright]\times\mleft[2^{-1},f\mleft(b\mright)\mright)\mright)-\delta\mleft(\left\{ x_{+}\right\} \times\mleft[f\mleft(b\mright),f\mleft(x_{+}\mright)\mright]\mright)
\end{equation}
and by the previous proposition (as $y\mapsto\mleft(g\mleft(y\mright),y\mright)$
also parametrizes part of the graph of $f$)
\begin{equation}
\left|\delta\mleft(B_{+}\mright)-\alpha_{+}-\beta_{+}\right|\leq C\frac{R_{2}}{R_{1}}R_{2}^{\frac{2}{3}}\log\mleft(1+2\sqrt{2R_{2}}\mright)^{\frac{2}{3}}
\end{equation}
where
\begin{align}
\alpha_{+} & :=\alpha_{x_{+}}\mleft(f\mleft(b\mright),f\mleft(x_{+}\mright)\mright)=\mleft(f\mleft(x_{+}\mright)-f\mleft(b\mright)\mright)\mleft(x_{+}-\frac{1}{2}\mright)-\mleft(\left\lfloor f\mleft(x_{+}\mright)\right\rfloor -\left\lceil f\mleft(b\mright)\right\rceil +1\mright)\mleft(\left\lceil x_{+}\right\rceil -1\mright)\\
\beta_{+} & :=\beta_{g}\mleft(f\mleft(b\mright),f\mleft(x_{+}\mright)\mright)=\mleft(\frac{1}{2}+f\mleft(b\mright)-\left\lceil f\mleft(b\mright)\right\rceil \mright)b+\mleft(\frac{1}{2}+\left\lfloor f\mleft(x_{+}\mright)\right\rfloor -f\mleft(x_{+}\mright)\mright)x_{+}.\nonumber 
\end{align}
Let
\begin{equation}
\gamma_{+}=\mleft(\frac{1}{2}+\left\lfloor x_{+}\right\rfloor -x_{+}\mright)f\mleft(x_{+}\mright),\quad\gamma_{b}=\mleft(\frac{1}{2}+\left\lfloor b\right\rfloor -b\mright)f\mleft(b\mright).
\end{equation}
We claim that
\begin{equation}
\phi_{+}:=\alpha_{+}+\beta_{+}+\gamma_{+}+\delta\mleft(\mleft(x_{+},b\mright]\times\mleft[2^{-1},f\mleft(b\mright)\mright)\mright)-\delta\mleft(\left\{ x_{+}\right\} \times\mleft[f\mleft(b\mright),f\mleft(x_{+}\mright)\mright]\mright)
\end{equation}
reduces up to an $O\mleft(1\mright)$ term to $\gamma_{b}$. Indeed,
we begin by noting that
\begin{align}
 & \quad\;\,\delta\mleft(\mleft(x_{+},b\mright]\times\mleft[2^{-1},f\mleft(b\mright)\mright)\mright)-\delta\mleft(\left\{ x_{+}\right\} \times\mleft[f\mleft(b\mright),f\mleft(x_{+}\mright)\mright]\mright)\\
 & =\delta\mleft(\mleft[x_{+},b\mright]\times\mleft[2^{-1},f\mleft(b\mright)\mright]\mright)-\delta\mleft(\mleft[x_{+},b\mright]\times\left\{ f\mleft(b\mright)\right\} \mright)-\delta\mleft(\left\{ x_{+}\right\} \times\mleft[2^{-1},f\mleft(x_{+}\mright)\mright]\mright)\nonumber 
\end{align}
and
\begin{align}
\delta\mleft(\mleft[x_{+},b\mright]\times\mleft[2^{-1},f\mleft(b\mright)\mright]\mright) & =\mleft(\left\lfloor b\right\rfloor -\left\lceil x_{+}\right\rceil +1\mright)\left\lfloor f\mleft(b\mright)\right\rfloor -\mleft(b-x_{+}\mright)\mleft(f\mleft(b\mright)-2^{-1}\mright)\nonumber \\
\delta\mleft(\mleft[x_{+},b\mright]\times\left\{ f\mleft(b\mright)\right\} \mright) & =\mleft(\left\lfloor b\right\rfloor -\left\lceil x_{+}\right\rceil +1\mright)1_{\mathbb{Z}}\mleft(f\mleft(b\mright)\mright)\\
\delta\mleft(\left\{ x_{+}\right\} \times\mleft[2^{-1},f\mleft(x_{+}\mright)\mright]\mright) & =\left\lfloor f\mleft(x_{+}\mright)\right\rfloor 1_{\mathbb{Z}}\mleft(x_{+}\mright),\nonumber 
\end{align}
so
\begin{align}
\phi_{+}-\gamma_{b} & =\mleft(x_{+}-\left\lceil x_{+}\right\rceil +1-1_{\mathbb{Z}}\mleft(x_{+}\mright)\mright)\left\lfloor f\mleft(x_{+}\mright)\right\rfloor +\mleft(\left\lfloor x_{+}\right\rfloor -x_{+}\mright)f\mleft(x_{+}\mright)\nonumber \\
 & +\mleft(1+f\mleft(b\mright)-\left\lceil f\mleft(b\mright)\right\rceil \mright)b+\mleft(\left\lfloor f\mleft(b\mright)\right\rfloor -f\mleft(b\mright)-1_{\mathbb{Z}}\mleft(f\mleft(b\mright)\mright)\mright)\left\lfloor b\right\rfloor \\
 & +\mleft(\left\lceil f\mleft(b\mright)\right\rceil -\left\lfloor f\mleft(b\mright)\right\rfloor -1+1_{\mathbb{Z}}\mleft(f\mleft(b\mright)\mright)\mright)\mleft(\left\lceil x_{+}\right\rceil -1\mright)\nonumber \\
 & =\mleft(x_{+}-\left\lfloor x_{+}\right\rfloor \mright)\mleft(\left\lfloor f\mleft(x_{+}\mright)\right\rfloor -f\mleft(x_{+}\mright)\mright)+\mleft(\left\lfloor f\mleft(b\mright)\right\rfloor -f\mleft(b\mright)-1_{\mathbb{Z}}\mleft(f\mleft(b\mright)\mright)\mright)\mleft(\left\lfloor b\right\rfloor -b\mright)\nonumber 
\end{align}
where we used that $\left\lceil x\right\rceil -\left\lfloor x\right\rfloor -1+1_{\mathbb{Z}}\mleft(x\mright)=0$
for all $x\in\mathbb{R}$. This implies that $\left|\phi_{+}-\gamma_{b}\right|\leq2$,
so
\begin{align}
\left|\delta\mleft(A_{+}\mright)+\gamma_{+}-\gamma_{b}\right| & =\left|\delta\mleft(B_{+}\mright)+\phi_{+}-\gamma_{b}\right|\leq C\frac{R_{2}}{R_{1}}R_{2}^{\frac{2}{3}}\log\mleft(1+2\sqrt{2R_{2}}\mright)^{\frac{2}{3}}+2\\
 & \leq C'\frac{R_{2}}{R_{1}}R_{2}^{\frac{2}{3}}\log\mleft(1+2\sqrt{2R_{2}}\mright)^{\frac{2}{3}}\nonumber 
\end{align}
(since $R_{2}\geq1$) hence if $A_{-}=\emptyset$
\begin{align}
\left|\delta\mleft(\Gamma_{f}^{c\leq}\mright)-\beta_{f}\mleft(a,b\mright)\right| & =\left|\delta\mleft(A_{0}\mright)-\beta_{f}\mleft(a,x_{+}\mright)\right|+\left|\delta\mleft(A_{+}\mright)+\gamma_{+}-\gamma_{b}\right|\\
 & \leq C''\frac{R_{2}}{R_{1}}R_{2}^{\frac{2}{3}}\log\mleft(1+2\sqrt{2R_{2}}\mright)^{\frac{2}{3}}.\nonumber 
\end{align}
If $A_{-}\neq\emptyset$, a similar argument as what we used for $A_{+}$
still establishes the claim.

$\hfill\square$

As a corollary we obtain a result on the number of point directly
on the graph of a function:
\begin{cor}
Let $f\in C^{2}\mleft(\mleft[a,b\mright]\mright)$ be such that $\Gamma_{f}$
has minimal and maximal radii of curvature $0<R_{1}\leq R_{2}$. Then
if $R_{2}\geq1$ it holds that
\[
\left|\Gamma_{f}\cap\mathbb{Z}^{2}\right|\leq C\frac{R_{2}}{R_{1}}R_{2}^{\frac{2}{3}}\log\mleft(1+2\sqrt{2R_{2}}\mright)^{\frac{2}{3}}
\]
for a constant $C>0$ independent of all quantities.
\end{cor}

\textbf{Proof:} We can assume without loss of generality that $f>2^{-1}$.
Furthermore, by discreteness there must exist some $\epsilon>0$ such
that $\left|\Gamma_{f}\cap\mathbb{Z}^{2}\right|=\left|\Gamma_{f}^{\delta}\cap\mathbb{Z}^{2}\right|$
for all $0<\delta<\epsilon$, where
\begin{equation}
\Gamma_{f}^{\delta}=\left\{ \mleft(x,y\mright)\in\mathbb{R}^{2}\mid a\leq x\leq b,\,f\mleft(x\mright)-\delta<y\leq f\mleft(x\mright)+\delta\right\} .
\end{equation}
As $\Gamma_{f}^{\delta}=\Gamma_{f+\delta}^{2^{-1}\leq}\backslash\Gamma_{f-\delta}^{2^{-1}\leq}$
we have by the previous proposition that
\begin{align}
 & \;\,\left|\Gamma_{f}\cap\mathbb{Z}^{2}\right|=\left|\Gamma_{f+\delta}^{2^{-1}\leq}\cap\mathbb{Z}^{2}\right|-\left|\Gamma_{f-\delta}^{2^{-1}\leq}\cap\mathbb{Z}^{2}\right|\nonumber \\
 & =\mathrm{Area}\mleft(\Gamma_{f+\delta}^{2^{-1}\leq}\mright)-\mathrm{Area}\mleft(\Gamma_{f-\delta}^{2^{-1}\leq}\mright)+\delta\mleft(\Gamma_{f+\delta}^{2^{-1}\leq}\mright)-\delta\mleft(\Gamma_{f-\delta}^{2^{-1}\leq}\mright)\\
 & \leq\mleft(\mathrm{Area}\mleft(\Gamma_{f+\delta}^{2^{-1}\leq}\mright)-\mathrm{Area}\mleft(\Gamma_{f-\delta}^{2^{-1}\leq}\mright)\mright)+\mleft(\beta_{f+\delta}\mleft(a,b\mright)-\beta_{f-\delta}\mleft(a,b\mright)\mright)+C'\frac{R_{2}}{R_{1}}R_{2}^{\frac{2}{3}}\log\mleft(1+2\sqrt{2R_{2}}\mright)^{\frac{2}{3}}.\nonumber 
\end{align}
The first two terms on the right-hand side vanish as $\delta\rightarrow0$,
yielding the claim.

$\hfill\square$

We can now conclude Theorem \ref{them:ConvexLatticePoints}:
\begin{thm*}[\ref{them:ConvexLatticePoints}]
Let $K\subset\mathbb{R}^{2}$ be a compact, strictly convex set with
$C^{2}$ boundary. Let $\partial K$ have minimal and maximal radii
of curvature $0<R_{1}\leq R_{2}$. If $R_{2}\geq1$ then
\[
\left|\delta\mleft(K\mright)\right|\leq C\frac{R_{2}}{R_{1}}R_{2}^{\frac{2}{3}}\log\mleft(1+2\sqrt{2R_{2}}\mright)^{\frac{2}{3}}
\]
for a constant $C>0$ independent of $K$, $R_{1}$ and $R_{2}$.
\end{thm*}
\textbf{Proof:} Let $f_{\pm}$ be as in Fact \ref{fact:CompactStrictlyConvexRegions},
and assume without loss of generality that $f_{-}>2^{-1}$. Define
for $\epsilon>0$ the sets
\begin{align}
K_{\pm}^{\epsilon} & =\left\{ \mleft(x,y\mright)\in\mathbb{R}^{2}\mid a+\epsilon\leq x\leq b-\epsilon,\,2^{-1}\leq y\leq f_{\pm}\mleft(x\mright)\right\} \\
\Gamma_{-}^{\epsilon} & =\left\{ \mleft(x,y\mright)\in\mathbb{R}^{2}\mid a+\epsilon\leq x\leq b-\epsilon,\,y=f_{-}\mleft(x\mright)\right\} .\nonumber 
\end{align}
Note that
\begin{align}
\liminf_{\epsilon\rightarrow0^{+}}\left|\delta\mleft(\mleft(K_{+}^{\epsilon}\backslash K_{-}^{\epsilon}\mright)\cup\Gamma_{-}^{\epsilon}\mright)\right| & \geq\left|\delta\mleft(\mleft(K_{+}\backslash K_{-}\mright)\cup\Gamma_{-}\mright)\right|-2=\left|\delta\mleft(K\mright)\right|-2\\
\liminf_{\epsilon\rightarrow0^{+}}\left|\Gamma_{-}^{\epsilon}\cap\mathbb{Z}^{2}\right| & \geq\left|\Gamma_{-}\cap\mathbb{Z}^{2}\right|-2\nonumber 
\end{align}
since $\lim_{\epsilon\rightarrow0^{+}}\mathrm{Area}\mleft(K_{\pm}^{\epsilon}\mright)=\mathrm{Area}\mleft(K_{\pm}\mright)$
and e.g. $\lim_{\epsilon\rightarrow0^{+}}\left|K_{+}^{\epsilon}\cap\mathbb{Z}^{2}\right|=\left|K_{+}\cap\mathbb{Z}^{2}\right|-1_{\mathbb{Z}}\mleft(b\mright)1_{\mathbb{Z}}\mleft(f\mleft(b\mright)\mright).$

Now, for any $\epsilon>0$ we can estimate
\begin{align}
 & \quad\,\left|\delta\mleft(\mleft(K_{+}^{\epsilon}\backslash K_{-}^{\epsilon}\mright)\cup\Gamma_{-}^{\epsilon}\mright)\right|\leq\left|\delta\mleft(K_{+}^{\epsilon}\mright)-\delta\mleft(K_{-}^{\epsilon}\mright)\right|+\left|\delta\mleft(\Gamma_{-}^{\epsilon}\mright)\right|\\
 & \leq\left|\beta_{f_{+}}\mleft(a+\epsilon,b-\epsilon\mright)-\beta_{f_{-}}\mleft(a+\epsilon,b-\epsilon\mright)\right|+2+C'\frac{R_{2}}{R_{1}}R_{2}^{\frac{2}{3}}\log\mleft(1+2\sqrt{2R_{2}}\mright)^{\frac{2}{3}}\nonumber 
\end{align}
and as $\epsilon\rightarrow0$, the first term on the right-hand side
vanishes, so (as $R_{2}\geq1$)
\begin{equation}
\left|\delta\mleft(K\mright)\right|\leq4+C'\frac{R_{2}}{R_{1}}R_{2}^{\frac{2}{3}}\log\mleft(1+2\sqrt{2R_{2}}\mright)^{\frac{2}{3}}\leq C''\frac{R_{2}}{R_{1}}R_{2}^{\frac{2}{3}}\log\mleft(1+2\sqrt{2R_{2}}\mright)^{\frac{2}{3}}.
\end{equation}
$\hfill\square$

%\appendix

\subsection{\label{sec:ProofsoftheAuxilliaryResults}Proofs of the Auxilliary
Results}
\begin{thm*}[\ref{them:DirichletApp}]
Let $a,\tau\in\mathbb{R}$ be given with $\tau\geq1$. Then there
exists coprime integers $p$ and $q$ such that
\[
\left|a-\frac{p}{q}\right|<\frac{1}{q\tau}\quad\text{with}\quad1\leq q\leq\tau.
\]
\end{thm*}
\textbf{Proof:} For any given $q\in\mathbb{N}$ it is obvious that
\begin{equation}
\min_{p\in\mathbb{Z}}\left|aq-p\right|=\min\mleft(\left\{ aq\right\} ,1-\left\{ aq\right\} \mright)
\end{equation}
and we claim that $1\leq q\leq N=:\left\lfloor \tau\right\rfloor $
can be chosen such that either $\left\{ aq\right\} \leq\mleft(N+1\mright)^{-1}$
or $N\mleft(N+1\mright)^{-1}\leq\left\{ aq\right\} $.

Indeed, consider the $N$ numbers $\mleft(\left\{ aq\right\} \mright)_{q=1}^{N}$.
These fall into the $N+1$ intervals
\begin{equation}
\mleft[0,\mleft(N+1\mright)^{-1}\mright),\mleft[\mleft(N+1\mright)^{-1},2\mleft(N+1\mright)^{-1}\mright),\ldots,\mleft[N\mleft(N+1\mright)^{-1},1\mright)
\end{equation}
and our claim is that some $\left\{ aq\right\} $ must fall into either
the first or last interval. Suppose otherwise. Then by the pigeonhole
principle, there must be some $1\leq k\leq N-1$ and $1\leq q_{1}<q_{2}\leq N$
such that $\left\{ aq_{1}\right\} ,\left\{ aq_{2}\right\} \in\mleft[k\mleft(N+1\mright)^{-1},\mleft(k+1\mright)\mleft(N+1\mright)^{-1}\mright)$.
But then
\begin{equation}
\frac{1}{N+1}\geq\left|\left\{ aq_{2}\right\} -\left\{ aq_{1}\right\} \right|=\left|a\mleft(q_{2}-q_{1}\mright)-\mleft(\left\lfloor aq_{2}\right\rfloor -\left\lfloor aq_{1}\right\rfloor \mright)\right|
\end{equation}
which shows that $q=q_{2}-q_{1}$ and $p=\left\lfloor aq_{2}\right\rfloor -\left\lfloor aq_{1}\right\rfloor $
obey $\left|aq-p\right|\leq\mleft(N+1\mright)^{-1}$, which was by assumption
not the case. Thus there exists $p,q$ such that $\left|aq-p\right|\leq\mleft(N+1\mright)^{-1}$,
hence
\begin{equation}
\left|a-\frac{p}{q}\right|\leq\frac{1}{q\mleft(N+1\mright)}=\frac{1}{q\mleft(\left\lfloor \tau\right\rfloor +1\mright)}\leq\frac{1}{q\tau}.
\end{equation}
That $p,q$ can furthermore be taken to be coprime is obvious, as
the fraction $\frac{p}{q}$ can be reduced to ensure this, and diminishing
$q$ only worsens the estimate.

$\hfill\square$
\begin{lem*}[\ref{lemma:GroupingLemma}]
Let $n_{1},\ldots,n_{k}\in\mathbb{N}$ be given with $1\leq n_{i}\leq N$,
$1\leq i\leq k$, for some $N\in\mathbb{N}$. Then there exists a
$k'\geq k-N$ such that $\mleft(n_{1},\ldots,n_{k'}\mright)$ admits
a consecutive grouping $\left\{ Q_{j}\right\} _{j=1}^{T}$ with the
property that
\[
\max\mleft(Q_{j}\mright)=\left|Q_{j}\right|
\]
for an integer $T\leq\sum_{i=1}^{k}n_{i}^{-1}$.
\end{lem*}
\textbf{Proof:} We first prove the existence of such a grouping. We
induct on the length of the sequence, $k$. If $k\leq N$ the statement
is trivial - in particular the base case $k=1$ certainly holds.

Suppose then the statement holds for sequences of length $\leq k$,
and let $\mleft(n_{1},\ldots,n_{k+1}\mright)$ be arbitrary. Define
subsequences $M_{1},\ldots,M_{N}\subset\mleft(n_{1},\ldots,n_{k+1}\mright)$
recursively by setting $M_{1}=\mleft(n_{1},\ldots,n_{n_{1}}\mright)$
and
\begin{equation}
M_{j}=\mleft(n_{1},\ldots,n_{\max\mleft(M_{j-1}\mright)}\mright),\quad2\leq j\leq N.
\end{equation}
Note that in fact $M_{1}\subset\cdots\subset M_{N}$. Indeed, it is
clear that $M_{1}\subset M_{2}$, and for $j\geq3$, the statement
$M_{j-1}\subset M_{j}$ is the claim that $\max\mleft(M_{j-2}\mright)\leq\max\mleft(M_{j-1}\mright)$,
which follows inductively from $M_{j-2}\subset M_{j-1}$.

We claim that $\max M_{j}=\left|M_{j}\right|$ for some $1\leq j\leq N$.
Indeed, by construction
\begin{equation}
\left|M_{1}\right|\leq\max\mleft(M_{1}\mright)=\left|M_{2}\right|\leq\max\mleft(M_{2}\mright)=\quad\cdots\quad=\left|M_{N}\right|\leq\max\mleft(M_{N}\mright)
\end{equation}
and the claim is that at least one of these inequalities is in fact
an equality. Since there are $N$ inequalities, if this was not the
case we would have $\max\mleft(M_{N}\mright)>N$, which contradicts
the uniform bound on the $n_{i}$.

We can then set $Q_{1}=M_{j}$ and apply the lemma for length $l=k+1-\left|Q_{1}\right|$
to the sequence $\mleft(n_{\left|Q_{1}\right|+1},\ldots,n_{k+1}\mright)$
to obtain admissible groupings $\left\{ Q_{j}\right\} _{j=2}^{T}$
of $\mleft(n_{\left|Q_{1}\right|+1},\ldots,n_{k'}\mright)$ with $k'-\left|Q_{1}\right|\geq l-N$.
Then $\left\{ Q_{j}\right\} _{j=1}^{T}$ is an admissible grouping
of $\mleft(n_{1},\ldots,n_{k'}\mright)$ with
\begin{equation}
k'\geq k+1-N
\end{equation}
which proves the claim for sequences of length $k+1$.

For the bound on $T$, note that for such a grouping $\left\{ Q_{j}\right\} _{j=1}^{T}$
\begin{equation}
T=\sum_{j=1}^{T}\sum_{n\in Q_{j}}\frac{1}{\left|Q_{j}\right|}\leq\sum_{j=1}^{T}\sum_{n\in Q_{j}}\frac{1}{n}=\sum_{i=1}^{k'}\frac{1}{n_{i}}\leq\sum_{i=1}^{k}\frac{1}{n_{i}}.
\end{equation}
$\hfill\square$
\begin{prop*}[\ref{prop:PerturbedSumEstimate}]
Let $n\in\mathbb{N}$ and $k\in\mathbb{Z}$ be coprime, and let for
some $M\in\mathbb{Z}$, 
$$P:\left\{ M+1,\ldots,M+n\right\} \rightarrow\mathbb{R}$$
be a function such that
\[
\max_{M+1\leq l,m\leq M+n}\left|P\mleft(l\mright)-P\mleft(m\mright)\right|\leq C
\]
for some $C>0$. Then
\[
\left|\sum_{m=M+1}^{M+n}\left\{ \frac{km+P\mleft(m\mright)}{n}\right\} -\frac{n}{2}\right|\leq C+\frac{1}{2}.
\]
\end{prop*}
\textbf{Proof:} If $\frac{n}{2}\leq C+\frac{1}{2}$ the claim is immediate,
so we suppose otherwise. Let $$P_{\min}=\min_{M+1\leq m\leq M+n}P\mleft(m\mright)$$
and define for convenience $Q$$\mleft(m\mright)=P\mleft(m\mright)-P_{\min}$.
Then the assumption on $P$ implies that
\begin{equation}
0\leq Q\mleft(m\mright)\leq C,\quad M+1\leq m\leq M+n.
\end{equation}
Now define $\mu\mleft(m\mright)=\mleft(km+\left\lfloor P_{\min}\right\rfloor \mod n\mright)$.
Since $k$ and $n$ are coprime, it holds that 
$$\left\{ \mu\mleft(M+1\mright),\ldots,\mu\mleft(M+n\mright)\right\} $$
is a permutation of $\left\{ 0,\ldots,n-1\right\} $. Defining $Q':\left\{ 0,\ldots,n-1\right\} \rightarrow\mathbb{R}$
by the relation $Q\mleft(m\mright)=Q'\mleft(\mu\mleft(M+m\mright)\mright)$
we can then write
\begin{align}
\sum_{m=M+1}^{M+n}\left\{ \frac{km+P\mleft(m\mright)}{n}\right\}  & =\sum_{m=M+1}^{M+n}\left\{ \frac{km+\left\lfloor P_{\min}\right\rfloor +\left\{ P_{\min}\right\} +Q\mleft(m\mright)}{n}\right\} \nonumber \\
 & =\sum_{m=M+1}^{M+n}\left\{ \frac{\mu\mleft(M+m\mright)+\left\{ P_{\min}\right\} +Q'\mleft(\mu\mleft(M+m\mright)\mright)}{n}\right\} \\
 & =\sum_{m=0}^{n-1}\left\{ \frac{m+\left\{ P_{\min}\right\} +Q'\mleft(m\mright)}{n}\right\} .\nonumber 
\end{align}
Now, note that for any $0\leq m\leq n-1$
\begin{equation}
0\leq\frac{m+\left\{ P_{\min}\right\} +Q'\mleft(m\mright)}{n}\leq1+\frac{C}{n}<2
\end{equation}
since $C+\frac{1}{2}<\frac{n}{2}$, hence $C<n$, by assumption. Defining
$$S=\left\{ m\mid n\leq m+\left\{ P_{\min}\right\} +Q'\mleft(m\mright)\right\} $$
and $S^{c}=\left\{ 0,\ldots,n-1\right\} \backslash S$ we then have
\begin{align}
 & \sum_{m=0}^{n-1}\left\{ \frac{m+\left\{ P_{\min}\right\} +Q'\mleft(m\mright)}{n}\right\} =\sum_{m\in S^{c}}\frac{m+\left\{ P_{\min}\right\} +Q'\mleft(m\mright)}{n}+\sum_{m\in S}\mleft(\frac{m+\left\{ P_{\min}\right\} +Q'\mleft(m\mright)}{n}-1\mright)\nonumber \\
 & =\sum_{m=0}^{n-1}\frac{m+\left\{ P_{\min}\right\} +Q'\mleft(m\mright)}{n}-\left|S\right|=\frac{n}{2}-\frac{1}{2}+\left\{ P_{\min}\right\} +\sum_{m=0}^{n-1}\frac{Q'\mleft(m\mright)}{n}-\left|S\right|.
\end{align}
Now, note that $\left|S\right|\leq C+\left\{ P_{\min}\right\} $.
Indeed, $m\in S$ implies that
\begin{equation}
n-1\geq m\geq n-Q'\mleft(m\mright)-\left\{ P_{\min}\right\} \geq n-C-\left\{ P_{\min}\right\} 
\end{equation}
i.e. $S\subset\mleft[n-C-\left\{ P_{\min}\right\} ,n-1\mright]\cap\mathbb{Z}$,
so
\begin{equation}
\left|S\right|\leq n-1-\left\lfloor n-C-\left\{ P_{\min}\right\} \right\rfloor +1\leq C+\left\{ P_{\min}\right\} .
\end{equation}
By the relations above we can thus estimate
\begin{align}
\sum_{m=M+1}^{M+n}\left\{ \frac{km+P\mleft(m\mright)}{n}\right\} -\frac{n}{2} & \leq\left\{ P_{\min}\right\} -\frac{1}{2}+\sum_{m=0}^{n-1}\frac{Q'\mleft(m\mright)}{n}\leq\frac{1}{2}+C\\
\sum_{m=M+1}^{M+n}\left\{ \frac{km+P\mleft(m\mright)}{n}\right\} -\frac{n}{2} & \geq\left\{ P_{\min}\right\} -\frac{1}{2}-\left|S\right|\geq-\frac{1}{2}-C\nonumber 
\end{align}
which is the claim.
$\hfill\square$

\end{document}